\documentclass[12pt,a4paper,twoside,titlepage]{book}
\usepackage{latexsym,amsmath,amssymb,amsfonts,amsthm}
\usepackage[mathscr]{eucal}
\usepackage{makeidx}
\newtheorem{proposition}{Proposition}[chapter]
\newtheorem{definition}{Definition}[chapter]
\newtheorem{corollary}{Corollary}[chapter]
\newtheorem{theorem}{Theorem}[chapter]
\def\nabla{\bigtriangledown}
\makeindex
\begin{document}
 \frontmatter
\begin{titlepage}
\vbox{\begin{center} {\Large Sergiu  Vacaru and Panayiotis
Stavrinos}
\end{center}}
\vspace*{2mm} \vskip100pt \vbox{\begin{center} \Huge \bf SPINORS
\\ and
\\ SPACE--TIME ANISOTROPY
\end{center}}
\vskip200pt \vbox{\begin{center} \Large \sf University of Athens,\
2002
\end{center}}
 \vskip50pt

---------------------------------------------------

{\copyright \quad Sergiu  Vacaru and Panyiotis Stavrinos}
\end{titlepage}

\newpage
...
\newpage

\section*{ABOUT THE BOOK}

This is the first monograph on the geometry of anisotropic spinor
spaces and its applications in modern physics. The main subjects
are the theory of gravity and matter fields in spaces provided
with off--diagonal metrics and associated anholonomic frames and
nonlinear connection structures, the algebra and geometry of
distinguished anisotropic Clifford and spinor spaces, their
extension to spaces of higher order anisotropy and the geometry
of gravity and gauge theories with anisotropic spinor variables.
The book summarizes the authors' results and can be also
considered as a pedagogical survey on the mentioned subjects.

\newpage - \newpage

\section*{ABOUT THE AUTHORS}

{\quad }\textbf{Sergiu Ion Vacaru} was born in 1958 in the
Republic of Moldova. He was educated at the Universities of the
former URSS (in Tomsk, Moscow, Dubna and Kiev) and reveived his
PhD in theoretical physics in 1994 at ''Al. I. Cuza'' University,
Ia\c{s}i, Romania. He was employed as principal senior
researcher, associate and full professor and obtained a number of
NATO/UNESCO grants and fellowships at various academic
institutions in R. Moldova, Romania, Germany, United Kingdom,
Italy, Portugal and USA. He has published in English two
scientific monographs, a university text--book and more than
hundred scientific works (in English, Russian and Romanian) on
(super) gravity and string theories, extra--dimension and brane
gravity, black hole physics and cosmolgy, exact solutions of
Einstein equations, spinors and twistors, anistoropic stochastic
and kinetic processes and thermodynamics in curved spaces,
generalized Finsler (super) geometry and gauge gravity, quantum
field and geometric methods in condensed matter physics. \vskip6pt

\textbf{Panayiotis Stavrinos} is Assistant Professor in the
University of Athens, where he obtained his Ph. D in 1990. He
teaches some courses on Differential Geometry in the Department
of Mathematics and Applications of Differential Geometry to
Physics in the Department of Physics. He also supervises
undergraduate these ans graduate researches at the University of
Athens. He is a Founding Member and Vice President of the Balkan
Society of Geometers, Member of the Editorial Board of the
Journal of the Balkan Society, Honorary Member to The Research
Board of Advisors of the American Biographical Institute
(U.S.A.), 1996, and Member of the Tensor Society, (Japan), 1981.
Dr. Stavrinos has published over 40 research papers in
international Journals on The topics of local differential
geometry, Finsler and Lagrange geometry, applications of Finsler
and Lagrange geometry to gravitation, gauge and spinor theory as
well as Einstein equations, deviation of geodesics, tidal forces,
weak gravitational fields, gravitational waves. He is co-author
of the monograph ''Introduction to the Physical Principles of
Differential Geometry'', in Russian, published in St. Petersburg
in 1996 (second edition in English, University of Athens Press,
2000). He has published two monographs in Greek for undergraduate
and graduate students in the Department of Mathematics and
Physics: ''Differential Geometry and its Applications'', Vol. I,
II (University of
Athens Press, 2000).

\newpage
---
 \newpage
 \tableofcontents
 \newpage
 ---
 \newpage

\section[Preface]{Preface}

\subsection{Historical remarks on spinor theory}

Spinors and Clifford algebras play a major role in contemporary
physics and mathematics. In their mathematical form spinors were
discovered by \`{E}lie Cartan in 1913 in his researches on
representation group theory \cite{car38} where he showed that
spinors furnish a linear representation of the groups of
rotations of a space of arbitrary dimensions. In 1927 Pauli
\cite{pauli} and Dirac \cite{dirac} (respectively, for the
three--dimensional and four--dimensional space--time) introduced
spinors for the representation of the wave functions.

The spinors studied by mathematicians and physicists are
connected with the general theory of Clifford spaces introduced
in 1876 \cite{clifford}.

In general relativity theory spinors and the Dirac equations on
(pseudo) Riemannian spaces were defined in 1929 by H. Weyl
\cite{weyl}, V. Fock \cite {foc} and E. Schr\"{o}dinger
\cite{schr}. The book by R. Penrose\cite{pen}, and the R. Penrose
and W. Rindler monograph \cite{penr1,penr2} summarize the spinor
and twistor methods in space--time geometry (see additional
references \cite{hladik,benn,morand,lue,tur,car} on Clifford
structures and spinor theory).

Spinor variables were introduced in Finsler geometries by Y.
Takano in 1983 \cite{t1} where he dismissed anisotropic
dependencies not only on vectors on the tangent bundle but on
some spinor variables in a spinor bundle on a space--time
manifold. That work was inspired by H. Yukawa's quantum theory of
non--local fields \cite{yuk}; it was suggested that
non--localization may be in Finsler-like space but on spinor
variables. There was also a similarity with supersymmetric models
(see, for instance, references \cite {wess,west}), where spinor
variables are also used. The approach of Y. Takano followed
standard concepts on Finsler geometries and was not concerned
with topics related to supersymmetries of interactions.

Generalized Finsler geometries, with spinor variables, were
developed by T. Ono and Y. Takano in a series of publications
during 1990--1993 \cite {ot1,ot2,ot3,ono}. The next steps were
investigations of anisotropic and deformed geometries with spinor
and vector variables and applications in gauge and gravity
theories elaborated by P. Stavrinosand his students, S.
Koutroubis, P. Manouselis, and Professor V. Balan beginning 1994
\cite {sk,sm2,sm3,sbmp,sbmp1}. In those works the authors assumed
that some spinor variables may be introduced in a Finsler-like
way, they did not relate the Finlser metric to a Clifford
structure and restricted the spinor--gauge Finsler constructions
only for antisymmetric spinor metrics on two--spinor fibers with
possible generalizations to four dimensional Dirac spinors.

Isotopic spinors, related with $SU(2)$ internal structural
groups, were considered in generalized Finsler gravity and gauge
theories also by G. Asanov and S. Ponomarenko \cite{asa88} in
1988. But in that book and in other papers on Finsler geometry
with spinor variables, the authors did not investigate the
possibility of introducing a rigorous mathematical definition of
spinors on spaces with generic local anisotropy.

An alternative approach to spinor differential geometry and
generalized Finsler spaces was elaborated, beginning 1994, in a
series of papers and communications by S. Vacaru with
participation of S. Ostaf \cite {vdeb,vo1,vrom,vbm}. This
direction originates from Clifford algebras and Clifford bundles
\cite{kar,tur} and Penrose's spinor and twistor space--time
geometry \cite{pen,penr1,penr2}, which were re--considered for
the case of nearly autoparallel maps (generalized conformal
transforms) in Refs. \cite {v87,vkaz,vit}. In the works
\cite{viasm1,vjmp,vsp2}, a rigorous definition of spinors for
Finsler spaces, and their generalizations, was given. It was
proven that a Finsler, or Lagrange, metric (in a tangent, or, more
generally, in a vector bundle) induces naturally a distinguished
Clifford (spinor) structure which is locally adapted to the
nonlinear connection structure. Such spinor spaces could be
defined for arbitrary dimensions of base and fiber subspaces,
their spinor metrics are symmetric, antisymmetric or
nonsymmetric, depending on the corresponding base and fiber
dimensions. This work resulted in the formation of spinor
differential geometry of generalized Finsler spaces and developed
a number of geometric applications to the theory of gravitational
and matter field interactions with generic local anisotropy.

Furthermore, the geometry of anisotropic spinors and
(distinguished by nonlinear connections) Clifford structures was
elaborated for higher order anisotropic spaces
\cite{vsp1,vhsp,vbook} and, more recently, for Hamilton and
Lagrange spaces \cite{vsv}.

Here it is appropriate to emphasize that the theory of
anisotropic spinors may be related not only to generalized
Finsler, Lagrange, Cartan and Hamilton spaces or their higher
order generalizations, but also to anholonomic frames with
associated nonlinear connections which appear naturally even in
(pseudo) Riemannian geometry if off--diagonal metrics are
considered \cite{vb,vbh,vkinet,vtheor,vtor}. In order to
construct exact solutions of the Einstein equations in general
relativity and extra dimension gravity (for lower dimensions see
\cite{vtherm,vsgab,vsgon}), it is more convenient to diagonalize
space--time metrics by using some anholonomic transforms. As a
result one induces locally anisotropic structures on space--time
which are related to anholonomic (anisotropic) spinor structures.

The main purpose of the present book is to present a detailed
summary and new results on spinor differential geometry for
generalized Finsler spaces and (pseudo) Riemannian space--times
provided with anholonomic frame and associated nonlinear
connection structure, to discuss and compare the existing
approaches and to consider applications to modern gravity and
gauge theories.

\subsection{Nonlinear connection geometry and physics}

It was \`{E}lie Cartan in the last thirty years of the previous
century in addition to his first monograph on spinors wrote
several fundamental books on the geometry of Riemannian, fibred
and Finsler spaces, where he developed the moving frame method
and the formalism of Pfaffian forms for systems of first order
partial differential equations \cite{car,car35,car45}. The first
examples of Finsler metrics and original definitions were given
by B. Riemann \cite{riemann} in 1854 and in P. Finsler's thesis
\cite{fin}, written under the supervision of Caratheodory in
1918. In these works one could find the origins of the notion of
locally trivial fiber bundle, which naturally generalize that of
the manifold; the theory of these bundles was developed, twenty
years later, especially by Gh. Ehresmann and of the nonlinear
connection (appearing as a set of coefficients in the book \cite
{car35} and in a more explicit form in several papers by A.
Kawaguchi \cite {kawaguchi}).

The global formulation of nonlinear connection is due to W.
Barthel \cite {barth};\ detailed investigations of nonlinear
connection geometry in vector bundles and higher order tangent
bundles, with applications to physics and mechanics, are
contained in the monographs and works \cite
{ma87,ma94,m1,m2,mirata,mhss} summarizing the investigations of
the R. Miron school on Finsler and Lagrange geometry and their
generalizations. The geometry of nonlinear connections was
developed in S. Vacaru's works and monograph for vector and
higher order \cite{vlasg,vbook} superbundles and anisotropic
Clifford/spinor fibrations \cite{vdeb,viasm1,vjmp,vsp1,vsp2},
with generalizations and applications in (super) gravity \cite
{vcv,vbh,vkinet,vd,vd,vg,vp,vsol} and string theories
\cite{vstr1,vstr2} and noncommutative gravity \cite{vsf}. There
are a number of results on nonlinear connections and Finsler
geometry, for instance in \cite {run,bao,mat}, with
generalizations and applications in mechanics, physics and
biology which can be found in references \cite
{ant,ai,al,am,az,asa,asa89,asa88,bej,bej1,bog}.

Finsler spaces and their generalizations have been also developed
with the view to applications in classical and higher order
mechanics, optics, generalized Kaluza--Klein theories and gauge
theories. But for a long time Finsler geometry was considered to
be very complicated and not compatible with the standard paradigm
of modern physics. The first objection was that on spaces with
local anisotropy there do not exist local groups of
authomorphisms which made impossible the definition of local
conservation laws and the development of a theory of anisotropic
random and kinetic processes by the introduction of spinor
fields. The second objection was based on an erroneous view that
in Finsler like gravity theories the local Lorentz symmetry is
broken, a fact which is not compatible with the modern paradigms
of particle physics and gravity \cite{will}. Nevertheless, it was
proven that there are no more conceptual problems with definition
of local conservation laws than in the usual theory of gravity on
pseudo--Riemannian spaces if Finsler like theories are formulated
with respect to local frames adapted to the nonlinear connection
structure\ a variant of the definition of conservation laws for
locally anisotropic gravitational and matter field interactions
was proposed by using chains of nearly autoparallel maps
generalizing conformal transforms \cite{vcl96,voa,vodg}. As to the
violations of the local Lorentz symmetries, one should mention
that in fact some classes of such Finsler like metrics were
investigated with the aim of revising the special and general
theories of relativity (see, for instance, Refs.
\cite{asa89b,ari,bog,gb}), but it is also possible to define
Finsler-like and other types of anisotropic structures, even in
the framework of general relativity theory. Such structures are
described by exact solutions of the Einstein equations if
off--diagonal frames and anholonomic frames are taken into
consideration \cite {vb,vbh,vkinet,vtheor,vtor,vd}. We conclude
that there are different classes of generalized Finsler like
metrics: some of them posses broken Lorentz symmetries and others
do not have such properties and are compatible with the general
relativity. Here, it should be emphasized that the violation of
Lorentz geometry is not already a subject not allowed in modern
physics, for instance, the effects induced by Lorentz violations
are analyzed in brane physics \cite{csaki} and non--commutative
field theories \cite{moc,carroll}.

The third objection was based on the ''absence'' of a
mathematical theory of stochastic processes and diffusion on
spaces with generic local anisotropy. But this problem was also
solved in a series of papers. The first results on diffusion
processes on Finsler manifolds were announced in 1992 by P.
Antonelli and T. Zastavniak \cite{az1,az2}; their formalism was
not yet
adapted to the nonlinear connection structure. In a communication at the Ia%
\c{s}i Academic Days (1994, Romania) \cite{vs1} S. Vacaru
developed the theory of stochastic differential equations as in
the Riemannian spaces but on vector bundles provided with
nonlinear connection structures. As a result the theory of
anisotropic processes was in parallel developed on vector bundles
by S. Vacaru \cite{vs1,vs2,vs3} (see Chapter 5 in \cite{vbook} for
supersymmetric anisotropic stochastic processes) and P.
Antonelli, T. Zastavniak and D. Hrimiuc \cite{az1,az2,h1,h2,ah1}
( the last three authors provide a number of applications in
biology and biophysics) following a theory of stochastic
differential equations formulated on bundles provided with
anholnomic frames and nonlinear connections. It was also possible
to formulate a theory of anisotropic kinetic processes and
thermodynamics \cite {vtherm,vanphl,vkinet} with applications in
modern cosmology and astrophysics. So, the third difficulty for
anisotropic physics, connected with the definition of random and
kinetic models on spaces with generic local anisotropy was
successfully met.

A fourth objection for the acceptance by ''the physics
community'' of Finsler spaces were the arguments like: ''it is
not clear how to supersymmetrize such theories and how to embed
them in a modern string theory because at low energies from
string theories one applies only (pseudo) Riemannian geometries
and their supersymmetric generalizations''. The problem on the
definition of nonlinear connections in superbundles was solved in
a series of preprints in 1996 \cite{vlasg} with the results
included in the paper \cite{vstr2} and monograph \cite{vbook}
where a new Finsler supergeometry with generalizations and
applications in (super) gravity and string theories \cite{vcv}
was formulated. The works \cite {vstr1,vstr2} contained explicit
proofs that we can embed in (super) string theories Finsler-like
geometries, if we are dealing with anholonomomic (super) frame
structures, at low energies we obtain anholonomic frames on
(pseudo) Riemannian space--times or, alternatively, different
type of Finsler like geometries.

The monograph \cite{vbook} summarizes the basic results on
anisotropic (in general, supersymmetric) field interactions,
stochastic processes and strings. It was the first book where the
basic directions in modern physics were reconsidered on (super)
spaces provided with nonlinear connection structure. It was
proven that following the E. Cartan geometrical ideas and methods
to vector bundles, spinors, moving frames, nonlinear connections,
Finsler and (pseudo) Riemannian spaces the modern physical
theories can be formulated in a unified manner both on spaces
with generic local anisotropy and on locally isotropic spaces if
local frames adapted to nonlinear connection structures are taken
into consideration.

The present book covers a more restricted area, compared with the
monograph [172],connected in the bulk with the spinor geometry
and physics, and is intended to provide the reader with a
thorough background for the theory of anisotropic spinors in
generalized Finsler spaces and for the theory of anholonomic
spinor structures in (pseudo) Riemannian spaces. The reader is
assumed to be familiar with basic concepts from the theory of
bundle spaces, spinor geometry, classical field theory and
general relativity at a standard level for graduate students
mathematics and theoretical in physics. The primary purpose of
the book is to introduce the new geometrical ideas in the
language of standard fiber bundle geometry and establish a working
familiarity with the modern applications of spinor geometry,
anholonomic frame method and nonlinear connections formalism in
physics. These techniques are subsequently generalized and
applied to gravity and gauge theories. The secondary purpose is
to consider and compare different approaches which deal with
spinors in Finsler like geometries.

\subsection{Anholonomic frames and N--connections in Einstein gravity}

Let us consider a $\left( n+m\right) $--dimensional (pseudo)
Riemannian spacetime \newline
$V^{(n+m)},$ being a paracompact and connected Hausdorff $C^\infty $%
--ma\-ni\-fold, enabled with a nonsigular metric
\[
ds^2={g}_{\alpha \beta }\ du^\alpha \otimes du^\beta
\]
with the coefficients
\begin{equation}
{g}_{\alpha \beta }=\left[
\begin{array}{cc}
g_{ij}+N_i^aN_j^bh_{ab} & N_j^eh_{ae} \\
N_i^eh_{be} & h_{ab}
\end{array}
\right]  \nonumber
\end{equation}
parametrized with respect to a local coordinate basis $du^\alpha
=\left( dx^i,dy^a\right) ,$ having its dual $\partial /u^\alpha
=\left( \partial /x^i,\partial /y^a\right) ,$ where the indices
of geometrical objects and local coordinate $u^\alpha =\left(
x^k,y^a\right) $ run correspondingly the values:\ (for Greek
indices)$\alpha ,\beta ,\ldots =n+m;$ for (Latin indices)
$i,j,k,...=1,2,...,n$ and $a,b,c,...=1,2,...,m.$ Such
off--diagonal ansatz for metric were considered, for instance, in
Salam--Strathdee--Percacci--Randjbar-Daemi works on Kaluza--Klein
theory \cite{salam,percacci,over} as well in four and five
dimensional gravity \cite
{vb,vbh,vkinet,vp,vtheor,vtor,vsgab1,vsgon1,vsol,vranc}.

The metric ansatz can be rewritten equivalently in a block
$(n\times n) + (m \times m) $ form
\begin{equation}
g_{\alpha \beta }=\left(
\begin{array}{cc}
g_{ij}(x^k,y^a) & 0 \\
0 & h_{ab}(x^k,y^a)
\end{array}
\right)  \nonumber
\end{equation}
with respect to a subclass of $n+m$ anholonomic frame basis (for
four dimensions one used terms tetrads, or vierbiends)
\index{tetrad} \index{vierbiend} defined
\begin{equation}
\delta _\alpha =(\delta _i,\partial _a)=\frac \delta {\partial
u^\alpha }=\left( \delta _i=\frac \delta {\partial x^i}=\frac
\partial {\partial x^i}-N_i^b\left( x^j,y^c\right) \frac \partial
{\partial y^b},\partial _a=\frac \partial {\partial y^a}\right)
\nonumber
\end{equation}
and
\begin{equation}
\delta ^\beta =\left( d^i,\delta ^a\right) =\delta u^\beta =\left(
d^i=dx^i,\delta ^a=\delta y^a=dy^a+N_k^a\left( x^j,y^b\right)
dx^k\right) , \nonumber
\end{equation}
called locally anisotropic bases (in brief, anisotropic bases)
adapted to the coefficients $N_j^a.$ The $n\times n$ matrix
$g_{ij}$ defines the so--called horizontal metric (in brief,
h--metric) and the $m\times m$ matrix $h_{ab}$ defines the
vertical (v--metric) with respect to the associated nonlinear
connection (N--connection) structure given by its coefficients
$N_j^a\left( u^\alpha \right) ,$ see for instance \cite{ma94}
where the geometry of N--connections is studied in detail for
generalized Finsler and Lagrange spaces (the $y$--coordinates
parametrizing fibers in a bundle).

Here we emphasize that a matter of principle we can consider that
our ansatz and N--elongated bases are defined on a (pseudo)
Riemannian manifold, and not on a bundle space. In this case we
can treat that the $x$--coordinates are holonomic ones given with
respect to a sub--basis not subjected to any constraints, but the
$y$--coordinates are those defined with respcect to an
anholonomic (constrained) sub--basis.

An anholonomic frame structure $\delta _\alpha $ on $V^{(n+m)}$ is
characterized by its anholonomy relations
\begin{equation}
\delta _\alpha \delta _\beta -\delta _\beta \delta _\alpha
=w_{~\alpha \beta }^\gamma \delta _\gamma .  \nonumber
\end{equation}
with anholonomy coefficients $w_{~\beta \gamma }^\alpha .$ The
elongation of partial derivatives (by N--coefficients) in the
locally adapted partial
derivatives reflects the fact that on the (pseudo) Riemannian space--time $%
V^{(n+m)}$ it is modeled a generic local anisotropy characterized
by some anholonomy relations when the anholonomy coefficients are
computed as follows
\begin{eqnarray}
w_{~ij}^k & = &
0,w_{~aj}^k=0,w_{~ia}^k=0,w_{~ab}^k=0,w_{~ab}^c=0,  \nonumber
\\
w_{~ij}^a & = & -\Omega _{ij}^a,w_{~aj}^b=-\partial
_aN_i^b,w_{~ia}^b=\partial _aN_i^b,  \nonumber
\end{eqnarray}
where
\[
\Omega _{ij}^a=\partial _iN_j^a-\partial _jN_i^a+N_i^b\partial
_bN_j^a-N_j^b\partial _b N_i^a
\]
defines the coefficients of the N--connection curvature, in brief,
N--curvature. On (pseudo) Riemannian space--times this is a
characteristic of a chosen anholonomic system of reference.

For generic off--diagonal metrics we have two alternatives: The
first one is to try to compute the connection coefficients and
components of the Einstein tensor directly with respect to a
usual coordinate basis. This is connected to a cumbersome tensor
calculus and off--diagonal systems of partial differential
equations which makes almost impossible to find exact solutions
of Einstein equations. But we may try do diagonalize the metric
by some anholonomic transforms to a suitable N--elongated
anholonomic basis. Even this modifies the low of partial
derivation (like in all tetradic theories of gravity) the
procedure of computing the non--trivial components of the Ricci
and Einstein tensor simplifies substantially, and for a very large
class of former off--diagonal ansatz of metric, anholonomically
diagonalized, the Einstein equations can be integrated in general
form \cite {vb,vbh,vkinet,vp,vtheor,vtor}.

So, we conclude that when generic off--diagonal metrics and
anholonomic frames are introduces into consideration on (pseudo)
Riemannian spaces the space--time geometry may be equivalently
modeled as the geometry of moving anholonomic frames with
associated nonlinear connection structure. In this case the
problem of definition of anholonomic (anisotropic) spinor
structures arises even in general relativity theory which points
to the fact that the topic of anisotropic spinor differential
geometry is not an exotic subject from Finsler differential
geometry but a physical important problem which must be solved in
order to give a spinor interpretation of space--times provided
with off--diagonal metrics and anholonomic gravitational and
matter field interactions.

\subsection{Metrics depending on spinor variables and gauge
 theories}


An interesting study of differential geometry of spaces whose metric tensor $%
g_{\mu \nu }$ depends on spinor variables $\xi $ and $\bar{\xi}$
(its adjoint) as well as coordinates $x^i$, has been proposed by
Y. Takano \cite{t1}.
 Then Y. Takano and T. Ono  \cite{ot1,ot2,ot3} 
studied the above-mentioned spaces and they gave a generalization
of these spaces in the case where the metric tensor depends on
spinor variables $\xi $ and $\bar{\xi}$ and vector variables
$y^i$ as well as coordinates $x^i$. Such spaces are considered as
a generalization of Finsler spaces.

Later P. Stavrinos and S. Koutroubis studied the Lorentz
transformations and the curvature of generalized spaces with
metric tensor $g_{\mu \nu }(x,y,\xi ,\bar{\xi})$ \cite{sk}. 

The gravitational field equations are derived in the framework of
these
spaces whose metric tensor depends also on spinor variables $\xi $ and $\bar{%
\xi}$. The attempt is to describe gravity by a tetrad field and
the Lorentz connection coefficients in a more generalized
framework than the one developed by P. Ramond (cf. eg. \cite{ramon}).  
An interesting case with generalized conformally flat spaces with
metric $g_{\mu \nu }(x,\xi ,\bar{\xi})=\exp [2\sigma (x,\xi
,\bar{\xi})]\eta _{\mu \nu }$ was studied and the deviation of
geodesic equation in this space was derived.

In Chapter 12 we study the differential structure of a spinor
bundle in spaces with metric tensor $g_{\mu \nu }(x,\xi
,\bar{\xi})$ of the base manifold. Notions such as gauge
covariant derivatives of tensors, connections, curvatures,
torsions and Bianchi identities are presented in the context of a
gauge approach due to the introduction of a Poincar\'{e} group is
also essential in our study. The gauge field equations are
derived. Also we give the Yang-Mills and the Yang-Mills-Higgs
equations in a form sufficiently generalized for our approach.

Using the Hilbert-Palatini technique for a Utiyama-type
Lagrangian density in the deformed spinor bundle $S^{(2)}M\times
R$, the explicit expression of the field equations are determined
generalizing previous results; also, the equivalence principle is
shown to represent an extension for the corresponding one from
$S^{(2)}M$.

In this chapter we study the spinor bundle of order two $\tilde{S}^{(2)}(M)$%
, which is a foliation of the structure of the spinor bundle
presented in \cite{st1,sm3}. 
 In the present approach the generalized
tetrads and the spin-tetrads define, by means of the relations
 (\ref{sx9}), 
a generalized principle of equivalence in the spinor
bundle $\tilde{S}^{(2)}(M)$. Also, employing the Miron - type
connections, we cover all the possibilities for the S - bundle
connections, which represent the gauge potential. These have, in
the framework of our considerations, the remarkable property of
isotopic spin conservation. The introduction of the internal
deformed system (as a fibre) in $\tilde{S}^{(2)}(M)$, is expected
to produce as a natural consequence the Higgs field, for a
definite value $\kappa $,$\varphi ^{\alpha }$, where $\kappa $ is
a constant and $\varphi ^{\alpha }$ a scalar field.

In chapter 14 the Bianchi equations are determined for a deformed
spinor bundle $\tilde{S}^{(2)}(M)=S^{(2)}M\times R$. Also the
Yang-Mills-Higgs equations are derived and a geometrical
interpretation of the Higgs field is given \cite{st2}. 

We proceed as follows:

\begin{enumerate}
\item We study the Bianchi identities, choosing a Lagrangian
density that
contains the component $\varphi $ of a g-valued spinor gauge field of mass $%
m\in R$. Also we derive the Yang-Mills-Higgs equations on $\tilde{S}%
^{(2)}M\times R$. When $m_{0}\in R$ the gauge symmetry is
spontaneously broken, a fact connected to the Higgs field.

\item We introduce d-connections in the internal (spinor)
structures on \\ $\ \tilde{S}^{(2)}M$--bundle, which provide the
presentation of parallelism of the spin component constraints,
satisfied by the field strengths.

\item We interpret the metric G (relation (\ref{eq1.1}) 
 of the bundle $\tilde{S}%
^{(2)}M$, where the term $g^{\alpha \beta }D\xi _{\alpha }D\xi
_{\beta }^{\ast }$ has a physical meaning since it expresses the
measure of the number of particles in the same point of the space.

\item The above mentioned approach can be combined with the phase
transformations of the fibre $U(1)$ on a bundle $S^{(2)}M\times
U(1)$ in the Higgs mechanism. This will be the subject of our
future study.
\end{enumerate}

In the last part of our monograph, we establish the relation
between spinors of the $SL(2,C)$ group and tensors in the
framework of Lagrange spaces. A geometrical extension to
generalized metric tangent bundles is developed by means of
spinors. Also, the spinorial equation of casuality for the unique
solution of the null-cone in Finsler or Lagrange spaces is given
explicitly \cite{sm4}. 

\subsection{The layout of the book}

This book is organized in four Parts: the first three Parts
consists of three Chapters, the fourth Part consists of six
Chapters.

Part I presents an introduction to the geometry of anisotropic
spaces while outlines original results on the geometry of
anholonomic frames with associated nonlinear connections
structures in (pseudo) Riemannian spaces. In Chapter 1, we give
the basic definitions from the theory of generalized Finsler,
Lagrange, Cartan and Hamilton spaces on vector and co--vector
(tangent and co--tangent spaces) and their generalizations for
higher order vector--covector bundles following the monographs
\cite{ma94,mhss,vbook}. The next two Chapters are devoted to a
discussion and explicit examples when anisotropic (Finsler like
and more general ones) structures can be modeled on
pseudo--Riemannian spacetimes and in gravitational theories.
[They are based on results of works ellaborated by S. Vacaru and
co--authors \cite {vb,vbh,vkinet,vtheor,vd,vp,vsol,vt}]

Part II covers the algebra (Chapter 4) and geometry (Chapter 5)
of Clifford and spinor structures in vector bundles provided with
nonlinear connection structure. A spinor formulation of
generalized Finsler gravity and anisotropic matter field
interactions is given in Chapter 6. [This Part originates from S.
Vacaru and co--authors \cite {vdeb,vrom,vbm,viasm1,vjmp,vsp1}.]

Part III is a generalization of results on Clifford and spinor
structures for higher order vector bundles (Chapters 7--9 extend
respectively the results of Chapters 4--6), which are based on S.
Vacaru's papers \cite {vsp2,vhsp}.

Part IV (Chapters 10--15) summarizes the basic results on various
extensions of Finsler like geometries by considering spinor
variables. [In the main, this Part originates from papers of Y.
Takano and T. Ono \cite {t1,ot1,ot2,ot3,ono} and explicates the
most important contributions by P. Stavrinos and co--authors \cite
{st1,st2,sbmp,sbmp1,si1,sk,sm1,sm2,sm3,sm4,spmb}.

In summing up, this monograph we investigate anholonomic
(anisotrop\-ic) spinor structures in space--times with generic
local anisotropy (i. e. in generaliz\-ed Finsler spaces) and in
(pseudo) Riemannian spaces provided with off--dia\-go\-nal
metrics and anholonomic frame bases. It is addressed primarily to
researches and other readers in theoretical and mathematical
physics and differential geometry, both at the graduate and more
advanced levels. The book was also communicated at some scientific
Conferences \cite{vvcook,vsv,vscook} with some chapters outlined
as lectures in the preprint \cite{vvl}.

\subsection{ Acknowledgments}

The authors also would like to express their graditude to the
Vice-Rector of the University of Athens Prof. Dr. Dermitzakis for
his kindness to support this monograph to publish it by the
University of Athens. The second author would like to express his
gratitude to the emeritous Professor Y. Takano, for the
encouragement and the valuable discussions. It is also a pleasure
for the authors to give many thanks especially to Professors
Douglas Singleton, Heinz Dehnen, J. P. S.  Lemos, R. Miron, M.
Anastasiei, Mihai Visinescu, Vladimir Balan and Bertfried Fauser
for valuable disclussions, collaboration and support of scientific
investigations. The warmest thanks are extended to Foivos
Diakogiannis for the collaboration and help in the text of the
manuscript, to Evgenii Gaburov, Denis Gontsa, Nadejda Vicol,
Ovidiu Tintareanu--Mircea and Florian Catalin Popa for their
collaboration and help. We should like to express our deep
gratitude to the publishers and to the NATO/Portugal science
officials.

The authors are grateful to their families for patience and
understanding enabled to write this book.


\vskip 20pt

\begin{tabular}{|c|c|c|}
\hline\hline
&  &  \\
Sergiu I. \textbf{Vacaru } &  & Panayiotis \textbf{Stavrinos } \\
&  &  \\
Physics Department, &  &  \\
California State University, &  &  \\
Fresno, CA 93740--8031, USA &  &  \\
and &  &  \\
Centro Multidisciplinar &  & Department of Mathematics, \\
de Astrofisica -- CENTRA, &  & University of Athens, \\
Departamento de Fisica &  & 15784 Panepistimiopolis, \\
Instituto Superior Tecnico, &  & Athens, Greece \\
Av. Rovisco Pais 1, Lisboa, &  &  \\
1049--001, Portugal &  &  \\
&  &  \\
E-mails: &  & E--mail: \\
vacaru@fisica.ist.utl.pt &  & pstavrin@cc.uoa.gr \\
sergiu$_{-}$vacaru@yahoo.com &  &  \\
&  &  \\ \hline\hline
\end{tabular}

\newpage
---

\newpage

\section[Notation]{Notation}

The reader is advised to refer as and when necessary to the list
below where there are set out the conventions that will be
followed in this book with regard to the presentation of the
various physical and mathematical expressions.

(1) {\it Equations.}\ For instance, equation (3.16) is the 16th
equation in Chapter 3.\

(2) {\it Indices.} It is impossible to satisfy everybody in matter
of choice of labels of geometrical objects and coordinates. In
general, we shall use Greek superscripts for labels on both vector
bundles and superbundles. The reader will have to consult the
first sections in every Chapter in order to understand the meaning
of various types of boldface and/or underlined Greek or Latin
letters for operators, distinguished spinors and tensors.

(3) {\it Differentiation.} Ordinary partial differentiation with
respect to a coordinate $x^i$ will either be denoted by the
operator ${\partial}_i$ or
by subscript $i$ following a comma, for instance, ${\frac{\partial A^i }{%
\partial x^j}} \equiv \partial _j A^i \equiv {A^i}_{,j} .$ We shall use the
denotation ${\frac{\delta A^i }{\delta x^j}} \equiv \delta _j A^i
$ for partial derivations locally adapted to a nonlinear
connection structure.

(4) {\it Summation convention.} We shall follow the Einstein
summation rule for spinor and tensor indices.

(5) {\it References. } In the bibliography we cite the scientific
journals in a generally accepted abbreviated form, give the
volume, the year and the first page of the authors' articles; the
monographs and collections of works are cited completely. For the
author's works and communications, a part of them been published
in not enough accessible issues, or being under consideration,
the extended form (with the titles of articles and communications)
is presented. We emphasize that the references are intended to
give a sense of the book's scopes. We ask kindly the readers they
do not feel offended by any omissions.

(6) {\it Introductions and Conclusions.} If it is considered
necessary a Chapter starts with an introduction into the subject
and ends with concluding remarks.

 \mainmatter

\pagenumbering{arabic}







\part{Space--Time Anisotropy}

\chapter[Vector Bundles and N--Connections ]{Vector/Covector Bundles and
Nonlinear Connections}

In this Chapter the space--time geometry is modeled not only on a
(pseudo) Riemannian manifold $V^{[n+m]}$ of dimension $n+m$ but
it is considered on a vector bundle (or its dual, covector
bundle) being, for simplicity, locally trivial with a base space
$M$ of dimension $n$ and a typical fiber $F $ (cofiber $F^{*}$)
of dimension $m,$ or as a higher order extended vector/covector
bundle (we follow the geometric constructions and definitions of
monographs \cite{ma94,ma87,mhss,m1,m2} which were generalized for
vector superbundles in Refs. \cite{vstr2,vbook}). Such fibered
space--times (in general, with extra dimensions and duality
relations) are supposed to be provided with compatible structures
of nonlinear and linear connections and (pseudo) Riemannian
metric. For the particular cases when:\ a) the total space of the
vector bundle is substituted by a pseudo--Riemannian manifold of
necessary signature we can model the usual pseudo--Riemannian
space--time from the Einstein gravity theory with field equations
and geometric objects defined with respect to some classes of
moving anholonomic frames with associated nonlinear connection
structure; b) if the dimensions of the base and fiber spaces are
identical, $n=m,$ for the first order anisotropy, we obtain the
tangent bundle $TM.$

Such both (pseudo) Riemanian spaces and vector/covector (in
particular cases, tangent/cotangent) bundles of metric signature
(-,+,...,+) enabled with compatible fibered and/or anholonomic
structures, the metric in the total space being a solution of the
Einstein equations, will be called \textbf{anisotropic
space--times}. If the anholonomic structure with associated
nonlinear connection is modeled on higher order vector/covector
bundles we shall use the term of \textbf{higher order anisotropic
space--time.}

The geometric constructions are outlined as to present the main
concepts and formulas in a unique way for both type of vector and
covector structures. In this part of the book we usually shall
omit proofs which can be found in the mentioned monographs
\cite{ma87,ma94,m1,m2,mhss,vbook}.

\section{Vector and Covector Bundles}

In this Section we introduce the basic definitions and
denotations for vector and tangent (and theirs dual spaces)
bundles and higher order vector/covector bundle geometry.

\subsection{Vector and tangent bundles}

A locally trivial \textbf{vector bundle}, in brief,
\textbf{v--bundle}, \index{vector bundle} \index{v--bundle}
$\mathcal{E}=\left( E,\pi ,M,Gr,F\right) $ is introduced as a set
of spaces and surjective map with the properties that a real
vector space $F=\mathcal{R}^m$ of dimension $m$ ($\dim F=m,$
$\mathcal{R\ }$ denotes the real number field) defines the
typical fibre, the structural group \index{v--bundle} is chosen
to be the group of automorphisms of $\mathcal{R}^m,$ i. e.
$Gr=GL\left( m,\mathcal{R}\right) ,\ $ and $\pi :E\rightarrow M$
is a differentiable surjection of a differentiable
\index{manifold} manifold $E$ (total space, $\dim E=n+m)$ to a
differentiable manifold $M$ $\left( \mbox{base
space, }\dim M=n\right) .$ Local coordinates on $\mathcal{E}$ are denoted $%
u^{{\alpha }}=\left( x^{{i}},y^{{a}}\right) ,$ or in brief
${u=\left( x,y\right) }$ (the Latin indices
${i,j,k,...}=1,2,...,n$ define coordinates of geometrical objects
with respect to a local frame on base space $M;$\ the Latin
indices ${a,b,c,...=}1,2,...,m$ define fibre coordinates of
geometrical objects and the Greek indices ${\alpha ,\beta ,\gamma
,...}$ are considered as cumulative ones for coordinates of
objects defined on the total space of a v-bundle).

Coordinate trans\-forms $u^{{\ \alpha ^{\prime }\ }} = u^{{\
\alpha ^{\prime }\ }}\left( u^{{\ \alpha }}\right) $ on a
v--bundle $\mathcal{E}$ are defined as
\[
\left(x^{{\ i}}, y^{{\ a}}\right) \rightarrow \left( x^{{\
i^{\prime }\ }}, y^{{\ a^{\prime }}}\right) ,
\]
where
\begin{equation}
x^{{i^{\prime }\ }} = x^{{\ i^{\prime }\ }}(x^{{\ i}}),\qquad
y^{{a^{\prime }\ }} = K_{{\ a\ }}^{{\ a^{\prime }}}(x^{{i\
}})y^{{a}}  \label{coordtr}
\end{equation}
and matrix $K_{{\ a\ }}^{{\ a^{\prime }}}(x^{{\ i\ }}) \in GL\left( m,%
\mathcal{R}\right) $ are functions of necessary smoothness class.

A local coordinate parametrization of v--bundle $\mathcal{E}$
naturally defines a coordinate basis
\begin{equation}
\partial _\alpha =\frac \partial {\partial u^\alpha }= \left( \partial _i =
\frac \partial {\partial x^i},\ \partial_a = \frac \partial
{\partial y^a}\right) ,  \label{pder}
\end{equation}
and the reciprocal to (\ref{pder}) 
coordinate basis
\begin{equation}  \label{pdif}
d^\alpha = du^\alpha =(d^i = dx^i,\ d^a = dy^a)
\end{equation}
which is uniquely defined from the equations
\[
d^\alpha \circ \partial _\beta =\delta _\beta ^\alpha ,
\]
where $\delta _\beta ^\alpha $ is the Kronecher symbol and by
''$\circ $" we denote the inner (scalar) product in the tangent
bundle $\mathcal{TE.}$

A \textbf{tangent bundle} \index{tangent bundle} (in brief,
\textbf{t--bundle}) \index{t--bundle} $(TM,\pi ,M)$ to a manifold
$M$ can be defined as a particular case of a v--bundle when the
dimension of the base and fiber spaces (the last one considered
as the
tangent subspace) are identic, $n=m.$ In this case both type of indices $%
i,k,...$ and $a,b,...$ take the same values $1,2,...n$. For
t--bundles the matrices of fiber coordinates transforms from
(\ref{coordtr}) can be written $K_{{\ i\ }}^{{\ i^{\prime }}} =
{\partial x^{i^{\prime }}}/ {\partial x^i}.$

We shall distinguish the base and fiber indices and values which
is necessary for our further geometric and physical applications.

\subsection{Covector and cotangent bundles}

We shall also use the concept of \textbf{covector bundle},
\index{covector bundle} (in brief, \textbf{cv--bundles)} \index{
cv--bundle}\newline $\breve {\mathcal{E}}=\left({\breve E},\pi
^{*} ,M,Gr,F^{*}\right) $, which is introduced as a dual vector
bundle for which the typical fiber $F^{*}$ (cofiber)
\index{cofiber} is considered to be the dual vector space
(covector space) \index{covector space}  to the vector space $F.$
The fiber coordinates $p_a$ of $\breve E$ are dual to $y^a$
in $E.$ The local coordinates on total space $\breve E$ are denoted $%
\breve{u}=(x,p)=(x^i,p_a).$ The coordinate transform on $\breve
E,$
\[
{\breve{u}}=(x^i,p_a)\to {\breve{u}}^{\prime }=(x^{i^{\prime
}},p_{a^{\prime }}),
\]
are written
\begin{equation}
x^{{i^{\prime }\ }} = x^{{\ i^{\prime }\ }}(x^{{\ i}}),\qquad
p_{{a^{\prime }\ }} = K^{{\ a\ }}_{{\ a^{\prime }}}(x^{{i\
}})p_{{a}}.  \label{coordtrd}
\end{equation}
The coordinate bases on $E^{*}$ are denoted
\begin{equation}
{\breve{\partial}}_\alpha =\frac{\breve{\partial}}{\partial u^\alpha }%
=\left( \partial _i=\frac \partial {\partial x^i},{{\breve {\partial}}^a}=%
\frac{\breve{\partial}}{\partial p_a}\right)  \label{pderct}
\end{equation}
and
\begin{equation}
{\breve{d}}^\alpha ={\breve{d}}u^\alpha =\left(
d^i=dx^i,{\breve{d}}_a= dp_a\right) .  \label{pdifct}
\end{equation}
We shall use ''breve'' symbols in order to distinguish the
geometrical
objects on a cv--bundle $\mathcal{E}^{*}$ from those on a v--bundle $%
\mathcal{E}$.

As a particular case with the same dimension of base space and
cofiber one obtains the \textbf{cotangent bundle}
\index{cotangent bundle} $(T^{*}M,\pi ^{*},M)$ , in brief,
\textbf{ct--bundle,} \index{ct--bundle} being dual to $TM.$ The fibre coordinates $p_i$ of $%
T^{*}M$ are dual to $y^i$ in $TM.$ The coordinate transforms (\ref{coordtrd}%
) on $T^{*}M$ are stated by some matrices $K^{k}_{{\ k^{\prime }}}(x^i)= {%
\partial x^k}/{\partial x^{k^{\prime }}}.$

In our further considerations we shall distinguish the base and
cofiber indices.

\subsection{Higher order vector/covector bundles}

The geometry of higher order tangent and cotangent bundles
provided with nonlinear connection structure was elaborated in
Refs. \cite {m1,m2,mirata,mhss} following the aim of
geometrization of higher order Lagrange and Hamilton mechanics.
In this case we have base spaces and fibers of the same
dimension. In order to develop the approach to modern high energy
physics (in superstring and Kaluza--Klein theories) one had to
introduce (in Refs \cite{vsp1,vhsp,vbook,vstr2}) the concept of
higher order vector bundle with the fibers consisting from finite
'shells" of vector, or covector, spaces of different dimensions
not obligatory coinciding with the base space dimension.

\begin{definition}
A distinguished vector/covector space, in brief dvc--space,
\index{dvc--space} of type
\begin{equation}
{\tilde{F}}=F[v(1),v(2),cv(3),...,cv(z-1),v(z)]  \label{orient}
\end{equation}
is a vector space decomposed into an invariant oriented direct
summ
\[
{\tilde{F}}=F_{(1)}\oplus F_{(2)}\oplus F_{(3)}^{*}\oplus
...\oplus F_{(z-1)}^{*}\oplus F_{(z)}
\]
of vector spaces $F_{(1)},F_{(2)},...,F_{(z)}$ of respective
dimensions
\[
dimF_{(1)}=m_1,dimF_{(2)}=m_2,...,dimF_{(z)}=m_z
\]
and of covector spaces $F_{(3)}^{*},...,F_{(z-1)}^{*}$ of
respective dimensions
\[
dimF_{(3)}^{*}=m_3^{*},...,dimF_{(z-1)}^{*}=m_{(z-1)}^{*}.
\]
\end{definition}

As a particular case we obtain a distinguished vector space, in
brief dv--space  \index{dv--space} (a distinguished covector
space, in brief dcv--space), \index{dcv--space} if all components
of the sum are vector (covector) spaces. We note that we have
fixed for simplicity an orientation of vector/covector subspaces like in (%
\ref{orient}); in general there are possible various type of
orientations, number of subspaces and dimensions of subspaces.

Coordinates on ${\tilde F}$ are denoted
\[
\tilde{y} = (y_{(1)},y_{(2)},p_{(3)},...,p_{(z-1)},y_{(z)})=
\{y^{<\alpha _z>}\} =
(y^{a_1},y^{a_2},p_{a_3},...,p_{a_{z-1}},y^{a_z}),
\]
where indices run corresponding values:\
\[
a_1 = 1,2,...,m_1;\ a_2 = 1,2,...,m_2,\ ..., a_z = 1,2,...,m_z.
\]

\begin{definition}
A higher order vector/covector bundle (in brief, hvc-\--bund\-le)
 \index{hvc--bundle} of type ${%
\tilde{\mathcal{E}}}={\tilde{\mathcal{E}}}[v(1),v(2),cv(3),...,cv(z-1),v(z)]$
is a vector bundle ${\tilde{\mathcal{E}}}=({\tilde{E}},p^{<d>},{\tilde{F}}%
,M) $ with corresponding total, ${\tilde{E}}$, and base, $M,$
spaces,
surjective projection $p^{<d>}:\ {\tilde{E}}\to M$ and typical fibre ${%
\tilde{F}}.$
\end{definition}

We define higher order vector (covector) bundles, in brief,
hv--bundles (in brief, hcv--bundles), if the typical fibre is a
dv--space (dcv--space) as particular cases of hvc--bundles.

A hvc--bundle is constructed as an oriented set of enveloping
'shell by shell' v--bundles and/or cv--bundles,
\[
p^{<s>}:\ {\tilde{E}}^{<s>}\to {\tilde{E}}^{<s-1>},
\]
where we use the index $<s>=0,1,2,...,z$ in order to enumerate
the shells, when ${\tilde{E}}^{<0>}=M.$ Local coordinates on
${\tilde{E}}^{<s>}$ are denoted
\begin{eqnarray}
{\tilde{u}}_{(s)} &=&(x,{\tilde{y}}%
_{<s>})=(x,y_{(1)},y_{(2)},p_{(3)},...,y_{(s)})  \nonumber \\
&=&(x^i,y^{a_1},y^{a_2},p_{a_3},...,y^{a_s}).  \nonumber
\end{eqnarray}
If $<s>=<z>$ we obtain a complete coordinate system on
${\tilde{\mathcal{E}}} $ denoted in brief
\[
\tilde{u}=(x,{\tilde{y}})=\tilde{u}^\alpha
=(x^i=y^{a_0},y^{a_1},y^{a_2},p_{a_3},...,p_{a_{z-1}},y^{a_z}).
\]
We shall use the general commutative indices $\alpha ,\beta ,...$
for objects on hvc---bundles which are marked by tilde, like
$\tilde{u},\tilde{u}^\alpha ,...,$ ${\tilde{E}}^{<s>},....$

The coordinate transforms for a hvc--bundle
${\tilde{\mathcal{E}},}$
\[
\tilde{u}=(x,{\tilde{y}})\rightarrow \tilde{u}^{\prime }=(x^{\prime },{%
\tilde{y}}^{\prime })
\]
are given by recurrent formulas
\begin{eqnarray*}
x^{i^{\prime }} &=&x^{i^{\prime }}\left( x^i\right) ,\ rank\left( \frac{%
\partial x^{i^{\prime }}}{\partial x^i}\right) =n; \\
y^{a_1^{\prime }} &=&K_{a_1}^{a_1^{^{\prime
}}}(x)y^{a_1},K_{a_1}^{a_1^{^{\prime }}}\in GL(m_1,\mathcal{R}); \\
y^{a_2^{\prime }} &=&K_{a_2}^{a_2^{^{\prime
}}}(x,y_{(1)})y^{a_2},K_{a_2}^{a_2^{^{\prime }}}\in GL(m_2,\mathcal{R}); \\
p_{a_3^{\prime }} &=&K_{a_3^{\prime
}}^{a_3}(x,y_{(1)},y_{(2)})p_{a_3},K_{a_3^{\prime }}^{a_3}\in GL(m_3,%
\mathcal{R}); \\
y^{a_4^{\prime }} &=&K_{a_4}^{a_4^{^{\prime
}}}(x,y_{(1)},y_{(2)},p_{(3)})y^{a_4},K_{a_4}^{a_4^{^{\prime }}}\in GL(m_4,%
\mathcal{R}); \\
&&................ \\
p_{a_{z-1}^{\prime }} &=&K_{a_{z-1}^{\prime
}}^{a_{z-1}}(x,y_{(1)},y_{(2)},p_{(3)},...,y_{(z-2)})p_{a_{z-1}},K_{a_{z-1}^{\prime }}^{a_{z-1}}\in GL(m_{z-1},%
\mathcal{R}); \\
y^{a_z^{\prime }}
&=&K_{a_z}^{a_z^{^{%
\prime}}}(x,y_{(1)},y_{(2)},p_{(3)},...,y_{(z-2)},p_{a_{z-1}})y^{a_z},K_{a_z}^{a_z^{^{\prime }}}\in GL(m_z,%
\mathcal{R}),
\end{eqnarray*}
where, for instance. by $GL(m_2,\mathcal{R})$ we denoted the
group of linear transforms of a real vector space of dimension
$m_2.$

The coordinate bases on ${\tilde{\mathcal{E}}}$ are denoted
\begin{eqnarray}
{\tilde \partial}_\alpha & =& \frac{{\tilde \partial}}{\partial
u^\alpha }
\label{pderho} \\
&=& \left( \partial _i=\frac \partial {\partial x^i},\partial
_{a_1}=\frac
\partial {\partial y^{a_1}},\partial _{a_2}=\frac \partial {\partial
y^{a_2}},{{\breve{\partial}}}^{a_3}=\frac{\breve{\partial}}{\partial p_{a_3}}%
,...,\partial _{a_z} =\frac \partial {\partial y^{a_z}}\right)
\nonumber
\end{eqnarray}
and
\begin{eqnarray}
{\tilde d}^\alpha & =& {\tilde d}u^\alpha  \label{pdifho} \\
&=& \left( d^i=dx^i,d^{a_1}=dy^{a_1},d^{a_2}=dy^{a_2},{\breve{d}}%
_{a_3}=dp_{a_3},...,d^{a_z}=dy^{a_z}\right) .  \nonumber
\end{eqnarray}

We end this subsection with two examples of higher order tangent /
co\-tan\-gent bundles (when the dimensions of fibers/cofibers
coincide with the dimension of bundle space, see Refs.
\cite{m1,m2,mirata,mhss}).

\subsubsection{Osculator bundle}
\index{osculator}
 \label{oscsect}

The $k$--osculator bundle \index{$k$--osculator} is identified
with the $k$--tangent bundle\newline $\left(
T^kM,p^{(k)},M\right) $ of a $n$--dimensional manifold $M.$ We
denote the local coordinates
\[
{\tilde{u}}^\alpha =\left( x^i,y_{(1)}^i,...,y_{(k)}^i\right) ,
\]
where we have identified $y_{(1)}^i\simeq
y^{a_1},...,y_{(k)}^i\simeq y^{a_k},k=z,$ in order to to have
similarity with denotations from \cite {mhss}. The coordinate
transforms
\[
{\tilde{u}}^{\alpha ^{\prime }}\rightarrow {\tilde{u}}^{\alpha
^{\prime }}\left( {\tilde{u}}^\alpha \right)
\]
preserving the structure of such higher order vector bundles are
parametrized
\begin{eqnarray*}
x^{i^{\prime }} &=&x^{i^{\prime }}\left( x^i\right) ,\det \left( \frac{%
\partial x^{i^{\prime }}}{\partial x^i}\right) \neq 0, \\
y_{(1)}^{i^{\prime }} &=&\frac{\partial x^{i^{\prime }}}{\partial x^i}%
y_{(1)}^i, \\
2y_{(2)}^{i^{\prime }} &=&\frac{\partial y_{(1)}^{i^{\prime }}}{\partial x^i}%
y_{(1)}^i+2\frac{\partial y_{(1)}^{i^{\prime }}}{\partial y^i}y_{(2)}^i, \\
&&................... \\
ky_{(k)}^{i^{\prime }} &=&\frac{\partial y_{(1)}^{i^{\prime }}}{\partial x^i}%
y_{(1)}^i+...+k\frac{\partial y_{(k-1)}^{i^{\prime }}}{\partial y_{(k-1)}^i}%
y_{(k)}^i,
\end{eqnarray*}
where the equalities
\[
\frac{\partial y_{(s)}^{i^{\prime }}}{\partial x^i}=\frac{\partial
y_{(s+1)}^{i^{\prime }}}{\partial y_{(1)}^i}=...=\frac{\partial
y_{(k)}^{i^{\prime }}}{\partial y_{(k-s)}^i}
\]
hold for $s=0,...,k-1$ and $y_{(0)}^i=x^i.$

The natural coordinate frame on $\left( T^kM,p^{(k)},M\right) $
is defined
\[
{\tilde{\partial}}_\alpha =\left( \frac \partial {\partial
x^i},\frac
\partial {\partial y_{(1)}^i},...,\frac \partial {\partial y_{(k)}^i}\right)
\]
and the coframe is
\[
{\tilde d}_\alpha =\left( dx^i,dy_{(1)}^i,...,dy_{(k)}^i\right) .
\]
These formulas are respectively some particular cases of $\left( \ref{pderho}%
\right) $ and $\left( \ref{pdifho}\right) .$

\subsubsection{The dual bundle of k--osculator bundle}

\label{doscsect}

This higher order vector/covector bundle, denoted as $\left(
T^{*k}M,p^{*k},M\right) ,$ is defined as the dual bundle to the
k--tangent bundle $\left( T^kM,p^k,M\right) .$ The local
coordinates (parametrized as in the previous paragraph) are
\[
\tilde u=\left( x,y_{(1)},...,y_{(k-1)},p\right) =\left(
x^i,y_{(1)}^i,...,y_{(k-1)}^i,p_i\right) \in T^{*k}M.
\]
The coordinate transforms on $\left( T^{*k}M,p^{*k},M\right) $ are
\begin{eqnarray*}
x^{i^{\prime }} &=&x^{i^{\prime }}\left( x^i\right) ,\det \left( \frac{%
\partial x^{i^{\prime }}}{\partial x^i}\right) \neq 0, \\
y_{(1)}^{i^{\prime }} &=&\frac{\partial x^{i^{\prime }}}{\partial x^i}%
y_{(1)}^i, \\
2y_{(2)}^{i^{\prime }} &=&\frac{\partial y_{(1)}^{i^{\prime }}}{\partial x^i}%
y_{(1)}^i+2\frac{\partial y_{(1)}^{i^{\prime }}}{\partial y^i}y_{(2)}^i, \\
&&................... \\
(k-1)y_{(k-1)}^{i^{\prime }} &=&\frac{\partial y_{(k-2)}^{i^{\prime }}}{%
\partial x^i}y_{(1)}^i+...+k\frac{\partial y_{(k-1)}^{i^{\prime }}}{\partial
y_{(k-2)}^i}y_{(k-1)}^i, \\
p_{i^{\prime }} &=&\frac{\partial x^i}{\partial x^{i^{\prime
}}}p_i,
\end{eqnarray*}
where the equalities
\[
\frac{\partial y_{(s)}^{i^{\prime }}}{\partial x^i}=\frac{\partial
y_{(s+1)}^{i^{\prime }}}{\partial y_{(1)}^i}=...=\frac{\partial
y_{(k-1)}^{i^{\prime }}}{\partial y_{(k-1-s)}^i}
\]
hold for $s=0,...,k-2$ and $y_{(0)}^i=x^i.$

The natural coordinate frame on $\left( T^{\ast k}M,p^{\ast
(k)},M\right) $ is defined
\[
{\tilde{\partial}}_{\alpha }=\left( \frac{\partial }{\partial x^{i}},\frac{%
\partial }{\partial y_{(1)}^{i}},...,\frac{\partial }{\partial y_{(k-1)}^{i}}%
,\frac{\partial }{\partial p_{i}}\right)
\]
and the coframe is
\[
{\tilde{d}}_{\alpha }=\left(
dx^{i},dy_{(1)}^{i},...,dy_{(k-1)}^{i},dp_{i}\right) .
\]
These formulas are respectively another particular cases of
$\left( \ref {pderho}\right) $ and $\left( \ref{pdifho}\right) .$

\section{Nonlinear Connections}

\label{subsnc}

The concept of \textbf{nonlinear connection,} \index{nonlinear
connection} in brief, N-connection, \index{N--connection} is
fundamental in the geometry of vector bundles and anisotropic
spaces (see a detailed study and basic references in
\cite{ma87,ma94}). A rigorous mathematical definition is possible
by using the formalism of exact sequences of vector bundles.

\subsection{N--connections in vector bundles}

Let $\mathcal{E=}=(E,p,M)$ be a v--bundle with typical fibre
$\mathcal{R}^m$ and $\pi ^T:\ TE\to TM$ being the differential of
the map $P$ which is a fibre--preserving morphism of the tangent
bundle $TE,\tau _E,E)\to E$ and of tangent bundle $(TM,\tau
,M)\to M.$ The kernel of the vector bundle morphism, denoted as
$(VE,\tau _V,E),$ is called the \textbf{vertical
subbundle} over $E,$ which is a vector subbundle of the vector bundle $%
(TE,\tau _E,E).$

A vector $X_u$ tangent to a point $u\in E$ is locally written as
\[
(x,y,X,Y)=(x^i,y^a,X^i,Y^a),
\]
where the coordinates $(X^i,Y^a)$ are defined by the equality
\[
X_u=X^i\partial _i+Y^a\partial _a.
\]
We have\ $\pi ^T(x,y,X,Y)=(x,X).$ Thus the submanifold $VE$
contains the elements which are locally represented as
$(x,y,0,Y).$

\begin{definition}
\label{ncon} A nonlinear connection $\mathbf{N}$ in a vector bundle $%
\mathcal{E}=(E,\pi ,M)$ is the splitting on the left of the exact
sequence
\[
0\mapsto VE\mapsto TE\mapsto TE/VE\mapsto 0
\]
where $TE/VE$ is the factor bundle.
\end{definition}

By definition (\ref{ncon}) it is defined a morphism of vector
bundles $C:\ TE\to VE$ such the superposition of maps $C\circ i$
is the identity on $VE,$ where $i:\ VE\mapsto VE.$ The kernel of
the morphism $C$ is a vector
subbundle of $(TE,\tau _E,E)$ which is the horizontal subbundle, denoted by $%
(HE,\tau _H,E).$ Consequently, we can prove that in a v-bundle
$\mathcal{E}$ a N--connection can be introduced as a distribution
\[
\{N:\ E_u\rightarrow H_uE,T_uE=H_uE\oplus V_uE\}
\]
for every point $u\in E$ defining a global decomposition, as a
Whitney sum, into horizontal,$H\mathcal{E\ },$ and vertical,
$V\mathcal{E},$ subbundles of the tangent bundle $T\mathcal{E}$
\begin{equation}
T\mathcal{E}=H\mathcal{E}\oplus V\mathcal{E}.  \label{whitney}
\end{equation}

Locally a N-connection in a v--bundle $\mathcal{E}$ is given by
its coefficients\newline $N_{{\ i}}^{{a}}({\
u})=N_{{i}}^{{a}}({x,y})$ with respect to bases
(\ref{pder}) and (\ref{pdif})
\[
\mathbf{N}=N_i^{~a}(u)d^i\otimes \partial _a.
\]

We note that a linear connection in a v--bundle $\mathcal{E}$ can
be
considered as a particular case of a N--connection when $%
N_i^{~a}(x,y)=K_{bi}^a\left( x\right) y^b,$ where functions
$K_{ai}^b\left( x\right) $ on the base $M$ are called the
Christoffel coefficients.

\subsection{N--connections in covector bundles:}

A nonlinear connection in a cv--bundle ${\breve{\mathcal{E}}}$
(in brief a \v N--connection) can be introduces in a similar
fashion as for v--bundles by reconsidering the corresponding
definitions for cv--bundles. For
instance, it is stated by a Whitney sum, into horizontal,$H{\breve{\mathcal{E%
}}},\ $ and vertical, $V{\breve{\mathcal{E}}},$ subbundles of the
tangent bundle $T{\breve{\mathcal{E}}}:$
\begin{equation}
T{\breve{\mathcal{E}}}=H{\breve{\mathcal{E}}}\oplus
V{\breve{\mathcal{E}}}. \label{whitneyc}
\end{equation}

Hereafter, for the sake of brevity we shall omit details on
definition of geometrical objects on cv--bundles if they are very
similar to those for v--bundles:\ we shall present only the basic
formulas by emphasizing the most important particularities and
differences.

\begin{definition}
\label{ctvn} A \v {N}--connection on ${\breve{\mathcal{E}}}$ is a
differentiable distribution
\[
\breve{N}:\ {\breve{\mathcal{E}}}\to {\breve{N}}_u\in T_u^{*}{\breve{%
\mathcal{E}}}
\]
which is suplimentary to the vertical distribution $V,$ i. e.
\[
T_u{\breve{\mathcal{E}}}={\breve{N}}_u\oplus {\breve{V}}_u,\forall {\breve{%
\mathcal{E}}}.
\]
\end{definition}

The same definition is true for \v N--connections in ct--bundles,
we have to change in the definition (\ref{ctvn}) the symbol
${\breve{\mathcal{E}}}$ into $T^*M.$

A \v N--connection in a cv--bundle ${\breve{\mathcal{E}}}$ is
given locally by its coefficients\newline ${\breve N}_{{\ ia}}({\
u})= {\breve N}_{{ia}}({x,p})$ with respect to bases (\ref{pder})
and (\ref{pdif})
\[
\mathbf{\breve N}={\breve N}_{ia}(u)d^i\otimes
{{\breve{\partial}}^a}.
\]

We emphasize that if a N--connection is introduced in a v--bundle
(cv--bundle) we have to adapt the geometric constructions to the
N--connection structure.

\subsection{N--connections in higher order bundles}

The concept of N--connection can be defined for higher order
vector / covec\-tor bundle in a standard manner like in the usual
vector bundles:

\begin{definition}
A nonlinear connection ${\tilde{\mathbf{N}}}$ in hvc--bundle
\index{N--connection}
\[
{\tilde{\mathcal{E}}}={\tilde{\mathcal{E}}}%
[v(1),v(2),cv(3),...,cv(z-1),v(z)]
\]
is a splitting of the left of the exact sequence
\begin{equation}
0\to V{\tilde{\mathcal{E}}}\to T{\tilde{\mathcal{E}}}\to T{\tilde{\mathcal{E}%
}}/V{\tilde{\mathcal{E}}}\to 0  \label{exacts}
\end{equation}
\end{definition}

We can associate sequences of type (\ref{exacts}) to every
mappings of intermediary subbundles. For simplicity, we present
here the Whitney decomposition
\[
T{\tilde{\mathcal{E}}}=H{\tilde{\mathcal{E}}}\oplus V_{v(1)}{\tilde{\mathcal{%
E}}}\oplus V_{v(2)}{\tilde{\mathcal{E}}}\oplus V_{cv(3)}^{*}{\tilde{\mathcal{%
E}}}\oplus ....\oplus V_{cv(z-1)}^{*}{\tilde{\mathcal{E}}}\oplus V_{v(z)}{%
\tilde{\mathcal{E}}}.
\]
Locally a N--connection ${\tilde{\mathbf{N}}}$ in
${\tilde{\mathcal{E}}}$ is given by its coefficients
\begin{equation}
\begin{array}{llllll}
N_i^{~a_1}, & N_i^{~a_2}, & N_{ia_3}, & ..., & N_{ia_{z-1}}, &
N_i^{~a_z},
\\
0, & N_{a_1}^{~a_2}, & N_{a_1a_3}, & ..., & N_{a_1a_{z-1}}, &
N_{a_1}^{~a_z},
\\
0, & 0, & N_{a_2a_3}, & ..., & N_{a_2a_{z-1}}, & N_{a_2}^{~a_z}, \\
..., & ..., & ..., & ..., & ..., & ..., \\
0, & 0, & 0, & ..., & N_{a_{z-2}~a_{z-1}}, & N_{a_{z-2}}^{~a_z}, \\
0, & 0, & 0, & ..., & 0, & N^{a_{z-1}a_z},
\end{array}
\label{nconho}
\end{equation}
which are given with respect to the components of bases $\left( \ref{pderho}%
\right) $ and $\left( \ref{pdifho}\right) .$

We end this subsection with two exemples of N--connections in
higher order vector/covector bundles:

\subsubsection{N--connection in osculator bundle}

\label{nconoscs}

Let us consider the second order of osculator bundle (see
subsection (\ref
{oscsect})) $T^2M=Osc^2M.$ A N--connection ${\tilde {\mathbf{N}}}$ in $%
Osc^2M $ is associated to a Whitney summ
\[
TT^2M=NT^2M\oplus VT^2M
\]
which defines in every point $\tilde u\in T^2M$ a distribution
\[
T_uT^2M=N_0\left( \tilde u\right) \oplus N_1\left( \tilde
u\right) \oplus VT^2M.
\]
We can parametrize ${\tilde {\mathbf{N}}}$ with respect to natural
coordinate bases as
\begin{equation}  \label{nconoscsf}
\begin{array}{ll}
N_i^{a_1}, & N_i^{a_2}, \\
0, & N_{a_1}^{a_2}.
\end{array}
\end{equation}
As a particular case we can consider $N_{a_1}^{a_2}=0.$

\subsubsection{N--connection in dual osculator bundle}

\label{nconoscds}

In a similar fashion we can take the bundle $\left(
T^{*2}M,p^{*2},M\right) $ being dual bundle to the $Osc^2M$ (see
subsection (\ref{doscsect})). We have
\[
T^{*2}M=TM\otimes T^{*}M.
\]
The local coefficients of a N--connection in $\left(
T^{*2}M,p^{*2},M\right) $ are parametrizied
\begin{equation}  \label{nconoscsdf}
\begin{array}{ll}
N_i^{~a_1}, & N_{ia_2}, \\
0, & N_{a_1a_2}.
\end{array}
\end{equation}
We can choose a particular case when $N_{a_1a_2}=0.$

\subsection{Anholonomic frames and N--connections}

Having defined a N--connection structure in a (vector, covector,
or higher order vector / covenctor) bundle we can adapt to this
structure, (by 'N--elonga\-ti\-on', the operators of partial
derivatives and differentials and to consider decompositions of
geometrical objects with respect to adapted bases and cobases.

\subsubsection{Anholonomic frames in v--bundles}

In a v--bunde $\mathcal{E}$ provided with a N-connection we can
adapt to this structure the geometric constructions by
introducing locally adapted basis (N--frame, or N--basis):
\begin{equation}
\delta _\alpha =\frac \delta {\delta u^\alpha }=\left( \delta
_i=\frac \delta {\delta x^i}=\partial _i-N_i^{~a}\left( u\right)
\partial _a,\partial _a=\frac \partial {\partial y^a}\right) ,
\label{dder}
\end{equation}
and its dual N--basis, (N--coframe, or N--cobasis),
\begin{equation}
\delta \ ^\alpha =\delta u^\alpha =\left( d^i=\delta
x^i=dx^i,\delta ^a=\delta y^a+N_i^{~a}\left( u\right) dx^i\right)
.  \label{ddif}
\end{equation}

The\textbf{\ anholonomic coefficients, } $\mathbf{w}=\{w_{\beta
\gamma }^\alpha \left( u\right) \},$ of N--frames are defined to
satisfy the relations \index{anholonomic }
\begin{equation}  \label{anhol}
\left[ \delta _\alpha ,\delta _\beta \right] =\delta _\alpha
\delta _\beta -\delta _\beta \delta _\alpha =w_{\beta \gamma
}^\alpha \left( u\right) \delta _\alpha .
\end{equation}


A frame bases is holonomic is all anholonomy coefficients vanish
(like for usual coordinate bases (\ref{pdif})), or anholonomic if
there are nonzero values of $w_{\beta \gamma }^\alpha.$

So, we conclude that a N--connection structure splitting
conventionally a
v--bundle $\mathcal{E}$ into some horizontal $H\mathcal{E}$ and vertical $V%
\mathcal{E} $ subbundles can be modelled by an anholonomic frame
structure with mixed holonomic $\{x^i\}$ and anholonomic
$\{y^a\}$ variables. This case differs from usual, for instance,
tetradic approach in general relativity when tetradic (frame)
fields are stated to have only for holonomic or only for
anholonomic variables. By using the N--connection formalism we
can investigate geometrical and physical systems when some degees
of freedoms (variables) are subjected to anholonomic constraints,
the rest of variables being holonomic.

The operators (\ref{dder}) and (\ref{ddif}) on a v--bundle
$\mathcal{E} $ en\-abled with a N--connecti\-on can be considered
as respective equivalents of the operators of partial derivations
and differentials:\ the existence of
a N--connection structure results in 'elongation' of partial derivations on $%
x$--variables and in 'elongation' of differentials on
$y$--variables.


The \textbf{algebra of tensorial distinguished fields} $DT\left( \mathcal{E}%
\right) $ (d--fields, d--ten\-sors, d--objects) on $\mathcal{E}$
is introduced as the tensor algebra $\mathcal{T} =\{
\mathcal{T}_{qs}^{pr}\}$ of the v--bundle
\[
\mathcal{E}_{\left( d\right) }= \left(H\mathcal{E}\oplus
V\mathcal{E}, p_d, \mathcal{E} \right),
\]
where $p_d:\ H\mathcal{E}\oplus V\mathcal{E}\rightarrow
\mathcal{E}.$

An element $\mathbf{t}\in \mathcal{T}_{qs}^{pr},$ d--tensor field of type $%
\left(
\begin{array}{cc}
p & r \\
q & s
\end{array}
\right) ,$ can be written in local form as
\begin{eqnarray}
\mathbf{t}&=&t_{j_1...j_qb_1...b_r}^{i_1...i_pa_1...a_r}\left(
u\right) \delta _{i_1}\otimes ...\otimes \delta _{i_p}\otimes
\partial _{a_1}\otimes
...\otimes \partial _{a_r}  \nonumber \\
& & \otimes d^{j_1}\otimes ...\otimes d^{j_q}\otimes \delta
^{b_1}...\otimes \delta ^{b_r}.  \nonumber
\end{eqnarray}

We shall respectively use the denotations $\mathcal{X\left( E\right) }$ (or $%
\mathcal{X\ } {\left( M\right) ),\ } \Lambda ^p\left(
\mathcal{E}\right) $ or \newline $\left( \Lambda ^p\left(
M\right) \right) $ and $\mathcal{F\left( E\right) }$
(or $\mathcal{F}$ $\left( M\right) $) for the module of d--vector fields on $%
\mathcal{E}$ (or\newline $M$), the exterior algebra of p--forms
on $\mathcal{E}$ (or $M$) and the set of real functi\-ons on
$\mathcal{E}$ (or $M).$

\subsubsection{Anholonomic frames in cv--bundles}

The anholnomic frames \index{anholonomic} adapted to the \v
N--connection structure are introduced similarly to (\ref{dder})
and (\ref{ddif}):

the locally adapted basis (\v N--basis, or \v N--frame):
\begin{equation}
{\breve \delta}_\alpha = \frac {\breve \delta}{\delta u^\alpha}=
\left(
\delta _i=\frac \delta {\delta x^i}= \partial _i + {\breve N}_{ia}\left({%
\breve u}\right) {\breve \partial}^a, {\breve \partial}^a = \frac
\partial {\partial p_a}\right) ,  \label{ddercv}
\end{equation}

and its dual (\v N--cobasis, or \v N--coframe) \index{N--cobasis}
\index{N--coframe} :
\begin{equation}
{\breve \delta}^\alpha ={\breve \delta} u^\alpha =\left(
d^i=\delta
x^i=dx^i,\ {\breve \delta}_a= {\breve \delta} p_a= d p_a - {\breve N}%
_{ia}\left({\breve u}\right) dx^i\right) .  \label{ddifcv}
\end{equation}

We note that for the signes of \v N--elongations are inverse to
those for N--elongations.

The\textbf{\ anholonomic coefficients, } $\mathbf{\breve w}= \{{\breve w}%
_{\beta \gamma }^\alpha \left({\breve u}\right) \},$ of \v
N--frames are defined by the relations
\begin{equation}  \label{anhola}
\left[{\breve \delta} _\alpha ,{\breve \delta} _\beta \right] ={\breve \delta%
}_\alpha {\breve \delta}_\beta - {\breve \delta}_\beta {\breve \delta}%
_\alpha = {\breve w}_{\beta \gamma }^\alpha \left({\breve
u}\right) {\breve \delta}_\alpha .
\end{equation}

The \textbf{algebra of tensorial distinguished fields} $DT\left({\breve {%
\mathcal{E}}}\right) $ (d--fields, d--tensors, d--objects) on ${\breve {%
\mathcal{E}}}$ is introduced as the tensor algebra ${\breve
{\mathcal{T}}} =\{ {\breve {\mathcal{T}}}_{qs}^{pr}\}$ of the
cv--bundle
\[
{\breve {\mathcal{E}}}_{\left( d\right) }= \left(H{\breve {\mathcal{E}}}%
\oplus V{\breve {\mathcal{E}}},{\breve p}_d, {\breve
{\mathcal{E}}} \right),
\]
where ${\breve p}_d:\ H{\breve {\mathcal{E}}}\oplus V{\breve {\mathcal{E}}}%
\rightarrow {\breve {\mathcal{E}}}.$

An element ${\breve {\mathbf{t}}}\in {\breve
{\mathcal{T}}}_{qs}^{pr},$ d--tensor field of type $\left(
\begin{array}{cc}
p & r \\
q & s
\end{array}
\right) ,$ can be written in local form as
\begin{eqnarray}
{\breve {\mathbf{t}}} &=& {\breve t}%
_{j_1...j_qb_1...b_r}^{i_1...i_pa_1...a_r} \left({\breve
u}\right){\breve \delta}_{i_1}\otimes ... \otimes {\breve
\delta}_{i_p}\otimes {\breve
\partial}_{a_1}\otimes ...\otimes {\breve \partial}_{a_r}  \nonumber \\
& & \otimes {\breve d}^{j_1}\otimes ...\otimes {\breve d}^{j_q}\otimes {%
\breve \delta}^{b_1}...\otimes {\breve \delta}^{b_r}.  \nonumber
\end{eqnarray}

We shall respectively use the denotations $\mathcal{X}\left({\breve E}%
\right) $ (or $\mathcal{X} \left( M\right) ),\ \Lambda ^p\left({\breve {%
\mathcal{E}}}\right) $ or\newline $\left(\Lambda ^p\left(
M\right) \right) $ and $\mathcal{F}\left( {\breve E} \right)$ (or
$\mathcal{F}$ $\left( M\right) $) for the module of d--vector
fields on ${\breve {\mathcal{E}}}$ (or\newline $M$), the exterior
algebra of p--forms on ${\breve {\mathcal{E}}}$\ (or $M)$ and the
set of real functions on ${\breve {\mathcal{E}}}$ (or $M).$

\subsubsection{Anholonomic frames in hvc--bundles} \index{anholonomic}

The anholnomic frames adapted to a N--connection in hvc--bundle $\tilde{%
\mathcal{E}}$ are defined by the set of coefficients
(\ref{nconho}); having restricted the constructions to a vector
(covector) shell we obtain some generalizations of the formulas
for corresponding N(or \v {N})--connection
elongations of partial derivatives defined by (\ref{dder}) (or (\ref{ddercv}%
)) and (\ref{ddif}) (or (\ref{ddifcv})).

We introduce the adapted partial derivatives (anholonomic
N--frames, or
N--bases) in $\tilde{\mathcal{E}}$ by applying the coefficients (\ref{nconho}%
)
\[
{\tilde{\delta}}_\alpha =\frac{{\tilde{\delta}}}{\delta \tilde{u}^\alpha }%
=\left( \delta _i,\delta _{a_1},\delta _{a_2},{\breve{\delta}}^{a_3},...,{%
\breve{\delta}}^{a_{z-1}},\partial _{a_z}\right) ,
\]
where 
\begin{eqnarray*}
&&\delta _i=\partial _i-N_i^{~a_1}\partial
_{a_1}-N_i^{~a_2}\partial
_{a_2}+N_{ia_3}{\breve{\partial}}^{a_3}-...+N_{ia_{z-1}}{\breve{\partial}}%
^{a_{z-1}}-N_i^{~a_z}\partial _{a_z}, \\
&&\delta _{a_1}=\partial _{a_1}-N_{a_1}^{~a_2}\partial _{a_2}+N_{a_1a_3}{%
\breve{\partial}}^{a_3}-...+N_{a_1a_{z-1}}{\breve{\partial}}%
^{a_{z-1}}-N_{a_1}^{~a_z}\partial _{a_z}, \\
&&\delta _{a_2}=\partial _{a_2}+N_{a_2a_3}{\breve{\partial}}%
^{a_3}-...+N_{a_2a_{z-1}}{\breve{\partial}}^{a_{z-1}}-N_{a_2}^{~a_z}\partial
_{a_z}, \\
&&{\breve{\delta}}^{a_3}={\tilde{\partial}}^{a_3}-N^{a_3a_4}\partial
_{a_4}-...+N_{~a_{z-1}}^{a_3}{\breve{\partial}}^{a_{z-1}}-N^{~a_3a_z}%
\partial _{a_z}, \\
&&................. \\
&&{\breve{\delta}}^{a_{z-1}}={\tilde{\partial}}^{a_{z-1}}-N^{~a_{z-1}a_z}%
\partial _{a_z}, \\
&&\partial _{a_z}=\partial /\partial y^{a_z}.
\end{eqnarray*}
These formulas can be written in the matrix form:
\begin{equation}
{\tilde{\delta}}_{_{\bullet }}=\widehat{\mathbf{N}}(u)\times {\tilde{\partial%
}}_{_{\bullet }}  \label{dderho}
\end{equation}
where
\begin{eqnarray}
{\tilde{\delta}}_{_{\bullet }} &=&\left(
\begin{array}{l}
\delta _i \\
\delta _{a_1} \\
\delta _{a_2} \\
{\breve{\delta}}^{a_3} \\
... \\
{\breve{\delta}}^{a_{z-1}} \\
\partial _{a_z}
\end{array}
\right) ,\quad {\tilde{\partial}}_{_{\bullet }}=\left(
\begin{array}{l}
\partial _i \\
\partial _{a_1} \\
\partial _{a_2} \\
{\tilde{\partial}}^{a_3} \\
... \\
{\tilde{\partial}}^{a_{z-1}} \\
\partial _{a_z}
\end{array}
\right) ,\quad  \label{rows}
\end{eqnarray}
and
\begin{eqnarray}
\widehat{\mathbf{N}} &=&\left(
\begin{array}{llllllll}
1 & -N_i^{~a_1} & -N_i^{~a_2} & N_{ia_3} & -N_i^{~a_4} & ... &
N_{ia_{z-1}}
& -N_i^{~a_z} \\
0 & 1 & -N_{a_1}^{~a_2} & N_{a_1a_3} & -N_{a_1}^{~a_4} & ... &
N_{a_1a_{z-1}}
& -N_{a_1}^{~a_z} \\
0 & 0 & 1 & N_{a_2a_3} & -N_{a_2}^{~a_4} & ... & N_{a_2a_{z-1}} &
-N_{a_2}^{~a_z} \\
0 & 0 & 0 & 1 & -N^{a_3a_4} & ... & N_{~a_{z-1}}^{a_3} & -N^{~a_3a_z} \\
... & ... & ... & ... & ... & ... & ... & ... \\
0 & 0 & 0 & 0 & 0 & ... & 1 & -N^{~a_{z-1}a_z} \\
0 & 0 & 0 & 0 & 0 & ... & 0 & 1
\end{array}
\right) .  \nonumber
\end{eqnarray}

The adapted differentials (anholonomic N--coframes, or N--cobases) in $%
\tilde{\mathcal{E}}$ are introduced in the symplest form by using
matrix formalism: The respective dual matrices to (\ref{rows})
\begin{eqnarray*}
{\tilde{\delta}}^{\bullet } &=&\{{\tilde{\delta}}^\alpha \}=\left(
\begin{array}{lllllll}
d^i & \delta ^{a_1} & \delta ^{a_2} & {\breve{\delta}}_{a_3} & ... & {\breve{%
\delta}}_{a_{z-1}} & \delta ^{a_z}
\end{array}
\right) , \\
{\tilde{d}}^{\bullet } &=&\{{\tilde{\partial}}^\alpha \}=\left(
\begin{array}{lllllll}
d^i & d^{a_1} & d^{a_2} & d_{a_3} & ... & {d}_{a_{z-1}} & d^{a_z}
\end{array}
\right)
\end{eqnarray*}
are related via a matrix relation
\begin{equation}
{\tilde{\delta}}^{\bullet }={\tilde{d}}^{\bullet
}\widehat{\mathbf{M}} \label{ddifho}
\end{equation}
which defines the formulas for anholonomic N--coframes. The matrix $\widehat{%
\mathbf{M}}$ from (\ref{ddifho}) is the inverse to
$\widehat{\mathbf{N}},$ i. e. satisfies the condition
\begin{equation}
\widehat{\mathbf{M}}\times \widehat{\mathbf{N}}=I. \label{invmatr}
\end{equation}

The\textbf{\ anholonomic coefficients, } $\widetilde{\mathbf{w}}=\{%
\widetilde{w}_{\beta \gamma }^\alpha \left( \widetilde{u}\right)
\},$ on hcv--bundle $\tilde{\mathcal{E}}$ are expressed via
coefficients of the matrix $\widehat{\mathbf{N}}$ and their
partial derivatives following the relations
\begin{equation}
\left[ \widetilde{\delta }_\alpha ,\widetilde{\delta }_\beta \right] =%
\widetilde{\delta }_\alpha \widetilde{\delta }_\beta -\widetilde{\delta }%
_\beta \widetilde{\delta }_\alpha =\widetilde{w}_{\beta \gamma
}^\alpha \left( \widetilde{u}\right) \widetilde{\delta }_\alpha
.  \label{anholho}
\end{equation}
We omit the explicit formulas on shells.

A d--tensor formalism can be also developed on the space $\tilde{\mathcal{E}}%
.$ In this case the indices have to be stipulated for every shell
separately, like for v--bunles or cv--bundles.

Let us consider some examples for particular cases of
hcv--bundles:

\subsubsection{Anholonomic frames in osculator bundle}
For the osculator bundle $T^2M=Osc^2M$ from subsection
(\ref{nconoscs}) the formulas (\ref{dderho}) and (\ref{ddifho})
are written respectively in the form
\[
{\tilde{\delta}}_\alpha =\left( \frac \delta {\delta x^i},\frac
\delta {\delta y_{(1)}^i},\frac \partial {\partial
y_{(2)}^i}\right) ,
\]
where
\begin{eqnarray*}
\frac \delta {\delta x^i} &=&\frac \partial {\partial
x^i}-N_{(1)i}^{\qquad j}\frac \partial {\partial
y_{(1)}^i}-N_{(2)i}^{\qquad j}\frac \partial
{\partial y_{(2)}^i}, \\
\frac \delta {\delta y_{(1)}^i} &=&\frac \partial {\partial
y_{(1)}^i}-N_{(2)i}^{\qquad j}\frac \partial {\partial y_{(2)}^j},
\end{eqnarray*}
and
\begin{equation}
{\tilde{\delta}}^\alpha =\left( dx^i,\delta y_{(1)}^i,\delta
y_{(2)}^i\right) ,  \label{ddifosc2}
\end{equation}
where
\begin{eqnarray*}
\delta y_{(1)}^i &=&dy_{(1)}^i+M_{(1)j}^idx^j, \\
\delta y_{(2)}^i
&=&dy_{(2)}^i+M_{(1)j}^idy_{(1)}^j+M_{(2)j}^idx^j,
\end{eqnarray*}
with the dual coefficients $M_{(1)j}^i$ and $M_{(2)j}^i$ (see (\ref{invmatr}%
)) expressed via primary coefficients $N_{(1)j}^i$ and
$N_{(2)j}^i$ as
\[
M_{(1)j}^i=N_{(1)j}^i,M_{(2)j}^i=N_{(2)j}^{\quad
i}+N_{(1)m}^{\quad i}N_{(1)j}^{\quad m}.
\]

\subsubsection{Anholonomic frames in dual osculator bundle}

\label{nconoscds1}

Following the definitions for dual osculator bundle
$(T^{*2}M,p^{*2},M)$ in subsection (\ref{nconoscds}) the formulas
(\ref{dderho}) and (\ref{ddifho}) are written respectively in the
form
\[
{\tilde{\delta}}_\alpha =\left( \frac \delta {\delta x^i},\frac
\delta {\delta y_{(1)}^i},\frac \partial {\partial
p_{(2)}i}\right) ,
\]
where
\begin{eqnarray*}
\frac \delta {\delta x^i} &=&\frac \partial {\partial
x^i}-N_{(1)i}^{\qquad j}\frac \partial {\partial
y_{(1)}^i}+N_{(2)ij}^{\quad }\frac \partial
{\partial p_{(2)}j}, \\
\frac \delta {\delta y_{(1)}^i} &=&\frac \partial {\partial
y_{(1)}^i}+N_{(2)ij}^{\qquad }\frac \partial {\partial p_{(2)j}},
\end{eqnarray*}
and
\begin{equation}
{\tilde{\delta}}^\alpha =\left( dx^i,\delta y_{(1)}^i,\delta
p_{(2)i}\right) ,  \label{ddifosc2d}
\end{equation}
where
\begin{eqnarray*}
\delta y_{(1)}^i &=&dy_{(1)}^i+N_{(1)j}^idx^j, \\
\delta p_{(2)i} &=&dp_{(2)i}-N_{(2)ij}dx^j,
\end{eqnarray*}
with the dual coefficients $M_{(1)j}^i$ and $M_{(2)j}^i$ (see (\ref{invmatr}%
)) were expressed via $N_{(1)j}^i$ and $N_{(2)j}^i$ like in Ref.
\cite{mhss}.

\section{Distinguished connections and metrics}

In general, distinguished objects (d--objects) \index{d--objects }
on a v--bundle $\mathcal{E}$ (or cv--bundle
${\breve{\mathcal{E}}}$) are introduced as geometric objects with
various group and coordinate transforms coordinated with the
N--connection structure on $\mathcal{E}$ (or
${\breve{\mathcal{E}}}$). For example, a distinguished connection
(in brief, \textbf{d--connection)} \index{d--connection}
 $D$ on $\mathcal{E}$ (or
${\breve{\mathcal{E}}}$) is defined as a linear connection $D$ on
$E$ (or $\breve{E}$) conserving under a parallelism the
global decomposition (\ref{whitney}) 
(or (\ref{whitneyc})) into horizontal and vertical subbundles of $T\mathcal{E%
}$ (or $T{\breve{\mathcal{E}}}).$ A covariant derivation
associated to a d--connection becomes d--covariant. We shall give
necessary formulas for cv--bundles in round backets.

\subsection{D--connections}

\subsubsection{D--connections in v--bundles (cv--bundles)}

A N--connection in a v--bundle $\mathcal{E}$ (cv--bundle ${\breve{\mathcal{E}%
}}$) induces a corresponding decomposition of d--tensors into
sums of horizontal and vertical parts, for example, for every
d--vector $X\in
\mathcal{X\left( E\right) }$ (${\breve X}\in \mathcal{X}\left( {\breve{%
\mathcal{E}}}\right) $ ) and 1--form $A\in \Lambda ^1\left( \mathcal{E}%
\right) $ ($\breve{A}\in \Lambda ^1\left(
{\breve{\mathcal{E}}}\right) $) we have respectively
\begin{eqnarray}
X &=&hX+vX\mathbf{\ \quad }\mbox{and \quad }A=hA+vA,  \label{vdecomp} \\
(\breve{X} &=&h\breve{X}+vX\mathbf{\ \quad }\mbox{and \quad }\breve{A}=h%
\breve{A}+v\breve{A})  \nonumber
\end{eqnarray}
where
\[
hX = X^i\delta _i,vX=X^a\partial _a \ (h\breve{X} = \breve{X}^i{\tilde \delta%
}_i,v\breve{X}=\breve{X}_a{\breve \partial}^a)
\]
and
\[
hA =A_i\delta ^i,vA=A_ad^a \ (h\breve{A} = \breve{A}_i{\breve \delta}^i,v%
\breve{A}= \breve{A}^a{\breve d}_a).
\]

In consequence, we can associate to every d--covariant derivation
along the
d--vector (\ref{vdecomp}), 
$D_X=X\circ D$ ($D_{\breve{X}}=\breve{X}\circ D$) two new
operators of h- and v--covariant derivations
\begin{eqnarray*}
D_X^{(h)}Y &=&D_{hX}Y\quad \mbox{ and \quad }D_X^{\left( v\right)
}Y=D_{vX}Y,\quad \forall Y{\ \in }\mathcal{X\left( E\right) } \\
(D_{\breve{X}}^{(h)}\breve{Y} &=&D_{h\breve{X}}\breve{Y}\quad
\mbox{ and
\quad }D_{\breve{X}}^{\left( v\right) }\breve{Y}=D_{v\breve{X}}\breve{Y}%
,\quad \forall \breve{Y}{\ \in }\mathcal{X}\left( {\breve{\mathcal{E}}}%
\right) )
\end{eqnarray*}
for which the following conditions hold:
\begin{eqnarray}
D_XY &=&D_X^{(h)}Y{\ +}D_X^{(v)}Y  \label{dcovpr} \\
(D_{\breve{X}}\breve{Y} &=&D_{\breve{X}}^{(h)}\breve{Y}{\ +}D_{\breve{X}%
}^{(v)}\breve{Y}),  \nonumber
\end{eqnarray}
where
\begin{eqnarray*}
D_X^{(h)}f &=&(hX\mathbf{)}f\mbox{ \quad and\quad }D_X^{(v)}f=(vX\mathbf{)}%
f,\quad X,Y\mathbf{\in }\mathcal{X\left( E\right) },f\in
\mathcal{F}\left(
M\right) \\
({\breve{D}}_{\breve{X}}^{(h)}f &=&(h\breve{X}\mathbf{)}f%
\mbox{ \quad
and\quad }{\breve{D}}_{\breve{X}}^{(v)}f=(v\breve{X}\mathbf{)}f,\quad \breve{%
X},\breve{Y}{\in }\mathcal{X}\left( {\breve{\mathcal{E}}}\right)
,f\in \mathcal{F}\left( M\right) ).
\end{eqnarray*}

The components $\Gamma _{\beta \gamma }^\alpha $ (
${\breve{\Gamma}}_{\beta
\gamma }^\alpha )$of a d--connection ${\breve{D}}_\alpha =({\breve{\delta}}%
_\alpha \circ D),$ locally adapted to the N---connection
structure with respect to the frames (\ref{dder}) and
(\ref{ddif}) ((\ref{ddercv}) and (\ref {ddifcv})), are defined by
the equations
\[
D_\alpha \delta _\beta =\Gamma _{\alpha \beta }^\gamma \delta _\gamma ~({%
\breve{D}}_\alpha {\breve{\delta}}_\beta
={\breve{\Gamma}}_{\alpha \beta }^\gamma {\breve{\delta}}_\gamma
~),
\]
from which one immediately follows
\begin{equation}
\Gamma _{\alpha \beta }^\gamma \left( u\right) =\left( D_\alpha
\delta _\beta \right) \circ \delta ^\gamma \quad
~({\breve{\Gamma}}_{\alpha \beta
}^\gamma \left( {\breve{u}}\right) =\left( {\breve{D}}_\alpha {\breve{\delta}%
}_\beta \right) \circ {\breve{\delta}}^\gamma ).  \label{gamma}
\end{equation}

The coefficients of operators of h- and v--covariant derivations,
\begin{eqnarray*}
D_k^{(h)} &=&\{L_{jk}^i,L^a_{bk\;}\}\mbox{ and }D_c^{(v)}=%
\{C_{jk}^i,C_{bc}^a\} \\
({\breve{D}}_k^{(h)} &=&\{{\breve{L}}_{jk}^i,{\breve{L}}_{a k}^{~b}\}%
\mbox{ and }{\breve{D}}^{(v)c}=\{{\breve C}_{~j}^{i~c},{\breve
C}_a^{~bc}\})
\end{eqnarray*}
(see (\ref{dcovpr})), are introduced as corresponding h- and
v--paramet\-ri\-za\-ti\-ons of (\ref{gamma})
\begin{eqnarray}
L_{jk}^i &=&\left( D_k\delta _j\right) \circ d^i,\quad
L_{bk}^a=\left(
D_k\partial _b\right) \circ \delta ^a  \label{hgamma} \\
({\breve{L}}_{jk}^i &=&\left(
{\breve{D}}_k{\breve{\delta}}_j\right) \circ
d^i,\quad {\breve{L}}_{a k}^{~b}=\left( {\breve{D}}_k{\breve \partial}%
^b\right) \circ {\breve{\delta}}_a)  \nonumber
\end{eqnarray}
and
\begin{eqnarray}
C_{jc}^i &=&\left( D_c\delta _j\right) \circ d^i,\quad
C_{bc}^a=\left(
D_c\partial _b\right) \circ \delta ^a  \label{vgamma} \\
({\breve C}_{~j}^{i~c} &=&\left(
{\breve{D}}^c{\breve{\delta}}_j\right)
\circ d^i,\quad {\breve C}_a^{~bc}=\left( {\breve{D}}^c{\breve \partial}%
^b\right) \circ {\breve{\delta}}_a).  \nonumber
\end{eqnarray}

A set of components (\ref{hgamma}) and (\ref{vgamma}) \
\[
\Gamma _{\alpha \beta }^\gamma
=[L_{jk}^i,L_{bk}^a,C_{jc}^i,C_{bc}^a]~\left(
{\breve{\Gamma}}_{\alpha \beta }^\gamma =[{\breve{L}}_{jk}^i,{\breve{L}}%
_{ak}^{~b},{\breve{C}}_{~j}^{i~c},{\breve{C}}_a^{~bc}]\right)
\]
completely defines the local action of a d---connection $D$ in
$\mathcal{E}$ (${\breve{D}}$ in ${\breve{\mathcal{E}}).}$

For instance, having taken on $\mathcal{E}$
(${\breve{\mathcal{E}})}$ a d---tensor field of type $\left(
\begin{array}{cc}
1 & 1 \\
1 & 1
\end{array}
\right) ,$
\begin{eqnarray*}
\mathbf{t} &=&t_{jb}^{ia}\delta _i\otimes \partial _a\otimes
d^j\otimes
\delta ^b, \\
{{\tilde{\mathbf{t}}}} &=&{\breve{t}}_{ja}^{ib}{\breve{\delta}}_i\otimes {%
\breve{\partial}}^a\otimes d^j\otimes {\breve{\delta}}_b,
\end{eqnarray*}
and a d--vector $\mathbf{X}$ ($\mathbf{{\breve{X}}})$ we obtain
\begin{eqnarray*}
D_X\mathbf{t} &=&D_X^{(h)}\mathbf{t+}D_X^{(v)}\mathbf{t=}\left( X^k{\breve{t}%
}_{jb|k}^{ia}+X^ct_{jb\perp c}^{ia}\right) \delta _i\otimes
\partial
_a\otimes d^j\otimes \delta ^b, \\
({\breve{D}}_{\breve{X}}{{\tilde{\mathbf{t}}}} &=&\breve{D}_{\breve{X}}^{(h)}%
{\tilde{\mathbf{t}}}+\breve{D}_{\breve{X}}^{(v)}{\tilde{\mathbf{t}}}=\left(
\breve{X}^k{\breve{t}}_{ja|k}^{ib}+\breve{X}_c{\breve{t}}_{ja}^{ib\perp
c}\right) {\breve{\delta}}_i\otimes {\breve{\partial}}^a\otimes d^j\otimes {%
\breve{\delta}}_b)
\end{eqnarray*}
where the h--covariant derivative is written
\begin{eqnarray*}
t_{jb|k}^{ia} &=&\delta
_kt_{jb}^{ia}+L_{hk}^it_{jb}^{ha}+L_{ck}^at_{jb}^{ic}-L_{jk}^ht_{hb}^{ia}-L_{bk}^ct_{jc}^{ia}
\\
({\breve{t}}_{ja|k}^{ib} &=&{\breve{\delta}}_k{\breve{t}}_{ja}^{ib}+{\breve{L%
}}_{hk}^i{\breve{t}}_{ja}^{hb}+{\breve{L}}_{ck}^{~b}{\breve{t}}_{ja}^{ic}-{%
\breve{L}}_{jk}^h{\breve{t}}_{ha}^{ib}-{\breve{L}}_{ck}^{~b}{\breve{t}}%
_{ja}^{ic})
\end{eqnarray*}
and the v-covariant derivative is written
\begin{eqnarray}
t_{jb\perp c}^{ia} &=&\partial
_ct_{jb}^{ia}+C_{hc}^it_{jb}^{ha}+C_{dc}^at_{jb}^{id}-C_{jc}^ht_{hb}^{ia}-C_{bc}^dt_{jd}^{ia}
\label{covder1} \\
({\breve{t}}_{ja}^{ib\perp c} &=&{\breve{\partial}}^c{\breve{t}}_{ja}^{ib}+{%
\breve{C}}_{~j}^{i~c}{\breve{t}}_{ja}^{hb}+{\breve{C}}_a^{~dc}{\breve{t}}%
_{jd}^{ib}-{\breve{C}}_{~j}^{i~c}{\breve{t}}_{ha}^{ib}-{\breve{C}}_d^{~bc}{%
\breve{t}}_{ja}^{id}).  \label{covder2}
\end{eqnarray}
For a scalar function $f\in \mathcal{F}\left(\mathcal{E}\right) $
( $f\in \mathcal{F}\left( {\breve{\mathcal{E}}}\right) $) we have
\begin{eqnarray*}
D_k^{(h)} &=&\frac{\delta f}{\delta x^k}=\frac{\partial f}{\partial x^k}%
-N_k^a\frac{\partial f}{\partial y^a}\mbox{ and }D_c^{(v)}f=\frac{\partial f%
}{\partial y^c} \\
(\breve{D}_k^{(h)} &=&\frac{{\breve{\delta}}f}{\delta x^k}=\frac{\partial f}{%
\partial x^k}+N_{ka}\frac{\partial f}{\partial p_a}\mbox{ and }\breve{D}%
^{(v)c}f=\frac{\partial f}{\partial p_c}).
\end{eqnarray*}

\subsubsection{D--connections in hvc--bundles} \index{d--connection}

The theory of connections in higher order anisotropic vector
superbundles and vector bundles was elaborated in Refs.
\cite{vstr2,vhsp,vbook}. Here we re--formulate that formalism for
the case when some shells of higher order anisotropy could be
covector spaces by stating the general rules of covariant
derivation compatible with the N--connection structure in
hvc--bundle $\tilde{\mathcal{E}}$ and omit details and combersome
formulas.

For a hvc--bundle of type ${\tilde{\mathcal{E}}}={\tilde{\mathcal{E}}}%
[v(1),v(2),cv(3),...,cv(z-1),v(z)]$ a d--connection ${\tilde
\Gamma}_{\alpha \beta }^\gamma $ has the next shell decomposition
of components (on induction being on the $p$-th shell, considered
as the base space, which in this case a hvc--bundle, we introduce
in a usual manner, like a vector or covector fibre, the
$(p+1)$-th shell)
\begin{eqnarray*}
{\tilde \Gamma}_{\alpha \beta }^\gamma &=&\{\Gamma _{\alpha
_1\beta _1}^{\gamma
_1}=[L_{j_1k_1}^{i_1},L_{b_1k_1}^{a_1},C_{j_1c_1}^{i_1},C_{b_1c_1}^{a_1}], \\
& & \Gamma _{\alpha _2\beta _2}^{\gamma _2} =
[L_{j_2k_2}^{i_2},L_{b_2k_2}^{a_2},C_{j_2c_2}^{i_2},C_{b_2c_2}^{a_2}], \\
& & {\breve{\Gamma}}_{\alpha _3\beta _3}^{\gamma _3} =[{\breve{L}}%
_{j_3k_3}^{i_3},{\breve{L}}_{a_3k_3}^{~b_3},{\breve{C}}_{~j_3}^{i_3~c_3},{%
\breve{C}}_{a_3}^{~b_3c_3}], \\
&& ...................................., \\
& & {\breve{\Gamma}}_{\alpha _{z-1}\beta _{z-1}}^{\gamma _{z-1}}= [{\breve{L}%
}_{j_{z-1}k_{z-1}}^{i_{z-1}},{\breve{L}}_{a_{z-1}k_{z-1}}^{~b_{z-1}},{\breve{%
C}}_{~j_{z-1}}^{i_{z-1}~c_{z-1}},{\breve{C}}_{a_{z-1}}^{~b_{z-1}c_{z-1}}], \\
& & \Gamma _{\alpha _z\beta _z}^{\gamma _z}
=[L_{j_zk_z}^{i_z},L_{b_zk_z}^{a_z},C_{j_zc_z}^{i_z},C_{b_zc_z}^{a_z}]\}.
\end{eqnarray*}
These coefficients determine the rules of a covariant derivation
$\tilde D$ on ${\tilde {\mathcal{E}}}.$

For example, let us consider a d--tensor ${\tilde {\mathbf{t}}}$
of type
\[
\left(
\begin{array}{llllll}
1 & 1_1 & 1_2 & {\breve 1}_3 & ... & 1_z \\
1 & 1_1 & 1_2 & {\breve 1}_3 & ... & 1_z
\end{array}
\right)
\]
with corresponding tensor product of components of anholonomic N--frames (%
\ref{dderho}) and (\ref{ddifho})
\begin{eqnarray*}
{\tilde {\mathbf{t}}} &=&{\tilde t}_{jb_1b_2{\breve a}_3...{\breve a}%
_{z-1}b_z}^{ia_1a_2{\breve b}_3...{\breve b}_{z-1}a_z}\delta
_i\otimes
\partial _{a_1}\otimes d^j\otimes \delta ^{b_1}\otimes \partial
_{a_2}\otimes \delta ^{b_2}\otimes {\breve{\partial}}^{a_3}\otimes {\breve{%
\delta}}_{b_3}, \\
&&{...}\otimes {\breve{\partial}}^{a_{z-1}}\otimes {\breve{\delta}}%
_{bz-1}\otimes \partial _{a_z}\otimes \delta ^{b_z}.
\end{eqnarray*}
The d--covariant derivation $\tilde D$ of ${\tilde {\mathbf{t}}}$
is to be performed separately for every shall according the rule
(\ref{covder1}) if a shell is defined by a vector subspace, or
according the rule (\ref{covder2}) if the shell is defined by a
covector subspace.

\subsection{Metric structure}

\subsubsection{D--metrics in v--bundles}

We define a \textbf{metric structure }$\mathbf{G\ }$
\index{metric} in the total space $E$ of a v--bundle
$\mathcal{E=}$ $\left( E,p,M\right) $ over a connected and
paracompact base $M$ as a symmetric covariant tensor field of
type $\left( 0,2\right) $,
\[
\mathbf{G} = G_{\alpha \beta } du^{\alpha}\otimes du^\beta
\]
being non degenerate and of constant signature on $E.$

Nonlinear connection $\mathbf{N}$ and metric $\mathbf{G}$ structures on $%
\mathcal{E}$ are mutually compatible it there are satisfied the
conditions:
\begin{equation}  \label{comp}
\mathbf{G}\left( \delta _i,\partial _a\right) =0,\mbox{or equivalently, }%
G_{ia}\left( u\right) -N_i^b\left( u\right) h_{ab}\left( u\right)
=0,
\end{equation}
where $h_{ab}=\mathbf{G}\left( \partial _a,\partial _b\right) $ and $G_{ia}=%
\mathbf{G}\left( \partial _i,\partial _a\right),$ which gives
\begin{equation}  \label{ncon1}
N_i^b\left( u\right) = h^{ab}\left( u\right) G_{ia}\left( u\right)
\end{equation}
( the matrix $h^{ab}$ is inverse to $h_{ab}).$ In consequence one
obtains the following decomposition of metric:
\begin{equation}  \label{metrdec}
\mathbf{G}(X,Y)\mathbf{=hG}(X,Y)+\mathbf{vG}(X,Y),
\end{equation}
where the d--tensor $\mathbf{hG}(X,Y)\mathbf{= G}(hX,hY)$ is of
type $\left(
\begin{array}{cc}
0 & 0 \\
2 & 0
\end{array}
\right) $ and the d--tensor $\mathbf{vG}(X,Y) =
\mathbf{G}(vX,vY)$ is of type $\left(
\begin{array}{cc}
0 & 0 \\
0 & 2
\end{array}
\right) .$ With respect to anholonomic basis (\ref{dder}) the
d--metric (\ref {metrdec}) is written
\begin{equation}  \label{dmetric}
\mathbf{G}=g_{\alpha \beta }\left( u\right) \delta ^\alpha
\otimes \delta ^\beta =g_{ij}\left( u\right) d^i\otimes
d^j+h_{ab}\left( u\right) \delta ^a\otimes \delta ^b,
\end{equation}
where $g_{ij}=\mathbf{G}\left( \delta _i,\delta _j\right) .$

A metric structure of type (\ref{metrdec}) (equivalently, of type
(\ref {dmetric})) or a metric on $E$ with components satisfying
constraints (\ref {comp}), (equivalently (\ref{ncon1})) defines
an adapted to the given N--connection inner (d--scalar) product
on the tangent bundle $\mathcal{TE}$.

We shall say that a d--connection $\widehat{D}_X$ is compatible
with the d-scalar product on $\mathcal{TE\ }$ (i. e. it is a
standard d--connection) if
\[
\widehat{D}_X\left( \mathbf{X\cdot Y}\right) =\left( \widehat{D}_X\mathbf{Y}%
\right) \cdot \mathbf{Z+Y\cdot }\left(
\widehat{D}_X\mathbf{Z}\right)
,\forall \mathbf{X,Y,Z}\mathbf{\in }\mathcal{X\left( E\right) }.
\]
An arbitrary d--connection $D_X$ differs from the standard one $\widehat{D}%
_X $ by an operator $\widehat{P}_X\left( u\right) =\{X^\alpha \widehat{P}%
_{\alpha \beta }^\gamma \left( u\right) \},$ called the
deformation d-tensor
with respect to $\widehat{D}_X,$ which is just a d-linear transform of $%
\mathcal{E}_u,$\ $\forall \ u\in \mathcal{E}$. The explicit form of $%
\widehat{P}_X $ can be found by using the corresponding axiom
defining linear connections \cite{lue}
\[
\left( D_X-\widehat{D}_X\right) fZ=f\left( D_X-\widehat{D}_X\right) Z\mathbf{%
,}
\]
written with respect to N--elongated bases (\ref{dder}) and
(\ref{ddif}). From the last expression we obtain
\[
\widehat{P}_X\left( u\right) =\left[ (D_X-\widehat{D}_X)\delta
_\alpha
\left( u\right) \right] \delta ^\alpha \left( u\right) ,
\]
therefore
\begin{equation}
D_XZ\mathbf{\ }=\widehat{D}_XZ\mathbf{\ +}\widehat{P}_XZ.
\label{deft}
\end{equation}

A d--connection $D_X$ is \textbf{metric} (or \textbf{compatible }
with metric $\mathbf{G}$) on $\mathcal{E}$ if
\[
D_X\mathbf{G} =0,\forall X\mathbf{\in }\mathcal{X\left( E\right)
}.
\]
With respect to anholonomic frames these conditions are written
\begin{equation}  \label{comatib}
D_\alpha g_{\beta\gamma}=0,
\end{equation}
where by $g_{\beta\gamma}$ we denote the coefficients in the
block form (\ref {dmetric}).

\subsubsection{D--metrics in cv-- and hvc--bundles}

The presented considerations on self--consisten definition of
N--connection, d--connection and metric structures in v--bundles
can reformulated in a similar fashion for another types of
anisotropic space--times, on cv--bundles and on shells of
hvc--bundles. For symplicity, we give here only the anagolous
formulas for the metric d--tensor (\ref{dmetric}):

\begin{itemize}
\item  On cv--bundle ${\breve{\mathcal{E}}}$ we write
\begin{equation}
{\breve{\mathbf{G}}}={\breve{g}}_{\alpha \beta }\left( {\breve{u}}\right) {%
\breve{\delta}}^\alpha \otimes {\breve{\delta}}^\beta ={\breve{g}}%
_{ij}\left( {\breve{u}}\right) d^i\otimes d^j+{\breve{h}}^{ab}\left( {\breve{%
u}}\right) {\breve{\delta}}_a\otimes {\breve{\delta}}_b,
\label{dmetricvc}
\end{equation}
where ${\breve{g}}_{ij}={\breve{\mathbf{G}}}\left( {\breve{\delta}}_i,{%
\breve{\delta}}_j\right) $ and ${\breve{h}}^{ab}={\breve{\mathbf{G}}}\left( {%
\breve{\partial}}^a,{\breve{\partial}}^b\right) $ and the
N--coframes are given by formulas (\ref{ddifcv}).

For simplicity, we shall consider that the metricity conditions
are satisfied, ${\breve{D}}_\gamma {\breve{g}}_{\alpha \beta}=0.$

\item  On hvc--bundle ${\tilde{\mathcal{E}}}$ we write
\begin{eqnarray}
{\tilde{\mathbf{G}}} &=&{\tilde{g}}_{\alpha \beta }\left(
{\tilde{u}}\right)
{\tilde{\delta}}^\alpha \otimes {\tilde{\delta}}^\beta ={\tilde{g}}%
_{ij}\left( {\tilde{u}}\right) d^i\otimes d^j+{\tilde{h}}_{a_1b_1}\left( {%
\tilde{u}}\right) {\delta }^{a_1}\otimes {\delta }^{b_1}
\label{dmetrichcv}
\\
&&+{\tilde{h}}_{a_2b_2}\left( {\tilde{u}}\right) {\delta }^{a_2}\otimes {%
\delta }^{b_2}+{\tilde{h}}^{a_3b_3}\left( {\tilde{u}}\right) {\breve{\delta}}%
_{a_3}\otimes {\breve{\delta}}_{b_3}+...  \nonumber \\
&&+{\tilde{h}}^{a_{z-1}b_{z-1}}\left( {\tilde{u}}\right) {\breve{\delta}}%
_{a_{z-1}}\otimes {\breve{\delta}}_{b_{z-1}}+{\tilde{h}}_{a_zb_z}\left( {%
\tilde{u}}\right) {\delta }^{a_z}\otimes {\delta }^{b_z},
\nonumber
\end{eqnarray}
where ${\tilde{g}}_{ij}={\tilde{\mathbf{G}}}\left( {\tilde{\delta}}_i,{%
\tilde{\delta}}_j\right) $ and ${\tilde{h}}_{a_1b_1}={\tilde{\mathbf{G}}}%
\left( \partial _{a_1},\partial _{b_1}\right) ,$ ${\tilde{h}}_{a_2b_2}={%
\tilde{\mathbf{G}}}\left( \partial _{a_2},\partial _{b_2}\right) ,$ ${\tilde{%
h}}^{a_3b_3}={\tilde{\mathbf{G}}}\left( {\breve{\partial}}^{a_3},{\breve{%
\partial}}^{b_3}\right) ,....$ and the N--coframes are given by formulas (%
\ref{ddifho}).

The metricity conditions are ${\tilde{D}}_\gamma
{\tilde{g}}_{\alpha \beta }=0.$

\item  On osculator bundle $T^2M=Osc^2M$ we have a particular case of (\ref
{dmetrichcv}) when
\begin{eqnarray}  \label{dmetrichosc2}
{\tilde{\mathbf{G}}}&=&{\tilde{g}}_{\alpha \beta }\left( {\tilde{u}}\right) {%
\tilde{\delta}}^\alpha \otimes {\tilde{\delta}}^\beta \\
&=& {\tilde{g}}_{ij}\left( {\tilde{u}}\right) d^i\otimes d^j+{\tilde{h}}%
_{ij}\left( {\tilde{u}}\right) {\delta y}_{(1)}^i\otimes {\delta y}_{(1)}^i+{%
\tilde{h}}_{ij}\left( {\tilde{u}}\right) {\delta y}_{(2)}^i\otimes {\delta y}%
_{(2)}^i  \nonumber
\end{eqnarray}
where the N--coframes are given by (\ref{ddifosc2}).

\item  On dual osculator bundle $\left( T^{*2}M,p^{*2},M\right) $ we have
another particular case of (\ref{dmetrichcv}) when
\begin{eqnarray}  \label{dmetrichosc2d}
{\tilde{\mathbf{G}}} &=& {\tilde{g}}_{\alpha \beta }\left( {\tilde{u}}%
\right) {\tilde{\delta}}^\alpha \otimes {\tilde{\delta}}^\beta \\
&=& {\tilde{g}}_{ij}\left( {\tilde{u}}\right) d^i\otimes d^j+{\tilde{h}}%
_{ij}\left( {\tilde{u}}\right) {\delta y}_{(1)}^i\otimes {\delta y}_{(1)}^i+{%
\tilde{h}}^{ij}\left( {\tilde{u}}\right) {\delta p}_i^{(2)}\otimes {\delta p}%
_i^{(2)}  \nonumber
\end{eqnarray}
where the N--coframes are given by (\ref{ddifosc2d}).
\end{itemize}


\subsection{Some remarkable d--connections}

We emphasize that the geometry of connections in a v--bundle
$\mathcal{E}$ is very reach. If a triple of fundamental geometric
objects $\left( N_i^a\left( u\right) ,\Gamma _{\beta \gamma
}^\alpha \left( u\right) ,g_{\alpha \beta }\left( u\right)
\right) $ is fixed on $\mathcal{E}$, a multi--connection
structure (with corresponding different rules of covariant
derivation, which are, or not, mutually compatible and with the
same, or not, induced d--scalar products in $\mathcal{TE)}$ is
defined on this v--bundle. We can give a priority to a connection
structure following some physical arguments, like the reduction
to the Christoffel symbols in the holonomic case, mutual
compatibility between metric and N--connection and d--connection
structures and so on.

In this subsection we enumerate some of the connections and
covariant derivations in v--bundle $\mathcal{E}$, cv--bundle
${\breve{\mathcal{E}}}$ and in some hvc--bundles which can
present interest in investigation of locally anisotropic
gravitational and matter field interactions :

\begin{enumerate}
\item  Every N--connection in $\mathcal{E}$ with coefficients $N_i^a\left(
x,y\right) $ being differentiable on y--variables, induces a
structure of linear connection $N_{\beta \gamma }^\alpha ,$ where
\begin{equation}
N_{bi}^a=\frac{\partial N_i^a}{\partial y^b}\mbox{ and
}N_{bc}^a\left( x,y\right) =0.  \label{nlinearized}
\end{equation}
For some $Y\left( u\right) =Y^i\left( u\right) \partial
_i+Y^a\left( u\right) \partial _a$ and $B\left( u\right)
=B^a\left( u\right) \partial _a$ one introduces a covariant
derivation as
\[
D_Y^{(\widetilde{N})}B=\left[ Y^i\left( \frac{\partial B^a}{\partial x^i}%
+N_{bi}^aB^b\right) +Y^b\frac{\partial B^a}{\partial y^b}\right]
\frac
\partial {\partial y^a}.
\]

\item  The d--connection of Berwald type \cite{berw} on v--bundle $\mathcal{E%
}$ (cv--bundle ${\breve{\mathcal{E}})}$
\begin{eqnarray}
\Gamma _{\beta \gamma }^{(B)\alpha } &=&\left( L_{jk}^i,\frac{\partial N_k^a%
}{\partial y^b},0,C_{bc}^a\right) ,  \label{berwald} \\
({\breve{\Gamma}}_{\beta \gamma }^{(B)\alpha } &=&\left( \breve{L}_{jk}^i,-%
\frac{\partial \breve{N}_{ka}}{\partial
p_b},0,{\breve{C}}_a^{~bc}\right) ) \nonumber
\end{eqnarray}
where 
\begin{eqnarray}
L_{.jk}^i\left( x,y\right) &=&\frac 12g^{ir}\left( \frac{\delta g_{jk}}{%
\delta x^k}+\frac{\delta g_{kr}}{\delta x^j}-\frac{\delta g_{jk}}{\delta x^r}%
\right) ,  \label{lccoef} \\
C_{.bc}^a\left( x,y\right) &=&\frac 12h^{ad}\left( \frac{\partial h_{bd}}{%
\partial y^c}+\frac{\partial h_{cd}}{\partial y^b}-\frac{\partial h_{bc}}{%
\partial y^d}\right)  \nonumber \\
(\breve{L}_{.jk}^i\left( x,p\right) &=&\frac 12\breve{g}^{ir}\left( \frac{{%
\breve{\delta}}\breve{g}_{jk}}{\delta x^k}+\frac{{\breve{\delta}}\breve{g}%
_{kr}}{\delta x^j}-\frac{{\breve{\delta}}\breve{g}_{jk}}{\delta
x^r}\right) ,
\nonumber \\
{\breve{C}}_a^{~bc}\left( x,p\right) &=&\frac 12\breve{h}_{ad}\left( \frac{%
\partial \breve{h}^{bd}}{\partial p_c}+\frac{\partial \breve{h}^{cd}}{%
\partial p_b}-\frac{\partial \breve{h}^{bc}}{\partial p_d}\right) ),
\nonumber
\end{eqnarray}

which is hv---metric, i.e. there are satisfied the conditions $%
D_k^{(B)}g_{ij}=0$ and $D_c^{(B)}h_{ab}=0$ ($\breve{D}_k^{(B)}\breve{g}%
_{ij}=0$ and $\breve{D}^{(B)c}\breve{h}^{ab}=0).$

\item  The canonical d--connection $\mathbf{\Gamma ^{(c)}}$ (or $\mathbf{%
\breve{\Gamma}^{(c)})}$ on a v--bundle (or cv--bundle) is
associated to a
metric $\mathbf{G}$ (or $\mathbf{\breve{G})}$ of type (\ref{dmetric}) (or (%
\ref{dmetricvc})),
\[
\Gamma _{\beta \gamma }^{(c)\alpha
}=[L_{jk}^{(c)i},L_{bk}^{(c)a},C_{jc}^{(c)i},C_{bc}^{(c)a}]~(\breve{\Gamma}%
_{\beta \gamma }^{(c)\alpha }=[\breve{L}_{jk}^{(c)i},\breve{L}%
_{~a~.k}^{(c).b},\breve{C}_{~j}^{(c)i\ c},{\breve{C}}_a^{(c)~bc}])
\]
with coefficients
\begin{eqnarray}
L_{jk}^{(c)i} &=&L_{.jk}^i,C_{bc}^{(c)a}=C_{.bc}^a~(\breve{L}_{jk}^{(c)i}=%
\breve{L}_{.jk}^i,{\breve{C}}_a^{(c)~bc}={\breve{C}}_a^{~bc}),%
\mbox{ (see
(\ref{lccoef})}  \nonumber \\
L_{bi}^{(c)a} &=&\frac{\partial N_i^a}{\partial y^b}+\frac
12h^{ac}\left(
\frac{\delta h_{bc}}{\delta x^i}-\frac{\partial N_i^d}{\partial y^b}h_{dc}-%
\frac{\partial N_i^d}{\partial y^c}h_{db}\right)  \nonumber \\
~(\breve{L}_{~a~.i}^{(c).b} &=&-\frac{\partial {\breve{N}}_i^a}{\partial p_b}%
+\frac 12\breve{h}_{ac}\left(
\frac{{\breve{\delta}}\breve{h}^{bc}}{\delta
x^i}+\frac{\partial {\breve{N}}_{id}}{\partial p_b}\breve{h}^{dc}+\frac{%
\partial {\breve{N}}_{id}}{\partial p_c}\breve{h}^{db}\right) ),  \nonumber
\\
~C_{jc}^{(c)i} &=&\frac 12g^{ik}\frac{\partial g_{jk}}{\partial y^c}~(\breve{%
C}_{~j}^{(c)i\ c}=\frac 12\breve{g}^{ik}\frac{\partial \breve{g}_{jk}}{%
\partial p_c}).  \label{inters}
\end{eqnarray}
This is a metric d--connection which satisfies conditions
\begin{eqnarray*}
D_k^{(c)}g_{ij} &=&0,D_c^{(c)}g_{ij}=0,D_k^{(c)}h_{ab}=0,D_c^{(c)}h_{ab}=0 \\
(\breve{D}_k^{(c)}\breve{g}_{jk} &=&0,\breve{D}^{(c)c}\breve{g}_{jk}=0,%
\breve{D}_k^{(c)}\breve{h}^{bc}=0,\breve{D}^{(c)c}\breve{h}^{ab}=0).
\end{eqnarray*}
In physical applications we shall use the canonical connection
and for
symplicity we chall omit the index $(c).$ The coefficients (\ref{inters})$\,$%
are to be extended to higher order if we are dealing with
derivations of geometrical objects with ''shell'' indices. In
this case the fiber indices are to be stipulated for every type
of shell into consideration.

\item  We can consider the N--adapted Christoffel d--symbols
\begin{equation}
\widetilde{\Gamma }_{\beta \gamma }^\alpha =\frac 12g^{\alpha
\tau }\left( \delta _\gamma g_{\tau \beta }+\delta _\beta g_{\tau
\gamma }-\delta g_{\beta \gamma }\right) ,  \label{dchrist}
\end{equation}
which have the components of d--connection $\widetilde{\Gamma
}_{\beta
\gamma }^\alpha =\left( L_{jk}^i,0,0,C_{bc}^a\right) ,$ with $L_{jk}^i$ and $%
C_{bc}^a$ as in (\ref{lccoef}) 
if $g_{\alpha \beta }$ is taken in the form (\ref{dmetric}). 
\end{enumerate}

Arbitrary linear connections on a v-bundle $\mathcal{E}$ can be
also
characterized by theirs deformation tensors (see (\ref{deft})) 
with respect, for instance, to the d--connect\-i\-on
(\ref{dchrist}):
\[
\Gamma _{\beta \gamma }^{(B)\alpha }=\widetilde{\Gamma }_{\beta
\gamma }^\alpha +P_{\beta \gamma }^{(B)\alpha },\Gamma _{\beta
\gamma }^{(c)\alpha }=\widetilde{\Gamma }_{\beta \gamma }^\alpha
+P_{\beta \gamma }^{(c)\alpha }
\]
or, in general,
\[
\Gamma _{\beta \gamma }^\alpha =\widetilde{\Gamma }_{\beta \gamma
}^\alpha +P_{\beta \gamma }^\alpha ,
\]
where $P_{\beta \gamma }^{(B)\alpha },P_{\beta \gamma }^{(c)\alpha }$ and $%
P_{\beta \gamma }^\alpha $ are respectively the deformation
d-tensors of
d--connect\-i\-ons (\ref{berwald}),\ (\ref{inters}) 
or of a general one. Similar deformation d--tensors can be
introduced for d--connections on cv--bundles and hvc--bundles. We
omit explicit formulas.

\subsection{Amost Hermitian anisotropic spaces} \index{almost Hermitian}

The are possible very interesting particular constructions \cite
{ma87,ma94,mhss} on t--bundle $TM$ provided with N--connection
which defines
a N--adapted frame structure $\delta _{\alpha }=(\delta _{i},\dot{\partial}%
_{i})$ (for the same formulas (\ref{dder}) and (\ref{ddif}) but
with identified fiber and base indices). We are using the 'dot'
symbol in order to distinguish the horizontal and vertical
operators because on t--bundles the indices could take the same
values both for the base and fiber objects. This allow us to
define an almost complex structure $\mathbf{J}=\{J_{\alpha }^{\
\beta }\}$ on $TM$ as follows
\begin{equation}
\mathbf{J}(\delta _{i})=-\dot{\partial}_{i},\ \mathbf{J}(\dot{\partial}%
_{i})=\delta _{i}.  \label{alcomp}
\end{equation}
It is obvious that $\mathbf{J}$ is well--defined and
$\mathbf{J}^{2}=-I.$

For d--metrics of type (\ref{dmetric}), on $TM,$ we can consider
the case when\newline $g_{ij}(x,y)=h_{ab}(x,y),$ i. e.
\begin{equation}  \label{dmetrict}
\mathbf{G}_{(t)}= g_{ij}(x,y)dx^i\otimes dx^j + g_{ij}(x,y)\delta
y^i\otimes \delta y^j,
\end{equation}
where the index $(t)$ denotes that we have geometrical object
defined on tangent space.

An almost complex \index{almost complex}  structure
$J_{\alpha}^{\ \beta}$ is compatible with a d--metric of type
(\ref{dmetrict}) and a d--connection $D$ on tangent bundle $TM$
if the conditions
\[
J_{\alpha}^{\ \beta} J_{\gamma}^{\ \delta} g_{\beta \delta} =
g_{\alpha \gamma} \ \mbox{ and } \ D_{\alpha} J^{\gamma}_{\
\beta}=0
\]
are satisfied.

The pair $(\mathbf{G}_{(t)},\mathbf{J})$ is an almost Hermitian
structure on $TM.$

One can introduce an almost sympletic 2--form associated to the
almost Hermitian structure $(\mathbf{G}_{(t)},\mathbf{J}),$
\begin{equation}  \label{hermit}
\theta = g_{ij}(x,y)\delta y^i\wedge dx^j.
\end{equation}

If the 2--form (\ref{hermit}), defined by the coefficients
$g_{ij},$ is closed, we obtain an almost K\"{a}hlerian structure
in $TM.$

\begin{definition}
\label{kahlcon} An almost K\"{a}hler metric connection is a linear
connection $D^{(H)}$ on $T{\tilde{M}}=TM\setminus \{0\}$ with the
properties:

\begin{enumerate}
\item  $D^{(H)}$ preserve by parallelism the vertical distribution defined
by the N--connection structure;

\item  $D^{(H)}$ is compatible with the almost K\"{a}hler structure $(%
\mathbf{G}_{(t)},\mathbf{J})$, i. e.
\[
D_{X}^{(H)}g=0,\ D_{X}^{(H)}J=0,\ \forall X\in \mathcal{X}\left( T{\tilde{M}}%
\right) .
\]
\end{enumerate}
\end{definition}

By straightforward calculation we can prove that a d--connection
$D\Gamma =\left(L_{jk}^i,L_{jk}^i,C_{jc}^i,C_{jc}^i\right) $ with
the coefficients defined by
\begin{eqnarray}  \label{kahlerconm}
D^{(H)}_{\delta _i}\delta _j & = & L_{jk}^i \delta _i,\ D^{(H)}_{\delta _i}%
\dot{\partial}_j = L_{jk}^i \dot{\partial}_i, \\
D^{(H)}_{\delta _i}\delta _j & = & C_{jk}^i \delta _i,\ D^{(H)}_{\delta _i}%
\dot{\partial}_j = C_{jk}^i \dot{\partial}_i,  \nonumber
\end{eqnarray}
where $L_{jk}^i$ and $C_{ab}^e \to C_{jk}^i,$\ on $TM$ are
defined by the formulas (\ref{lccoef}), define a torsionless (see
the next section on torsion structures) metric d--connection
which satisfy the compatibility conditions (\ref{comatib}).

Almost complex structures and almost K\"{a}hler models of
Finsler, Lagrange, Hamilton and Cartan geometries (of first an
higher orders) are investigated in details in Refs.
\cite{m1,m2,mhss,vbook}.

\section{ Torsions and Curvatures} \index{torsion} \index{curvature}

In this section we outline the basic definitions and formulas for
the torsion and curvature structures in v--bundles and
cv--bundles provided with N--connection structure.

\subsection{N--connection curvature}

\begin{enumerate}
\item  The curvature $\mathbf{\Omega }$$\,$ of a nonlinear connection $%
\mathbf{N}$ in a v--bundle $\mathcal{E}$ can be defined in local form as $%
\mathbf{\ }$ \cite{ma87,ma94}:
\[
\mathbf{\Omega }=\frac 12\Omega _{ij}^ad^i\bigwedge d^j\otimes
\partial _a,
\]
where
\begin{eqnarray}
\Omega _{ij}^a &=&\delta _jN_i^a-\delta _iN_j^a  \label{ncurv} \\
&=&\partial _jN_i^a-\partial
_iN_j^a+N_i^bN_{bj}^a-N_j^bN_{bi}^a,  \nonumber
\end{eqnarray}
$N_{bi}^a$ being that from (\ref{nlinearized}).

\item  For the curvature $\mathbf{\breve{\Omega}},$ of a nonlinear
connection $\mathbf{\breve{N}}$ in a cv--bundle
$\mathcal{\breve{E}}$ we introduce
\[
\mathbf{\breve{\Omega}}\,=\frac 12\breve{\Omega}_{ija}d^i\bigwedge
d^j\otimes \breve{\partial}^a,
\]
where
\begin{eqnarray}
\breve{\Omega}_{ija} &=&-\breve{\delta}_j\breve{N}_{ia}+\breve{\delta}_i%
\breve{N}_{ja}  \label{ncurvcv} \\
&=&-\partial _j\breve{N}_{ia}+\partial _i\breve{N}_{ja}+\breve{N}_{ib}\breve{%
N}_{ja}^{\quad b}-\breve{N}_{jb}\breve{N}_{ja}^{\quad b},  \nonumber \\
\breve{N}_{ja}^{\quad b} &=&\breve{\partial}^b\breve{N}_{ja}=\partial \breve{%
N}_{ja}/\partial p_b.  \nonumber
\end{eqnarray}

\item  Curvatures $\mathbf{\tilde{\Omega}}$$\,$ of different type of
nonlinear connections $\mathbf{\tilde{N}}$ in higher order
anisotropic bundles were analyzed for different type of higher
order tangent/dual tangent bundles and higher order prolongations
of generalized Finsler, Lagrange and Hamiloton spaces in Refs.
\cite{m1,m2,mhss} and for higher order anisotropic superspaces
and spinor bundles in Refs. \cite {vbook,vsp1,vhsp,vstr2}: For
every higher order anisotropy shell we shall define the
coefficients (\ref{ncurv}) or (\ref{ncurvcv}) in dependence of
the fact with type of subfiber we are considering (a vector or
covector fiber).
\end{enumerate}

\subsection{d--Torsions in v- and cv--bundles}

The torsion $\mathbf{T}$ of a d--connection $\mathbf{D\ }$ in v--bundle $%
\mathcal{E}$ (cv--bundle $\mathcal{\breve{E})}$ is defined by the
equation
\begin{equation}
\mathbf{T\left( X,Y\right) =XY_{\circ }^{\circ }T\doteq }D_X\mathbf{Y-}D_Y%
\mathbf{X\ -\left[ X,Y\right] .}  \label{torsion}
\end{equation}
One holds the following h- and v--decompositions
\[
\mathbf{T\left( X,Y\right) =T\left( hX,hY\right) +T\left(
hX,vY\right) +T\left( vX,hY\right) +T\left( vX,vY\right) .}
\]
We consider the projections:
\[
\mathbf{hT\left( X,Y\right) ,vT\left( hX,hY\right) ,hT\left(
hX,hY\right) ,...}
\]
and say that, for instance, $\mathbf{hT\left( hX,hY\right) }$ is
the h(hh)--torsion of $\mathbf{D}$ ,\newline $\mathbf{vT\left(
hX,hY\right) }$ is the v(hh)--torsion of $\mathbf{D}$ and so on.

The torsion (\ref{torsion}) in v-bundle is locally determined by
five d--tensor fields, torsions, defined as
\begin{eqnarray}
T_{jk}^i &=&\mathbf{hT}\left( \delta _k,\delta _j\right) \cdot
d^i,\quad T_{jk}^a=\mathbf{vT}\left( \delta _k,\delta _j\right)
\cdot \delta ^a,
\label{dtorsions} \\
P_{jb}^i &=&\mathbf{hT}\left( \partial _b,\delta _j\right) \cdot
d^i,\quad P_{jb}^a=\mathbf{vT}\left( \partial _b,\delta _j\right)
\cdot \delta ^a,
\nonumber \\
S_{bc}^a &=&\mathbf{vT}\left( \partial _c,\partial _b\right)
\cdot \delta ^a. \nonumber
\end{eqnarray}
Using formulas (\ref{dder}), (\ref{ddif}), (\ref{ncurv})
and (\ref{torsion}) 
we can computer \cite{ma87,ma94} in explicit form the components
of torsions
(\ref{dtorsions}) 
for a d--connection of type (\ref{hgamma}) and (\ref{vgamma}):
\begin{eqnarray}
T_{.jk}^i &=&T_{jk}^i=L_{jk}^i-L_{kj}^i,\quad
T_{ja}^i=C_{.ja}^i,T_{aj}^i=-C_{ja}^i,  \label{dtorsc} \\
T_{.ja}^i &=&0,\qquad T_{.bc}^a=S_{.bc}^a=C_{bc}^a-C_{cb}^a,  \nonumber \\
T_{.ij}^a &=&\delta _jN_i^a-\delta _jN_j^a,\quad
T_{.bi}^a=P_{.bi}^a=\partial _bN_i^a-L_{.bj}^a,\quad
T_{.ib}^a=-P_{.bi}^a. \nonumber
\end{eqnarray}

Formulas similar to (\ref{dtorsions}) and (\ref{dtorsc}) hold for
cv--bundles:
\begin{eqnarray}
\check{T}_{jk}^i &=&\mathbf{hT}\left( \delta _k,\delta _j\right)
\cdot d^i,\quad \check{T}_{jka}=\mathbf{vT}\left( \delta
_k,\delta _j\right) \cdot
\check{\delta}_a,  \label{torsionsa} \\
\check{P}_j^{i\quad b} &=&\mathbf{hT}\left(
\check{\partial}^b,\delta
_j\right) \cdot d^i,\quad \check{P}_{aj}^{\quad b}=\mathbf{vT}\left( \check{%
\partial}^b,\delta _j\right) \cdot \check{\delta}_a,  \nonumber \\
\check{S}_a^{\quad bc} &=&\mathbf{vT}\left( \check{\partial}^c,\check{%
\partial}^b\right) \cdot \check{\delta}_a.  \nonumber
\end{eqnarray}
and
\begin{eqnarray}
\check{T}_{.jk}^i &=&\check{T}_{jk}^i=L_{jk}^i-L_{kj}^i,\quad \check{T}%
_j^{ia}=\check{C}_{.j}^{i\ a},\check{T}_{~\
j}^{ia}=-\check{C}_j^{i~a},
\label{dtorsca} \\
\check{T}_{.j}^{i~a} &=&0,\qquad \check{T}_a^{~bc}=\check{S}_a^{~bc}=\check{C%
}_a^{~bc}-\check{C}_a^{~cb},  \nonumber \\
\check{T}_{.ija} &=&-\delta _j\check{N}_{ia}+\delta
_j\check{N}_{ja},\quad
\check{T}_a^{~bi}=\check{P}_a^{~bi}=-\check{\partial}^b\check{N}_{ia}-\check{%
L}_a^{~bi},\quad \check{T}_{a~b}^{~j}=-\check{P}_{a\ b}^{~j}.
\nonumber
\end{eqnarray}

The formulas for torsion can be generalized for hvc--bundles (on
every shell we must write (\ref{dtorsc}) or (\ref{dtorsca}) in
dependence of the type of shell, vector or co-vector one, we are
dealing).

\subsection{d--Curvatures in v- and cv--bundles} \index{d--curvatures}

The curvature $\mathbf{R}$ of a d--connection in v--bundle
$\mathcal{E}$ is defined by the equation
\[
\mathbf{R}\left( X,Y\right) Z = XY_{\bullet }^{\bullet }R\bullet
Z =D_XD_Y{\
Z} - D_Y D_X Z - D_{[X,Y]}{Z.} 
\]
One holds the next properties for the h- and v--decompositions of
curvature:
\begin{eqnarray}  \label{curvaturehv}
\mathbf{vR}\left( X,Y\right) hZ &=& 0,\ \mathbf{hR}\left( X,Y\right) vZ=0, \\
\mathbf{R}\left( X,Y\right) Z & = & \mathbf{hR} \left( X,Y\right)
hZ + \mathbf{vR} \left( X,Y\right) vZ.  \nonumber
\end{eqnarray}

From (\ref{curvaturehv}) 
and the equation $\mathbf{R\left( X,Y\right) =-R\left( Y,X\right)
}$ we get that the curvature of a d--con\-nec\-ti\-on
$\mathbf{D}$ in $\mathcal{E}$ is completely determined by the
following six d--tensor fields:
\begin{eqnarray}
R_{h.jk}^{.i} &=& d^i\cdot \mathbf{R}\left( \delta _k,\delta
_j\right) \delta _h,~R_{b.jk}^{.a}=\delta ^a\cdot
\mathbf{R}\left( \delta _k,\delta
_j\right) \partial _b,  \label{rps} \\
P_{j.kc}^{.i} &=& d^i\cdot \mathbf{R}\left( \partial _c,\partial
_k\right) \delta _j,~P_{b.kc}^{.a}=\delta ^a\cdot
\mathbf{R}\left( \partial
_c,\partial _k\right) \partial _b,  \nonumber \\
S_{j.bc}^{.i} &=& d^i\cdot \mathbf{R}\left( \partial _c,\partial
_b\right) \delta _j,~S_{b.cd}^{.a}=\delta ^a\cdot
\mathbf{R}\left( \partial _d,\partial _c\right) \partial _b.
\nonumber
\end{eqnarray}
By a direct computation, using 
(\ref{dder}),(\ref{ddif}),(\ref{hgamma}),(\ref{vgamma}) and
(\ref{rps}) we get:
\begin{eqnarray}
R_{h.jk}^{.i} &=& {\delta}_h L_{.hj}^i - {\delta}_j L_{.hk}^i
+L_{.hj}^m
L_{mk}^i-L_{.hk}^m L_{mj}^i+C_{.ha}^i R_{.jk}^a,  \label{dcurvatures} \\
R_{b.jk}^{.a} &=& {\delta}_k L_{.bj}^a - {\delta}_j L_{.bk}^a
+L_{.bj}^c
L_{.ck}^a - L_{.bk}^c L_{.cj}^a + C_{.bc}^a R_{.jk}^c,  \nonumber \\
P_{j.ka}^{.i} &=& {\partial}_a L_{.jk}^i - \left({\delta}_k
C_{.ja}^i + L_{.lk}^i C_{.ja}^l - L_{.jk}^l C_{.la}^i - L_{.ak}^c
C_{.jc}^i\right)
+C_{.jb}^i P_{.ka}^b,  \nonumber \\
P_{b.ka}^{.c} &=& {\partial}_a L_{.bk}^c -\left( {\delta}_k
C_{.ba}^c + L_{.dk}^{c} C_{.ba}^d - L_{.bk}^d C_{.da}^c -
L_{.ak}^d C_{.bd}^c \right)
+C_{.bd}^c P_{.ka}^d,  \nonumber \\
S_{j.bc}^{.i} &=& {\partial}_c C_{.jb}^i - {\partial}_b C_{.jc}^i
+
C_{.jb}^h C_{.hc}^i - C_{.jc}^h C_{hb}^i,  \nonumber \\
S_{b.cd}^{.a} & = & {\partial}_d C_{.bc}^a - {\partial}_c
C_{.bd}^a + C_{.bc}^e C_{.ed}^a - C_{.bd}^eC_{.ec}^a.  \nonumber
\end{eqnarray}

We note that d--torsions (\ref{dtorsc}) and d--curvatures (\ref{dcurvatures}%
) 
are computed in explicit form by particular cases of
d--connections (\ref {berwald}), (\ref{inters}) and
(\ref{dchrist}).

For cv--bundles we have
\begin{eqnarray}
\check{R}_{h.jk}^{.i} &=&d^{i}\cdot \mathbf{R}\left( \delta
_{k},\delta _{j}\right) \delta _{h},~\check{R}_{\
a.jk}^{b}=\check{\delta}_{a}\cdot \mathbf{R}\left( \delta
_{k},\delta _{j}\right) \check{\partial}^{b},
\label{rpsa} \\
\check{P}_{j.k}^{.i\quad c} &=&d^{i}\cdot \mathbf{R}\left( \check{\partial}%
^{c},\partial _{k}\right) \delta _{j},~\check{P}_{\quad a.k}^{b\quad c}=%
\check{\delta}_{a}\cdot \mathbf{R}\left(
\check{\partial}^{c},\partial
_{k}\right) \check{\partial}^{b},  \nonumber \\
\check{S}_{j.}^{.ibc} &=&d^{i}\cdot \mathbf{R}\left( \check{\partial}^{c},%
\check{\partial}^{b}\right) \delta _{j},~\check{S}_{.a}^{b.cd}=\check{\delta}%
_{a}\cdot \mathbf{R}\left(
\check{\partial}^{d},\check{\partial}^{c}\right)
\check{\partial}^{b}.  \nonumber
\end{eqnarray}
and
\begin{eqnarray}
\check{R}_{h.jk}^{.i} &=&{\check{\delta}}_{h}L_{.hj}^{i}-{\check{\delta}}%
_{j}L_{.hk}^{i}+L_{.hj}^{m}L_{mk}^{i}-L_{.hk}^{m}L_{mj}^{i}+C_{.h}^{i~a}%
\check{R}_{.ajk},  \label{dcurvaturesa} \\
\check{R}_{.ajk}^{b.} &=&{\check{\delta}}_{k}\check{L}_{a.j}^{~b}-{\check{%
\delta}}_{j}\check{L}_{~b~k}^{a}+\check{L}_{cj}^{~b}\check{L}_{.ak}^{~c}-%
\check{L}_{ck}^{b}\check{L}_{a.j}^{~c}+\check{C}_{a}^{~bc}\check{R}_{c.jk},
\nonumber \\
\check{P}_{j.k}^{.i~a} &=&{\check{\partial}}^{a}L_{.jk}^{i}-\left( {\check{%
\delta}}_{k}\check{C}_{.j}^{i~a}+L_{.lk}^{i}\check{C}_{.j}^{l~a}-L_{.jk}^{l}%
\check{C}_{.l}^{i~a}-\check{L}_{ck}^{~a}\check{C}_{.j}^{i~c}\right) +\check{C%
}_{.j}^{i~b}\check{P}_{bk}^{~\quad a},  \nonumber \\
\check{P}_{ck}^{b~a} &=&{\check{\partial}}^{a}\check{L}_{c.k}^{~b}-({\check{%
\delta}}_{k}\check{C}_{c.}^{~ba}+\check{L}_{c.k}^{bd}\check{C}_{d}^{\ ba}-%
\check{L}_{d.k}^{\quad b}\check{C}_{c.}^{\ ad}),  \nonumber \\
&&-\check{L}_{dk}^{\quad a}\check{C}_{c.}^{\ bd})+\check{C}_{c.}^{\ bd}%
\check{P}_{d.k}^{\quad a},  \nonumber \\
\check{S}_{j.}^{.ibc} &=&{\check{\partial}}^{c}\check{C}_{.j}^{i\ b}-{\check{%
\partial}}^{b}\check{C}_{.j}^{i\ c}+\check{C}_{.j}^{h\ b}\check{C}_{.h}^{i\
c}-\check{C}_{.j}^{h\ c}\check{C}_{h}^{i\ b},  \nonumber \\
\check{S}_{\ a.}^{b\ cd} &=&{\check{\partial}}^{d}\check{C}_{a.}^{\ bc}-{%
\check{\partial}}^{c}\check{C}_{a.}^{\ bd}+\check{C}_{e.}^{\ bc}\check{C}%
_{a.}^{\ ed}-\check{C}_{e.}^{\ bd}\check{C}_{.a}^{\ ec}. \nonumber
\end{eqnarray}

The formulas for curvature can be also generalized for
hvc--bundles (on every shell we must write (\ref{dtorsc}) or
(\ref{torsionsa}) in dependence of the type of shell, vector or
co-vector one, we are dealing).

\section{Generalizations of Finsler Spaces}

We outline the basic definitions and formulas for Finsler,
Lagrange and generalized Lagrange spaces (constructed on tangent
bundle) and for Cartan, Hamilton and generalized Hamilton spaces
(constructed on cotangent bundle). The original results are given
in details in monographs \cite{ma87,ma94,mhss}

\subsection{Finsler Spaces} \index{Finsler space}

The Finsler geometry is modeled on tangent bundle $TM.$

\begin{definition}
A Finsler space (manifold) is a pair $F^{n}=\left(
M,F(x,y)\right) $ \ where $M$ is a real $n$--dimensional
differentiable manifold and $F:TM\rightarrow \mathcal{R}$ \ is a
scalar function which satisfy the following conditions:

\begin{enumerate}
\item  $F$ is a differentiable function on the manifold $\widetilde{TM}$ $%
=TM\backslash \{0\}$ and $F$ is continous on the null section of
the projection $\pi :TM\rightarrow M;$

\item  $F$ is a positive function, homogeneous on the fibers of the $TM,$ i.
e. $F(x,\lambda y)=\lambda F(x,y),\lambda \in \mathcal{R};$

\item  The Hessian of $F^{2}$ with elements
\begin{equation}
g_{ij}^{(F)}(x,y)=\frac{1}{2}\frac{\partial ^{2}F^{2}}{\partial
y^{i}\partial y^{j}}  \label{finm}
\end{equation}
is positively defined on $\widetilde{TM}.$
\end{enumerate}
\end{definition}

The function $F(x,y)$ and $g_{ij}(x,y)$ are called respectively
the fundamental function and the fundamental (or metric) tensor
of the Finsler space $F.$

One considers ''anisotropic'' (depending on directions $y^{i})$
Christoffel symbols, for simplicity we write
$g_{ij}^{(F)}=g_{ij},$
\[
\gamma _{~jk}^{i}(x,y)=\frac{1}{2}g^{ir}\left( \frac{\partial g_{rk}}{%
\partial x^{j}}+\frac{\partial g_{jr}}{\partial x^{k}}-\frac{\partial g_{jk}%
}{\partial x^{r}}\right) ,
\]
which are used for definition of the Cartan N--connection,
\begin{equation}
N_{(c)~j}^{i}=\frac{1}{2}\frac{\partial }{\partial y^{j}}\left[
\gamma _{~nk}^{i}(x,y)y^{n}y^{k}\right] .  \label{ncartan}
\end{equation}
This N--connection can be used for definition of an almost
complex structure like in (\ref{alcomp}) and to define on $TM$ a
d--metric
\begin{equation}
\mathbf{G}_{(F)}=g_{ij}(x,y)dx^{i}\otimes dx^{j}+g_{ij}(x,y)\delta
y^{i}\otimes \delta y^{j},  \label{dmfin}
\end{equation}
with $g_{ij}(x,y)$ taken as (\ref{finm}).

Using the Cartan N--connection (\ref{ncartan}) and Finsler metric tensor (%
\ref{finm}) (or, equivalently, the d--metric (\ref{dmfin})) we
can introduce the canonical d--connection
\[
D\Gamma \left( N_{(c)}\right) =\Gamma _{(c)\beta \gamma }^{\alpha
}=\left( L_{(c)~jk}^{i},C_{(c)~jk}^{i}\right)
\]
with the coefficients computed like in (\ref{kahlerconm}) and
(\ref{lccoef}) with $h_{ab}\rightarrow g_{ij}.$ The d--connection
$D\Gamma \left( N_{(c)}\right) $ has the unique property that it
is torsionless and satisfies the metricity conditions both for
the horizontal and vertical components, i. e. $D_{\alpha
}g_{\beta \gamma }=0.$

The d--curvatures
\[
\check{R}_{h.jk}^{.i}=\{\check{R}_{h.jk}^{.i},\check{P}_{j.k}^{.i\quad
l},S_{(c)j.kl}^{.i}\}
\]
on a Finsler space provided with Cartan N--connection and Finsler
metric
structures are computed following the formulas (\ref{dcurvatures}) when the $%
a,b,c...$ indices are identified with $i,j,k,...$ indices. It
should be emphasized that in this case all values $g_{ij,}\Gamma
_{(c)\beta \gamma }^{\alpha }$ and $R_{(c)\beta .\gamma \delta
}^{.\alpha }$ are defined by a fundamental function $F\left(
x,y\right) .$

In general, we can consider that a Finsler space is provided with a metric $%
g_{ij}=\partial ^{2}F^{2}/2\partial y^{i}\partial y^{j},$ but the
N--connection and d--connection are be defined in a different
manner, even not be determined by $F.$

\subsection{Lagrange and Generalized Lagrange Spaces} \index{Lagrange space}

The notion of Finsler spaces was extended by J. Kern \cite{ker}
and R. Miron \cite{mironlg}. It is widely developed in monographs
\cite{ma87,ma94} and exteded to superspaces by S. Vacaru
\cite{vlasg,vstr2,vbook}.

The idea of extension was to consider instead of the homogeneous
fundamental
function $F(x,y)$ in a Finsler space a more general one, a Lagrangian $%
L\left( x,y\right) $, defined as a differentiable mapping
$L:(x,y)\in
TM\rightarrow L(x,y)\in \mathcal{R},$ of class $C^{\infty }$ on manifold $%
\widetilde{TM}$ and continous on the null section $0:M\rightarrow
TM$ of the projection $\pi :TM\rightarrow M.$ A Lagrangian is
regular if it is differentiable and the Hessian
\begin{equation}
g_{ij}^{(L)}(x,y)=\frac{1}{2}\frac{\partial ^{2}L^{2}}{\partial
y^{i}\partial y^{j}}  \label{lagm}
\end{equation}
is of rank $n$ on $M.$

\begin{definition}
A Lagrange space is a pair $L^{n}=\left( M,L(x,y)\right) $ where
$M$ is a
smooth real $n$--dimensional manifold provided with regular Lagrangian \ $%
L(x,y)$ structure $L:TM\rightarrow \mathcal{R}$ $\ $for which
$g_{ij}(x,y)$
from (\ref{lagm}) has a constant signature over the manifold $\widetilde{TM}%
. $
\end{definition}

The fundamental Lagrange function $L(x,y)$ defines a canonical
N--con\-nec\-ti\-on
\[
N_{(cL)~j}^{i}=\frac{1}{2}\frac{\partial }{\partial y^{j}}\left[
g^{ik}\left( \frac{\partial ^{2}L^{2}}{\partial y^{k}\partial y^{h}}y^{h}-%
\frac{\partial L}{\partial x^{k}}\right) \right]
\]
as well a d-metric
\begin{equation}
\mathbf{G}_{(L)}=g_{ij}(x,y)dx^{i}\otimes dx^{j}+g_{ij}(x,y)\delta
y^{i}\otimes \delta y^{j},  \label{dmlag}
\end{equation}
with $g_{ij}(x,y)$ taken as (\ref{lagm}). As well we can
introduce an almost K\"{a}hlerian structure and an almost
Hermitian model of $L^{n},$ denoted as $H^{2n}$ as in the case of
Finsler spaces but with a proper fundamental Lagange function and
metric tensor $g_{ij}.$ The canonical metric d--connection
$D\Gamma \left( N_{(cL)}\right) =\Gamma _{(cL)\beta \gamma
}^{\alpha }=\left( L_{(cL)~jk}^{i},C_{(cL)~jk}^{i}\right) $ is to
computed
by the same formulas (\ref{kahlerconm}) and (\ref{lccoef}) with $%
h_{ab}\rightarrow g_{ij}^{(L)},$ for $N_{(cL)~j}^{i}.$ The
d--torsions (\ref {dtorsc}) and d--curvatures (\ref{dcurvatures})
are defined, in this case,
by $L_{(cL)~jk}^{i}$ and $C_{(cL)~jk}^{i}.$ We also note that instead of $%
N_{(cL)~j}^{i}$ and $\Gamma _{(cL)\beta \gamma }^{\alpha }$ one
can consider
on a $L^{n}$--space arbitrary N--connections $N_{~j}^{i},$ d--connections $%
\Gamma _{\beta \gamma }^{\alpha }$ which are not defined only by
$L(x,y)$ and $g_{ij}^{(L)}$ but can be metric, or non--metric
with respect to the Lagrange metric.

The next step of generalization is to consider an arbitrary metric $%
g_{ij}\left( x,y\right) $ on $TM$ instead of (\ref{lagm}) which
is the second derivative of ''anisotropic'' coordinates $y^{i}$
of a Lagrangian \cite{mironlg,mironlgg}.

\begin{definition}
A generalized Lagrange space is a pair $GL^{n}=\left(
M,g_{ij}(x,y)\right) $ where $g_{ij}(x,y)$ is a covariant,
symmetric d--tensor field, of rank $n$ and of constant signature
on $\widetilde{TM}.$
\end{definition}

\bigskip One can consider different classes of N-- and d--connections on $%
TM, $ which are compatible (metric) or non compatible with
(\ref{dmlag}) for arbitrary $g_{ij}(x,y).$ We can apply all
formulas for d--connections, N-curvatures, d-torsions and
d-curvatures as in a v--bundle $\mathcal{E},$ but reconsidering
them on $TM,$ by changing \ $h_{ab}\rightarrow g_{ij}(x,y)$ and
$N_{i}^{a}\rightarrow N_{~i}^{k}.$

\subsection{Cartan Spaces} \index{Cartan space}

The theory of Cartan spaces (see, for instance, \cite{run,kaw1})
\ was formulated in a new fashion in R. Miron's works
\cite{mironc1,mironc2} by considering them as duals to the
Finsler spaces (see details  and references in \cite{mhss}).
Roughly, a Cartan space is constructed on a cotangent bundle
$T^{\ast }M$ like a Finsler space on the corresponding tangent
bundle $TM.$

Consider a real smooth manifold $M,$ the cotangent bundle $\left(
T^{\ast }M,\pi ^{\ast },M\right) $ and the manifold
$\widetilde{T^{\ast }M}=T^{\ast }M\backslash \{0\}.$

\begin{definition}
A Cartan space is a pair $C^{n}=\left( M,K(x,p)\right) $ \ such that $%
K:T^{\ast }M\rightarrow \mathcal{R}$ is a scalar function which
satisfy the following conditions:

\begin{enumerate}
\item  $K$ is a differentiable function on the manifold $\widetilde{T^{\ast
}M}$ $=T^{\ast }M\backslash \{0\}$ and  continous on the null
section of the projection $\pi ^{\ast }:T^{\ast }M\rightarrow M;$

\item  $K$ is a positive function, homogeneous on the fibers of the $T^{\ast
}M,$ i. e. $K(x,\lambda p)=\lambda F(x,p),\lambda \in
\mathcal{R};$

\item  The Hessian of $K^{2}$ with elements
\begin{equation}
\check{g}_{(K)}^{ij}(x,p)=\frac{1}{2}\frac{\partial
^{2}K^{2}}{\partial p_{i}\partial p_{j}}  \label{carm}
\end{equation}
is positively defined on $\widetilde{T^{\ast }M}.$
\end{enumerate}
\end{definition}

The function $K(x,y)$ and $\check{g}^{ij}(x,p)$ are called \
respectively the fundamental function and the fundamental (or
metric) tensor of the Cartan space $C^{n}.$ We use symbols like
$"\check{g}"$ as to emphasize that the geometrical objects are
defined on a dual space.

One considers ''anisotropic'' (depending on directions, momenta,
$p_{i})$
\newline
Christoffel symbols, for symplicty, we write the inverse to (\ref{carm}) as $%
g_{ij}^{(K)}=\check{g}_{ij},$
\[
\check{\gamma}_{~jk}^{i}(x,p)=\frac{1}{2}\check{g}^{ir}\left(
\frac{\partial
\check{g}_{rk}}{\partial x^{j}}+\frac{\partial \check{g}_{jr}}{\partial x^{k}%
}-\frac{\partial \check{g}_{jk}}{\partial x^{r}}\right) ,
\]
which are used for definition of the canonical N--connection,
\begin{equation}
\check{N}_{ij}=\check{\gamma}_{~ij}^{k}p_{k}-\frac{1}{2}\gamma
_{~nl}^{k}p_{k}p^{l}{\breve{\partial}}^{n}\check{g}_{ij},~{\breve{\partial}}%
^{n}=\frac{\partial }{\partial p_{n}}.  \label{nccartan}
\end{equation}
This N--connection can be used for definition of an almost
complex structure like in (\ref{alcomp}) and to define on
$T^{\ast }M$ a d--metric
\begin{equation}
\mathbf{\check{G}}_{(k)}=\check{g}_{ij}(x,p)dx^{i}\otimes dx^{j}+\check{g}%
^{ij}(x,p)\delta p_{i}\otimes \delta p_{j},  \label{dmcar}
\end{equation}
with $\check{g}^{ij}(x,p)$ taken as (\ref{carm}).

Using the canonical N--connection (\ref{nccartan}) and Finsler
metric tensor (\ref{carm}) (or, equivalently, the d--metric
(\ref{dmcar}) we can introduce the canonical d--connection
\[
D\check{\Gamma}\left( \check{N}_{(k)}\right)
=\check{\Gamma}_{(k)\beta \gamma }^{\alpha }=\left(
\check{H}_{(k)~jk}^{i},\check{C}_{(k)~i}^{\quad jk}\right)
\]
with the coefficients \ are computed
\begin{eqnarray*}
\check{H}_{(k)~jk}^{i} &=&\frac{1}{2}\check{g}^{ir}\left( \check{\delta}_{j}%
\check{g}_{rk}+\check{\delta}_{k}\check{g}_{jr}-\check{\delta}_{r}\check{g}%
_{jk}\right) , \\
\check{C}_{(k)~i}^{\quad jk} &=&\check{g}_{is}{\breve{\partial}}^{s}\check{g}%
^{jk},
\end{eqnarray*}
The d--connection $D\check{\Gamma}\left( \check{N}_{(k)}\right) $
has the unique property that it is torsionless and satisfies the
metricity
conditions both for the horizontal and vertical components, i. e. $\check{D}%
_{\alpha }\check{g}_{\beta \gamma }=0.$

The d--curvatures
\[
\check{R}_{(k)\beta .\gamma \delta }^{.\alpha
}=\{R_{(k)h.jk}^{.i},P_{(k)j.km}^{.i},\check{S}_{j.}^{.ikl}\}
\]
on a Finsler space provided with Cartan N--connection and Finsler
metric structures are computed following the formulas
(\ref{dcurvaturesa}) when the $a,b,c...$ indices are identified
with $i,j,k,...$ indices. It should be
emphasized that in this case all values $\check{g}_{ij,}\check{\Gamma}%
_{(k)\beta \gamma }^{\alpha }$ and $\check{R}_{(k)\beta .\gamma
\delta }^{.\alpha }$ are defined by a fundamental function
$K\left( x,p\right) .$

In general, we can consider that a Cartan space is provided with a metric $%
\check{g}^{ij}=\partial ^{2}K^{2}/2\partial p_{i}\partial p_{j},$
but the N--connection and d--connection could be defined in a
different manner, even not be determined by $K.$

\subsection{ Generalized Hamilton and Hamilton Spaces} \index{Hamilton space}

The geometry  of Hamilton spaces was defined and investigated by
R. Miron in the papers \cite{mironh1,mironh2,mironh3} (see
details and references in \cite{mhss}). It was developed on the
cotangent bundel as a dual geometry to the geometry of Lagrange
spaces. \ Here we start with the definition of generalized
Hamilton spaces and then consider the particular case.

\begin{definition}
A generalized Hamilton space is a pair\\ $GH^{n}=\left( M,\check{g}%
^{ij}(x,p)\right) $ where $M$ is a real $n$--dimensional manifold and  $%
\check{g}^{ij}(x,p)$ is a contravariant, symmetric, nondegenerate
of rank $n$ and of constant signature on $\widetilde{T^{\ast }M}.$
\end{definition}

\bigskip The value $\check{g}^{ij}(x,p)$ is called the fundamental (or
metric) tensor of the space $GH^{n}.$ One can define such values
for every paracompact manifold $M.$ In general, a N--connection
on $GH^{n}$ is not determined by $\check{g}^{ij}.$ Therefore we
can consider arbitrary coefficients $\check{N}_{ij}\left(
x,p\right) $ and define on $T^{\ast }M$ a d--metric like
(\ref{dmetricvc})
\begin{equation}
{\breve{\mathbf{G}}}={\breve{g}}_{\alpha \beta }\left( {\breve{u}}\right) {%
\breve{\delta}}^{\alpha }\otimes {\breve{\delta}}^{\beta }={\breve{g}}%
_{ij}\left( {\breve{u}}\right) d^{i}\otimes d^{j}+{\check{g}}^{ij}\left( {%
\breve{u}}\right) {\breve{\delta}}_{i}\otimes
{\breve{\delta}}_{j}, \label{dmghs}
\end{equation}
This N--coefficients $\check{N}_{ij}\left( x,p\right) $ and
d--metric structure (\ref{dmghs}) allow to define an almost
K\"{a}hler model of generalized Hamilton spaces and to define
canonical d--connections, d--torsions and d-curvatures (see
respectively the formulas (\ref{lccoef}), (\ref{inters}),
(\ref{dtorsca}) and (\ref{dcurvatures}) with the fiber
coefficients redefined for the cotangent bundle $T^{\ast }M$ ).

A generalized Hamilton space $GH^{n}=\left(
M,\check{g}^{ij}(x,p)\right) $
is called reducible to a Hamilton one if there exists a Hamilton function $%
H\left( x,p\right) $ on $T^{\ast }M$ such that
\begin{equation}
\check{g}^{ij}(x,p)=\frac{1}{2}\frac{\partial ^{2}H}{\partial
p_{i}\partial p_{j}}.  \label{hsm}
\end{equation}

\begin{definition}
A Hamilton space is a pair $H^{n}=\left( M,H(x,p)\right) $ \ such that $%
H:T^{\ast }M\rightarrow \mathcal{R}$ is a scalar function which
satisfy the following conditions:

\begin{enumerate}
\item  $H$ is a differentiable function on the manifold $\widetilde{T^{\ast
}M}$ $=T^{\ast }M\backslash \{0\}$ and  continous on the null
section of the projection $\pi ^{\ast }:T^{\ast }M\rightarrow M;$

\item  The Hessian of $H$ with elements (\ref{hsm}) is positively defined on
$\widetilde{T^{\ast }M}$ and $\check{g}^{ij}(x,p)$ is
nondegenerate matrix of rank $n$ and of constant signature.
\end{enumerate}
\end{definition}

For Hamilton spaces the canonical N--connection (defined by $H$
and its Hessian) exists,
\[
\check{N}_{ij}=\frac{1}{4}\{\check{g}_{ij},H\}-\frac{1}{2}\left( \check{g}%
_{ik}\frac{\partial ^{2}H}{\partial p_{k}\partial x^{j}}+\check{g}_{jk}\frac{%
\partial ^{2}H}{\partial p_{k}\partial x^{i}}\right) ,
\]
where the Poisson brackets, for arbitrary functions $f$ and $g$
on $T^{\ast }M,$ act as
\[
\{f,g\}=\frac{\partial f}{\partial p_{i}}\frac{\partial g}{\partial x^{i}}-%
\frac{\partial g}{\partial p_{i}}\frac{\partial p}{\partial
x^{i}}.
\]

The canonical d--connection $D\check{\Gamma}\left( \check{N}_{(c)}\right) =%
\check{\Gamma}_{(c)\beta \gamma }^{\alpha }=\left( \check{H}_{(c)~jk}^{i},%
\check{C}_{(c)~i}^{\quad jk}\right) $is defined by the
coefficients
\begin{eqnarray*}
\check{H}_{(c)~jk}^{i} &=&\frac{1}{2}\check{g}^{is}\left( \check{\delta}_{j}%
\check{g}_{sk}+\check{\delta}_{k}\check{g}_{js}-\check{\delta}_{s}\check{g}%
_{jk}\right) , \\
\check{C}_{(c)~i}^{\quad jk} &=&-\frac{1}{2}\check{g}_{is}\check{\partial}%
^{j}\check{g}^{sk}.
\end{eqnarray*}
In result we can compute the d--torsions and d--curvatures like on
cv--bundle \ or on Cartan spaces. On Hamilton spaces all such
objects are defined by the Hamilton function $H(x,p)$ and indeces
have to be reconsidered for co--fibers of the co-tangent bundle.

\section{Gravity on Vector Bundles}

The components of the Ricci d--tensor \index{Ricci d--tensor}
\[
R_{\alpha \beta }=R_{\alpha .\beta \tau }^{.\tau }
\]
with respect to a locally adapted frame (\ref{ddif}) are as
follows:
\begin{eqnarray}
R_{ij} &=&R_{i.jk}^{.k},\quad R_{ia}=-^2P_{ia}=-P_{i.ka}^{.k},
\label{2.33}
\\
R_{ai} &=&^1P_{ai}=P_{a.ib}^{.b},\quad R_{ab}=S_{a.bc}^{.c}.
\nonumber
\end{eqnarray}
We point out that because, in general, $^1P_{ai}\neq ~^2P_{ia}$
the Ricci d-tensor is non symmetric.

Having defined a d-metric of type in $\mathcal{E}$ we can
introduce the scalar curvature of d--connection $\mathbf{D}$:
\begin{equation}
{\overleftarrow{R}}=G^{\alpha \beta }R_{\alpha \beta }=R+S,
\label{2.34}
\end{equation}
where $R=g^{ij}R_{ij}$ and $S=h^{ab}S_{ab}.$

For our further considerations it will be also useful to use an
alternative way of definition torsion (\ref{torsion}) and
curvature (\ref{curvaturehv}) by using the commutator
\[
\Delta _{\alpha \beta }\doteq \bigtriangledown _\alpha
\bigtriangledown _\beta -\bigtriangledown _\beta \bigtriangledown
_\alpha =2\bigtriangledown _{[\alpha }\bigtriangledown _{\beta ]}.
\]
For components of d--torsion we have
\[
\Delta _{\alpha \beta }f=T_{.\alpha \beta }^\gamma
\bigtriangledown _\gamma f
\]
for every scalar function $f\,\,$ on $\mathcal{E}$. Curvature can
be introduced as an operator acting on arbitrary d-vector
$V^\delta :$
\begin{equation}
(\Delta _{\alpha \beta }-T_{.\alpha \beta }^\gamma
\bigtriangledown _\gamma )V^\delta =R_{~\gamma .\alpha \beta
}^{.\delta }V^\gamma   \label{2.37}
\end{equation}
(we note that in this section we shall follow conventions of
Miron and Anastasiei \cite{ma87,ma94} on d-tensors; we can obtain
corresponding Penrose and Rindler abstract index formulas
\cite{penr1,penr2} just for a trivial N-connection structure and
by changing denotations for components of torsion and curvature
in this manner:\ $T_{.\alpha \beta }^\gamma \rightarrow T_{\alpha
\beta }^{\quad \gamma }$ and $R_{~\gamma .\alpha \beta }^{.\delta
}\rightarrow R_{\alpha \beta \gamma }^{\qquad \delta }).$

Here we also note that torsion and curvature of a d-connection on $\mathcal{E%
}$ satisfy generalized for locally anisotropic spaces Ricci and
Bianchi identities \cite{ma87,ma94} which in terms of components
(\ref{2.37}) are written respectively as
\begin{equation}
R_{~[\gamma .\alpha \beta ]}^{.\delta }+\bigtriangledown _{[\alpha
}T_{.\beta \gamma ]}^\delta +T_{.[\alpha \beta }^\nu T_{.\gamma
]\nu }^\delta =0  \label{2.38}
\end{equation}
and
\begin{equation}
\bigtriangledown _{[\alpha }R_{|\nu |\beta \gamma ]}^{\cdot \sigma
}+T_{\cdot [\alpha \beta }^\delta R_{|\nu |.\gamma ]\delta
}^{\cdot \sigma }=0.  \label{2.39}
\end{equation}
Identities (\ref{2.38}) and (\ref{2.39}) can be proved similarly
as in \cite{penr1} by taking into account that indices play a
distinguished character.

We can also consider a la-generalization of the so-called
conformal Weyl tensor (see, for instance, \cite{penr1}) which can
be written as a d-tensor in this form:
\begin{eqnarray}
C_{\quad \alpha \beta }^{\gamma \delta } &=&R_{\quad \alpha \beta
}^{\gamma \delta }-\frac 4{n+m-2}R_{\quad [\alpha }^{[\gamma
}~\delta _{\quad \beta
]}^{\delta ]}  \label{2.40} \\
&&+\frac 2{(n+m-1)(n+m-2)}{\overleftarrow{R}~\delta _{\quad
[\alpha }^{[\gamma }~\delta _{\quad \beta ]}^{\delta ]}.}
\nonumber
\end{eqnarray}
This object is conformally invariant on locally anisotropic
spaces provided with d-connection generated by d-metric
structures.

The Einstein equations and conservation laws on v-bundles
provided with N-connection structures are studied in detail in
\cite {ma87,ma94,ana86,ana87}. In Ref. \cite{vg} we proved that
the locally anisotropic gravity can be formulated in a gauge like
manner and analyzed the conditions when the Einstein locally
anisotropic gravitational field equations are equivalent to a
corresponding form of Yang-Mills equations. In this subsection
we  write the locally anisotropic gravitational field equations
in a form more convenient for theirs equivalent reformulation in
locally anisotropic spinor variables.

We define d-tensor $\Phi _{\alpha \beta }$ as to satisfy
conditions
\begin{equation}
-2\Phi _{\alpha \beta }\doteq R_{\alpha \beta }-\frac 1{n+m}\overleftarrow{R}%
g_{\alpha \beta }
 \label{2.41}
\end{equation}
which is the torsionless part of the Ricci tensor for locally
isotropic spaces \cite{penr1,penr2}, i.e. $\Phi _\alpha
^{~~\alpha }\doteq 0$.\ The Einstein equations on locally
anisotropic spaces
\begin{equation}
\overleftarrow{G}_{\alpha \beta }+\lambda g_{\alpha \beta }=
\kappa E_{\alpha \beta },
\label{2.42}
\end{equation}
where
\begin{equation}
\overleftarrow{G}_{\alpha \beta }=R_{\alpha \beta }-\frac 12\overleftarrow{R}%
g_{\alpha \beta }
  \label{2.43}
\end{equation}
is the Einstein d-tensor, $\lambda $ and $\kappa $ are
correspondingly the cosmological and gravitational constants and
by $E_{\alpha \beta }$ is denoted the locally anisotropic
energy-momentum d-tensor \cite{ma87,ma94}, can be rewritten in
equivalent form:
\begin{equation}
\Phi _{\alpha \beta }=-\frac \kappa 2(E_{\alpha \beta }-\frac
1{n+m}E_\tau
^{~\tau }~g_{\alpha \beta }).
 \label{2.44}
\end{equation}
Because the locally anisotropic
spaces generally have nonzero torsions we shall add to (\ref{2.44}%
) (equivalently to (\ref{2.42})) a system of algebraic d-field
equations with the source $S_{~\beta \gamma }^\alpha $ being the
locally anisotropic spin density of matter (if we consider a
variant of locally anisotropic Einstein-Cartan theory):
\begin{equation}
T_{~\alpha \beta }^\gamma +2\delta _{~[\alpha }^\gamma T_{~\beta
]\delta
}^\delta =\kappa S_{~\alpha \beta .}^\gamma 
 \label{2.45}
\end{equation}
From (\ref{2.38}) and (\ref{2.45}) one follows the conservation
law of locally anisotropic spin matter:
\[
\bigtriangledown _\gamma S_{~\alpha \beta }^\gamma -T_{~\delta
\gamma }^\delta S_{~\alpha \beta }^\gamma =E_{\beta \alpha
}-E_{\alpha \beta }.
\]

Finally, in this section, we remark that all presented geometric
constructions contain those elaborated for generalized Lagrange
spaces \cite {ma87,ma94} (for which a tangent bundle $TM$ is
considered instead of a v-bundle $\mathcal{E}$ ). We also note
that the Lagrange (Finsler) geometry is characterized by a metric
with components parametized as $g_{ij}=\frac 12\frac{\partial
^2\mathcal{L}}{\partial
y^i\partial y^j}$ $\left( g_{ij}=\frac 12\frac{\partial ^2\Lambda ^2}{%
\partial y^i\partial y^j}\right) $ and $h_{ij}=g_{ij},$ where $\mathcal{L=L}$
$(x,y)$ $\left( \Lambda =\Lambda \left( x,y\right) \right) $ is a
Lagrangian $\left( \mbox{Finsler metric}\right) $ on $TM$ (see
details in \cite{ma87,ma94,mat,bej}).





\chapter[Anholonomic Einstein and Gauge Gravity]{Anholonomic Einstein and
Gauge Gravity}

We analyze local anisotropies induced by anholonomic frames and
associated nonlinear connections in general relativity and
extensions to affine--Poincar\'e and de Sitter gauge gravity and
different types of Kaluza--Klein theories. We construct some new
classes of cosmological solutions of gravitational field
equations describing Friedmann--Robertson--Walker like universes
with  rotation (ellongated and flattened) ellipsoidal or torus
symmetry \cite{vd}.

\section{Introduction}

The search for exact solutions with generic local anisotropy in
general relativity, gauge gravity and non--Riemannian extensions
has its motivation from low energy limits in modern string and
Kaluza--Klein theories. Such classes of solutions constructed by
using moving anholonomic frame fields (tetrads, or vierbeins; we
shall use the term frames for higher dimensions) reflect a new
type of constrained dynamics and locally anisotropic interactions
of gravitational and matter fields \cite{vbh}.

What are the requirements of such constructions and their physical
treatment? We believe that such solutions should have the
properties:\ (i) they satisfy the Einstein equations in general
relativity and are locally anisotropic generalizations of some
known solutions in isotropic limits with a well posed Cauchy
problem;\ (ii) the corresponding geometrical and physical values
are defined, as a rule, with respect to an anholonomic system of
reference which reflects the imposed constraints and supposed
symmetry of locally anisotropic interactions; the reformulation
of results for a coordinate frame is also possible;\ (iii) by
applying the method of moving frames of reference, we can
generalize the solutions to some analogous in metric--affine
and/or gauge gravity, in higher dimension and string theories.

Comparing with the previos results
\cite{vjmp,vstr1,vhsp,vbook,vg} on definition of self--consistent
field theories incorporating various possible anisotropic,
inhomogeneous and sto\-chas\-tic manifestations of classical and
quantum interactions on locally anisotropic and higher order
anisotropic spaces, we emphasize that, in this Chapter, we shall
be interested not in some extensions of the well known gravity
theories with locally isotropic space--times ((pseudo) Riemannian
or Riemanian--Cartan--Weyl ones, in brief, RCW space--times) to
Finsler geometry and its generalizations. We shall present a
proof that locally an\-iso\-trop\-ic structures (Finsler, Lagrange
and higher order developments
\cite{fin,car35,run,mat,asa,ma94,m1,bej,gb}) could be induced by
anholonomic frames on locally isotropic spaces, even in general
relativity and its metric--affine and gauge like modifications
\cite {hgmn,tseyt,p,pd,mielke,deh,vg,pbo,wal}.

To evolve some new (frame anholonomy) features of locally
isotropic gravity theories we shall apply the methods of the
geometry of anholonomic frames and associated nonlinear
connection (in brief, N--connection) structures elaborated in
details for bundle spaces and generalized Finsler spaces in
monographs \cite{ma94,m1,bej} with further developments for spinor
differential geometry, superspaces and stochastic calculus in
\cite {vjmp,vstr2,vhsp,vbook}. The first rigorous global
definition of N--connections is due to W. Barthel \cite{barth}
but the idea and some rough constructions could be found in the
E. Cartan's works \cite{car35}. We note that the point of this
paper is to emphasize the generic locally anisotropic geometry
and physics and apply the N--connection method for
\'{}%
non--Finslerian%
\'{}%
(pseudo) Riemannian and RCW spacetimes. Here, it should be
mentioned that anholonomic frames are considered in detail, for
instance, in monographs \cite{eh,mtw,penr1,penr2} and with
respect to geometrization of gauge theories in \cite{mielke,pbo}
but not concerning the topic on associated N--connection
structures which grounds our geometric approach to anisotropies
in physical theories and developing of a new method of
integrating gravitational field equations.

\section[Anholonomic Frames]{ Anholonomic Frames on (Pseudo) Riemannian
Spaces}

For definiteness, we consider a $\left( n+m\right) $--dimensional
(pseudo) Riemannian spacetime $V^{(n+m)},$ being a paracompact
and connected Hausdorff $C^\infty $--ma\-ni\-fold, enabled with a
nonsigular metric
\[
ds^2=\widetilde{g}_{\alpha \beta }\ du^\alpha \otimes du^\beta
\]
with the coefficients
\begin{equation}  \label{4ansatz}
\widetilde{g}_{\alpha \beta }=\left[
\begin{array}{cc}
g_{ij}+N_i^aN_j^bh_{ab} & N_j^eh_{ae} \\
N_i^eh_{be} & h_{ab}
\end{array}
\right]
\end{equation}
parametrized with respect to a local coordinate basis $du^\alpha
=\left( dx^i,dy^a\right) ,$ having its dual $\partial /u^\alpha
=\left( \partial /x^i,\partial /y^a\right) ,$ where the indices
of geometrical objects and local coordinate $u^\alpha =\left(
x^k,y^a\right) $ run correspondingly the values:\ (for Greek
indices)$\alpha ,\beta ,\ldots =n+m;$ for (Latin indices)
$i,j,k,...=1,2,...,n$ and $a,b,c,...=1,2,...,m$ . We shall use
'tilds' if would be necessary to emphasize that a value is
defined with respect to a coordinate basis.

The metric (\ref{4ansatz}) can be rewritten in a block $(n\times
n) + (m \times m) $ form
\begin{equation}  \label{4dm}
g_{\alpha \beta }=\left(
\begin{array}{cc}
g_{ij}(x^k,y^a) & 0 \\
0 & h_{ab}(x^k,y^a)
\end{array}
\right)
\end{equation}
with respect to a subclass of $n+m$ anholonomic frame basis (for
four dimensions one used terms tetrads, or vierbiends)
\index{tetrad} \index{vierbiend}  defined
\begin{equation}  \label{4dder}
\delta _\alpha =(\delta _i,\partial _a)=\frac \delta {\partial
u^\alpha }=\left( \delta _i=\frac \delta {\partial x^i}=\frac
\partial {\partial x^i}-N_i^b\left( x^j,y^c\right) \frac \partial
{\partial y^b},\partial _a=\frac \partial {\partial y^a}\right)
\end{equation}
and
\begin{equation}  \label{4ddif}
\delta ^\beta =\left( d^i,\delta ^a\right) =\delta u^\beta =\left(
d^i=dx^i,\delta ^a=\delta y^a=dy^a+N_k^a\left( x^j,y^b\right)
dx^k\right) ,
\end{equation}
called the locally anisotropic bases (in brief, la--bases)
adapted to the coefficients $N_j^a.$ The $n\times n$ matrix
$g_{ij}$ defines the so--called horizontal metric (in brief,
h--metric) and the $m\times m$ matrix $h_{ab}$ defines the
vertical (v--metric) with respect to the associated nonlinear
connection (N--connection) structure given by its coefficients
$N_j^a\left( u^\alpha \right) $ from (\ref{4dder}) and (\ref
{4ddif}). The geometry of N--connections is studied in detail in
\cite {barth,ma94};\ here we shall consider its applications with
respect to anholonomic frames in general relativity and its
locally isotropic generalizations.

A frame structure $\delta _\alpha $ (\ref{4dder}) on $V^{(n+m)}$
is characterized by its anholonomy relations
\begin{equation}  \label{4anholon}
\delta _\alpha \delta _\beta -\delta _\beta \delta _\alpha
=w_{~\alpha \beta }^\gamma \delta _\gamma .
\end{equation}
with anholonomy coefficients $w_{~\beta \gamma }^\alpha .$The
elongation of partial derivatives (by N--coefficients) in the
locally adapted partial derivatives (\ref{4dder}) reflects the
fact that on the (pseudo) Riemannian space--time $V^{(n+m)}$ it
is modeled a generic local anisotropy characterized by the
anholonomy relations (\ref{4anholon}) when the anholonomy
coefficients are computed as follows
\begin{eqnarray}
w_{~ij}^k & = &
0,w_{~aj}^k=0,w_{~ia}^k=0,w_{~ab}^k=0,w_{~ab}^c=0,  \nonumber
\\
w_{~ij}^a & = & -\Omega _{ij}^a,w_{~aj}^b=-\partial
_aN_i^b,w_{~ia}^b=\partial _aN_i^b,  \nonumber
\end{eqnarray}
where
\[
\Omega _{ij}^a=\partial _iN_j^a-\partial _jN_i^a+N_i^b\partial
_bN_j^a-N_j^b\partial _bN_i^a
\]
defines the coefficients of the N--connection curvature, in brief,
N--curvature. On (pseudo) Riemannian space--times this is a
characteristic of a chosen anholonomic system of reference.

A N--connection $N$ defines a global decomposition,
\[
N:V^{(n+m)}=H^{(n)}\oplus V^{(m)},
\]
of spacetime $V^{(n+m)}$ into a $n$--dimensional horizontal subspace $%
H^{(n)} $ (with holonomic $x$--coordinates) and into a
$m$--dimensional
vertical subspace $V^{(m)}$ (with anisotropic, anholonomic, $y$%
--coordinates). This form of parametrizations of sets of mixt
holonomic--anholonomic frames is very useful for investigation,
for instance, of kinetic and thermodynamic systems in general
relativity, spinor and gauge field interactions in curved
space--times and for definition of non--trivial reductions from
higher dimension to lower dimension ones in Kaluza--Klein
theories. In the last case the N--connection could be treated as
a 'splitting' field into base's and extra dimensions with the
anholonomic (equivalently, anisotropic) structure defined from
some prescribed types of symmetries and constraints (imposed on a
physical system) or, for a different class of theories, with some
dynamical field equations following in the low energy limit of
string theories \cite{vstr1,vstr2} or from Einstein equations on
a higher dimension space.

The locally anisotropic spacetimes, anisoropic spacetimes, to be
investigated in this section  are considered to be some (pseudo)
Riemannian manifolds $V^{(n+m)}$ enabled with a frame, in
general, anholonomic structures of basis vector fields, $\delta
^\alpha =(\delta ^i,\delta ^a)$ and theirs duals $\delta _\alpha
=(\delta _i,\delta _a)$ (equivalently to an
associated N--connection structure), adapted to a symmetric metric field $%
g_{\alpha \beta }$ (\ref{4dm}) of necessary signature and to a
linear, in general nonsymmetric, connection $\Gamma _{~\beta
\gamma }^\alpha $ defining
the covariant derivation $D_\alpha $ satisfying the metricity conditions $%
D_\alpha g_{\beta \gamma }=0.$ The term anisotropic points to a
prescribed type of anholonomy structure. As a matter of
principle, on a (pseudo) Riemannian space--time, we can always,
at least locally, remove our considerations with respect to a
coordinate basis. In this case the geometric anisotopy is
modelled by metrics of type (\ref{4ansatz}). Such ansatz for
metrics are largely applied in modern Kaluza--Klein theory \cite
{over} where the N--conection structures have been not pointed
out because in the simplest approximation on topological
compactification of extra dimensions the N--connection geometry
is trivial. A rigorous analysis of systems with mixed
holonomic--anholonomic variables was not yet provided for general
relativity, extra dimension and gauge like gravity theories..

A $n+m$ anholonomic structure distinguishes (d) the geometrical
objects into h-- and v--components. Such objects are briefly
called d--tensors, d--metrics and/or d--connections. Their
components are defined with respect
to a locally anisotropic basis of type (\ref{4dder}), its dual (\ref{4ddif}%
), or their tensor products (d--linear or d--affine transforms of
such frames could also be considered). For instance, a covariant
and contravariant d--tensor $Z,$ is expressed
\begin{equation}
Z = Z_{~\beta }^\alpha \delta _\alpha \otimes \delta ^\beta =
Z_{~j}^i\delta _i\otimes d^j+Z_{~a}^i\delta _i\otimes \delta
^a+Z_{~j}^b\partial _b\otimes d^j+Z_{~a}^b\partial _b\otimes
\delta ^a.  \nonumber
\end{equation}

A linear d--connection $D$ on locally anisotropic space--time
$V^{(n+m)},$
\[
D_{\delta _\gamma }\delta _\beta =\Gamma _{~\beta \gamma }^\alpha
\left( x,y\right) \delta _\alpha ,
\]
is parametrized by non--trivial h--v--com\-po\-nents,
\begin{equation}  \label{4dcon}
\Gamma _{~\beta \gamma }^\alpha =\left(
L_{~jk}^i,L_{~bk}^a,C_{~jc}^i,C_{~bc}^a\right) .
\end{equation}

A metric on $V^{(n+m)}$ with $(m \times m)+ (n \times n)$  block
coefficients (\ref{4dm}) is written in distinguished form, as a
metric
d--tensor (in brief, d--metric), with respect to a locally anisotropic base (%
\ref{4ddif})
\begin{equation}  \label{4dmetric}
\delta s^2 = g_{\alpha \beta }\left( u\right) \delta ^\alpha
\otimes \delta^\beta = g_{ij}(x,y)dx^idx^j+h_{ab}(x,y)\delta
y^a\delta y^b.
\end{equation}

Some d--connection and d--metric structures are compatible if
there are satisfied the conditions
\[
D_\alpha g_{\beta \gamma }=0.
\]
For instance, a canonical compatible d--connection
\[
^c\Gamma _{~\beta \gamma }^\alpha =\left(
^cL_{~jk}^i,^cL_{~bk}^a,^cC_{~jc}^i,^cC_{~bc}^a\right)
\]
is defined by the coefficients of d--metric (\ref{4dmetric}),
$g_{ij}\left( x,y\right) $ and $h_{ab}\left( x,y\right) ,$ and by
the N--coefficients,
\begin{eqnarray}
^cL_{~jk}^i & = & \frac 12g^{in}\left( \delta _kg_{nj}+\delta
_jg_{nk}-\delta _ng_{jk}\right) ,  \label{4cdcon} \\
^cL_{~bk}^a & = & \partial _bN_k^a+\frac 12h^{ac}\left( \delta
_kh_{bc}-h_{dc}\partial _bN_i^d-h_{db}\partial _cN_i^d\right) ,  \nonumber \\
^cC_{~jc}^i & = & \frac 12g^{ik}\partial _cg_{jk},  \nonumber \\
^cC_{~bc}^a & = & \frac 12h^{ad}\left( \partial _ch_{db}+\partial
_bh_{dc}-\partial _dh_{bc}\right)  \nonumber
\end{eqnarray}
The coefficients of the canonical d--connection generalize for
locally an\-iso\-trop\-ic space--times the well known Christoffel
symbols; on a (pseudo) Riemannian spacetime with a fixed
anholonomic frame the d--connection coefficients transform
exactly into the metric connection coefficients.

For a d--connection (\ref{4dcon}) the components of torsion,
\begin{eqnarray}
&T\left( \delta _\gamma ,\delta _\beta \right) &=T_{~\beta \gamma
}^\alpha
\delta _\alpha ,  \nonumber \\
&T_{~\beta \gamma }^\alpha &= \Gamma _{~\beta \gamma }^\alpha
-\Gamma _{~\gamma \beta }^\alpha +w_{~\beta \gamma }^\alpha
\nonumber
\end{eqnarray}
are expressed via d--torsions \index{d--torsions}
\begin{eqnarray}
T_{.jk}^i & = & - T_{.kj}^i=L_{jk}^i-L_{kj}^i,\quad
T_{ja}^i=C_{.ja}^i,T_{aj}^i=-C_{ja}^i,  \nonumber \\
T_{.ab}^i & = & 0,\quad T_{.bc}^a=S_{.bc}^a=C_{bc}^a-C_{cb}^a,
\label{4dtors} \\
T_{.ij}^a & = & -\Omega _{ij}^a,\quad T_{.bi}^a= \partial _b N_i^a
-L_{.bj}^a,\quad T_{.ib}^a=-T_{.bi}^a.  \nonumber
\end{eqnarray}
We note that for symmetric linear connections the d--torsions are
induced as a pure anholonomic effect. They vanish with respect to
a coordinate frame of reference.

In a similar manner, putting non--vanishing coefficients
(\ref{4dcon}) into the formula for curvature,
\begin{eqnarray}
&&R\left( \delta _{\tau },\delta _{\gamma }\right) \delta _{\beta
}=R_{\beta
~\gamma \tau }^{~\alpha }\delta _{\alpha },  \nonumber \\
&&R_{\beta ~\gamma \tau }^{~\alpha }=\delta _{\tau }\Gamma
_{~\beta \gamma }^{\alpha }-\delta _{\gamma }\Gamma _{~\beta
\delta }^{\alpha }+\Gamma _{~\beta \gamma }^{\varphi }\Gamma
_{~\varphi \tau }^{\alpha }-\Gamma _{~\beta \tau }^{\varphi
}\Gamma _{~\varphi \gamma }^{\alpha }+\Gamma _{~\beta \varphi
}^{\alpha }w_{~\gamma \tau }^{\varphi },  \nonumber
\end{eqnarray}
we can compute the components of d--curvatures
\index{d--curvatures}
\begin{eqnarray}
R_{h.jk}^{.i} &=&\delta _{k}L_{.hj}^{i}-\delta
_{j}L_{.hk}^{i}+L_{.hj}^{m}L_{mk}^{i}-L_{.hk}^{m}L_{mj}^{i}-C_{.ha}^{i}%
\Omega _{.jk}^{a},  \nonumber \\
R_{b.jk}^{.a} &=&\delta _{k}L_{.bj}^{a}-\delta
_{j}L_{.bk}^{a}+L_{.bj}^{c}L_{.ck}^{a}-L_{.bk}^{c}L_{.cj}^{a}-C_{.bc}^{a}%
\Omega _{.jk}^{c},  \nonumber \\
P_{j.ka}^{.i} &=&\partial
_{k}L_{.jk}^{i}+C_{.jb}^{i}T_{.ka}^{b}-(\delta
_{k}C_{.ja}^{i}+L_{.lk}^{i}C_{.ja}^{l}-L_{.jk}^{l}C_{.la}^{i}-L_{.ak}^{c}C_{.jc}^{i}),
\nonumber \\
P_{b.ka}^{.c} &=&\partial
_{a}L_{.bk}^{c}+C_{.bd}^{c}T_{.ka}^{d}-(\delta
_{k}C_{.ba}^{c}+L_{.dk}^{c%
\,}C_{.ba}^{d}-L_{.bk}^{d}C_{.da}^{c}-L_{.ak}^{d}C_{.bd}^{c}),  \nonumber \\
S_{j.bc}^{.i} &=&\partial _{c}C_{.jb}^{i}-\partial
_{b}C_{.jc}^{i}+C_{.jb}^{h}C_{.hc}^{i}-C_{.jc}^{h}C_{hb}^{i},  \nonumber \\
S_{b.cd}^{.a} &=&\partial _{d}C_{.bc}^{a}-\partial
_{c}C_{.bd}^{a}+C_{.bc}^{e}C_{.ed}^{a}-C_{.bd}^{e}C_{.ec}^{a}.
\nonumber
\end{eqnarray}

The Ricci tensor \index{Ricci d--tensor}
\[
R_{\beta \gamma }=R_{\beta ~\gamma \alpha }^{~\alpha }
\]
has the d--components
\begin{eqnarray}
R_{ij} & = & R_{i.jk}^{.k},\quad R_{ia}=-^2P_{ia}=-P_{i.ka}^{.k},
\label{4dricci} \\
R_{ai} &= & ^1P_{ai}=P_{a.ib}^{.b},\quad R_{ab}=S_{a.bc}^{.c}.
\nonumber
\end{eqnarray}
We point out that because, in general, $^1P_{ai}\neq ~^2P_{ia},$
the Ricci d--tensor is non symmetric.

Having defined a d-metric of type (\ref{4dmetric}) in $V^{(n+m)}$
we can compute the scalar curvature
\[
\overleftarrow{R}=g^{\beta \gamma }R_{\beta \gamma }
\]
of a d-connection $D,$%
\begin{equation}  \label{4dscalar}
{\overleftarrow{R}}=\widehat{R}+S,
\end{equation}
where $\widehat{R}=g^{ij}R_{ij}$ and $S=h^{ab}S_{ab}.$

Now, by introducing the values (\ref{4dricci}) and
(\ref{4dscalar}) into the Einstein's equations
\[
R_{\beta \gamma }-\frac 12g_{\beta \gamma
}\overleftarrow{R}=k\Upsilon _{\beta \gamma },
\]
we can write down the system of field equations for locally
anisotropic gravity with  anholonomic (N--connection) structure:
\begin{eqnarray}
R_{ij}-\frac 12\left( \widehat{R}+S\right) g_{ij} & = & k\Upsilon
_{ij},
\label{4einsteq2} \\
S_{ab}-\frac 12\left( \widehat{R}+S\right) h_{ab} & = & k\Upsilon
_{ab},
\nonumber \\
^1P_{ai} & = & k\Upsilon _{ai},  \nonumber \\
^2P_{ia} & = & -k\Upsilon _{ia},  \nonumber
\end{eqnarray}
where $\Upsilon _{ij},\Upsilon _{ab},\Upsilon _{ai}$ and
$\Upsilon _{ia}$ are the components of the energy--momentum
d--tensor field $\Upsilon _{\beta \gamma }$ (which includes
possible cosmological constants, contributions of anholonomy
d--torsions (\ref{4dtors}) and matter) and $k$ is the coupling
constant.

The h- v- decomposition of gravitational field equations
(\ref{4einsteq2}) was introduced by Miron and Anastasiei
\cite{ma94} in their N--connection approach to generalized
Finsler and Lagrange spaces. It holds true as well on (pseudo)
Riemannian spaces, in general gravity; in this case we obtain the
usual form of Einstein equations if we transfer considerations
with respect to coordinate frames. If the N--coefficients are
prescribed by fixing the anholonomic frame of reference,
different classes of solutions
are to be constructed by finding the h-- and v--components, $g_{ij}$ and $%
h_{ab},$ of metric (\ref{4ansatz}), or its equivalent
(\ref{4dm}). A more general approach is to consider the
N--connection as 'free' but subjected to the condition that its
coefficients along with the d--metric components are chosen to
solve the Einsten equations in the form (\ref{4einsteq2}) for some
suggested symmetries, configurations of horizons and type of
singularities and well defined Cauchy problem. This way one can
construct new classes of metrics with generic local anisotropy
(see \cite{vbh}).

\section{Higher Order Anisotropic Structures}

Miron and Atanasiu \cite{mirata,m1,m2} developed the higher order
Lagrange and Finsler geometry with applications in mechanics in
order to geometrize the concepts of classical mechanics on higher
order tangent bundles. The work \cite{vstr2} was a proof that
higher order anisotropies (in brief, one writes abbreviations
like ha--, ha--superspace, ha--spacetime, ha--geometry and so on)
can be induced alternatively in low energy limits of (super)
string theories and a higher order superbundle N--connection
formalism was proposed. There were developed the theory of
spinors \cite{vhsp}, proposed models of ha--(super)gravity and
matter interactions on ha--spaces and defined the supersymmetric
stochastic calculus in ha--superspaces which were summarized in
the monograph \cite{vbook} containing a local (super) geometric
approach to so called ha--superstring and generalized
Finsler--Kaluza--Klein (super) gravities.

The aim of this section is to proof that higher order anisotropic
(ha--structu\-res) are induced by respective anholonomic frames
in higher dimension Einstein gravity, to present the basic
geometric background for a such moving frame formalism and
associated N--connections and to deduce the system of
gravitational field equations with respect to ha--frames.

\subsection{Ha--frames and corresponding N--connections}

Let us consider a (pseudo) Riemannian spacetime $V^{(\overline{n})}=V^{(n+%
\overline{m})}$ where the anisot\-rop\-ic dimension
$\overline{m}$ is split
into $z$ sub--dimensions $m_p,$ $(p=1,2,...,z),$ i. e. $\overline{m}%
=m_1+m_2+...+m_z.$ The local coordinates on a such higher
dimension curved space--time will be denoted as to take into
account the $m$--decomposition,
\begin{eqnarray}
u&=&\{u^{\overline \alpha}\equiv u^{\alpha
_z}=(x^i,y^{a_1},y^{a_2},\ldots
,y^{a_p},\ldots y^{a_z})\},  \nonumber \\
u^{\alpha _p}&=&(x^i,y^{a_1},y^{a_2},\ldots ,y^{a_p})=\left(
u^{\alpha _{p-1}},y^{a_p}\right) .  \nonumber
\end{eqnarray}
The la--constructions from the previous Section are considered to
describe anholonomic structures of first order; for $z=1$ we put
$u^{\alpha _1}=\left( x^i,y^{a_1}\right) =u^\alpha =\left(
x^i,y^{a_1}\right) .$ The higher order anisotropies are defined
inductively, 'shell by shell', starting from the first order to
the higher order, $z$--anisotropy. In order
to distinguish the components of geometrical objects with respect to a $p$%
--shell we provide both Greek and Latin indices with a
corresponding subindex like ${\alpha _p}=({\alpha
_{p-1}},{a_p}),$ and $a_p=(1,2,...,m_p),$ i. e. one holds a shell
parametrization for coordinates,
\[
y^{a_p}=(y_{(p)}^1=y^1,y_{(p)}^2=y^2,...,y_{(p)}^{m_p}=y^{m_p}).
\]
We shall overline some indices, for instance, $\overline{\alpha }$ and $%
\overline{a},$ if would be necessary to point that it could be
split into shell components and omit the $p$--shell mark $(p)$ if
this does not lead to misunderstanding. Such decompositions of
indices and geometrical and physical values are introduced with
the aim for a further modelling of (in general, dynamical)
spllittings of higher dimension spacetimes, step by step, with
'interior' subspaces being of different dimension, to lower
dimensions, with nontrival topology and anholonomic (anisotropy)
structures in generalized Kaluza--Klein theories.

The coordinate frames are denoted
\[
\partial _{\overline{\alpha }}=\partial /u^{\overline{\alpha }}=\left(
\partial /x^i,\partial /y^{a_1},...,\partial /y^{a_z}\right)
\]
with the dual ones
\[
d^{\overline{\alpha }}=du^{\overline{\alpha }}=\left(
dx^i,dy^{a_1},...,dy^{a_z}\right) ,
\]
when
\[
\partial _{\alpha _p}=\partial /u^{\alpha _p}=\left( \partial /x^i,\partial
/y^{a_1},...,\partial /y^{a_p}\right)
\]
and
\[
d^{\alpha _p}=du^{\alpha _p}=\left(
dx^i,dy^{a_1},...,dy^{a_p}\right)
\]
if considerations are limited to the $p$-th shell.

With respect to a coordinate frame a nonsigular metric
\[
ds^2=\widetilde{g}_{\overline{\alpha }\overline{\beta }}\ du^{\overline{%
\alpha }}\otimes du^{\overline{\beta }}
\]
with coefficients $\widetilde{g}_{\overline{\alpha
}\overline{\beta }}$ defined on induction,
\begin{eqnarray}  \label{4ansatz1}
\widetilde{g}_{\alpha _1\beta _1}&=&\left[
\begin{array}{cc}
g_{ij}+M_i^{a_1}M_j^{b_1}h_{a_1b_1} & M_j^{e_1}h_{a_1e_1} \\
M_i^{e_1}h_{b_1e_1} & h_{a_1b_1}
\end{array}
\right] , \\
& \vdots &  \nonumber \\
\widetilde{g}_{\alpha _p\beta _p} &=&\left[
\begin{array}{cc}
g_{\alpha _{p-1}\beta _{p-1}}+M_{\alpha _{p-1}}^{a_p}M_{\beta
_{p-1}}^{b_p}h_{a_pb_p} & M_{\beta _{p-1}}^{e_p}h_{a_pe_p} \\
M_{\alpha _{p-1}}^{e_p}h_{b_pe_p} & h_{a_pb_p}
\end{array}
\right] ,  \nonumber \\
& \vdots &  \nonumber \\
\widetilde{g}_{{\overline \alpha}{\overline \beta}} =
\widetilde{g}_{\alpha _z\beta _z} &=&\left[
\begin{array}{cc}
g_{\alpha _{z-1}\beta _{z-1}}+M_{\alpha _{z-1}}^{a_z}M_{\beta
_{z-1}}^{b_z}h_{a_zb_z} & M_{\beta _{z-1}}^{e_z}h_{a_ze_z} \\
M_{\alpha _{z-1}}^{e_z}h_{b_ze_z} & h_{a_zb_z}
\end{array}
\right] ,  \nonumber
\end{eqnarray}
where indices are split as $\alpha _1=\left( i_1,a_1\right) ,$
$\alpha _2=\left( \alpha _1,a_2\right) ,$ $\alpha _p=\left(
\alpha _{p-1},a_p\right) ;\ p=1,2,...z.$

The metric (\ref{4ansatz1}) on $V^{(\overline{n})}$ splits into
symmetric blocks of matrices of dimensions
\[
\left( n\times n\right) \oplus \left( m_1\times m_1\right) \oplus
...\oplus \left( m_z\times m_z\right) ,
\]
$n+m$ form
\begin{equation}  \label{4dm1}
g_{\alpha \beta }=\left(
\begin{array}{cccc}
g_{ij}(u) & 0 & \ldots & 0 \\
0 & h_{a_1b_1} & \ldots & 0 \\
\ldots & \ldots & \cdots & \ldots \\
0 & 0 & \ldots & h_{a_zb_z}
\end{array}
\right)
\end{equation}
with respect to an anholonomic frame basis defined on induction
\begin{eqnarray}  \label{4dder1}
\delta _{\alpha _p}&=&(\delta _{\alpha _{p-1}},\partial
_{a_p})=\left(
\delta _i,\delta _{a_1},...,\delta _{a_{p-1}},\partial _{a_p}\right) \\
&=&\frac \delta {\partial u^{\alpha _p}}= \left( \frac \delta
{\partial u^{\alpha _{p-1}}}=\frac \partial {\partial u^{\alpha
_{p-1}}}-N_{\alpha _{p-1}}^{b_p}\left( u\right) \frac \partial
{\partial y^{b_p}},\frac
\partial {\partial y^{a_p}}\right) ,  \nonumber
\end{eqnarray}
and
\begin{eqnarray}  \label{4ddif1}
\delta ^{\beta _p} &= &\left( d^i,\delta ^{{\overline
a}_p}\right) =\left(
d^i,\delta ^{a_1},...,\delta ^{a_{p-1}},\delta ^{a_p}\right) \\
&=& \delta u^{\beta _p} = \left( d^i=dx^i,\delta ^{{\overline
a}_p}=\delta
y^{{\overline a}_p}=dy^{{\overline a}_p}+ M_{\alpha _{p-1}}^{{\overline a}%
_p}\left( u\right) du^{\alpha _{p-1}}\right) ,  \nonumber
\end{eqnarray}
where $\overline{a}_p=\left( a_1,a_2,...,a_p\right) ,$ are called
the locally anisotropic bases (in brief la--bases) adapted
respectively to the N--coefficients
\[
N_{\alpha _{p-1}}^{a_p}=\left\{
N_i^{a_p},N_{a_1}^{a_p},...,N_{a_{p-2}}^{a_p},N_{a_{p-1}}^{a_p}\right\}
\]
and M--coefficients
\[
M_{\alpha _{p-1}}^{a_p}=\left\{
M_i^{a_p},M_{a_1}^{a_p},...,M_{a_{p-2}}^{a_p},M_{a_{p-1}}^{a_p}\right\}
;
\]
the coefficients $M_{\alpha _{p-1}}^{a_p}$ are related via some
algebraic relations with $N_{\alpha _{p-1}}^{a_p}$ in order to be
satisfied the locally anisotropic basis duality conditions
\[
\delta _{\alpha _p}\otimes \delta ^{\beta _p}={\delta }_{\alpha
_p}^{\beta _p},
\]
where ${\delta }_{\alpha _p}^{\beta _p}$ is the Kronecker symbol,
for every shell.

The geometric structure of N-- and M--coefficients of a higher
order nonlinear connection becomes more explicit if we write the
relations (\ref {4dder1}) and (\ref{4ddif1}) in matrix form,
respectively,
\[
\delta _{\bullet }=\widehat{N}\left( u\right) \times \partial
_{\bullet }
\]
and
\[
\delta ^{\bullet }=d^{\bullet }\times M\left( u\right) ,
\]
where
\[
\delta _{\bullet }=\delta _{\overline \alpha}=\left(
\begin{array}{c}
\delta _i \\
\delta _{a_1} \\
\delta _{a_2} \\
\cdots \\
\delta _{a_z}
\end{array}
\right) =\left(
\begin{array}{c}
\delta /\partial x^i \\
\delta /\partial y^{a_1} \\
\delta /\partial y^{a_2} \\
\cdots \\
\delta /\partial y^{a_z}
\end{array}
\right) ,\partial _{\bullet }=\partial _{\overline \alpha}=\left(
\begin{array}{c}
\partial _i \\
\partial _{a_1} \\
\partial _{a_2} \\
\cdots \\
\partial _{a_z}
\end{array}
\right) =\left(
\begin{array}{c}
\partial /\partial x^i \\
\partial /\partial y^{a_1} \\
\partial /\partial y^{a_2} \\
\cdots \\
\partial /\partial y^{a_z}
\end{array}
\right) ,
\]
\[
\delta ^{\bullet }=\left(
\begin{array}{ccccc}
dx^i & \delta y^{a_1} & \delta y^{a_2} & \ldots & \delta y^{a_z}
\end{array}
\right) ,\ d^{\bullet }=\left(
\begin{array}{ccccc}
dx^i & dy^{a_1} & dy^{a_2} & \ldots & dy^{a_z}
\end{array}
\right) ,
\]
and
\[
\widehat{N}=\left(
\begin{array}{ccccc}
1 & -N_i^{a_1} & -N_i^{a_2} & \ldots & -N_i^{a_z} \\
0 & 1 & -N_{a_1}^{a_2} & \ldots & -N_{a_1}^{a_z} \\
0 & 0 & 1 & \ldots & -N_{a_2}^{a_z} \\
\ldots & \ldots & \ldots & \ldots & \ldots \\
0 & 0 & 0 & \ldots & 1
\end{array}
\right) ,
\]
\[
\ M=\left(
\begin{array}{ccccc}
1 & M_i^{a_1} & M_i^{a_2} & \ldots & M_i^{a_z} \\
0 & 1 & M_{a_1}^{a_2} & \ldots & M_{a_1}^{a_z} \\
0 & 0 & 1 & \ldots & M_{a_2}^{a_z} \\
\ldots & \ldots & \ldots & \ldots & \ldots \\
0 & 0 & 0 & \ldots & 1
\end{array}
\right) .
\]

The $n\times n$ matrix $g_{ij}$ defines the horizontal metric (in brief, $h$%
--metric) and the $m_p\times m_p$ matrices $h_{a_pb_p}$ defines
the vertical, $v_p$--metrics with respect to the associated
nonlinear connection (N--connection) structure given by its
coefficients $N_{\alpha _{p-1}}^{a_p}$ from (\ref{4dder1}). The
geometry of N--connections on higher order tangent bundles is
studied in detail in \cite{mirata,m1,m2}, for vector
(super)bundles there it was proposed the approach from
\cite{vstr2,vbook}; the approach and denotations elaborated in
this work is adapted to further applications in higher dimension
Einstein gravity and its non--Riemannian locally anisotropic
extensions.

A ha--basis $\delta _{\overline \alpha }$ (\ref{4ddif}) on $V^{(\overline{n}%
)} $ is characterized by its anholonomy relations
\begin{equation}  \label{4anholon1}
\delta _{\overline{\alpha }}\delta _{\overline{\beta }}-\delta _{\overline{%
\beta }}\delta _{\overline{\alpha }}=w_{~\overline{\alpha }\overline{\beta }%
}^{\overline{\gamma }}\delta _{\overline{\gamma }}.
\end{equation}
with anholonomy coefficients $w_{~\overline{\alpha }\overline{\beta }}^{%
\overline{\gamma }}.$ The anholonomy  \index{anholonomy}
coefficients are computed
\begin{eqnarray}
w_{~ij}^k & = & 0;w_{~{a_p}j}^k=0;w_{~i{a_p}}^k=0;w_{~{a_p}{b_p}}^k=0; w_{~{%
a_p}{b_p}}^{c_p}=0;  \nonumber \\
w_{~ij}^{a_p} & = & -\Omega _{ij}^{a_p};
w_{~{a_p}j}^{b_p}=-\delta _{a_p}
N_i^{b_p}; w_{~i{a_p}}^{b_p}=\delta _{a_p} N_i^{b_p};  \nonumber \\
w_{~{a_p}{b_p}}^{k_p} & = & 0;w_{~{a_p}{b_f}}^{c_f}=0, f<p; w_{~{b_f}{a_p}%
}^{c_f}=0, f<p; w_{~{a_p}{b_p}}^{c_f}=0, f<p;  \nonumber \\
w_{~{c_f}{d_s}}^{a_p} & = & -\Omega _{{c_f}{d_s}}^{a_p},
(f,s<p);  \nonumber
\\
w_{~{a_p}{c_f}}^{b_p} &=& -\delta _{a_p} N_{c_f}^{b_p}, f<p; w_{~{c_f}{a_p}%
}^{b_p}=\delta _{a_p} N_{c_f}^{b_p}, f<p;  \nonumber
\end{eqnarray}
where
\begin{eqnarray}
\Omega _{ij}^{a_p}&=&\partial _iN_j^{a_p}-\partial
_jN_i^{a_p}+N_i^{b_p}\delta _{b_p}N_j^{a_p}- N_j^{b_p}\delta
_{b_p}N_i^{a_p},
\label{4ncurv1} \\
\Omega _{\alpha _f\beta _s}^{a_p} &=& \partial _{\alpha
_f}N_{\beta _s}^{a_p}-\partial _{\beta _s}N_{\alpha
_f}^{a_p}+N_{\alpha _f}^{b_p}\delta _{b_p}N_{\beta
_s}^{a_p}-N_{\beta_s}^{b_p} \delta _{b_p}N_{\alpha _f}^{a_p},
\nonumber
\end{eqnarray}
for $1\leq s,f<p,$ are the coefficients of higher order
N--connection curvature (N--curvature).

A higher order N--connection $N$ defines a global decomposition
\[
N:\ V^{(\overline{n})}=H^{(n)}\oplus V^{(m_1)}\oplus
V^{(m_2)}\oplus ...\oplus V^{(m_z)},
\]
of space--time $V^{(\overline{n})}$ into a $n$--dimensional
horizontal
subspace $H^{(n)}$ (with holonomic $x$--components) and into $m_p$%
--dimensional vertical subspaces $V^{(m_p)}$\newline (with
anisotropic, anholonomic, $y_{(p)}$--components).

\subsection{Distinguished linear connections}

In this section we consider fibered (pseudo) Riemannian manifolds $V^{(%
\overline{n})}$ enabled with anholonomic frame structures of
basis vector fields,\newline $\delta ^{\overline{\alpha
}}=(\delta ^i,\delta ^{\overline{a}})$ and theirs duals $\delta
_{\overline{\alpha }}=(\delta _i,\delta _{\overline{a}})$ with
associated N--connection structure, adapted to a symmetric metric field $g_{%
\overline{\alpha }\overline{\beta }}$ (\ref{4dm1}) and to a
linear, in
general nonsymmetric, connection $\Gamma _{~\overline{\beta }\overline{%
\gamma }}^{\overline{\alpha }}$ defining the covariant derivation $D_{%
\overline{\alpha }}$ satisfying the metricity conditions $D_{\overline{%
\alpha }}g_{\overline{\beta }\overline{\gamma }}=0.$ Such
space--times are provided with anholonomic higher order
anisotropic structures and, in brief, are called ha--spacetimes.

A higher order N--connection distinguishes (d) the geometrical
objects into h-- and $v_p$--components (d--tensors, d--metrics
and/or d--connections). For instance, a d-tensor field of type
$\left(
\begin{array}{cccccc}
p & r_1 & ... & r_p & ... & r_z \\
q & s_1 & ... & s_p & ... & s_z
\end{array}
\right) $ is written in local form as
\begin{eqnarray}
\mathbf{t} &=&t_{j_1...j_qb_1^{(1)}...b_{r_1}^{(1)}...b_1^{(p)}
...b_{r_p}^{(p)}...b_1^{(z)}...
b_{r_z}^{(z)}}^{i_1...i_pa_1^{(1)}...a_{r_1}^{(1)}...
a_1^{(p)}...a_{r_p}^{(p)}...a_1^{(z)}...a_{r_z}^{(z)}}\left(
u\right) \delta _{i_1}\otimes ...\otimes \delta _{i_p}\otimes
d^{j_1}\otimes ... \otimes
d^{j_q}\otimes  \nonumber \\
&{}& \delta _{a_1^{(1)}}\otimes ...\otimes \delta
_{a_{r_1}^{(1)}}\otimes \delta ^{b_1^{(1)}}...\otimes \delta
^{b_{s_1}^{(1)}}\otimes ...\otimes \delta _{a_1^{(p)}}\otimes
...\otimes \delta _{a_{r_p}^{(p)}}\otimes
...\otimes  \nonumber \\
&{}& \delta ^{b_1^{(p)}}...\otimes \delta ^{b_{s_p}^{(p)}}\otimes
\delta _{a_1^{(z)}}\otimes ...\otimes \delta
_{a_{rz}^{(z)}}\otimes \delta ^{b_1^{(z)}}...\otimes \delta
^{b_{s_z}^{(z)}}.  \nonumber
\end{eqnarray}

A linear d--connection $D$ on ha--spacetime $V^{(\overline{n})},$
\index{d--connection}
\[
D_{\delta _{\overline{\gamma }}}\delta _{\overline{\beta }}=\Gamma _{~%
\overline{\beta }\overline{\gamma }}^{\overline{\alpha }}\left(
u\right) \delta _{\overline{\alpha }},
\]
is defined by its non--trivial h--v--components,
\begin{equation}  \label{4dcon1}
\Gamma _{~\overline{\beta }\overline{\gamma }}^{\overline{\alpha
}}=\left(
L_{~jk}^i,L_{~\overline{b}k}^{\overline{a}},C_{~j\overline{c}}^i,C_{~%
\overline{b}\overline{c}}^{\overline{a}%
},K_{~b_pc_p}^{a_p},K_{~b_sc_f}^{a_p},Q_{~b_fc_p}^{a_f}\right) ,
\end{equation}
for $f<p,s.$

A metric with block coefficients (\ref{4dm1}) is written as a
d--metric, with respect to a la--base (\ref{4ddif1})
\begin{equation}  \label{4dmetric1}
\delta s^2=g_{\overline{\alpha }\overline{\beta }}\left( u\right) \delta ^{%
\overline{\alpha }}\otimes \delta ^{\overline{\beta }%
}=g_{ij}(u)dx^idx^j+h_{a_pb_p}(u)\delta y^{a_p}\delta y^{b_p},
\end{equation}
where $p=1,2,...,z.$

A d--connection and a d--metric structure are compatible if there
are satisfied the conditions
\[
D_{\overline{\alpha }}g_{\overline{\beta }\overline{\gamma }}=0.
\]

The canonical d--connection $^c\Gamma _{~\overline{\beta }\overline{\gamma }%
}^{\overline{\alpha }}$ is defined by the coefficients of
d--metric (\ref {4dmetric1}), and by the higher order
N--coefficients,
\begin{eqnarray}
^cL_{~jk}^i & = & \frac 12g^{in}\left( \delta _kg_{nj}+\delta
_jg_{nk}-\delta _ng_{jk}\right) ,  \label{4cdcon1} \\
^cL_{~{\overline b}k}^{\overline a} & = & \delta _{\overline
b}N_k^{\overline a}+ \frac 12h^{{\overline a}{\overline c}}
\left( \partial _k h_{{\overline b}{\overline c}}- h_{{\overline
d}{\overline c}} \delta _{\overline b} N_k^{\overline d}
-h_{{\overline d}{\overline b}}\delta
_{\overline c}N_k^{\overline d}\right),  \nonumber \\
^cC_{~j{\overline c}}^i & = & \frac 12g^{ik}\delta _{\overline
c}g_{jk},
\nonumber \\
^cC_{~{{\overline b}{\overline c}}}^{\overline a} & = & \frac 12h^{{%
\overline a}{\overline d}}\left( \delta _{\overline c}h_{{\overline d}{%
\overline b}}+ \delta_{\overline b}h_{{\overline d}{\overline
c}}-\delta
_{\overline d} h_{{\overline b}{\overline c}}\right),  \nonumber \\
^cK_{~{b_p}{c_p}}^{a_p} & = & \frac 12 g^{{a_p}{e_p}} \left(
\delta _{c_p} g_{{e_p}{b_p}}+ \delta _{b_p}g_{e_p c_p}-
\delta_{e_p} g_{b_p c_p}\right) ,
\nonumber \\
^cK_{~{b_s}e_f}^{a_p} & = & \delta _{b_s}N_{e_f}^{a_p}+ \frac
12h^{a_p c_p}\left( \partial _{e_f} h_{b_s c_p} -h_{d_p
c_p}\delta _{b_s}
N_{e_f}^{d_p} -h_{d_s b_s}\delta _{c_p}N_{e_f}^{d_s}\right) ,  \nonumber \\
^cQ_{~b_f c_p}^{a_f} & = & \frac 12 h^{a_f e_f}\delta
_{c_p}h_{b_f e_f}, \nonumber
\end{eqnarray}
where $f<p,s.$ They transform into usual Christoffel symbols with
respect to a coordinate base.

\subsection{Ha--torsions and ha--curvatures} \index{ha--torsion}
 \index{ha--curvature}

For a higher order anisotropic d--connection (\ref{4dcon1}) the
components of torsion,
\begin{eqnarray}
&T\left( \delta _{\overline \gamma} ,\delta _{\overline \beta}
\right) &= T_{~{\overline \beta}{\overline \gamma} }^{\overline
\alpha} \delta
_{\overline \alpha} ,  \nonumber \\
&T_{{\overline \beta} {\overline \gamma} }^{\overline \alpha} &= \Gamma _{~{%
\overline \beta}{\overline \gamma}}^{\overline \alpha} - \Gamma _{~{%
\overline \gamma}{\overline \beta} }^{\overline \alpha} +
w_{~{\overline \beta}{\overline \gamma} }^{\overline \alpha}
\nonumber
\end{eqnarray}
are expressed via d--torsions
\begin{eqnarray}
T_{.jk}^i&=&-T_{.kj}^i=L_{jk}^i-L_{kj}^i, \quad T_{j\overline{a}}^i=-T_{%
\overline{a}j}^i=C_{.j\overline{a}}^i,  \nonumber \\
T_{.\overline{a}\overline{b}}^i&=& 0, \quad T_{.\overline{b}\overline{c}}^{%
\overline{a}}= S_{.\overline{b}\overline{c}}^{\overline{a}}=C_{\overline{b}%
\overline{c}}^{\overline{a}}-C_{\overline{c}\overline{b}}^{\overline{a}},
\label{4dtors1} \\
T_{.ij}^{\overline{a}}& =& -\Omega _{ij}^{\overline{a}},\quad T_{.\overline{b%
}i}^{\overline{a}}= -T_{.\overline{b}i}^{\overline{a}}= \delta _{\overline{b}%
}N_i^{\overline{a}}-L_{.\overline{b}j}^{\overline{a}},   \nonumber \\
T_{.b_fc_f}^{a_f}& =&
-T_{.c_fb_f}^{a_f}=K_{.b_fc_f}^{a_f}-K_{.c_fb_f}^{a_f},
\nonumber \\
T_{.a_pb_s}^{a_f}&=& 0,\quad
T_{.b_fa_p}^{a_f}=-T_{.a_pb_f}^{a_f}=Q_{.b_fa_p}^{a_f},  \nonumber \\
T_{.a_fb_f}^{a_p}& = & -\Omega _{.a_fb_f}^{a_p},\quad
T_{.b_sa_f}^{a_p}=-T_{.a_fb_s}^{a_p}=\delta
_{b_s}N_{a_f}^{a_p}-K_{.b_sa_f}^{a_p}.  \nonumber
\end{eqnarray}

We note that for symmetric linear connections the d--torsion is
induced as a pure anholonomic effect.

In a similar manner, putting non--vanishing coefficients
(\ref{4dcon}) into the formula for curvature,
\begin{eqnarray}
&&R\left( \delta _{\overline{\tau }},\delta _{\overline{\gamma
}}\right)
\delta _{\overline{\beta }}=R_{\overline{\beta }~\overline{\gamma }\overline{%
\tau }}^{~\overline{\alpha }}\delta _{\overline{\alpha }},  \nonumber \\
&&R_{\overline{\beta }~\overline{\gamma }\overline{\tau }}^{~\overline{%
\alpha }}=\delta _{\overline{\tau }}\Gamma _{~\overline{\beta }\overline{%
\gamma }}^{\overline{\alpha }}-\delta _{\overline{\gamma }}\Gamma _{~%
\overline{\beta }\overline{\delta }}^{\overline{\alpha }}+\Gamma _{~%
\overline{\beta }\overline{\gamma }}^{\overline{\varphi }}\Gamma _{~%
\overline{\varphi }\overline{\tau }}^{\overline{\alpha }}-\Gamma _{~%
\overline{\beta }\overline{\tau }}^{\overline{\varphi }}\Gamma _{~\overline{%
\varphi }\overline{\gamma }}^{\overline{\alpha }}+\Gamma _{~\overline{\beta }%
\overline{\varphi }}^{\overline{\alpha }}w_{~\overline{\gamma }\overline{%
\tau }}^{\overline{\varphi }},  \nonumber
\end{eqnarray}
we can compute the components of d--curvatures
\begin{eqnarray}
R_{h.jk}^{.i} &=&\delta _{k}L_{.hj}^{i}-\delta
_{j}L_{.hk}^{i}+L_{.hj}^{m}L_{mk}^{i}-L_{.hk}^{m}L_{mj}^{i}-C_{.h\overline{a}%
}^{i}\Omega _{.jk}^{\overline{a}},  \label{4curvaturesha} \\
R_{\overline{b}.jk}^{.\overline{a}} &=&\delta _{k}L_{.\overline{b}j}^{%
\overline{a}}-\delta _{j}L_{.\overline{b}k}^{\overline{a}}+L_{.\overline{b}%
j}^{\overline{c}}L_{.\overline{c}k}^{\overline{a}}-L_{.\overline{b}k}^{%
\overline{c}}L_{.\overline{c}j}^{\overline{a}}-C_{.\overline{b}\overline{c}%
}^{\overline{a}}\Omega _{.jk}^{\overline{c}},  \nonumber\\
P_{j.k\overline{a}}^{.i} &=&\partial _{k}L_{.jk}^{i}+C_{.j\overline{b}%
}^{i}T_{.k\overline{a}}^{\overline{b}}-(\delta _{k}C_{.j\overline{a}%
}^{i}+L_{.lk}^{i}C_{.j\overline{a}}^{l}-L_{.jk}^{l}C_{.l\overline{a}%
}^{i}-L_{.\overline{a}k}^{\overline{c}}C_{.j\overline{c}}^{i}),  \nonumber \\
P_{\overline{b}.k\overline{a}}^{.\overline{c}} &=&\delta _{\overline{a}}L_{.%
\overline{b}k}^{\overline{c}}+C_{.\overline{b}\overline{d}}^{\overline{c}%
}T_{.k\overline{a}}^{\overline{d}}-(\delta _{k}C_{.\overline{b}\overline{a}%
}^{\overline{c}}+L_{.\overline{d}k}^{\overline{c}\,}C_{.\overline{b}%
\overline{a}}^{\overline{d}}-L_{.\overline{b}k}^{\overline{d}}C_{.\overline{d%
}\overline{a}}^{\overline{c}}-L_{.\overline{a}k}^{\overline{d}}C_{.\overline{%
b}\overline{d}}^{\overline{c}}),  \nonumber \\
S_{j.\overline{b}\overline{c}}^{.i} &=&\delta _{\overline{c}}C_{.j\overline{b%
}}^{i}-\delta _{\overline{b}}C_{.j\overline{c}}^{i}+C_{.j\overline{b}%
}^{h}C_{.h\overline{c}}^{i}-C_{.j\overline{c}}^{h}C_{h\overline{b}}^{i},
\nonumber \\
S_{\overline{b}.\overline{c}\overline{d}}^{.\overline{a}} &=&\delta _{%
\overline{d}}C_{.\overline{b}\overline{c}}^{\overline{a}}-\delta _{\overline{%
c}}C_{.\overline{b}\overline{d}}^{\overline{a}}+C_{.\overline{b}\overline{c}%
}^{\overline{e}}C_{.\overline{e}\overline{d}}^{\overline{a}}-C_{.\overline{b}%
\overline{d}}^{\overline{e}}C_{.\overline{e}\overline{c}}^{\overline{a}},
\nonumber
\end{eqnarray}
\begin{eqnarray}
 W_{b_{f}.c_{f}e_{f}}^{.a_{f}} &=&\delta
_{e_{f}}K_{.b_{f}c_{f}}^{a_{f}}-\delta
_{c_{f}}K_{.b_{f}e_{f}}^{a_{f}}+K_{.b_{f}c_{f}}^{h_{f}}K_{h_{f}e_{f}}^{a_{f}}
\nonumber \\
&{}&-K_{.b_{f}e_{f}}^{h_{f}}K_{h_{f}c_{f}}^{a_{f}}-Q_{.b_{f}a_{p}}^{a_{f}}%
\Omega _{.c_{f}e_{f}}^{a_{p}},  \nonumber \\
W_{b_{s}.c_{f}e_{f}}^{.a_{p}} &=&\delta
_{e_{f}}K_{.b_{s}c_{f}}^{a_{p}}-\delta
_{c_{f}}K_{.b_{s}e_{f}}^{a_{p}}+K_{.b_{s}c_{f}}^{c_{p}}K_{.c_{p}e_{f}}^{a_{p}}
\nonumber \\
&{}&-K_{.b_{s}e_{f}}^{c_{p}}L_{.c_{p}c_{f}}^{a_{p}}-K_{.b_{s}c_{p}}^{a_{p}}%
\Omega _{.c_{f}e_{f}}^{c_{p}},  \nonumber \\
Z_{b_{f}.c_{f}e_{f}}^{.a_{f}} &=&\partial
_{e_{p}}K_{.b_{f}c_{f}}^{a_{f}}+Q_{.b_{f}b_{p}}^{a_{f}}T_{.c_{f}e_{p}}^{b_{p}}
\nonumber \\
&{}&-(\delta
_{c_{f}}Q_{.b_{f}e_{p}}^{a_{f}}+K_{.h_{f}c_{f}}^{a_{f}}Q_{.b_{f}c_{p}}^{h_{f}}- %
K_{.b_{f}c_{f}}^{h_{f}}Q_{.h_{f}e_{p}}^{a_{f}}- %
K_{.e_{p}c_{f}}^{c_{p}}C_{.b_{f}c_{p}}^{a_{f}}), %
\nonumber \\
Z_{b_{r}.c_{f}e_{p}}^{.c_{s}} &=&\delta
_{e_{p}}K_{.b_{r}c_{f}}^{c_{s}}+K_{.b_{r}d_{f}}^{c_{s}}T_{.c_{f}e_{p}}^{d_{f}}
\nonumber \\
&{}&-(\delta
_{c_{f}}C_{.b_{r}e_{p}}^{c_{s}}+K_{.d_{f}c_{f}}^{c_{s}%
\,}C_{.b_{r}e_{p}}^{d_{f}}-K_{.b_{r}c_{f}}^{d_{t}}C_{.d_{t}e_{p}}^{c_{s}}- %
K_{.e_{p}c_{f}}^{d_{t}}C_{.b_{r}d_{t}}^{c_{s}}), %
\nonumber \\
Y_{b_{f}.c_{p}e_{p}}^{.a_{f}} &=&\delta
_{e_{p}}Q_{.b_{f}c_{p}}^{a_{f}}-\delta
_{c_{p}}Q_{.b_{f}e_{p}}^{a_{f}}+Q_{.b_{f}c_{p}}^{d_{f}}Q_{.d_{f}e_{p}}^{a_{f}}- %
Q_{.b_{f}e_{p}}^{d_{f}}Q_{d_{f}c_{p}}^{a_{f}}. \nonumber %
\end{eqnarray}
where $f<p,s,r,t.$

\subsection{Einstein equations with respect to ha--frames}

The Ricci tensor
\[
R_{\overline{\beta }\overline{\gamma }}=R_{\overline{\beta }~\overline{%
\gamma }\overline{\alpha }}^{~\overline{\alpha }}
\]
has the d--components
\begin{eqnarray}
R_{ij} & = & R_{i.jk}^{.k},\quad R_{i\overline a}=-^2P_{i
\overline
a}=-P_{i.k \overline a}^{.k},  \label{4dricci1} \\
R_{{\overline a}i} &= & ^1P_{{\overline a}i}= P_{{\overline a}.i{\overline b}%
}^{.\overline b}, \quad R_{{\overline a}{\overline b}}=S_{{\overline a}.{%
\overline b} {\overline c}}^{.{\overline c}}  \nonumber \\
R_{b_f c_f} & = & W_{b_f . c_f a_f}^{.a_f},\quad R_{e_p
b_f}=-^2P_{b_f
e_p}=- Z_{b_f . a_f e_p}^{.a_f},  \nonumber \\
R_{b_r c_f} &= & ^1P_{b_r c_f}= Z_{b_r . c_f e_s}^{.e_s}.
\nonumber
\end{eqnarray}
The Ricci d-tensor is non symmetric.

If a higher order d-metric of type (\ref{4dmetric1}) is defined in $V^{(%
\overline{n})},$ we can compute the scalar curvature
\[
\overline{R}=g^{\overline{\beta }\overline{\gamma }}R_{\overline{\beta }%
\overline{\gamma }}.
\]
of a d-connection $D,$%
\begin{equation}  \label{4dscalar1}
{\overline{R}}=\widehat{R}+\overline{S},
\end{equation}
where $\widehat{R}=g^{ij}R_{ij}$ and $\overline{S}=h^{\overline{a}\overline{b%
}}S_{\overline{a}\overline{b}}.$

The h-v parametrization of the gravitational field equations in
ha--spa\-ce\-ti\-mes is obtained by introducing the values
(\ref{4dricci1}) and (\ref{4dscalar1}) into the Einstein's
equations
\[
R_{\overline{\beta }\overline{\gamma }}-\frac 12g_{\overline{\beta }%
\overline{\gamma }}\overline{R}=k\Upsilon _{\overline{\beta }\overline{%
\gamma }},
\]
and written
\begin{eqnarray}
R_{ij}-\frac 12\left( \widehat{R}+ \overline S\right) g_{ij} & =
& k\Upsilon
_{ij},  \label{4einsteq3} \\
S_{{\overline a} \overline b}- \frac 12\left(
\widehat{R}+\overline S\right) h_{{\overline a}{\overline b}} & =
& k\Upsilon _{{\overline a}{\overline b}},
\nonumber \\
^1P_{{\overline a}i} = k\Upsilon _{{\overline a}i}, \quad
^1P_{a_p b_f} & =
& k \Upsilon _{a_p b_f}  \nonumber \\
^2P_{i \overline a} = -k\Upsilon _{i \overline a},\quad ^2P_{a_s
b_f} & = & - k \Upsilon _{a_f b_p},  \nonumber
\end{eqnarray}
where $\Upsilon _{ij},\Upsilon _{\overline{a}\overline{b}},\Upsilon _{%
\overline{a}i},\Upsilon _{i\overline{a}},\Upsilon
_{a_pb_f},\Upsilon _{a_fb_p}$ are the h-v--components of the
energy--mo\-men\-tum d--tensor field $\Upsilon _{\overline{\beta
}\overline{\gamma }}$ (which includes
possible cosmological constants, contributions of anholonomy d--torsions (%
\ref{4dtors1}) and matter) and $k$ is the coupling constant.

We note that, in general, the ha--torsions are not vanishing.
Nevetheless, for a (pseudo)--Riemannian spacetime with induced
anholonomic anisotropies it is not necessary to consider an
additional to (\ref{4einsteq3}) system of equations for torsion
becouse in this case the torsion structure is an anholonomic
effect wich becames trivial with respect to holonomic frames of
reference.

If a ha--spacetime structure is associated to a generic nonzero
torsion, we could consider additionally, for instance, as in
\cite{vg}, a system of
algebraic d--field equations with a source $S_{~\overline{\beta }\overline{%
\gamma }}^{\overline{\alpha }}$ for a locally anisotropic spin
density of matter (if we construct a variant of higher order
anisotropic Einstein--Cartan theory):
\[
T_{~\overline{\alpha }\overline{\beta }}^{\overline{\gamma }}+2\delta _{~[%
\overline{\alpha }}^{\overline{\gamma }}T_{~\overline{\beta }]\overline{%
\delta }}^{\overline{\delta }}=\kappa S_{~\overline{\alpha }\overline{\beta }%
.}^{\overline{\gamma }}
\]
In a more general case we have to introduce some new constraints
and/or dynamical equations for torsions and nonlinear connections
which are induced from (super) string theory and/ or higher order
anisotropic supergravity \cite{vstr1,vstr2}. Two variants of
gauge dynamical field equations with both frame like and torsion
variables will be considered in the Section 5 and 6 of this paper.

\section{Gauge Fields on Ha--Spaces}

This section is devoted to gauge field theories on spacetimes
provided with higher order anisotropic anholonomic frame
structures.  \index{ha--spaces}

\subsection{Bundles on ha--spaces}

Let us consider a principal bundle $\left( \mathcal{P},\pi ,Gr,V^{(\overline{%
n})}\right) $ over a ha--spacetime $V^{(\overline{n})}$ ($\mathcal{P}$ and $%
V^{(\overline{n})}$ are called respectively the base and total
spaces) with the structural group $Gr$ and surjective map $\pi
:\mathcal{P}\rightarrow V^{(\overline{n})}$ (on geometry of
bundle spaces see, for instance, \cite
{bishop,ma94,p}). At every point $u=(x,y_{(1)},...$ $,y_{(z)})$ $\in V^{(%
\overline{n})}$ there is a vicinity $\mathcal{U}\subset V^{(\overline{n}%
)},u\in \mathcal{U,}$ with trivializing $\mathcal{P}$
diffeomorphisms $f$
and $\varphi :$%
\begin{eqnarray}
f_{\mathcal{U}}:\ \pi ^{-1}\left( \mathcal{U}\right) &\rightarrow& \mathcal{%
U\times }Gr,\qquad f\left( p\right) =\left( \pi \left( p\right)
,\varphi
\left( p\right) \right) ,  \nonumber \\
\varphi _{\mathcal{U}}:\ \pi ^{-1}\left( \mathcal{U}\right)
&\rightarrow& Gr,\varphi (pq)=\varphi \left( p\right) q  \nonumber
\end{eqnarray}
for every group element $q\in Gr$ and point $~p\in \mathcal{P}.$
We remark that in the general case for two open regions
\[
\mathcal{U,V}\subset V^{(\overline{n})}\mathcal{,U\cap V}\neq \emptyset ,f_{%
\mathcal{U|}_p}\neq f_{\mathcal{V|}_p},\mbox{ even }p\in
\mathcal{U\cap V.}
\]

Transition functions $g_{\mathcal{UV}}$ are defined
\[
g_{\mathcal{UV}}:\mathcal{U\cap V\rightarrow
}Gr,g_{\mathcal{UV}}\left(
u\right) =\varphi _{\mathcal{U}}\left( p\right) \left( \varphi _{\mathcal{V}%
}\left( p\right) ^{-1}\right) ,\pi \left( p\right) =u.
\]

Hereafter we shall omit, for simplicity, the specification of
trivializing regions of maps and denote, for example, $f\equiv
f_{\mathcal{U}},\varphi \equiv \varphi _{\mathcal{U}},$ $s\equiv
s_{\mathcal{U}},$ if this will not give rise to ambiguities.

Let $\theta \,$ be the canonical left invariant 1-form on $Gr$
with values in algebra Lie $\mathcal{G}$ of group $Gr$ uniquely
defined from the relation $\theta \left( q\right) =q,$ for every
$q\in \mathcal{G,}$ and consider a 1-form $\omega $ on
$\mathcal{U}\subset V^{(\overline{n})}$ with values in
$\mathcal{G.}$ Using $\theta $ and $\omega ,$ we can locally
define the connection form $\Theta $ in $\mathcal{P}$ as a 1-form:
\begin{equation}  \label{4bundcon}
\Theta =\varphi ^{*}\theta +Ad\ \varphi ^{-1}\left( \pi
^{*}\omega \right)
\end{equation}
where $\varphi ^{*}\theta $ and $\pi ^{*}\omega $ are,
respectively, 1--forms induced on $\pi ^{-1}\left(
\mathcal{U}\right) $ and $\mathcal{P}$ by maps $\varphi $ and
$\pi $ and $\omega =s^{*}\Theta .$ The adjoint action on a form
$\lambda $ with values in $\mathcal{G}$ is defined as
\[
\left( Ad~\varphi ^{-1}\lambda \right) _p=\left( Ad~\varphi
^{-1}\left( p\right) \right) \lambda _p
\]
where $\lambda _p$ is the value of form $\lambda $ at point $p\in \mathcal{P}%
.$

Introducing a basis $\{\Delta _{\widehat{a}}\}$ in $\mathcal{G}$ (index $%
\widehat{a}$ enumerates the generators making up this basis), we
write the 1-form $\omega $ on $V^{(\overline{n})}$ as
\begin{equation}  \label{4form1}
\omega =\Delta _{\widehat{a}}\omega ^{\widehat{a}}\left( u\right) ,~\omega ^{%
\widehat{a}}\left( u\right) =\omega _{\overline{\mu
}}^{\widehat{a}}\left( u\right) \delta u^{\overline{\mu }}
\end{equation}
where $\delta u^{\overline{\mu }}=\left( dx^i,\delta
y^{\overline{a}}\right)
$ and the Einstein summation rule on indices $\widehat{a}$ and $\overline{%
\mu }$ is used. Functions $\omega _{\overline{\mu
}}^{\widehat{a}}\left( u\right) $ from (\ref{4form1}) are called
the components of Yang-Mills fields on ha-spacetime
$V^{(\overline{n})}.$ Gauge transforms of $\omega $
can be interpreted as transition relations for $\omega _{\mathcal{U}}$ and $%
\omega _{\mathcal{V}},$ when $u\in \mathcal{U\cap V,}$%
\begin{equation}  \label{4transf}
\left( \omega _{\mathcal{U}}\right) _u=\left(
g_{\mathcal{UV}}^{*}\theta
\right) _u+Ad~g_{\mathcal{UV}}\left( u\right) ^{-1}\left( \omega _{\mathcal{V%
}}\right) _u.
\end{equation}

To relate $\omega _{\overline{\mu }}^{\widehat{a}}$ with a
covariant
derivation we shall consider a vector bundle $\Xi $ associated to $\mathcal{P%
}.$ Let $\rho :Gr\rightarrow GL\left( \mathcal{R}^s\right) $ and
$\rho ^{\prime }:\mathcal{G}\rightarrow End\left( E^s\right) $
be, respectively,
linear representations of group $Gr$ and Lie algebra $\mathcal{G}$ (where $%
\mathcal{R}$ is the real number field$).$ Map $\rho $ defines a
left action on $Gr$ and associated vector bundle $\Xi =P\times
\mathcal{R}^s/Gr,~\pi _E:
E\rightarrow V^{(\overline{n})}.$ Introducing the standard basis $\xi _{%
\underline{i}}=\{\xi _{\underline{1}},\xi _{\underline{2}},...,\xi _{%
\underline{s}}\}$ in $\mathcal{R}^s,$ we can define the right action on $%
P\times $ $\mathcal{R}^s,\left( \left( p,\xi \right) q=\left(
pq,\rho \left(
q^{-1}\right) \xi \right) ,q\in Gr\right) ,$ the map induced from $\mathcal{P%
}$%
\[
p:\mathcal{R}^s\rightarrow \pi _E^{-1}\left( u\right) ,\quad
\left( p\left( \xi \right) =\left( p\xi \right) Gr,\xi \in
\mathcal{R}^s,\pi \left( p\right) =u\right)
\]
and a basis of local sections $e_{\underline{i}}:U\rightarrow \pi
_E^{-1}\left( U\right) ,~e_{\underline{i}}\left( u\right)
=s\left( u\right)
\xi _{\underline{i}}.$ Every section $\varsigma :V^{(\overline{n}%
)}\rightarrow \Xi $ can be written locally as $\varsigma
=\varsigma ^ie_i,\varsigma ^i\in C^\infty \left(
\mathcal{U}\right) .$ To every vector
field $X$ on $V^{(\overline{n})}$ and Yang-Mills field $\omega ^{\widehat{a}%
} $ on $\mathcal{P}$ we associate operators of covariant
derivations:
\begin{equation}  \label{4operat}
\bigtriangledown _X\zeta =e_{\underline{i}}\left[ X\zeta ^{\underline{i}%
}+B\left( X\right) _{\underline{j}}^{\underline{i}}\zeta ^{\underline{j}}%
\right] ,\ B\left( X\right) =\left( \rho ^{\prime }X\right) _{\widehat{a}%
}\omega ^{\widehat{a}}\left( X\right) .
\end{equation}
The transform (\ref{4transf}) and operators (\ref{4operat}) are
inter--related
by these transition transforms for values $e_{\underline{i}},\zeta ^{%
\underline{i}},$ and $B_{\overline{\mu }}:$
\begin{eqnarray}
e_{\underline{i}}^{\mathcal{V}}\left( u\right) & = & \left[ \rho g_{\mathcal{%
UV}}\left(u\right) \right] _{\underline{i}}^{\underline{j}}e_{\underline{i}%
}^{\mathcal{U}}, ~\zeta _{\mathcal{U}}^{\underline{i}}\left( u\right) = %
\left[ \rho g_{\mathcal{UV}}\left( u\right) \right] _{\underline{i}}^{%
\underline{j}} \zeta _{\mathcal{V}}^{\underline{i}},  \label{4transf1} \\
B_{\overline{\mu }}^{\mathcal{V}}\left( u\right) &=& \left[ \rho g_{\mathcal{%
UV}}\left( u\right) \right] ^{-1} \delta _{\overline{\mu }}\left[ \rho g_{%
\mathcal{UV}} \left( u\right) \right] +\left[ \rho
g_{\mathcal{UV}}\left( u\right) \right] ^{-1}B_{\overline{\mu
}}^{\mathcal{U}}\left( u\right) \left[ \rho
g_{\mathcal{UV}}\left( u\right) \right] ,  \nonumber
\end{eqnarray}
where $B_{\overline{\mu }}^{\mathcal{U}}\left( u\right) =B^{\overline{\mu }%
}\left( \delta /du^{\overline{\mu }}\right) \left( u\right) .$

Using (\ref{4transf1}), we can verify that the operator
$\bigtriangledown
_X^{\mathcal{U}}, $ acting on sections of $\pi _\Xi :\Xi \rightarrow V^{(%
\overline{n})}$ according to definition (\ref{4operat}),
satisfies the properties
\begin{eqnarray}
\bigtriangledown _{f_1X+f_2Y}^{\mathcal{U}} &= & f_1\bigtriangledown _X^{%
\mathcal{U}}+f_2\bigtriangledown _X^{\mathcal{U}},~\bigtriangledown _X^{%
\mathcal{U}}\left( f\zeta \right) =f\bigtriangledown
_X^{\mathcal{U}}\zeta
+\left( Xf\right) \zeta ,  \nonumber \\
\bigtriangledown _X^{\mathcal{U}}\zeta &=& \bigtriangledown _X^{\mathcal{V}%
}\zeta ,\quad u\in \mathcal{U\cap V,}f_1,f_2\in C^\infty \left( \mathcal{U}%
\right) .  \nonumber
\end{eqnarray}

So, we can conclude that the Yang--Mills connection in the vector bundle $%
\pi _\Xi :\Xi \rightarrow V^{(\overline{n})}$ is not a general
one, but is
induced from the principal bundle $\pi :\mathcal{P}\rightarrow V^{(\overline{%
n})}$ with structural group $Gr.$

The curvature $\mathcal{K}$ of connection $\Theta $ from
(\ref{4bundcon}) is defined as
\begin{equation}  \label{4bundcurv}
\mathcal{K}=D\Theta ,~D=\widehat{H}\circ d
\end{equation}
where $d$ is the operator of exterior derivation acting on $\mathcal{G}$%
-valued forms as
\[
d\left( \Delta _{\widehat{a}}\otimes \chi ^{\widehat{a}}\right) =\Delta _{%
\widehat{a}}\otimes d\chi ^{\widehat{a}}
\]
and $\widehat{H}\,$ is the horizontal projecting operator acting,
for example, on the 1-form $\lambda $ as $\left(
\widehat{H}\lambda \right) _P\left( X_p\right) =\lambda _p\left(
H_pX_p\right) ,$ where $H_p$ projects on the horizontal subspace
\[
\mathcal{H}_p\in P_p\left[ X_p\in \mathcal{H}_p\mbox{ is equivalent to }%
\Theta _p\left( X_p\right) =0\right] .
\]
We can express (\ref{4bundcurv}) locally as
\begin{equation}  \label{4bundcurv1}
\mathcal{K}=Ad~\varphi _{\mathcal{U}}^{-1}\left( \pi ^{*}\mathcal{K}_{%
\mathcal{U}}\right)
\end{equation}
where
\begin{equation}  \label{4bundcurv2}
\mathcal{K}_{\mathcal{U}}=d\omega _{\mathcal{U}}+\frac 12\left[ \omega _{%
\mathcal{U}},\omega _{\mathcal{U}}\right] .
\end{equation}
The exterior product of $\mathcal{G}$-valued form
(\ref{4bundcurv2}) is defined as
\[
\left[ \Delta _{\widehat{a}}\otimes \lambda ^{\widehat{a}},\Delta _{\widehat{%
b}}\otimes \xi ^{\widehat{b}}\right] =\left[ \Delta _{\widehat{a}},\Delta _{%
\widehat{b}}\right] \otimes \lambda ^{\widehat{a}}\bigwedge \xi ^{\widehat{b}%
},
\]
where the anti--symmetric tensorial product is denoted $\lambda ^{\widehat{a}%
}\bigwedge \xi ^{\widehat{b}}=\lambda ^{\widehat{a}}\xi ^{\widehat{b}}-\xi ^{%
\widehat{b}}\lambda ^{\widehat{a}}.$

Introducing structural coefficients
$f_{\widehat{b}\widehat{c}}^{\quad \widehat{a}}$ of $\mathcal{G}$
satisfying
\[
\left[ \Delta _{\widehat{b}},\Delta _{\widehat{c}}\right] =f_{\widehat{b}%
\widehat{c}}^{\quad \widehat{a}}\Delta _{\widehat{a}}
\]
we can rewrite (\ref{4bundcurv2}) in a form more convenient for
local considerations:

\begin{equation}  \label{4bundcurv3}
\mathcal{K}_{\mathcal{U}}=\Delta _{\widehat{a}}\otimes \mathcal{K}_{%
\overline{\mu }\overline{\nu }}^{\widehat{a}}\delta u^{\overline{\mu }%
}\bigwedge \delta u^{\overline{\nu }}
\end{equation}
where
\[
\mathcal{K}_{\overline{\mu }\overline{\nu
}}^{\widehat{a}}=\frac{\delta
\omega _{\overline{\nu }}^{\widehat{a}}}{\partial u^{\overline{\mu }}}-\frac{%
\delta \omega _{\overline{\mu }}^{\widehat{a}}}{\partial u^{\overline{\nu }}}%
+\frac 12f_{\widehat{b}\widehat{c}}^{\quad \widehat{a}}\left( \omega _{%
\overline{\mu }}^{\widehat{b}}\omega _{\overline{\nu
}}^{\widehat{c}}-\omega
_{\overline{\nu }}^{\widehat{b}}\omega _{\overline{\mu }}^{\widehat{c}%
}\right) .
\]

This subsection ends by considering the problem of reduction of
the local an\-i\-sot\-rop\-ic gauge symmetries and gauge fields
to isotropic ones. For local trivial considerations we can
consider that with respect to holonomic frames the higher order
anisotropic Yang-Mills fields reduce to usual ones on (pseudo)
Riemannian spaces.

\subsection{Yang-Mills equations on ha-spaces} \index{Yang--Mills}

Interior gauge symmetries are associated to semisimple structural
groups. On the principal bundle $\left( \mathcal{P},\pi
,Gr,V^{({\overline{n}})}\right) $ with nondegenerate Killing form
for semisimple group $Gr$ we can define the generalized bundle
metric
\begin{equation}  \label{4tmetric}
h_p\left( X_p,Y_p\right) =G_{\pi \left( p\right) }\left( d\pi
_PX_P,d\pi _PY_P\right) +K\left( \Theta _P\left( X_P\right)
,\Theta _P\left( X_P\right) \right) ,
\end{equation}
where $d\pi _P$ is the differential of map $\pi :\mathcal{P}\rightarrow V^{({%
\overline{n}})}\mathcal{,}$ $G_{\pi \left( p\right) }$ is locally
generated
as the ha-metric (\ref{4dmetric1}), and $K$ is the Killing form on $\mathcal{%
G:}$%
\[
K\left( \Delta _{\widehat{a}},\Delta _{\widehat{b}}\right) =f_{\widehat{b}%
\widehat{d}}^{\quad \widehat{c}}f_{\widehat{a}\widehat{c}}^{\quad \widehat{d}%
}=K_{\widehat{a}\widehat{b}}.
\]

Using the metric $g_{\overline{\alpha }\overline{\beta }}$ on $V^{({%
\overline{n}})}$ (respectively, $h_P\left( X_P,Y_P\right) $ on
$\mathcal{P)}$ we can introduce operators $*_G$ and
$\widehat{\delta }_G$ acting in the space of forms on
$V^{({\overline{n}})}$ ($*_H$ and $\widehat{\delta }_H$ acting on
forms on $\mathcal{P)}).$ Let $e_{\overline{\mu }}$ be an
orthonormalized frame on $\mathcal{U\subset
}V^{({\overline{n}})},$ locally adapted to the N--connection
structure, i. .e. being related via some local
distinguisherd linear transforms to a ha--frame (\ref{4dder1}) and $e^{%
\overline{\mu }}$ be the adjoint coframe. Locally
\[
G=\sum\limits_{\overline{\mu }}\eta \left( \overline{\mu }\right) e^{%
\overline{\mu }}\otimes e^{\overline{\mu }},
\]
where $\eta _{\overline{\mu }\overline{\mu }}=\eta \left( \overline{\mu }%
\right) =\pm 1,$ $\overline{\mu }=1,2,...,\overline{n},$ and the
Hodge \index{Hodge operator}
operator $*_G$ can be defined as $*_G:\Lambda ^{\prime }\left( V^{({%
\overline{n}})}\right) \rightarrow \Lambda ^{\overline{n}}\left( V^{({%
\overline{n}})}\right) ,$ or, in explicit form, as
\begin{equation}  \label{4hodge}
*_G\left( e^{\overline{\mu }_1}\bigwedge ...\bigwedge e^{\overline{\mu }%
_r}\right) =\eta \left( \overline{\nu }_1\right) ...\eta \left( \overline{%
\nu }_{\overline{n}-r}\right) \times
\end{equation}
\[
sign\left(
\begin{array}{ccccccc}
1 & 2 & \ldots & r & r+1 & \ldots & \overline{n} \\
\overline{\mu }_1 & \overline{\mu }_2 & \ldots & \overline{\mu
}_r & \overline{\nu }_1 & \ldots & \overline{\nu
}_{\overline{n}-r}
\end{array}
\right) e^{\overline{\nu }_1}\bigwedge ...\bigwedge e^{\overline{\nu }_{%
\overline{n}-r}}.
\]
Next, we define the operator
\[
*_G^{-1}=\eta \left( 1\right) ...\eta \left( \overline{n}\right)
\left( -1\right) ^{r\left( \overline{n}-r\right) }*_G
\]
and introduce the scalar product on forms $\beta _1,\beta
_2,...\subset \Lambda ^r\left( V^{({\overline{n}})}\right) $ with
compact carrier:
\[
\left( \beta _1,\beta _2\right) =\eta \left( 1\right) ...\eta
\left( n_E\right) \int \beta _1\bigwedge *_G\beta _2.
\]
The operator $\widehat{\delta }_G$ is defined as the adjoint to
$d$ associated to the scalar product for forms, specified for
$r$-forms as
\begin{equation}  \label{4adjoint}
\widehat{\delta }_G=\left( -1\right) ^r*_G^{-1}\circ d\circ *_G.
\end{equation}

We remark that operators $*_H$ and $\delta _H$ acting in the total space of $%
\mathcal{P}$ can be defined similarly to (\ref{4hodge}) and (\ref{4adjoint}%
), but by using metric (\ref{4tmetric}). Both these operators
also act in the space of $\mathcal{G}$-valued forms:
\[
*\left( \Delta _{\widehat{a}}\otimes \varphi ^{\widehat{a}}\right) =\Delta _{%
\widehat{a}}\otimes (*\varphi ^{\widehat{a}}),
\]
\[
\widehat{\delta }\left( \Delta _{\widehat{a}}\otimes \varphi ^{\widehat{a}%
}\right) =\Delta _{\widehat{a}}\otimes (\widehat{\delta }\varphi ^{\widehat{a%
}}).
\]

The form $\lambda $ on $\mathcal{P}$ with values in $\mathcal{G}$
is called horizontal if $\widehat{H}\lambda =\lambda $ and
equivariant if $R^{*}\left( q\right) \lambda =Ad~q^{-1}\varphi
,~\forall g\in Gr,R\left( q\right) $ being the right shift on
$\mathcal{P}.$ We can verify that equivariant and horizontal
forms also satisfy the conditions
\[
\lambda =Ad~\varphi _{\mathcal{U}}^{-1}\left( \pi ^{*}\lambda
\right) ,\qquad \lambda _{\mathcal{U}}=S_{\mathcal{U}}^{*}\lambda
,
\]
\[
\left( \lambda _{\mathcal{V}}\right) _{\mathcal{U}}=Ad~\left( g_{\mathcal{UV}%
}\left( u\right) \right) ^{-1}\left( \lambda
_{\mathcal{U}}\right) _u.
\]

Now, we can define the field equations for curvature
(\ref{4bundcurv1}) and connection (\ref{4bundcon}):
\begin{equation}  \label{4ym1}
\Delta \mathcal{K}=0,
\end{equation}
\begin{equation}  \label{4ym2}
\bigtriangledown \mathcal{K}=0,
\end{equation}
where $\Delta =\widehat{H}\circ \widehat{\delta }_H.$ Equations
(\ref{4ym1}) are similar to the well-known Maxwell equations and
for non-Abelian gauge
fields are called Yang-Mills equations. The structural equations (\ref{4ym2}%
) are called the Bianchi identities.

The field equations (\ref{4ym1}) do not have a physical meaning
because they are written in the total space of the bundle $\Xi $
and not on the base anisotropic spacetime $V^{({\overline{n}})}.$
But this difficulty may be
obviated by projecting the mentioned equations on the base. The 1-form $%
\Delta \mathcal{K}$ is horizontal by definition and its
equivariance follows from the right invariance of metric
(\ref{4tmetric}). So, there is a unique form $(\Delta
\mathcal{K})_{\mathcal{U}}$ satisfying
\[
\Delta \mathcal{K=}Ad~\varphi _{\mathcal{U}}^{-1}\pi ^{*}(\Delta \mathcal{K}%
)_{\mathcal{U}}.
\]
The projection of (\ref{4ym1}) on the base can be written as
$(\Delta
\mathcal{K})_{\mathcal{U}}=0.$ To calculate $(\Delta \mathcal{K})_{\mathcal{U%
}},$ we use the equality \cite{bishop,pd}
\[
d\left( Ad~\varphi _{\mathcal{U}}^{-1}\lambda \right) =Ad~~\varphi _{%
\mathcal{U}}^{-1}~d\lambda -\left[ \varphi
_{\mathcal{U}}^{*}\theta ,Ad~\varphi _{\mathcal{U}}^{-1}\lambda
\right]
\]
where $\lambda $ is a form on $\mathcal{P}$ with values in
$\mathcal{G.}$ For $r$-forms we have
\[
\widehat{\delta }\left( Ad~\varphi _{\mathcal{U}}^{-1}\lambda
\right) =Ad~\varphi _{\mathcal{U}}^{-1}\widehat{\delta }\lambda
-\left( -1\right)
^r*_H\{\left[ \varphi _{\mathcal{U}}^{*}\theta ,*_HAd~\varphi _{\mathcal{U}%
}^{-1}\lambda \right]
\]
and, as a consequence,
\begin{equation}  \label{4aux1}
\widehat{\delta }\mathcal{K}=Ad~\varphi
_{\mathcal{U}}^{-1}\{\widehat{\delta
}_H\pi ^{*}\mathcal{K}_{\mathcal{U}}+*_H^{-1}[\pi ^{*}\omega _{\mathcal{U}%
},*_H\pi ^{*}\mathcal{K}_{\mathcal{U}}]\} -*_H^{-1}\left[ \Theta
,Ad~\varphi _{\mathcal{U}}^{-1}*_H\left( \pi
^{*}\mathcal{K}\right) \right] .
\end{equation}
By using straightforward calculations in the adapted dual basis
on $\pi ^{-1}\left( \mathcal{U}\right) $ we can verify the
equalities
\begin{equation}  \label{4aux2}
\left[ \Theta ,Ad~\varphi _{\mathcal{U}}^{-1}~*_H\left( \pi ^{*}\mathcal{K}_{%
\mathcal{U}}\right) \right] =0,\widehat{H}\delta _H\left( \pi ^{*}\mathcal{K}%
_{\mathcal{U}}\right) =\pi ^{*}\left( \widehat{\delta
}_G\mathcal{K}\right) ,
\end{equation}
\[
*_H^{-1}\left[ \pi ^{*}\omega _{\mathcal{U}},*_H\left( \pi ^{*}\mathcal{K}_{%
\mathcal{U}}\right) \right] =\pi ^{*}\{*_G^{-1}\left[ \omega _{\mathcal{U}%
},*_G\mathcal{K}_{\mathcal{U}}\right] \}.
\]
From (\ref{4aux1}) and (\ref{4aux2}) one follows that
\begin{equation}  \label{4aux3}
\left( \Delta \mathcal{K}\right) _{\mathcal{U}}=\widehat{\delta }_G\mathcal{K%
}_{\mathcal{U}}+*_G^{-1}\left[ \omega _{\mathcal{U}},*_G\mathcal{K}_{%
\mathcal{U}}\right] .
\end{equation}

Taking into account (\ref{4aux3}) and (\ref{4adjoint}), we prove
that projection on the base of equations (\ref{4ym1}) and
(\ref{4ym2}) can be expressed respectively as
\begin{equation}  \label{4ym3}
*_G^{-1}\circ d\circ *_G\mathcal{K}_{\mathcal{U}}+*_G^{-1}\left[ \omega _{%
\mathcal{U}},*_G\mathcal{K}_{\mathcal{U}}\right] =0.
\end{equation}
\[
d\mathcal{K}_{\mathcal{U}}+\left[ \omega _{\mathcal{U}},\mathcal{K}_{%
\mathcal{U}}\right] =0.
\]

Equations (\ref{4ym3}) (see (\ref{4aux3})) are gauge--invariant
because
\[
\left( \Delta \mathcal{K}\right) _{\mathcal{U}}=Ad~g_{\mathcal{UV}%
}^{-1}\left( \Delta \mathcal{K}\right) _{\mathcal{V}}.
\]

By using formulas (\ref{4bundcurv3})-(\ref{4adjoint}) we can
rewrite (\ref {4ym3}) in coordinate form
\begin{equation}  \label{4ym4}
D_{\overline{\nu }}\left( G^{\overline{\nu }\overline{\lambda }}\mathcal{K}%
_{~\overline{\lambda }\overline{\mu }}^{\widehat{a}}\right) +f_{\widehat{b}%
\widehat{c}}^{\quad \widehat{a}}g^{\overline{v}\overline{\lambda }}\omega _{%
\overline{\lambda }}^{~\widehat{b}}\mathcal{K}_{~\overline{\nu }\overline{%
\mu }}^{\widehat{c}}=0,
\end{equation}
where $D_{\overline{\nu }}$ is a compatible with metric covariant
derivation on ha-spacetime (\ref{4ym4}).

We point out that for our bundles with semisimple structural
groups the Yang-Mills equations (\ref{4ym1}) (and, as a
consequence, their horizontal projections (\ref{4ym3}), or
(\ref{4ym4})) can be obtained by variation of the action
\begin{equation}  \label{4action}
I=\int \mathcal{K}_{~\overline{\mu }\overline{\nu }}^{\widehat{a}}\mathcal{K}%
_{~\overline{\alpha }\overline{\beta }}^{\widehat{b}}G^{\overline{\mu }%
\overline{\alpha }}g^{\overline{\nu }\overline{\beta }}K_{\widehat{a}%
\widehat{b}}\left| g_{\overline{\alpha }\overline{\beta }}\right|
^{1/2}dx^1...dx^n\delta y_{(1)}^1...\delta y_{(1)}^{m_1}...\delta
y_{(z)}^1...\delta y_{(z)}^{m_z}.
\end{equation}
Equations for extremals of (\ref{4action}) have the form
\[
K_{\widehat{r}\widehat{b}}g^{\overline{\lambda }\overline{\alpha }}g^{%
\overline{\kappa }\overline{\beta }}D_{\overline{\alpha }}\mathcal{K}_{~%
\overline{\lambda }\overline{\beta }}^{\widehat{b}}-K_{\widehat{a}\widehat{b}%
}g^{\overline{\kappa }\overline{\alpha }}g^{\overline{\nu }\overline{\beta }%
}f_{\widehat{r}\widehat{l}}^{\quad \widehat{a}}\omega _{\overline{\nu }}^{%
\widehat{l}}\mathcal{K}_{~\overline{\alpha }\overline{\beta }}^{\widehat{b}%
}=0,
\]
which are equivalent to ''pure'' geometric equations (\ref{4ym4})
(or (\ref
{4ym3})) due to nondegeneration of the Killing form $K_{\widehat{r}\widehat{b%
}}$ for semisimple groups.

To take into account gauge interactions with matter fields
(sections of vector bundle $\Xi $ on $V^{({\overline{n}})}$) we
have to introduce a source 1--form $\mathcal{J}$ in equations
(\ref{4ym1}) and to write them
\begin{equation}  \label{4yms}
\Delta \mathcal{K}=\mathcal{J}
\end{equation}

Explicit constructions of $\mathcal{J}$ require concrete
definitions of the bundle $\Xi ;$ for example, for spinor fields
an invariant formulation of the Dirac equations on ha--spaces is
necessary. We omit spinor considerations in this paper (see
\cite{vjmp,vhsp}).

\section{Gauge Ha-gravity} \index{gauge ha--gravity}

A considerable body of work on the formulation of gauge
gravitational models on isotropic spaces is based on application
of nonsemisimple groups, for example, of Poincare and affine
groups, as structural gauge groups (see critical analysis and
original results in \cite {deh,vg,mielke,hgmn,wal,tseyt,pbo}).
The main impediment to developing such models is caused by the
degeneration of Killing forms for nonsemisimple groups, which
make it impossible to construct consistent variational gauge
field theories (functional (\ref{4action}) and extremal equations
are degenerate in these cases). There are at least two
possibilities to get around the mentioned difficulty.\ The first
is to realize a minimal extension of the nonsemisimple group to a
semisimple one, similar to the extension of the Poincare group to
the de Sitter group considered in \cite {p,pd,tseyt}. The second
possibility is to introduce into consideration the
bundle of adapted affine frames on locally anisotropic space $V^{({\overline{%
n}})},$ to use an auxiliary nondegenerate bilinear form $a_{\widehat{a}%
\widehat{b}}$ instead of the degenerate Killing form $K_{\widehat{a}\widehat{%
b}}$ and to consider a ''pure'' geometric method, illustrated in
the previous section, of definition of gauge field equations.
Projecting on the base $V^{({\overline{n}})},$ we shall obtain
gauge gravitational field equations on a ha--space having a form
similar to Yang-Mills equations.

The goal of this section is to prove that a specific
parametrization of components of the Cartan connection in the
bundle of adapted affine frames on $V^{({\overline{n}})}$
establishes an equivalence between Yang-Mills equations
(\ref{4yms}) and Einstein equations (\ref{4einsteq3}) on
ha--spaces.

\subsection{Bundles of linear ha--frames} \index{ha--frames}

Let $\left( X_{\overline{\alpha }}\right) _u=\left( X_i,X_{\overline{a}%
}\right) _u=\left( X_i,X_{a_1},...,X_{a_z}\right) _u$ be a frame
locally adapted to the N--connection structure at a point $u\in
V^{({\overline{n}})}.$ We consider a local right distinguished
action of matrices
\[
A_{\overline{\alpha }^{\prime }}^{\quad \overline{\alpha }}=\left(
\begin{array}{cccc}
A_{i^{\prime }}^{\quad i} & 0 & ... & 0 \\
0 & B_{a_1^{\prime }}^{\quad a_1} & ... & 0 \\
... & ... & ... & ... \\
0 & 0 & ... & B_{a_z^{\prime }}^{\quad a_z}
\end{array}
\right) \subset GL_{\overline{n}}=
\]
\[
GL\left( n,\mathcal{R}\right) \oplus GL\left(
m_1,\mathcal{R}\right) \oplus ...\oplus GL\left(
m_z,\mathcal{R}\right) .
\]
Nondegenerate matrices $A_{i^{\prime }}^{\quad i}$ and
$B_{j^{\prime }}^{\quad j},$ respectively, transform linearly
$X_{i|u}$ into $X_{i^{\prime
}|u}=A_{i^{\prime }}^{\quad i}X_{i|u}$ and $X_{a_p^{\prime }|u}$ into $%
X_{a_p^{\prime }|u}=B_{a_p^{\prime }}^{\quad a_p}X_{a_p|u},$ where $X_{%
\overline{\alpha }^{\prime }|u}=A_{\overline{\alpha }^{\prime
}}^{\quad \overline{\alpha }}X_{\overline{\alpha }}$ is also an
adapted frame at the
same point $u\in V^{({\overline{n}})}.$ We denote by $La\left( V^{({%
\overline{n}})}\right) $ the set of all adapted frames $X_{\overline{\alpha }%
}$ at all points of $V^{({\overline{n}})}$ and consider the surjective map $%
\pi $ from $La\left( V^{({\overline{n}})}\right) $ to
$V^{({\overline{n}})}$ transforming every adapted frame
$X_{\overline{\alpha }|u}$ and point $u$ into the point $u.$
Every $X_{\overline{\alpha }^{\prime }|u}$ has a unique
representation as $X_{\overline{\alpha }^{\prime }}=A_{\overline{\alpha }%
^{\prime }}^{\quad \overline{\alpha }}X_{\overline{\alpha
}}^{\left( 0\right) },$ where $X_{\overline{\alpha }}^{\left(
0\right) }$ is a fixed distinguished basis in tangent space
$T\left( V^{({\overline{n}})}\right) .$ It is obvious that $\pi
^{-1}\left( \mathcal{U}\right) ,\mathcal{U}\subset
V^{({\overline{n}})},$ is bijective to $\mathcal{U}\times GL_{\overline{n}%
}\left( \mathcal{R}\right) .$ We can transform $La\left( V^{({\overline{n}}%
)}\right) $ in a differentiable manifold taking $\left( u^{\overline{\beta }%
},A_{\overline{\alpha }^{\prime }}^{\quad \overline{\alpha
}}\right) $ as a local coordinate system on $\pi ^{-1}\left(
\mathcal{U}\right) .$ Now, it is easy to verify that
\[
\mathcal{L}a(V^{({\overline{n}})})=(La(V^{({\overline{n}})}),V^{({\overline{n%
}})},GL_{\overline n}(\mathcal{R}))
\]
is a principal bundle. We call
$\mathcal{L}a(V^{({\overline{n}})})$ the bundle of linear adapted
frames on $V^{({\overline{n}})}.$

The next step is to identify the components of, for simplicity,
compatible
d-connection $\Gamma _{\overline{\beta }\overline{\gamma }}^{\overline{%
\alpha }}$ on $V^{({\overline{n}})},$ with the connection in $\mathcal{L}%
a(V^{({\overline{n}})})$%
\begin{equation}  \label{4conb}
\Theta _{\mathcal{U}}^{\widehat{a}}=\omega
^{\widehat{a}}=\{\omega _{\quad
\overline{\lambda }}^{\widehat{\alpha }\widehat{\beta }}\doteq \Gamma _{%
\overline{\beta }\overline{\gamma }}^{\overline{\alpha }}\}.
\end{equation}
Introducing (\ref{4conb}) in (\ref{4aux3}), we calculate the
local 1-form
\begin{equation}  \label{4aux4}
\left( \Delta \mathcal{R}^{(\Gamma )}\right) _{\mathcal{U}}=\Delta _{%
\widehat{\alpha }\widehat{\alpha }_1}\otimes (g^{\overline{\nu }\overline{%
\lambda }}D_{\overline{\lambda }}\mathcal{R}_{\quad \overline{\nu }\overline{%
\mu }}^{\widehat{\alpha }\widehat{\gamma }}+f_{\quad \widehat{\beta }%
\widehat{\delta }\widehat{\gamma }\widehat{\varepsilon }}^{\widehat{\alpha }%
\widehat{\gamma }}g^{\overline{\nu }\overline{\lambda }}\omega
_{\quad
\lambda }^{\widehat{\beta }\widehat{\delta }}\mathcal{R}_{\quad \overline{%
\nu }\overline{\mu }}^{\widehat{\gamma }\widehat{\varepsilon }})\delta u^{%
\overline{\mu }},
\end{equation}
where
\[
\Delta _{\widehat{\alpha }\widehat{\beta }}=\left(
\begin{array}{cccc}
\Delta _{\widehat{i}\widehat{j}} & 0 & ... & 0 \\
0 & \Delta _{\widehat{a}_1\widehat{b}_1} & ... & 0 \\
... & ... & ... & ... \\
0 & 0 & ... & \Delta _{\widehat{a}_z\widehat{b}_z}
\end{array}
\right)
\]
is the standard distinguished basis in the Lie algebra of matrices ${\mathcal%
{G}l}_{\overline{n}}\left( \mathcal{R}\right) $ with $(\Delta
_{\widehat{i}
\widehat{k}}) _{jl}$ $= \delta _{ij}\delta _{kl}$ and $\left( \Delta _{%
\widehat{a}_p\widehat{c}_p}\right) _{b_pd_p} = \delta
_{a_pb_p}\delta
_{c_pd_p} $ defining the stand\-ard bas\-es in $\mathcal{G}l\left( \mathcal{R%
}^{\overline{n}}\right) .$ We have denoted the curvature of
connection (\ref {4conb}), considered in (\ref{4aux4}), as
\[
\mathcal{R}_{\mathcal{U}}^{(\Gamma )}=\Delta _{\widehat{\alpha }\widehat{%
\alpha }_1}\otimes \mathcal{R}_{\quad \overline{\nu }\overline{\mu }}^{%
\widehat{\alpha }\widehat{\alpha }_1}X^{\overline{\nu }}\bigwedge X^{%
\overline{\mu }},
\]
where $\mathcal{R}_{\quad \overline{\nu }\overline{\mu }}^{\widehat{\alpha }%
\widehat{\alpha }_1}=R_{\overline{\alpha }_1\quad \overline{\nu }\overline{%
\mu }}^{\quad \overline{\alpha }}$ (see curvatures
(\ref{4curvaturesha})).

\subsection{Bundles of affine ha--frames and Einstein equations} \index{affine ha--frames}

Besides the bundles $\mathcal{L}a\left(
V^{({\overline{n}})}\right) $ on ha-spacetime
$V^{({\overline{n}})},$ there is another bundle, the bundle of
adapted affine frames with structural group $Af_{n_E}\left( \mathcal{R}%
\right) =GL_{n_E}\left( V^{({\overline{n}})}\right) $ $\otimes \mathcal{R}^{%
\overline{n}},$ which can be naturally related to the gravity
models on
(pseudo) Riemannian spaces. Because as a linear space the Lie Algebra $af_{%
\overline{n}}\left( \mathcal{R}\right) $ is a direct sum of ${\mathcal{G}l}_{%
\overline{n}}\left( \mathcal{R}\right) $ and
$\mathcal{R}^{\overline{n}},$
we can write forms on $\mathcal{A}a\left( V^{({\overline{n}})}\right) $ as $%
\Phi =\left( \Phi _1,\Phi _2\right) ,$ where $\Phi _1$ is the ${\mathcal{G}l}%
_{\overline{n}}\left( \mathcal{R}\right) $ component and $\Phi _2$ is the $%
\mathcal{R}^{\overline{n}}$ component of the form $\Phi .$ The connection (%
\ref{4conb}), $\Theta $ in ${\mathcal{L}a}\left(
V^{({\overline{n}})}\right)
,$ induces the Cartan connection $\overline{\Theta }$ in ${\mathcal{A}a}%
\left( V^{({\overline{n}})}\right) ;$ see the isotropic case in
\cite
{p,pd,bishop}. There is only one connection on ${\mathcal{A}a}\left( V^{({%
\overline{n}})}\right) $ represented as $i^{*}\overline{\Theta
}=\left(
\Theta ,\chi \right) ,$ where $\chi $ is the shifting form and $i:{\mathcal{A%
}a}\rightarrow {\mathcal{L}a}$ is the trivial reduction of bundles. If $s_{%
\mathcal{U}}^{(a)}$ is a local adapted frame in ${\mathcal{L}a}\left( V^{({%
\overline{n}})}\right) ,$ then
$\overline{s}_{\mathcal{U}}^{\left( 0\right)
}=i\circ s_{\mathcal{U}}$ is a local section in ${\mathcal{A}a}\left( V^{({%
\overline{n}})}\right) $ and
\begin{equation}  \label{4curvbaf}
\left( \overline{\Theta }_{\mathcal{U}}\right)
=s_{\mathcal{U}}\Theta =\left( \Theta _{\mathcal{U}},\chi
_{\mathcal{U}}\right) ,
\end{equation}
where $\chi =e_{\widehat{\alpha }}\otimes \chi _{~\overline{\mu }}^{\widehat{%
\alpha }}X^{\overline{\mu }},$ $g_{\overline{\alpha
}\overline{\beta }}=\chi
_{~\overline{\alpha }}^{\widehat{\alpha }}\chi _{~\overline{\beta }}^{%
\widehat{\beta }}\eta _{\widehat{\alpha }\widehat{\beta }}\quad (\eta _{%
\widehat{\alpha }\widehat{\beta }}$ is diagonal with $\eta
_{\widehat{\alpha }\widehat{\alpha }}=\pm 1)$ is a frame
decomposition of metric (\ref {4dmetric1}) on
$V^{({\overline{n}})},e_{\widehat{\alpha }}$ is the standard
distinguished basis on $\mathcal{R}^{\overline{n}},$ and the
projection of torsion , $T_{\mathcal{U}},$ on the base
$V^{({\overline{n}})}$ is defined as
\begin{equation}  \label{4torsbaf}
T_{\mathcal{U}}=d\chi _{\mathcal{U}}+\Omega _{\mathcal{U}}\bigwedge \chi _{%
\mathcal{U}}+\chi _{\mathcal{U}}\bigwedge \Omega _{\mathcal{U}}=e_{\widehat{%
\alpha }}\otimes \sum\limits_{\overline{\mu }\overline{\nu }}T_{~\overline{%
\mu }\overline{\nu }}^{\widehat{\alpha }}X^{\overline{\mu }}\bigwedge X^{%
\overline{\nu }}.
\end{equation}
For a fixed locally adapted basis on $\mathcal{U}\subset
V^{({\overline{n}})}
$ we can identify components $T_{~\overline{\mu }\overline{\nu }}^{\widehat{a%
}} $ of torsion (\ref{4torsbaf}) with components of torsion
(\ref{4dtors1})
on $V^{({\overline{n}})},$ i.e. $T_{~\overline{\mu }\overline{\nu }}^{%
\widehat{\alpha }}=T_{~\overline{\mu }\overline{\nu
}}^{\overline{\alpha }}.$ By straightforward calculation we obtain
\begin{equation}  \label{4aux5}
{(\Delta \overline{\mathcal{R}})}_{\mathcal{U}}=[{(\Delta \mathcal{R}%
^{(\Gamma )})}_{\mathcal{U}},\ {(R\tau )}_{\mathcal{U}}+{(Ri)}_{\mathcal{U}%
}],
\end{equation}
where
\[
\left( R\tau \right) _{\mathcal{U}}=\widehat{\delta }_GT_{\mathcal{U}%
}+*_G^{-1}\left[ \Omega _{\mathcal{U}},*_GT_{\mathcal{U}}\right]
,\quad
\left( Ri\right) _{\mathcal{U}}=*_G^{-1}\left[ \chi _{\mathcal{U}},*_G%
\mathcal{R}_{\mathcal{U}}^{(\Gamma )}\right] .
\]
Form $\left( Ri\right) _{\mathcal{U}}$ from (\ref{4aux5}) is
locally constructed by using components of the Ricci tensor (see
(\ref{4dricci1}))
as follows from decomposition on the local adapted basis $X^{\overline{\mu }%
}=\delta u^{\overline{\mu }}:$
\[
\left( Ri\right) _{\mathcal{U}}=e_{\widehat{\alpha }}\otimes
\left(
-1\right) ^{\overline{n}+1}R_{\overline{\lambda }\overline{\nu }}g^{\widehat{%
\alpha }\overline{\lambda }}\delta u^{\overline{\mu }}.
\]

We remark that for isotropic torsionless pseudo-Riemannian spaces
the
requirement that $\left( \Delta \overline{\mathcal{R}}\right) _{\mathcal{U}%
}=0,$ i.e., imposing the connection (\ref{4conb}) to satisfy
Yang-Mills equations (\ref{4ym1}) (equivalently (\ref{4ym3}) or
(\ref{4ym4})) we obtain \cite{p,pd} the equivalence of the
mentioned gauge gravitational equations with the vacuum Einstein
equations $R_{ij}=0.\,$ In the case of ha--spaces with arbitrary
given torsion, even considering vacuum gravitational fields, we
have to introduce a source for gauge gravitational equations in
order to compensate for the contribution of torsion and to obtain
equivalence with the Einstein equations.

Considerations presented in this section constitute the proof of
the following result:

\begin{theorem}
The Einstein equations (\ref{4einsteq3}) for ha--gravity are
equivalent to the Yang-Mills equations
\begin{equation}  \label{4yme}
\left( \Delta \overline{\mathcal{R}}\right)
=\overline{\mathcal{J}}
\end{equation}
for the induced Cartan connection $\overline{\Theta }$ (see
(\ref{4conb})
and (\ref{4curvbaf})) in the bundle of locally adapted affine frames $%
\mathcal{A}a\left( V^{({\overline{n}})}\right) $ with the source $\overline{%
\mathcal{J}}_{\mathcal{U}}$ constructed locally by using the same formulas (%
\ref{4aux5}) for $\left( \Delta \overline{\mathcal{R}}\right) $, but where $%
R_{\overline{\alpha }\overline{\beta }}$ is changed by the matter source ${E}%
_{\overline{\alpha }\overline{\beta }}-\frac 12g_{\overline{\alpha }%
\overline{\beta }}{E}$ with ${E}_{\overline{\alpha }\overline{\beta }%
}=k\Upsilon _{\overline{\alpha }\overline{\beta }}-\lambda g_{\overline{%
\alpha }\overline{\beta }}.$
\end{theorem}

We note that this theorem is an extension for higher order
anisotropic spacetimes of the Popov and Daikhin result \cite{pd}
with respect to a possible gauge like treatment of the Einstein
gravity. Similar theorems have been proved for locally
anisotropic gauge gravity \cite{vg} and in the framework of some
variants of locally (and higher order) anisotropic supergravity
\cite{vbook}.

\section{Nonlinear De Sitter Gauge Ha--Gravity} \index{de Sitter}

The equivalent reexpression of the Einstein theory as a gauge
like theory implies, for both locally isotropic and anisotropic
space--times, the nonsemisimplicity of the gauge group, which
leads to a nonvariational theory in the total space of the bundle
of locally adapted affine frames. A variational gauge
gravitational theory can be formulated by using a minimal
extension of the affine structural group ${\mathcal{A}f}_{\overline{n}%
}\left( \mathcal{R}\right) $ to the de Sitter gauge group $S_{\overline{n}%
}=SO\left( \overline{n}\right) $ acting on distinguished $\mathcal{R}^{%
\overline{n}+1}$ space.

\subsection{Nonlinear gauge theories of de Sitter group}

Let us consider the de Sitter space $\Sigma ^{\overline{n}}$ as a
hypersurface given by the equations $\eta _{AB}u^Au^B=-l^2$ in the flat $%
\left( \overline{n}+1\right) $--dimensional space enabled with
diagonal
metric $\eta _{AB},\eta _{AA}=\pm 1$ (in this subsection $A,B,C,...=1,2,...,%
\overline{n}+1),(\overline{n}=n+m_1+...+m_z),$ where $\{u^A\}$
are global Cartesian coordinates in
$\mathcal{R}^{\overline{n}+1};l>0$ is the curvature of de Sitter
space. The de Sitter group $S_{\left( \eta \right) }=SO_{\left(
\eta \right) }\left( \overline{n}+1\right) $ is defined as the
isometry group of $\Sigma ^{\overline{n}}$--space with
$\frac{\overline{n}}2\left( \overline{n}+1\right) $ generators of
Lie algebra ${\mathit{s}o}_{\left( \eta \right) }\left(
\overline{n}+1\right) $ satisfying the commutation relations
\begin{equation}  \label{4gener}
\left[ M_{AB},M_{CD}\right] =\eta _{AC}M_{BD}-\eta
_{BC}M_{AD}-\eta _{AD}M_{BC}+\eta _{BD}M_{AC}.
\end{equation}

Decomposing indices $A,B,...$ as $A=\left( \widehat{\alpha },\overline{n}%
+1\right) ,B=\left( \widehat{\beta },\overline{n}+1\right) ,$
$...,$ the
metric $\eta _{AB}$ as $\eta _{AB}=\left( \eta _{\widehat{\alpha }\widehat{%
\beta }},\eta _{\left( \overline{n}+1\right) \left(
\overline{n}+1\right)
}\right) ,$ and operators $M_{AB}$ as $M_{\widehat{\alpha }\widehat{\beta }}=%
\mathcal{F}_{\widehat{\alpha }\widehat{\beta }}$ and $P_{\widehat{\alpha }%
}=l^{-1}M_{\overline{n}+1,\widehat{\alpha }},$ we can write
(\ref{4gener}) as
\[
\left[ \mathcal{F}_{\widehat{\alpha }\widehat{\beta }},\mathcal{F}_{\widehat{%
\gamma }\widehat{\delta }}\right] =\eta _{\widehat{\alpha }\widehat{\gamma }}%
\mathcal{F}_{\widehat{\beta }\widehat{\delta }}-\eta _{\widehat{\beta }%
\widehat{\gamma }}\mathcal{F}_{\widehat{\alpha }\widehat{\delta }}+\eta _{%
\widehat{\beta }\widehat{\delta }}\mathcal{F}_{\widehat{\alpha }\widehat{%
\gamma }}-\eta _{\widehat{\alpha }\widehat{\delta }}\mathcal{F}_{\widehat{%
\beta }\widehat{\gamma }},
\]
\[
\left[ P_{\widehat{\alpha }},P_{\widehat{\beta }}\right] =-l^{-2}\mathcal{F}%
_{\widehat{\alpha }\widehat{\beta }},\quad \left[ P_{\widehat{\alpha }},%
\mathcal{F}_{\widehat{\beta }\widehat{\gamma }}\right] =\eta _{\widehat{%
\alpha }\widehat{\beta }}P_{\widehat{\gamma }}-\eta _{\widehat{\alpha }%
\widehat{\gamma }}P_{\widehat{\beta }},
\]
where we have indicated the possibility to decompose
${\mathit{s}o}_{\left(
\eta \right) }\left( \overline{n}+1\right) $ into a direct sum, ${\mathit{s}o%
}_{\left( \eta \right) }\left( \overline{n}+1\right)
={\mathit{s}o}_{\left( \eta \right) }(\overline{n})\oplus
v_{\overline{n}},$ where $v_{\overline{n}} $ is the vector space
stretched on vectors $P_{\widehat{\alpha }}.$ We remark that
$\Sigma ^{\overline{n}}=S_{\left( \eta \right) }/L_{\left( \eta
\right) },$ where $L_{\left( \eta \right) }=SO_{\left( \eta
\right) }\left(
\overline{n}\right) .$ For $\eta _{AB}=diag\left( 1,-1,-1,-1\right) $ and $%
S_{10}=SO\left( 1,4\right) ,L_6=SO\left( 1,3\right) $ is the
group of Lorentz rotations.

Let $W\left( \mathcal{E},\mathcal{R}^{\overline{n}+1},S_{\left(
\eta \right) },\mathcal{P}\right) $ be the vector bundle
associated with the principal bundle $\mathcal{P}\left( S_{\left(
\eta \right) },\mathcal{E}\right) $ on ha-spacetime
$v_{\overline{n}},$ where $S_{\left( \eta \right) }$ is taken to
be the structural group and by $\mathcal{E}$ it is denoted the
total
space. The action of the structural group $S_{\left( \eta \right) }$ on $%
\mathcal{E}\,$ can be realized by using $\overline{n}\times
\overline{n}$ matrices with a parametrization distinguishing
subgroup $L_{\left( \eta
\right) }:$%
\begin{equation}  \label{4aux6}
B=bB_L,
\end{equation}
where
\[
B_L=\left(
\begin{array}{cc}
L & 0 \\
0 & 1
\end{array}
\right) ,
\]
$L\in L_{\left( \eta \right) }$ is the de Sitter bust matrix
transforming the vector $\left( 0,0,...,\rho \right) \in
\mathcal{R}^{\overline{n}+1}$ into the point $\left(
v^1,v^2,...,v^{\overline{n}+1}\right) \in \Sigma _\rho
^{\overline{n}}\subset \mathcal{R}^{\overline{n}+1}$ for which
\[
v_Av^A=-\rho^2,v^A=t^A\rho .
\]
Matrix $b$ can be expressed
\[
b=\left(
\begin{array}{cc}
\delta _{\quad \widehat{\beta }}^{\widehat{\alpha }}+\frac{t^{\widehat{%
\alpha }}t_{\widehat{\beta }}}{\left( 1+t^{\overline{n}+1}\right) } & t^{%
\widehat{\alpha }} \\
t_{\widehat{\beta }} & t^{\overline{n}+1}
\end{array}
\right) .
\]

The de Sitter gauge field is associated with a linear connection
in $W$,
i.e., with a ${\mathit{s}o}_{\left( \eta \right) }\left( \overline{n}%
+1\right) $-valued connection 1--form on $V^{({\overline{n}})}:$%
\begin{equation}  \label{4convsit}
\breve \Theta=\left(
\begin{array}{cc}
\omega _{\quad \widehat{\beta }}^{\widehat{\alpha }} & \breve \theta^{%
\widehat{\alpha }} \\
\breve \theta_{\widehat{\beta }} & 0
\end{array}
\right) ,
\end{equation}
where $\omega _{\quad \widehat{\beta }}^{\widehat{\alpha }}\in so(\overline{n%
})_{\left( \eta \right) },$ $\breve \theta^{\widehat{\alpha }}\in \mathcal{R}%
^{\overline{n}},\breve \theta_{\widehat{\beta }}\in \eta _{\widehat{\beta }%
\widehat{\alpha }}\breve \theta^{\widehat{\alpha }}.$

Because $S_{\left( \eta \right) }$-transforms mix $\omega _{\quad \widehat{%
\beta }}^{\widehat{\alpha }}$ and $\breve \theta
^{\widehat{\alpha }}$ fields in (\ref{4convsit}) (the introduced
para\-met\-ri\-za\-ti\-on is invariant on action on $SO_{\left(
\eta \right) }\left( \overline{n}\right) $
group we cannot identify $\omega _{\quad \widehat{\beta }}^{\widehat{\alpha }%
}$ and $\breve \theta ^{\widehat{\alpha }},$ respectively, with
the connection $\Gamma _{~\overline{\beta }\overline{\gamma
}}^{\overline{\alpha
}}$ and the fundamental form $\chi ^{\overline{\alpha }}$ in $V^{({\overline{%
n}})}$ (as we have for (\ref{4conb}) and (\ref{4curvbaf})). To
avoid this difficulty we consider \cite{tseyt,pbo} a nonlinear
gauge realization of the de Sitter group $S_{\left( \eta \right)
}$ by introducing the nonlinear gauge field
\begin{equation}  \label{4convsitn}
\Theta =b^{-1}\breve \Theta b+b^{-1}db=\left(
\begin{array}{cc}
\Gamma _{~\widehat{\beta }}^{\widehat{\alpha }} & \theta
^{\widehat{\alpha }}
\\
\theta _{\widehat{\beta }} & 0
\end{array}
\right) ,
\end{equation}
where
\[
\Gamma _{\quad \widehat{\beta }}^{\widehat{\alpha }}=\omega _{\quad \widehat{%
\beta }}^{\widehat{\alpha }}-\left( t^{\widehat{\alpha }}Dt_{\widehat{\beta }%
}-t_{\widehat{\beta }}Dt^{\widehat{\alpha }}\right) /\left( 1+t^{\overline{n}%
+1}\right) ,
\]
\[
\theta ^{\widehat{\alpha }}=t^{\overline{n}+1}\breve \theta ^{\widehat{%
\alpha }}+Dt^{\widehat{\alpha }}-t^{\widehat{\alpha }}\left( dt^{\overline{n}%
+1}+\breve \theta _{\widehat{\gamma }}t^{\widehat{\gamma
}}\right) /\left( 1+t^{\overline{n}+1}\right) ,
\]
\[
Dt^{\widehat{\alpha }}=dt^{\widehat{\alpha }}+\omega _{\quad \widehat{\beta }%
}^{\widehat{\alpha }}t^{\widehat{\beta }}.
\]

The action of the group $S\left( \eta \right) $ is nonlinear,
yielding transforms
\[
\Gamma ^{\prime }=L^{\prime }\Gamma \left( L^{\prime }\right)
^{-1}+L^{\prime }d\left( L^{\prime }\right) ^{-1},\theta ^{\prime
}=L\theta ,
\]
where the nonlinear matrix-valued function $L^{\prime }=L^{\prime }\left( t^{%
\overline{\alpha }},b,B_T\right) $ is defined from $B_b=b^{\prime
}B_{L^{\prime }}$ (see the parametrization (\ref{4aux6})).

Now, we can identify components of (\ref{4convsitn}) with components of $%
\Gamma _{~\overline{\beta }\overline{\gamma }}^{\overline{\alpha }}$ and $%
\chi _{~\overline{\alpha }}^{\widehat{\alpha }}$ on
$V^{({\overline{n}})}$
and induce in a consistent manner on the base of bundle $W( \mathcal{E},%
\mathcal{R}^{\overline{n}+1},S_{( \eta )},\mathcal{P})$ the
ha--geometry.

\subsection{Dynamics of the nonlinear de Sitter ha--gravity}

Instead of the gravitational potential (\ref{4conb}), we
introduce the gravitational connection (similar to
(\ref{4convsitn}))
\begin{equation}  \label{4convsitn1}
\Gamma =\left(
\begin{array}{cc}
\Gamma _{~\widehat{\beta }}^{\widehat{\alpha }} & l_0^{-1}\chi ^{\widehat{%
\alpha }} \\
l_0^{-1}\chi _{\widehat{\beta }} & 0
\end{array}
\right)
\end{equation}
where
\[
\Gamma _{~\widehat{\beta }}^{\widehat{\alpha }}=\Gamma _{~\widehat{\beta }%
\overline{\mu }}^{\widehat{\alpha }}\delta u^{\overline{\mu }},
\]
\[
\Gamma _{\quad \widehat{\beta }\overline{\mu }}^{\widehat{\alpha
}}=\chi
_{\quad \overline{\alpha }}^{\widehat{\alpha }}\chi _{\quad \overline{\beta }%
}^{\widehat{\beta }}\Gamma _{\quad \overline{\beta }\overline{\gamma }}^{%
\overline{\alpha }}+\chi _{\quad \overline{\alpha }}^{\widehat{\alpha }%
}\delta _{\overline{\mu }}\chi _{\quad \widehat{\beta }}^{\overline{\alpha }%
},
\]
$\chi ^{\widehat{\alpha }}=\chi _{\quad \overline{\mu }}^{\widehat{\alpha }%
}\delta u^{\overline{\mu }},$ and $g_{\overline{\alpha }\overline{\beta }%
}=\chi _{\quad \overline{\alpha }}^{\widehat{\alpha }}\chi _{\quad \overline{%
\beta }}^{\widehat{\beta }}\eta _{\widehat{\alpha }\widehat{\beta }},$ and $%
\eta _{\widehat{\alpha }\widehat{\beta }}$ is parametrized as
\[
\eta _{\widehat{\alpha }\widehat{\beta }}=\left(
\begin{array}{cccc}
\eta _{ij} & 0 & ... & 0 \\
0 & \eta _{a_1b_1} & ... & 0 \\
... & ... & ... & ... \\
0 & 0 & ... & \eta _{a_zb_z}
\end{array}
\right) ,
\]
$\eta _{ij}=\left( 1,-1,...,-1\right) ,...\eta _{ij}=\left( \pm
1,\pm 1,...,\pm 1\right) ,...,l_0$ is a dimensional constant.

The curvature of (\ref{4convsitn1}), $\mathcal{R}^{(\Gamma
)}=d\Gamma +\Gamma \bigwedge \Gamma ,$ can be written as
\begin{equation}  \label{4curvdsit}
\mathcal{R}^{(\Gamma )}=\left(
\begin{array}{cc}
\mathcal{R}_{\quad \widehat{\beta }}^{\widehat{\alpha }}+l_0^{-1}\pi _{%
\widehat{\beta }}^{\widehat{\alpha }} & l_0^{-1}T^{\widehat{\alpha }} \\
l_0^{-1}T^{\widehat{\beta }} & 0
\end{array}
\right) ,
\end{equation}
where
\[
\pi _{\widehat{\beta }}^{\widehat{\alpha }}=\chi ^{\widehat{\alpha }%
}\bigwedge \chi _{\widehat{\beta }},\mathcal{R}_{\quad \widehat{\beta }}^{%
\widehat{\alpha }}=\frac 12\mathcal{R}_{\quad \widehat{\beta }\overline{\mu }%
\overline{\nu }}^{\widehat{\alpha }}\delta u^{\overline{\mu
}}\bigwedge \delta u^{\overline{\nu }},
\]
and
\[
\mathcal{R}_{\quad \widehat{\beta }\overline{\mu }\overline{\nu }}^{\widehat{%
\alpha }}=\chi _{\widehat{\beta }}^{\quad \overline{\beta }}\chi _{\overline{%
\alpha }}^{\quad \widehat{\alpha }}R_{\overline{\beta }.\overline{\mu }%
\overline{\nu }}^{~\overline{\alpha }}
\]
(see (\ref{4curvaturesha}) for components of d-curvatures). The
de Sitter gauge group is semisimple and we are able to construct
a variational gauge
gravitational locally an\-iso\-trop\-ic theory (bundle metric (\ref{4tmetric}%
) is nondegenerate). The Lagrangian of the theory is postulated as
\[
L=L_{\left( G\right) }+L_{\left( m\right) }
\]
where the gauge gravitational Lagrangian is defined as
\[
L_{\left( G\right) }=\frac 1{4\pi }Tr\left( \mathcal{R}^{(\Gamma
)}\bigwedge *_G\mathcal{R}^{(\Gamma )}\right)
=\mathcal{L}_{\left( G\right) }\left| g\right| ^{1/2}\delta
^{\overline{n}}u,
\]
\begin{equation}  \label{4lagrsit}
\mathcal{L}_{\left( G\right) }=\frac 1{2l^2}T_{\quad \overline{\mu }%
\overline{\nu }}^{\widehat{\alpha }}T_{\widehat{\alpha }}^{\quad \overline{%
\mu }\overline{\nu }}+\frac 1{8\lambda }\mathcal{R}_{\quad \widehat{\beta }%
\overline{\mu }\overline{\nu }}^{\widehat{\alpha
}}\mathcal{R}_{\quad
\widehat{\alpha }}^{\widehat{\beta }\quad \overline{\mu }\overline{\nu }%
}-\frac 1{l^2}\left( {\overleftarrow{R}}\left( \Gamma \right)
-2\lambda _1\right) ,
\end{equation}
$T_{\quad \overline{\mu }\overline{\nu }}^{\widehat{\alpha
}}=\chi _{\quad \overline{\alpha }}^{\widehat{\alpha }}T_{\quad
\overline{\mu }\overline{\nu
}}^{\overline{\alpha }}$ (the gravitational constant $l^2$ in (\ref{4lagrsit}%
) satisfies the relations $l^2=2l_0^2\lambda ,\lambda
_1=-3/l_0],$ $\quad Tr$ denotes the trace on $\widehat{\alpha
},\widehat{\beta }$ indices, and the matter field Lagrangian is
defined as
\[
L_{\left( m\right) }=\frac 12Tr\left( \Gamma \bigwedge
*_G\mathcal{I}\right)
=\mathcal{L}_{\left( m\right) }\left| g\right| ^{1/2}\delta ^{\overline{n}%
}u,
\]
\begin{equation}  \label{4lagrsitm}
\mathcal{L}_{\left( m\right) }=\frac 12\Gamma _{\quad \widehat{\beta }%
\overline{\mu }}^{\widehat{\alpha }}S_{\quad \overline{\alpha }}^{\widehat{%
\beta }\quad \overline{\mu }}-t_{\quad \widehat{\alpha }}^{\overline{\mu }%
}l_{\quad \overline{\mu }}^{\widehat{\alpha }}.
\end{equation}
The matter field source $\mathcal{I}$ is obtained as a variational
derivation of $\mathcal{L}_{\left( m\right) }$ on $\Gamma $ and is
parametrized as
\begin{equation}  \label{4sourcesit}
\mathcal{I}=\left(
\begin{array}{cc}
S_{\quad \widehat{\beta }}^{\widehat{\alpha }} &
-l_0t^{\widehat{\alpha }}
\\
-l_0t_{\widehat{\beta }} & 0
\end{array}
\right)
\end{equation}
with $t^{\widehat{\alpha }}=t_{\quad \overline{\mu }}^{\widehat{\alpha }%
}\delta u^{\overline{\mu }}$ and $S_{\quad \widehat{\beta }}^{\widehat{%
\alpha }}=S_{\quad \widehat{\beta }\overline{\mu }}^{\widehat{\alpha }%
}\delta u^{\overline{\mu }}$ being respectively the canonical
tensors of energy-momentum and spin density. Because of the
contraction of the
''interior'' indices $\widehat{\alpha },\widehat{\beta }$ in (\ref{4lagrsit}%
) and (\ref{4lagrsitm}) we used the Hodge operator $*_G$ instead
of $*_H$ (hereafter we consider $*_G=*).$

Varying the action
\[
S=\int \left| g\right| ^{1/2}\delta ^{\overline{n}}u\left( \mathcal{L}%
_{\left( G\right) }+\mathcal{L}_{\left( m\right) }\right)
\]
on the $\Gamma $-variables (\ref{4convsitn1}), we obtain the
gauge--gravitational field equations:
\begin{equation}  \label{4dsitteq1}
d\left( *\mathcal{R}^{(\Gamma )}\right) +\Gamma \bigwedge \left( *\mathcal{R}%
^{(\Gamma )}\right) -\left( *\mathcal{R}^{(\Gamma )}\right)
\bigwedge \Gamma =-\lambda \left( *\mathcal{I}\right) .
\end{equation}

Specifying the variations on $\Gamma _{\quad \widehat{\beta }}^{\widehat{%
\alpha }}$ and $l^{\widehat{\alpha }}$-variables, we rewrite (\ref{4dsitteq1}%
) as
\begin{equation}  \label{4dsitteq2}
\widehat{\mathcal{D}}\left( *\mathcal{R}^{(\Gamma )}\right)
+\frac{2\lambda }{l^2}\left( \widehat{\mathcal{D}}\left( *\pi
\right) +\chi \bigwedge \left( *T^T\right) -\left( *T\right)
\bigwedge \chi ^T\right) =-\lambda \left( *S\right) ,
\end{equation}
\begin{equation}  \label{4dsitteq3}
\widehat{\mathcal{D}}\left( *T\right) -\left(
*\mathcal{R}^{(\Gamma )}\right) \bigwedge \chi -\frac{2\lambda
}{l^2}\left( *\pi \right) \bigwedge \chi =\frac{l^2}2\left(
*t+\frac 1\lambda *\tau \right) ,
\end{equation}
where
\[
T^t=\{T_{\widehat{\alpha }}=\eta _{\widehat{\alpha }\widehat{\beta }}T^{%
\widehat{\beta }},~T^{\widehat{\beta }}=\frac 12T_{\quad \overline{\mu }%
\overline{\nu }}^{\widehat{\beta }}\delta u^{\overline{\mu
}}\bigwedge \delta u^{\overline{\nu }}\},
\]
\[
\chi ^T=\{\chi _{\widehat{\alpha }}=\eta _{\widehat{\alpha }\widehat{\beta }%
}\chi ^{\widehat{\beta }},~\chi ^{\widehat{\beta }}=\chi _{\quad \overline{%
\mu }}^{\widehat{\beta }}\delta u^{\overline{\mu }}\},\qquad \widehat{%
\mathcal{D}}=d+\widehat{\Gamma }
\]
($\widehat{\Gamma }$ acts as $\Gamma _{\quad \widehat{\beta }\overline{\mu }%
}^{\widehat{\alpha }}$ on indices $\widehat{\gamma
},\widehat{\delta },...$
and as $\Gamma _{\quad \overline{\beta }\overline{\mu }}^{\overline{\alpha }%
} $ on indices $\overline{\gamma },$ $\overline{\delta },...).$
In (\ref {4dsitteq3}), $\tau $ defines the energy--momentum
tensor of the $S_{\left(
\eta \right) }$--gauge gravitational field $\widehat{\Gamma }:$%
\begin{equation}  \label{4sourcesit1}
\tau _{\overline{\mu }\overline{\nu }}\left( \widehat{\Gamma
}\right) =\frac 12Tr\left( \mathcal{R}_{\overline{\mu
}\overline{\alpha }}\mathcal{R}_{\quad
\overline{\nu }}^{\overline{\alpha }}-\frac 14\mathcal{R}_{\overline{\alpha }%
\overline{\beta }}\mathcal{R}^{\overline{\alpha }\overline{\beta }} g_{%
\overline{\mu }\overline{\nu }}\right) .
\end{equation}

Equations (\ref{4dsitteq1}) (or, equivalently, (\ref{4dsitteq2})
and (\ref {4dsitteq3})) make up the complete system of
variational field equations for nonlinear de Sitter gauge gravity
with higher order anisotropy. They can be interpreted as a
variant of gauge like equations for ha--gravity \cite{vg} when
the (pseudo) Riemannian base frames and torsions are considered
to be induced by an anholonomic frame structure with associated
N--connection

A. Tseytlin \cite{tseyt} presented a quantum analysis of the
isotropic version of equations (\ref{4dsitteq2}) and
(\ref{4dsitteq3}). Of course, the problem of quantizing
gravitational interactions is unsolved for both variants of
locally anisotropic and isotropic gauge de Sitter gravitational
theories, but we think that the generalized Lagrange version of
$S_{\left( \eta \right) }$-gravity is more adequate for studying
quantum radiational and statistical gravitational processes. This
is a matter for further investigations.

Finally, we remark that we can obtain a nonvariational Poincare
gauge gravitational theory on ha--spaces if we consider the
contraction of the gauge potential (\ref{4convsitn1}) to a
potential with values in the Poincare Lie algebra
\[
\Gamma =\left(
\begin{array}{cc}
\Gamma _{\quad \widehat{\beta }}^{\widehat{\alpha }} & l_0^{-1}\chi ^{%
\widehat{\alpha }} \\
l_0^{-1}\chi _{\widehat{\beta }} & 0
\end{array}
\right) \rightarrow \Gamma =\left(
\begin{array}{cc}
\Gamma _{\quad \widehat{\beta }}^{\widehat{\alpha }} & l_0^{-1}\chi ^{%
\widehat{\alpha }} \\
0 & 0
\end{array}
\right) .
\]
Isotropic Poincare gauge gravitational theories are studied in a
number of papers (see, for example, \cite{wal,tseyt,pbo}). In a
manner similar to considerations presented in this work, we can
generalize Poincare gauge models for spaces with local anisotropy.

\section{An Ansatz for 4D d--Metrics}

We consider a 4D space--time $V^{(3+1)}$ provided with a
d--metric (\ref {dmetric}) when $g_i = g_i (x^k)$ and $h_a = h_a
(x^k, z)$  for $y^a = (z, y^4).$ The N--connection coefficients
are some  functions on three
coordinates $(x^i,z),$%
\begin{eqnarray}
N_1^3&=&q_1(x^i,z),\ N_2^3=q_2(x^i,z),  \label{4ncoef} \\
N_1^4&=&n_1(x^i,z),\ N_2^4=n_2(x^i,z).  \nonumber
\end{eqnarray}
For simplicity, we shall use brief denotations of partial
derivatives,  like
\begin{eqnarray*}
\dot a &=&\partial a/\partial x^1,a^{\prime }= \partial a/\partial x^2, \\
a^{*}&=&\partial a/\partial z \dot a^{\prime } = \partial
^2a/\partial
x^1\partial x^2, \\
a^{**}&=& \partial ^2a/ \partial z\partial z.
\end{eqnarray*}

The non--trivial components of the Ricci  d--tensor
(\ref{4dricci}),  for the mentioned type of d--metrics depending
on three variables, are\newpage
\begin{eqnarray}  \label{4ricci2}
&&R_1^1=R_2^2=\frac 1{2g_1g_2} [-(g_1^{^{\prime \prime }}+{\ddot
g}_2)+ \frac 1{2g_2}\left( {\dot g}_2^2+g_1^{\prime }g_2^{\prime
}\right) + \frac
1{2g_1}\left( g_1^{\prime \ 2}+\dot g_1\dot g_2\right) ];  \label{4ricci1} \\
&&S_3^3=S_4^4= \frac 1{h_3h_4}[-h_4^{**}+\frac
1{2h_4}(h_4^{*})^2+\frac
1{2h_3}h_3^{*}h_4^{*}]; \nonumber \\
&&P_{31}=\frac{q_1}2[\left( \frac{h_3^{*}}{h_3}\right) ^2- \frac{h_3^{**}}{%
h_3}+\frac{h_4^{*}}{2h_4^{\ 2}}-\frac{h_3^{*}h_4^{*}}{2h_3h_4}]
\nonumber \\ & & {\qquad} + \frac
1{2h_4}[\frac{\dot h_4}{2h_4}h_4^{*}-\dot h_4^{*}+ \frac{\dot h_3}{2h_3}%
h_4^{*}],  \label{4ricci3} \\
&&P_{32}=\frac{q_2}2[\left( \frac{h_3^{*}}{h_3}\right) ^2- \frac{h_3^{**}}{%
h_3}+\frac{h_4^{*}}{2h_4^{\ 2}}-\frac{h_3^{*}h_4^{*}}{2h_3h_4}]
\nonumber \\
&& {\qquad} +\frac 1{2h_4}[\frac{h_4^{\prime }}{2h_4}h_4^{*}-h_4^{\prime \ *}+ \frac{%
h_3^{\prime }}{2h_3}h_4^{*}];  \nonumber \\
&&  \nonumber \\
&&P_{41}=-\frac{h_4}{2h_3}n_1^{**} +
\frac{1}{4h_3}(\frac{h_4}{h_3} h^*_3 -
3 h^*_4) n^*_1 ,  \label{4ricci4} \\
&&P_{42}= -\frac{h_4}{2h_3}n_2^{**} +
\frac{1}{4h_3}(\frac{h_4}{h_3} h^*_3 - 3 h^*_4) n^*_2.  \nonumber
\end{eqnarray}

The curvature scalar $\overleftarrow{R}$ (\ref{4dscalar}) is
defined by  the sum of two non-trivial components
$\widehat{R}=2R_1^1$ and $S=2S_3^3.$

The system of Einstein equations (\ref{4einsteq2}) transforms into
\begin{eqnarray}
R_1^1&=&-\kappa \Upsilon _3^3=-\kappa \Upsilon _4^4,  \label{4einsteq3a} \\
S_3^3&=&-\kappa \Upsilon _1^1=-\kappa \Upsilon _2^2,  \label{4einsteq3b} \\
P_{3i}&=& \kappa \Upsilon _{3i},  \label{4einsteq3c} \\
P_{4i}&=& \kappa \Upsilon _{4i},  \label{4einsteq3d}
\end{eqnarray}
where the values of $R_1^1,S_3^3,P_{ai},$ are taken respectively
from (\ref {4ricci1}), (\ref{4ricci2}), (\ref{4ricci3}),
(\ref{4ricci4}).

By using the equations (\ref{4einsteq3c}) and (\ref{4einsteq3d})
we can define the N--coefficients (\ref{4ncoef}), $q_i(x^k,z)$
and $n_i(x^k,z),$ if the functions $g_i(x^k)$ and $h_i(x^k,z)$
are known as respective solutions of the equations
(\ref{4einsteq3a}) and (\ref{4einsteq3b}).  Let consider an
ansatz for a 4D d--metric of type
\begin{equation}
{\delta s}^2=g_1(x^k)(dx^1)^2+(dx^2)^2+h_3(x^i,t)(\delta t)^2+
h_4(x^i,t)(\delta y^4)^2,  \label{4dmetr4}
\end{equation}
where the $z$--parameter is considered to be the time like
coordinate and the energy momentum d--tensor is taken
\[
\Upsilon _\alpha^\beta = [p_1,p_2,-\varepsilon, p_4=p].
\]
The aim of this section is to analyze the system of partial
differential equations following from the Einsteni field equations
for these d--metric and energy--momentum d--tensor.

\subsection{The h--equations}

The Einstein equations (\ref{4einsteq3a}), with the Ricci
h--tensor (\ref {4ricci1}), for the d--metric (\ref{4dmetr4})
transform into
\begin{equation}  \label{4hbh1}
\frac{\partial ^2g_1}{\partial (x^1)^2}-\frac 1{2g_1}\left(
\frac{\partial g_1}{\partial x^1}\right) ^2+2\kappa \varepsilon
g_1=0.
\end{equation}
By introducing the coordinates $\chi ^i=x^i/\sqrt{\kappa
\varepsilon}$ and the variable
\begin{equation}
q=g_1^{\prime }/g_1,  \label{4eq4c}
\end{equation}
where by 'prime' in this Section is considered the partial derivative $%
\partial /\chi ^2,$ the equation (\ref{4hbh1}) transforms into
\begin{equation}  \label{4hbh2}
q^{\prime }+\frac{q^2}2+2\epsilon =0,
\end{equation}
where the vacuum case should be parametrized for $\epsilon =0$
with $\chi ^i=x^i$ and $\epsilon =-1$ for a matter state with
$\varepsilon = - p$.

The integral curve of (\ref{4hbh2}), intersecting a point $\left(
\chi _{(0)}^2,q_{(0)}\right) ,$ considered as a differential
equation on $\chi ^2$ is defined by the functions \cite{kamke}
\begin{eqnarray}
q &=& \frac{q_{(0)}}{1+\frac{q_{(0)}}2 \left( \chi ^2-\chi _{(0)}^2\right) }%
,\qquad \epsilon =0;  \label{4eq3a} \\
q & = & \frac{q_{(0)}-2\tan \left( \chi ^2-\chi _{(0)}^2\right) } {1+\frac{%
q_{(0)}}2\tan \left( \chi ^2-\chi _{(0)}^2\right) },\qquad
\epsilon <0. \label{4eq5c}
\end{eqnarray}

Because the function $q$ depends also parametrically on variable
$\chi ^1$ we can consider functions $\chi _{(0)}^2=\chi
_{(0)}^2\left( \chi ^1\right) $ and $q_{(0)}=q_{(0)}\left( \chi
^1\right) .$ We elucidate the non--vacuum case with $\epsilon <0.$
The general formula for the non--trivial component of h--metric
is to be obtained after integration on $\chi ^1$ of (\ref{4eq4c})
by using the solution (\ref{4eq5c})
\[
g_1\left( \chi ^1,\chi ^2\right) = g_{1(0)} \left( \chi ^1\right)
\left\{ \sin [\chi ^2-\chi _{(0)}^2 \left( \chi ^1\right)
]+\arctan \frac 2{q_{(0)}\left( \chi ^1\right) }\right\} ^2,
\]
for $q_{(0)}\left( \chi ^1\right) \neq 0,$ and
\begin{equation}  \label{4btzlh3}
g_1\left( \chi ^1,\chi ^2 \right) = g_{1(0)}\left( \chi ^1\right)
\ \cos ^2[\chi ^2- \chi _{(0)}^2\left( \chi ^1\right) ]
\end{equation}
for $q_{(0)}(\chi ^1) =0,$ where $g_{1(0)}(\chi^1), \chi
_{(0)}^2(\chi ^1) $ and $q_{(0)}(\chi^1) $ are some functions of
necessary smoothness class on variable $\chi ^1.$ For simplicity,
in our further considerations we shall apply the solution
(\ref{4btzlh3}).

\subsection{The v--equations}

For the ansatz (\ref{4dmetr4}) the Einstein equations
(\ref{4einsteq3b}) with the Ricci h--tensor (\ref{4ricci2})
transforms into
\begin{equation}  \label{4heq}
\frac{\partial ^2h_4}{\partial t^2} - \frac 1{2h_4}\left( \frac{\partial h_4%
}{\partial t}\right) ^2 -\frac 1{2h_3}\left( \frac{\partial h_3}{\partial t}%
\right) \left( \frac{\partial h_4}{\partial t}\right) - \frac
\kappa 2\Upsilon _1h_3h_4=0   \nonumber
\end{equation}
(here we write down the partial derivatives on $t$ in  explicit
form) which relates some first and second order partial on $z$
derivatives of diagonal components $h_a(x^i,t)$ of a v--metric
with a source
\[
\Upsilon_1(x^i,z)=\kappa \Upsilon _1^1=\kappa \Upsilon _2^2 =
p_1=p_2
\]
in the h--subspace. We can consider as  unknown the function
$h_3(x^i,t)$ (or, inversely, $h_4(x^i,t))$  for some compatible
values of $h_4(x^i,t)$ (or $h_3(x^i,t))$ and  source
$\Upsilon_1(x^i,t).$ By introducing a new variable $\beta
=h_4^{*}/h_4$ the equation (\ref{4heq}) transforms into
\begin{equation}  \label{4heq1}
\beta ^{*}+\frac 12\beta ^2-\frac{\beta h_3^{*}}{2h_3}- 2\kappa
\Upsilon _1h_3=0
\end{equation}
which relates two functions $\beta \left( x^i,t\right) $ and
$h_3\left(
x^i,t\right) .$ There are two possibilities: 1) to define $\beta $ (i. e. $%
h_4)$ when $\kappa \Upsilon _1$ and $h_3$ are  prescribed and,
inversely 2) to find $h_3$ for given $\kappa \Upsilon _1$ and
$h_4$ (i. e. $\beta );$ in both cases one considers only ''*''
derivatives on $t$--variable with coordinates $x^i$ treated as
parameters.

\begin{enumerate}
\item  In the first case the explicit solutions of  (\ref{4heq1}) have to be
constructed by using the integral varieties  of the general
Riccati equation
\cite{kamke} which by a corresponding  redefinition of variables, $%
t\rightarrow t\left( \varsigma \right) $ and $\beta \left(
t\right) \rightarrow \eta \left( \varsigma \right) $  (for
simplicity, we omit dependencies on $x^i)$ could be written in
the canonical form
\[
\frac{\partial \eta }{\partial \varsigma }+ \eta ^2+\Psi \left(
\varsigma \right) =0
\]
where $\Psi $ vanishes for vacuum gravitational fields. In vacuum
cases the Riccati equation reduces to a Bernoulli equation which
(we can use the former variables) for $s(t)=\beta ^{-1}$
transforms into a linear differential (on $t)$ equation,
\begin{equation}  \label{4heq1a}
s^{*}+\frac{h_3^{*}}{2h_3}s-\frac 12=0.
\end{equation}

\item  In the second (inverse) case when $h_3$ is to be found for some
prescribed $\kappa \Upsilon _1$ and $\beta $ the equation
(\ref{4heq1})  is to be treated as a Bernoulli type equation,
\begin{equation}  \label{4heq2}
h_3^{*}=-\frac{4\kappa \Upsilon _1}\beta (h_3)^2+ \left( \frac{ 2\beta ^{*}}%
\beta +\beta \right) h_3
\end{equation}
which can be solved by standard methods. In the vacuum case the squared on $%
h_3$ term vanishes and we obtain a linear differential (on $t)$
equation.
\end{enumerate}

Finally, in this Section we conclude that  the system of
equations (\ref {4einsteq3b}) is satisfied by arbitrary  functions
\[
\nonumber h_3=a_3(\chi ^i)\mbox{ and }h_4=a_4(\chi ^i).
\]
If v--metrics depending on three coordinates are introduced,
$h_a=h_a(\chi ^i,t),$ the v--components of the Einstein
equations  transforms into (\ref {4heq}) which reduces to
(\ref{4heq1}) for prescribed  values of $h_3(\chi ^i,t),\,$ and,
inversely, to (\ref{4heq2}) if $h_4(\chi ^i,t)$ is prescribed.

\subsection{H--v equations}

For the ansatz (\ref{4dmetr4}) with $h_4 = h_4 (x^i)$ and a
diagonal
energy--momentum d--tensor the h--v--com\-po\-nents of Einstein equations (%
\ref{4einsteq3c}) and (\ref{4einsteq3d}) are written respectively
as
\begin{equation}  \label{4einsteq5c}
P_{5i}=\frac{q_i}{2h_3} [ {({\frac{\partial h_3}{\partial t} })}^2 - \frac{{%
\partial} ^2 h_3}{{\partial t}^2} ]=0,
\end{equation}
and
\begin{equation}  \label{4einsteq5d}
P_{6i}= {\frac{h_4}{4 (h_3)^2}} {\frac{\partial n_i}{\partial t}}{\frac{%
\partial h_3}{\partial t}} - {\frac{h_4}{2h_3}}{\frac{\partial ^2 n_i}{{%
\partial t}^2}} = 0.
\end{equation}
The equations (\ref{4einsteq5c}) are satisfied by arbitrary coefficients $%
q_i(x^k,t)$ if the d--metric coefficient $h_3$ is a solution of
\begin{equation}  \label{4ncomp}
{({\frac{\partial h_3}{\partial t} })}^2 - \frac{{\partial} ^2 h_3}{{%
\partial t}^2} = 0
\end{equation}
and the $q$--coefficients must vanish if this  condition is not
satisfied. In the last case we obtain a $3+1$ anisotropy.  The
general solution of equations (\ref{4einsteq5d}) are written in
the form
\[
n_i= l^{(0)}_i (x^k) \int \sqrt {|h_3 (x^k,t)|} dt + n^{(0)}_i
(x^k)
\]
where $l^{(0)}_i (x^k)$ and $n^{(0)}_i (x^k)$ are arbitrary
functions on $x^k $ which have to be defined by some boundary
conditions.

\section{Anisotropic Cos\-mo\-lo\-gi\-cal So\-lu\-ti\-ons}

The aim of this section is to construct two classes of solutions
of Einstein equations describing Friedman--Robertson--Walker
(FRW) like universes with corresponding symmetries or rotational
ellipsoid  (elongated and flattened) and torus.

\subsection{Rotation ellipsoid FRW universes}

We proof that there are cosmological solutions constructed as
locally an\-iso\-trop\-ic deformations of the FRW spherical
symmetric solution  to the rotation ellipsoid configuration.
There are two types of rotation ellipsoids, elongated and
flattened ones.  We examine both cases of such horizon
configurations.

\subsubsection{Rotation elongated ellipsoid configuration}

An elongated rotation ellipsoid hypersurface is given by the
formula \cite {korn}
\begin{equation}  \label{4relhor}
\frac{{x}^2+{y}^2}{\sigma ^2-1}+ \frac{{z}^2}{\sigma ^2}={\rho
}^2,
\end{equation}
where $\sigma \geq 1,$ $x,y,z$ are Cartezian coordinates  and
${\rho }$ is similar to the radial coordinate in the spherical
symmetric case. The 3D special coordinate system is defined
\begin{eqnarray}
{x} &=&{\rho}\sinh u\sin v\cos \varphi ,\ {y}={\rho}\sinh u\sin
v\sin
\varphi ,  \nonumber \\
{z}&=& {\rho}\ \cosh u \cos v,  \nonumber
\end{eqnarray}
where $\sigma =\cosh u,(0\leq u<\infty ,\ 0\leq v\leq \pi ,\
0\leq \varphi <2\pi ). $\ The hypersurface metric (\ref{4relhor})
is
\begin{eqnarray}
g_{uu} &=& g_{vv}={\rho}^2\left( \sinh ^2u+\sin ^2v\right) ,
\label{4hsuf1}
\\
g_{\varphi \varphi } &=&{\rho}^2\sinh ^2u\sin ^2v.  \nonumber
\end{eqnarray}
Let us introduce a d--metric of class (\ref{4dmetr4})
\begin{equation}  \label{4rel1}
\delta s^2 = g_1(u,v)du^2+dv^2 + h_3\left( u,v,\tau \right)
\left( \delta \tau \right) ^2+ h_4\left( u,v\right) \left(
\delta\varphi \right) ^2,
\end{equation}
where $x^1=u, x^2=v,$ $y^4= \varphi,$ $y^3=\tau$ is the time like
cosmological coordinate and $\delta \tau$ and $\delta \varphi $
are N--elongated differentials. As a particular solution of
(\ref{4rel1})  for the h--metric we choose (see (\ref{4btzlh3}))
the  coefficient
\begin{equation}  \label{4relh1h}
g_1(u,v)=\cos ^2v
\end{equation}
and set for the v--metric components
\begin{equation}  \label{4relh1}
h_3(u,v,\tau)=-\frac 1{\rho ^2(\tau) (\sinh ^2u+\sin ^2v)}
\end{equation}
and
\begin{equation}
h_4(u,v,\tau)=\frac {\sinh ^2u \sin ^2v}{(\sinh ^2u+\sin ^2v)}.
\label{4relh2}
\end{equation}
The set of coefficients (\ref{4relh1h}),(\ref{4relh1}), and
(\ref{4relh2}), for the d--metric (\ref{4rel1}, and of $q_i=0$
and $n_i$ being solutions  of (\ref{4ncomp}), for the
N--connection, defines a solution of the Einstein equations
(\ref{4einsteq2}).  The physical treatment of the obtained
solutions follows from the  locally isotropic limit of a
conformal transform of this d--metric:  Multiplying (\ref{4rel1})
on
\[
{\rho ^2(\tau) (\sinh ^2u+\sin ^2v)},
\]
and considering $cos ^2v \simeq 1$ and $n_i \simeq = 0$ for
locally isotropic spacetimes we get the interval
\begin{eqnarray}
ds^2 &= &- d \tau ^2 + \rho ^2 (\tau) [(\sinh ^2u+\sin ^2v)(du^2 + dv^2) + {%
\sinh ^2u} {\sin ^2v} d\varphi ^2]  \nonumber \\
{\ }&{\ }& \mbox{for ellipsoidal coordinates on hypersurface
 (\ref{4hsuf1})};  \nonumber \\
{\ } &=& - d \tau ^2 + \rho ^2 (\tau) [dx^2 + dy^2 + dz^2]\
\mbox{for Cartezian coordinates},  \nonumber
\end{eqnarray}
which defines just the Robertson--Walker metric. So, the
d--metric (\ref {4rel1}), the coefficients of N--connection being
solutions of (\ref {4einsteq3c}) and (\ref{4einsteq3d}),
describes a 4D  cosmological solution of the Einstein equations
when instead of a  spherical symmetry one has a locally
anisotropic deformation to the  symmetry of rotation elongated
ellipsoid. The explicit dependence on  time $\tau$ of the
cosmological factor $\rho$ must be constructed by  using
additionally the matter state equations for a cosmological model
with local anisotropy.

\subsubsection{Flattened rotation ellipsoid coordinates}

In a similar fashion we can construct a locally anisotropic
deformation of the FRW metric with the symmetry of flattened
rotation ellipsoid \index{ellipsoid}.  The parametric equation
for a such hypersurface is \cite{korn}
\[
\frac{{x}^2+{y}^2}{1+\sigma ^2}+\frac{{z}^2}{\sigma ^2}={\rho }^2,
\]
where $\sigma \geq 0$ and $\sigma =\sinh u.$ The proper for
ellipsoid  3D space coordinate system is defined
\begin{eqnarray}
{x} &=&{\rho}\cosh u\sin v\cos \varphi , y = {\rho} \cosh u\sin v
\sin
\varphi  \nonumber \\
{z} &=& {\rho} \sinh u\cos v,  \nonumber
\end{eqnarray}
where $0\leq u<\infty ,\ 0\leq v\leq \pi ,\ 0\leq \varphi <2\pi
.$ The hypersurface metric is
\begin{eqnarray}
g_{uu} &=& g_{vv}={\rho}^2\left( \sinh ^2u+\cos ^2v\right) ,  \nonumber \\
g_{\varphi \varphi } &=&{\rho}^2\sinh ^2u\cos ^2v.  \nonumber
\end{eqnarray}
In the rest the cosmological la--solution is described by the
same formulas as in the previous subsection but with respect to
new canonical  coordinates for flattened rotation ellipsoid.

\subsection{Toroidal FRW universes} \index{Toroidal universe}

Let us construct a cosmological solution of the Einstein
equations with to\-ro\-id\-al symmetry. The hypersurface formula
of a torus \index{torus}  is \cite{korn}
\[
\left( \sqrt{{x}^2+{y}^2}-{\rho }\ c\tanh \sigma \right) ^2+{z}^2= \frac{{%
\rho }^2}{\sinh ^2\sigma }.
\]
The 3D space coordinate system is defined
\begin{eqnarray}
{x} &=& \frac{{\rho}\sinh \alpha \cos \varphi } {\cosh \alpha
-\cos\sigma }, \qquad {y} = \frac{{\rho}\sin \sigma \sin \varphi
}{\cosh \alpha -\cos
\sigma },  \nonumber \\
{z} &=& \frac{{\rho}\sinh \sigma }{\cosh \tau -\cos \sigma },  \nonumber \\
&{}& \left( -\pi <\sigma <\pi , 0\leq \alpha <\infty , 0\leq
\varphi <2\pi \right) .  \nonumber
\end{eqnarray}
The hypersurface metric is
\begin{equation}  \label{4mtor}
g_{\sigma \sigma }= g_{\alpha \alpha }=\frac{{\rho }^2} {\left(
\cosh \alpha -\cos \sigma \right) ^2}, g_{\varphi \varphi }=
\frac{{\rho }^2\sin ^2\sigma } {\left( \cosh \alpha -\cos \sigma
\right) ^2}.
\end{equation}
The d--metric of class (\ref{4dmetr4}) is chosen
\begin{equation}  \label{4mtora}
\delta s^2 = g_1(\alpha)d\sigma ^2+d \alpha ^2 + h_3\left(\sigma,
\alpha, \tau \right) \left( \delta \tau \right) ^2+ h_4\left(
\sigma \right) \left( \delta\varphi \right) ^2,
\end{equation}
where $x^1= \sigma , x^2= \alpha ,$ $y^4= \varphi,$ $y^3=\tau$ is
the time like cosmological coordinate and $\delta \tau$ and
$\delta \varphi $ are N--elongated differentials.  As a
particular solution of (\ref{4mtor}) for the h--metric we choose
(see (\ref{4btzlh3})) the coefficient
\begin{equation}  \label{4relh2h}
g_1(\alpha)=\cos ^2\alpha
\end{equation}
and set for the v--metric components
\begin{eqnarray}
h_3(\sigma, \alpha,\tau) &=& -\frac {{(\cosh\alpha - \cos \sigma
)}^2 }{\rho
^2(\tau) }  \nonumber \\
h_4(\sigma) & = &\sin ^2 \sigma.  \label{4relh2a}
\end{eqnarray}
Multiplying (\ref{4mtora}) on
\[
\frac{\rho ^2(\tau)}{{(\cosh \alpha - \cos \sigma)}^2},
\]
and considering $cos \alpha \simeq 1$ and $n_i \simeq = 0$ in the
locally isotropic limit we get the interval
\[
ds^2 = - d \tau ^2 + \frac {\rho ^2 (\tau)} {{(\cosh\alpha - \cos \sigma )}%
^2 } [(d \sigma ^2 + d\alpha ^2 + {\sin ^2 \sigma } d\varphi ^2]
\]
where the space part is just the torus hypersurface metric
(\ref{4mtor}). So, the set of coefficients (\ref{4relh2h}) and
(\ref{4relh2a}), for the d--metric (\ref{4mtora}, and of $q_i=0$
and $n_i$ being solutions of  (\ref {4ncomp}), for the
N--connection, defines a cosmological  solution of the Einstein
equations (\ref{4einsteq2}) with the torus  symmetry, when the
explicit form of the function $\rho (\tau)$ is  to be defined by
considering some additional equations  for the matter state (for
instance, with a scalar field defining  the torus inflation).

\section{ Concluding Remarks}

In this Chapter we have developed the method of anholonomic frames
on (pseudo) Riemannian  spacetimes by considering associated
nonlinear connection  (N--con\-nec\-ti\-on) strucutres. We
provided a rigorous geometric  background for description of
gravitational systems with mixed holonomic and anholonomic
(anisotropic) degrees of freedom by considering first and higher
order anisotropies induced by anholonomic constraints  and
corresponding frame bases.

 The first key result of this paper is
the proof that generic  anisotropic structures of different order
are contained in the Einstein theory. We reformulated the tensor
and linear connection  formalism for (pseudo) Riemannian spaces
enables with N--connections and computed the horizonal--vertical
splitting, with respect to  anholonomic frames with associated
N--connections,  of the Einstein equations. The (pseudo)
Riemannian spaces enabled  with compatible  anholonomic frame and
associated N--connection structures  and the metric being a
solution of the Einstein equations were  called as locally
anisotropic spacetimes (in brief, anisotropic spaces). The next
step was the definition of gauge field interactions on such
spacetimes. We have applied the bundle formalism and extended it
to the case of locally anisotropic bases
 and considered a 'pure' geometric method of deriving the Yang--Mills
  equations for generic  locally anisotropic gauge interactions, by genalizing
the absolut differential calculus and dual forms symmetries for
anisotropic spaces.

  The second key result was the proof by
geometric methods that the  Yang--Mills equations for a
correspondingly defined Cartan connection in the bundle of affine
frames on locally anisotropic spacetimes are equivalent to the
Einstein equations with anholonomic (N--connection) structures
(the original Popov--Daikhin papers \cite{p,pd}  were for the
locally isotropic spaces). The result was obtained by applying an
auxiliary bilinear form on the typical fiber because of
degeneration of the Killing form for the affine groups. After
projection on base spacetimes the dependence on auxiliar values
is eliminated.  We analyzed also a variant of variational gauge
locally anisotropic gauge theory by considering a minimal
extension of the affine structural group to the de Sitter one,
with a nonlinear realization for the gauge group as one was
performed in a locally isotropic version in Tseytlin's paper
\cite {tseyt}. If some former our works \cite{vg,vbook} where
devoted to extensions of some models of gauge gravity to
generalized Lagrange and Finsler spaces, in this paper  we
demonstrated which manner we could manage with anisotropies
arrising in locally isotropic,  but with anholonomic structures,
variants of gauge gravity. Here it should be emphasized  that
anisotropies of different type (Finsler like, or more general
ones) could be induced  in all variants of gravity theories
dealing with frame (tetrad, vierbiend, in four dimensions) fields
and decompisitions of geometrical and physical objects in
comonents with respect  to such frames and associated
N--connections. In a similar fashion anisotropies could arise
under nontrivial reductions from higher to lower dimensions in
Kaluza--Klein theories;  in this case the N--connection should be
treated as a splitting field  modeling the anholonomic
(anisotropic) character of some degrees of freedom.

The third basic result is the construction of a new class of
solutions, with generic local anisotropy, of the Einstein
equations. For simplicity, we defined these solutions in the
framework of general relativity, but they can be removed to
various variants of gauge and spinor gravity by using
corresponding decompositions of the metric into the frame fields.
We note that the obtained class of solutions also holds true for
the gauge models of gravity which, in this paper, were
constructed to b e equivalent to the Einstein theory. In explicit
form we considered the metric ansatz
\[
ds^{2}=g_{\alpha \beta }\ du^{\alpha }du^{\beta }
\]
when $g_{\alpha \beta }$ are parametrized by matrices of type
\begin{equation}
\left[
\begin{array}{cccc}
g_{1}+q_{1}{}^{2}h_{3}+n_{1}{}^{2}h_{4} &
q_{1}{}q_{2}h_{3}+n_{1}{}n_{2}h_{4}
& q_{1}h_{3} & n_{1}h_{4} \\
q_{1}{}q_{2}h_{3}+n_{1}{}n_{2}h_{4} &
g_{2}+q_{2}{}^{2}h_{3}+n_{2}{}^{2}h_{4}
& q_{2}h_{3} & n_{2}h_{4} \\
q_{1}h_{3} & q_{2}h_{3} & h_{3} & 0 \\
n_{1}h_{4} & n_{2}h_{4} & 0 & h_{4}
\end{array}
\right]   \label{4ansatz2}
\end{equation}
with coefficients being some functions of necessary smo\-oth class
\[
g_{i}=g_{i}(x^{j}),q_{i}=q_{i}(x^{j},t),n_{i}=n_{i}(x^{j},t),h_{a}=h_{a}(x^{j},t).
\]
Latin indices run respectively $i,j,k,...$ $=1,2$ and
$a,b,c,...=3,4$ and the local coordinates are denoted $u^{\alpha
}=(x^{i},y^{3}=t,y^{4}),$ where $t$ is treated as a time like
coordinate. A metric (\ref{4ansatz2}) can be diagonalized,
\begin{equation}
\delta s^{2}=g_{i}(x^{j})\left( dx^{i}\right)
^{2}+h_{a}(x^{j},t)\left( \delta y^{a}\right) ^{2},  \label{4diag}
\end{equation}
with respect to anholonomic frames (\ref{4dder}) and
(\ref{4ddif}), here we write down only the 'elongated'
differentials
\[
\delta t=dz+q_{i}(x^{j},t)dx^{i},\ \delta
y^{4}=dy^{4}+n_{i}(x^{j},t)dx^{i}.
\]

The ansatz (\ref{4ansatz2}) was formally introduced in \cite{vbh}
in order to construct locally anisotropic black hole solutions;
in this paper we applied it to cosmological locally anisotropic
space--times. In result, we get new metrics which describe
locally anisotropic Friedman--Robertson--Walker like universes
with the spherical symmetry de\-form\-ed to that of rotation
(elongated and/or flattened) ellipsoid and torus. Such solutions
are contained in general relativity: in the simplest diagonal
form they are parametrized by distinguished metrics of type (\ref
{4diag}), given with respect to anholonomic bases, but could be
also described equivalently with respect to a coordinate base by
matrices of type (\ref{4ansatz2}). The topic of construction of
cosmological models with generic spacetime and matter field
distribution and fluctuation anisotropies is under consideration.

Now, we point the item of definition of reference frames in
gravity theories: The form of basic field equations and
fundamental laws in general relativity do not depend on choosing
of coordinate systems and frame bases. Nevertheless, the problem
of fixing of an adequate system of reference is also a very
important physical task which is not solved by any dynamical
equations but following some arguments on measuring of physical
observables, imposed symmetry of interactions, types of horizons
and singularities, and by taken into consideration the posed
Cauchy problem. Having fixed a class of frame variables, the
frame coefficients being presented in the Einstein equations, the
type of constructed solution depends on the chosen holonomic or
anholonomic frame structure. As a result one could model various
forms of anisotropies in the framework of the Einsten theory
(roughly, on (pseudo) Riemannian spacetimes with corresponding
anholonomic frame structures it is possible to model Finsler like
metrics, or more general ones with anisotropies). Finally, it
should be noted that such questions on stability of obtained
solutions, analysis of energy--momentum conditions should be
performed in the simplest form with respect to the chosen class
of anholonomic frames.

\chapter[Anisotropic Taub NUT -- Dirac Spaces]{Anisotropic Taub NUT -- Dirac
Spaces}

The aim of this chapter is to outline the theory of gravity on
vector bundles provided with nonlinear connection structures
\cite{ma87,ma94} and to proof that anholonomic frames with
associated nonlinear connection structures can be introduced in
general relativity and in low dimensional and extra dimension
models of gravity on (pseudo) Riemannian space--times
\cite{vbh,vkinet}.

\section[N--connections in General Relativity]{Anholonomic Frames and
Nonlinear Connections in General Relativity}

The geometry of nonlinear connections on vector and higher order
vector bundles can be reformulated for anholonomic frames given
on a (pseudo) Riemannian spacetime of dimension $n+m$, or
$n+m_1+m_2+...+m_z,$ and provided with a d-metric structure which
induces on space--time a canonical d--connection structure
(\ref{inters}). In this case we can consider a formal splitting
of indices with respect to some holonomic and anholonomic frame
basis vectors. This approach was developed in references \cite
{vbh,vkinet} with the aim to construct exact solutions with
generic local anisotropy in general relativity and its low and
extra dimension modifications. For simplicity, in the further
sections of this chapter we shall restrict our constructions only
to first order anisotropic structures.

Recently one has proposed a new method of construction of exact
solutions of the Einstein equations on (pseudo) Riemannian spaces
of three, four and extra dimensions (in brief, 3D, 4D,...), by
applying the formalism of anholonomic moving frames \cite{vsol}.
There were constructed static solutions for black holes / tori,
soliton--dilaton systems and wormhole / flux tube configurations
and for anisotropic generalizations of the Taub NUT metric
\cite{vp}; all such solutions being, in general, with generic
local anisotropy. The method was elaborated following the
geometry of anholonomic frame (super) bundles and associated
nonlinear connections (in brief, N--connection) \cite{vsf} which
has a number of applications in generalized Finsler and Lagrange
geometry, anholonomic spinor geometry, (super) gravity and
strings with anisotropic (anholonomic) frame structures.

In this chapter we restrict our considerations for the 5D
Einstein gravity. In this case the N--connection coefficients are
defined by some particular parametrizations of funfbein, or
pentadic, coefficients defining a frame structure on (pseudo)
Riemannian spacetime and describing a gravitational and matter
field dynamics with mixed holonomic (unconstrained) and
anholonomic (constrained) variables. We emphasize that the
Einstein gravity theory in arbitrary dimensions can be
equivalently formulated with respect to both holonomic
(coordinate) and anholonomic frames. In the anholonomic cases the
rules of partial and covariant derivation are modified by some
pentad transforms. The point is to find such values of the
anholonomic frame (and associated N--connection) coefficients
when the metric is diagonalized and the Einstein equations are
written in a simplified form admitting exact solutions.

The class of new exact solutions of vacuum Einstein equations
describing anisotropic Taub NUT like spacetimes \cite{vt} is
defined by off--diagonal metrics if they are given with respect
to usual coordinate bases. Such metrics can be anholonomically
transformed into diagonal ones with coefficients being very
similar to the coefficients of the isotropic Taub NUT solution
but having additional dependencies on the 5th coordinate and
angular parameters.

We shall use the term locally anisotropic (spacetime) space (in
brief, anisotropic space) for a (pseudo) Riemannian space
provided with an anholonomic frame structure induced by a
procedure of anholonomic diagonalization of a off--diagonal
metric.

The Hawking's \cite{Ha} suggestion that the Euclidean Taub--NUT
metric might give rise to the gravitational analogue of the
Yang--Mills instanton holds true on anisotropic spaces but in
this case both the metric and instanton have some anisotropically
renormalized parameters being of higher dimension gravitational
vacuum polarization origin. The anisotropic Euclidean Taub-NUT
metric also satisfies the vacuum Einstein's equations with zero
cosmological constant when the spherical symmetry is deformed,
for instance, into ellipsoidal or even toroidal configuration.
Such anisotropic Taub-NUT metrics can be used for generation of
deformations of the space part of the line element defining an
anisotropic modification of the Kaluza-Klein monopole solutions
proposed by Gross and Perry \cite{GP} and Sorkin \cite{So}.

In the long-distance limit, neglecting radiation, the relative
motion of two such anisotropic monopoles can be also described by
geodesic motions, like in Ref. \cite{G1,G11,ah}, but these motions
are some anholonomic ones with associated nonlinear connection
structure and effective torsion induced by the anholonomy of the
systems of reference used for modeling anisotropies. The torsion
and N--connection corrections vanish if the geometrical objects
are transferred with respect to holonomic (coordinate) frames.

From the mathematical point of view, the new anholonomic geometry
of anisotropic Taub-NUT spaces is also very interesting. In the
locally isotropic Taub-NUT geometry there are four Killing-Yano
tensors \cite{GR}. Three of them form a complex structure
realizing the quaternionic algebra and the Taub-NUT manifold is
hyper-K\"{a}hler. In addition to such three vector-like
Killing-Yano tensors, there is a scalar one which exists by
virtue of the metric being of class $D,$ according to Petrov's
classification. Anisotropic deformations of metrics to
off--diagonal components introduce substantial changes in the
geometrical picture. Nevertheless, working with respect to
anholonomic frames with associated nonlinear connection structure
the basic properties and relations, even being anisotropically
modified, are preserved and transformed to similar ones for
deformed symmetries \cite{vt}.

The Schr\"{o}dinger quantum modes in the Euclidean Taub-NUT
geometry were analyzed using algebraic and analytical methods
\cite{GR,GR1,feher,cordani,gibm,CV,vh}. The Dirac equation was
studied in such locally isotropic curved backgrounds \cite
{DIRAC,kobs,bais}. One of the aims of this paper is to prove that
this approach can be developed as to include into consideration
anisotropic Taub-NUT backgrounds in the context of the standard
relativistic gauge-invariant theory \cite {W,BD} of the Dirac
field.

The purpose of the present work is to develop a general $SO(4,1)$
gauge-invariant theory of the Dirac fermions \cite{DKK} which can
be considered for locally anisotropic spaces, for instance, in
the external field of the Kaluza-Klein monopole
\cite{DIRAC,kobs,bais} which is anisotropically deformed.

Our goal is also to point out new features of the Einstein theory
in higher dimension spacetime when the locally anisotropic
properties, induced by anholonomic constraints and extra
dimension gravity, are emphasized. We shall analyze such effects
by constructing new classes of exact solutions of the
Einstein--Dirac equations defining 3D soliton--spinor
configurations propagating self--consistently in an anisotropic
5D Taub NUT spacetime.

We note that in this paper the 5D spacetime is modeled as a
direct time extension of a 4D Riemannian space provided with a
corresponding spinor structure, i. e. our spinor constructions
are not defined by some Clifford algebra associated to a 5D
bilinear form but, for simplicity, they are considered to be
extended from a spinor geometry defined for a 4D Riemannian space.

\subsection{Anholonomic Einstein--Dirac Equations} \index{Einsten--Dirac}

In this Section we introduce an ansatz for pseudo Riemannian
off--diagonal metrics and consider the anholonomic transforms
diagonalizing such metrics. The system of field Einstein
equations with the spinor matter energy--momentum tensor and of
Dirac equations are formulated on 5D pseudo--Riemannian
spacetimes constructed as a trivial extension by the time
variable of a 4D Riemannian space (an anisotropic deformation of
the Taub NUT instanton \cite{vt}).

\subsubsection{Ansatz for metrics}

We consider a 5D pseudo--Riemannian spacetime of signature
$(+,-,-,-,$ $-)$, with local coordinates
\[
u^\alpha =(x^i,y^a)=(x^0=t,x^1=r,x^2=\theta ,y^3=s,y^4=p),
\]
-- or more compactly $u=(x,y)$ -- where the Greek indices are
conventionally split into two subsets $x^i$ and $y^a$ labeled
respectively by Latin indices of type $i,j,k,...=0,1,2$ and
$a,b,...=3,4.$ The 5D (pseduo) Riemannian metric
\begin{equation}
ds^2=g_{\alpha \beta }du^\alpha du^\beta  \label{metric1}
\end{equation}
is given by a metric ansatz parametrized in the form
\begin{equation}
g_{\alpha \beta }=\left[
\begin{array}{ccccc}
1 & 0 & 0 & 0 & 0 \\
0 & g_1+w_1^{\ 2}h_3+n_1^{\ 2}h_4 & w_1w_2h_3+n_1n_2h_4 & w_1h_3 & n_1h_4 \\
0 & w_2w_1h_3+n_1n_2h_4 & g_2+w_2^{\ 2}h_3+n_2^{\ 2}h_4 & w_2h_3 & n_2h_4 \\
0 & w_1h_3 & w_2h_3 & h_3 & 0 \\
0 & n_1h_4 & n_2h_4 & 0 & h_4
\end{array}
\right] ,  \label{ansatz0}
\end{equation}
where the coefficients are some functions of type
\begin{eqnarray}
g_{1,2} &=&g_{1,2}(x^1,x^2),h_{3,4}=h_{3,4}(x^1,x^2,s),  \label{qvar} \\
w_{1,2} &=&w_{1,2}(x^1,x^2,s),n_{1,2}=n_{1,2}(x^1,x^2,s).
\nonumber
\end{eqnarray}
Both the inverse matrix (metric) as well the metric
(\ref{ansatz0}) is off--diagonal with respect to the coordinate
basis
\begin{equation}
\partial _\alpha \equiv \frac \partial {du^\alpha }=(\partial _i=\frac
\partial {dx^i},\partial _a=\frac \partial {dy^a})  \label{pder2}
\end{equation}
and, its dual basis,
\begin{equation}
d^\alpha \equiv du^\alpha =(d^i=dx^i,d^a=dy^a).  \label{pdif2}
\end{equation}

The metric (\ref{metric1}) with coefficients (\ref{ansatz0}) can
be equivalently rewritten in the diagonal form
\begin{eqnarray}
\delta s^2 &=&dt^2+g_1\left( x\right) (dx^1)^2+g_2\left( x\right)
(dx^2)^2
\label{dmetric2} \\
&{}&+h_3\left( x,s\right) (\delta y^3)^2+h_4\left( x,s\right)
(\delta y^4)^2, \nonumber
\end{eqnarray}
if instead the coordinate bases (\ref{pder2}) and (\ref{pdif2})
we introduce the anholonomic frames (anisotropic bases)
\begin{equation}
{\delta }_\alpha \equiv \frac \delta {du^\alpha }=(\delta
_i=\partial _i-N_i^b(u)\ \partial _b,\partial _a=\frac \partial
{dy^a})  \label{dder2}
\end{equation}
and
\begin{equation}
\delta ^\alpha \equiv \delta u^\alpha =(\delta ^i=dx^i,\delta
^a=dy^a+N_k^a(u)\ dx^k)  \label{ddif2}
\end{equation}
where the $N$--coefficients are parametrized
\[
N_0^a=0,\ N_{1,2}^3=w_{1,2}\mbox{ and }N_{1,2}^4=n_{1,2}
\]
and define the associated nonlinear connection (N--connection)
structure, see details in Refs \cite{vsol,vt,vsf}.

\subsubsection{Einstein equations with anholonomic variables}

The metric (\ref{metric1}) with coefficients (\ref{ansatz0})
(equivalently, the d--metric (\ref{dmetric2})) is assumed to
solve the 5D Einstein equations
\begin{equation}
R_{\alpha \beta }-\frac 12g_{\alpha \beta }R=\kappa \Upsilon
_{\alpha \beta },  \label{5einstein}
\end{equation}
where $\kappa $ and $\Upsilon _{\alpha \beta }$ are respectively
the coupling constant and the energy--momentum tensor.

The nontrivial components of the Ricci tensor for the metric
(\ref{metric1}) with coefficients (\ref{ansatz0}) (equivalently,
the d--metric (\ref {dmetric2})) are
\begin{eqnarray}
R_{1}^{1} &=&R_{2}^{2}=-\frac{1}{2g_{1}g_{2}}[g_{2}^{\bullet \bullet }-\frac{%
g_{1}^{\bullet }g_{2}^{\bullet }}{2g_{1}}-\frac{(g_{2}^{\bullet })^{2}}{%
2g_{2}}+g_{1}^{^{\prime \prime }}-\frac{g_{1}^{^{\prime }}g_{2}^{^{\prime }}%
}{2g_{2}}-\frac{(g_{1}^{^{\prime }})^{2}}{2g_{1}}],  \label{ricci1a} \\
R_{3}^{3} &=&R_{4}^{4}=-\frac{\beta }{2h_{3}h_{4}},  \label{ricci1b} \\
&{}&  \nonumber \\
R_{31} &=&-w_{1}\frac{\beta }{2h_{4}}-\frac{\alpha _{1}}{2h_{4}},
\label{ricci1c} \\
R_{32} &=&-w_{2}\frac{\beta }{2h_{4}}-\frac{\alpha
_{2}}{2h_{4}},  \nonumber
\\
&{}&  \nonumber \\
R_{41} &=&-\frac{h_{4}}{2h_{3}}\left[ n_{1}^{\ast \ast }+\gamma n_{1}^{\ast }%
\right] ,  \label{ricci1d} \\
R_{42} &=&-\frac{h_{4}}{2h_{3}}\left[ n_{2}^{\ast \ast }+\gamma n_{2}^{\ast }%
\right] ,  \nonumber
\end{eqnarray}
where, for simplicity, the partial derivatives are denoted
$h^{\bullet
}=\partial h/\partial x^{1},f^{\prime }=\partial f/\partial x^{2}$ and $%
f^{\ast }=\partial f/\partial s.$

The scalar curvature is computed
\[
R=2\left( R_1^1+R_3^3\right) .
\]

In result of the obtained equalities for some Ricci and Einstein
tensor components, we conclude that for the metric ansatz
(\ref{ansatz0}) the Einstein equations with matter sources are
compatible if the coefficients of the energy--momentum d--tensor
give with respect to anholonomic bases satisfy the conditions
\begin{equation}
\Upsilon _0^0=\Upsilon _1^1+\Upsilon _3^3,\Upsilon _1^1=\Upsilon
_2^2=\Upsilon _1,\Upsilon _3^3=\Upsilon _4^4=\Upsilon _3,
\label{spinorem}
\end{equation}
and could be written in the form
\begin{eqnarray}
R_1^1 &=&-\kappa \Upsilon _3,  \label{einsteq2a} \\
R_3^3 &=&-\kappa \Upsilon _1,  \label{einsteq2b} \\
R_{3\widehat{i}} &=&\kappa \Upsilon _{3\widehat{i}},  \label{einsteq2c} \\
R_{4\widehat{i}} &=&\kappa \Upsilon _{4\widehat{i}},
\label{einsteq2d}
\end{eqnarray}
where $\widehat{i}=1,2$ and the left parts are given by the
components of the Ricci tensor (\ref{ricci1a})-(\ref{ricci1d}).

The Einstein equations (\ref{5einstein}), equivalently (\ref{einsteq2a})--(%
\ref{einsteq2d}), reduce to this system of second order partial
derivation equations:
\begin{eqnarray}
g_{2}^{\bullet \bullet }-\frac{g_{1}^{\bullet }g_{2}^{\bullet }}{2g_{1}}-%
\frac{(g_{2}^{\bullet })^{2}}{2g_{2}}+g_{1}^{^{\prime \prime }}-\frac{%
g_{1}^{^{\prime }}g_{2}^{^{\prime }}}{2g_{2}}-\frac{(g_{1}^{^{\prime }})^{2}%
}{2g_{1}} &=&-2g_{1}g_{2}\Upsilon _{3},  \label{einsteq3a} \\
h_{4}^{\ast \ast }-\frac{(h_{4}^{\ast
})^{2}}{2h_{4}}-\frac{h_{4}^{\ast
}h_{3}^{\ast }}{2h_{3}} &=&-2h_{3}h_{4}\Upsilon _{1},  \label{einsteq3b} \\
\beta w_{i}+\alpha _{i} &=&-2h_{4}\kappa \Upsilon _{3i},
\label{einsteq3c}
\\
n_{i}^{\ast \ast }+\gamma n_{i}^{\ast }
&=&-\frac{2h_{3}}{h_{4}}\kappa \Upsilon _{4i},  \label{einsteq3d}
\end{eqnarray}
where
\begin{eqnarray}
\alpha _{1} &=&{h_{4}^{\ast }}^{\bullet }-\frac{{h_{4}^{\ast
}}}{2}\left( \frac{h_{3}^{\bullet }}{h_{3}}+\frac{h_{4}^{\bullet
}}{h_{4}}\right) ,
\label{alpha1} \\
\alpha _{2} &=&{h_{4}^{\ast }}^{\prime }-\frac{{h_{4}^{\ast
}}}{2}\left( \frac{h_{3}^{\prime }}{h_{3}}+\frac{h_{4}^{\prime
}}{h_{4}}\right) ,
\label{alpha2} \\
\beta  &=&h_{4}^{\ast \ast }-\frac{(h_{4}^{\ast })^{2}}{2h_{4}}-\frac{%
h_{4}^{\ast }h_{3}^{\ast }}{2h_{3}},  \label{beta} \\
\gamma  &=&\frac{3}{2}\frac{h_{4}}{h_{4}}^{\ast
}-\frac{h_{3}}{h_{3}}^{\ast },  \label{gamma2}
\end{eqnarray}
and the partial derivatives are denoted, for instance,
\begin{eqnarray*}
g_{2}^{\bullet } &=&\partial g_{2}/\partial x^{1}=\partial
g_{2}/\partial
r,g_{1}^{^{\prime }}=\partial g_{1}/\partial x^{2}=\partial g_{1}/\theta , \\
h_{3}^{\ast } &=&\partial h_{3}/\partial s=\partial
h_{3}/\partial \varphi \ (\mbox{or }\partial h_{3}/\partial
y^{4},\mbox{ for }s=y^{4}).
\end{eqnarray*}

\subsubsection{Dirac equations in anisotropic space--times}

The problem of definition of spinors in locally anisotropic
spaces and in spaces with higher order anisotropy was solved in
Refs. \cite{vsf}. In this paper we consider locally anisotropic
Dirac spinors given with respect to anholonomic frames with
associated N--connection structure on a 5D (pseudo) Riemannian
space $V^{(1,2,2)}$ constructed by a direct time extension of a
4D Riemannian space with two holonomic and two anholonomic
variables.

Having an anisotropic d--metric
\begin{eqnarray*}
g_{\alpha \beta }(u)
&=&(g_{ij}(u),h_{ab}(u))=(1,g_{\widehat{i}}(u),h_a(u)),
\\
\widehat{i} &=&1,2;i=0,1,2;a=3,4,
\end{eqnarray*}
defined with respect to an anholonomic basis (\ref{dder}) we can
easily define the funfbein (pentad) fields
\begin{eqnarray}
f_{\underline{\mu }} &=&f_{\underline{\mu }}^\mu \delta _\mu =\{f_{%
\underline{i}}=f_{\underline{i}}^i\delta _i,f_{\underline{a}}=f_{\underline{a%
}}^a\partial _a\},  \label{pentad1} \\
\ f^{\underline{\mu }} &=&f_\mu ^{\underline{\mu }}\delta ^\mu =\{f^{%
\underline{i}}=f_i^{\underline{i}}d^i,,f^{\underline{a}}=f_a^{\underline{a}%
}\delta ^a\}  \nonumber
\end{eqnarray}
satisfying the conditions
\begin{eqnarray*}
g_{ij} &=&f_i^{\underline{i}}f_j^{\underline{j}}g_{\underline{i}\underline{j}%
}\mbox{ and }h_{ab}=f_a^{\underline{a}}f_b^{\underline{b}}h_{\underline{a}%
\underline{b}}, \\
g_{\underline{i}\underline{j}} &=&diag[1,-1-1]\mbox{ and }h_{\underline{a}%
\underline{b}}=diag[-1,-1].
\end{eqnarray*}
For a diagonal d-metric of type (\ref{dmetric}) we have
\[
f_i^{\underline{i}}=\sqrt{\left| g_i\right| }\delta
_i^{\underline{i}}\mbox{ and }f_a^{\underline{a}}=\sqrt{\left|
h_a\right| }\delta _a^{\underline{a}},
\]
where $\delta _i^{\underline{i}}$ and $\delta _a^{\underline{a}}$
are Kronecker's symbols.

We can also introduce the corresponding funfbiends which are
related with
the off--diagonal metric ansatz (\ref{ansatz0}) for $g_{\alpha \beta },$%
\begin{equation}
e_{\underline{\mu }}=e_{\underline{\mu }}^\mu \partial _\mu \mbox{ and }e^{%
\underline{\mu }}=e_\mu ^{\underline{\mu }}\partial ^\mu
\label{pentad2}
\end{equation}
satisfying the conditions
\begin{eqnarray*}
g_{\alpha \beta } &=&e_\alpha ^{\underline{\alpha }}e_\beta ^{\underline{%
\beta }}g_{\underline{\alpha }\underline{\beta }}\mbox{ for }g_{\underline{%
\alpha }\underline{\beta }}=diag[1,-1,-1,-1,-1], \\
e_\alpha ^{\underline{\alpha }}e_{\underline{\alpha }}^\mu
&=&\delta _\alpha ^\mu \mbox{ and }e_\alpha ^{\underline{\alpha
}}e_{\underline{\mu }}^\alpha =\delta _{\underline{\mu
}}^{\underline{\alpha }}.
\end{eqnarray*}

The Dirac spinor fields on locally anisotropic deformations of
Taub NUT spaces,
\[
\Psi \left( u\right) =[\Psi ^{\overline{\alpha }}\left( u\right) ]=[\psi ^{%
\widehat{I}}\left( u\right) ,\chi _{\widehat{I}}\left( u\right) ],
\]
where $\widehat{I}=0,1,$ are defined with respect to the 4D
Euclidean tangent subspace belonging the tangent space to
$V^{(1,2,2)}.$ The $4\times 4 $ dimensional gamma matrices
$\gamma ^{\underline{\alpha }^{\prime }}=[\gamma
^{\underline{1}^{\prime }},\gamma ^{\underline{2}^{\prime
}},\gamma ^{\underline{3}^{\prime }},\gamma
^{\underline{4}^{\prime }}]$ are defined as to satisfy the
relation
\begin{equation}
\left\{ \gamma ^{\underline{\alpha }^{\prime }},\,\gamma ^{\underline{\beta }%
^{\prime }}\right\} =2g^{\underline{\alpha }\underline{^{\prime }\beta }%
^{\prime }},  \label{gammarel}
\end{equation}
where $\left\{ \gamma ^{\underline{\alpha }^{\prime }}\,\gamma ^{\underline{%
\beta }^{\prime }}\right\} $ is a symmetric commutator, $g^{\underline{%
\alpha }\underline{^{\prime }\beta }^{\prime }}=(-1,-1,-1,-1),$
which generates a Clifford algebra distinguished on two holonomic
and two anholonomic directions (hereafter the primed indices will
run values on the Euclidean and/or Riemannian, 4D component of
the 5D pseudo--Riemannian spacetime). In order to extend the
(\ref{gammarel}) relations for unprimed indices $\alpha ,\beta
...$ we conventionally complete the set of primed gamma matrices
with a matrix $\gamma ^{\underline{0}},$ i. .e. write $\gamma
^{\underline{\alpha }}=[\gamma ^{\underline{0}},\gamma ^{\underline{1}%
},\gamma ^{\underline{2}},\gamma ^{\underline{3}},\gamma
^{\underline{4}}]$ when
\[
\left\{ \gamma ^{\underline{\alpha }},\,\gamma ^{\underline{\beta
}}\right\} =2g^{\underline{\alpha }\underline{\beta }}.
\]

The coefficients of gamma matrices can be computed with respect to
coordinate bases (\ref{pder}) or with respect to anholonomic
bases (\ref
{dder}) by using respectively the funfbein coefficients (\ref{pentad1}) and (%
\ref{pentad2}),
\[
\gamma ^\alpha (u)=e_{\underline{\alpha }}^\alpha (u)\gamma ^{\underline{%
\alpha }}\mbox{ and }\widehat{\gamma }^\beta (u)=f_{\underline{\beta }%
}^\beta (u)\gamma ^{\underline{\beta }},
\]
were by $\gamma ^\alpha (u)$ we denote the curved spacetime gamma
matrices and by $\widehat{\gamma }^\beta (u)$ we denote the gamma
matrices adapted to the N--connection structure.

The covariant derivation of Dirac spinor field $\Psi \left( u\right) ,$ $%
\bigtriangledown _\alpha \Psi ,$ can be defined with respect to a
pentad decomposition of the off--diagonal metric (\ref{ansatz0})
\begin{equation}
\bigtriangledown _\alpha \Psi =\left[ \partial _\alpha +\frac 14C_{%
\underline{\alpha }\underline{\beta }\underline{\gamma }}\left(
u\right)
~e_\alpha ^{\underline{\alpha }}\left( u\right) \gamma ^{\underline{\beta }%
}\gamma ^{\underline{\gamma }}\right] \Psi ,  \label{covspinder}
\end{equation}
where the coefficients
\[
C_{\underline{\alpha }\underline{\beta }\underline{\gamma
}}\left( u\right)
=\left( D_\gamma e_{\underline{\alpha }}^\alpha \right) e_{\underline{\beta }%
\alpha }e_{\underline{\gamma }}^\gamma
\]
are called the rotation Ricci coefficients; the covariant derivative $%
D_\gamma $ is defined by the usual Christoffel symbols for the
off--diagonal metric.

We can also define an equivalent covariant derivation of the
Dirac spinor field, $\overrightarrow{\bigtriangledown }_\alpha
\Psi ,$ by using pentad decompositions of the diagonalized
d--metric (\ref{dmetric}),
\begin{equation}
\overrightarrow{\bigtriangledown }_\alpha \Psi =\left[ \delta
_\alpha +\frac 14C_{\underline{\alpha }\underline{\beta
}\underline{\gamma }}^{[\delta
]}\left( u\right) ~f_\alpha ^{\underline{\alpha }}\left( u\right) \gamma ^{%
\underline{\beta }}\gamma ^{\underline{\gamma }}\right] \Psi ,
\label{covspindder}
\end{equation}
where there are introduced N--elongated partial derivatives and
the coefficients
\[
C_{\underline{\alpha }\underline{\beta }\underline{\gamma
}}^{[\delta
]}\left( u\right) =\left( D_\gamma ^{[\delta ]}f_{\underline{\alpha }%
}^\alpha \right) f_{\underline{\beta }\alpha
}f_{\underline{\gamma }}^\gamma
\]
are transformed into rotation Ricci d--coefficients which
together with the d--covariant derivative $D_\gamma ^{[\delta ]}$
are defined by anholonomic pentads and anholonomic transforms of
the Christoffel symbols.

For diagonal d--metrics the funfbein coefficients can be taken in
their turn
in diagonal form and the corresponding gamma matrix $\widehat{\gamma }%
^\alpha \left( u\right) $ for anisotropic curved spaces are
proportional to the usual gamma matrix in flat spaces $\gamma
^{\underline{\gamma }}$. The Dirac equations for locally
anisotropic spacetimes are written in the simplest form with
respect to anholonomic frames,
\begin{equation}
(i\widehat{\gamma }^\alpha \left( u\right)
\overrightarrow{\bigtriangledown _\alpha }-\mu )\Psi =0,
\label{diraceq}
\end{equation}
where $\mu $ is the mass constant of the Dirac field. The Dirac
equations are the Euler equations for the Lagrangian
\begin{eqnarray}
&\mathcal{L}^{(1/2)}\left( u\right) &= \sqrt{\left| g\right|
}\{[\Psi
^{+}\left( u\right) \widehat{\gamma }^\alpha \left( u\right) \overrightarrow{%
\bigtriangledown _\alpha }\Psi \left( u\right)  \label{direq} \\
&{}&-(\overrightarrow{\bigtriangledown _\alpha }\Psi ^{+}\left( u\right) )%
\widehat{\gamma }^\alpha \left( u\right) \Psi \left( u\right)
]-\mu \Psi ^{+}\left( u\right) \Psi \left( u\right) \},  \nonumber
\end{eqnarray}
where by $\Psi ^{+}\left( u\right) $ we denote the complex
conjugation and transposition of the column$~\Psi \left( u\right)
.$

Varying $\mathcal{L}^{(1/2)}$ on d--metric (\ref{direq}) we
obtain the symmetric energy--mo\-ment\-um d--tensor
\begin{eqnarray}
\Upsilon _{\alpha \beta }\left( u\right)  &=&\frac{i}{4}[\Psi
^{+}\left(
u\right) \widehat{\gamma }_{\alpha }\left( u\right) \overrightarrow{%
\bigtriangledown _{\beta }}\Psi \left( u\right) +\Psi ^{+}\left(
u\right) \widehat{\gamma }_{\beta }\left( u\right)
\overrightarrow{\bigtriangledown
_{\alpha }}\Psi \left( u\right)   \nonumber \\
&{}&-(\overrightarrow{\bigtriangledown _{\alpha }}\Psi ^{+}\left( u\right) )%
\widehat{\gamma }_{\beta }\left( u\right) \Psi \left( u\right) -(%
\overrightarrow{\bigtriangledown _{\beta }}\Psi ^{+}\left( u\right) )%
\widehat{\gamma }_{\alpha }\left( u\right) \Psi \left( u\right) ].
\label{diracemd}
\end{eqnarray}
We choose such spinor field configurations in curved spacetime as
to be satisfied the conditions (\ref{spinorem}).

One can introduce similar formulas to
(\ref{diraceq})--(\ref{diracemd}) for spacetimes provided with
off-diagonal metrics with respect to holonomic frames by changing
of operators $\widehat{\gamma }_\alpha \left( u\right)
\rightarrow \gamma _\alpha \left( u\right) $ and $\overrightarrow{%
\bigtriangledown _\beta }\rightarrow \bigtriangledown _\beta .$

\subsection{Anisotropic Taub NUT -- Dirac Spinor Solutions}

By straightforward calculations we can verify that because the conditions $%
D_\gamma ^{[\delta ]}f_{\underline{\alpha }}^\alpha =0$ are
satisfied the Ricci rotation coefficients vanishes,
\[
C_{\underline{\alpha }\underline{\beta }\underline{\gamma
}}^{[\delta
]}\left( u\right) =0\mbox{ and }\overrightarrow{\bigtriangledown _\alpha }%
\Psi =\delta _\alpha \Psi ,
\]
and the anisotropic Dirac equations (\ref{diraceq}) transform into
\begin{equation}
(i\widehat{\gamma }^\alpha \left( u\right) \delta _\alpha -\mu
)\Psi =0. \label{diraceq1}
\end{equation}

Further simplifications are possible for Dirac fields depending
only on coordinates $(t,x^1=r,x^2=\theta )$, i. e. $\Psi =\Psi
(x^k)$ when the equation (\ref{diraceq1}) transforms into
\[
(i\gamma ^{\underline{0}}\partial _t+i\gamma ^{\underline{1}}\frac 1{\sqrt{%
\left| g_1\right| }}\partial _1+i\gamma ^{\underline{2}}\frac
1{\sqrt{\left| g_2\right| }}\partial _2-\mu )\Psi =0.
\]
The equation (\ref{diraceq1}) simplifies substantially in $\zeta $%
--coordinates
\[
\left( t,\zeta ^1=\zeta ^1(r,\theta ),\zeta ^2=\zeta ^2(r,\theta
)\right) ,
\]
defined as to be satisfied the conditions
\begin{equation}
\frac \partial {\partial \zeta ^1}=\frac 1{\sqrt{\left| g_1\right| }%
}\partial _1\mbox{ and }\frac \partial {\partial \zeta ^2}=\frac 1{\sqrt{%
\left| g_2\right| }}\partial _2  \label{zetacoord}
\end{equation}
We get
\begin{equation}
(-i\gamma _{\underline{0}}\frac \partial {\partial t}+i\gamma _{\underline{1}%
}\frac \partial {\partial \zeta ^1}+i\gamma _{\underline{2}}\frac
\partial {\partial \zeta ^2}-\mu )\Psi (t,\zeta ^1,\zeta ^2)=0.
\label{diraceq2}
\end{equation}
The equation (\ref{diraceq2}) describes the wave function of a
Dirac particle of mass $\mu $ propagating in a three dimensional
Minkowski flat plane which is imbedded as an anisotropic
distribution into a 5D pseudo--Riemannian spacetime.

The solution $\Psi = \Psi (t,\zeta ^1,\zeta ^2)$ of
(\ref{diraceq2}) can be written
\[
\Psi =\left\{
\begin{array}{rcl}
\Psi ^{(+)}(\zeta ) & = & \exp {[-i(k_0t+k_1\zeta ^1+k_2\zeta
^2)]}\varphi
^0(k) \\
&  & \mbox{for positive energy;} \\
\Psi ^{(-)}(\zeta ) & = & \exp {[i(k_0t+k_1\zeta ^1+k_2\zeta
^2)]}\chi ^0(k)
\\
&  & \mbox{for negative energy,}
\end{array}
\right.
\]
with the condition that $k_0$ is identified with the positive energy and $%
\varphi ^0(k)$ and $\chi ^0(k)$ are constant bispinors. To
satisfy the Klein--Gordon equation we must have
\[
k^2={k_0^2-k_1^2-k_2^2}=\mu ^2.
\]
The Dirac equations implies
\[
(\sigma ^ik_i-\mu )\varphi ^0(k)\mbox{ and }(\sigma ^ik_i+\mu
)\chi ^0(k),
\]
where $\sigma ^i(i=0,1,2)$ are Pauli matrices corresponding to a
realization of gamma matrices as to a form of splitting to usual
Pauli equations for the bispinors $\varphi ^0(k)$ and $\chi
^0(k).$

In the rest frame for the horizontal plane parametrized by coordinates $%
\zeta =\{t,\zeta ^1,\zeta ^2\}$ there are four independent
solutions of the Dirac equations,
\[
\varphi _{(1)}^0(\mu ,0)=\left(
\begin{array}{c}
1 \\
0 \\
0 \\
0
\end{array}
\right) ,\ \varphi _{(2)}^0(\mu ,0)=\left(
\begin{array}{c}
0 \\
1 \\
0 \\
0
\end{array}
\right) ,\
\]
\[
\chi _{(1)}^0(\mu ,0)=\left(
\begin{array}{c}
0 \\
0 \\
1 \\
0
\end{array}
\right) ,\ \chi _{(2)}^0(\mu ,0)=\left(
\begin{array}{c}
0 \\
0 \\
0 \\
1
\end{array}
\right) .
\]
In order to satisfy the conditions (\ref{spinorem}) for
compatibility of the equations
(\ref{einsteq3a})--(\ref{einsteq3d}) we must consider wave packets
of type (for simplicity, we can use only superpositions of
positive energy solutions)
\begin{eqnarray}
\Psi ^{(+)}(\zeta ) &=&\int \frac{d^3p}{2\pi ^3}\frac \mu
{\sqrt{\mu
^2+(k^2)^2}}  \nonumber \\
&{}&\times \sum_{[\alpha ]=1,2,3}b(p,[\alpha ])\varphi ^{[\alpha ]}(k)\exp {%
[-ik_i\zeta ^i]}  \label{packet}
\end{eqnarray}
when the coefficients $b(p,[\alpha ])$ define a current (the
group velocity)
\begin{eqnarray}
&&J^2=\sum_{[\alpha ]=1,2,3}\int \frac{d^3p}{2\pi ^3}\frac \mu
{\sqrt{\mu
^2+(k^2)^2}}|b(p,[\alpha ])|^2\frac{p^2}{\sqrt{\mu ^2+(k^2)^2}}  \nonumber \\
&&{}\equiv <\frac{p^2}{\sqrt{\mu ^2+(k^2)^2}}>  \nonumber
\end{eqnarray}
with $|p^2|\sim \mu $ and the energy--momentum d--tensor
(\ref{diracemd}) has the next nontrivial coefficients
\begin{eqnarray}
\Upsilon _0^0 &=&2\Upsilon (\zeta ^1,\zeta ^2)=k_0\Psi ^{+}\gamma
_0\Psi ,
\nonumber \\
\Upsilon _1^1 &=&-k_1\Psi ^{+}\gamma _1\Psi ,\Upsilon
_2^2=-k_2\Psi ^{+}\gamma _2\Psi \   \label{compat}
\end{eqnarray}
where the holonomic coordinates can be reexpressed $\zeta
^i=\zeta ^i(x^i).$
We must take two or more waves in the packet and choose such coefficients $%
b(p,[\alpha ]),$ satisfying corresponding algebraic equations, as
to have in (\ref{compat}) the equalities
\begin{equation}
\Upsilon _1^1=\Upsilon _2^2=\Upsilon (\zeta ^1,\zeta ^2)=\Upsilon
(x^1,x^2), \label{compat1}
\end{equation}
required by the conditions (\ref{diracemd}).

\section[Anisotropic Taub NUT Solutions]{Taub NUT Solutions with Generic
Local Anisotropy} \index{Taub NUT}

The Kaluza-Klein monopole \cite{GP,So} was obtained by embedding
the Taub-NUT gravitational instanton into five-dimensional
theory, adding the time coordinate in a trivial way. There are
anisotropic variants of such solutions \cite{vt} when
anisotropies are modelled by effective polarizations of the
induced magnetic field. The aim of this Section is to analyze
such Taub--NUT solutions for both cases of locally isotropic and
locally anisotropic configurations.

\subsection{A conformal transform of the Taub NUT metric}

We consider the Taub NUT solutions and introduce a conformal
transformation and a such redefinition of variables which will be
useful for further generalizations to anisotropic vacuum
solutions.

\subsubsection{The Taub NUT solution}

This locally isotropic solution of the 5D vacuum Einstein
equations is expressed by the line element
\begin{eqnarray}
ds_{(5D)}^{2} &=&dt^{2}+ds_{(4D)}^{2};  \label{nut} \\
ds_{(4D)}^{2} &=&-V^{-1}(dr^{2}+r^{2}d\theta ^{2}+\sin ^{2}\theta
d\varphi ^{2})-V(dx^{4}+A_{i}dx^{i})^{2}\,\   \nonumber
\end{eqnarray}
where
\[
V^{-1}=1+\frac{m_{0}}{r},m_{0}=const.
\]
The functions $A_{i}$ are static ones associated to the
electromagnetic potential,
\[
A_{r}=0,A_{\theta }=0,A_{\varphi }=4m_{0}\left( 1-\cos \theta
\right)
\]
resulting into ''pure'' magnetic field
\begin{equation}
\vec{B}\,=\mathrm{rot}\,\vec{A}=m_{0}\frac{\overrightarrow{r}}{r^{3}}\,.
\label{magnetic}
\end{equation}
of a Euclidean instanton; $\overrightarrow{r}$ is the spherical
coordinate's unity vector. The spacetime defined by (\ref{nut})
has the {\ global} symmetry of the group $G_{s}=SO(3)\otimes
U_{4}(1)\otimes T_{t}(1)$ since the line element is invariant
under the global rotations of the Cartesian
space coordinates and $y^{4}$ and $t$ translations of the Abelian groups $%
U_{4}(1)$ and $T_{t}(1)$ respectively. We note that the
$U_{4}(1)$ symmetry eliminates the so called NUT singularity if
$y^{4}$ has the period $4\pi m_{0}$.

\subsubsection{Conformally transformed Taub NUT metrics}

With the aim to construct anisotropic generalizations it is more
convenient to introduce a new 5th coordinate,
\begin{equation}
y^{4}\rightarrow \varsigma =y^{4}-\int \mu ^{-1}(\theta ,\varphi
)d\xi (\theta ,\varphi ),  \label{coordch}
\end{equation}
with the property that
\[
d\varsigma +4m_{0}(1-\cos \theta )d\theta =dy^{4}+4m_{0}(1-\cos
\theta )d\varphi ,
\]
which holds for
\[
d\xi =\mu (\theta ,\varphi )d(\varsigma -y^{4})=\frac{\partial \xi }{%
\partial \theta }d\theta +\frac{\partial \xi }{\partial \varphi }d\varphi ,
\]
when
\[
\frac{\partial \xi }{\partial \theta }=4m_{0}(1-\cos \theta )\mu
,\quad \frac{\partial \xi }{\partial \varphi }=-4m_{0}(1-\cos
\theta )\mu ,
\]
and, for instance,
\[
\mu =\left( 1-\cos \theta \right) ^{-2}\exp [\theta -\varphi ].
\]
The changing of coordinate (\ref{coordch}) describe a
re--orientation of the 5th coordinate in a such way as we could
have only one nonvanishing component of the electromagnetic
potential
\[
A_{\theta }=4m_{0}\left( 1-\cos \theta \right) .
\]

The next step is to perform a conformal transform,
\[
ds_{(4D)}^2\rightarrow d\widehat{s}_{(4D)}^2=Vds_{(4D)}^2
\]
and to consider the 5D metric

\begin{eqnarray}
ds_{(5D)}^{2} &=&dt^{2}+d\widehat{s}_{(4D)}^{2};  \label{conf4d} \\
d\widehat{s}_{(4D)}^{2} &=&-(dr^{2}+r^{2}d\theta ^{2})-r^{2}\sin
^{2}\theta d\varphi ^{2}-V^{2}(d\zeta +A_{\theta }d\theta
)^{2},\,  \nonumber
\end{eqnarray}
(not being an exact solution of the Einstein equations) which
will transform into some exact solutions after corresponding
anholonomic transforms.

Here, we emphasize that we chose the variant of transformation of
a locally isotropic non--Einsteinian metrics into an anisotropic
one solving the vacuum Einstein equations in order to illustrate
a more simple procedure of construction of 5D vacuum metrics with
generic local anisotropy. As a metter of principle we could
remove vacuum isotropic solutions into vacuum anisotropic ones,
but the formula in this case would became very combersome.. The
fact of selection as an isotropic 4D Riemannian background just
the metric from the linear interval $d\widehat{s}_{(4D)}^2$ can be
treated as a conformal transformation of an instanton solution
which is anisotropically deformed and put trivially (by extension
to the time like coordinate) into a 5D metric as to generate a
locally isotropic vacuum gravitational field.

\subsection{Anisotropic Taub NUT solutions with mag\-netic polarization}

We outline two classes of exact solutions of 5D vacuum Einstein
equations with generic anisotropies (see details in Ref.
\cite{vt}) which will be extended to configurations with spinor
matter field source.

\subsubsection{Solutions with angular polarization}

The ansatz for a d--metric (\ref{dmetric}), with a distinguished
anisot\-rop\-ic dependence on the angular coordinate $\varphi ,$ when $%
s=\varphi ,$ is taken in the form
\begin{eqnarray*}
\delta s^{2} &=&dt^{2}-\delta s_{(4D)}^{2}, \\
\delta s_{(4D)}^{2} &=&-(dr^{2}+r^{2}d\theta ^{2})-r^{2}\sin
^{2}\theta d\varphi ^{2}-V^{2}(r)\eta _{4}^{2}(\theta ,\varphi
)\delta \varsigma ^{2},
\\
\delta \varsigma  &=&d\varsigma +n_{2}(\theta ,\varphi )d\theta ,
\end{eqnarray*}
where the values $\eta _{4}^{2}(\theta ,\varphi )$ (we use
non--negative
values $\eta _{4}^{2}$ not changing the signature of metrics) and $%
n_{2}(\theta ,\varphi )$ must be found as to satisfy the vacuum
Einstein equations in the form
(\ref{einsteq3a})--(\ref{einsteq3d}). We can verify that the data
\begin{eqnarray}
x^{0} &=&t,x^{1}=r,x^{2}=\theta ,y^{3}=s=\varphi ,y^{4}=\varsigma
,
\label{sol1} \\
g_{0} &=&1,g_{1}=-1,g_{2}=-r^{2},h_{3}=-r^{2}\sin ^{2}\theta ,  \nonumber \\
h_{4} &=&V^{2}\left( r\right) \eta _{(\varphi )}^{2},\eta
_{(\varphi
)}^{2}=[1+\varpi (r,\theta )\varphi ]^{2},w_{i}=0;  \nonumber \\
n_{0,1} &=&0;n_{2}=n_{2[0]}\left( r,\theta \right)
+n_{2[1]}\left( r,\theta \right) /[1+\varpi (r,\theta )\varphi
]^{2}.  \nonumber
\end{eqnarray}
give an exact solution. If we impose the condition to obtain in
the locally isotropic limit just the metric (\ref{conf4d}), we
have to choose the arbitrary functions from the general solution
of (\ref{einsteq3b}) as to have
\[
\eta _{(\varphi )}^{2}=[1+\varpi (r,\theta )\varphi
]^{2}\rightarrow 1\mbox{ for }\varpi (r,\theta )\varphi
\rightarrow 0.
\]
For simplicity, we can analyze only angular anisotropies with
$\varpi =\varpi (\theta ),$ when
\[
\eta _{(\varphi )}^{2}=\eta _{(\varphi )}^{2}(\theta ,\varphi
)=[1+\varpi (\theta )\varphi ]^{2}.
\]

In the locally isotropic limit of the solution for $n_2\left(
r,\theta ,\varphi \right) $, when $\varpi \varphi \rightarrow 0,$
we could obtain the particular magnetic configuration contained
in the metric (\ref{conf4d}) if we impose the condition that
\[
n_{2[0]}\left( r,\theta \right) +n_{2[1]}\left( r,\theta \right)
=A_\theta =4m_0\left( 1-\cos \theta \right) ,
\]
which defines only one function from two unknown values
$n_{2[0]}\left( r,\theta \right) $ and $n_{2[1]}\left( r,\theta
\right) .$ This could have a corresponding physical motivation.
From the usual Kaluza--Klein procedure we induce the 4D
gravitational field (metric) and 4D electro--magnetic field
(potentials $A_i),$ which satisfy the Maxwell equations in 4D
pseudo--Riemannian space--time. For the case of spherical, locally
isotropic, symmetries the Maxwell equations can be written for
vacuum magnetic fields without any polarizations. When we
introduce into consideration anholonomic constraints and locally
anisotropic gravitational configurations the effective magnetic
field could be effectively renormalized by higher dimension
gravitational field. This effect, for some classes of
anisotropies, can be modeled by considering that the constant $m_0
$ is polarized,
\[
m_0\rightarrow m\left( r,\theta ,\varphi \right) =m_0\eta
_m\left( r,\theta ,\varphi \right)
\]
for the electro--magnetic potential and resulting magnetic field.
For ''pure" angular anisotropies we write that
\begin{eqnarray}
n_2\left( \theta ,\varphi \right)& = &n_{2[0]}\left( \theta
\right)
+n_{2[1]}\left( \theta \right) /[1+\varpi (\theta )\varphi ]^2  \nonumber \\
& {} & =4m_0\eta _m\left( \theta ,\varphi \right) \left( 1-\cos
\theta \right) ,  \nonumber
\end{eqnarray}
for
\[
\eta _{(\varphi )}^2\left( \theta ,\varphi \right) =\eta
_{(\varphi )[0]}^2\left( \theta \right) +\eta _{(\varphi
)[1]}^2\left( \theta \right) /[1+\varpi (\theta )\varphi ]^2.
\]
This could result in a constant angular renormalization even
$\varpi (\theta )\varphi \rightarrow 0.$

\subsubsection{Solutions with extra--dimension induced polarization}

Another class of solutions is constructed if we consider a
d--metric of the type (\ref{dmetric}), when $s=\varsigma ,$ with
anisotropic dependence on the 5th coordinate $\varsigma ,$
\begin{eqnarray*}
\delta s^{2} &=&dt^{2}-\delta s_{(4D)}^{2}, \\
\delta s_{(4D)}^{2} &=&-(dr^{2}+r^{2}d\theta ^{2})-r^{2}\sin
^{2}\theta d\varphi ^{2}-V^{2}(r)\eta _{(\varsigma )}^{2}(\theta
,\varsigma )\delta
\varsigma ^{2}, \\
\delta \varsigma  &=&d\varsigma +w_{3}(\theta ,\varsigma )d\theta
,
\end{eqnarray*}
where, for simplicity, we omit possible anisotropies on variable
$r,$ i. e. we state that $\eta _{(\varsigma )}$ and $w_{2}$ are
not functions on $r.$

The data for a such solution are
\begin{eqnarray}
x^0 &=&t,x^1=r,x^2=\theta ,y^3=s=\varsigma ,y^4=\varphi ,  \label{sol2} \\
g_0 &=&1,g_1=-1,g_2=-r^2,h_4=-r^2\sin ^2\theta ,  \nonumber \\
h_3 &=&V^2\left( r\right) \eta _{(\varsigma )}^2,\eta
_{(\varsigma )}^2=\eta
_{(\varsigma )}^2(r,\theta ,\varsigma ),n_{0,1}=0;  \nonumber \\
w_{0,1} &=&0,w_2=4m_0\eta _m\left( \theta ,\varsigma \right)
\left( 1-\cos
\theta \right) ,n_0=0,  \nonumber \\
n_{1,2} &=&n_{1,2[0]}\left( r,\theta \right) +n_{1,2[1]}\left(
r,\theta \right) \int \eta _{(\varsigma )}^{-3}(r,\theta
,\varsigma )d\varsigma , \nonumber
\end{eqnarray}
where the function $\eta _{(\varsigma )}=\eta _{(\varsigma
)}(r,\theta ,\varsigma )$ is an arbitrary one as follow for the
case $h_4^{*}=0,$ for angular polarizations we state, for
simplicity, that $\eta _{(\varsigma )}$ does not depend on $r,$
i. e. $\eta _{(\varsigma )}=\eta _{(\varsigma )}(\theta
,\varsigma ).$ We chose the coefficient
\[
w_4=4m_0\eta _m\left( \theta ,\varsigma \right) \left( 1-\cos
\theta \right)
\]
as to have compatibility with the locally isotropic limit when
$w_2\simeq A_\theta $ with a ''polarization'' effect modeled by
$\eta _m\left( \theta ,\varsigma \right) ,$ which could have a
constant component $\eta _m\simeq \eta _{m[0]}=const$ for small
anisotropies. In the simplest cases we can fix the conditions
$n_{1,2[0,1]}\left( r,\theta \right) =0.$ All functions $\eta
_{(\varsigma )}^2,\eta _m$ and $n_{1,2[0,1]}$ can be treated as
some possible induced higher dimensional polarizations.

\section{Anisotropic Taub NUT--Dirac Fields}

In this Section we construct two new classes of solutions of the
5D Einstein--Dirac fields in a manner as to extend the locally
anisotropic Taub NUT metrics defined by data (\ref{sol1}) and
(\ref{sol2}) as to be solutions of the Einstein equations
(\ref{einsteq3a})--(\ref{einsteq3d}) with a nonvanishing diagonal
energy momentum d--tensor
\[
\Upsilon _\beta ^\alpha =\{2\Upsilon (r,\theta ),\Upsilon
(r,\theta ),\Upsilon (r,\theta ),0,0\}
\]
for a Dirac wave packet satisfying the conditions (\ref{compat})
and (\ref {compat1}).

\subsection{Dirac fields and angular polarizations}

In order to generate from the data (\ref{sol1}) a new solution
with Dirac spinor matter field we consider instead of a linear
dependence of polarization,
\[
\eta _{(\varphi )}\sim [1+\varpi \left( r,\theta \right) \varphi
],
\]
an arbitrary function $\eta _{(\varphi )}\left( r,\theta ,\varphi
\right) $ for which
\[
h_4=V^2(r)\eta _{(\varphi )}^2\left( r,\theta ,\varphi \right)
\]
is an exact solution of the equation (\ref{einsteq3b}) with
$\Upsilon _1=\Upsilon \left( r,\theta \right) .$ With respect to
the variable $\eta _{(\varphi )}^2\left( r,\theta ,\varphi
\right) $ this component of the Einstein equations becomes linear
\begin{equation}
\eta _{(\varphi )}^{**}+r^2\sin ^2\theta \Upsilon \eta _{(\varphi
)}=0 \label{eqaux11}
\end{equation}
which is a second order linear differential equation on variable
$\varphi $ with parametric dependencies of the coefficient
$r^2\sin ^2\theta \Upsilon $ on coordinates $\left( r,\theta
\right) .$ The solution of equation (\ref
{eqaux11}) is to be found following the method outlined in Ref. \cite{kamke}%
:
\begin{eqnarray}
\eta _{(\varphi )} &=&C_1\left( r,\theta \right) \cosh [\varphi
r\sin \theta \sqrt{\left| \Upsilon \left( r,\theta \right)
\right| }+C_2\left( r,\theta
\right) ],  \nonumber \\
&{}&\Upsilon \left( r,\theta \right) <0;  \label{solaux21} \\
&=&C_1\left( r,\theta \right) +C_2\left( r,\theta \right) \varphi
,\Upsilon
\left( r,\theta \right) =0;  \label{solaux22} \\
&=&C_1\left( r,\theta \right) \cos [\varphi r\sin \theta
\sqrt{\Upsilon
\left( r,\theta \right) }+C_2\left( r,\theta \right) ],  \nonumber \\
&{}&\Upsilon \left( r,\theta \right) >0,  \label{solaux23}
\end{eqnarray}
where $C_{1,2}\left( r,\theta \right) $ are some functions to be
defined from some boundary conditions. The first solution
(\ref{solaux21}), for negative densities of energy should be
excluded as unphysical, the second solution (\ref{solaux22}) is
just that from (\ref{sol1}) for the vacuum case. A new
interesting physical situation is described by the solution (\ref
{solaux23}) when we obtain a Taub NUT anisotropic metric with
periodic anisotropic dependencies on the angle $\varphi $ where
the periodicity could variate on coordinates $\left( r,\theta
\right) $ as it is defined by the energy density $\Upsilon \left(
r,\theta \right) .$ For simplicity, we can consider a package of
spinor waves with constant value of $\Upsilon
=\Upsilon _0$ and fix some boundary and coordinate conditions when $%
C_{1,2}=C_{1,2[0]}$ are constant. This type of anisotropic Taub
NUT solutions are described by a d--metric coefficient
\begin{equation}
h_4=V^2(r)C_{1[0]}^2\cos ^2[\varphi r\sin \theta \sqrt{\Upsilon _0}%
+C_{2[0]}].  \label{aux31}
\end{equation}
Putting this value into the formulas (\ref{alpha1}),
(\ref{alpha2}) and (\ref
{beta}) for coefficients in equations (\ref{einsteq3c}) we can express $%
\alpha _{1,2}=\alpha _{1,2}[h_3,h_4,\Upsilon _0]$ and $\beta
=\beta [h_3,h_4,\Upsilon _0]$ (we omit these rather simple but
cumbersome formulas) and in consequence we can define the values
$w_{1,2}$ by solving linear algebraic equations:
\[
w_{1,2}\left( r,\theta ,\varphi \right) =\alpha _{1,2}\left(
r,\theta ,\varphi \right) /\beta \left( r,\theta ,\varphi \right)
.
\]

Having defined the values (\ref{aux31}) it is a simple task of two
integrations on $\varphi $ in order to define
\begin{eqnarray}
n_2 &=&n_{2[0]}\left( r,\theta \right) \left[ \ln \frac{1+\cos {\tilde{\kappa%
}}}{1-\cos {\tilde{\kappa}}}+\frac 1{1-\cos
{\tilde{\kappa}}}+\frac 1{1-\sin
{\tilde{\kappa}}}\right]  \nonumber \\
&{}&+n_{2[1]}\left( r,\theta \right) ,  \label{n2conf}
\end{eqnarray}
were
\[
{\tilde{\kappa}}=\varphi r\sin \theta \sqrt{\Upsilon _0}+C_{2[0]},
\]
$n_{2[0,1]}\left( r,\theta \right) $ are some arbitrary functions
to be defined by boundary conditions. We put $n_{0,1}=0$ to
obtain in the vacuum limit the solution (\ref{sol1}).

Finally, we can summarize the data defining an exact solution for
an anisotropic (on angle $\varphi $) Dirac wave packet -- Taub NUT
configuration:
\begin{eqnarray}
x^0 &=&t,x^1=r,x^2=\theta ,y^3=s=\varphi ,y^4=\varsigma ,  \label{sol1a} \\
g_0 &=&1,g_1=-1,g_2=-r^2,h_3=-r^2\sin ^2\theta ,  \nonumber \\
h_4 &=&V^2\left( r\right) \eta _{(\varphi )}^2,\eta _{(\varphi
)}=C_1\left(
r,\theta \right) \cos {\tilde {\kappa}} (r,\theta ,\varphi ),  \nonumber \\
w_i &=&0,n_{0,1}=0,n_2=n_2\left( r,\theta ,{\tilde {\kappa}}
(r,\theta
,\varphi )\right) \mbox{ see  (\ref{n2conf})},  \nonumber \\
\Psi &=&\Psi ^{(+)}\left( \zeta ^{1,2}(x^1,x^2)\right) \mbox{ see
(\ref{packet})},  \nonumber \\
\Upsilon &=&\Upsilon \left( \zeta ^{1,2}(x^1,x^2)\right) \mbox{
see (\ref{compat})}.  \nonumber
\end{eqnarray}
This solution will be extended to additional soliton anisotropic
\index{soliton} configurations in the next Section.

\subsection{Dirac fields and extra dimension polarizations}

Now we consider a generalization of the data (\ref{sol2}) for
generation of a new solution, with generic local anisotropy on
extra dimension 5th coordinate, of the Einstein -- Dirac
equations. Following the equation (\ref {einsteq3c}) we conclude
that there are not nonvacuum solutions of the Einstein equations
(with $\Upsilon \neq 0)$ if $h_4^{*}=0$ which impose the
condition $\Upsilon =0$ for $h_3,h_4\neq 0.$ So, we have to
consider that the d--metric component $h_4=-r^2\sin ^2\theta $
from the data (\ref{sol2}) is generalized to a function
$h_4\left( r,\theta ,\varsigma \right) $
satisfying a second order nonlinear differential equation on variable $%
\varsigma $ with coefficients depending parametrically on coordinates $%
\left( r,\theta \right) .$ The equation (i. e. (\ref{einsteq3c}))
can be
linearized (see Ref. \cite{kamke}) if we introduce a new variable $h_4=h^2,$%
\[
h^{**}-\frac{h_3^{*}}{2h_3}h^{*}+h_3\Upsilon h=0,
\]
which, in its turn, can be transformed :

a) to a Riccati form if we introduce a new variable $v,$ for which $%
h=v^{*}/v,$%
\begin{equation}
v^{*}+v^2-\frac{h_3^{*}}{2h_3}v+h_3\Upsilon =0;  \label{riccati1}
\end{equation}

b) to the so--called normal form \cite{kamke},
\begin{equation}
\lambda ^{**}+I\lambda =0,  \label{normal}
\end{equation}
obtained by a redefinition of variables like
\[
\lambda =h\exp \left[ -\frac 14\int \frac{h_3^{*}}{h_3}d\varsigma
\right] =h\ h_3^{-1/4}
\]
where
\[
I=h_3\Upsilon -\frac 1{16}\frac{h_3^{*}}{h_3}+\frac 14\left( \frac{h_3^{*}}{%
h_3}\right) ^{*}.
\]
We can construct explicit series and/or numeric solutions (for
instance, by using Mathematica or Maple programs) of both type of
equations (\ref {riccati1}) and normal (\ref{normal}) for some
stated boundary conditions and type of polarization of the
coefficient $h_3\left( r,\theta ,\varsigma \right) =V^2\left(
r\right) \eta _{(\varsigma )}^2(r,\theta ,\varsigma )$
and, in consequence, to construct different classes of solutions for $%
h_4\left( r,\theta ,\varsigma \right) .$ In order to have
compatibility with the data (\ref{sol2}) we must take $h_4$ in
the form
\[
h_4\left( r,\theta ,\varsigma \right) =-r^2\sin ^2\theta
+h_{4(\varsigma )}\left( r,\theta ,\varsigma \right) ,
\]
where $h_{4(\varsigma )}\left( r,\theta ,\varsigma \right) $ vanishes for $%
\Upsilon \rightarrow 0.$

Having defined a value of $h_4\left( r,\theta ,\varsigma \right)
$ we can compute the coefficients (\ref{alpha1}), (\ref{alpha2})
and (\ref{beta}) and find from the equations (\ref{einsteq3c})
\[
w_{1,2}\left( r,\theta ,\varsigma \right) =\alpha _{1,2}\left(
r,\theta ,\varsigma \right) /\beta \left( r,\theta ,\varsigma
\right) .
\]

From the equations (\ref{einsteq3d}), after two integrations on variable $%
\varsigma $ one obtains the values of $n_{1,2}\left( r,\theta
,\varsigma \right) .$ Two integrations of equations
(\ref{einsteq3d}) define
\[
n_i(r,\theta ,\varsigma )=n_{i[0]}(r,\theta )\int_0^\varsigma
dz\int_0^zdsP(r,\theta ,s)+n_{i[1]}(r,\theta ),
\]
where
\[
P\equiv \frac 12(\frac{h_3^{*}}{h_3}-3\frac{h_4^{*}}{h_4})
\]
and the functions $n_{i[0]}(r,\theta )$ and $n_{i[1]}(r,\theta )$ on $%
(r,\theta )$ have to be defined by solving the Cauchy problem.
The boundary conditions of both type of coefficients $w_{1,2}$
and $n_{1,2}$ should be
expressed in some forms transforming into corresponding values for the data (%
\ref{sol2}) if the source $\Upsilon \rightarrow 0.$ We omit
explicit
formulas for exact Einstein--Dirac solutions with $\varsigma $%
--polarizations because their forms depend very strongly on the
type of polarizations and vacuum solutions.

\section{Anholonomic Dirac--Taub NUT Solitons} \index{soliton}

In the next subsections we analyze two explicit examples when the
spinor field induces two dimensional, depending on three
variables, solitonic anisotropies.

\subsection{Kadomtsev--Petviashvili type solitons}

By straightforward verification we conclude that the d--metric component $%
h_4(r,\theta ,s)$ could be a solution of Kadomtsev--Petviashvili
(KdP) \index{KdP soliton} equation \cite{kad} (the first methods
of integration of 2+1 dimensional soliton equations where
developed by Dryuma \cite{dryuma} and Zakharov and Shabat
\cite{zakhsh})
\begin{equation}
h_4^{**}+\epsilon \left( \dot{h}_4+6h_4h_4^{\prime }+h_4^{\prime
\prime \prime }\right) ^{\prime }=0,\epsilon =\pm 1,  \label{kdp}
\end{equation}
if the component $h_3(r,\theta ,s)$ satisfies the Bernoulli
equations \cite {kamke}
\begin{equation}
h_3^{*}+Y\left( r,\theta ,s\right) (h_3)^2+F_\epsilon \left(
r,\theta ,s\right) h_3=0,  \label{bern1}
\end{equation}
where, for $h_4^{*}\neq 0,$%
\begin{equation}
Y\left( r,\theta ,s\right) =\kappa \Upsilon \frac{h_4}{h_4^{*}},
\label{sourse1}
\end{equation}
and
\[
F_\epsilon \left( r,\theta ,s\right) =\frac{h_4^{*}}{h_4}+\frac{2\epsilon }{%
h_4^{*}}\left( \dot{h}_4+6h_4h_4^{\prime }+h_4^{\prime \prime
\prime }\right) ^{\prime }.
\]
The three dimensional integral variety of (\ref{bern1}) is
defined by formulas
\[
h_3^{-1}\left( r,\theta ,s\right) =h_{3(x)}^{-1}\left( r,\theta
\right) E_\epsilon \left( x^i,s\right) \times \int \frac{Y\left(
r,\theta ,s\right) }{E_\epsilon \left( r,\theta ,s\right) }ds,
\]
where
\[
E_\epsilon \left( r,\theta ,s\right) =\exp \int F_\epsilon \left(
r,\theta ,s\right) ds
\]
and $h_{3(x)}\left( r,\theta \right) $ is a nonvanishing function.

In the vacuum case $Y\left( r,\theta ,s\right) =0$ and we can
write the integral variety of (\ref{bern1})
\[
h_3^{(vac)}\left( r,\theta ,s\right) =h_{3(x)}^{(vac)}\left(
r,\theta \right) \exp \left[ -\int F_\epsilon \left( r,\theta
,s\right) ds\right] .
\]

We conclude that a solution of KdP equation (\ref{bern1}) could
be generated by a non--perturbative component $h_4(r,\theta ,s)$
of a diagonal h--metric if the second component $h_3\left(
r,\theta ,s\right) $ is a solution of Bernoulli equations
(\ref{bern1}) with coefficients determined both by $h_4$ and its
partial derivatives and by the $\Upsilon _1^1$ component of the
energy--momentum d--tensor (see (\ref{compat1})). The parameters
(coefficients) of (2+1) dimensional KdV solitons are induced by
gravity and spinor constants and spinor field configuration
defining locally anisotropic interactions of packets of Dirac's
spinor waves.

\subsection{(2+1) sine--Gordon type solitons} \index{sine--Gordon}

In a similar manner we can prove that solutions $h_4(r,\theta
,s)$ of (2+1) sine--Gordon equation (see, for instance,
\cite{har,lieb,whith})
\[
h_4^{**}+h_4^{^{\prime \prime }}-\ddot{h}_4=\sin (h_4)
\]
also induce solutions for $h_3\left( r,\theta ,s\right) $
following from the Bernoulli equation
\[
h_3^{*}+\kappa E(r,\theta )\frac{h_4}{h_4^{*}}(h_3)^2+F\left(
r,\theta ,s\right) h_3=0,h_4^{*}\neq 0,
\]
where
\[
F\left( r,\theta ,s\right) =\frac{h_4^{*}}{h_4}+\frac
2{h_4^{*}}\left[ h_4^{^{\prime \prime }}-\ddot{h}_4-\sin
(h_4)\right] .
\]
The general solutions (with energy--momentum sources and in
vacuum cases) are constructed by a corresponding redefinition of
coefficients in the formulas from the previous subsection. We
note that we can consider both
type of anisotropic solitonic polarizations, depending on angular variable $%
\varphi $ or on extra dimension coordinate $\varsigma .$ Such
classes of solutions of the Einstein--Dirac equations describe
three dimensional spinor wave packets induced and moving
self--consistently on solitonic gravitational locally anisotropic
configurations. In a similar manner, we can consider Dirac wave
packets generating and propagating on locally anisotropic black
hole (with rotation ellipsoid horizons), black tori, anisotropic
disk and two or three dimensional black hole anisotropic
gravitational structures \cite{vsol}. Finally, we note that such
gravitational solitons are induced by Dirac field matter sources
and are different from those soliton solutions of vacuum Einstein
equations originally considered by Belinski and Zakharov
\cite{belinski}.

Finally, we conclude that we have argued that the anholonomic
frame method can be applied for construction on new classes of
Einstein--Dirac equations in five dimensional (5D) space--times.
Subject to a form of metric ansatz with dependencies of
coefficients on two holonomic and one anholonomic variables we
obtained a very simplified form of field equations which admit
exact solutions. We have identified two classes of solutions
describing Taub NUT like metrics with anisotropic dependencies on
angular parameter or on the fifth coordinate. We have shown that
both classes of anisotropic vacuum solutions can be generalized
to matter sources with the energy--momentum tensor defined by
some wave packets of Dirac fields. Although the Dirac equation is
a quantum one, in the quasi--classical approximation we can
consider such spinor fields as some spinor waves propagating in a
three dimensional Minkowski plane which is imbedded in a
self--consistent manner in a Taub--NUT anisotropic space--time.
At the classical level it should be emphasized that the results
of this paper are very general in nature, depending in a crucial
way only on the locally Lorentzian nature of 5D space--time and
on the supposition that this space--time is constructed as a
trivial time extension of 4D space--times. We have proved that
the new classes of solutions admit generalizations to nontrivial
topological configurations of 3D dimensional solitons (induced by
anisotropic spinor matter) defined as solutions
Kadomtsev--Petviashvili or sine--Gordon equations.










\part[Anisotropic Spinors]{Anisotropic Spinors}

Spinor variables and interactions of spinor fields on Finsler
spaces were used in a heuristic manner, for instance, in works
\cite{asa88,ono}, where the problem of a rigorous definition of
spinors for locally anisotropic spaces was not considered. Here
we note that, in general, the nontrivial nonlinear connection and
torsion structures and possible incompatibility of metric and
connections makes the solution of the mentioned problem very
sophisticate. The geometric definition of locally anisotropic
spinors and a detailed study of the relationship between
Clifford, spinor and nonlinear and distinguished connections
structures in vector bundles, generalized Lagrange and Finsler
spaces are presented in Refs. \cite{vjmp,viasm1,vsp1}.

The purpose of this Part is to summarize our investigations \cite
{vjmp,viasm1,vsp1,vg,vlasg} on formulation of the theory of
classical and quantum field interactions on locally anisotropic
spaces. We receive primary attention to the development of the
necessary geometric framework: to propose an abstract spinor
formalism and formulate the differential geometry of locally
anisotropic spaces (the second step after the definition of
locally anisotropic spinors in \cite{vjmp,viasm1}). The next step
is the investigation of locally anisotropic interactions of
fundamental fields on generic locally anisotropic spaces
\cite{vsp1}.

For our considerations on the locally anisotropic spinor theory
it will be convenient to extend the Penrose and Rindler abstract
index formalism \cite {pen,penr1,penr2} (see also the Luehr and
Rosenbaum index free methods \cite {lue}) proposed for spinors on
locally isotropic spaces. We note that in order to formulate the
locally anisotropic physics usually we have dimensions $d>4$ for
the fundamental locally anisotropic space-time. In this case the
2-spinor calculus does not play a preferential role.

\chapter{Anisotropic Clifford Structures}

If a nonlinear connection structure is defined on a vector
(covector, or higher order vector--covector) bundle, or on a
pseudo--Riemannian spacetime, the geometrical objects on this
space are distinguished into some ''horizontal'' and ''vertical''
(co-vertical, or higher order vertical--covertical) invariant
components. Our idea on definition of Clifford and spinor
structure on such locally anisotropic spaces is to consider
distinguished Cliffor algebras, which consists from blocks of
usual Clifford algebras for every horizontal and vertical
subspace (for every ''shall'' of higher order anisotropies). For
symplicity, we restrict our constructions only to vector bundles
(the covector bundles with respective Clifford co-algebras are
similar dual constructions \cite{vsv}, we can for instance to
develop a respective theory fo Clifford co--structures on
Hamilton and Cartan spaces).

\section{Distinguished Clifford Algebras} \index{Clifford algebras}

The typical fiber of a vector bunde (v-bundle) $\xi _{d}\ ,\ \pi
_{d}:\
HE\oplus VE\rightarrow E$ is a d-vector space, $\mathcal{F}=h\mathcal{F}%
\oplus v\mathcal{F},$ split into horizontal $h\mathcal{F}$ and vertical $v%
\mathcal{F}$ subspaces, with metric $G(g,h)$ induced by v-bundle
metric (\ref {dmetric}). Clifford algebras (see, for example,
Refs. \cite{kar,tur,penr2}) formulated for d-vector spaces will
be called Clifford d-algebras \cite {vjmp,viasm1} \index{Clifford
d--algebras}. In this section we shall consider the main
properties of Clifford d-algebras. The proof of theorems will be
based on the technique developed in Ref. \cite{kar}
correspondingly adapted to the distinguished character of spaces
in consideration.

Let $k$ be a number field (for our purposes $k=\mathcal{R}$ or $k=\mathcal{C}%
,\mathcal{R}$ and $\mathcal{C},$ are, respectively real and
complex number fields) and define $\mathcal{F},$ as a d-vector
space on $k$ provided with nondegenerate symmetric quadratic form
(metric)\ $G.$ Let $C$ be an algebra on $k$ (not necessarily
commutative) and $j\ :\ \mathcal{F}$ $\rightarrow C$ a
homomorphism of underlying vector spaces such that
$j(u)^2=\;G(u)\cdot 1\ (1$ is the unity in algebra $C$ and
d-vector $u\in \mathcal{F}).$ We are interested in definition of
the pair $\left( C,j\right) $ satisfying the next universitality
conditions. For every $k$-algebra $A$ and arbitrary homomorphism
$\varphi :\mathcal{F}\rightarrow A$ of the underlying d-vector
spaces, such that $\left( \varphi (u)\right) ^2\rightarrow
G\left( u\right) \cdot 1,$ there is a unique homomorphism of
algebras $\psi \ :\ C\rightarrow A$ transforming the diagram 1
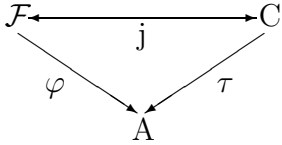
\begin{figure}[htbp]
\begin{center}
\begin{picture}(100,50) \setlength{\unitlength}{1pt}
\thinlines \put(0,45){${\cal F}$} \put(96,45){C} \put(48,2){A}
\put(50,38){j} \put(50,48){ \vector(-1,0){45}}
\put(50,48){\vector(1,0){45}} \put(5,42){\vector(3,-2){45}}
\put(98,42){\vector(-3,-2){45}} \put(15,20){$\varphi$}
\put(80,20){$\tau$}
\end{picture}
\end{center}
\caption{Diagram 1}
\end{figure}
into a commutative one.\ The algebra solving this problem will be
denoted as
$C\left( \mathcal{F},A\right) $ [equivalently as $C\left( G\right) $ or $%
C\left( \mathcal{F}\right) ]$ and called as Clifford d-algebra
associated with pair $\left( \mathcal{F},G\right) .$

\begin{theorem}
\label{2.1t} The above-presented diagram has a unique solution
$\left( C,j\right) $ up to isomorphism.
\end{theorem}

\textbf{Proof:} (We adapt for d-algebras that of Ref. \cite{kar},
p. 127.) For a universal problem the uniqueness is obvious if we
prove the existence
of solution $C\left( G\right) $ . To do this we use tensor algebra $\mathcal{%
L}^{(F)}=\oplus \mathcal{L}_{qs}^{pr}\left( \mathcal{F}\right) $
=$\oplus
_{i=0}^\infty T^i\left( \mathcal{F}\right) ,$ where $T^0\left( \mathcal{F}%
\right) =k$ and $T^i\left( \mathcal{F}\right) =k$ and $T^i\left( \mathcal{F}%
\right) =\mathcal{F}\otimes ...\otimes \mathcal{F}$ for $i>0.$
Let $I\left( G\right) $ be the bilateral ideal generated by
elements of form $\epsilon \left( u\right) =u\otimes u-G\left(
u\right) \cdot 1$ where $u\in \mathcal{F}
$ and $1$ is the unity element of algebra $\mathcal{L}\left( \mathcal{F}%
\right) .$ Every element from $I\left( G\right) $ can be written as $%
\sum\nolimits_i\lambda _i\epsilon \left( u_i\right) \mu _i,$
where $\lambda
_{i},\mu _i\in \mathcal{L}(\mathcal{F})$ and $u_i\in \mathcal{F}.$ Let $%
C\left( G\right) $ =$\mathcal{L}(\mathcal{F})/I\left( G\right) $ and define $%
j:\mathcal{F}\rightarrow C\left( G\right) $ as the composition of
monomorphism $i:{\mathcal{F}\rightarrow L}^1 (\mathcal{F})\subset \mathcal{L}%
(\mathcal{F})$ and projection $p:\mathcal{L}\left(
\mathcal{F}\right) \rightarrow C\left( G\right) .$ In this case
pair $\left( C\left( G\right) ,j\right) $ is the solution of our
problem. From the general properties of tensor algebras the
homomorphism $\varphi :\mathcal{F}\rightarrow A$ can be extended
to $\mathcal{L}(\mathcal{F})$ , i.e., the diagram 2
\begin{figure}[htbp]
\begin{center}
\begin{picture}(100,50) \setlength{\unitlength}{1pt}
\thinlines \put(0,45){${\cal F}$} \put(96,45){${\cal L}({\cal
F})$} \put(48,2){A} \put(50,38){i} \put(50,48){ \line(-1,0){45}}
\put(50,48){\vector(1,0){45}} \put(5,42){\vector(3,-2){45}}
\put(98,42){\vector(-3,-2){45}} \put(15,20){$\varphi$}
\put(80,20){$\rho$}
\end{picture}
\end{center}
\caption{Diagram 2}
\end{figure}
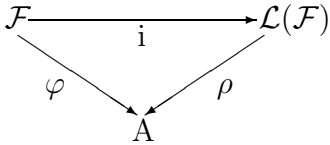
is commutative, where $\rho $ is a monomorphism of algebras.
Because $\left( \varphi \left( u\right) \right) ^2=G\left(
u\right) \cdot 1,$ then $\rho $ vanishes on ideal $I\left(
G\right) $ and in this case the necessary homomorphism $\tau $ is
defined. As a consequence of uniqueness of $\rho ,$ the
homomorphism $\tau $ is unique.

Tensor d-algebra $\mathcal{L}(\mathcal{F )}$ can be considered as a $%
\mathcal{Z}/2$ graded algebra. Really, let us in\-tro\-duce $\mathcal{L}%
^{(0)}(\mathcal{F}) = \sum_{i=1}^\infty T^{2i}\left(
\mathcal{F}\right) $ and $\mathcal{L}^{(1)}(\mathcal{F})
=\sum_{i=1}^\infty T^{2i+1}\left( \mathcal{F}\right) .$ Setting
$I^{(\alpha )}\left( G\right) =I\left( G\right) \cap
\mathcal{L}^{(\alpha )}(\mathcal{F}).$ Define $C^{(\alpha
)}\left( G\right) $ as $p\left( \mathcal{L}^{(\alpha )}
(\mathcal{F})\right) ,$ where $p:\mathcal{L}\left(
\mathcal{F}\right) \rightarrow C\left( G\right) $ is the
canonical projection. Then $C\left( G\right) =C^{(0)}\left(
G\right) \oplus C^{(1)}\left( G\right) $ and in consequence we
obtain that the Clifford d-algebra is $\mathcal{Z}/2$ graded.

It is obvious that Clifford d-algebra functorially depends on
pair $\left( \mathcal{F},G\right) .$ If
$f:\mathcal{F}\rightarrow\mathcal{F}^{\prime }$ is a homomorphism
of k-vector spaces, such that $G^{\prime }\left( f(u)\right)
=G\left( u\right) ,$ where $G$ and $G^{\prime }$ are,
respectively, metrics on $\mathcal{F}$ and $\mathcal{F}^{\prime
},$ then $f$ induces an homomorphism of d-algebras
\[
C\left( f\right) :C\left( G\right) \rightarrow C\left( G^{\prime
}\right)
\]
with identities $C\left( \varphi \cdot f\right) =C\left( \varphi
\right)
C\left( f\right) $ and $C\left( Id_{\mathcal{F}}\right) =Id_{C(\mathcal{F)}%
}. $

If $\mathcal{A}^{\alpha}$ and $\mathcal{B}^{\beta}$ are $\mathcal{Z}/2$%
--graded d--algebras, then their graded tensorial product $\mathcal{A}%
^\alpha \otimes \mathcal{B}^\beta $ is defined as a d-algebra for
k-vector d-space $\mathcal{A}^\alpha \otimes \mathcal{B}^\beta $
with the graded product induced as $\left( a\otimes b\right)
\left( c\otimes d\right)
=\left( -1\right) ^{\alpha \beta }ac\otimes bd,$ where $b\in \mathcal{B}%
^\alpha $ and $c\in \mathcal{A}^\alpha \quad \left( \alpha ,\beta
=0,1\right) .$

Now we reformulate for d--algebras the Chevalley theorem
\cite{chev}:

\begin{theorem}
\label{2.2t} The Clifford d-algebra $C\left( h\mathcal{F}\oplus v\mathcal{F}%
,g+h\right) $ is naturally isomorphic to $C(g)\otimes C\left(
h\right) .$
\end{theorem}

\textbf{Proof. }Let $n:h\mathcal{F}\rightarrow C\left( g\right) $ and $%
n^{\prime }:v\mathcal{F}\rightarrow C\left( h\right) $ be
canonical maps and map\newline $m:h\mathcal{F}\oplus
v\mathcal{F}\rightarrow C(g)\otimes C\left( h\right) $
is defined as $m(x,y)=n(x)\otimes 1+1\otimes n^{\prime }(y),$ $x\in h%
\mathcal{F}, y\in v\mathcal{F}.$ We have $\left( m(x,y)\right)
^2=\left[ \left( n\left( x\right) \right) ^2+\left( n^{\prime
}\left( y\right) \right) ^2\right] \cdot 1=[g\left( x\right)
+h\left( y\right) ].$ \ Taking into account the universality
property of Clifford d-algebras we conclude that $m$ induces the
homomorphism
\[
\begin{array}{c}
\zeta :C\left( h\mathcal{F}\oplus v\mathcal{F}, g+h\right)
\rightarrow
C\left( h\mathcal{F},g\right) \widehat{\otimes }C\left( v\mathcal{F}%
,h\right) .
\end{array}
\]
We also can define a homomorphism
\[
\upsilon :C\left( h\mathcal{F},g\right) \widehat{\otimes }C\left( v\mathcal{F%
},h\right) \rightarrow C\left(h\mathcal{F}\oplus v\mathcal{F},
g+h\right)
\]
by using formula $\upsilon \left( x\otimes y\right) =\delta
\left( x\right)
\delta ^{\prime }\left( y\right) ,$ where homomorphysms $\delta $ and $%
\delta ^{\prime }$ are, respectively, induced by imbeddings of
$h\mathcal{F}$
and $v\mathcal{F}$ into $h\mathcal{F}\oplus v\mathcal{F} :$%
\[
\delta :C\left( h\mathcal{F},g\right) \rightarrow C\left(
h\mathcal{F}\oplus v\mathcal{F} , g+h\right) ,
\]
\[
\begin{array}{c}
\delta ^{\prime }:C\left( v\mathcal{F},h\right) \rightarrow C\left( h%
\mathcal{F}\oplus v\mathcal{F} , g+h\right) .
\end{array}
\]
Because $x\in C^{(\alpha )}\left( g\right) $ and $y\in C^{(\alpha
)}\left( g\right) ,$ we have
\[
\delta \left( x\right) \delta ^{\prime }\left( y\right) =\left(
-1\right) ^{\left( \alpha \right) }\delta ^{\prime }\left(
y\right) \delta \left( x\right) .
\]

Superpositions of homomorphisms $\zeta $ and $\upsilon $ lead to
identities
\begin{equation}
\upsilon \zeta =Id_{C\left( h\mathcal{F},g\right) \widehat{\otimes }C\left( v%
\mathcal{F},h\right) },\zeta \upsilon =Id_{C\left(
h\mathcal{F},g\right) \widehat{\otimes }C\left(
v\mathcal{F},h\right) }.  \label{2.46}
\end{equation}
Really, the d-dalgebra $C\left( h\mathcal{F}\oplus
v\mathcal{F},g+h\right) $ is generated by elements of type
$m(x,y).$ Calculating
\begin{eqnarray}
\upsilon \zeta \left( m\left( x,y\right) \right) & =& \upsilon
\left( n\left( x\right) \otimes 1+1\otimes n^{\prime }\left(
y\right) \right)
\nonumber \\
&=& \delta \left( n\left( x\right) \right) \delta \left(
n^{\prime }\left( y\right) \right) = m\left( x,0\right)
+m(0,y)=m\left( x,y\right) ,  \nonumber
\end{eqnarray}
we prove the first identity in (\ref{2.46}).

On the other hand, d-algebra $C\left( h\mathcal{F},g\right)
\widehat{\otimes }C\left( v\mathcal{F},h\right) $ is generated by
elements of type $n\left( x\right) \otimes 1$ and $1\otimes
n^{\prime }\left( y\right) ,$ we prove the second identity in
(\ref{2.46}).

Following from the above -mentioned properties of homomorphisms
$\zeta $ and $\upsilon $ we can assert that the natural
isomorphism is explicitly constructed.$\Box $ \vskip20pt

In consequence of theorem \ref{2.2t} we conclude that all
operations with
Clifford d-algebras can be reduced to calculations for $C\left( h\mathcal{F}%
,g\right) $ and $C\left( v\mathcal{F},h\right) $ which are usual
Clifford algebras of dimension $2^n$ and, respectively, $2^m$
\cite{kar,ati}.

Of special interest is the case when $k=\mathcal{R}$ and
$\mathcal{F}$ is isomorphic to vector space
$\mathcal{R}^{p+q,a+b}$ provided with quadratic form
$-x_1^2-...-x_p^2+x_{p+q}^2-y_1^2-...-y_a^2+...+y_{a+b}^2.$ In
this case, the Clifford algebra, denoted as $\left(
C^{p,q},C^{a,b}\right) ,\,$
is generated by symbols $%
e_1^{(x)},e_2^{(x)},...,e_{p+q}^{(x)},e_1^{(y)},e_2^{(y)},...,e_{a+b}^{(y)}$
satisfying properties $\left( e_i\right) ^2=-1~\left( 1\leq i\leq
p\right) ,\left( e_j\right) ^2=-1~\left( 1\leq j\leq a\right)
,\left( e_k\right) ^2=1~(p+1\leq k\leq p+q),$

$\left( e_j\right) ^2=1~(n+1\leq s\leq
a+b),~e_ie_j=-e_je_i,~i\neq j.\,$ Explicit calculations of
$C^{p,q}$ and $C^{a,b}$ are possible by using isomorphisms
\cite{kar,penr2}
\[
C^{p+n,q+n}\simeq C^{p,q}\otimes M_2\left( \mathcal{R}\right)
\otimes ...\otimes M_2\left( \mathcal{R}\right) \cong
C^{p,q}\otimes M_{2^n}\left( \mathcal{R}\right) \cong
M_{2^n}\left( C^{p,q}\right) ,
\]
where $M_s\left( A\right) $ denotes the ring of quadratic matrices of order $%
s$ with coefficients in ring $A.$ Here we write the simplest isomorphisms $%
C^{1,0}\simeq \mathcal{C},C^{0,1}\simeq \mathcal{R}\oplus \mathcal{R,}$ and $%
C^{2,0}=\mathcal{H},$ where by $\mathcal{H}$ is denoted the body
of quaternions. We summarize this calculus as (as in Ref.
\cite{ati})
\begin{eqnarray*}
C^{0,0}
&=&\mathcal{R},C^{1,0}=\mathcal{C},C^{0,1}=\mathcal{R}\oplus
\mathcal{R},C^{2,0}=\mathcal{H},C^{0,2}=M_2\left( \mathcal{R}\right) , \\
C^{3,0} &=&\mathcal{H}\oplus \mathcal{H},C^{0,3}=M_2\left( \mathcal{R}%
\right) , \\
C^{4,0} &=&M_2\left( \mathcal{H}\right) ,C^{0,4}=M_2\left( \mathcal{H}%
\right) ,C^{5,0}=M_4\left( \mathcal{C}\right) , \\
~C^{0,5} &=&M_2\left( \mathcal{H}\right) \oplus M_2\left(
\mathcal{H}\right) ,C^{6,0}=M_8\left( \mathcal{R}\right)
,C^{0,6}=M_4\left( \mathcal{H}\right) ,
\\
~C^{7,0} &=&M_8\left( \mathcal{R}\right) \oplus M_8\left(
\mathcal{R}\right)
,~C^{0,7}=M_8\left( \mathcal{C}\right) , \\
~C^{8,0} &=&M_{16}\left( \mathcal{R}\right) ,~C^{0,8}=M_{16}\left( \mathcal{R%
}\right) .
\end{eqnarray*}
One of the most important properties of real algebras
$C^{0,p}~\left( C^{0,a}\right) $ and\newline $C^{p,0}~\left(
C^{a,0}\right) $ is eightfold periodicity of $p(a).$

Now, we emphasize that $H^{2n}$-spaces admit locally a structure
of Clifford algebra on complex vector spaces. Really, by using
almost \ Hermitian
structure $J_\alpha ^{\quad \beta }$ and considering complex space $\mathcal{%
C}^n$ with nondegenarate quadratic form
\[
\sum_{a=1}^n\left| z_a\right| ^2,~z_a\in \mathcal{C}^2
\]
induced locally by metric (\ref{dmetric}) (rewritten in complex coordinates $%
z_a=x_a+iy_a)$ we define Clifford algebra \index{Clifford algebra}
\[
\overleftarrow{C}^n=\underbrace{\overleftarrow{C}^1\otimes
...\otimes \overleftarrow{C}^1}_n,
\]
where $\overleftarrow{C}^1=\mathcal{C\otimes
}_R\mathcal{C=C\oplus C}$ or in
consequence, $\overleftarrow{C}^n\simeq C^{n,0}\otimes _{\mathcal{R}}%
\mathcal{C}\approx C^{0,n}\otimes _{\mathcal{R}}\mathcal{C}.$
Explicit
calculations lead to isomorphisms $\overleftarrow{C}^2=C^{0,2}\otimes _{%
\mathcal{R}}\mathcal{C}\approx M_2\left( \mathcal{R}\right) \otimes _{%
\mathcal{R}}\mathcal{C}\approx M_2\left(
\overleftarrow{C}^n\right)
,~C^{2p}\approx M_{2^p}\left( \mathcal{C}\right) $ and $\overleftarrow{C}%
^{2p+1}\approx M_{2^p}\left( \mathcal{C}\right) \oplus
M_{2^p}\left( \mathcal{C}\right) ,$ which show that complex
Clifford algebras, defined locally for $H^{2n}$-spaces, have
periodicity 2 on $p.$

Considerations presented in the proof of theorem \ref{2.2t} show that map $j:%
\mathcal{F}\rightarrow C\left( \mathcal{F}\right) $ is
monomorphic, so we
can identify space $\mathcal{F}$ with its image in $C\left( \mathcal{F}%
,G\right) ,$ denoted as $u\rightarrow \overline{u},$ if $u\in
C^{(0)}\left( \mathcal{F},G\right) ~\left( u\in C^{(1)}\left(
\mathcal{F},G\right) \right) ;$ then $u=\overline{u}$ (
respectively, $\overline{u}=-u).$

\begin{definition} \index{Clifford d--group}
\label{2.1d} The set of elements $u\in C\left( G\right) ^{*},$ where $%
C\left( G\right) ^{*}$ denotes the multiplicative group of
invertible
elements of $C\left( \mathcal{F},G\right) $ satisfying $\overline{u}\mathcal{%
F}u^{-1}\in \mathcal{F},$ is called the twisted Clifford d-group,
denoted as $\widetilde{\Gamma }\left( \mathcal{F}\right) .$
\end{definition}

Let $\widetilde{\rho }:\widetilde{\Gamma }\left(
\mathcal{F}\right)
\rightarrow GL\left( \mathcal{F}\right) $ be the homorphism given by $%
u\rightarrow \rho \widetilde{u},$ where $\widetilde{\rho }_u\left( w\right) =%
\overline{u}wu^{-1}.$ We can verify that $\ker \widetilde{\rho }=\mathcal{R}%
^{*}$is a subgroup in $\widetilde{\Gamma }\left(
\mathcal{F}\right) .$

Canonical map $j:\mathcal{F}\rightarrow C\left(
\mathcal{F}\right) $ can be
interpreted as the linear map $\mathcal{F}\rightarrow C\left( \mathcal{F}%
\right) ^0 $ satisfying the universal property of Clifford
d-algebras. This leads to a homomorphism of algebras, $C\left(
\mathcal{F}\right) \rightarrow C\left( \mathcal{F}\right) ^t,$
considered by an anti-involution of $C\left(
\mathcal{F}\right) $ and denoted as $u\rightarrow ~^tu.$ More exactly, if $%
u_1...u_n\in \mathcal{F,}$ then $t_u=u_n...u_1$ and $^t\overline{u}=%
\overline{^tu}=\left( -1\right) ^nu_n...u_1.$

\begin{definition} \index{spinor norm}
\label{2.2d} The spinor norm of arbitrary $u\in C\left(
\mathcal{F}\right) $ is defined as\newline $S\left( u\right)
=~^t\overline{u}\cdot u\in C\left( \mathcal{F}\right) .$
\end{definition}

It is obvious that if $u,u^{\prime },u^{\prime \prime }\in
\widetilde{\Gamma }\left( \mathcal{F}\right) ,$ then
$S(u,u^{\prime })=S\left( u\right) S\left( u^{\prime }\right) $
and \newline $S\left( uu^{\prime }u^{\prime \prime }\right)
=S\left( u\right) S\left( u^{\prime }\right) S\left( u^{\prime
\prime }\right) .$ For $u,u^{\prime }\in \mathcal{F} S\left(
u\right) =-G\left( u\right) $ and $S\left( u,u^{\prime }\right)
=S\left( u\right) S\left( u^{\prime }\right) =S\left( uu^{\prime
}\right) .$

Let us introduce the orthogonal group $O\left( G\right) \subset
GL\left( G\right) $ defined by metric $G$ on $\mathcal{F}$ and
denote sets $SO\left( G\right) =\{u\in O\left( G\right) ,\det
\left| u\right| =1\},~Pin\left( G\right) =\{u\in
\widetilde{\Gamma }\left( \mathcal{F}\right) ,S\left( u\right)
=1\}$ and $Spin\left( G\right) =Pin\left( G\right) \cap C^0\left(
\mathcal{F}\right) .$ For ${\mathcal{F}\cong \mathcal{R}}^{n+m}$ we write $%
Spin\left( n+m\right) .$ By straightforward calculations (see
similar considerations in Ref. \cite{kar}) we can verify the
exactness of these sequences:
\begin{eqnarray*}
1 &\rightarrow &\mathcal{Z}/2\rightarrow Pin\left( G\right)
\rightarrow
O\left( G\right) \rightarrow 1, \\
1 &\rightarrow &\mathcal{Z}/2\rightarrow Spin\left( G\right)
\rightarrow
SO\left( G\right) \rightarrow 0, \\
1 &\rightarrow &\mathcal{Z}/2\rightarrow Spin\left( n+m\right)
\rightarrow SO\left( n+m\right) \rightarrow 1.
\end{eqnarray*}
We conclude this section by emphasizing that the spinor norm was
defined with respect to a quadratic form induced by a metric in
v-bundle $\xi _d$ (or by an $H^{2n}$-metric in the case of
generalized Lagrange spaces). This approach differs from those
presented in Refs. \cite{asa88} and \cite{ono}.

\section[Anisotropic Clifford Bundles]{Anisotropic Clifford Bundles and
Spinor Structures}

There are two possibilities for generalizing our spinor
constructions defined for d-vector spaces to the case of vector
bundle spaces enabled with the structure of N-connection. The
first is to use the extension to the category of vector bundles.
The second is to define the Clifford fibration associated with
compatible linear d-connection and metric $G$ on a vector bundle
(or with an $H^{2n}$-metric on GL-space). Let us consider both
variants.

\subsection{Clifford d-module structure} \index{Clifford d--module}

Because functor $\mathcal{F}\to C(\mathcal{F})$ is smooth we can
extend it to the category of vector bundles of type $\xi _d=\{\pi
_d:HE\oplus VE\rightarrow E\}.$ Recall that by $\mathcal{F}$ we
denote the typical fiber
of such bundles. For $\xi _d$ we obtain a bundle of algebras, denoted as $%
C\left( \xi _d\right) ,\,$ such that $C\left( \xi _d\right)
_u=C\left( \mathcal{F}_u\right) .$ Multiplication in every fibre
defines a continuous map $C\left( \xi _d\right) \times C\left(
\xi _d\right) \rightarrow C\left(
\xi _d\right) .$ If $\xi _d$ is a vector bundle on number field $k$%
,\thinspace \thinspace the structure of the $C\left( \xi
_d\right) $-module,
the d-module, the d-module, on $\xi _d$ is given by the continuous map $%
C\left( \xi _d\right) \times _E\xi _d\rightarrow \xi _d$ with every fiber $%
\mathcal{F}_u$ provided with the structure of the $C\left( \mathcal{F}%
_u\right) -$module, correlated with its $k$-module structure, Because $%
\mathcal{F}\subset C\left( \mathcal{F}\right) ,$ we have a fiber
to fiber map $\mathcal{F}\times _E\xi _d\rightarrow \xi _d,$
inducing on every fiber the map $\mathcal{F}_u\times _E\xi
_{d\left( u\right) }\rightarrow \xi
_{d\left( u\right) }$ ($\mathcal{R}$-linear on the first factor and $k$%
-linear on the second one ). Inversely, every such bilinear map defines on $%
\xi _d$ the structure of the $C\left( \xi _d\right) $-module by
virtue of universal properties of Clifford d-algebras.
Equivalently, the above-mentioned bilinear map defines a morphism
of v-bundles $m:\xi _d\rightarrow HOM\left( \xi _d,\xi _d\right)
\quad [HOM\left( \xi _d,\xi _d\right) $ denotes the bundles of
homomorphisms] when $\left( m\left( u\right) \right) ^2=G\left(
u\right) $ on every point.

Vector bundles $\xi _d$ provided with $C\left( \xi _d\right)
$-structures are objects of the category with morphisms being
morphisms of v-bundles,
which induce on every point $u\in \xi $ morphisms of $C\left( \mathcal{F}%
_u\right) -$modules. This is a Banach category contained in the
category of finite-dimensional d-vector spaces on filed $k.$ We
shall not use category formalism in this work, but point to its
advantages in further formulation of new directions of K-theory
(see , for example, an introduction in Ref. \cite{kar}) concerned
with locally anisotorpic spaces.

Let us denote by $H^s\left( \xi ,GL_{n+m}\left(
\mathcal{R}\right) \right) \, $ the s-dimensional cohomology
group of the algebraic sheaf of germs of
continuous maps of v-bundle $\xi $ with group $GL_{n+m}\left( \mathcal{R}%
\right) $ the group of automorphisms of $\mathcal{R}^{n+m}\,$
(for the language of algebraic topology see, for example, Refs.
\cite{kar} and \cite {god}). We shall also use the group
$SL_{n+m}\left( \mathcal{R}\right) =\{A\subset GL_{n+m}\left(
\mathcal{R}\right) ,\det A=1\}.\,$ Here we point out that
cohomologies $H^s(M,Gr)$ characterize the class of a principal
bundle $\pi :P\rightarrow M$ on $M$ with structural group $Gr.$
Taking into account that we deal with bundles distinguished by an
N-connection we introduce into consideration cohomologies
$H^s\left( \xi ,GL_{n+m}\left( \mathcal{R}\right) \right) $ as
distinguished classes (d-classes) of bundles $\xi $ provided with
a global N-connection structure.

For a real vector bundle $\xi _d$ on compact base $\xi $ we can
define the orientation on $\xi _d$ as an element $\alpha _d\in
H^1\left( \xi ,GL_{n+m}\left( \mathcal{R}\right) \right) $ whose
image on map
\[
H^1\left( \xi ,SL_{n+m}\left( \mathcal{R}\right) \right)
\rightarrow H^1\left( \xi ,GL_{n+m}\left( \mathcal{R}\right)
\right)
\]
is the d-class of bundle $\xi .$

\begin{definition} \index{spinor structure}
\label{2.3d} The spinor structure on $\xi _d$ is defined as an element%
\newline
$\beta _d\in H^1\left( \xi ,Spin\left( n+m\right) \right) $ whose
image in the composition
\[
H^1\left( \xi ,Spin\left( n+m\right) \right) \rightarrow
H^1\left( \xi ,SO\left( n+m\right) \right) \rightarrow H^1\left(
\xi ,GL_{n+m}\left( \mathcal{R}\right) \right)
\]
is the d-class of $\xi .$
\end{definition}

The above definition of spinor structures can be reformulated in
terms of principal bundles. Let $\xi _d$ be a real vector bundle
of rank n+m on a compact base $\xi .$ If there is a principal
bundle $P_d$ with structural
group $SO\left( n+m\right) ~[$ or $Spin\left( n+m\right) ],$ this bundle $%
\xi _d$ can be provided with orientation (or spinor) structure. The bundle $%
P_d$ is associated with element $\alpha _d\in H^1\left( \xi
,SO(n+m)\right) $ [or $\beta _d\in H^1\left( \xi ,Spin\left(
n+m\right) \right) .$

We remark that a real bundle is oriented if and only if its first
Stiefel-Whitney d-class vanishes,
\[
w_1\left( \xi _d\right) \in H^1\left( \xi ,\mathcal{Z}/2\right)
=0,
\]
where $H^1\left( \xi ,\mathcal{Z}/2\right) $ is the first group
of Chech cohomology with coefficients in $\mathcal{Z}/2 ,$
Considering the second
Stiefel-Whitney class $w_2\left( \xi _d\right) \in H^{21}\left( \xi ,%
\mathcal{Z}/2\right) $ it is well known that vector bundle $\xi
_d$ admits the spinor structure if and only if $w_2\left( \xi
_d\right) =0.$ Finally,
in this subsection, we emphasize that taking into account that base space $%
\xi $ is also a v-bundle, $p:E\rightarrow M,$ we have to make
explicit calculations in order to express cohomologies $H^s\left(
\xi ,GL_{n+m}\right) \,$ and $^{}H^s\left( \xi ,SO\left(
n+m\right) \right) $ through cohomologies $H^s\left(
M,GL_n\right) ,H^s\left( M,SO\left( m\right) \right) $ , which
depends on global topological structures of spaces $M$ and $\xi
.$ For general bundle and base spaces this requires a cumbersome
cohomological calculus.

\subsection{Anisotropic Clifford fibration} \index{Clifford fibration}

Another way of defining the spinor structure is to use Clifford
fibrations. Consider the principal bundle with the structural
group $Gr$ being a subgroup of orthogonal group $O\left( G\right)
,$ where $G$ is a quadratic nondegenerate form
(see(\ref{dmetric})) defined on the base (also being a
bundle space) space $\xi .$ The fibration associated to principal fibration $%
P\left( \xi ,Gr\right) \left[ \mbox{or }P\left( H^{2n},Gr\right)
\right] $ with a typical fiber having Clifford algebra $C\left(
G\right) $ is, by definition, the Clifford fibration $PC\left(
\xi ,Gr\right) .$ We can always
define a metric on the Clifford fibration if every fiber is isometric to $%
PC\left( \xi ,G\right) $ (this result is proved for arbitrary
quadratic
forms $G$ on pseudo-Riemannian bases \cite{tur}). If, additionally, $%
Gr\subset SO\left( G\right) $ a global section can be defined on
$PC\left( G\right) .$

Let $\mathcal{P}\left( \xi ,Gr\right) $ be the set of principal
bundles with differentiable base $\xi $ and structural group
$Gr.$ If $g:Gr\rightarrow Gr^{\prime }$ is an homomorphism of Lie
groups and $P\left( \xi ,Gr\right) \subset \mathcal{P}\left( \xi
,Gr\right) $ (for simplicity in this section we shall denote
mentioned bundles and sets of bundles as $P,P^{\prime }$ and
respectively, $\mathcal{P},\mathcal{P}^{\prime }),$ we can always
construct
a principal bundle with the property that there is as homomorphism $%
f:P^{\prime }\rightarrow P$ of principal bundles which can be
projected to the identity map of $\xi $ and corresponds to
isomorphism $g:Gr\rightarrow Gr^{\prime }.$ If the inverse
statement also holds, the bundle $P^{\prime }$ is called as the
extension of $P$ associated to $g$ and $f$ is called the
extension homomorphism denoted as $\widetilde{g.}$

Now we can define distinguished spinor structures on bundle
spaces (compare with definition 2.3 ).

\begin{definition}
\label{2.4d} Let $P\in \mathcal{P}\left( \xi ,O\left( G\right)
\right) $ be a principal bundle. A distinguished spinor structure
of $P,$ equivalently a ds-structure of $\xi $ is an extension
$\widetilde{P}$ of $P$ associated to homomorphism
$h:PinG\rightarrow O\left( G\right) $ where $O\left( G\right) $
is the group of orthogonal rotations, generated by metric $G,$ in bundle $%
\xi .$
\end{definition}

So, if $\widetilde{P}$ is a spinor structure of the space $\xi ,$ then $%
\widetilde{P}\in \mathcal{P}\left( \xi ,PinG\right) .$

The definition of spinor structures on varieties was given in Ref.\cite{cru1}%
. In Refs. \cite{cru2} and \cite{cru2} it is proved that a
necessary and sufficient condition for a space time to be
orientable is to admit a global field of orthonormalized frames.
We mention that spinor structures can be also defined on
varieties modeled on Banach spaces \cite{ana77}. As we have shown
in this subsection, similar constructions are possible for the
cases when space time has the structure of a v-bundle with an
N-connection.

\begin{definition}
\label{2.5d} A special distinguished spinor structure,
ds-structure, of principal bundle $P=P\left( \xi ,SO\left(
G\right) \right) $ is a principal bundle
$\widetilde{P}=\widetilde{P}\left( \xi ,SpinG\right) $ for which a
homomorphism of principal bundles
$\widetilde{p}:\widetilde{P}\rightarrow P,$ projected on the
identity map of $\xi $ (or of $H^{2n})$ and corresponding to
representation
\[
R:SpinG\rightarrow SO\left( G\right) ,
\]
is defined.
\end{definition}

In the case when the base space variety is oriented, there is a
natural bijection between tangent spinor structures with a common
base. For special ds-structures we can define, as for any spinor
structure, the concepts of spin tensors, spinor connections, and
spinor covariant derivations (see Refs. \cite{viasm1,vdeb,vsp1}).

\section[Almost Complex Spinors]{Almost Complex Anisotropic Spinor \newline
Structures}

Almost complex structures are an important characteristic of $H^{2n}$%
-spaces. We can rewrite the almost Hermitian metric
\cite{ma87,ma94}, in complex form \cite{vjmp}:

\begin{equation}
G=H_{ab}\left( z,\xi \right) dz^a\otimes dz^b,  \label{2.47}
\end{equation}
where
\[
z^a=x^a+iy^a,~\overline{z^a}=x^a-iy^a,~H_{ab}\left(
z,\overline{z}\right) =g_{ab}\left( x,y\right) \mid _{y=y\left(
z,\overline{z}\right) }^{x=x\left( z,\overline{z}\right) },
\]
and define almost complex spinor structures. For given metric
(\ref{2.47}) on $H^{2n}$-space there is always a principal bundle
$P^U$ with unitary structural group U(n) which allows us to
transform $H^{2n}$-space into v-bundle $\xi ^U\approx P^U\times
_{U\left( n\right) }\mathcal{R}^{2n}.$ This statement will be
proved after we introduce complex
\begin{figure}[tbph]
\label{2.48}
\par
\begin{center}
\begin{picture}(255,50) \setlength{\unitlength}{1pt}
\thinlines \put(0,45){$U(n)$} \put(116,45){$SO(2n)$}
\put(53,2){${Spin}^c (2n)$} \put(70,38){i} \put(70,48){
\line(-1,0){45}} \put(70,48){\vector(1,0){45}}
\put(25,42){\vector(3,-2){45}} \put(78,13) {\vector(3,2){45}}
\put(245,10){
}
\put(35,20){$\sigma$} \put(100,20){${\rho}^c$}
\end{picture}
\end{center}
\caption{Diagram 3}
\end{figure}
spinor structures on oriented real vector bundles \cite{kar}.

Let us consider momentarily $k=\mathcal{C}$ and introduce into
consideration
[instead of the group $Spin(n)]$ the group $Spin^c\times _{\mathcal{Z}%
/2}U\left( 1\right) $ being the factor group of the product
$Spin(n)\times U\left( 1\right) $ with the respect to equivalence
\[
\left( y,z\right) \sim \left( -y,-a\right) ,\quad y\in Spin(m).
\]
This way we define the short exact sequence
\[
1\rightarrow U\left( 1\right) \rightarrow Spin^c\left( n\right) \stackrel{S^c%
}{\to }SO\left( n\right) \rightarrow 1,
\]
where $\rho ^c\left( y,a\right) =\rho ^c\left( y\right) .$ If
$\lambda $ is oriented , real, and rank $n,$ $\gamma $-bundle
$\pi :E_\lambda \rightarrow
M^n,$ with base $M^n,$ the complex spinor structure, spin structure, on $%
\lambda $ is given by the principal bundle $P$ with structural group $%
Spin^c\left( m\right) $ and isomorphism $\lambda \approx P\times
_{Spin^c\left( n\right) }\mathcal{R}^n.$ For such bundles the
categorial equivalence can be defined as
\begin{equation}
\epsilon ^c:\mathcal{E}_{\mathcal{C}}^T\left( M^n\right)
\rightarrow \mathcal{E}_{\mathcal{C}}^\lambda \left( M^n\right)
,  \label{2.49}
\end{equation}
where $\epsilon ^c\left( E^c\right) =P\bigtriangleup
_{Spin^c\left( n\right)
}E^c$ is the category of trivial complex bundles on $M^n,\mathcal{E}_{%
\mathcal{C}}^\lambda \left( M^n\right) $ is the category of
complex v-bundles on $M^n$ with action of Clifford bundle
$C\left( \lambda \right) ,P\bigtriangleup _{Spin^c(n)}$ and $E^c$
is the factor space of the bundle product $P\times _ME^c$ with
respect to the equivalence $\left( p,e\right)
\sim \left( p\widehat{g}^{-1},\widehat{g}e\right) ,p\in P,e\in E^c,$ where $%
\widehat{g}\in Spin^c\left( n\right) $ acts on $E$ by via the imbedding $%
Spin\left( n\right) \subset C^{0,n}$ and the natural action
$U\left( 1\right) \subset \mathcal{C}$ on complex v-bundle $\xi
^c,E^c=tot\xi ^c,$ for bundle $\pi ^c:E^c\rightarrow M^n.$

Now we return to the bundle $\xi .$ A real v-bundle (not being a
spinor bundle) admits a complex spinor structure if and only if
there exist a homomorphism $\sigma :U\left( n\right) \rightarrow
Spin^c\left( 2n\right) $ making the diagram 3 commutative. The
explicit construction of $\sigma $ for
arbitrary $\gamma $-bundle is given in Refs. \cite{kar} and \cite{ati}. For $%
H^{2n}$-spaces it is obvious that a diagram similar to
(\ref{2.48}) can be defined for the tangent bundle $TM^n,$ which
directly points to the possibility of defining the
$^cSpin$-structure on $H^{2n}$-spaces.

Let $\lambda $ be a complex, rank\thinspace $n,$ spinor bundle
with
\begin{equation}
\tau :Spin^c\left( n\right) \times _{\mathcal{Z}/2}U\left(
1\right) \rightarrow U\left( 1\right)  \label{2.50}
\end{equation}
the homomorphism defined by formula $\tau \left( \lambda ,\delta
\right) =\delta ^2.$ For $P_s$ being the principal bundle with
fiber $Spin^c\left( n\right) $ we introduce the complex linear
bundle $L\left( \lambda ^c\right)
=P_S\times _{Spin^c(n)}\mathcal{C}$ defined as the factor space of $%
P_S\times \mathcal{C}$ on equivalence relation
\[
\left( pt,z\right) \sim \left( p,l\left( t\right) ^{-1}z\right) ,
\]
where $t\in Spin^c\left( n\right) .$ This linear bundle is
associated to complex spinor structure on $\lambda ^c.$

If $\lambda ^c$ and $\lambda ^{c^{\prime }}$ are complex spinor
bundles, the Whitney sum $\lambda ^c\oplus \lambda ^{c^{\prime
}}$ is naturally provided with the structure of the complex
spinor bundle. This follows from the holomorphism
\begin{equation}
\omega ^{\prime }:Spin^c\left( n\right) \times Spin^c\left(
n^{\prime }\right) \rightarrow Spin^c\left( n+n^{\prime }\right)
,  \label{2.51}
\end{equation}
given by formula $\left[ \left( \beta ,z\right) ,\left( \beta
^{\prime },z^{\prime }\right) \right] \rightarrow \left[ \omega
\left( \beta ,\beta ^{\prime }\right) ,zz^{\prime }\right] ,$
where $\omega $ is the homomorphism making the following diagram
4 commutative.
\begin{figure}[tbph]
\begin{center}
\begin{picture}(190,50) \setlength{\unitlength}{1pt}
\thinlines \put(0,45){$Spin(n)\times Spin(n')$}
\put(160,45){$Spin(n+n')$} \put(15,2){$O(n)\times O(n')$}
\put(168,02){$O(n+n')$} \put(100,45){ \vector(1,0){55}}
\put(48,40){\vector(0,-1){25}} \put(90,5){\vector(1,0){70}}
\put(185,40) {\vector(0,-1){25}}
\end{picture}
\end{center}
\caption{Diagram 4}
\end{figure}
Here, $z,z^{\prime }\in U\left( 1\right) .$ It is obvious that
$L\left( \lambda ^c\oplus \lambda ^{c^{\prime }}\right) $ is
isomorphic to $L\left( \lambda ^c\right) \otimes L\left( \lambda
^{c^{\prime }}\right) .$

We conclude this section by formulating our main result on
complex spinor structures for $H^{2n}$-spaces:

\begin{theorem}
\label{2.3t} Let $\lambda ^c$ be a complex spinor bundle of rank $n$ and $%
H^{2n}$-space considered as a real vector bundle $\lambda
^c\oplus \lambda ^{c^{\prime }}$ provided with almost complex
structure $J_{\quad \beta }^\alpha ;$ multiplication on $i$ is
given by $\left(
\begin{array}{cc}
0 & -\delta _j^i \\
\delta _j^i & 0
\end{array}
\right) $. Then, the diagram 5 is commutative up to isomorphisms
$\epsilon ^c $ and $\widetilde{\epsilon }^c$ defined as in
(2.49), $\mathcal{H}$ is
functor $E^c\rightarrow E^c\otimes L\left( \lambda ^c\right) $ and $\mathcal{%
E}_{\mathcal{C}}^{0,2n}\left( M^n\right) $ is defined by functor $\mathcal{E}%
_{\mathcal{C}}\left( M^n\right) \rightarrow \mathcal{E}_{\mathcal{C}%
}^{0,2n}\left( M^n\right) $ given as correspondence
$E^c\rightarrow \Lambda \left( \mathcal{C}^n\right) \otimes E^c$
(which is a categorial equivalence), $\Lambda \left(
\mathcal{C}^n\right) $ is the exterior algebra on
$\mathcal{C}^n.$ $W$ is the real bundle $\lambda ^c\oplus \lambda
^{c^{\prime }}$ provided with complex structure.
\end{theorem}

\begin{figure}[tbph]
\begin{center}
\begin{picture}(150,50) \setlength{\unitlength}{1pt}
\thinlines \put(0,45){${\cal E}_{\cal C}^{0,2n} (M^{2n})$}
\put(145,45){${\cal E}_{\cal C}^{{\lambda}^c \oplus {\lambda}^c }
(M^n )$} \put(81,0){${\cal E}_{\cal C}^W (M^n )$}
\put(99,38){${\epsilon}^c$} \put(99,48){ \line(-1,0){45}}
\put(99,48){\vector(1,0){45}} \put(54,42){\vector(3,-2){45}}
\put(147,42){\vector(-3,-2){45}}
\put(64,20){${\tilde{\varepsilon}}^c$} \put(129,20){$\cal H$}
\end{picture}

\end{center}
\caption{Diagram 5}
\end{figure}
\textbf{Proof: }We use composition of homomorphisms
\[
\mu :Spin^c\left( 2n\right) \stackrel{\pi }{\to }SO\left( n\right) \stackrel{%
r}{\to }U\left( n\right) \stackrel{\sigma }{\to }Spin^c\left(
2n\right) \times _{\mathcal{Z}/2}U\left( 1\right) ,
\]
commutative diagram 6 and introduce composition of homomorphisms
\[
\mu :Spin^c\left( n\right) \stackrel{\Delta }{\to }Spin^c\left(
n\right) \times Spin^c\left( n\right) \stackrel{{\omega }^c}{\to
}Spin^c\left( n\right) ,
\]
where $\Delta $ is the diagonal homomorphism and $\omega ^c$ is
defined as in (\ref{2.51}). Using homomorphisms (\ref{2.50}) and
(\ref{2.51}) we obtain formula $\mu \left( t\right) =\mu \left(
t\right) r\left( t\right) .$

Now consider bundle $P\times _{Spin^c\left( n\right)
}Spin^c\left( 2n\right) $ as the principal $Spin^c\left(
2n\right) $-bundle, associated to $M\oplus M $ being the factor
space of the product $P\times Spin^c\left( 2n\right) $ on the
equivalence relation $\left( p,t,h\right) \sim \left( p,\mu \left(
t\right) ^{-1}h\right) .$ In this case the categorial equivalence (\ref{2.49}%
) can be rewritten as
\[
\epsilon ^c\left( E^c\right) =P\times _{Spin^c\left( n\right)
}Spin^c\left( 2n\right) \Delta _{Spin^c\left( 2n\right) }E^c
\]
and seen as factor space of $P\times Spin^c\left( 2n\right)
\times _ME^c$ on equivalence relation
\[
\left( pt,h,e\right) \sim \left( p,\mu \left( t\right) ^{-1}h,e\right) %
\mbox{and}\left( p,h_1,h_2,e\right) \sim \left(
p,h_1,h_2^{-1}e\right)
\]
(projections of elements $p$ and $e$ coincides on base $M).$
Every element of $\epsilon ^c\left( E^c\right) $ can be
represented as $P\Delta _{Spin^c\left( n\right) }E^c,$ i.e., as a
factor space $P\Delta E^c$ on equivalence relation $\left(
pt,e\right) \sim \left( p,\mu ^c\left( t\right) ,e\right) ,$ when
$t\in Spin^c\left( n\right) .$
\begin{figure}[tbph]
\begin{center}
\begin{picture}(140,70) \setlength{\unitlength}{1pt}
\thinlines \put(0,60){$Spin(2n)$} \put(100,60){${Spin}^c (2n)$}
\put(6,0){$SO(n)$} \put(106,0){$SO(2n)$} \put(36,2){
\vector(1,0){65}} \put(20,13){\vector(0,1){40}}
\put(126,13) {\vector(0,1){40}}
\put(8,30){$\beta$} \put(70,60){$\subset$}
\end{picture}
\end{center}
\caption{Diagram 6}
\end{figure}
The complex line bundle $L\left( \lambda ^c\right) $ can be
interpreted as the factor space of\newline
$P\times _{Spin^c\left( n\right) }\mathcal{C}$ on equivalence relation $%
\left( pt,\delta \right) \sim \left( p,r\left( t\right)
^{-1}\delta \right) . $

Putting $\left( p,e\right) \otimes \left( p,\delta \right) \left(
p,\delta e\right) $ we introduce morphism
\[
\epsilon ^c\left( E\right) \times L\left( \lambda ^c\right)
\rightarrow \epsilon ^c\left( \lambda ^c\right)
\]
with properties $\left( pt,e\right) \otimes \left( pt,\delta
\right) \rightarrow \left( pt,\delta e\right) =\left( p,\mu
^c\left( t\right) ^{-1}\delta e\right) ,$

$\left( p,\mu ^c\left( t\right) ^{-1}e\right) \otimes \left(
p,l\left( t\right) ^{-1}e\right) \rightarrow \left( p,\mu
^c\left( t\right) r\left( t\right) ^{-1}\delta e\right) $
pointing to the fact that we have defined
the isomorphism correctly and that it is an isomorphism on every fiber. $%
\Box $ \vskip20pt

\chapter{Spinors and Anisotropic Spaces}

The purpose of this Chapter is to show how a corresponding
abstract spinor technique entailing notational and calculations
advantages can be developed
for arbitrary splits of dimensions of a d-vector space $\mathcal{F}=h%
\mathcal{F}\oplus v\mathcal{F}$, where $\dim h\mathcal{F}=n$ and $\dim v%
\mathcal{F}=m.$ For convenience we shall also present some
necessary coordinate expressions.

The problem of a rigorous definition of spinors on locally
anisotropic spaces (anisotropic spinors, d--spinors) was posed
and solved \cite {vjmp,viasm1,vdeb} in the framework of the
formalism of Clifford and spinor structures on v--bundles
provided with compatible nonlinear and distinguished connections
and metric \index{d--spinor}. We introduced d--spinors as
corresponding objects of the Clifford d--algebra $\mathcal{C}\left( \mathcal{%
F},G\right) $, defined for a d--vector space $\mathcal{F}$ in a
standard
manner (see, for instance, \cite{kar}) and proved that operations with $%
\mathcal{C}\left( \mathcal{F},G\right) \ $ can be reduced to
calculations
for $\mathcal{C}\left( h\mathcal{F},g\right) $ and $\mathcal{C}\left( v%
\mathcal{F},h\right) ,$ which are usual Clifford algebras of
respective dimensions 2$^n$ and 2$^m$ (if it is necessary we can
use quadratic forms $g$ and $h$ correspondingly induced on
$h\mathcal{F}$ and $v\mathcal{F}$ by a
metric $\mathbf{G}$ (\ref{dmetric})). Considering the orthogonal subgroup $O%
\mathbf{\left( G\right) }\subset GL\mathbf{\left( G\right) }$
defined by a metric $\mathbf{G}$ we can define the d-spinor norm
and parametrize
d-spinors by ordered pairs of elements of Clifford algebras $\mathcal{C}%
\left( h\mathcal{F},g\right) $ and $\mathcal{C}\left(
v\mathcal{F},h\right) . $ We emphasize that the splitting of a
Clifford d-algebra associated to a v-bundle $\mathcal{E}$ is a
straightforward consequence of the global decomposition defining
a N-connection structure in $\mathcal{E}$.

In this Chapter, as a rule, we shall omit proofs which in most
cases are mechanical but rather tedious. We can apply the methods
developed in \cite {pen,penr1,penr2,lue} in a straightforward
manner on h- and v-subbundles in order to verify the correctness
of affirmations.

\section[Anisotropic Spinors and Twistors]{Anisotropic Clifford Algebras,
\newline
Spinors and Twistors} \index{twistor}

In order to relate the succeeding constructions with Clifford
d-algebras \cite{vjmp,viasm1} we consider a locally anisotropic
frame decomposition of the metric (\ref{dmetric}):
\begin{equation}
G_{\alpha \beta }\left( u\right) =l_\alpha ^{\widehat{\alpha
}}\left(
u\right) l_\beta ^{\widehat{\beta }}\left( u\right) G_{\widehat{\alpha }%
\widehat{\beta }},  \label{2.52}
\end{equation}
where the frame d-vectors and constant metric matrices are
distinguished as

\begin{equation}
l_\alpha ^{\widehat{\alpha }}\left( u\right) =\left(
\begin{array}{cc}
l_j^{\widehat{j}}\left( u\right) & 0 \\
0 & l_a^{\widehat{a}}\left( u\right)
\end{array}
\right) ,G_{\widehat{\alpha }\widehat{\beta }}\left(
\begin{array}{cc}
g_{\widehat{i}\widehat{j}} & 0 \\
0 & h_{\widehat{a}\widehat{b}}
\end{array}
\right) ,  \label{2.53}
\end{equation}
$g_{\widehat{i}\widehat{j}}$ and $h_{\widehat{a}\widehat{b}}$ are
diagonal matrices with $g_{\widehat{i}\widehat{i}}=$
$h_{\widehat{a}\widehat{a}}=\pm 1.$

To generate Clifford d--algebras we start with matrix equations
\begin{equation}
\sigma _{\widehat{\alpha }}\sigma _{\widehat{\beta }}+\sigma _{\widehat{%
\beta }}\sigma _{\widehat{\alpha }}=-G_{\widehat{\alpha
}\widehat{\beta }}I, \label{2.54}
\end{equation}
where $I$ is the identity matrix, matrices $\sigma _{\widehat{\alpha }%
}\,(\sigma $-objects) act on a d-vector space $\mathcal{F}=h\mathcal{F}%
\oplus v\mathcal{F}$ and theirs components are distinguished as
\begin{equation}
\sigma _{\widehat{\alpha }}\,=\left\{ (\sigma _{\widehat{\alpha }})_{%
\underline{\beta }}^{\cdot \underline{\gamma }}=\left(
\begin{array}{cc}
(\sigma _{\widehat{i}})_{\underline{j}}^{\cdot \underline{k}} & 0 \\
0 & (\sigma _{\widehat{a}})_{\underline{b}}^{\cdot \underline{c}}
\end{array}
\right) \right\} ,  \label{2.55}
\end{equation}
indices \underline{$\beta $},\underline{$\gamma $},... refer to
spin spaces of type $\mathcal{S}=S_{(h)}\oplus S_{(v)}$ and
underlined Latin indices \underline{$j$},$\underline{k},...$ and
$\underline{b},\underline{c},...$
refer respectively to a h-spin space $\mathcal{S}_{(h)}$ and a v-spin space $%
\mathcal{S}_{(v)},\ $which are correspondingly associated to a h-
and v-decomposition of a v-bundle $\mathcal{E}_{(d)}.$ The
irreducible algebra of matrices $\sigma _{\widehat{\alpha }}$ of
minimal dimension $N\times N,$ where $N=N_{(n)}+N_{(m)},$ $\dim
\mathcal{S}_{(h)}$=$N_{(n)}$ and $\dim
\mathcal{S}_{(v)}$=$N_{(m)},$ has these dimensions
\begin{eqnarray*}
{N_{(n)}} &=&{\left\{
\begin{array}{rl}
{\ 2^{(n-1)/2},} & n=2k+1 \\
{2^{n/2},\ } & n=2k;
\end{array}
\right. }\quad \quad \\
{N_{(m)}} &=&{\left\{
\begin{array}{rl}
{2^{(m-1)/2},} & m=2k+1 \\
{2^{m/2},\ } & m=2k,
\end{array}
\right. }
\end{eqnarray*}
where $k=1,2,...$ .

The Clifford d-algebra is generated by sums on $n+1$ elements of
form
\[
A_1I+B^{\widehat{i}}\sigma _{\widehat{i}}+C^{\widehat{i}\widehat{j}}\sigma _{%
\widehat{i}\widehat{j}}+D^{\widehat{i}\widehat{j}\widehat{k}}\sigma _{%
\widehat{i}\widehat{j}\widehat{k}}+...
\]
and sums of $m+1$ elements of form
\[
A_2I+B^{\widehat{a}}\sigma _{\widehat{a}}+C^{\widehat{a}\widehat{b}}\sigma _{%
\widehat{a}\widehat{b}}+D^{\widehat{a}\widehat{b}\widehat{c}}\sigma _{%
\widehat{a}\widehat{b}\widehat{c}}+...
\]
with antisymmetric coefficients $C^{\widehat{i}\widehat{j}}=C^{[\widehat{i}%
\widehat{j}]},C^{\widehat{a}\widehat{b}}=C^{[\widehat{a}\widehat{b]}},D^{%
\widehat{i}\widehat{j}\widehat{k}}=D^{[\widehat{i}\widehat{j}\widehat{k}%
]},D^{\widehat{a}\widehat{b}\widehat{c}}=D^{[\widehat{a}\widehat{b}\widehat{c%
}]},...$ and matrices $\sigma _{\widehat{i}\widehat{j}}=\sigma _{[\widehat{i}%
}\sigma _{\widehat{j}]},\sigma _{\widehat{a}\widehat{b}}=\sigma _{[\widehat{a%
}}\sigma _{\widehat{b}]},\sigma
_{\widehat{i}\widehat{j}\widehat{k}}=\sigma _{[\widehat{i}}\sigma
_{\widehat{j}}\sigma _{\widehat{k}]},...$ . Really, we
have 2$^{n+1}$ coefficients $\left( A_1,C^{\widehat{i}\widehat{j}},D^{%
\widehat{i}\widehat{j}\widehat{k}},...\right) $ and 2$^{m+1}$ coefficients%
\newline
$\left( A_2,C^{\widehat{a}\widehat{b}},D^{\widehat{a}\widehat{b}\widehat{c}%
},...\right) $ of the Clifford algebra on $\mathcal{F}$.

For simplicity, in this subsection, we shall present the
necessary geometric
constructions only for h-spin spaces $\mathcal{S}_{(h)}$ of dimension $%
N_{(n)}.$ Considerations for a v-spin space $\mathcal{S}_{(v)}$
are similar but with proper characteristics for a dimension
$N_{(m)}.$

In order to define the scalar (spinor) product on
$\mathcal{S}_{(h)}$ we introduce into consideration this finite
sum (because of a finite number of elements $\sigma
_{[\widehat{i}\widehat{j}...\widehat{k}]}):$
\begin{equation}
^{(\pm
)}E_{\underline{k}\underline{m}}^{\underline{i}\underline{j}}=\delta
_{\underline{k}}^{\underline{i}}\delta _{\underline{m}}^{\underline{j}%
}+\frac 2{1!}(\sigma
_{\widehat{i}})_{\underline{k}}^{.\underline{i}}(\sigma
^{\widehat{i}})_{\underline{m}}^{.\underline{j}}+\frac{2^2}{2!}(\sigma _{%
\widehat{i}\widehat{j}})_{\underline{k}}^{.\underline{i}}(\sigma ^{\widehat{i%
}\widehat{j}})_{\underline{m}}^{.\underline{j}}+\frac{2^3}{3!}(\sigma _{%
\widehat{i}\widehat{j}\widehat{k}})_{\underline{k}}^{.\underline{i}}(\sigma
^{\widehat{i}\widehat{j}\widehat{k}})_{\underline{m}}^{.\underline{j}}+...
\label{2.56}
\end{equation}
which can be factorized as
\begin{equation}
^{(\pm )}E_{\underline{k}\underline{m}}^{\underline{i}\underline{j}}=N_{(n)}{%
\ }^{(\pm )}\epsilon _{\underline{k}\underline{m}}{\ }^{(\pm )}\epsilon ^{%
\underline{i}\underline{j}}\mbox{ for }n=2k  \label{2.57}
\end{equation}
and
\begin{eqnarray}
^{(+)}E_{\underline{k}\underline{m}}^{\underline{i}\underline{j}}
&=&2N_{(n)}\epsilon _{\underline{k}\underline{m}}\epsilon ^{\underline{i}%
\underline{j}},{\ }^{(-)}E_{\underline{k}\underline{m}}^{\underline{i}%
\underline{j}}=0\mbox{ for }n=3(mod4),  \label{2.58} \\
^{(+)}E_{\underline{k}\underline{m}}^{\underline{i}\underline{j}} &=&0,{\ }%
^{(-)}E_{\underline{k}\underline{m}}^{\underline{i}\underline{j}%
}=2N_{(n)}\epsilon _{\underline{k}\underline{m}}\epsilon ^{\underline{i}%
\underline{j}}\mbox{ for }n=1(mod4).  \nonumber
\end{eqnarray}

Antisymmetry of $\sigma _{\widehat{i}\widehat{j}\widehat{k}...}$
and the construction of the objects (\ref{2.56}),(\ref{2.57}) and
(\ref{2.58})
define the properties of $\epsilon $-objects $^{(\pm )}\epsilon _{\underline{%
k}\underline{m}}$ and $\epsilon _{\underline{k}\underline{m}}$
which have an eight-fold periodicity on $n$ (see details in
\cite{penr2} and, with respect to locally anisotropic spaces,
\cite{vjmp}).

For even values of $n$ it is possible the decomposition of every
h-spin space $\mathcal{S}_{(h)}$into irreducible h-spin spaces
$\mathbf{S}_{(h)}$ and $\mathbf{S}_{(h)}^{\prime }$ (one
considers splitting of h-indices, for instance,
\underline{$l$}$=L\oplus L^{\prime },\underline{m}=M\oplus
M^{\prime },...;$ for v-indices we shall write
$\underline{a}=A\oplus
A^{\prime },\underline{b}=B\oplus B^{\prime },...)$ and defines new $%
\epsilon $-objects
\begin{equation}
\epsilon ^{\underline{l}\underline{m}}=\frac 12\left( ^{(+)}\epsilon ^{%
\underline{l}\underline{m}}+^{(-)}\epsilon ^{\underline{l}\underline{m}%
}\right) \mbox{ and }\widetilde{\epsilon }^{\underline{l}\underline{m}%
}=\frac 12\left( ^{(+)}\epsilon
^{\underline{l}\underline{m}}-^{(-)}\epsilon
^{\underline{l}\underline{m}}\right) .  \label{2.59}
\end{equation}
We shall omit similar formulas for $\epsilon $-objects with lower
indices.

We can verify, by using expressions (\ref{2.58}) and
straightforward
calculations, these para\-met\-ri\-za\-ti\-ons on symmetry properties of $%
\epsilon $-objects (\ref{2.59})
\begin{eqnarray}
\epsilon ^{\underline{l}\underline{m}}=\left(
\begin{array}{cc}
\epsilon ^{LM}=\epsilon ^{ML} & 0 \\
0 & 0
\end{array}
\right) \mbox{ and } \widetilde{\epsilon }^{\underline{l}\underline{m}%
}=\left(
\begin{array}{cc}
0 & 0 \\
0 & \widetilde{\epsilon }^{LM}=\widetilde{\epsilon }^{ML}
\end{array}
\right) & &  \nonumber \\
\mbox{ for } n=0(mod8); & &  \nonumber \\
\epsilon ^{\underline{l}\underline{m}}=-\frac 12{}^{(-)}\epsilon ^{%
\underline{l}\underline{m}}=\epsilon ^{\underline{m}\underline{l}},%
\mbox{where }^{(+)}\epsilon ^{\underline{l}\underline{m}}=0,
\mbox{ and }& &
\label{2.60} \\
\widetilde{\epsilon }^{\underline{l}\underline{m}}=-\frac
12{}^{(-)}\epsilon
^{\underline{l}\underline{m}}=\widetilde{\epsilon }^{\underline{m}\underline{%
l}} \mbox{for } n=1(mod8); & &  \nonumber \\
\epsilon ^{\underline{l}\underline{m}}=\left(
\begin{array}{cc}
0 & 0 \\
\epsilon ^{L^{\prime }M} & 0
\end{array}
\right) \mbox{ and } \widetilde{\epsilon }^{\underline{l}\underline{m}%
}=\left(
\begin{array}{cc}
0 & \widetilde{\epsilon }^{LM^{\prime }}=-\epsilon ^{M^{\prime }L} \\
0 & 0
\end{array}
\right) & &  \nonumber \\
\mbox{ for } n=2(mod8); & &  \nonumber \\
\epsilon ^{\underline{l}\underline{m}}=-\frac 12{}^{(+)}\epsilon ^{%
\underline{l}\underline{m}}=-\epsilon ^{\underline{m}\underline{l}},%
\mbox{ where }^{(-)}\epsilon ^{\underline{l}\underline{m}}=0,
\mbox{ and } & &
\nonumber \\
\widetilde{\epsilon }^{\underline{l}\underline{m}}=\frac
12{}^{(+)}\epsilon
^{\underline{l}\underline{m}}=-\widetilde{\epsilon }^{\underline{m}%
\underline{l}} \mbox{for } n=3(mod8); & &  \nonumber \\
\epsilon ^{\underline{l}\underline{m}}=\left(
\begin{array}{cc}
\epsilon ^{LM}=-\epsilon ^{ML} & 0 \\
0 & 0
\end{array}
\right) \mbox{ and }\widetilde{\epsilon }^{\underline{l}\underline{m}%
}=\left(
\begin{array}{cc}
0 & 0 \\
0 & \widetilde{\epsilon }^{LM}=-\widetilde{\epsilon }^{ML}
\end{array}
\right) & &  \nonumber \\
\mbox{ for } n=4(mod8); & &  \nonumber \\
\epsilon ^{\underline{l}\underline{m}}=-\frac 12{}^{(-)}\epsilon ^{%
\underline{l}\underline{m}}=-\epsilon ^{\underline{m}\underline{l}},%
\mbox{ where }^{(+)}\epsilon ^{\underline{l}\underline{m}}=0,
\mbox{ and } & &
\nonumber \\
\widetilde{\epsilon }^{\underline{l}\underline{m}}=-\frac
12{}^{(-)}\epsilon
^{\underline{l}\underline{m}}=-\widetilde{\epsilon }^{\underline{m}%
\underline{l}} \mbox{for } n=5(mod8); & &  \nonumber \\
\epsilon ^{\underline{l}\underline{m}}=\left(
\begin{array}{cc}
0 & 0 \\
\epsilon ^{L^{\prime }M} & 0
\end{array}
\right) \mbox{ and }\widetilde{\epsilon }^{\underline{l}\underline{m}%
}=\left(
\begin{array}{cc}
0 & \widetilde{\epsilon }^{LM^{\prime }}=\epsilon ^{M^{\prime }L} \\
0 & 0
\end{array}
\right) & &  \nonumber \\
\mbox{ for } n=6(mod8); & &  \nonumber \\
\epsilon ^{\underline{l}\underline{m}}=\frac 12{}^{(-)}\epsilon ^{\underline{%
l}\underline{m}}=\epsilon ^{\underline{m}\underline{l}},\mbox{
where }{}^{(+)}\epsilon ^{\underline{l}\underline{m}}=0, \mbox{
and } & &
\nonumber \\
\widetilde{\epsilon }^{\underline{l}\underline{m}}=-\frac
12{}^{(-)}\epsilon
^{\underline{l}\underline{m}}=\widetilde{\epsilon }^{\underline{m}\underline{%
l}} \mbox{ for } n=7(mod8). & &  \nonumber
\end{eqnarray}

Let denote reduced and irreducible h-spinor spaces
\index{h--spinor} in a form pointing to the
symmetry of spinor inner products in dependence of values $n=8k+l$ ($%
k=0,1,2,...;l=1,2,...7)$ of the dimension of the horizontal
subbundle (we shall write respectively $\bigtriangleup $ and
$\circ $ for antisymmetric and symmetric inner products of
reduced spinors and $\diamondsuit =(\bigtriangleup ,\circ )$ and
$\widetilde{\diamondsuit }=(\circ ,\bigtriangleup )$ for
corresponding parametrizations of inner products, in brief
\textit{i.p.}, of irreducible spinors; properties of scalar
products
of spinors are defined by $\epsilon $-objects (\ref{2.60}); we shall use $%
\diamond $ for a general \textit{i.p.} when the symmetry is not
pointed out): \newpage
\begin{eqnarray}
\mathcal{S}_{(h)}{\ }\left( 8k\right) &=&\mathbf{S_{\circ }\oplus
S_{\circ
}^{\prime };\quad }  \label{2.61} \\
\mathcal{S}_{(h)}{\ }\left( 8k+1\right) &=&\mathcal{S}_{\circ
}^{(-)}\ \mbox{({\it
i.p.} is defined by an }^{(-)}\epsilon \mbox{-object);}  \nonumber \\
\mathcal{S}_{(h)}{\ }\left( 8k+2\right) &=&\{
\begin{array}{c}
\mathcal{S}_{\diamond }=(\mathbf{S}_{\diamond },\mathbf{S}_{\diamond }),%
\mbox{ or} \\
\mathcal{S}_{\diamond }^{\prime }=(\mathbf{S}_{\widetilde{\diamond }%
}^{\prime },\mathbf{S}_{\widetilde{\diamond }}^{\prime });
\end{array}
\nonumber \\
\mathcal{S}_{(h)}\left( 8k+3\right) &=&\mathcal{S}_{\bigtriangleup }^{(+)}\ %
\mbox{({\it i.p.} is defined by an }^{(+)}\epsilon
\mbox{-object);}
\nonumber \\
\mathcal{S}_{(h)}\left( 8k+4\right) &=&\mathbf{S}_{\bigtriangleup
}\oplus
\mathbf{S}_{\bigtriangleup }^{\prime };\quad  \nonumber \\
\mathcal{S}_{(h)}\left( 8k+5\right)
&=&\mathcal{S}_{\bigtriangleup }^{(-)}\ \mbox{({\it i.p. }is
defined
by an }^{(-)}\epsilon \mbox{-object),}  \nonumber \\
\mathcal{S}_{(h)}\left( 8k+6\right) &=&\{
\begin{array}{c}
\mathcal{S}_{\diamond }=(\mathbf{S}_{\diamond },\mathbf{S}_{\diamond }),%
\mbox{ or} \\
\mathcal{S}_{\diamond }^{\prime }=(\mathbf{S}_{\widetilde{\diamond }%
}^{\prime },\mathbf{S}_{\widetilde{\diamond }}^{\prime });
\end{array}
\nonumber \\
\mathcal{S}_{(h)}\left( 8k+7\right) &=&\mathcal{S}_{\circ }^{(+)}\
\mbox{({\it i.p. } is defined by an }^{(+)}\epsilon
\mbox{-object)}.  \nonumber
\end{eqnarray}
\qquad We note that by using corresponding $\epsilon $-objects we
can lower and rise indices of reduced and irreducible spinors
(for $n=2,6(mod4)$ we can exclude primed indices, or inversely,
see details in \cite {pen,penr1,penr2}).

The similar v-spinor \index{v--spinor}  spaces are denoted by the same symbols as in (\ref{2.61}%
) provided with a left lower mark ''$|"$ and parametrized with
respect to the values $m=8k^{\prime }+l$ (k'=0,1,...;
l=1,2,...,7) of the dimension of the vertical subbundle, for
example, as
\begin{equation}
\mathcal{S}_{(v)}(8k^{\prime })=\mathbf{S}_{|\circ }\oplus \mathbf{S}%
_{|\circ }^{\prime },\mathcal{S}_{(v)}\left( 8k+1\right) =\mathcal{S}%
_{|\circ }^{(-)},...  \label{2.62}
\end{equation}
We use '' $\widetilde{}$ ''-overlined symbols,
\begin{equation}
{\widetilde{\mathcal{S}}}_{(h)}\left( 8k\right) ={\widetilde{\mathbf{S}}}%
_{\circ }\oplus \widetilde{S}_{\circ }^{\prime },{\widetilde{\mathcal{S}}}%
_{(h)}\left( 8k+1\right) ={\widetilde{\mathcal{S}}}_{\circ
}^{(-)},... \label{2.63}
\end{equation}
and
\begin{equation}
{\widetilde{\mathcal{S}}}_{(v)}(8k^{\prime })={\widetilde{\mathbf{S}}}%
_{|\circ }\oplus {\widetilde{S}}_{|\circ }^{\prime },{\widetilde{\mathcal{S}}%
}_{(v)}\left( 8k^{\prime }+1\right)
={\widetilde{\mathcal{S}}}_{|\circ }^{(-)},...  \label{2.64}
\end{equation}
respectively for the dual to (\ref{2.61}) and (\ref{2.62}) spinor
spaces.

The spinor spaces (\ref{2.61})-(\ref{2.64}) are called the prime
spinor spaces, in brief p-spinors \index{p--spinor}. They are
considered as building blocks of distinguished (n,m)--spinor
spaces constructed in this manner:
$$
\eqno(2.65)
$$
\begin{eqnarray}
\mathcal{S}(_{\circ \circ ,\circ \circ }) &=&\mathbf{S_{\circ
}\oplus
S_{\circ }^{\prime }\oplus S_{|\circ }\oplus S_{|\circ }^{\prime },}\mathcal{%
S}(_{\circ \circ ,\circ }\mid ^{\circ })=\mathbf{S_{\circ }\oplus
S_{\circ }^{\prime }\oplus S_{|\circ }\oplus
\widetilde{S}_{|\circ }^{\prime },}
\nonumber \\
\mathcal{S}(_{\circ \circ ,} &\mid &^{\circ \circ
})=\mathbf{S_{\circ
}\oplus S_{\circ }^{\prime }\oplus \widetilde{S}_{|\circ }\oplus \widetilde{S%
}_{|\circ }^{\prime },}\mathcal{S}(_{\circ }\mid ^{\circ \circ \circ })=%
\mathbf{S_{\circ }\oplus \widetilde{S}_{\circ }^{\prime }\oplus \widetilde{S}%
_{|\circ }\oplus \widetilde{S}_{|\circ }^{\prime },}  \nonumber \\
&&...............................................  \nonumber \\
\mathcal{S}(_{\triangle },_{\triangle }) &=&\mathcal{S}_{\triangle
}^{(+)}\oplus S_{|\bigtriangleup }^{(+)},S(_{\triangle },^{\triangle })=%
\mathcal{S}_{\triangle }^{(+)}\oplus \widetilde{S}_{|\triangle
}^{(+)},
\label{2.65} \\
&&................................  \nonumber \\
\mathcal{S}(_{\triangle }|^{\circ },_{\diamondsuit }) &=&\mathbf{S}%
_{\triangle }\oplus \widetilde{S_{\circ }}^{\prime }\oplus \mathcal{S}%
_{|\diamondsuit },\mathcal{S}(_{\triangle }|^{\circ },^{\diamondsuit })=%
\mathbf{S}_{\triangle }\oplus \widetilde{S_{\circ }}^{\prime
}\oplus
\mathcal{\widetilde{S}}_{|}^{\diamondsuit },  \nonumber \\
&&................................  \nonumber
\end{eqnarray}
Considering the operation of dualisation of prime components in
(\ref{2.65}) we can generate different isomorphic variants of
distinguished (n,m)--spi\-nor spa\-ces.

We define a d--spinor space \index{d--spinor}
$\mathcal{S}_{(n,m)}\ $ as a direct sum of a horizontal and a
vertical spinor spaces of type (\ref{2.65}), for instance,
\begin{eqnarray*}
\mathcal{S}_{(8k,8k^{\prime })} &=&\mathbf{S}_{\circ }\oplus \mathbf{S}%
_{\circ }^{\prime }\oplus \mathbf{S}_{|\circ }\oplus
\mathbf{S}_{|\circ }^{\prime },\mathcal{S}_{(8k,8k^{\prime }+1)}\
=\mathbf{S}_{\circ }\oplus
\mathbf{S}_{\circ }^{\prime }\oplus \mathcal{S}_{|\circ }^{(-)},..., \\
\mathcal{S}_{(8k+4,8k^{\prime }+5)} &=&\mathbf{S}_{\triangle }\oplus \mathbf{%
S}_{\triangle }^{\prime }\oplus \mathcal{S}_{|\triangle
}^{(-)},...
\end{eqnarray*}
The scalar product on a $\mathcal{S}_{(n,m)}\ $ is induced by
(corresponding to fixed values of $n$ and $m$ ) $\epsilon
$-objects (\ref{2.60}) considered for h- and v-components.

Having introduced d-spinors for dimensions $\left( n,m\right) $
we can write out the generalization for locally anisotropic
spaces of twistor \index{twistor} equations \cite{penr1} by using
the distinguished $\sigma $-objects (\ref{2.55}):
\begin{equation}
(\sigma _{(\widehat{\alpha }})_{|\underline{\beta }|}^{..\underline{\gamma }%
}\quad \frac{\delta \omega ^{\underline{\beta }}}{\delta u^{\widehat{\beta }%
)}}=\frac 1{n+m}\quad G_{\widehat{\alpha }\widehat{\beta }}(\sigma ^{%
\widehat{\epsilon }})_{\underline{\beta }}^{..\underline{\gamma
}}\quad \frac{\delta \omega ^{\underline{\beta }}}{\delta
u^{\widehat{\epsilon }}}, \label{2.66}
\end{equation}
where $\left| \underline{\beta }\right| $ denotes that we do not
consider symmetrization on this index. The general solution of
(\ref{2.66}) on the d-vector space $\mathcal{F}$ looks like as
\begin{equation}
\omega ^{\underline{\beta }}=\Omega ^{\underline{\beta
}}+u^{\widehat{\alpha
}}(\sigma _{\widehat{\alpha }})_{\underline{\epsilon }}^{..\underline{\beta }%
}\Pi ^{\underline{\epsilon }},  \label{2.67}
\end{equation}
where $\Omega ^{\underline{\beta }}$ and $\Pi
^{\underline{\epsilon }}$ are constant d-spinors. For fixed
values of dimensions $n$ and $m$ we mast analyze the reduced and
irreducible components of h- and v-parts of equations
(\ref{2.66}) and their solutions (\ref{2.67}) in order to find the
symmetry properties of a d-twistor \index{d--twistor}
$\mathbf{Z^\alpha \ }$ defined as a pair of d--spinors
\[
\mathbf{Z}^\alpha =(\omega ^{\underline{\alpha }},\pi _{\underline{\beta }%
}^{\prime }),
\]
where $\pi _{\underline{\beta }^{\prime }}=\pi _{\underline{\beta
}^{\prime }}^{(0)}\in {\widetilde{\mathcal{S}}}_{(n,m)}$ is a
constant dual d-spinor. The problem of definition of spinors and
twistors on locally anisotropic spaces was firstly considered in
\cite{vdeb} (see also \cite{v87}) in connection with the
possibility to extend the equations (\ref{2.66}) and their
solutions (\ref{2.67}), by using nearly autoparallel maps, on
curved, locally isotropic or anisotropic, spaces.

\section{ Mutual Transforms of Tensors and Spi\-nors}

The spinor algebra for spaces of higher dimensions can not be
considered as a real alternative to the tensor algebra as for
locally isotropic spaces of dimensions $n=3,4$
\cite{pen,penr1,penr2}. The same holds true for locally
anisotropic spaces and we emphasize that it is not quite
convenient to perform a spinor calculus for dimensions $n,m>>4$.
Nevertheless, the concept of spinors is important for every type
of spaces, we can deeply understand the fundamental properties of
geometical objects on locally anisotropic spaces, and we shall
consider in this subsection some questions concerning transforms
of d-tensor objects into d-spinor ones.

\subsection{ Transformation of d-tensors into d-spinors}

In order to pass from d-tensors to d-spinors we must use $\sigma $-objects (%
\ref{2.55}) written in reduced or irreduced form \quad (in
dependence of fixed values of dimensions $n$ and $m$ ):

\begin{equation}
(\sigma _{\widehat{\alpha }})_{\underline{\beta }}^{\cdot \underline{\gamma }%
},~(\sigma ^{\widehat{\alpha }})^{\underline{\beta }\underline{\gamma }%
},~(\sigma ^{\widehat{\alpha }})_{\underline{\beta }\underline{\gamma }%
},...,(\sigma _{\widehat{a}})^{\underline{b}\underline{c}},...,(\sigma _{%
\widehat{i}})_{\underline{j}\underline{k}},...,(\sigma _{\widehat{a}%
})^{AA^{\prime }},...,(\sigma ^{\widehat{i}})_{II^{\prime }},....
\label{2.68}
\end{equation}
It is obvious that contracting with corresponding $\sigma
$-objects (\ref {2.68}) we can introduce instead of d-tensors
indices the d-spinor ones, for instance,
\[
\omega ^{\underline{\beta }\underline{\gamma }}=(\sigma ^{\widehat{\alpha }%
})^{\underline{\beta }\underline{\gamma }}\omega
_{\widehat{\alpha }},\quad
\omega _{AB^{\prime }}=(\sigma ^{\widehat{a}})_{AB^{\prime }}\omega _{%
\widehat{a}},\quad ...,\zeta _{\cdot
\underline{j}}^{\underline{i}}=(\sigma
^{\widehat{k}})_{\cdot \underline{j}}^{\underline{i}}\zeta _{\widehat{k}%
},....
\]
For d-tensors containing groups of antisymmetric indices there is
a more simple procedure of theirs transforming into d-spinors
because the objects
\begin{equation}
(\sigma _{\widehat{\alpha }\widehat{\beta }...\widehat{\gamma }})^{%
\underline{\delta }\underline{\nu }},\quad (\sigma ^{\widehat{a}\widehat{b}%
...\widehat{c}})^{\underline{d}\underline{e}},\quad ...,(\sigma ^{\widehat{i}%
\widehat{j}...\widehat{k}})_{II^{\prime }},\quad ...  \label{2.69}
\end{equation}
can be used for sets of such indices into pairs of d-spinor
indices. Let us enumerate some properties of $\sigma $-objects of
type (\ref{2.69}) (for
simplicity we consider only h-components having q indices $\widehat{i},%
\widehat{j},\widehat{k},...$ taking values from 1 to $n;$ the
properties of
v-components can be written in a similar manner with respect to indices $%
\widehat{a},\widehat{b},\widehat{c}...$ taking values from 1 to
$m$):
\begin{equation}
(\sigma _{\widehat{i}...\widehat{j}})^{\underline{k}\underline{l}}%
\mbox{
 is\ }\left\{ \
\begin{array}{c}
\mbox{symmetric on }\underline{k},\underline{l}\mbox{ for
}n-2q\equiv
1,7~(mod~8); \\
\mbox{antisymmetric on }\underline{k},\underline{l}\mbox{ for
}n-2q\equiv 3,5~(mod~8)
\end{array}
\right\}  \label{2.70}
\end{equation}
for odd values of $n,$ and an object
\[
(\sigma _{\widehat{i}...\widehat{j}})^{IJ}~\left( (\sigma _{\widehat{i}...%
\widehat{j}})^{I^{\prime }J^{\prime }}\right)
\]
\begin{equation}
\mbox{ is\ }\left\{
\begin{array}{c}
\mbox{symmetric on }I,J~(I^{\prime },J^{\prime })\mbox{ for
}n-2q\equiv
0~(mod~8); \\
\mbox{antisymmetric on }I,J~(I^{\prime },J^{\prime })\mbox{ for
}n-2q\equiv 4~(mod~8)
\end{array}
\right\}  \label{2.71}
\end{equation}
or
\begin{equation}
(\sigma _{\widehat{i}...\widehat{j}})^{IJ^{\prime }}=\pm (\sigma _{\widehat{i%
}...\widehat{j}})^{J^{\prime }I}\{
\begin{array}{c}
n+2q\equiv 6(mod8); \\
n+2q\equiv 2(mod8),
\end{array}
\label{2.72}
\end{equation}
with vanishing of the rest of reduced components of the d-tensor $(\sigma _{%
\widehat{i}...\widehat{j}})^{\underline{k}\underline{l}}$ with
prime/unprime sets of indices.

\subsection{ Fundamental d--spinors}

We can transform every d-spinor $\xi ^{\underline{\alpha }}=\left( \xi ^{%
\underline{i}},\xi ^{\underline{a}}\right) $ into a corresponding
d-tensor.
For simplicity, we consider this construction only for a h-component $\xi ^{%
\underline{i}}$ on a h-space being of dimension $n$. The values
\begin{equation}
\xi ^{\underline{\alpha }}\xi ^{\underline{\beta }}(\sigma ^{\widehat{i}...%
\widehat{j}})_{\underline{\alpha }\underline{\beta }}\quad \left( n%
\mbox{ is odd}\right)  \label{2.73}
\end{equation}
or
\begin{equation}
\xi ^I\xi ^J(\sigma ^{\widehat{i}...\widehat{j}})_{IJ}~\left(
\mbox{or }\xi
^{I^{\prime }}\xi ^{J^{\prime }}(\sigma ^{\widehat{i}...\widehat{j}%
})_{I^{\prime }J^{\prime }}\right) ~\left( n\mbox{ is even}\right)
\label{2.74}
\end{equation}
with a different number of indices $\widehat{i}...\widehat{j},$
taken together, defines the h-spinor $\xi ^{\underline{i}}\,$ to
an accuracy to the sign. We emphasize that it is necessary to
choose only those h-components of d-tensors (\ref{2.73}) (or
(\ref{2.74})) which are symmetric
on pairs of indices $\underline{\alpha }\underline{\beta }$ (or $IJ\,$ (or $%
I^{\prime }J^{\prime }$ )) and the number $q$ of indices $\widehat{i}...%
\widehat{j}$ satisfies the condition (as a respective consequence
of the properties (\ref{2.70}) and/or (\ref{2.71}), (\ref{2.72}))
\begin{equation}
n-2q\equiv 0,1,7~(mod~8).  \label{2.75}
\end{equation}
Of special interest is the case when
\begin{equation}
q=\frac 12\left( n\pm 1\right) ~\left( n\mbox{ is odd}\right)
\label{2.76}
\end{equation}
or
\begin{equation}
q=\frac 12n~\left( n\mbox{ is even}\right) .  \label{2.77}
\end{equation}
If all expressions (\ref{2.73}) and/or (\ref{2.74}) are zero for
all values of $q\,$ with the exception of one or two ones defined
by the condition (\ref {2.76}) (or (\ref{2.77})), the value $\xi
^{\widehat{i}}$ (or $\xi ^I$ ($\xi ^{I^{\prime }}))$ is called a
fundamental h-spinor. Defining in a similar manner the
fundamental v-spinors we can introduce fundamental d-spinors as
pairs of fundamental h- and v-spinors. Here we remark that a h(v)-spinor $%
\xi ^{\widehat{i}}~(\xi ^{\widehat{a}})\,$ (we can also consider
reduced components) is always a fundamental one for $n(m)<7,$
which is a consequence of (\ref{2.75})).

Finally, in this section, we note that the geometry of
fundamental h- and v-spinors is similar to that of usual
fundamental spinors (see Appendix to the monograph \cite{penr2}).
We omit such details in this work, but emphasize that
constructions with fundamental d-spinors, for a locally
anisotropic space, must be adapted to the corresponding global
splitting by N-connection of the space.

\section{Anisotropic Spinor Differential Geometry}

The goal of the section is to formulate the differential geometry
of d-spinors for locally anisotropic spaces.

We shall use denotations of type
\[
v^\alpha =(v^i,v^a)\in {\mathcal{\sigma }^\alpha }=({\mathcal{\sigma }^i,%
\mathcal{\sigma }^a})\,\mbox{ and }\zeta ^{\underline{\alpha }}=(\zeta ^{%
\underline{i}},\zeta ^{\underline{a}})\in {\mathcal{\sigma }^{\underline{%
\alpha }}}=({\mathcal{\sigma }^{\underline{i}},\mathcal{\sigma }^{\underline{%
a}}})\,
\]
for, respectively, elements of modules of d-vector and irreduced
d-spinor fields (see details in \cite{vjmp}). D-tensors and
d-spinor tensors
(irreduced or reduced) will be interpreted as elements of corresponding $%
\mathcal{\sigma }$ -modules, for instance,
\[
q_{~\beta ...}^\alpha \in \mathcal{\sigma ^\alpha {}_\beta },\psi _{~%
\underline{\beta }\quad ...}^{\underline{\alpha }\quad \underline{\gamma }%
}\in \mathcal{\sigma }_{~\underline{\beta }\quad ...}^{\underline{\alpha }%
\quad \underline{\gamma }}~,\xi _{\quad JK^{\prime }N^{\prime
}}^{II^{\prime }}\in \mathcal{\sigma }_{\quad JK^{\prime
}N^{\prime }}^{II^{\prime }}~,...
\]

We can establish a correspondence between the d--metric $g_{\alpha \beta }$ (%
\ref{dmetric}) and d-spinor metric $\epsilon _{\underline{\alpha }\underline{%
\beta }}$ ( $\epsilon $-objects (\ref{2.60}) for both h- and v-subspaces of $%
\mathcal{E\,}$ ) of a locally anisotropic space $\mathcal{E}$ by
using the relation
\begin{equation}
g_{\alpha \beta }=-\frac 1{N(n)+N(m)}((\sigma _{(\alpha }(u))^{\underline{%
\alpha }_1\underline{\beta }_1}(\sigma _{\beta )}(u))^{\underline{\beta }_2%
\underline{\alpha }_2})\epsilon _{\underline{\alpha }_1\underline{\alpha }%
_2}\epsilon _{\underline{\beta }_1\underline{\beta }_2},
\label{2.78}
\end{equation}
where
\begin{equation}
(\sigma _\alpha (u))^{\underline{\nu }\underline{\gamma }}=l_\alpha ^{%
\widehat{\alpha }}(u)(\sigma _{\widehat{\alpha }})^{\underline{\nu }%
\underline{\gamma }},  \label{2.79}
\end{equation}
which is a consequence of formulas (\ref{2.52})-(\ref{2.57}). In
brief we can write (\ref{2.78}) as
\begin{equation}
g_{\alpha \beta }=\epsilon _{\underline{\alpha }_1\underline{\alpha }%
_2}\epsilon _{\underline{\beta }_1\underline{\beta }_2}
\label{2.80}
\end{equation}
if the $\sigma $-objects are considered as a fixed structure, whereas $%
\epsilon $-objects are treated as caring the metric ''dynamics ''
, on locally anisotropic space--times. This variant is used, for
instance, in the so-called 2-spinor geometry \cite{penr1,penr2}
and should be preferred if we have to make explicit the algebraic
symmetry properties of d-spinor objects.
An alternative way is to consider as fixed the algebraic structure of $%
\epsilon $-objects and to use variable components of $\sigma
$-objects of type (\ref{2.79}) for developing a variational
d-spinor approach to gravitational and matter field interactions
on locally anisotropic spaces ( the spinor Ashtekar variables
\cite{ash} are introduced in this manner).

We note that a d--spinor metric \index{d--spinor metric}
\[
\epsilon _{\underline{\nu }\underline{\tau }}=\left(
\begin{array}{cc}
\epsilon _{\underline{i}\underline{j}} & 0 \\
0 & \epsilon _{\underline{a}\underline{b}}
\end{array}
\right)
\]
on the d-spinor space
$\mathcal{S}=(\mathcal{S}_{(h)},\mathcal{S}_{(v)})$
can have symmetric or antisymmetric h (v) -components $\epsilon _{\underline{%
i}\underline{j}}$ ($\epsilon _{\underline{a}\underline{b}})$ ,
see $\epsilon $-objects (\ref{2.60}). For simplicity, in this
section (in order to avoid
cumbersome calculations connected with eight-fold periodicity on dimensions $%
n$ and $m$ of a locally anisotropic space $\mathcal{E\ }$) we
shall develop
a general d-spinor formalism only by using irreduced spinor spaces $\mathcal{%
S}_{(h)}$ and $\mathcal{S}_{(v)}.$

\section{ D-covariant derivation}

Let $\mathcal{E}$ be a locally anisotropic space. We define the
action on a d-spinor of a d-covariant operator
\[
\bigtriangledown _\alpha =\left( \bigtriangledown
_i,\bigtriangledown
_a\right) =(\sigma _\alpha )^{\underline{\alpha }_1\underline{\alpha }%
_2}\bigtriangledown _{^{\underline{\alpha }_1\underline{\alpha
}_2}}=\left(
(\sigma _i)^{\underline{i}_1\underline{i}_2}\bigtriangledown _{^{\underline{i%
}_1\underline{i}_2}},~(\sigma _a)^{\underline{a}_1\underline{a}%
_2}\bigtriangledown _{^{\underline{a}_1\underline{a}_2}}\right)
\]
(in brief, we shall write
\[
\bigtriangledown _\alpha =\bigtriangledown _{^{\underline{\alpha }_1%
\underline{\alpha }_2}}=\left( \bigtriangledown _{^{\underline{i}_1%
\underline{i}_2}},~\bigtriangledown _{^{\underline{a}_1\underline{a}%
_2}}\right) )
\]
as a map
\[
\bigtriangledown _{{\underline{\alpha }}_1{\underline{\alpha
}}_2}\ :\
\mathcal{\sigma }^{\underline{\beta }}\rightarrow \sigma _\alpha ^{%
\underline{\beta }}=\sigma _{{\underline{\alpha }}_1{\underline{\alpha }}%
_2}^{\underline{\beta }}
\]
satisfying conditions
\[
\bigtriangledown _\alpha (\xi ^{\underline{\beta }}+\eta ^{\underline{\beta }%
})=\bigtriangledown _\alpha \xi ^{\underline{\beta
}}+\bigtriangledown _\alpha \eta ^{\underline{\beta }},
\]
and
\[
\bigtriangledown _\alpha (f\xi ^{\underline{\beta
}})=f\bigtriangledown _\alpha \xi ^{\underline{\beta }}+\xi
^{\underline{\beta }}\bigtriangledown _\alpha f
\]
for every $\xi ^{\underline{\beta }},\eta ^{\underline{\beta }}\in \mathcal{%
\sigma ^{\underline{\beta }}}$ and $f$ being a scalar field on $\mathcal{E}$.%
$\mathcal{\ }$ It is also required that one holds the Leibnitz
rule
\[
(\bigtriangledown _\alpha \zeta _{\underline{\beta }})\eta ^{\underline{%
\beta }}=\bigtriangledown _\alpha (\zeta _{\underline{\beta }}\eta ^{%
\underline{\beta }})-\zeta _{\underline{\beta }}\bigtriangledown
_\alpha \eta ^{\underline{\beta }}
\]
and that $\bigtriangledown _\alpha \,$ is a real operator, i.e.
it commuters with the operation of complex conjugation:
\[
\overline{\bigtriangledown _\alpha \psi _{\underline{\alpha }\underline{%
\beta }\underline{\gamma }...}}=\bigtriangledown _\alpha (\overline{\psi }_{%
\underline{\alpha }\underline{\beta }\underline{\gamma }...}).
\]

Let now analyze the question on uniqueness of action on d-spinors
of an operator $\bigtriangledown _\alpha $ satisfying necessary
conditions . Denoting by $\bigtriangledown _\alpha ^{(1)}$ and
$\bigtriangledown _\alpha $ two such d-covariant operators we
consider the map
\begin{equation}
(\bigtriangledown _\alpha ^{(1)}-\bigtriangledown _\alpha
):\mathcal{\sigma
^{\underline{\beta }}\rightarrow \sigma _{\underline{\alpha }_1\underline{%
\alpha }_2}^{\underline{\beta }}}.  \label{2.81}
\end{equation}
Because the action on a scalar $f$ of both operators
$\bigtriangledown _\alpha ^{(1)}$ and $\bigtriangledown _\alpha $
must be identical, i.e.
\begin{equation}
\bigtriangledown _\alpha ^{(1)}f=\bigtriangledown _\alpha f,
\label{2.82}
\end{equation}
the action (\ref{2.81}) on $f=\omega _{\underline{\beta }}\xi ^{\underline{%
\beta }}$ must be written as
\[
(\bigtriangledown _\alpha ^{(1)}-\bigtriangledown _\alpha )(\omega _{%
\underline{\beta }}\xi ^{\underline{\beta }})=0.
\]
In consequence we conclude that there is an element $\Theta _{\underline{%
\alpha }_1\underline{\alpha }_2\underline{\beta }}^{\quad \quad \underline{%
\gamma }}\in \mathcal{\sigma }_{\underline{\alpha }_1\underline{\alpha }_2%
\underline{\beta }}^{\quad \quad \underline{\gamma }}$ for which
\begin{equation}
\bigtriangledown _{\underline{\alpha }_1\underline{\alpha }_2}^{(1)}\xi ^{%
\underline{\gamma }}=\bigtriangledown _{\underline{\alpha }_1\underline{%
\alpha }_2}\xi ^{\underline{\gamma }}+\Theta _{\underline{\alpha }_1%
\underline{\alpha }_2\underline{\beta }}^{\quad \quad \underline{\gamma }%
}\xi ^{\underline{\beta }}  \label{2.83}
\end{equation}
and
\[
\bigtriangledown _{\underline{\alpha }_1\underline{\alpha }_2}^{(1)}\omega _{%
\underline{\beta }}=\bigtriangledown _{\underline{\alpha }_1\underline{%
\alpha }_2}\omega _{\underline{\beta }}-\Theta _{\underline{\alpha }_1%
\underline{\alpha }_2\underline{\beta }}^{\quad \quad \underline{\gamma }%
}\omega _{\underline{\gamma }}~.
\]
The action of the operator (\ref{2.81}) on a d-vector $v^\beta =v^{%
\underline{\beta }_1\underline{\beta }_2}$ can be written by using formula (%
\ref{2.83}) for both indices $\underline{\beta }_1$ and $\underline{\beta }%
_2 $ :
\begin{eqnarray*}
(\bigtriangledown _\alpha ^{(1)}-\bigtriangledown _\alpha )v^{\underline{%
\beta }_1\underline{\beta }_2} &=&\Theta _{\alpha \underline{\gamma }%
}^{\quad \underline{\beta }_1}v^{\underline{\gamma }\underline{\beta }%
_2}+\Theta _{\alpha \underline{\gamma }}^{\quad \underline{\beta }_2}v^{%
\underline{\beta }_1\underline{\gamma }} \\
&=&(\Theta _{\alpha \underline{\gamma }_1}^{\quad \underline{\beta }%
_1}\delta _{\underline{\gamma }_2}^{\quad \underline{\beta
}_2}+\Theta
_{\alpha \underline{\gamma }_1}^{\quad \underline{\beta }_2}\delta _{%
\underline{\gamma }_2}^{\quad \underline{\beta }_1})v^{\underline{\gamma }_1%
\underline{\gamma }_2}=Q_{\ \alpha \gamma }^\beta v^\gamma ,
\end{eqnarray*}
where
\begin{equation}
Q_{\ \alpha \gamma }^\beta =Q_{\qquad \underline{\alpha
}_1\underline{\alpha
}_2~\underline{\gamma }_1\underline{\gamma }_2}^{\underline{\beta }_1%
\underline{\beta }_2}=\Theta _{\alpha \underline{\gamma
}_1}^{\quad \underline{\beta }_1}\delta _{\underline{\gamma
}_2}^{\quad \underline{\beta
}_2}+\Theta _{\alpha \underline{\gamma }_1}^{\quad \underline{\beta }%
_2}\delta _{\underline{\gamma }_2}^{\quad \underline{\beta }_1}.
\label{2.84}
\end{equation}
The d-commutator $\bigtriangledown _{[\alpha }\bigtriangledown
_{\beta ]}$ defines the d--torsion. So, applying operators
$\bigtriangledown _{[\alpha }^{(1)}\bigtriangledown _{\beta
]}^{(1)}$ and $\bigtriangledown _{[\alpha
}\bigtriangledown _{\beta ]}$ on $f=\omega _{\underline{\beta }}\xi ^{%
\underline{\beta }}$ we can write
\[
T_{\quad \alpha \beta }^{(1)\gamma }-T_{~\alpha \beta }^\gamma
=Q_{~\beta \alpha }^\gamma -Q_{~\alpha \beta }^\gamma
\]
with $Q_{~\alpha \beta }^\gamma $ from (\ref{2.84}).

The action of operator $\bigtriangledown _\alpha ^{(1)}$ on
d-spinor tensors
of type $\chi _{\underline{\alpha }_1\underline{\alpha }_2\underline{\alpha }%
_3...}^{\qquad \quad \underline{\beta }_1\underline{\beta
}_2...}$ must be
constructed by using formula (\ref{2.83}) for every upper index $\underline{%
\beta }_1\underline{\beta }_2...$ and formula (\ref{2.84}) for
every lower index $\underline{\alpha }_1\underline{\alpha
}_2\underline{\alpha }_3...$ .

\section{Infeld - van der Waerden coefficients}

Let
\[
\delta _{\underline{\mathbf{\alpha }}}^{\quad \underline{\alpha
}}=\left(
\delta _{\underline{\mathbf{1}}}^{\quad \underline{i}},\delta _{\underline{%
\mathbf{2}}}^{\quad \underline{i}},...,\delta _{\underline{\mathbf{N(n)}}%
}^{\quad \underline{i}},\delta _{\underline{\mathbf{1}}}^{\quad \underline{a}%
},\delta _{\underline{\mathbf{2}}}^{\quad \underline{a}},...,\delta _{%
\underline{\mathbf{N(m)}}}^{\quad \underline{i}}\right)
\]
be a d-spinor basis. The dual to it basis is denoted as
\[
\delta _{\underline{\alpha }}^{\quad \underline{\mathbf{\alpha
}}}=\left(
\delta _{\underline{i}}^{\quad \underline{\mathbf{1}}},\delta _{\underline{i}%
}^{\quad \underline{\mathbf{2}}},...,\delta
_{\underline{i}}^{\quad
\underline{\mathbf{N(n)}}},\delta _{\underline{i}}^{\quad \underline{\mathbf{%
1}}},\delta _{\underline{i}}^{\quad \underline{\mathbf{2}}},...,\delta _{%
\underline{i}}^{\quad \underline{\mathbf{N(m)}}}\right) .
\]
A d-spinor $\kappa ^{\underline{\alpha }}\in \mathcal{\sigma }$ $^{%
\underline{\alpha }}$ has components $\kappa ^{\underline{\mathbf{\alpha }}%
}=\kappa ^{\underline{\alpha }}\delta _{\underline{\alpha
}}^{\quad \underline{\mathbf{\alpha }}}.$ Taking into account that
\[
\delta _{\underline{\mathbf{\alpha }}}^{\quad \underline{\alpha }}\delta _{%
\underline{\mathbf{\beta }}}^{\quad \underline{\beta }}\bigtriangledown _{%
\underline{\alpha }\underline{\beta }}=\bigtriangledown _{\underline{\mathbf{%
\alpha }}\underline{\mathbf{\beta }}},
\]
we write out the components $\bigtriangledown _{\underline{\alpha }%
\underline{\beta }}$ $\kappa ^{\underline{\gamma }}$ as
\begin{eqnarray}
\delta _{\underline{\mathbf{\alpha }}}^{\quad \underline{\alpha }}~\delta _{%
\underline{\mathbf{\beta }}}^{\quad \underline{\beta }}~\delta _{\underline{%
\gamma }}^{\quad \underline{\mathbf{\gamma }}}~\bigtriangledown _{\underline{%
\alpha }\underline{\beta }}\kappa ^{\underline{\gamma }} &=&\delta _{%
\underline{\mathbf{\epsilon }}}^{\quad \underline{\tau }}~\delta _{%
\underline{\tau }}^{\quad \underline{\mathbf{\gamma }}}~\bigtriangledown _{%
\underline{\mathbf{\alpha }}\underline{\mathbf{\beta }}}\kappa ^{\underline{%
\mathbf{\epsilon }}}+\kappa ^{\underline{\mathbf{\epsilon }}}~\delta _{%
\underline{\epsilon }}^{\quad \underline{\mathbf{\gamma
}}}~\bigtriangledown
_{\underline{\mathbf{\alpha }}\underline{\mathbf{\beta }}}\delta _{%
\underline{\mathbf{\epsilon }}}^{\quad \underline{\epsilon }}  \nonumber \\
&=&\bigtriangledown _{\underline{\mathbf{\alpha }}\underline{\mathbf{\beta }}%
}\kappa ^{\underline{\mathbf{\gamma }}}+\kappa
^{\underline{\mathbf{\epsilon
}}}\gamma _{~\underline{\mathbf{\alpha }}\underline{\mathbf{\beta }}%
\underline{\mathbf{\epsilon }}}^{\underline{\mathbf{\gamma }}},
\label{2.85}
\end{eqnarray}
where the coordinate components of the d-spinor connection $\gamma _{~%
\underline{\mathbf{\alpha }}\underline{\mathbf{\beta }}\underline{\mathbf{%
\epsilon }}}^{\underline{\mathbf{\gamma }}}$ are defined as
\begin{equation}
\gamma _{~\underline{\mathbf{\alpha }}\underline{\mathbf{\beta }}\underline{%
\mathbf{\epsilon }}}^{\underline{\mathbf{\gamma }}}\doteq \delta _{%
\underline{\tau }}^{\quad \underline{\mathbf{\gamma }}}~\bigtriangledown _{%
\underline{\mathbf{\alpha }}\underline{\mathbf{\beta }}}\delta _{\underline{%
\mathbf{\epsilon }}}^{\quad \underline{\tau }}.  \label{2.86}
\end{equation}
We call the Infeld - van der Waerden d-symbols a set of $\sigma $-objects ($%
\sigma _{\mathbf{\alpha }})^{\underline{\mathbf{\alpha }}\underline{\mathbf{%
\beta }}}$ paramet\-ri\-zed with respect to a coordinate d-spinor
basis. Defining
\[
\bigtriangledown _{\mathbf{\alpha }}=(\sigma _{\mathbf{\alpha }})^{%
\underline{\mathbf{\alpha }}\underline{\mathbf{\beta }}}~\bigtriangledown _{%
\underline{\mathbf{\alpha }}\underline{\mathbf{\beta }}},
\]
introducing denotations
\[
\gamma ^{\underline{\mathbf{\gamma }}}{}_{\mathbf{\alpha \underline{\tau }}%
}\doteq \gamma ^{\underline{\mathbf{\gamma }}}{}_{\mathbf{\underline{\alpha }%
\underline{\beta }\underline{\tau }}}~(\sigma _{\mathbf{\alpha }})^{%
\underline{\mathbf{\alpha }}\underline{\mathbf{\beta }}}
\]
and using properties (\ref{2.85}) we can write relations
\begin{equation}
l_{\mathbf{\alpha }}^\alpha ~\delta _{\underline{\beta }}^{\quad \underline{%
\mathbf{\beta }}}~\bigtriangledown _\alpha \kappa ^{\underline{\beta }%
}=\bigtriangledown _{\mathbf{\alpha }}\kappa ^{\underline{\mathbf{\beta }}%
}+\kappa ^{\underline{\mathbf{\delta }}}~\gamma _{~\mathbf{\alpha }%
\underline{\mathbf{\delta }}}^{\underline{\mathbf{\beta }}}
\label{2.87}
\end{equation}
and
\begin{equation}
l_{\mathbf{\alpha }}^\alpha ~\delta _{\underline{\mathbf{\beta
}}}^{\quad
\underline{\beta }}~\bigtriangledown _\alpha ~\mu _{\underline{\beta }%
}=\bigtriangledown _{\mathbf{\alpha }}~\mu _{\underline{\mathbf{\beta }}%
}-\mu _{\underline{\mathbf{\delta }}}\gamma _{~\mathbf{\alpha }\underline{%
\mathbf{\beta }}}^{\underline{\mathbf{\delta }}}  \label{2.88}
\end{equation}
for d-covariant derivations $~\bigtriangledown
_{\underline{\alpha }}\kappa
^{\underline{\beta }}$ and $\bigtriangledown _{\underline{\alpha }}~\mu _{%
\underline{\beta }}.$

We can consider expressions similar to (\ref{2.87}) and
(\ref{2.88}) for values having both types of d-spinor and
d-tensor indices, for instance,
\[
l_{\mathbf{\alpha }}^\alpha ~l_\gamma ^{\mathbf{\gamma }}~\delta _{%
\underline{\mathbf{\delta }}}^{\quad \underline{\delta
}}~\bigtriangledown
_\alpha \theta _{\underline{\delta }}^{~\gamma }=\bigtriangledown _{\mathbf{%
\alpha }}\theta _{\underline{\mathbf{\delta }}}^{~\mathbf{\gamma }}-\theta _{%
\underline{\mathbf{\epsilon }}}^{~\mathbf{\gamma }}\gamma _{~\underline{%
\mathbf{\alpha }}\underline{\mathbf{\delta }}}^{\underline{\mathbf{\epsilon }%
}}+\theta _{\underline{\mathbf{\delta }}}^{~\mathbf{\tau
}}~\Gamma _{\quad \mathbf{\alpha \tau }}^{~\mathbf{\gamma }}
\]
(we can prove this by a straightforward calculation of the
derivation
\newline
$\bigtriangledown _\alpha (\theta _{\underline{\mathbf{\epsilon }}}^{~%
\mathbf{\tau }}$ $~\delta _{\underline{\delta }}^{\quad \underline{\mathbf{%
\epsilon }}}~l_{\mathbf{\tau }}^\gamma )).$

Now we shall consider some possible relations between components
of
d-connec\-ti\-ons $\gamma _{~\underline{\mathbf{\alpha }}\underline{\mathbf{%
\delta }}}^{\underline{\mathbf{\epsilon }}}$ and $\Gamma _{\quad \mathbf{%
\alpha \tau }}^{~\mathbf{\gamma }}$ and derivations of $(\sigma _{\mathbf{%
\alpha }})^{\underline{\mathbf{\alpha }}\underline{\mathbf{\beta
}}}$ . We can write
\begin{eqnarray*}
\Gamma _{~\mathbf{\beta \gamma }}^{\mathbf{\alpha }} &=&l_\alpha ^{\mathbf{%
\alpha }}\bigtriangledown _{\mathbf{\gamma }}l_{\mathbf{\beta
}}^\alpha
=l_\alpha ^{\mathbf{\alpha }}\bigtriangledown _{\mathbf{\gamma }}(\sigma _{%
\mathbf{\beta }})^{\underline{\epsilon }\underline{\tau }}=l_\alpha ^{%
\mathbf{\alpha }}\bigtriangledown _{\mathbf{\gamma }}((\sigma _{\mathbf{%
\beta }})^{\underline{\mathbf{\epsilon }}\underline{\mathbf{\tau }}}\delta _{%
\underline{\mathbf{\epsilon }}}^{~\underline{\epsilon }}\delta _{\underline{%
\mathbf{\tau }}}^{~\underline{\tau }}) \\
&=&l_\alpha ^{\mathbf{\alpha }}\delta _{\underline{\mathbf{\alpha }}}^{~%
\underline{\alpha }}\delta _{\underline{\mathbf{\epsilon }}}^{~\underline{%
\epsilon }}\bigtriangledown _{\mathbf{\gamma }}(\sigma _{\mathbf{\beta }})^{%
\underline{\mathbf{\alpha }}\underline{\mathbf{\epsilon }}}+l_\alpha ^{%
\mathbf{\alpha }}(\sigma _{\mathbf{\beta }})^{\underline{\mathbf{\epsilon }}%
\underline{\mathbf{\tau }}}(\delta _{\underline{\mathbf{\tau }}}^{~%
\underline{\tau }}\bigtriangledown _{\mathbf{\gamma }}\delta _{\underline{%
\mathbf{\epsilon }}}^{~\underline{\epsilon }}+\delta _{\underline{\mathbf{%
\epsilon }}}^{~\underline{\epsilon }}\bigtriangledown _{\mathbf{\gamma }%
}\delta _{\underline{\mathbf{\tau }}}^{~\underline{\tau }}) \\
&=&l_{\underline{\mathbf{\epsilon }}\underline{\mathbf{\tau }}}^{\mathbf{%
\alpha }}~\bigtriangledown _{\mathbf{\gamma }}(\sigma _{\mathbf{\beta }})^{%
\underline{\mathbf{\epsilon }}\underline{\mathbf{\tau }}}+l_{\underline{%
\mathbf{\mu }}\underline{\mathbf{\nu }}}^{\mathbf{\alpha }}\delta _{%
\underline{\epsilon }}^{~\underline{\mathbf{\mu }}}\delta _{\underline{\tau }%
}^{~\underline{\mathbf{\nu }}}(\sigma _{\mathbf{\beta }})^{\underline{%
\epsilon }\underline{\tau }}(\delta _{\underline{\mathbf{\tau }}}^{~%
\underline{\tau }}\bigtriangledown _{\mathbf{\gamma }}\delta _{\underline{%
\mathbf{\epsilon }}}^{~\underline{\epsilon }}+\delta _{\underline{\mathbf{%
\epsilon }}}^{~\underline{\epsilon }}\bigtriangledown _{\mathbf{\gamma }%
}\delta _{\underline{\mathbf{\tau }}}^{~\underline{\tau }}),
\end{eqnarray*}
where $l_\alpha ^{\mathbf{\alpha }}=(\sigma _{\underline{\mathbf{\epsilon }}%
\underline{\mathbf{\tau }}})^{\mathbf{\alpha }}$ , from which it
follows
\[
(\sigma _{\mathbf{\alpha }})^{\underline{\mathbf{\mu }}\underline{\mathbf{%
\nu }}}(\sigma _{\underline{\mathbf{\alpha }}\underline{\mathbf{\beta }}})^{%
\mathbf{\beta }}\Gamma _{~\mathbf{\gamma \beta }}^{\mathbf{\alpha
}}=(\sigma
_{\underline{\mathbf{\alpha }}\underline{\mathbf{\beta }}})^{\mathbf{\beta }%
}\bigtriangledown _{\mathbf{\gamma }}(\sigma _{\mathbf{\alpha }})^{%
\underline{\mathbf{\mu }}\underline{\mathbf{\nu }}}+\delta _{\underline{%
\mathbf{\beta }}}^{~\underline{\mathbf{\nu }}}\gamma
_{~\mathbf{\gamma
\underline{\alpha }}}^{\underline{\mathbf{\mu }}}+\delta _{\underline{%
\mathbf{\alpha }}}^{~\underline{\mathbf{\mu }}}\gamma
_{~\mathbf{\gamma \underline{\beta }}}^{\underline{\mathbf{\nu
}}}.
\]
Connecting the last expression on \underline{$\mathbf{\beta }$}
and \underline{$\mathbf{\nu }$} and using an orthonormalized
d-spinor basis when
$\gamma _{~\mathbf{\gamma \underline{\beta }}}^{\underline{\mathbf{\beta }}%
}=0$ (a consequence from (\ref{2.86})) we have
\begin{equation}
\gamma _{~\mathbf{\gamma \underline{\alpha }}}^{\underline{\mathbf{\mu }}%
}=\frac 1{N(n)+N(m)}(\Gamma _{\quad \mathbf{\gamma ~\underline{\alpha }%
\underline{\beta }}}^{\underline{\mathbf{\mu }}\underline{\mathbf{\beta }}%
}-(\sigma _{\underline{\mathbf{\alpha }}\underline{\mathbf{\beta }}})^{%
\mathbf{\beta }}\bigtriangledown _{\mathbf{\gamma }}(\sigma _{\mathbf{\beta }%
})^{\underline{\mathbf{\mu }}\underline{\mathbf{\beta }}}),
\label{2.89}
\end{equation}
where
\begin{equation}
\Gamma _{\quad \mathbf{\gamma ~\underline{\alpha }\underline{\beta }}}^{%
\underline{\mathbf{\mu }}\underline{\mathbf{\beta }}}=(\sigma _{\mathbf{%
\alpha }})^{\underline{\mathbf{\mu }}\underline{\mathbf{\beta }}}(\sigma _{%
\underline{\mathbf{\alpha }}\underline{\mathbf{\beta }}})^{\mathbf{\beta }%
}\Gamma _{~\mathbf{\gamma \beta }}^{\mathbf{\alpha }}.
\label{2.90}
\end{equation}
We also note here that, for instance, for the canonical and
Berwald connections, Christoffel d-symbols we can express
d-spinor connection (\ref {2.90}) through corresponding locally
adapted derivations of components of metric and N-connection by
introducing respectively the coefficients of the Barwald,
canonical or another type d--connections.

\section{ D-spinors of Anisotropic Curvature and Torsion}

The d-tensor indices of the commutator $\Delta _{\alpha \beta }$
can be transformed into d-spinor ones:
\begin{equation}
\Box _{\underline{\alpha }\underline{\beta }}=(\sigma ^{\alpha \beta })_{%
\underline{\alpha }\underline{\beta }}\Delta _{\alpha \beta }=(\Box _{%
\underline{i}\underline{j}},\Box _{\underline{a}\underline{b}}),
\label{2.91}
\end{equation}
with h- and v-components,
\[
\Box _{\underline{i}\underline{j}}=(\sigma ^{\alpha \beta })_{\underline{i}%
\underline{j}}\Delta _{\alpha \beta }\mbox{ and }\Box _{\underline{a}%
\underline{b}}=(\sigma ^{\alpha \beta
})_{\underline{a}\underline{b}}\Delta _{\alpha \beta },
\]
being symmetric or antisymmetric in dependence of corresponding
values of dimensions $n\,$ and $m$ (see eight-fold
parametizations (\ref{2.69})--(\ref {2.71})). Considering the
actions of operator (\ref{2.91}) on d-spinors $\pi
^{\underline{\gamma }}$ and $\mu _{\underline{\gamma }}$ we
introduce the
d-spinor curvature $X_{\underline{\delta }\quad \underline{\alpha }%
\underline{\beta }}^{\quad \underline{\gamma }}\,$ as to satisfy
equations
\begin{equation}
\Box _{\underline{\alpha }\underline{\beta }}\ \pi ^{\underline{\gamma }}=X_{%
\underline{\delta }\quad \underline{\alpha }\underline{\beta
}}^{\quad \underline{\gamma }}\pi ^{\underline{\delta }}
\label{2.92}
\end{equation}
and
\[
\Box _{\underline{\alpha }\underline{\beta }}\ \mu _{\underline{\gamma }}=X_{%
\underline{\gamma }\quad \underline{\alpha }\underline{\beta
}}^{\quad \underline{\delta }}\mu _{\underline{\delta }}.
\]
The gravitational d-spinor $\Psi _{\underline{\alpha }\underline{\beta }%
\underline{\gamma }\underline{\delta }}$ is defined by a
corresponding symmetrization of d-spinor indices:
\begin{equation}
\Psi _{\underline{\alpha }\underline{\beta }\underline{\gamma }\underline{%
\delta }}=X_{(\underline{\alpha }|\underline{\beta }|\underline{\gamma }%
\underline{\delta })}.  \label{2.93}
\end{equation}
We note that d-spinor tensors $X_{\underline{\delta }\quad
\underline{\alpha
}\underline{\beta }}^{\quad \underline{\gamma }}$ and $\Psi _{\underline{%
\alpha }\underline{\beta }\underline{\gamma }\underline{\delta
}}\,$ are transformed into similar 2-spinor objects on locally
isotropic spaces \cite {penr1,penr2} if we consider vanishing of
the N-connection structure and a limit to a locally isotropic
space.

Putting $\delta _{\underline{\gamma }}^{\quad
\mathbf{\underline{\gamma }}}$
instead of $\mu _{\underline{\gamma }}$ in (\ref{2.92}) and using (\ref{2.93}%
) we can express respectively the curvature and gravitational
d-spinors as
\[
X_{\underline{\gamma }\underline{\delta }\underline{\alpha
}\underline{\beta
}}=\delta _{\underline{\delta }\underline{\mathbf{\tau }}}\Box _{\underline{%
\alpha }\underline{\beta }}\delta _{\underline{\gamma }}^{\quad \mathbf{%
\underline{\tau }}}
\]
and
\[
\Psi _{\underline{\gamma }\underline{\delta }\underline{\alpha }\underline{%
\beta }}=\delta _{\underline{\delta }\underline{\mathbf{\tau }}}\Box _{(%
\underline{\alpha }\underline{\beta }}\delta _{\underline{\gamma
})}^{\quad \mathbf{\underline{\tau }}}.
\]

The d-spinor torsion $T_{\qquad \underline{\alpha }\underline{\beta }}^{%
\underline{\gamma }_1\underline{\gamma }_2}$ is defined similarly
as for d-tensors) by using the d-spinor commutator (\ref{2.91})
and equations
\[
\Box _{\underline{\alpha }\underline{\beta }}f=T_{\qquad \underline{\alpha }%
\underline{\beta }}^{\underline{\gamma }_1\underline{\gamma }%
_2}\bigtriangledown _{\underline{\gamma }_1\underline{\gamma
}_2}f.
\]

The d-spinor components $R_{\underline{\gamma
}_1\underline{\gamma }_2\qquad
\underline{\alpha }\underline{\beta }}^{\qquad \underline{\delta }_1%
\underline{\delta }_2}$ of the curvature d-tensor $R_{\gamma
\quad \alpha
\beta }^{\quad \delta }$ can be computed by using the relations (\ref{2.90}%
), (\ref{2.91}) and (\ref{2.93}) as to satisfy the equations (the
d-spinor analogous of equations (\ref{2.37}) )
\begin{equation}
(\Box _{\underline{\alpha }\underline{\beta }}-T_{\qquad \underline{\alpha }%
\underline{\beta }}^{\underline{\gamma }_1\underline{\gamma }%
_2}\bigtriangledown _{\underline{\gamma }_1\underline{\gamma }_2})V^{%
\underline{\delta }_1\underline{\delta }_2}=R_{\underline{\gamma }_1%
\underline{\gamma }_2\qquad \underline{\alpha }\underline{\beta
}}^{\qquad
\underline{\delta }_1\underline{\delta }_2}V^{\underline{\gamma }_1%
\underline{\gamma }_2},  \label{2.94}
\end{equation}
here d-vector $V^{\underline{\gamma }_1\underline{\gamma }_2}$ is
considered
as a product of d-spinors, i.e. $V^{\underline{\gamma }_1\underline{\gamma }%
_2}=\nu ^{\underline{\gamma }_1}\mu ^{\underline{\gamma }_2}$. We
find

\begin{eqnarray}
R_{\underline{\gamma }_1\underline{\gamma }_2\qquad \underline{\alpha }%
\underline{\beta }}^{\qquad \underline{\delta
}_1\underline{\delta }_2}
&=&\left( X_{\underline{\gamma }_1~\underline{\alpha }\underline{\beta }%
}^{\quad \underline{\delta }_1}+T_{\qquad \underline{\alpha }\underline{%
\beta }}^{\underline{\tau }_1\underline{\tau }_2}\quad \gamma
_{\quad
\underline{\tau }_1\underline{\tau }_2\underline{\gamma }_1}^{\underline{%
\delta }_1}\right) \delta _{\underline{\gamma }_2}^{\quad \underline{\delta }%
_2}  \label{2.95} \\
&&+\left( X_{\underline{\gamma }_2~\underline{\alpha }\underline{\beta }%
}^{\quad \underline{\delta }_2}+T_{\qquad \underline{\alpha }\underline{%
\beta }}^{\underline{\tau }_1\underline{\tau }_2}\quad \gamma
_{\quad
\underline{\tau }_1\underline{\tau }_2\underline{\gamma }_2}^{\underline{%
\delta }_2}\right) \delta _{\underline{\gamma }_1}^{\quad \underline{\delta }%
_1}.  \nonumber
\end{eqnarray}

It is convenient to use this d-spinor expression for the
curvature d-tensor
\begin{eqnarray*}
R_{\underline{\gamma }_1\underline{\gamma }_2\qquad \underline{\alpha }_1%
\underline{\alpha }_2\underline{\beta }_1\underline{\beta
}_2}^{\qquad
\underline{\delta }_1\underline{\delta }_2} &=&\left( X_{\underline{\gamma }%
_1~\underline{\alpha }_1\underline{\alpha }_2\underline{\beta }_1\underline{%
\beta }_2}^{\quad \underline{\delta }_1}+T_{\qquad \underline{\alpha }_1%
\underline{\alpha }_2\underline{\beta }_1\underline{\beta }_2}^{\underline{%
\tau }_1\underline{\tau }_2}~\gamma _{\quad \underline{\tau }_1\underline{%
\tau }_2\underline{\gamma }_1}^{\underline{\delta }_1}\right) \delta _{%
\underline{\gamma }_2}^{\quad \underline{\delta }_2} \\
&&+\left( X_{\underline{\gamma }_2~\underline{\alpha }_1\underline{\alpha }_2%
\underline{\beta }_1\underline{\beta }_2}^{\quad \underline{\delta }%
_2}+T_{\qquad \underline{\alpha }_1\underline{\alpha }_2\underline{\beta }_1%
\underline{\beta }_2~}^{\underline{\tau }_1\underline{\tau
}_2}\gamma
_{\quad \underline{\tau }_1\underline{\tau }_2\underline{\gamma }_2}^{%
\underline{\delta }_2}\right) \delta _{\underline{\gamma
}_1}^{\quad \underline{\delta }_1}
\end{eqnarray*}
in order to get the d-spinor components of the Ricci d-tensor
\begin{eqnarray}
R_{\underline{\gamma }_1\underline{\gamma }_2\underline{\alpha }_1\underline{%
\alpha }_2} &=&R_{\underline{\gamma }_1\underline{\gamma }_2\qquad
\underline{\alpha }_1\underline{\alpha }_2\underline{\delta }_1\underline{%
\delta }_2}^{\qquad \underline{\delta }_1\underline{\delta }_2}=X_{%
\underline{\gamma }_1~\underline{\alpha }_1\underline{\alpha }_2\underline{%
\delta }_1\underline{\gamma }_2}^{\quad \underline{\delta
}_1}+T_{\qquad
\underline{\alpha }_1\underline{\alpha }_2\underline{\delta }_1\underline{%
\gamma }_2}^{\underline{\tau }_1\underline{\tau }_2}~\gamma
_{\quad
\underline{\tau }_1\underline{\tau }_2\underline{\gamma }_1}^{\underline{%
\delta }_1}  \nonumber \\
&&+X_{\underline{\gamma }_2~\underline{\alpha }_1\underline{\alpha }_2%
\underline{\delta }_1\underline{\gamma }_2}^{\quad \underline{\delta }%
_2}+T_{\qquad \underline{\alpha }_1\underline{\alpha }_2\underline{\gamma }_1%
\underline{\delta }_2~}^{\underline{\tau }_1\underline{\tau
}_2}\gamma
_{\quad \underline{\tau }_1\underline{\tau }_2\underline{\gamma }_2}^{%
\underline{\delta }_2}  \label{2.96}
\end{eqnarray}
and this d-spinor decomposition of the scalar curvature:
\begin{eqnarray}
q\overleftarrow{R} &=&R_{\qquad \underline{\alpha }_1\underline{\alpha }_2}^{%
\underline{\alpha }_1\underline{\alpha }_2}=X_{\quad ~\underline{~\alpha }%
_1\quad \underline{\delta }_1\underline{\alpha }_2}^{\underline{\alpha }_1%
\underline{\delta }_1~~\underline{\alpha }_2}+T_{\qquad ~~\underline{\alpha }%
_2\underline{\delta }_1}^{\underline{\tau }_1\underline{\tau }_2\underline{%
\alpha }_1\quad ~\underline{\alpha }_2}~\gamma _{\quad \underline{\tau }_1%
\underline{\tau }_2\underline{\alpha }_1}^{\underline{\delta }_1}
\label{2.97} \\
&&+X_{\qquad \quad \underline{\alpha }_2\underline{\delta }_2\underline{%
\alpha }_1}^{\underline{\alpha }_2\underline{\delta }_2\underline{\alpha }%
_1}+T_{\qquad \underline{\alpha }_1\quad ~\underline{\delta }_2~}^{%
\underline{\tau }_1\underline{\tau }_2~~\underline{\alpha }_2\underline{%
\alpha }_1}\gamma _{\quad \underline{\tau }_1\underline{\tau }_2\underline{%
\alpha }_2}^{\underline{\delta }_2}.  \nonumber
\end{eqnarray}

Putting (2.96) and (2.97) into (2.43) and, correspondingly,
(2.41) we find the d-spinor components of the Einstein and $\Phi
_{\alpha \beta }$ d-tensors:
\begin{eqnarray}
\overleftarrow{G}_{\gamma \alpha } &=&\overleftarrow{G}_{\underline{\gamma }%
_1\underline{\gamma }_2\underline{\alpha }_1\underline{\alpha }_2}=X_{%
\underline{\gamma }_1~\underline{\alpha }_1\underline{\alpha }_2\underline{%
\delta }_1\underline{\gamma }_2}^{\quad \underline{\delta
}_1}+T_{\qquad
\underline{\alpha }_1\underline{\alpha }_2\underline{\delta }_1\underline{%
\gamma }_2}^{\underline{\tau }_1\underline{\tau }_2}~\gamma
_{\quad
\underline{\tau }_1\underline{\tau }_2\underline{\gamma }_1}^{\underline{%
\delta }_1}  \label{2.98} \\
&&+X_{\underline{\gamma }_2~\underline{\alpha }_1\underline{\alpha }_2%
\underline{\delta }_1\underline{\gamma }_2}^{\quad \underline{\delta }%
_2}+T_{\qquad \underline{\alpha }_1\underline{\alpha }_2\underline{\gamma }_1%
\underline{\delta }_2~}^{\underline{\tau }_1\underline{\tau
}_2}\gamma
_{\quad \underline{\tau }_1\underline{\tau }_2\underline{\gamma }_2}^{%
\underline{\delta }_2}  \nonumber \\
&&-\frac 12\varepsilon _{\underline{\gamma }_1\underline{\alpha }%
_1}\varepsilon _{\underline{\gamma }_2\underline{\alpha }_2}[X_{\quad ~%
\underline{~\beta }_1\quad \underline{\mu }_1\underline{\beta }_2}^{%
\underline{\beta }_1\underline{\mu }_1~~\underline{\beta }_2}+T_{\qquad ~~%
\underline{\beta }_2\underline{\mu }_1}^{\underline{\tau }_1\underline{\tau }%
_2\underline{\beta }_1\quad ~\underline{\beta }_2}~\gamma _{\quad \underline{%
\tau }_1\underline{\tau }_2\underline{\beta }_1}^{\underline{\mu
}_1}
\nonumber \\
&&+X_{\qquad \quad \underline{\beta }_2\underline{\mu }_2\underline{\nu }%
_1}^{\underline{\beta }_2\underline{\mu }_2\underline{\nu
}_1}+T_{\qquad
\underline{\beta }_1\quad ~\underline{\delta }_2~}^{\underline{\tau }_1%
\underline{\tau }_2~~\underline{\beta }_2\underline{\beta
}_1}\gamma _{\quad
\underline{\tau }_1\underline{\tau }_2\underline{\beta }_2}^{\underline{%
\delta }_2}]  \nonumber
\end{eqnarray}
and
\begin{eqnarray}
\Phi _{\gamma \alpha } &=&\Phi _{\underline{\gamma }_1\underline{\gamma }_2%
\underline{\alpha }_1\underline{\alpha }_2}=\frac 1{2(n+m)}\varepsilon _{%
\underline{\gamma }_1\underline{\alpha }_1}\varepsilon _{\underline{\gamma }%
_2\underline{\alpha }_2}[X_{\quad ~\underline{~\beta }_1\quad
\underline{\mu
}_1\underline{\beta }_2}^{\underline{\beta }_1\underline{\mu }_1~~\underline{%
\beta }_2}  \label{2.99} \\
&& +T_{\qquad ~~\underline{\beta }_2\underline{\mu }_1}^{\underline{\tau }_1%
\underline{\tau }_2\underline{\beta }_1\quad ~\underline{\beta
}_2}~\gamma
_{\quad \underline{\tau }_1\underline{\tau }_2\underline{\beta }_1}^{%
\underline{\mu }_1} +X_{\qquad \quad \underline{\beta }_2\underline{\mu }_2%
\underline{\nu }_1}^{\underline{\beta }_2\underline{\mu }_2\underline{\nu }%
_1}+T_{\qquad \underline{\beta }_1\quad ~\underline{\delta }_2~}^{\underline{%
\tau }_1\underline{\tau }_2~~\underline{\beta }_2\underline{\beta
}_1}\gamma
_{\quad \underline{\tau }_1\underline{\tau }_2\underline{\beta }_2}^{%
\underline{\delta }_2}]  \nonumber \\
&&-\frac 12[X_{\underline{\gamma }_1~\underline{\alpha }_1\underline{\alpha }%
_2\underline{\delta }_1\underline{\gamma }_2}^{\quad \underline{\delta }%
_1}+T_{\qquad \underline{\alpha }_1\underline{\alpha }_2\underline{\delta }_1%
\underline{\gamma }_2}^{\underline{\tau }_1\underline{\tau
}_2}~\gamma
_{\quad \underline{\tau }_1\underline{\tau }_2\underline{\gamma }_1}^{%
\underline{\delta }_1}  \nonumber \\
&&+X_{\underline{\gamma }_2~\underline{\alpha }_1\underline{\alpha }_2%
\underline{\delta }_1\underline{\gamma }_2}^{\quad \underline{\delta }%
_2}+T_{\qquad \underline{\alpha }_1\underline{\alpha }_2\underline{\gamma }_1%
\underline{\delta }_2~}^{\underline{\tau }_1\underline{\tau
}_2}\gamma
_{\quad \underline{\tau }_1\underline{\tau }_2\underline{\gamma }_2}^{%
\underline{\delta }_2}].  \nonumber
\end{eqnarray}

The components of the conformal Weyl d-spinor can be computed by
putting d-spinor values of the curvature (\ref{2.95}) and Ricci
(\ref{2.95}) d-tensors into corresponding expression for the
d-tensor (\ref{2.40}). We omit this calculus in this work.

\chapter{Anisotropic Spinors and Field Equations}

The problem of formulation gravitational and gauge field
equations on different types of locally anisotropic spaces is
considered, for instance, in \cite{ma94,bej,asa88} and \cite{vg}.
In this section we shall introduce the basic field equations for
gravitational and matter field locally anisotropic interactions
in a generalized form for generic locally anisotropic spaces.

\section{ Anisotropic Scalar Field Interactions}

Let $\varphi \left( u\right) =(\varphi _1\left( u\right) ,\varphi
_2\left( u\right) \dot{,}...,\varphi _k\left( u\right) )$ be a
complex k-component scalar field of mass $\mu $ on locally
anisotropic space $\mathcal{E}$. The d-covariant generalization
of the conformally invariant (in the massless case) scalar field
equation \cite{penr1,penr2} can be defined by using the
d'Alambert locally anisotropic operator \cite{ana94,vst96} $\Box
=D^\alpha D_\alpha $, where $D_\alpha $ is a d-covariant
derivation on $\mathcal{E}$ constructed, for simplicity, by using
Christoffel d--simbols (all formulas for field equations and
conservation values can be deformed by using corresponding
defrormation d--tensors $P_{\beta \gamma }^\alpha $ from the
Cristoffel d--simbols, or the canonical d--connection to a general
d-connection into consideration):

\begin{equation}
(\Box +\frac{n+m-2}{4(n+m-1)}\overleftarrow{R}+\mu ^2)\varphi
\left( u\right) =0.  \label{2.100}
\end{equation}
We must change d-covariant derivation $D_\alpha $ into
$^{\diamond }D_\alpha =D_\alpha +ieA_\alpha $ and take into
account the d-vector current
\[
J_\alpha ^{(0)}\left( u\right) =i(\left( \overline{\varphi
}\left( u\right) D_\alpha \varphi \left( u\right) -D_\alpha
\overline{\varphi }\left( u\right) )\varphi \left( u\right)
\right)
\]
if interactions between locally anisotropic electromagnetic field
( d-vector potential $A_\alpha $ ), where $e$ is the
electromagnetic constant, and charged scalar field $\varphi $ are
considered. The equations (\ref{2.100}) are (locally adapted to
the N-connection structure) Euler equations for the Lagrangian
\begin{equation}
\mathcal{L}^{(0)}\left( u\right) =\sqrt{|g|}\left[ g^{\alpha
\beta }\delta _\alpha \overline{\varphi }\left( u\right) \delta
_\beta \varphi \left(
u\right) -\left( \mu ^2+\frac{n+m-2}{4(n+m-1)}\right) \overline{\varphi }%
\left( u\right) \varphi \left( u\right) \right] ,  \label{2.101}
\end{equation}
where $|g|=detg_{\alpha \beta }.$

The locally adapted variations of the action with Lagrangian
(\ref{2.101}) on variables $\varphi \left( u\right) $ and
$\overline{\varphi }\left( u\right) $ leads to the locally
anisotropic generalization of the energy-momentum tensor,
\begin{equation}
E_{\alpha \beta }^{(0,can)}\left( u\right) =\delta _\alpha
\overline{\varphi }\left( u\right) \delta _\beta \varphi \left(
u\right) +\delta _\beta \overline{\varphi }\left( u\right) \delta
_\alpha \varphi \left( u\right) -\frac 1{\sqrt{|g|}}g_{\alpha
\beta }\mathcal{L}^{(0)}\left( u\right) , \label{2.102}
\end{equation}
and a similar variation on the components of a d-metric
(\ref{dmetric}) leads to a symmetric metric energy-momentum
d-tensor,
\begin{equation}
E_{\alpha \beta }^{(0)}\left( u\right) =E_{(\alpha \beta
)}^{(0,can)}\left( u\right) -\frac{n+m-2}{2(n+m-1)}\left[
R_{(\alpha \beta )}+D_{(\alpha }D_{\beta )}-g_{\alpha \beta }\Box
\right] \overline{\varphi }\left( u\right) \varphi \left(
u\right) .  \label{2.103}
\end{equation}
Here we note that we can obtain a nonsymmetric energy-momentum
d-tensor if we firstly vary on $G_{\alpha \beta }$ and than
impose the constraints of compatibility with the N-connection
structure. We also conclude that the existence of a N-connection
in v-bundle $\mathcal{E}$ results in a nonequivalence of
energy-momentum d-tensors (\ref{2.102}) and (\ref{2.103}),
nonsymmetry of the Ricci tensor (see (\ref{2.33})), nonvanishing
of the d-covariant derivation of the Einstein d-tensor
(\ref{2.43}), $D_\alpha \overleftarrow{G}^{\alpha \beta }\neq 0$
and, in consequence, a corresponding breaking of conservation
laws on locally anisotropic spaces when $D_\alpha E^{\alpha \beta
}\neq 0\,$ \cite{ma87,ma94}. The problem of formulation of
conservation laws on locally anisotropic spaces is discussed in
details and two variants of its solution (by using nearly
autoparallel maps and tensor integral formalism on locally
anisotropic multispaces) are proposed in \cite{vst96}. In this
section we shall present only straightforward generalizations of
field equations and necessary formulas for energy-momentum
d-tensors of matter fields on $\mathcal{E}$ considering that it
is naturally that the conservation laws (usually being
consequences of global, local and/or intrinsic symmetries of the
fundamental space-time and of the type of field interactions)
have to be broken on locally anisotropic spaces.

\section{ Anisotropic Proca equations} \index{Proca}

Let consider a d-vector field $\varphi _\alpha \left( u\right) $ with mass $%
\mu ^2$ (locally anisotropic Proca field ) interacting with
exterior locally anisotropic gravitational field. From the
Lagrangian
\begin{equation}
\mathcal{L}^{(1)}\left( u\right) =\sqrt{\left| g\right| }\left[ -\frac 12{%
\overline{f}}_{\alpha \beta }\left( u\right) f^{\alpha \beta
}\left( u\right) +\mu ^2{\overline{\varphi }}_\alpha \left(
u\right) \varphi ^\alpha \left( u\right) \right] ,  \label{2.104}
\end{equation}
where $f_{\alpha \beta }=D_\alpha \varphi _\beta -D_\beta \varphi
_\alpha ,$ one follows the Proca equations on locally anisotropic
spaces
\begin{equation}
D_\alpha f^{\alpha \beta }\left( u\right) +\mu ^2\varphi ^\beta
\left( u\right) =0.  \label{2.105}
\end{equation}
Equations (\ref{2.105}) are a first type constraints for $\beta
=0.$ Acting with $D_\alpha $ on (\ref{2.105}), for $\mu \neq 0$
we obtain second type constraints
\begin{equation}
D_\alpha \varphi ^\alpha \left( u\right) =0.  \label{2.106}
\end{equation}

Putting (\ref{2.106}) into (\ref{2.105}) we obtain second order
field equations with respect to $\varphi _\alpha $ :
\begin{equation}
\Box \varphi _\alpha \left( u\right) +R_{\alpha \beta }\varphi
^\beta \left( u\right) +\mu ^2\varphi _\alpha \left( u\right)
=0.  \label{2.107}
\end{equation}
The energy-momentum d-tensor and d-vector current following from
the (\ref {2.107}) can be written as
\[
E_{\alpha \beta }^{(1)}\left( u\right) =-g^{\varepsilon \tau }\left( {%
\overline{f}}_{\beta \tau }f_{\alpha \varepsilon
}+{\overline{f}}_{\alpha
\varepsilon }f_{\beta \tau }\right) +\mu ^2\left( {\overline{\varphi }}%
_\alpha \varphi _\beta +{\overline{\varphi }}_\beta \varphi _\alpha \right) -%
\frac{g_{\alpha \beta }}{\sqrt{\left| g\right|
}}\mathcal{L}^{(1)}\left( u\right) .
\]
and
\[
J_\alpha ^{\left( 1\right) }\left( u\right) =i\left(
{\overline{f}}_{\alpha
\beta }\left( u\right) \varphi ^\beta \left( u\right) -{\overline{\varphi }}%
^\beta \left( u\right) f_{\alpha \beta }\left( u\right) \right) .
\]

For $\mu =0$ the d-tensor $f_{\alpha \beta }$ and the Lagrangian (\ref{2.104}%
) are invariant with respect to locally anisotropic gauge
transforms of type
\[
\varphi _\alpha \left( u\right) \rightarrow \varphi _\alpha
\left( u\right) +\delta _\alpha \Lambda \left( u\right) ,
\]
where $\Lambda \left( u\right) $ is a d-differentiable scalar
function, and we obtain a locally anisot\-rop\-ic variant of
Maxwell theory.

\section{ Anisotropic Gravitons and Backgrounds}

Let a massless d-tensor field $h_{\alpha \beta }\left( u\right) $
is interpreted as a small perturbation of the locally anisotropic
background metric d-field $g_{\alpha \beta }\left( u\right) .$
Considering, for simplicity, a torsionless background we have
locally anisotropic Fierz-Pauli equations
\begin{equation}
\Box h_{\alpha \beta }\left( u\right) +2R_{\tau \alpha \beta \nu
}\left( u\right) ~h^{\tau \nu }\left( u\right) =0  \label{2.108}
\end{equation}
and d-gauge conditions
\begin{equation}
D_\alpha h_\beta ^\alpha \left( u\right) =0,\quad h\left(
u\right) \equiv h_\beta ^\alpha (u)=0,  \label{2.109}
\end{equation}
where $R_{\tau \alpha \beta \nu }\left( u\right) $ is curvature
d-tensor of the locally anisotropic background space (these
formulae can be obtained by using a perturbation formalism with
respect to $h_{\alpha \beta }\left( u\right) $ developed in
\cite{gri}; in our case we must take into account the
distinguishing of geometrical objects and operators on locally
anisotropic spaces).

We note that we can rewrite d-tensor formulas
(\ref{2.100})-(\ref{2.109})
into similar d-spinor ones by using formulas (\ref{2.78})--(\ref{2.80}), (%
\ref{2.90}), (\ref{2.92}) and (\ref{2.96})--(\ref{2.105}) (for
simplicity, we omit these considerations in this work).

\section{ Anisotropic Dirac Equations} \index{Dirac equations}

Let denote the Dirac d-spinor field on $\mathcal{E}$ as $\psi
\left( u\right) =\left( \psi ^{\underline{\alpha }}\left(
u\right) \right) $ and consider as the generalized Lorentz
transforms the group of automorphysm of the metric
$G_{\widehat{\alpha }\widehat{\beta }}$ (see the locally
anisotropic frame decomposition of d-metric (\ref{2.54})). The
d-covariant derivation of field $\psi \left( u\right) $ is
written as
\begin{equation}
\overrightarrow{\bigtriangledown _\alpha }\psi =\left[ \delta
_\alpha +\frac 14C_{\widehat{\alpha }\widehat{\beta
}\widehat{\gamma }}\left( u\right)
~l_\alpha ^{\widehat{\alpha }}\left( u\right) \sigma ^{\widehat{\beta }%
}\sigma ^{\widehat{\gamma }}\right] \psi ,  \label{2.110}
\end{equation}
where coefficients $C_{\widehat{\alpha }\widehat{\beta }\widehat{\gamma }%
}=\left( D_\gamma l_{\widehat{\alpha }}^\alpha \right) l_{\widehat{\beta }%
\alpha }l_{\widehat{\gamma }}^\gamma $ generalize for locally
anisotropic spaces the corresponding Ricci coefficients on
Riemannian spaces \cite{foc}. Using $\sigma $-objects $\sigma
^\alpha \left( u\right) $ (see (\ref{2.79}) and (\ref{2.55})) we
define the Dirac equations on locally anisotropic spaces:
\begin{equation}
(i\sigma ^\alpha \left( u\right) \overrightarrow{\bigtriangledown _\alpha }%
-\mu )\psi =0,  \label{2.111}
\end{equation}
which are the Euler equations for the Lagrangian
\begin{eqnarray}
\mathcal{L}^{(1/2)}\left( u\right) &=&\sqrt{\left| g\right|
}\{[\psi
^{+}\left( u\right) \sigma ^\alpha \left( u\right) \overrightarrow{%
\bigtriangledown _\alpha }\psi \left( u\right)  \label{2.112} \\
&&-(\overrightarrow{\bigtriangledown _\alpha }\psi ^{+}\left(
u\right) )\sigma ^\alpha \left( u\right) \psi \left( u\right)
]-\mu \psi ^{+}\left( u\right) \psi \left( u\right) \},  \nonumber
\end{eqnarray}
where $\psi ^{+}\left( u\right) $ is the complex conjugation and
transposition of the column $\psi \left( u\right).$

From (\ref{2.112}) we obtain the d-metric energy-momentum d-tensor
\begin{eqnarray*}
E_{\alpha \beta }^{(1/2)}\left( u\right) &=&\frac i4[\psi
^{+}\left( u\right) \sigma _\alpha \left( u\right)
\overrightarrow{\bigtriangledown _\beta }\psi \left( u\right)
+\psi ^{+}\left( u\right) \sigma _\beta \left(
u\right) \overrightarrow{\bigtriangledown _\alpha }\psi \left( u\right) \\
&&-(\overrightarrow{\bigtriangledown _\alpha }\psi ^{+}\left(
u\right)
)\sigma _\beta \left( u\right) \psi \left( u\right) -(\overrightarrow{%
\bigtriangledown _\beta }\psi ^{+}\left( u\right) )\sigma _\alpha
\left( u\right) \psi \left( u\right) ]
\end{eqnarray*}
and the d-vector source
\[
J_\alpha ^{(1/2)}\left( u\right) =\psi ^{+}\left( u\right) \sigma
_\alpha \left( u\right) \psi \left( u\right) .
\]
We emphasize that locally anisotropic interactions with exterior
gauge fields can be introduced by changing the locally
anisotropic partial derivation from (\ref{2.110}) in this manner:
\begin{equation}
\delta _\alpha \rightarrow \delta _\alpha +ie^{\star }B_\alpha ,
\label{2.113}
\end{equation}
where $e^{\star }$ and $B_\alpha $ are respectively the constant
d-vector potential of locally anisotropic gauge interactions on
locally anisotropic spaces (see \cite{vg} and the next section).

\section{Yang-Mills Equations in Anisotropic Spi\-nor Form}

We consider a v-bundle $\mathcal{B}_E,~\pi
_B:\mathcal{B\rightarrow E,}$ on locally anisotropic space
$\mathcal{E}$.$\mathcal{\,}$ Additionally to d-tensor and
d-spinor indices we shall use capital Greek letters, $\Phi
,\Upsilon ,\Xi ,\Psi ,...$ for fibre (of this bundle) indices
(see details in \cite{penr1,penr2} for the case when the base
space of the v-bundle $\pi
_B$ is a locally isotropic space-time). Let $\underline{\bigtriangledown }%
_\alpha $ be, for simplicity, a torsionless, linear connection in $\mathcal{B%
}_E$ satisfying conditions:
\begin{eqnarray*}
\underline{\bigtriangledown }_\alpha &:&{\ \Upsilon }^\Theta
\rightarrow {\ \Upsilon }_\alpha ^\Theta \quad \left[ \mbox{or
}{\ \Xi }^\Theta \rightarrow
{\ \Xi }_\alpha ^\Theta \right] , \\
\underline{\bigtriangledown }_\alpha \left( \lambda ^\Theta +\nu
^\Theta
\right) &=&\underline{\bigtriangledown }_\alpha \lambda ^\Theta +\underline{%
\bigtriangledown }_\alpha \nu ^\Theta , \\
\underline{\bigtriangledown }_\alpha ~(f\lambda ^\Theta )
&=&\lambda ^\Theta \underline{\bigtriangledown }_\alpha
f+f\underline{\bigtriangledown }_\alpha \lambda ^\Theta ,\quad
f\in {\ \Upsilon }^\Theta ~[\mbox{or }{\ \Xi }^\Theta ],
\end{eqnarray*}
where by ${\ \Upsilon }^\Theta ~\left( {\ \Xi }^\Theta \right) $
we denote the module of sections of the real (complex) v-bundle
$\mathcal{B}_E$
provided with the abstract index $\Theta .$ The curvature of connection $%
\underline{\bigtriangledown }_\alpha $ is defined as
\[
K_{\alpha \beta \Omega }^{\qquad \Theta }\lambda ^\Omega =\left( \underline{%
\bigtriangledown }_\alpha \underline{\bigtriangledown }_\beta -\underline{%
\bigtriangledown }_\beta \underline{\bigtriangledown }_\alpha
\right) \lambda ^\Theta .
\]

For Yang-Mills fields \index{Yang--Mills}  as a rule one considers
that $\mathcal{B}_E$ is enabled with a unitary (complex) structure
(complex conjugation changes mutually the upper and lower Greek
indices). It is useful to introduce
instead of $K_{\alpha \beta \Omega }^{\qquad \Theta }$ a Hermitian matrix $%
F_{\alpha \beta \Omega }^{\qquad \Theta }=i$ $K_{\alpha \beta
\Omega }^{\qquad \Theta } $ connected with components of the
Yang-Mills d-vector potential $B_{\alpha \Xi }^{\quad \Phi }$
according the formula:

\begin{equation}
\frac 12F_{\alpha \beta \Xi }^{\qquad \Phi }=\underline{\bigtriangledown }%
_{[\alpha }B_{\beta ]\Xi }^{\quad \Phi }-iB_{[\alpha |\Lambda
|}^{\quad \Phi }B_{\beta ]\Xi }^{\quad \Lambda },  \label{2.114}
\end{equation}
where the locally anisotropic space indices commute with capital
Greek indices. The gauge transforms are written in the form:

\begin{eqnarray*}
B_{\alpha \Theta }^{\quad \Phi } &\mapsto &B_{\alpha \widehat{\Theta }%
}^{\quad \widehat{\Phi }}=B_{\alpha \Theta }^{\quad \Phi }~s_\Phi
^{\quad \widehat{\Phi }}~q_{\widehat{\Theta }}^{\quad \Theta
}+is_\Theta ^{\quad
\widehat{\Phi }}\underline{\bigtriangledown }_\alpha ~q_{\widehat{\Theta }%
}^{\quad \Theta }, \\
F_{\alpha \beta \Xi }^{\qquad \Phi } &\mapsto &F_{\alpha \beta \widehat{\Xi }%
}^{\qquad \widehat{\Phi }}=F_{\alpha \beta \Xi }^{\qquad \Phi
}s_\Phi ^{\quad \widehat{\Phi }}q_{\widehat{\Xi }}^{\quad \Xi },
\end{eqnarray*}
where matrices $s_\Phi ^{\quad \widehat{\Phi }}$ and $q_{\widehat{\Xi }%
}^{\quad \Xi }$ are mutually inverse (Hermitian conjugated in the
unitary case). The Yang-Mills equations on torsionless locally
anisotropic spaces \cite{vg} (see details in the next Chapter)
are written in this form:
\begin{eqnarray}
\underline{\bigtriangledown }^\alpha F_{\alpha \beta \Theta
}^{\qquad \Psi }
&=&J_{\beta \ \Theta }^{\qquad \Psi },  \label{2.115} \\
\underline{\bigtriangledown }_{[\alpha }F_{\beta \gamma ]\Theta
}^{\qquad \Xi } &=&0.  \label{2.116}
\end{eqnarray}
We must introduce deformations of connection of type,\newline
$\underline{\bigtriangledown }_\alpha ^{\star }~\longrightarrow \underline{%
\bigtriangledown }_\alpha +P_\alpha ,$ (the deformation d-tensor
$P_\alpha $ is induced by the torsion in v-bundle
$\mathcal{B}_E)$ into the definition of the curvature of locally
anisotropic gauge fields (\ref{2.114}) and motion equations
(\ref{2.115}) and (\ref{2.116}) if interactions are modeled on a
generic locally anisotropic space.

Now we can write out the field equations of the Einstein-Cartan
theory in the d-spinor form. So, for the Einstein equations
(\ref{2.42}) we have

\[
\overleftarrow{G}_{\underline{\gamma }_1\underline{\gamma }_2\underline{%
\alpha }_1\underline{\alpha }_2}+\lambda \varepsilon _{\underline{\gamma }_1%
\underline{\alpha }_1}\varepsilon _{\underline{\gamma }_2\underline{\alpha }%
_2}=\kappa E_{\underline{\gamma }_1\underline{\gamma }_2\underline{\alpha }_1%
\underline{\alpha }_2},
\]
with $\overleftarrow{G}_{\underline{\gamma }_1\underline{\gamma }_2%
\underline{\alpha }_1\underline{\alpha }_2}$ from (\ref{2.98}),
or, by using the d-tensor (\ref{2.99}),

\[
\Phi _{\underline{\gamma }_1\underline{\gamma }_2\underline{\alpha }_1%
\underline{\alpha }_2}+(\frac{\overleftarrow{R}}4-\frac \lambda
2)\varepsilon _{\underline{\gamma }_1\underline{\alpha }_1}\varepsilon _{%
\underline{\gamma }_2\underline{\alpha }_2}=-\frac \kappa 2E_{\underline{%
\gamma }_1\underline{\gamma }_2\underline{\alpha
}_1\underline{\alpha }_2},
\]
which are the d-spinor equivalent of the equations (\ref{2.44}).
These equations must be completed by the algebraic equations
(\ref{2.45}) for the d-torsion and d-spin density with d-tensor
indices changed into corresponding d-spinor ones.












\part[Higher Order Anisotropic Spinors]{Higher Order Anisotropic Spinors}

The theory of anisotropic spinors formulated in the Part II is
extended for higher order anisotropic (ha) spaces. In brief, such
spinors will be called ha--spinors which are defined as some
Clifford ha--structures defined with respect to a distinguished
quadratic form (\ref{dmetrichcv}) on a hvc--bundle. For
simplicity,  the bulk of formulas will be given with respect to
higher order vector bundles. To rewrite such formulas for
hvc--bundles is to consider for the ''dual'' shells of higher
order anisotropy some dual vector spaces and associated dual
spinors.

\chapter{ Clifford Ha--Structures} \index{Clifford ha--sructure}

\section{Distinguished Clifford Algebras}

The typical fiber of dv--bundle $\xi _d\ ,\ \pi _d:\ HE\oplus
V_1E\oplus
...\oplus V_zE\rightarrow E$ is a d-vector space, $\mathcal{F}=h\mathcal{F}%
\oplus v_1\mathcal{F\oplus }...\oplus v_z\mathcal{F},$ split into
horizontal $h\mathcal{F}$ and verticals
$v_p\mathcal{F},p=1,...,z$ subspaces, with a
bilinear quadratic form $G(g,h)$ induced by a hvc--bundle metric (%
\ref{dmetrichcv}). Clifford algebras (see, for example, Refs.
\cite {kar,tur,penr2}) formulated for d-vector spaces will be
called Clifford d-algebras \cite{vjmp,viasm1,vdeb}. We shall
consider the main properties of Clifford d--algebras. The proof
of theorems will be based on the technique developed in Ref.
\cite{kar} correspondingly adapted to the distinguished character
of spaces in consideration.

Let $k$ be a number field (for our purposes $k=\mathcal{R}$ or $k=\mathcal{C}%
,\mathcal{R}$ and $\mathcal{C},$ are, respectively real and
complex number fields) and define $\mathcal{F},$ as a d-vector
space on $k$ provided with nondegenerate symmetric quadratic form
(metric)\ $G.$ Let $C$ be an algebra on $k$ (not necessarily
commutative) and $j\ :\ \mathcal{F}$ $\rightarrow C$ a
homomorphism of underlying vector spaces such that
$j(u)^2=\;G(u)\cdot 1\ (1$ is the unity in algebra $C$ and
d-vector $u\in \mathcal{F}).$ We are interested in definition of
the pair $\left( C,j\right) $ satisfying the next universitality
conditions. For every $k$-algebra $A$ and arbitrary homomorphism
$\varphi :\mathcal{F}\rightarrow A$ of the underlying d-vector
spaces, such that $\left( \varphi (u)\right) ^2\rightarrow
G\left( u\right) \cdot 1,$ there is a unique homomorphism of
algebras $\psi \ :\ C\rightarrow A$ transforming the diagram 1
into a commutative one.

The algebra solving this problem will be denoted as $C\left( \mathcal{F}%
,A\right) $ [equivalently as $C\left( G\right) $ or $C\left( \mathcal{F}%
\right) ]$ and called as Clifford d--algebra associated with pair
$\left( \mathcal{F},G\right) .$

\begin{theorem}
The above-presented diagram has a unique solution $\left(
C,j\right) $ up to isomorphism.
\end{theorem}

\textbf{Proof:} (We adapt for d-algebras that of Ref. \cite{kar},
p. 127 and extend for higher order anisotropies a similar proof
presented in  the Part II).  For a universal problem the
uniqueness is obvious if we prove the existence of solution
$C\left( G\right) $ . To do this we use tensor algebra
$\mathcal{L}^{(F)}=\oplus \mathcal{L}_{qs}^{pr}\left( \mathcal{F}\right) $ =$%
\oplus _{i=0}^\infty T^i\left( \mathcal{F}\right) ,$ where
$T^0\left( \mathcal{F}\right) =k$ and $T^i\left(
\mathcal{F}\right) =k$ and $T^i\left( \mathcal{F}\right)
=\mathcal{F}\otimes ...\otimes \mathcal{F}$ for $i>0.$ Let
$I\left( G\right) $ be the bilateral ideal generated by elements
of form $\epsilon \left( u\right) =u\otimes u-G\left( u\right)
\cdot 1$ where $u\in \mathcal{F}$ and $1$ is the unity element of
algebra $\mathcal{L}\left( \mathcal{F}\right) .$ Every element
from $I\left( G\right) $ can be written
as $\sum\nolimits_i\lambda _i\epsilon \left( u_i\right) \mu _i,$ where $%
\lambda _i,\mu _i\in \mathcal{L}(\mathcal{F})$ and $u_i\in
\mathcal{F}.$ Let $C\left( G\right) $
=$\mathcal{L}(\mathcal{F})/I\left( G\right) $ and define
$j:\mathcal{F}\rightarrow C\left( G\right) $ as the composition of
monomorphism $i:{\mathcal{F}\rightarrow L}^1(\mathcal{F})\subset \mathcal{L}(%
\mathcal{F})$ and projection $p:\mathcal{L}\left(
\mathcal{F}\right) \rightarrow C\left( G\right) .$ In this case
pair $\left( C\left( G\right) ,j\right) $ is the solution of our
problem. From the general properties of tensor algebras the
homomorphism $\varphi :\mathcal{F}\rightarrow A$ can be extended
to $\mathcal{L}(\mathcal{F})$ , i.e., the diagram 2 is
commutative, where $\rho $ is a monomorphism of algebras. Because
$\left( \varphi \left( u\right) \right) ^2=G\left( u\right) \cdot
1,$ then $\rho $ vanishes on ideal $I\left( G\right) $ and in
this case the necessary homomorphism $\tau $
is defined. As a consequence of uniqueness of $\rho ,$ the homomorphism $%
\tau $ is unique.

Tensor d--algebra $\mathcal{L}(\mathcal{F)}$ can be considered as a $%
\mathcal{Z}/2$ graded algebra. Really, let us in\-tro\-duce $\mathcal{L}%
^{(0)}(\mathcal{F})=\sum_{i=1}^\infty T^{2i}\left( \mathcal{F}\right) $ and $%
\mathcal{L}^{(1)}(\mathcal{F})=\sum_{i=1}^\infty T^{2i+1}\left( \mathcal{F}%
\right) .$ Setting $I^{(\alpha )}\left( G\right) =I\left(
G\right) \cap \mathcal{L}^{(\alpha )}(\mathcal{F}).$ Define
$C^{(\alpha )}\left( G\right) $
as $p\left( \mathcal{L}^{(\alpha )}(\mathcal{F})\right) ,$ where $p:\mathcal{%
L}\left( \mathcal{F}\right) \rightarrow C\left( G\right) $ is the
canonical projection. Then $C\left( G\right) =C^{(0)}\left(
G\right) \oplus C^{(1)}\left( G\right) $ and in consequence we
obtain that the Clifford d--algebra is $\mathcal{Z}/2$ graded.

It is obvious that Clifford d-algebra \index{Clifford d--algebra}
functorially depends on pair $\left( \mathcal{F},G\right) .$ If
$f:\mathcal{F}\rightarrow\mathcal{F}^{\prime }$ is a homomorphism
of k-vector spaces, such that $G^{\prime }\left( f(u)\right)
=G\left( u\right) ,$ where $G$ and $G^{\prime }$ are,
respectively, metrics on $\mathcal{F}$ and $\mathcal{F}^{\prime
},$ then $f$ induces an homomorphism of d-algebras
\[
C\left( f\right) :C\left( G\right) \rightarrow C\left( G^{\prime
}\right)
\]
with identities $C\left( \varphi \cdot f\right) =C\left( \varphi
\right)
C\left( f\right) $ and $C\left( Id_{\mathcal{F}}\right) =Id_{C(\mathcal{F)}%
}. $

If $\mathcal{A}^{\alpha}$ and $\mathcal{B}^{\beta}$ are $\mathcal{Z}/2$%
--graded d--algebras, then their graded tensorial product $\mathcal{A}%
^\alpha \otimes \mathcal{B}^\beta $ is defined as a d-algebra for
k-vector d-space $\mathcal{A}^\alpha \otimes \mathcal{B}^\beta $
with the graded product induced as $\left( a\otimes b\right)
\left( c\otimes d\right)
=\left( -1\right) ^{\alpha \beta }ac\otimes bd,$ where $b\in \mathcal{B}%
^\alpha $ and $c\in \mathcal{A}^\alpha \quad \left( \alpha ,\beta
=0,1\right) .$

Now we re--formulate for d--algebras the Chevalley theorem
\cite{chev}:
\begin{theorem}
The Clifford d-algebra
\[
C\left( h\mathcal{F}\oplus v_1\mathcal{F}\oplus ...\oplus v_z\mathcal{F}%
,g+h_1+...+h_z\right)
\]
is naturally isomorphic to $C(g)\otimes C\left( h_1\right) \otimes
...\otimes C\left( h_z\right) .$
\end{theorem}

\textbf{Proof. }Let $n:h\mathcal{F}\rightarrow C\left( g\right) $ and $%
n_{(p)}^{\prime }:v_{(p)}\mathcal{F}\rightarrow C\left(
h_{(p)}\right) $ be canonical maps and map
\[
m:h\mathcal{F}\oplus v_1\mathcal{F}\oplus ...\oplus v_z\mathcal{F}%
\rightarrow C(g)\otimes C\left( h_1\right) \otimes ...\otimes
C\left( h_z\right)
\]
is defined as \begin{eqnarray}
& & m(x,y_{(1)},...,y_{(z)})= \nonumber \\
& & n(x)\otimes 1\otimes ...\otimes 1+1\otimes n^{\prime
}(y_{(1)})\otimes ...\otimes 1+1\otimes ...\otimes 1\otimes
n^{\prime }(y_{(z)}), \nonumber
\end{eqnarray}
$x\in h\mathcal{F},y_{(1)}\in v_{(1)}\mathcal{F},...,y_{(z)}\in v_{(z)}%
\mathcal{F}.$ We have
\begin{eqnarray*}
\left( m(x,y_{(1)},...,y_{(z)})\right) ^2 &=&\left[ \left(
n\left( x\right) \right) ^2+\left( n^{\prime }\left(
y_{(1)}\right) \right) ^2+...+\left(
n^{\prime }\left( y_{(z)}\right) \right) ^2\right] \cdot 1 \\
&=&[g\left( x\right) +h\left( y_{(1)}\right) +...+h\left(
y_{(z)}\right) ].
\end{eqnarray*}
\ Taking into account the universality property of Clifford
d-algebras we conclude that $m_1+...+m_z$ induces the homomorphism
\begin{eqnarray}
 & &
\zeta :C\left( h\mathcal{F}\oplus v_1\mathcal{F}\oplus ...\oplus v_z\mathcal{%
F},g+h_1+...+h_z\right) \rightarrow \nonumber \\
 & &
C\left( h\mathcal{F},g\right) \widehat{\otimes }C\left( v_1\mathcal{F}%
,h_1\right) \widehat{\otimes }...C\left(
v_z\mathcal{F},h_z\right) . \nonumber
\end{eqnarray}
 We also can define a homomorphism
\begin{eqnarray}
 & &
\upsilon :C\left( h\mathcal{F},g\right) \widehat{\otimes }C\left( v_1%
\mathcal{F},h_{(1)}\right) \widehat{\otimes }...\widehat{\otimes }C\left( v_z%
\mathcal{F},h_{(z)}\right) \rightarrow \nonumber \\
 & &
C\left( h\mathcal{F}\oplus v_1\mathcal{F}\oplus ...\oplus v_z\mathcal{F}%
,g+h_{(1)}+...+h_{(z)}\right) \nonumber
\end{eqnarray}
by using formula $\upsilon \left( x\otimes y_{(1)}\otimes
...\otimes y_{(z)}\right) =\delta \left( x\right) \delta
_{(1)}^{\prime }\left( y_{(1)}\right) ...\delta _{(z)}^{\prime
}\left( y_{(z)}\right) ,$ where homomorphysms $\delta $ and
$\delta _{(1)}^{\prime },...,\delta _{(z)}^{\prime }$ are,
respectively, induced by imbeddings of $h\mathcal{F}$ and
$v_1\mathcal{F}$ into $h\mathcal{F}\oplus v_1\mathcal{F}\oplus
...\oplus
v_z\mathcal{F}:$%
\begin{eqnarray*}
\delta  &:&C\left( h\mathcal{F},g\right) \rightarrow C\left( h\mathcal{F}%
\oplus v_1\mathcal{F}\oplus ...\oplus v_z\mathcal{F},g+h_{(1)}+...+h_{(z)}%
\right) , \\
\delta _{(1)}^{\prime } &:&C\left( v_1\mathcal{F},h_{(1)}\right)
\rightarrow
C\left( h\mathcal{F}\oplus v_1\mathcal{F}\oplus ...\oplus v_z\mathcal{F}%
,g+h_{(1)}+...+h_{(z)}\right) , \\
&&................................... \\
\delta _{(z)}^{\prime } &:&C\left( v_z\mathcal{F},h_{(z)}\right)
\rightarrow
C\left( h\mathcal{F}\oplus v_1\mathcal{F}\oplus ...\oplus v_z\mathcal{F}%
,g+h_{(1)}+...+h_{(z)}\right) .
\end{eqnarray*}

Superpositions of homomorphisms $\zeta $ and $\upsilon $ lead to
identities
\begin{eqnarray}
\upsilon \zeta  &=&Id_{C\left( h\mathcal{F},g\right) \widehat{\otimes }%
C\left( v_1\mathcal{F},h_{(1)}\right) \widehat{\otimes }...\widehat{\otimes }%
C\left( v_z\mathcal{F},h_{(z)}\right) },  \label{2.37a} \\
\zeta \upsilon  &=&Id_{C\left( h\mathcal{F},g\right) \widehat{\otimes }%
C\left( v_1\mathcal{F},h_{(1)}\right) \widehat{\otimes }...\widehat{\otimes }%
C\left( v_z\mathcal{F},h_{(z)}\right) }.  \nonumber
\end{eqnarray}
Really, d-algebra $C\left( h\mathcal{F}\oplus
v_1\mathcal{F}\oplus ...\oplus
v_z\mathcal{F},g+h_{(1)}+...+h_{(z)}\right) $ is generated by
elements of type $m(x,y_{(1)},...y_{(z)}).$ Calculating
\begin{eqnarray*}
& & \upsilon \zeta \left( m\left( x,y_{(1)},...y_{(z)}\right)
\right) =\upsilon (n\left( x\right) \otimes 1\otimes ...\otimes
1+1\otimes
n_{(1)}^{\prime }\left( y_{(1)}\right) \otimes ...\otimes 1 \\
& & + ...+1\otimes ....\otimes n_{(z)}^{\prime }\left(
y_{(z)}\right) ) =\delta \left( n\left( x\right) \right) \delta
\left( n_{(1)}^{\prime }\left( y_{(1)}\right) \right) ...\delta
\left( n_{(z)}^{\prime }\left(
y_{(z)}\right) \right)  \\
& &=m\left( x,0,...,0\right) +m(0,y_{(1)},...,0)+...+m(0,0,...,y_{(z)}) \\
& & =m\left( x,y_{(1)},...,y_{(z)}\right) ,
\end{eqnarray*}
we prove the first identity in (\ref{2.37a}).

On the other hand, d-algebra
\[
C\left( h\mathcal{F},g\right) \widehat{\otimes }C\left( v_1\mathcal{F}%
,h_{(1)}\right) \widehat{\otimes }...\widehat{\otimes }C\left( v_z\mathcal{F}%
,h_{(z)}\right)
\]
is generated by elements of type
\[
n\left( x\right) \otimes 1\otimes ...\otimes ,1\otimes
n_{(1)}^{\prime }\left( y_{(1)}\right) \otimes ...\otimes
1,...1\otimes ....\otimes n_{(z)}^{\prime }\left( y_{(z)}\right) ,
\]
we prove the second identity in  (\ref{2.37a}).

Following from the above--mentioned properties of homomorphisms
$\zeta $ and $\upsilon $ we can assert that the natural
isomorphism is explicitly constructed.$\Box $

In consequence of the presented in this section Theorems  we
conclude that
all operations with Clifford d-algebras can be reduced to calculations for $%
C\left( h\mathcal{F},g\right) $ and $C\left( v_{(p)}\mathcal{F}%
,h_{(p)}\right) $ which are usual Clifford algebras of dimension
$2^n$ and, respectively, $2^{m_p}$ \cite{kar,ati}.

Of special interest is the case when $k=\mathcal{R}$ and
$\mathcal{F}$ is isomorphic to vector space
$\mathcal{R}^{p+q,a+b}$ provided with quadratic form
\[
-x_1^2-...-x_p^2+x_{p+q}^2-y_1^2-...-y_a^2+...+y_{a+b}^2.
\]
In this case, the Clifford algebra, denoted as $\left(
C^{p,q},C^{a,b}\right) ,\,$ is generated by symbols $%
e_1^{(x)},e_2^{(x)},...,e_{p+q}^{(x)},e_1^{(y)},e_2^{(y)},...,e_{a+b}^{(y)}$
satisfying properties
\begin{eqnarray*}
\left( e_i\right) ^2 &=&-1~\left( 1\leq i\leq p\right) ,\left(
e_j\right)
^2=-1~\left( 1\leq j\leq a\right) , \\
\left( e_k\right) ^2 &=&1~(p+1\leq k\leq p+q), \\
\left( e_j\right) ^2 &=&1~(n+1\leq s\leq
a+b),~e_ie_j=-e_je_i,~i\neq j.\,
\end{eqnarray*}

Explicit calculations of $C^{p,q}$ and $C^{a,b}$ are possible by
using isomorphisms \cite{kar,penr2}
\begin{eqnarray*}
C^{p+n,q+n} &\simeq &C^{p,q}\otimes M_2\left( \mathcal{R}\right)
\otimes
...\otimes M_2\left( \mathcal{R}\right)  \\
&\cong &C^{p,q}\otimes M_{2^n}\left( \mathcal{R}\right) \cong
M_{2^n}\left( C^{p,q}\right) ,
\end{eqnarray*}
where $M_s\left( A\right) $ denotes the ring of quadratic matrices of order $%
s$ with coefficients in ring $A.$ Here we write the simplest isomorphisms $%
C^{1,0}\simeq \mathcal{C},C^{0,1}\simeq \mathcal{R}\oplus \mathcal{R}$ and $%
C^{2,0}=\mathcal{H},$ where by $\mathcal{H}$ is denoted the body
of quaternions.

Now, we emphasize that higher order Lagrange and Finsler spaces, denoted  $%
H^{2n}$--spaces, admit locally a structure of Clifford algebra on
complex vector spaces. Really, by using almost \ Hermitian
structure $J_\alpha ^{\quad \beta }$ and considering complex
space $\mathcal{C}^n$ with nondegenarate quadratic form
$\sum_{a=1}^n\left| z_a\right| ^2,~z_a\in \mathcal{C}^2$ induced
locally by metric (\ref{dmetrichcv}) (rewritten in complex
coordinates $z_a=x_a+iy_a)$ we define Clifford algebra $\overleftarrow{C}^n=%
\underbrace{\overleftarrow{C}^1\otimes ...\otimes
\overleftarrow{C}^1}_n,$ where
$\overleftarrow{C}^1=\mathcal{C\otimes }_R\mathcal{C=C\oplus C}$
or in
consequence, $\overleftarrow{C}^n\simeq C^{n,0}\otimes _{\mathcal{R}}%
\mathcal{C}\approx C^{0,n}\otimes _{\mathcal{R}}\mathcal{C}.$
Explicit calculations lead to isomorphisms
\[
\overleftarrow{C}^2=C^{0,2}\otimes
_{\mathcal{R}}\mathcal{C}\approx M_2\left( \mathcal{R}\right)
\otimes _{\mathcal{R}}\mathcal{C}\approx
M_2\left( \overleftarrow{C}^n\right) ,~C^{2p}\approx M_{2^p}\left( \mathcal{C%
}\right)
\]
and
\[
\overleftarrow{C}^{2p+1}\approx M_{2^p}\left( \mathcal{C}\right)
\oplus M_{2^p}\left( \mathcal{C}\right) ,
\]
which show that complex Clifford algebras, defined locally for $H^{2n}$%
-spaces, have periodicity 2 on $p.$

Considerations presented in the proof of theorem 2.2 show that map $j:%
\mathcal{F}\rightarrow C\left( \mathcal{F}\right) $ is
monomorphic, so we
can identify space $\mathcal{F}$ with its image in $C\left( \mathcal{F}%
,G\right) ,$ denoted as $u\rightarrow \overline{u},$ if $u\in
C^{(0)}\left( \mathcal{F},G\right) ~\left( u\in C^{(1)}\left(
\mathcal{F},G\right) \right) ;$ then $u=\overline{u}$ (
respectively, $\overline{u}=-u).$

\begin{definition} \index{Clifford d--group}
The set of elements $u\in C\left( G\right) ^{*},$ where $C\left(
G\right) ^{*}$ denotes the multiplicative group of invertible
elements of $C\left(
\mathcal{F},G\right) $ satisfying $\overline{u}\mathcal{F}u^{-1}\in \mathcal{%
F},$ is called the twisted Clifford d-group, denoted as $\widetilde{\Gamma }%
\left( \mathcal{F}\right) .$
\end{definition}

Let $\widetilde{\rho }:\widetilde{\Gamma }\left(
\mathcal{F}\right)
\rightarrow GL\left( \mathcal{F}\right) $ be the homorphism given by $%
u\rightarrow \rho \widetilde{u},$ where $\widetilde{\rho }_u\left( w\right) =%
\overline{u}wu^{-1}.$ We can verify that $\ker \widetilde{\rho }=\mathcal{R}%
^{*}$is a subgroup in $\widetilde{\Gamma }\left(
\mathcal{F}\right) .$

The canonical map $j:\mathcal{F}\rightarrow C\left(
\mathcal{F}\right) $ can
be interpreted as the linear map $\mathcal{F}\rightarrow C\left( \mathcal{F}%
\right) ^0$ satisfying the universal property of Clifford
d-algebras. This leads to a homomorphism of algebras, $C\left(
\mathcal{F}\right) \rightarrow C\left( \mathcal{F}\right) ^t,$
considered by an anti-involution of $C\left(
\mathcal{F}\right) $ and denoted as $u\rightarrow ~^tu.$ More exactly, if $%
u_1...u_n\in \mathcal{F,}$ then $t_u=u_n...u_1$ and $^t\overline{u}=%
\overline{^tu}=\left( -1\right) ^nu_n...u_1.$

\begin{definition} \index{spinor norm}
The spinor norm of arbitrary $u\in C\left( \mathcal{F}\right) $ is defined as%
\newline
$S\left( u\right) =~^t\overline{u}\cdot u\in C\left(
\mathcal{F}\right) .$
\end{definition}

It is obvious that if $u,u^{\prime },u^{\prime \prime }\in
\widetilde{\Gamma }\left( \mathcal{F}\right) ,$ then
$S(u,u^{\prime })=S\left( u\right) S\left( u^{\prime }\right) $
and \newline $S\left( uu^{\prime }u^{\prime \prime }\right)
=S\left( u\right) S\left( u^{\prime }\right) S\left( u^{\prime
\prime }\right) .$ For $u,u^{\prime }\in \mathcal{F} S\left(
u\right) =-G\left( u\right) $ and $S\left( u,u^{\prime }\right)
=S\left( u\right) S\left( u^{\prime }\right) =S\left( uu^{\prime
}\right) .$

Let us introduce the orthogonal group $O\left( G\right) \subset
GL\left( G\right) $ defined by metric $G$ on $\mathcal{F}$ and
denote sets
\[
SO\left( G\right) =\{u\in O\left( G\right) ,\det \left| u\right|
=1\},~Pin\left( G\right) =\{u\in \widetilde{\Gamma }\left( \mathcal{F}%
\right) ,S\left( u\right) =1\}
\]
and $Spin\left( G\right) =Pin\left( G\right) \cap C^0\left( \mathcal{F}%
\right) .$ For ${\mathcal{F}\cong \mathcal{R}}^{n+m}$ we write
$Spin\left( n_E\right) .$ By straightforward calculations (see
similar considerations in Ref. \cite{kar}) we can verify the
exactness of these sequences:
\begin{eqnarray*}
1 &\rightarrow &\mathcal{Z}/2\rightarrow Pin\left( G\right)
\rightarrow
O\left( G\right) \rightarrow 1, \\
1 &\rightarrow &\mathcal{Z}/2\rightarrow Spin\left( G\right)
\rightarrow
SO\left( G\right) \rightarrow 0, \\
1 &\rightarrow &\mathcal{Z}/2\rightarrow Spin\left( n_E\right)
\rightarrow SO\left( n_E\right) \rightarrow 1.
\end{eqnarray*}
We conclude this subsection by emphasizing that the spinor norm
was defined
with respect to a quadratic form induced by a metric in dv-bundle $\mathcal{E%
}^{<z>}$. This approach differs from those presented in Refs.
\cite{asa88} and \cite{ono}.

\section{Clifford Ha--Bundles}

We shall consider two variants of generalization of spinor
constructions defined for d-vector spaces to the case of
distinguished vector bundle spaces enabled with the structure of
N-connection. The first is to use the extension to the category
of vector bundles. The second is to define the Clifford fibration
associated with compatible linear d-connection and metric $G$ on
a dv--bundle. We shall analyze both variants.

\subsection{Clifford d--module structure in dv--bundles} \index{Clifford d--module}

Because functor $\mathcal{F}\to C(\mathcal{F})$ is smooth we can
extend it to the category of vector bundles of type
\[
\xi ^{<z>}=\{\pi _d:HE^{<z>}\oplus V_1E^{<z>}\oplus ...\oplus
V_zE^{<z>}\rightarrow E^{<z>}\}.
\]
Recall that by $\mathcal{F}$ we denote the typical fiber of such
bundles. For $\xi ^{<z>}$ we obtain a bundle of algebras, denoted
as $C\left( \xi
^{<z>}\right) ,\,$ such that $C\left( \xi ^{<z>}\right) _u=C\left( \mathcal{F%
}_u\right) .$ Multiplication in every fibre defines a continuous
map
\[
C\left( \xi ^{<z>}\right) \times C\left( \xi ^{<z>}\right)
\rightarrow C\left( \xi ^{<z>}\right) .
\]
If $\xi ^{<z>}$ is a distinguished vector bundle on number field $k$%
,\thinspace \thinspace the structure of the $C\left( \xi ^{<z>}\right) $%
-module, the d-module, the d-module, on $\xi ^{<z>}$ is given by
the continuous map $C\left( \xi ^{<z>}\right) \times _E\xi
^{<z>}\rightarrow \xi
^{<z>}$ with every fiber $\mathcal{F}_u$ provided with the structure of the $%
C\left( \mathcal{F}_u\right) -$module, correlated with its
$k$-module structure, Because $\mathcal{F}\subset C\left(
\mathcal{F}\right) ,$ we have a fiber to fiber map
$\mathcal{F}\times _E\xi ^{<z>}\rightarrow \xi ^{<z>},$ inducing
on every fiber the map $\mathcal{F}_u\times _E\xi
_{(u)}^{<z>}\rightarrow \xi _{(u)}^{<z>}$ ($\mathcal{R}$-linear
on the first factor and $k$-linear on the second one ).
Inversely, every such bilinear
map defines on $\xi ^{<z>}$ the structure of the $C\left( \xi ^{<z>}\right) $%
-module by virtue of universal properties of Clifford d--algebras.
Equivalently, the above--mentioned bilinear map defines a
morphism of v--bundles $$m:\xi ^{<z>}\rightarrow HOM\left( \xi
^{<z>},\xi ^{<z>}\right) \quad [HOM\left( \xi ^{<z>},\xi
^{<z>}\right) $$ denotes the bundles of homomorphisms] when
$\left( m\left( u\right) \right) ^2=G\left( u\right) $ on every
point.

Vector bundles $\xi ^{<z>}$ provided with $C\left( \xi ^{<z>}\right) $%
--structures are objects of the category with morphisms being
morphisms of
dv-bundles, which induce on every point $u\in \xi ^{<z>}$ morphisms of $%
C\left( \mathcal{F}_u\right) -$modules. This is a Banach category
contained in the category of finite-dimensional d-vector spaces
on filed $k.$

Let us denote by $H^s\left( \mathcal{E}^{<z>},GL_{n_E}\left( \mathcal{R}%
\right) \right) ,\,$ where $n_E=n+m_1+...+m_z,\,$ the
s-dimensional cohomology group of the algebraic sheaf of germs of
continuous maps of dv-bundle $\mathcal{E}^{<z>}$ with group
$GL_{n_E}\left( \mathcal{R}\right) $ the group of automorphisms
of $\mathcal{R}^{n_E}\,$ (for the language of algebraic topology
see, for example, Refs. \cite{kar} and \cite{god}). We shall also
use the group $SL_{n_E}\left( \mathcal{R}\right) =\{A\subset
GL_{n_E}\left( \mathcal{R}\right) ,\det A=1\}.\,$ Here we point
out that cohomologies  $H^s(M,Gr)$ characterize the class of a
principal bundle $\pi :P\rightarrow M $ on $M$ with structural
group $Gr.$ Taking into account that we deal with bundles
distinguished by an N-connection we introduce into consideration
cohomologies $H^s\left( \mathcal{E}^{<z>},GL_{n_E}\left(
\mathcal{R}\right) \right) $ as distinguished classes (d-classes)
of bundles $\mathcal{E}^{<z>}$ provided with a global
N-connection structure.

For a real vector bundle $\xi ^{<z>}$ on compact base
$\mathcal{E}^{<z>}$ we can define the orientation on $\xi ^{<z>}$
as an element $\alpha _d\in H^1\left(
\mathcal{E}^{<z>},GL_{n_E}\left( \mathcal{R}\right) \right) $
whose image on map
\[
H^1\left( \mathcal{E}^{<z>},SL_{n_E}\left( \mathcal{R}\right)
\right) \rightarrow H^1\left( \mathcal{E}^{<z>},GL_{n_E}\left(
\mathcal{R}\right) \right)
\]
is the d-class of bundle $\mathcal{E}^{<z>}.$

\begin{definition}
The spinor structure on $\xi ^{<z>}$ is defined as an
element\newline $\beta _d\in H^1\left(
\mathcal{E}^{<z>},Spin\left( n_E\right) \right) $ whose image in
the composition
\[
H^1\left( \mathcal{E}^{<z>},Spin\left( n_E\right) \right)
\rightarrow H^1\left( \mathcal{E}^{<z>},SO\left( n_E\right)
\right) \rightarrow H^1\left( \mathcal{E}^{<z>},GL_{n_E}\left(
\mathcal{R}\right) \right)
\]
is the d-class of $\mathcal{E}^{<z>}.$
\end{definition}

The above definition of spinor structures can be re--formulated in
terms of principal bundles. Let $\xi ^{<z>}$ be a real vector
bundle of rank n+m on a compact base $\mathcal{E}^{<z>}.$ If
there is a principal bundle $P_d$ with structural group $SO( n_E
) $  or $Spin( n_E ) ],$ this bundle $\xi ^{<z>}$ can be provided
with orientation (or spinor) structure. The bundle $P_d$ is
associated with element\\ $\alpha _d\in
H^1\left(\mathcal{E}^{<z>},SO(n_{<z>})\right) $
 [or $\beta _d\in H^1\left( \mathcal{E}^{<z>},
 Spin\left( n_E\right) \right) .$

We remark that a real bundle is oriented if and only if its first
Stiefel--Whitney d--class vanishes,
\[
w_1\left( \xi _d\right) \in H^1\left( \xi ,\mathcal{Z}/2\right)
=0,
\]
where $H^1\left( \mathcal{E}^{<z>},\mathcal{Z}/2\right) $ is the
first group of Chech cohomology with coefficients in
$\mathcal{Z}/2,$ Considering the second Stiefel--Whitney class
$w_2\left( \xi ^{<z>}\right) \in H^2\left(
\mathcal{E}^{<z>},\mathcal{Z}/2\right) $ it is well known that
vector bundle $\xi ^{<z>}$ admits the spinor structure if and
only if $w_2\left( \xi ^{<z>}\right) =0.$ Finally, we emphasize
that taking into account that base space $\mathcal{E}^{<z>}$ is
also a v-bundle, $p:E^{<z>}\rightarrow M,$ we
have to make explicit calculations in order to express cohomologies $%
H^s\left( \mathcal{E}^{<z>},GL_{n+m}\right) \,$ and $H^s\left( \mathcal{E}%
^{<z>},SO\left( n+m\right) \right) $ through cohomologies
\[
H^s\left( M,GL_n\right) ,H^s\left( M,SO\left( m_1\right) \right)
,...H^s\left( M,SO\left( m_z\right) \right) ,
\]
which depends on global topological structures of spaces $M$ and $\mathcal{E}%
^{<z>}$ $.$ For general bundle and base spaces this requires a
cumbersome cohomological calculus.

\subsection{Clifford fibration} \index{Clifford fibration}

Another way of defining the spinor structure is to use Clifford
fibrations. Consider the principal bundle with the structural
group $Gr$ being a subgroup of orthogonal group $O\left( G\right)
,$ where $G$ is a quadratic
nondegenerate form ) defined on the base (also being a bundle space) space $%
\mathcal{E}^{<z>}.$ The fibration associated to principal
fibration $P\left(
\mathcal{E}^{<z>},Gr\right) $ with a typical fiber having Clifford algebra $%
C\left( G\right) $ is, by definition, the Clifford fibration
$PC\left( \mathcal{E}^{<z>},Gr\right) .$ We can always define a
metric on the Clifford
fibration if every fiber is isometric to $PC\left( \mathcal{E}%
^{<z>},G\right) $ (this result is proved for arbitrary quadratic
forms $G$ on pseudo--Riemannian bases \cite{tur}). If,
additionally, $Gr\subset SO\left( G\right) $ a global section can
be defined on $PC\left( G\right) .$

Let $\mathcal{P}\left( \mathcal{E}^{<z>},Gr\right) $ be the set
of principal
bundles with differentiable base $\mathcal{E}^{<z>}$ and structural group $%
Gr.$ If $g:Gr\rightarrow Gr^{\prime }$ is an homomorphism of Lie groups and $%
P\left( \mathcal{E}^{<z>},Gr\right) \subset \mathcal{P}\left( \mathcal{E}%
^{<z>},Gr\right) $ (for simplicity in this subsection we shall
denote
mentioned bundles and sets of bundles as $P,P^{\prime }$ and respectively, $%
\mathcal{P},\mathcal{P}^{\prime }),$ we can always construct a
principal bundle with the property that there is an homomorphism
$f:P^{\prime }\rightarrow P$ of principal bundles which can be
projected to the identity map of $\mathcal{E}^{<z>}$ and
corresponds to isomorphism $g:Gr\rightarrow Gr^{\prime }.$ If the
inverse statement also holds, the bundle $P^{\prime }$ is called
as the extension of $P$ associated to $g$ and $f$ is called the
extension homomorphism denoted as $\widetilde{g.}$

Now we can define distinguished spinor structures on bundle
spaces \index{spinor structure}.

\begin{definition}
Let $P\in \mathcal{P}\left( \mathcal{E}^{<z>},O\left( G\right)
\right) $ be a principal bundle. A distinguished spinor structure
of $P,$ equivalently a ds-structure of $\mathcal{E}^{<z>}$ is an
extension $\widetilde{P}$ of $P$
associated to homomorphism $h:PinG\rightarrow O\left( G\right) $ where $%
O\left( G\right) $ is the group of orthogonal rotations,
generated by metric $G,$ in bundle $\mathcal{E}^{<z>}.$
\end{definition}

So, if $\widetilde{P}$ is a spinor structure of the space $\mathcal{E}%
^{<z>}, $ then $\widetilde{P}\in \mathcal{P}\left( \mathcal{E}%
^{<z>},PinG\right) .$

The definition of spinor structures on varieties was given in Ref.\cite{cru1}%
. In Refs. \cite{cru2} and \cite{cru2} it is proved that a
necessary and sufficient condition for a space time to be
orientable is to admit a global field of orthonormalized frames.
We mention that spinor structures can be also defined on
varieties modeled on Banach spaces \cite{ana77}. As we have shown
similar constructions are possible for the cases when space time
has the structure of a v-bundle with an N-connection.

\begin{definition}
A special distinguished spinor structure, ds-structure, \index{ds--structure}  of principal bundle $%
P=P\left( \mathcal{E}^{<z>},SO\left( G\right) \right) $ is a
principal bundle\\
 $\widetilde{P}=\widetilde{P}\left(
\mathcal{E}^{<z>},SpinG\right) $
for which a homomorphism of principal bundles $\widetilde{p}:\widetilde{P}%
\rightarrow P,$ projected on the identity map of
$\mathcal{E}^{<z>}$ and corresponding to representation
\[
R:SpinG\rightarrow SO\left( G\right) ,
\]
is defined.
\end{definition}

In the case when the base space variety is oriented, there is a
natural bijection between tangent spinor structures with a common
base. For special ds--structures we can define, as for any spinor
structure, the concepts of spin tensors, spinor connections, and
spinor covariant derivations (see Refs. \cite{viasm1,vdeb,vsp1}).

\section{Almost Complex Spinor Structures} \index{complex spinors}

Almost complex structures are an important characteristic of
$H^{2n}$-spaces and of osculator bundles $Osc^{k=2k_1}(M),$ where
$k_1=1,2,...$ . For simplicity in this subsection we restrict our
analysis to the case of $H^{2n}
$-spaces. We can rewrite the almost Hermitian metric \cite{ma87,ma94}, $%
H^{2n}$-metric  in complex form \cite{vjmp}:

\begin{equation}
G=H_{ab}\left( z,\xi \right) dz^a\otimes dz^b,  \label{2.38a}
\end{equation}
where
\[
z^a=x^a+iy^a,~\overline{z^a}=x^a-iy^a,~H_{ab}\left(
z,\overline{z}\right) =g_{ab}\left( x,y\right) \mid _{y=y\left(
z,\overline{z}\right) }^{x=x\left( z,\overline{z}\right) },
\]
and define almost complex spinor structures. For given metric
(\ref{2.38a}) on $H^{2n}$-space there is always a principal
bundle $P^U$ with unitary structural group U(n) which allows us
to transform $H^{2n}$-space into v-bundle $\xi ^U\approx
P^U\times _{U\left( n\right) }\mathcal{R}^{2n}.$ This statement
will be proved after we introduce complex spinor structures on
oriented real vector bundles \cite{kar}.

Let us consider momentarily $k=\mathcal{C}$ and introduce into
consideration
[instead of the group $Spin(n)]$ the group $Spin^c\times _{\mathcal{Z}%
/2}U\left( 1\right) $ being the factor group of the product
$Spin(n)\times U\left( 1\right) $ with the respect to equivalence
\[
\left( y,z\right) \sim \left( -y,-a\right) ,\quad y\in Spin(m).
\]
This way we define the short exact sequence
\begin{equation}
1\rightarrow U\left( 1\right) \rightarrow Spin^c\left( n\right) \stackrel{S^c%
}{\to }SO\left( n\right) \rightarrow 1,  \label{2.39a}
\end{equation}
where $\rho ^c\left( y,a\right) =\rho ^c\left( y\right) .$ If
$\lambda $ is oriented , real, and rank $n,$ $\gamma $-bundle
$\pi :E_\lambda \rightarrow
M^n,$ with base $M^n,$ the complex spinor structure, spin structure, on $%
\lambda $ is given by the principal bundle $P$ with structural group $%
Spin^c\left( m\right) $ and isomorphism $\lambda \approx P\times
_{Spin^c\left( n\right) }\mathcal{R}^n$ (see (\ref{2.39a})). For
such bundles the categorial equivalence can be defined as
\begin{equation}
\epsilon ^c:\mathcal{E}_{\mathcal{C}}^T\left( M^n\right)
\rightarrow \mathcal{E}_{\mathcal{C}}^\lambda \left( M^n\right)
,  \label{2.40a}
\end{equation}
where $\epsilon ^c\left( E^c\right) =P\bigtriangleup
_{Spin^c\left( n\right)
}E^c$ is the category of trivial complex bundles on $M^n,\mathcal{E}_{%
\mathcal{C}}^\lambda \left( M^n\right) $ is the category of
complex v-bundles on $M^n$ with action of Clifford bundle
$C\left( \lambda \right) ,P\bigtriangleup _{Spin^c(n)}$ and $E^c$
is the factor space of the bundle product $P\times _ME^c$ with
respect to the equivalence $\left( p,e\right)
\sim \left( p\widehat{g}^{-1},\widehat{g}e\right) ,p\in P,e\in E^c,$ where $%
\widehat{g}\in Spin^c\left( n\right) $ acts on $E$ by via the imbedding $%
Spin\left( n\right) \subset C^{0,n}$ and the natural action
$U\left( 1\right) \subset \mathcal{C}$ on complex v-bundle $\xi
^c,E^c=tot\xi ^c,$ for bundle $\pi ^c:E^c\rightarrow M^n.$

Now we return to the bundle $\xi =\mathcal{E}^{<1>}.$ A real
v-bundle (not being a spinor bundle) admits a complex spinor
structure if and only if there exist a homomorphism $\sigma
:U\left( n\right) \rightarrow Spin^c\left( 2n\right) $ making the
diagram 3 commutative. The explicit construction of $\sigma $ for
arbitrary $\gamma $-bundle is given in Refs. \cite{kar} and
\cite{ati}. For $H^{2n}$-spaces it is obvious that a diagram
similar to (\ref{2.40a}) can be defined for the tangent bundle
$TM^n,$ which
directly points to the possibility of defining the $^cSpin$-structure on $%
H^{2n}$--spaces.

Let $\lambda $ be a complex, rank\thinspace $n,$ spinor bundle
with
\begin{equation}
\tau :Spin^c\left( n\right) \times _{\mathcal{Z}/2}U\left(
1\right) \rightarrow U\left( 1\right)   \label{2.41a}
\end{equation}
the homomorphism defined by formula $\tau \left( \lambda ,\delta
\right) =\delta ^2.$ For $P_s$ being the principal bundle with
fiber $Spin^c\left( n\right) $ we introduce the complex linear
bundle $L\left( \lambda ^c\right)
=P_S\times _{Spin^c(n)}\mathcal{C}$ defined as the factor space of $%
P_S\times \mathcal{C}$ on equivalence relation

\[
\left( pt,z\right) \sim \left( p,l\left( t\right) ^{-1}z\right) ,
\]
where $t\in Spin^c\left( n\right) .$ This linear bundle is
associated to complex spinor structure on $\lambda ^c.$

If $\lambda ^c$ and $\lambda ^{c^{\prime }}$ are complex spinor
bundles, the Whitney sum $\lambda ^c\oplus \lambda ^{c^{\prime
}}$ is naturally provided with the structure of the complex
spinor bundle. This follows from the holomorphism
\begin{equation}
\omega ^{\prime }:Spin^c\left( n\right) \times Spin^c\left(
n^{\prime }\right) \rightarrow Spin^c\left( n+n^{\prime }\right)
,  \label{2.42a}
\end{equation}
given by formula $\left[ \left( \beta ,z\right) ,\left( \beta
^{\prime },z^{\prime }\right) \right] \rightarrow \left[ \omega
\left( \beta ,\beta ^{\prime }\right) ,zz^{\prime }\right] ,$
where $\omega $ is the homomorphism making the diagram 4
commutative. Here, $z,z^{\prime }\in U\left( 1\right) .$ It is
obvious that $L\left( \lambda ^c\oplus \lambda ^{c^{\prime
}}\right) $ is isomorphic to $L\left( \lambda ^c\right) \otimes
L\left( \lambda ^{c^{\prime }}\right) .$

We conclude this subsection by formulating our main result on
complex spinor structures for $H^{2n}$-spaces:

\begin{theorem}
Let $\lambda ^c$ be a complex spinor bundle of rank $n$ and
$H^{2n}$-space considered as a real vector bundle $\lambda
^c\oplus \lambda ^{c^{\prime }}$ provided with almost complex
structure $J_{\quad \beta }^\alpha ;$ multiplication on $i$ is
given by $\left(
\begin{array}{cc}
0 & -\delta _j^i \\
\delta _j^i & 0
\end{array}
\right) $. Then, the diagram 5 is commutative up to isomorphisms
$\epsilon ^c $ and $\widetilde{\epsilon }^c$ defined as in
(\ref{2.40a}), $\mathcal{H}$
is functor $E^c\rightarrow E^c\otimes L\left( \lambda ^c\right) $ and $%
\mathcal{E}_{\mathcal{C}}^{0,2n}\left( M^n\right) $ is defined by functor $%
\mathcal{E}_{\mathcal{C}}\left( M^n\right) \rightarrow \mathcal{E}_{\mathcal{%
C}}^{0,2n}\left( M^n\right) $ given as correspondence
$E^c\rightarrow \Lambda \left( \mathcal{C}^n\right) \otimes E^c$
(which is a categorial equivalence), $\Lambda \left(
\mathcal{C}^n\right) $ is the exterior algebra on
$\mathcal{C}^n.$ $W$ is the real bundle $\lambda ^c\oplus \lambda
^{c^{\prime }}$ provided with complex structure.
\end{theorem}

\textbf{Proof: }We use composition of homomorphisms
\[
\mu :Spin^c\left( 2n\right) \stackrel{\pi }{\to }SO\left( n\right) \stackrel{%
r}{\to }U\left( n\right) \stackrel{\sigma }{\to }Spin^c\left(
2n\right) \times _{\mathcal{Z}/2}U\left( 1\right) ,
\]
commutative diagram 6 and introduce composition of homomorphisms
\[
\mu :Spin^c\left( n\right) \stackrel{\Delta }{\to }Spin^c\left(
n\right) \times Spin^c\left( n\right) \stackrel{{\omega }^c}{\to
}Spin^c\left( n\right) ,
\]
where $\Delta $ is the diagonal homomorphism and $\omega ^c$ is
defined as in (\ref{2.42a}). Using homomorphisms (\ref{2.41a})
and ((\ref{2.42a})) we obtain formula $\mu \left( t\right) =\mu
\left( t\right) r\left( t\right) .$

Now consider bundle $P\times _{Spin^c\left( n\right)
}Spin^c\left( 2n\right) $ as the principal $Spin^c\left(
2n\right) $-bundle, associated to $M\oplus M $ being the factor
space of the product $P\times Spin^c\left( 2n\right) $ on the
equivalence relation $\left( p,t,h\right) \sim \left( p,\mu \left(
t\right) ^{-1}h\right) .$ In this case the categorial equivalence (%
\ref{2.40a}) can be rewritten as
\[
\epsilon ^c\left( E^c\right) =P\times _{Spin^c\left( n\right)
}Spin^c\left( 2n\right) \Delta _{Spin^c\left( 2n\right) }E^c
\]
and seen as factor space of $P\times Spin^c\left( 2n\right)
\times _ME^c$ on equivalence relation
\[
\left( pt,h,e\right) \sim \left( p,\mu \left( t\right) ^{-1}h,e\right) %
\mbox{and}\left( p,h_1,h_2,e\right) \sim \left(
p,h_1,h_2^{-1}e\right)
\]
(projections of elements $p$ and $e$ coincides on base $M).$
Every element of $\epsilon ^c\left( E^c\right) $ can be
represented as $P\Delta _{Spin^c\left( n\right) }E^c,$ i.e., as a
factor space $P\Delta E^c$ on equivalence relation $\left(
pt,e\right) \sim \left( p,\mu ^c\left( t\right)
,e\right) ,$ when $t\in Spin^c\left( n\right) .$ The complex line bundle $%
L\left( \lambda ^c\right) $ can be interpreted as the factor
space of\newline
$P\times _{Spin^c\left( n\right) }\mathcal{C}$ on equivalence relation $%
\left( pt,\delta \right) \sim \left( p,r\left( t\right)
^{-1}\delta \right) . $

Putting $\left( p,e\right) \otimes \left( p,\delta \right) \left(
p,\delta e\right) $ we introduce morphism
\[
\epsilon ^c\left( E\right) \times L\left( \lambda ^c\right)
\rightarrow \epsilon ^c\left( \lambda ^c\right)
\]
with properties
\begin{eqnarray*}
\left( pt,e\right) \otimes \left( pt,\delta \right)  &\rightarrow
&\left(
pt,\delta e\right) =\left( p,\mu ^c\left( t\right) ^{-1}\delta e\right) , \\
\left( p,\mu ^c\left( t\right) ^{-1}e\right) \otimes \left(
p,l\left( t\right) ^{-1}e\right)  &\rightarrow &\left( p,\mu
^c\left( t\right) r\left( t\right) ^{-1}\delta e\right)
\end{eqnarray*}
pointing to the fact that we have defined the isomorphism
correctly and that it is an isomorphism on every fiber. $\Box $

\chapter{Spinors and Ha--Spaces}

\section{ D--Spinor Techniques}

The purpose of this section is to show how a corresponding
abstract spinor technique entailing notational and calculations
advantages can be developed
for arbitrary splits of dimensions of a d-vector space $\mathcal{F}=h%
\mathcal{F}\oplus v_1\mathcal{F}\oplus ...\oplus v_z\mathcal{F}$, where $%
\dim h\mathcal{F}=n$ and $\dim v_p\mathcal{F}=m_p.$ For
convenience we shall also present some necessary coordinate
expressions.

The problem of a rigorous definition of spinors on la-spaces
(la-spinors, d-spinors) was posed and solved
\cite{vjmp,viasm1,vsp1}  in the framework of the formalism of
Clifford and spinor structures on v-bundles provided with
compatible nonlinear and distinguished connections and metric. We
introduced
d-spinors as corresponding objects of the Clifford d-algebra $\mathcal{C}%
\left( \mathcal{F},G\right) $, defined for a d-vector space
$\mathcal{F}$ in a standard manner (see, for instance,
\cite{kar}) and proved that operations with $\mathcal{C}\left(
\mathcal{F},G\right) \ $ can be reduced to calculations for
$\mathcal{C}\left( h\mathcal{F},g\right) ,\mathcal{C}\left(
v_1\mathcal{F},h_1\right) ,...$ and $\mathcal{C}\left( v_z\mathcal{F}%
,h_z\right) ,$ which are usual Clifford algebras of respective dimensions $%
2^n,2^{m_1},...$ and $2^{m_z}$ (if it is necessary we can use
quadratic
forms $g$ and $h_p$ correspondingly induced on $h\mathcal{F}$ and $v_p%
\mathcal{F}$ by a metric $\mathbf{G}$ (\ref{dmetrichcv})).
Considering the orthogonal subgroup $O\mathbf{\left( G\right)
}\subset GL\mathbf{\left( G\right) }$ defined by a metric
$\mathbf{G}$ we can define the d-spinor norm
and parametrize d-spinors by ordered pairs of elements of Clifford algebras $%
\mathcal{C}\left( h\mathcal{F},g\right) $ and $\mathcal{C}\left( v_p\mathcal{%
F},h_p\right) ,p=1,2,...z.$ We emphasize that the splitting of a
Clifford d-algebra associated to a dv-bundle $\mathcal{E}^{<z>}$
is a straightforward consequence of the global decomposition
defining a N-connection structure in $\mathcal{E}^{<z>}$.

In this subsection we shall omit detailed proofs which in most
cases are mechanical but rather tedious. We can apply the methods
developed in \cite {pen,penr1,penr2,lue} in a straightforward
manner on h- and v-subbundles in order to verify the correctness
of affirmations.

\subsection{Clifford d--algebra, d--spinors and d--twistors}
\index{d--spinor} \index{d--twistor}

In order to relate the succeeding constructions with Clifford
d-algebras
\cite{vjmp,viasm1} we consider a la-frame decomposition of the metric (%
\ref{dmetrichcv}):
\[
G_{<\alpha ><\beta >}\left( u\right) =l_{<\alpha >}^{<\widehat{\alpha }%
>}\left( u\right) l_{<\beta >}^{<\widehat{\beta }>}\left( u\right) G_{<%
\widehat{\alpha }><\widehat{\beta }>},
\]
where the frame d-vectors and constant metric matrices are
distinguished as

\begin{eqnarray*}
l_{<\alpha >}^{<\widehat{\alpha }>}\left( u\right)  &=&\left(
\begin{array}{cccc}
l_j^{\widehat{j}}\left( u\right)  & 0 & ... & 0 \\
0 & l_{a_1}^{\widehat{a}_1}\left( u\right)  & ... & 0 \\
... & ... & ... & ... \\
0 & 0 & .. & l_{a_z}^{\widehat{a}_z}\left( u\right)
\end{array}
\right) , \\
G_{<\widehat{\alpha }><\widehat{\beta }>} &=&\left(
\begin{array}{cccc}
g_{\widehat{i}\widehat{j}} & 0 & ... & 0 \\
0 & h_{\widehat{a}_1\widehat{b}_1} & ... & 0 \\
... & ... & ... & ... \\
0 & 0 & 0 & h_{\widehat{a}_z\widehat{b}_z}
\end{array}
\right) ,
\end{eqnarray*}
$g_{\widehat{i}\widehat{j}}$ and $h_{\widehat{a}_1\widehat{b}_1},...,h_{%
\widehat{a}_z\widehat{b}_z}$ are diagonal matrices with $g_{\widehat{i}%
\widehat{i}}=$ $h_{\widehat{a}_1\widehat{a}_1}=...=h_{\widehat{a}_z\widehat{b%
}_z}=\pm 1.$

To generate Clifford d-algebras we start with matrix equations
\begin{equation}
\sigma _{<\widehat{\alpha }>}\sigma _{<\widehat{\beta }>}+\sigma _{<\widehat{%
\beta }>}\sigma _{<\widehat{\alpha }>}=-G_{<\widehat{\alpha }><\widehat{%
\beta }>}I,  \label{2.43a}
\end{equation}
where $I$ is the identity matrix, matrices $\sigma _{<\widehat{\alpha }%
>}\,(\sigma $-objects) act on a d-vector space $\mathcal{F}=h\mathcal{F}%
\oplus v_1\mathcal{F}\oplus ...\oplus v_z\mathcal{F}$ and theirs
components are distinguished as
\begin{equation}
\sigma _{<\widehat{\alpha }>}\,=\left\{ (\sigma _{<\widehat{\alpha }>})_{%
\underline{\beta }}^{\cdot \underline{\gamma }}=\left(
\begin{array}{cccc}
(\sigma _{\widehat{i}})_{\underline{j}}^{\cdot \underline{k}} & 0
& ... & 0
\\
0 & (\sigma _{\widehat{a}_1})_{\underline{b}_1}^{\cdot
\underline{c}_1} & ...
& 0 \\
... & ... & ... & ... \\
0 & 0 & ... & (\sigma _{\widehat{a}_z})_{\underline{b}_z}^{\cdot \underline{c%
}_z}
\end{array}
\right) \right\} ,  \label{2.44a}
\end{equation}
indices \underline{$\beta $},\underline{$\gamma $},... refer to
spin spaces of type $\mathcal{S}=S_{(h)}\oplus S_{(v_1)}\oplus
...\oplus S_{(v_z)}$ and
underlined Latin indices \underline{$j$},$\underline{k},...$ and $\underline{%
b}_1,\underline{c}_1,...,\underline{b}_z,\underline{c}_z...$ refer
respectively to h-spin space $\mathcal{S}_{(h)}$ and v$_p$-spin space $%
\mathcal{S}_{(v_p)},(p=1,2,...,z)\ $which are correspondingly
associated to a h- and v$_p$-decomposition of a dv-bundle
$\mathcal{E}^{<z>}.$ The irreducible algebra of matrices $\sigma
_{<\widehat{\alpha }>}$ of minimal dimension $N\times N,$ where
$N=N_{(n)}+N_{(m_1)}+...+N_{(m_z)},$ $\dim
\mathcal{S}_{(h)}$=$N_{(n)}$ and $\dim
\mathcal{S}_{(v_p)}$=$N_{(m_p)},$ has these dimensions
\begin{eqnarray*}
{N_{(n)}} &=&{\left\{
\begin{array}{rl}
{\ 2^{(n-1)/2},} & n=2k+1 \\
{2^{n/2},\ } & n=2k;
\end{array}
\right. }\quad , \\
\quad {N}_{(m_p)}{} &=&{}\left|
\begin{array}{cc}
2^{(m_p-1)/2}, & m_p=2k_p+1 \\
2^{m_p}, & m_p=2k_p
\end{array}
\right| ,
\end{eqnarray*}
where $k=1,2,...,k_p=1,2,....$

The Clifford d-algebra is generated by sums on $n+1$ elements of
form
\begin{equation}
A_1I+B^{\widehat{i}}\sigma _{\widehat{i}}+C^{\widehat{i}\widehat{j}}\sigma _{%
\widehat{i}\widehat{j}}+D^{\widehat{i}\widehat{j}\widehat{k}}\sigma _{%
\widehat{i}\widehat{j}\widehat{k}}+...  \label{2.45a}
\end{equation}
and sums of $m_p+1$ elements of form
\[
A_{2(p)}I+B^{\widehat{a}_p}\sigma _{\widehat{a}_p}+C^{\widehat{a}_p\widehat{b%
}_p}\sigma _{\widehat{a}_p\widehat{b}_p}+D^{\widehat{a}_p\widehat{b}_p%
\widehat{c}_p}\sigma
_{\widehat{a}_p\widehat{b}_p\widehat{c}_p}+...
\]
with antisymmetric coefficients
\[
C^{\widehat{i}\widehat{j}}=C^{[\widehat{i}\widehat{j}]},C^{\widehat{a}_p%
\widehat{b}_p}=C^{[\widehat{a}_p\widehat{b}_p]},D^{\widehat{i}\widehat{j}%
\widehat{k}}=D^{[\widehat{i}\widehat{j}\widehat{k}]},D^{\widehat{a}_p%
\widehat{b}_p\widehat{c}_p}=D^{[\widehat{a}_p\widehat{b}_p\widehat{c}_p]},...
\]
and matrices
\[
\sigma _{\widehat{i}\widehat{j}}=\sigma _{[\widehat{i}}\sigma _{\widehat{j}%
]},\sigma _{\widehat{a}_p\widehat{b}_p}=\sigma _{[\widehat{a}_p}\sigma _{%
\widehat{b}_p]},\sigma _{\widehat{i}\widehat{j}\widehat{k}}=\sigma _{[%
\widehat{i}}\sigma _{\widehat{j}}\sigma _{\widehat{k}]},....
\]
Really, we have 2$^{n+1}$ coefficients $\left( A_1,C^{\widehat{i}\widehat{j}%
},D^{\widehat{i}\widehat{j}\widehat{k}},...\right) $ and
2$^{m_p+1}$
coefficients $(A_{2(p)},C^{\widehat{a}_p\widehat{b}_p},D^{\widehat{a}_p%
\widehat{b}_p\widehat{c}_p},...)$ of the Clifford algebra on
$\mathcal{F}$.

For simplicity, we shall present the necessary geometric
constructions only for h-spin spaces $\mathcal{S}_{(h)}$ of
dimension $N_{(n)}.$ Considerations for a v-spin space
$\mathcal{S}_{(v)}$ are similar but with proper characteristics
for a dimension $N_{(m)}.$

In order to define the scalar (spinor) product on
$\mathcal{S}_{(h)}$ we introduce into consideration this finite
sum (because of a finite number of elements $\sigma
_{[\widehat{i}\widehat{j}...\widehat{k}]}$):
\begin{eqnarray}
^{(\pm
)}E_{\underline{k}\underline{m}}^{\underline{i}\underline{j}}
&=&\delta _{\underline{k}}^{\underline{i}}\delta _{\underline{m}}^{%
\underline{j}}+\frac 2{1!}(\sigma _{\widehat{i}})_{\underline{k}}^{.%
\underline{i}}(\sigma ^{\widehat{i}})_{\underline{m}}^{.\underline{j}}+\frac{%
2^2}{2!}(\sigma _{\widehat{i}\widehat{j}})_{\underline{k}}^{.\underline{i}%
}(\sigma
^{\widehat{i}\widehat{j}})_{\underline{m}}^{.\underline{j}}
\nonumber \\
&&+\frac{2^3}{3!}(\sigma _{\widehat{i}\widehat{j}\widehat{k}})_{\underline{k}%
}^{.\underline{i}}(\sigma ^{\widehat{i}\widehat{j}\widehat{k}})_{\underline{m%
}}^{.\underline{j}}+...  \label{2.46a}
\end{eqnarray}
which can be factorized as
\begin{equation}
^{(\pm )}E_{\underline{k}\underline{m}}^{\underline{i}\underline{j}}=N_{(n)}{%
\ }^{(\pm )}\epsilon _{\underline{k}\underline{m}}{\ }^{(\pm )}\epsilon ^{%
\underline{i}\underline{j}}\mbox{ for }n=2k  \label{2.47a}
\end{equation}
and
\begin{eqnarray}
^{(+)}E_{\underline{k}\underline{m}}^{\underline{i}\underline{j}}
&=&2N_{(n)}\epsilon _{\underline{k}\underline{m}}\epsilon ^{\underline{i}%
\underline{j}},{\ }^{(-)}E_{\underline{k}\underline{m}}^{\underline{i}%
\underline{j}}=0\mbox{ for }n=3(mod4),  \label{2.48a} \\
^{(+)}E_{\underline{k}\underline{m}}^{\underline{i}\underline{j}} &=&0,{\ }%
^{(-)}E_{\underline{k}\underline{m}}^{\underline{i}\underline{j}%
}=2N_{(n)}\epsilon _{\underline{k}\underline{m}}\epsilon ^{\underline{i}%
\underline{j}}\mbox{ for }n=1(mod4).  \nonumber
\end{eqnarray}

Antisymmetry of $\sigma _{\widehat{i}\widehat{j}\widehat{k}...}$
and the construction of the objects (\ref{2.45a})--(\ref{2.48a})
define the
properties of $\epsilon $-objects $^{(\pm )}\epsilon _{\underline{k}%
\underline{m}}$ and $\epsilon _{\underline{k}\underline{m}}$
which have an eight-fold periodicity on $n$ (see details in
\cite{penr2} and, with respect to locally anisotropic spaces,
\cite{vjmp}).

For even values of $n$ it is possible the decomposition of every
h-spin space $\mathcal{S}_{(h)}$ into irreducible h-spin spaces
$\mathbf{S}_{(h)}$ and $\mathbf{S}_{(h)}^{\prime }$ (one
considers splitting of h-indices, for instance,
\underline{$l$}$=L\oplus L^{\prime },\underline{m}=M\oplus
M^{\prime },...;$ for v$_p$-indices we shall write $\underline{a}%
_p=A_p\oplus A_p^{\prime },\underline{b}_p=B_p\oplus B_p^{\prime
},...)$ and defines new $\epsilon $-objects
\begin{equation}
\epsilon ^{\underline{l}\underline{m}}=\frac 12\left( ^{(+)}\epsilon ^{%
\underline{l}\underline{m}}+^{(-)}\epsilon ^{\underline{l}\underline{m}%
}\right) \mbox{ and }\widetilde{\epsilon }^{\underline{l}\underline{m}%
}=\frac 12\left( ^{(+)}\epsilon
^{\underline{l}\underline{m}}-^{(-)}\epsilon
^{\underline{l}\underline{m}}\right)   \label{2.49a}
\end{equation}
We shall omit similar formulas for $\epsilon $-objects with lower
indices.

In general,  the spinor  $\epsilon $-objects should be defined
for every shell of an\-iso\-tro\-py according the formulas
(\ref{2.60}) where instead of dimension $n$ we shall consider the
dimensions $m_p$, $1\leq p\leq z,$ of shells.

We define a d-spinor space $\mathcal{S}_{(n,m_1)}\ $ as a direct
sum of a horizontal and a vertical spinor spaces of type
(\ref{2.55}), for instance,
\[
\mathcal{S}_{(8k,8k^{\prime })}=\mathbf{S}_{\circ }\oplus
\mathbf{S}_{\circ
}^{\prime }\oplus \mathbf{S}_{|\circ }\oplus \mathbf{S}_{|\circ }^{\prime },%
\mathcal{S}_{(8k,8k^{\prime }+1)}\ =\mathbf{S}_{\circ }\oplus \mathbf{S}%
_{\circ }^{\prime }\oplus \mathcal{S}_{|\circ }^{(-)},...,
\]
\[
\mathcal{S}_{(8k+4,8k^{\prime }+5)}=\mathbf{S}_{\triangle }\oplus \mathbf{S}%
_{\triangle }^{\prime }\oplus \mathcal{S}_{|\triangle }^{(-)},...
\]
The scalar product on a $\mathcal{S}_{(n,m_1)}\ $ is induced by
(corresponding to fixed values of $n$ and $m_1$ ) $\epsilon $-objects (%
\ref{2.60}) considered for h- and v$_1$-components. We present
also an
example for $\mathcal{S}_{(n,m_1+...+m_z)}:$%
\begin{eqnarray}
 & &
\mathcal{S}_{(8k+4,8k_{(1)}+5,...,8k_{(p)}+4,...8k_{(z)})}=
\nonumber \\
 & &
 [ \mathbf{S}_{\triangle }\oplus \mathbf{S}_{\triangle }^{\prime
}\oplus \mathcal{S}_{|(1)\triangle }^{(-)}\oplus ...\oplus
\mathbf{S}_{|(p)\triangle
}\oplus \mathbf{S}_{|(p)\triangle }^{\prime }\oplus ...\oplus \mathbf{S}%
_{|(z)\circ }\oplus \mathbf{S}_{|(z)\circ }^{\prime }. \nonumber
\end{eqnarray}

Having introduced d-spinors for dimensions $\left(
n,m_1+...+m_z\right) $ we can write out the generalization for
ha--spaces of twistor equations \cite {penr1} by using the
distinguished $\sigma $-objects (\ref{2.44a}):
\begin{equation}
(\sigma _{(<\widehat{\alpha }>})_{|\underline{\beta
}|}^{..\underline{\gamma }}\quad \frac{\delta \omega
^{\underline{\beta }}}{\delta u^{<\widehat{\beta
}>)}}=\frac 1{n+m_1+...+m_z}\quad G_{<\widehat{\alpha }><\widehat{\beta }%
>}(\sigma ^{\widehat{\epsilon }})_{\underline{\beta }}^{..\underline{\gamma }%
}\quad \frac{\delta \omega ^{\underline{\beta }}}{\delta u^{\widehat{%
\epsilon }}},  \label{2.56a}
\end{equation}
where $\left| \underline{\beta }\right| $ denotes that we do not
consider symmetrization on this index. The general solution of
(\ref{2.56a}) on the d-vector space $\mathcal{F}$ looks like as
\begin{equation}
\omega ^{\underline{\beta }}=\Omega ^{\underline{\beta }}+u^{<\widehat{%
\alpha }>}(\sigma _{<\widehat{\alpha }>})_{\underline{\epsilon }}^{..%
\underline{\beta }}\Pi ^{\underline{\epsilon }},  \label{2.57a}
\end{equation}
where $\Omega ^{\underline{\beta }}$ and $\Pi
^{\underline{\epsilon }}$ are constant d-spinors. For fixed
values of dimensions $n$ and $m=m_1+...m_z$ we mast analyze the
reduced and irreducible components of h- and v$_p$-parts of
equations (\ref{2.56a}) and their solutions (\ref{2.57a}) in
order to find the symmetry properties of a d-twistor
$\mathbf{Z^\alpha \ }$ defined as a pair of d-spinors
\[
\mathbf{Z}^\alpha =(\omega ^{\underline{\alpha }},\pi _{\underline{\beta }%
}^{\prime }),
\]
where $\pi _{\underline{\beta }^{\prime }}=\pi _{\underline{\beta
}^{\prime }}^{(0)}\in
{\widetilde{\mathcal{S}}}_{(n,m_1,...,m_z)}$ is a constant dual
d-spinor. The problem of definition of spinors and twistors on
ha-spaces was firstly considered in \cite{vdeb} (see also
\cite{v87}) in connection with the possibility to extend the
equations (\ref{2.57a}) and theirs solutions (\ref{2.58a}), by
using nearly autoparallel maps, on curved, locally isotropic or
anisotropic, spaces. We note that the definition of twistors have
been extended to higher order anisotropic spaces with trivial N--
and d--connections.

\subsection{ Mutual transforms of d-tensors and d-spinors}

The spinor algebra for spaces of higher dimensions can not be
considered as a real alternative to the tensor algebra as for
locally isotropic spaces of dimensions $n=3,4$
\cite{pen,penr1,penr2}. The same holds true for ha--spaces and we
emphasize that it is not quite convenient to perform a spinor
calculus for dimensions $n,m>>4$. Nevertheless, the concept of
spinors is important for every type of spaces, we can deeply
understand the fundamental properties of geometical objects on
ha--spaces, and we shall consider in this subsection some
questions concerning transforms of d-tensor objects into d-spinor
ones.

\subsection{ Transformation of d-tensors into d-spinors}

In order to pass from d-tensors to d-spinors we must use $\sigma $-objects (%
\ref{2.44a}) written in reduced or irreduced form \quad (in
dependence of fixed values of dimensions $n$ and $m$ ):

\begin{eqnarray}
&&(\sigma _{<\widehat{\alpha }>})_{\underline{\beta }}^{\cdot \underline{%
\gamma }},~(\sigma ^{<\widehat{\alpha }>})^{\underline{\beta }\underline{%
\gamma }},~(\sigma ^{<\widehat{\alpha }>})_{\underline{\beta }\underline{%
\gamma }},...,(\sigma
_{<\widehat{a}>})^{\underline{b}\underline{c}},...,
\label{2.58a} \\
&&(\sigma _{\widehat{i}})_{\underline{j}\underline{k}},...,(\sigma _{<%
\widehat{a}>})^{AA^{\prime }},...,(\sigma
^{\widehat{i}})_{II^{\prime }},.... \nonumber
\end{eqnarray}
It is obvious that contracting with corresponding $\sigma $-objects (%
\ref{2.58a}) we can introduce instead of d-tensors indices the
d-spinor ones, for instance,
\[
\omega ^{\underline{\beta }\underline{\gamma }}=(\sigma ^{<\widehat{\alpha }%
>})^{\underline{\beta }\underline{\gamma }}\omega _{<\widehat{\alpha }%
>},\quad \omega _{AB^{\prime }}=(\sigma ^{<\widehat{a}>})_{AB^{\prime
}}\omega _{<\widehat{a}>},\quad ...,\zeta _{\cdot \underline{j}}^{\underline{%
i}}=(\sigma ^{\widehat{k}})_{\cdot \underline{j}}^{\underline{i}}\zeta _{%
\widehat{k}},....
\]
For d-tensors containing groups of antisymmetric indices there is
a more simple procedure of theirs transforming into d-spinors
because the objects
\begin{equation}
(\sigma _{\widehat{\alpha }\widehat{\beta }...\widehat{\gamma }})^{%
\underline{\delta }\underline{\nu }},\quad (\sigma ^{\widehat{a}\widehat{b}%
...\widehat{c}})^{\underline{d}\underline{e}},\quad ...,(\sigma ^{\widehat{i}%
\widehat{j}...\widehat{k}})_{II^{\prime }},\quad ... \label{2.59a}
\end{equation}
can be used for sets of such indices into pairs of d-spinor
indices. Let us enumerate some properties of $\sigma $-objects of
type (\ref{2.59a}) (for
simplicity we consider only h-components having q indices $\widehat{i},%
\widehat{j},\widehat{k},...$ taking values from 1 to $n;$ the properties of v%
$_p$-components can be written in a similar manner with respect to indices $%
\widehat{a}_p,\widehat{b}_p,\widehat{c}_p...$ taking values from
1 to $m$):
\begin{equation}
(\sigma _{\widehat{i}...\widehat{j}})^{\underline{k}\underline{l}}%
\mbox{
 is\ }\left\{ \
\begin{array}{c}
\mbox{symmetric on }\underline{k},\underline{l}\mbox{ for
}n-2q\equiv 1,7~(mod~8); \\
\mbox{antisymmetric on }\underline{k},\underline{l}\mbox{ for
}n-2q\equiv 3,5~(mod~8)
\end{array}
\right\}   \label{2.60a}
\end{equation}
for odd values of $n,$ and an object
\[
(\sigma _{\widehat{i}...\widehat{j}})^{IJ}~\left( (\sigma _{\widehat{i}...%
\widehat{j}})^{I^{\prime }J^{\prime }}\right)
\]
\begin{equation}
\mbox{ is\ }\left\{
\begin{array}{c}
\mbox{symmetric on }I,J~(I^{\prime },J^{\prime })\mbox{ for
}n-2q\equiv 0~(mod~8); \\
\mbox{antisymmetric on }I,J~(I^{\prime },J^{\prime })\mbox{ for
}n-2q\equiv 4~(mod~8)
\end{array}
\right\}   \label{2.61a}
\end{equation}
or
\begin{equation}
(\sigma _{\widehat{i}...\widehat{j}})^{IJ^{\prime }}=\pm (\sigma _{\widehat{i%
}...\widehat{j}})^{J^{\prime }I}\{
\begin{array}{c}
n+2q\equiv 6(mod8); \\
n+2q\equiv 2(mod8),
\end{array}
\label{2.62a}
\end{equation}
with vanishing of the rest of reduced components of the d-tensor $(\sigma _{%
\widehat{i}...\widehat{j}})^{\underline{k}\underline{l}}$ with
prime/ unprime sets of indices.

\subsection{  Fundamental d--spinors}

We can transform every d--spinor $\xi ^{\underline{\alpha }}=\left( \xi ^{%
\underline{i}},\xi ^{\underline{a}_1},...,\xi
^{\underline{a}_z}\right) $ into a corresponding d-tensor. For
simplicity, we consider this construction only for a h-component
$\xi ^{\underline{i}}$ on a h-space being of dimension $n$. The
values
\begin{equation}
\xi ^{\underline{\alpha }}\xi ^{\underline{\beta }}(\sigma ^{\widehat{i}...%
\widehat{j}})_{\underline{\alpha }\underline{\beta }}\quad \left( n%
\mbox{ is odd}\right)   \label{2.63a}
\end{equation}
or
\begin{equation}
\xi ^I\xi ^J(\sigma ^{\widehat{i}...\widehat{j}})_{IJ}~\left(
\mbox{or }\xi
^{I^{\prime }}\xi ^{J^{\prime }}(\sigma ^{\widehat{i}...\widehat{j}%
})_{I^{\prime }J^{\prime }}\right) ~\left( n\mbox{ is even}\right)
\label{2.64a}
\end{equation}
with a different number of indices $\widehat{i}...\widehat{j},$
taken together, defines the h-spinor \index{h--spinor}  $\xi
^{\underline{i}}\,$ to an accuracy to the sign. We emphasize that
it is necessary to choose only those h-components of d-tensors
(\ref{2.63a}) (or (\ref{2.64a})) which are
symmetric on pairs of indices $\underline{\alpha }\underline{\beta }$ (or $%
IJ\,$ (or $I^{\prime }J^{\prime }$ )) and the number $q$ of indices $%
\widehat{i}...\widehat{j}$ satisfies the condition (as a
respective
consequence of the properties (\ref{2.60a}) and/ or (\ref{2.61a}), (%
\ref{2.62a}))
\begin{equation}
n-2q\equiv 0,1,7~(mod~8).  \label{2.65a}
\end{equation}
Of special interest is the case when
\begin{equation}
q=\frac 12\left( n\pm 1\right) ~\left( n\mbox{ is odd}\right)
\label{2.66a}
\end{equation}
or
\begin{equation}
q=\frac 12n~\left( n\mbox{ is even}\right) .  \label{2.67a}
\end{equation}
If all expressions (\ref{2.63a}) and/or (\ref{2.64a}) are zero
for all values of $q\,$ with the exception of one or two ones
defined by the
conditions (\ref{2.65a}), (\ref{2.66a}) (or (\ref{2.67a})), the value $\xi ^{%
\widehat{i}}$ (or $\xi ^I$ ($\xi ^{I^{\prime }}))$ is called a
fundamental h-spinor. Defining in a similar manner the
fundamental v-spinors \index{v--spinor} we can introduce
fundamental d-spinors as pairs of fundamental h- and v-spinors.
Here we remark that a h(v$_p$)-spinor $\xi ^{\widehat{i}}~(\xi ^{\widehat{a}%
_p})\,$ (we can also consider reduced components) is always a
fundamental one for $n(m)<7,$ which is a consequence of
(\ref{2.67a})).

\section{ Differential Geometry of Ha--Spinors}

This subsection is devoted to the differential geometry of
d--spinors in higher order anisotropic spaces. We shall use
denotations of type
\[
v^{<\alpha >}=(v^i,v^{<a>})\in \sigma ^{<\alpha >}=(\sigma
^i,\sigma ^{<a>})
\]
and
\[
\zeta ^{\underline{\alpha }_p}=(\zeta ^{\underline{i}_p},\zeta ^{\underline{a%
}_p})\in \sigma ^{\alpha _p}=(\sigma ^{i_p},\sigma ^{a_p})\,
\]
for, respectively, elements of modules of d-vector and irreduced
d-spinor fields (see details in \cite{vjmp}). D-tensors and
d-spinor tensors
(irreduced or reduced) will be interpreted as elements of corresponding $%
\mathcal{\sigma }$--modules, for instance,
\[
q_{~<\beta >...}^{<\alpha >}\in \mathcal{\sigma ^{<\alpha >}\mathbf{/}}%
^{\prime };[-0\mathcal{_{<\beta >}},\psi _{~\underline{\beta }_p\quad ...}^{%
\underline{\alpha }_p\quad \underline{\gamma }_p}\in \mathcal{\sigma }_{~%
\underline{\beta _p}\quad ...}^{\underline{\alpha }_p\quad
\underline{\gamma }_p}~,\xi _{\quad J_pK_p^{\prime }N_p^{\prime
}}^{I_pI_p^{\prime }}\in \mathcal{\sigma }_{\quad J_pK_p^{\prime
}N_p^{\prime }}^{I_pI_p^{\prime }}~,...
\]

We can establish a correspondence between the higher order
anisotropic adapted to the N--connection metric $g_{\alpha \beta
}$ (\ref{dmetrichcv})
and d-spinor metric $\epsilon _{\underline{\alpha }\underline{\beta }}$ ($%
\epsilon $-objects (\ref{2.60}) for both h- and v$_p$-subspaces of $\mathcal{%
E}^{<z>}$) of a ha--space $\mathcal{E}^{<z>}$ by using the
relation
\begin{equation}
g_{<\alpha ><\beta >}=-\frac 1{N(n)+N(m_1)+...+N(m_z)}\times
\label{2.68a}
\end{equation}
\[
((\sigma _{(<\alpha >}(u))^{\underline{\alpha }\underline{\beta
}}(\sigma
_{<\beta >)}(u))^{\underline{\delta }\underline{\gamma }})\epsilon _{%
\underline{\alpha }\underline{\gamma }}\epsilon _{\underline{\beta }%
\underline{\delta }},
\]
where
\begin{equation}
(\sigma _{<\alpha >}(u))^{\underline{\nu }\underline{\gamma
}}=l_{<\alpha
>}^{<\widehat{\alpha }>}(u)(\sigma _{<\widehat{\alpha }>})^{<\underline{\nu }%
><\underline{\gamma }>},  \label{2.69a}
\end{equation}
which is a consequence of formulas (\ref{2.43a})--(\ref{2.49a}).
In brief we can write (\ref{2.68a}) as
\begin{equation}
g_{<\alpha ><\beta >}=\epsilon _{\underline{\alpha }_1\underline{\alpha }%
_2}\epsilon _{\underline{\beta }_1\underline{\beta }_2}
\label{2.70a}
\end{equation}
if the $\sigma $-objects are considered as a fixed structure, whereas $%
\epsilon $-objects are treated as caring the metric ''dynamics ''
, on higher order anisotropic space. This variant is used, for
instance, in the so-called 2-spinor geometry \cite{penr1,penr2}
and should be preferred if we have to make explicit the algebraic
symmetry properties of d-spinor objects by using metric
decomposition (\ref{2.70a}). An alternative way is to consider as
fixed the algebraic structure of $\epsilon $-objects and to use
variable components of $\sigma $-objects of type (\ref{2.69a}) for
developing a variational d-spinor approach to gravitational and
matter field interactions on ha-spaces ( the spinor Ashtekar
variables \cite{ash} are introduced in this manner).

We note that a d--spinor metric
\[
\epsilon _{\underline{\nu }\underline{\tau }}=\left(
\begin{array}{cccc}
\epsilon _{\underline{i}\underline{j}} & 0 & ... & 0 \\
0 & \epsilon _{\underline{a}_1\underline{b}_1} & ... & 0 \\
... & ... & ... & ... \\
0 & 0 & ... & \epsilon _{\underline{a}_z\underline{b}_z}
\end{array}
\right)
\]
on the d-spinor space $\mathcal{S}=(\mathcal{S}_{(h)},\mathcal{S}%
_{(v_1)},...,\mathcal{S}_{(v_z)})$ can have symmetric or antisymmetric h (v$%
_p$) -components $\epsilon _{\underline{i}\underline{j}}$ ($\epsilon _{%
\underline{a}_p\underline{b}_p})$ , see $\epsilon $-objects
(\ref{2.60}). For simplicity, in order to avoid cumbersome
calculations connected with
eight-fold periodicity on dimensions $n$ and $m_p$ of a ha-space $\mathcal{E}%
^{<z>},$ we shall develop a general d-spinor formalism only by
using irreduced spinor spaces $\mathcal{S}_{(h)}$ and
$\mathcal{S}_{(v_p)}.$

\subsection{ D-covariant derivation on ha--spaces}

Let $\mathcal{E}^{<z>}$ be a ha-space. We define the action on a
d-spinor of a d-covariant operator
\begin{eqnarray*}
\bigtriangledown _{<\alpha >} &=&\left( \bigtriangledown
_i,\bigtriangledown _{<a>}\right) \\
 & =&(\sigma _{<\alpha
>})^{\underline{\alpha }_1\underline{\alpha }_2}\bigtriangledown
_{^{\underline{\alpha }_1\underline{\alpha }_2}}=\left(
(\sigma _i)^{\underline{i}_1\underline{i}_2}\bigtriangledown _{^{\underline{i%
}_1\underline{i}_2}},~(\sigma _{<a>})^{\underline{a}_1\underline{a}%
_2}\bigtriangledown _{^{\underline{a}_1\underline{a}_2}}\right)  \\
&=&((\sigma _i)^{\underline{i}_1\underline{i}_2}\bigtriangledown _{^{%
\underline{i}_1\underline{i}_2}},~(\sigma _{a_1})^{\underline{a}_1\underline{%
a}_2}\bigtriangledown _{(1)^{\underline{a}_1\underline{a}_2}},..., \\
&&(\sigma _{a_p})^{\underline{a}_1\underline{a}_2}\bigtriangledown _{(p)^{%
\underline{a}_1\underline{a}_2}},...,(\sigma _{a_z})^{\underline{a}_1%
\underline{a}_2}\bigtriangledown
_{(z)^{\underline{a}_1\underline{a}_2}})
\end{eqnarray*}
(in brief, we shall write
\[
\bigtriangledown _{<\alpha >}=\bigtriangledown _{^{\underline{\alpha }_1%
\underline{\alpha }_2}}=\left( \bigtriangledown _{^{\underline{i}_1%
\underline{i}_2}},~\bigtriangledown _{(1)^{\underline{a}_1\underline{a}%
_2}},...,\bigtriangledown _{(p)^{\underline{a}_1\underline{a}%
_2}},...,\bigtriangledown
_{(z)^{\underline{a}_1\underline{a}_2}}\right) )
\]
as maps
\[
\bigtriangledown _{{\underline{\alpha }}_1{\underline{\alpha
}}_2}\ :\
\mathcal{\sigma }^{\underline{\beta }}\rightarrow \sigma _{<\alpha >}^{%
\underline{\beta }}=\sigma _{{\underline{\alpha }}_1{\underline{\alpha }}%
_2}^{\underline{\beta }}=
\]
\[
\left( \sigma _i^{\underline{\beta }}=\sigma _{{\underline{i}}_1{\underline{i%
}}_2}^{\underline{\beta }},\sigma _{(1)a_1}^{\underline{\beta }}=\sigma _{(1)%
{\underline{\alpha }}_1{\underline{\alpha }}_2}^{\underline{\beta }%
},...,\sigma _{(p)a_p}^{\underline{\beta }}=\sigma _{(p){\underline{\alpha }}%
_1{\underline{\alpha }}_2}^{\underline{\beta }},...,\sigma _{(z)a_z}^{%
\underline{\beta }}=\sigma _{(z){\underline{\alpha }}_1{\underline{\alpha }}%
_2}^{\underline{\beta }}\right)
\]
satisfying conditions
\[
\bigtriangledown _{<\alpha >}(\xi ^{\underline{\beta }}+\eta ^{\underline{%
\beta }})=\bigtriangledown _{<\alpha >}\xi ^{\underline{\beta }%
}+\bigtriangledown _{<\alpha >}\eta ^{\underline{\beta }},
\]
and
\[
\bigtriangledown _{<\alpha >}(f\xi ^{\underline{\beta
}})=f\bigtriangledown
_{<\alpha >}\xi ^{\underline{\beta }}+\xi ^{\underline{\beta }%
}\bigtriangledown _{<\alpha >}f
\]
for every $\xi ^{\underline{\beta }},\eta ^{\underline{\beta }}\in \mathcal{%
\sigma ^{\underline{\beta }}}$ and $f$ being a scalar field on $\mathcal{E}%
^{<z>}.\mathcal{\ }$ It is also required that one holds the
Leibnitz rule
\[
(\bigtriangledown _{<\alpha >}\zeta _{\underline{\beta }})\eta ^{\underline{%
\beta }}=\bigtriangledown _{<\alpha >}(\zeta _{\underline{\beta }}\eta ^{%
\underline{\beta }})-\zeta _{\underline{\beta }}\bigtriangledown
_{<\alpha
>}\eta ^{\underline{\beta }}
\]
and that $\bigtriangledown _{<\alpha >}\,$ is a real operator,
i.e. it commuters with the operation of complex conjugation:
\[
\overline{\bigtriangledown _{<\alpha >}\psi _{\underline{\alpha }\underline{%
\beta }\underline{\gamma }...}}=\bigtriangledown _{<\alpha
>}(\overline{\psi }_{\underline{\alpha }\underline{\beta
}\underline{\gamma }...}).
\]

Let now analyze the question on uniqueness of action on d-spinors
of an operator $\bigtriangledown _{<\alpha >}$ satisfying
necessary conditions . Denoting by $\bigtriangledown _{<\alpha
>}^{(1)}$ and $\bigtriangledown _{<\alpha >}$ two such
d-covariant operators we consider the map
\begin{equation}
(\bigtriangledown _{<\alpha >}^{(1)}-\bigtriangledown _{<\alpha >}):\mathcal{%
\sigma ^{\underline{\beta }}\rightarrow \sigma _{\underline{\alpha }_1%
\underline{\alpha }_2}^{\underline{\beta }}}.  \label{2.71a}
\end{equation}
Because the action on a scalar $f$ of both operators
$\bigtriangledown _\alpha ^{(1)}$ and $\bigtriangledown _\alpha $
must be identical, i.e.
\[
\bigtriangledown _{<\alpha >}^{(1)}f=\bigtriangledown _{<\alpha
>}f,
\]
the action (\ref{2.71a}) on $f=\omega _{\underline{\beta }}\xi ^{\underline{%
\beta }}$ must be written as
\[
(\bigtriangledown _{<\alpha >}^{(1)}-\bigtriangledown _{<\alpha >})(\omega _{%
\underline{\beta }}\xi ^{\underline{\beta }})=0.
\]
In consequence we conclude that there is an element $\Theta _{\underline{%
\alpha }_1\underline{\alpha }_2\underline{\beta }}^{\quad \quad \underline{%
\gamma }}\in \mathcal{\sigma }_{\underline{\alpha }_1\underline{\alpha }_2%
\underline{\beta }}^{\quad \quad \underline{\gamma }}$ for which
\begin{eqnarray}
\bigtriangledown _{\underline{\alpha }_1\underline{\alpha }_2}^{(1)}\xi ^{%
\underline{\gamma }} &=&\bigtriangledown _{\underline{\alpha }_1\underline{%
\alpha }_2}\xi ^{\underline{\gamma }}+\Theta _{\underline{\alpha }_1%
\underline{\alpha }_2\underline{\beta }}^{\quad \quad \underline{\gamma }%
}\xi ^{\underline{\beta }},  \nonumber \\
\bigtriangledown _{\underline{\alpha }_1\underline{\alpha }_2}^{(1)}\omega _{%
\underline{\beta }} &=&\bigtriangledown _{\underline{\alpha }_1\underline{%
\alpha }_2}\omega _{\underline{\beta }}-\Theta _{\underline{\alpha }_1%
\underline{\alpha }_2\underline{\beta }}^{\quad \quad \underline{\gamma }%
}\omega _{\underline{\gamma }}~.  \label{2.72a}
\end{eqnarray}
The action of the operator (\ref{2.71a}) on a d-vector $v^{<\beta >}=v^{%
\underline{\beta }_1\underline{\beta }_2}$ can be written by using formula (%
\ref{2.72a}) for both indices $\underline{\beta }_1$ and $\underline{\beta }%
_2$ :
\begin{eqnarray*}
(\bigtriangledown _{<\alpha >}^{(1)}-\bigtriangledown _{<\alpha >})v^{%
\underline{\beta }_1\underline{\beta }_2} &=&\Theta _{<\alpha >\underline{%
\gamma }}^{\quad \underline{\beta }_1}v^{\underline{\gamma
}\underline{\beta
}_2}+\Theta _{<\alpha >\underline{\gamma }}^{\quad \underline{\beta }_2}v^{%
\underline{\beta }_1\underline{\gamma }} \\
&=&(\Theta _{<\alpha >\underline{\gamma }_1}^{\quad \underline{\beta }%
_1}\delta _{\underline{\gamma }_2}^{\quad \underline{\beta
}_2}+\Theta
_{<\alpha >\underline{\gamma }_1}^{\quad \underline{\beta }_2}\delta _{%
\underline{\gamma }_2}^{\quad \underline{\beta }_1})v^{\underline{\gamma }_1%
\underline{\gamma }_2} \\
 &=& Q_{\ <\alpha ><\gamma >}^{<\beta >}v^{<\gamma >},
\end{eqnarray*}
where
\begin{equation}
Q_{\ <\alpha ><\gamma >}^{<\beta >}=Q_{\qquad \underline{\alpha }_1%
\underline{\alpha }_2~\underline{\gamma }_1\underline{\gamma }_2}^{%
\underline{\beta }_1\underline{\beta }_2}=\Theta _{<\alpha >\underline{%
\gamma }_1}^{\quad \underline{\beta }_1}\delta _{\underline{\gamma }%
_2}^{\quad \underline{\beta }_2}+\Theta _{<\alpha >\underline{\gamma }%
_1}^{\quad \underline{\beta }_2}\delta _{\underline{\gamma
}_2}^{\quad \underline{\beta }_1}.  \label{2.73a}
\end{equation}
The d-commutator $\bigtriangledown _{[<\alpha >}\bigtriangledown
_{<\beta >]} $ defines the d-torsion.  So, applying operators
$\bigtriangledown
_{[<\alpha >}^{(1)}\bigtriangledown _{<\beta >]}^{(1)}$ and $%
\bigtriangledown _{[<\alpha >}\bigtriangledown _{<\beta >]}$ on $f=\omega _{%
\underline{\beta }}\xi ^{\underline{\beta }}$ we can write
\[
T_{\quad <\alpha ><\beta >}^{(1)<\gamma >}-T_{~<\alpha ><\beta
>}^{<\gamma
>}=Q_{~<\beta ><\alpha >}^{<\gamma >}-Q_{~<\alpha ><\beta >}^{<\gamma >}
\]
with $Q_{~<\alpha ><\beta >}^{<\gamma >}$ from (\ref{2.73a}).

The action of operator $\bigtriangledown _{<\alpha >}^{(1)}$ on
d-spinor
tensors of type $\chi _{\underline{\alpha }_1\underline{\alpha }_2\underline{%
\alpha }_3...}^{\qquad \quad \underline{\beta }_1\underline{\beta
}_2...}$
must be constructed by using formula (\ref{2.72a}) for every upper index $%
\underline{\beta }_1\underline{\beta }_2...$ and formula
(\ref{2.73a}) for
every lower index $\underline{\alpha }_1\underline{\alpha }_2\underline{%
\alpha }_3...$ .

\subsection{Infeld--van der Waerden co\-ef\-fi\-ci\-ents }

Let
\[
\delta _{\underline{\mathbf{\alpha }}}^{\quad \underline{\alpha
}}=\left(
\delta _{\underline{\mathbf{1}}}^{\quad \underline{i}},\delta _{\underline{%
\mathbf{2}}}^{\quad \underline{i}},...,\delta _{\underline{\mathbf{N(n)}}%
}^{\quad \underline{i}},\delta _{\underline{\mathbf{1}}}^{\quad \underline{a}%
},\delta _{\underline{\mathbf{2}}}^{\quad \underline{a}},...,\delta _{%
\underline{\mathbf{N(m)}}}^{\quad \underline{i}}\right)
\]
be a d--spinor basis. The dual to it basis is denoted as
\[
\delta _{\underline{\alpha }}^{\quad \underline{\mathbf{\alpha
}}}=\left(
\delta _{\underline{i}}^{\quad \underline{\mathbf{1}}},\delta _{\underline{i}%
}^{\quad \underline{\mathbf{2}}},...,\delta
_{\underline{i}}^{\quad
\underline{\mathbf{N(n)}}},\delta _{\underline{i}}^{\quad \underline{\mathbf{%
1}}},\delta _{\underline{i}}^{\quad \underline{\mathbf{2}}},...,\delta _{%
\underline{i}}^{\quad \underline{\mathbf{N(m)}}}\right) .
\]
A d-spinor $\kappa ^{\underline{\alpha }}\in \mathcal{\sigma }$ $^{%
\underline{\alpha }}$ has components $\kappa ^{\underline{\mathbf{\alpha }}%
}=\kappa ^{\underline{\alpha }}\delta _{\underline{\alpha
}}^{\quad \underline{\mathbf{\alpha }}}.$ Taking into account that
\[
\delta _{\underline{\mathbf{\alpha }}}^{\quad \underline{\alpha }}\delta _{%
\underline{\mathbf{\beta }}}^{\quad \underline{\beta }}\bigtriangledown _{%
\underline{\alpha }\underline{\beta }}=\bigtriangledown _{\underline{\mathbf{%
\alpha }}\underline{\mathbf{\beta }}},
\]
we write out the components $\bigtriangledown _{\underline{\alpha }%
\underline{\beta }}$ $\kappa ^{\underline{\gamma }}$ as
\begin{eqnarray}
\delta _{\underline{\mathbf{\alpha }}}^{\quad \underline{\alpha }}~\delta _{%
\underline{\mathbf{\beta }}}^{\quad \underline{\beta }}~\delta _{\underline{%
\gamma }}^{\quad \underline{\mathbf{\gamma }}}~\bigtriangledown _{\underline{%
\alpha }\underline{\beta }}\kappa ^{\underline{\gamma }} &=&\delta _{%
\underline{\mathbf{\epsilon }}}^{\quad \underline{\tau }}~\delta _{%
\underline{\tau }}^{\quad \underline{\mathbf{\gamma }}}~\bigtriangledown _{%
\underline{\mathbf{\alpha }}\underline{\mathbf{\beta }}}\kappa ^{\underline{%
\mathbf{\epsilon }}}+\kappa ^{\underline{\mathbf{\epsilon }}}~\delta _{%
\underline{\epsilon }}^{\quad \underline{\mathbf{\gamma
}}}~\bigtriangledown
_{\underline{\mathbf{\alpha }}\underline{\mathbf{\beta }}}\delta _{%
\underline{\mathbf{\epsilon }}}^{\quad \underline{\epsilon }}  \nonumber \\
&=&\bigtriangledown _{\underline{\mathbf{\alpha }}\underline{\mathbf{\beta }}%
}\kappa ^{\underline{\mathbf{\gamma }}}+\kappa
^{\underline{\mathbf{\epsilon
}}}\gamma _{~\underline{\mathbf{\alpha }}\underline{\mathbf{\beta }}%
\underline{\mathbf{\epsilon }}}^{\underline{\mathbf{\gamma }}},
\label{2.74a}
\end{eqnarray}
where the coordinate components of the d--spinor connection $\gamma _{~%
\underline{\mathbf{\alpha }}\underline{\mathbf{\beta }}\underline{\mathbf{%
\epsilon }}}^{\underline{\mathbf{\gamma }}}$ are defined as
\begin{equation}
\gamma _{~\underline{\mathbf{\alpha }}\underline{\mathbf{\beta }}\underline{%
\mathbf{\epsilon }}}^{\underline{\mathbf{\gamma }}}\doteq \delta _{%
\underline{\tau }}^{\quad \underline{\mathbf{\gamma }}}~\bigtriangledown _{%
\underline{\mathbf{\alpha }}\underline{\mathbf{\beta }}}\delta _{\underline{%
\mathbf{\epsilon }}}^{\quad \underline{\tau }}.  \label{2.75a}
\end{equation}
We call the Infeld - van der Waerden d-symbols a set of $\sigma $-objects ($%
\sigma _{\mathbf{\alpha }})^{\underline{\mathbf{\alpha }}\underline{\mathbf{%
\beta }}}$ parametri\-zed with respect to a coordinate d-spinor
basis. Defining
\[
\bigtriangledown _{<\mathbf{\alpha >}}=(\sigma _{<\mathbf{\alpha >}})^{%
\underline{\mathbf{\alpha }}\underline{\mathbf{\beta }}}~\bigtriangledown _{%
\underline{\mathbf{\alpha }}\underline{\mathbf{\beta }}},
\]
introducing denotations
\[
\gamma ^{\underline{\mathbf{\gamma }}}{}_{<\mathbf{\alpha >\underline{\tau }}%
}\doteq \gamma ^{\underline{\mathbf{\gamma }}}{}_{\mathbf{\underline{\alpha }%
\underline{\beta }\underline{\tau }}}~(\sigma _{<\mathbf{\alpha >}})^{%
\underline{\mathbf{\alpha }}\underline{\mathbf{\beta }}}
\]
and using properties (\ref{2.74a}) we can write relations
\begin{eqnarray}
l_{<\mathbf{\alpha >}}^{<\alpha >}~\delta _{\underline{\beta
}}^{\quad
\underline{\mathbf{\beta }}}~\bigtriangledown _{<\alpha >}\kappa ^{%
\underline{\beta }} &=&\bigtriangledown _{<\mathbf{\alpha >}}\kappa ^{%
\underline{\mathbf{\beta }}}+\kappa ^{\underline{\mathbf{\delta
}}}~\gamma _{~<\mathbf{\alpha >}\underline{\mathbf{\delta
}}}^{\underline{\mathbf{\beta
}}},  \label{2.76a} \\
l_{<\mathbf{\alpha >}}^{<\alpha >}~\delta _{\underline{\mathbf{\beta }}%
}^{\quad \underline{\beta }}~\bigtriangledown _{<\alpha >}~\mu _{\underline{%
\beta }} &=&\bigtriangledown _{<\mathbf{\alpha >}}~\mu _{\underline{\mathbf{%
\beta }}}-\mu _{\underline{\mathbf{\delta }}}\gamma _{~<\mathbf{\alpha >}%
\underline{\mathbf{\beta }}}^{\underline{\mathbf{\delta }}}.
\nonumber
\end{eqnarray}
for d-covariant derivations $~\bigtriangledown
_{\underline{\alpha }}\kappa
^{\underline{\beta }}$ and $\bigtriangledown _{\underline{\alpha }}~\mu _{%
\underline{\beta }}.$

We can consider expressions similar to (\ref{2.76a}) for values
having both types of d-spinor and d-tensor indices, for instance,
\begin{eqnarray}
& &
 l_{<\mathbf{\alpha >}}^{<\alpha >}~l_{<\gamma
>}^{<\mathbf{\gamma
>}}~\delta _{\underline{\mathbf{\delta }}}^{\quad
\underline{\delta }}~\bigtriangledown _{<\alpha >}\theta
_{\underline{\delta }}^{~<\gamma >}= \nonumber \\
 & & \bigtriangledown _{<%
\mathbf{\alpha >}}\theta _{\underline{\mathbf{\delta }}}^{~<\mathbf{\gamma >}%
}-\theta _{\underline{\mathbf{\epsilon }}}^{~<\mathbf{\gamma >}}\gamma _{~<%
\mathbf{\alpha >}\underline{\mathbf{\delta }}}^{\underline{\mathbf{\epsilon }%
}}+\theta _{\underline{\mathbf{\delta }}}^{~<\mathbf{\tau
>}}~\Gamma _{\quad <\mathbf{\alpha ><\tau >}}^{~<\mathbf{\gamma
>}}
 \nonumber
\end{eqnarray}
(we can prove this by a straightforward calculation).

Now we shall consider some possible relations between components
of
d-connec\-ti\-ons $\gamma _{~<\mathbf{\alpha >}\underline{\mathbf{\delta }}%
}^{\underline{\mathbf{\epsilon }}}$ and $\Gamma _{\quad
<\mathbf{\alpha
><\tau >}}^{~<\mathbf{\gamma >}}$ and derivations of $(\sigma _{<\mathbf{%
\alpha >}})^{\underline{\mathbf{\alpha }}\underline{\mathbf{\beta
}}}$ .  We can write
\begin{eqnarray*}
 \Gamma _{~<\mathbf{\beta ><\gamma >}}^{<\mathbf{\alpha >}}
&=& l_{<\alpha
>}^{<\mathbf{\alpha >}}\bigtriangledown _{<\mathbf{\gamma >}}l_{<\mathbf{%
\beta >}}^{<\alpha >} \\
& &=l_{<\alpha >}^{<\mathbf{\alpha >}}\bigtriangledown _{<%
\mathbf{\gamma >}}(\sigma _{<\mathbf{\beta >}})^{\underline{\epsilon }%
\underline{\tau }}l_{<\alpha >}^{<\mathbf{\alpha >}}\bigtriangledown _{<%
\mathbf{\gamma >}}((\sigma _{<\mathbf{\beta >}})^{\underline{\mathbf{%
\epsilon }}\underline{\mathbf{\tau }}}\delta _{\underline{\mathbf{\epsilon }}%
}^{~\underline{\epsilon }}\delta _{\underline{\mathbf{\tau }}}^{~\underline{%
\tau }}) \\
& &=l_{<\alpha >}^{<\mathbf{\alpha >}}\delta _{\underline{\mathbf{\alpha }}%
}^{~\underline{\alpha }}\delta _{\underline{\mathbf{\epsilon }}}^{~%
\underline{\epsilon }}\bigtriangledown _{<\mathbf{\gamma >}}(\sigma _{<%
\mathbf{\beta >}})^{\underline{\mathbf{\alpha }}\underline{\mathbf{\epsilon }%
}} \\
& & +l_{<\alpha >}^{<\mathbf{\alpha >}}(\sigma _{<\mathbf{\beta >}})^{%
\underline{\mathbf{\epsilon }}\underline{\mathbf{\tau }}}(\delta _{%
\underline{\mathbf{\tau }}}^{~\underline{\tau }}\bigtriangledown _{<\mathbf{%
\gamma >}}\delta _{\underline{\mathbf{\epsilon }}}^{~\underline{\epsilon }%
}+\delta _{\underline{\mathbf{\epsilon }}}^{~\underline{\epsilon }%
}\bigtriangledown _{<\mathbf{\gamma >}}\delta _{\underline{\mathbf{\tau }}%
}^{~\underline{\tau }}) \\
& &=l_{\underline{\mathbf{\epsilon }}\underline{\mathbf{\tau }}}^{<\mathbf{%
\alpha >}}~\bigtriangledown _{<\mathbf{\gamma >}}(\sigma _{<\mathbf{\beta >}%
})^{\underline{\mathbf{\epsilon }}\underline{\mathbf{\tau }}} \\
& &+l_{\underline{%
\mathbf{\mu }}\underline{\mathbf{\nu }}}^{<\mathbf{\alpha >}}\delta _{%
\underline{\epsilon }}^{~\underline{\mathbf{\mu }}}\delta _{\underline{\tau }%
}^{~\underline{\mathbf{\nu }}}(\sigma _{<\mathbf{\beta >}})^{\underline{%
\epsilon }\underline{\tau }}(\delta _{\underline{\mathbf{\tau }}}^{~%
\underline{\tau }}\bigtriangledown _{<\mathbf{\gamma >}}\delta _{\underline{%
\mathbf{\epsilon }}}^{~\underline{\epsilon }}+\delta _{\underline{\mathbf{%
\epsilon }}}^{~\underline{\epsilon }}\bigtriangledown _{<\mathbf{\gamma >}%
}\delta _{\underline{\mathbf{\tau }}}^{~\underline{\tau }}),
\end{eqnarray*}
where $l_{<\alpha >}^{<\mathbf{\alpha >}}=(\sigma _{\underline{\mathbf{%
\epsilon }}\underline{\mathbf{\tau }}})^{<\mathbf{\alpha >}}$ ,
from which one follows
\begin{eqnarray}
& & (\sigma _{<\mathbf{\alpha >}})^{\underline{\mathbf{\mu }}
\underline{\mathbf{%
\nu }}}(\sigma _{\underline{\mathbf{\alpha }}\underline{\mathbf{\beta }}})^{<%
\mathbf{\beta >}}\Gamma _{~<\mathbf{\gamma ><\beta >}}^{<\mathbf{\alpha >}%
}= \nonumber \\ & &
(\sigma _{\underline{\mathbf{\alpha }}\underline{\mathbf{\beta }}})^{<%
\mathbf{\beta >}}\bigtriangledown _{<\mathbf{\gamma >}}(\sigma _{<\mathbf{%
\alpha >}})^{\underline{\mathbf{\mu }}\underline{\mathbf{\nu }}}+\delta _{%
\underline{\mathbf{\beta }}}^{~\underline{\mathbf{\nu }}}\gamma _{~<\mathbf{%
\gamma >\underline{\alpha }}}^{\underline{\mathbf{\mu }}}+\delta _{%
\underline{\mathbf{\alpha }}}^{~\underline{\mathbf{\mu }}}\gamma _{~<\mathbf{%
\gamma >\underline{\beta }}}^{\underline{\mathbf{\nu }}}.
\nonumber
\end{eqnarray}
Connecting the last expression on \underline{$\mathbf{\beta }$}
and \underline{$\mathbf{\nu }$} and using an orthonormalized
d-spinor basis when
$\gamma _{~<\mathbf{\gamma >\underline{\beta }}}^{\underline{\mathbf{\beta }}%
}=0$ (a consequence from (\ref{2.75a})) we have
\begin{eqnarray}
& &
\gamma _{~<\mathbf{\gamma >\underline{\alpha }}}^{\underline{\mathbf{\mu }}%
}=\frac 1{N(n)+N(m_1)+...+N(m_z)}(\Gamma _{\quad <\mathbf{\gamma >~%
\underline{\alpha }\underline{\beta }}}^{\underline{\mathbf{\mu }}\underline{%
\mathbf{\beta }}}   \label{2.78a}  \\ & &
-(\sigma _{\underline{\mathbf{\alpha }}\underline{\mathbf{%
\beta }}})^{<\mathbf{\beta >}}\bigtriangledown _{<\mathbf{\gamma
>}}(\sigma
_{<\mathbf{\beta >}})^{\underline{\mathbf{\mu }}\underline{\mathbf{\beta }}%
}), \nonumber
\end{eqnarray}
where
\begin{equation}
\Gamma _{\quad <\mathbf{\gamma >~\underline{\alpha }\underline{\beta }}}^{%
\underline{\mathbf{\mu }}\underline{\mathbf{\beta }}}=(\sigma _{<\mathbf{%
\alpha >}})^{\underline{\mathbf{\mu }}\underline{\mathbf{\beta }}}(\sigma _{%
\underline{\mathbf{\alpha }}\underline{\mathbf{\beta }}})^{\mathbf{\beta }%
}\Gamma _{~<\mathbf{\gamma ><\beta >}}^{<\mathbf{\alpha >}}.
\label{2.79a}
\end{equation}
We also note here that, for instance, for the canonical and
Berwald
connections and Christoffel d-symbols  we can express d-spinor connection (%
\ref{2.79a}) through corresponding locally adapted derivations of
components of metric and N-connection by introducing
corresponding coefficients instead
of $\Gamma _{~<\mathbf{\gamma ><\beta >}}^{<\mathbf{\alpha >}}$ in (%
\ref{2.79a}) and than in (\ref{2.78a}).

\subsection{ D--spinors of ha--space curvature and torsion}

The d-tensor indices of the commutator  $\Delta _{<\alpha ><\beta
>}$ can be transformed into d-spinor ones:
\begin{eqnarray}
\Box _{\underline{\alpha }\underline{\beta }} &=&(\sigma
^{<\alpha ><\beta
>})_{\underline{\alpha }\underline{\beta }}\Delta _{\alpha \beta }=(\Box _{%
\underline{i}\underline{j}},\Box _{\underline{a}\underline{b}})
\label{2.80a} \\
&=&(\Box _{\underline{i}\underline{j}},\Box _{\underline{a}_1\underline{b}%
_1},...,\Box _{\underline{a}_p\underline{b}_p},...,\Box _{\underline{a}_z%
\underline{b}_z}),  \nonumber
\end{eqnarray}
with h- and v$_p$-components,
\[
\Box _{\underline{i}\underline{j}}=(\sigma ^{<\alpha ><\beta >})_{\underline{%
i}\underline{j}}\Delta _{<\alpha ><\beta >}\mbox{ and }\Box _{\underline{a}%
\underline{b}}=(\sigma ^{<\alpha ><\beta >})_{\underline{a}\underline{b}%
}\Delta _{<\alpha ><\beta >},
\]
being symmetric or antisymmetric in dependence of corresponding
values of
dimensions $n\,$ and $m_p$ (see eight-fold parametizations (\ref{2.60}) and (%
\ref{2.61})). Considering the actions of operator (\ref{2.80a})
on d-spinors $\pi ^{\underline{\gamma }}$ and $\mu
_{\underline{\gamma }}$ we introduce
the d-spinor curvature $X_{\underline{\delta }\quad \underline{\alpha }%
\underline{\beta }}^{\quad \underline{\gamma }}\,$ as to satisfy
equations
\begin{equation}
\Box _{\underline{\alpha }\underline{\beta }}\ \pi ^{\underline{\gamma }}=X_{%
\underline{\delta }\quad \underline{\alpha }\underline{\beta
}}^{\quad
\underline{\gamma }}\pi ^{\underline{\delta }}\mbox{ and }\Box _{\underline{%
\alpha }\underline{\beta }}\ \mu _{\underline{\gamma
}}=X_{\underline{\gamma }\quad \underline{\alpha
}\underline{\beta }}^{\quad \underline{\delta }}\mu
_{\underline{\delta }}.  \label{2.81a}
\end{equation}
The gravitational d-spinor $\Psi _{\underline{\alpha }\underline{\beta }%
\underline{\gamma }\underline{\delta }}$ is defined by a
corresponding symmetrization of d-spinor indices:
\begin{equation}
\Psi _{\underline{\alpha }\underline{\beta }\underline{\gamma }\underline{%
\delta }}=X_{(\underline{\alpha }|\underline{\beta }|\underline{\gamma }%
\underline{\delta })}.  \label{2.82a}
\end{equation}
We note that d-spinor tensors $X_{\underline{\delta }\quad
\underline{\alpha
}\underline{\beta }}^{\quad \underline{\gamma }}$ and $\Psi _{\underline{%
\alpha }\underline{\beta }\underline{\gamma }\underline{\delta
}}\,$ are transformed into similar 2-spinor objects on locally
isotropic spaces \cite {penr1,penr2} if we consider vanishing of
the N-connection structure and a limit to a locally isotropic
space.

Putting $\delta _{\underline{\gamma }}^{\quad
\mathbf{\underline{\gamma }}}$
instead of $\mu _{\underline{\gamma }}$ in (\ref{2.81a}) and using (%
\ref{2.82a}) we can express respectively the curvature and
gravitational d-spinors as
\[
X_{\underline{\gamma }\underline{\delta }\underline{\alpha
}\underline{\beta
}}=\delta _{\underline{\delta }\underline{\mathbf{\tau }}}\Box _{\underline{%
\alpha }\underline{\beta }}\delta _{\underline{\gamma }}^{\quad \mathbf{%
\underline{\tau }}}\mbox{ and }\Psi _{\underline{\gamma }\underline{\delta }%
\underline{\alpha }\underline{\beta }}=\delta _{\underline{\delta }%
\underline{\mathbf{\tau }}}\Box _{(\underline{\alpha }\underline{\beta }%
}\delta _{\underline{\gamma })}^{\quad \mathbf{\underline{\tau
}}}.
\]

The d-spinor torsion $T_{\qquad \underline{\alpha }\underline{\beta }}^{%
\underline{\gamma }_1\underline{\gamma }_2}$ is defined similarly
as for d-tensors by using the d-spinor commutator (\ref{2.80a})
and equations
\[
\Box _{\underline{\alpha }\underline{\beta }}f=T_{\qquad \underline{\alpha }%
\underline{\beta }}^{\underline{\gamma }_1\underline{\gamma }%
_2}\bigtriangledown _{\underline{\gamma }_1\underline{\gamma
}_2}f.
\]

The d-spinor components $R_{\underline{\gamma
}_1\underline{\gamma }_2\qquad
\underline{\alpha }\underline{\beta }}^{\qquad \underline{\delta }_1%
\underline{\delta }_2}$ of the curvature d-tensor $R_{\gamma
\quad \alpha \beta }^{\quad \delta }$ can be computed by using
relations (\ref{2.79a}), and (\ref{2.80a}) and (\ref{2.82a}) as
to satisfy the equations
\[
(\Box _{\underline{\alpha }\underline{\beta }}-T_{\qquad \underline{\alpha }%
\underline{\beta }}^{\underline{\gamma }_1\underline{\gamma }%
_2}\bigtriangledown _{\underline{\gamma }_1\underline{\gamma }_2})V^{%
\underline{\delta }_1\underline{\delta }_2}=R_{\underline{\gamma }_1%
\underline{\gamma }_2\qquad \underline{\alpha }\underline{\beta
}}^{\qquad
\underline{\delta }_1\underline{\delta }_2}V^{\underline{\gamma }_1%
\underline{\gamma }_2},
\]
here d-vector $V^{\underline{\gamma }_1\underline{\gamma }_2}$ is
considered
as a product of d-spinors, i.e. $V^{\underline{\gamma }_1\underline{\gamma }%
_2}=\nu ^{\underline{\gamma }_1}\mu ^{\underline{\gamma }_2}$. We
find

\begin{eqnarray}
R_{\underline{\gamma }_1\underline{\gamma }_2\qquad \underline{\alpha }%
\underline{\beta }}^{\qquad \underline{\delta
}_1\underline{\delta }_2}
&=&\left( X_{\underline{\gamma }_1~\underline{\alpha }\underline{\beta }%
}^{\quad \underline{\delta }_1}+T_{\qquad \underline{\alpha }\underline{%
\beta }}^{\underline{\tau }_1\underline{\tau }_2}\quad \gamma
_{\quad
\underline{\tau }_1\underline{\tau }_2\underline{\gamma }_1}^{\underline{%
\delta }_1}\right) \delta _{\underline{\gamma }_2}^{\quad \underline{\delta }%
_2}  \label{2.83a} \\
&&+\left( X_{\underline{\gamma }_2~\underline{\alpha }\underline{\beta }%
}^{\quad \underline{\delta }_2}+T_{\qquad \underline{\alpha }\underline{%
\beta }}^{\underline{\tau }_1\underline{\tau }_2}\quad \gamma
_{\quad
\underline{\tau }_1\underline{\tau }_2\underline{\gamma }_2}^{\underline{%
\delta }_2}\right) \delta _{\underline{\gamma }_1}^{\quad \underline{\delta }%
_1}.  \nonumber
\end{eqnarray}

It is convenient to use this d-spinor expression for the
curvature d-tensor
\begin{eqnarray*}
R_{\underline{\gamma }_1\underline{\gamma }_2\qquad \underline{\alpha }_1%
\underline{\alpha }_2\underline{\beta }_1\underline{\beta
}_2}^{\qquad
\underline{\delta }_1\underline{\delta }_2} &=&\left( X_{\underline{\gamma }%
_1~\underline{\alpha }_1\underline{\alpha }_2\underline{\beta }_1\underline{%
\beta }_2}^{\quad \underline{\delta }_1}+T_{\qquad \underline{\alpha }_1%
\underline{\alpha }_2\underline{\beta }_1\underline{\beta }_2}^{\underline{%
\tau }_1\underline{\tau }_2}~\gamma _{\quad \underline{\tau }_1\underline{%
\tau }_2\underline{\gamma }_1}^{\underline{\delta }_1}\right) \delta _{%
\underline{\gamma }_2}^{\quad \underline{\delta }_2} \\
&&+\left( X_{\underline{\gamma }_2~\underline{\alpha }_1\underline{\alpha }_2%
\underline{\beta }_1\underline{\beta }_2}^{\quad \underline{\delta }%
_2}+T_{\qquad \underline{\alpha }_1\underline{\alpha }_2\underline{\beta }_1%
\underline{\beta }_2~}^{\underline{\tau }_1\underline{\tau
}_2}\gamma
_{\quad \underline{\tau }_1\underline{\tau }_2\underline{\gamma }_2}^{%
\underline{\delta }_2}\right) \delta _{\underline{\gamma
}_1}^{\quad \underline{\delta }_1}
\end{eqnarray*}
in order to get the d--spinor components of the Ricci d-tensor
\begin{eqnarray}
& &R_{\underline{\gamma }_1\underline{\gamma }_2\underline{\alpha }_1\underline{%
\alpha }_2} = R_{\underline{\gamma }_1\underline{\gamma }_2\qquad
\underline{\alpha }_1\underline{\alpha }_2\underline{\delta }_1\underline{%
\delta }_2}^{\qquad \underline{\delta }_1\underline{\delta }_2}=X_{%
\underline{\gamma }_1~\underline{\alpha }_1\underline{\alpha }_2\underline{%
\delta }_1\underline{\gamma }_2}^{\quad \underline{\delta }_1}+
\label{2.84a} \\
&&T_{\qquad \underline{\alpha }_1\underline{\alpha }_2\underline{\delta }_1%
\underline{\gamma }_2}^{\underline{\tau }_1\underline{\tau
}_2}~\gamma
_{\quad \underline{\tau }_1\underline{\tau }_2\underline{\gamma }_1}^{%
\underline{\delta }_1}+X_{\underline{\gamma }_2~\underline{\alpha }_1%
\underline{\alpha }_2\underline{\delta }_1\underline{\gamma
}_2}^{\quad
\underline{\delta }_2}+T_{\qquad \underline{\alpha }_1\underline{\alpha }_2%
\underline{\gamma }_1\underline{\delta }_2~}^{\underline{\tau }_1\underline{%
\tau }_2}\gamma _{\quad \underline{\tau }_1\underline{\tau }_2\underline{%
\gamma }_2}^{\underline{\delta }_2}  \nonumber
\end{eqnarray}
and this d-spinor decomposition of the scalar curvature:
\begin{eqnarray}
q\overleftarrow{R} &=&R_{\qquad \underline{\alpha }_1\underline{\alpha }_2}^{%
\underline{\alpha }_1\underline{\alpha }_2}=X_{\quad ~\underline{~\alpha }%
_1\quad \underline{\delta }_1\underline{\alpha }_2}^{\underline{\alpha }_1%
\underline{\delta }_1~~\underline{\alpha }_2}+T_{\qquad ~~\underline{\alpha }%
_2\underline{\delta }_1}^{\underline{\tau }_1\underline{\tau }_2\underline{%
\alpha }_1\quad ~\underline{\alpha }_2}~\gamma _{\quad \underline{\tau }_1%
\underline{\tau }_2\underline{\alpha }_1}^{\underline{\delta }_1}
\label{2.85a} \\
&&+X_{\qquad \quad \underline{\alpha }_2\underline{\delta }_2\underline{%
\alpha }_1}^{\underline{\alpha }_2\underline{\delta }_2\underline{\alpha }%
_1}+T_{\qquad \underline{\alpha }_1\quad ~\underline{\delta }_2~}^{%
\underline{\tau }_1\underline{\tau }_2~~\underline{\alpha }_2\underline{%
\alpha }_1}\gamma _{\quad \underline{\tau }_1\underline{\tau }_2\underline{%
\alpha }_2}^{\underline{\delta }_2}.  \nonumber
\end{eqnarray}

Putting (\ref{2.84a}) and (\ref{2.85a}) into (\ref{2.34}) and,
correspondingly, (\ref{2.35a}) we find the d--spinor components
of the Einstein and $\Phi _{<\alpha ><\beta >}$ d--tensors:
\begin{eqnarray}
\overleftarrow{G}_{<\gamma ><\alpha >} &=&\overleftarrow{G}_{\underline{%
\gamma }_1\underline{\gamma }_2\underline{\alpha }_1\underline{\alpha }%
_2}=X_{\underline{\gamma }_1~\underline{\alpha }_1\underline{\alpha }_2%
\underline{\delta }_1\underline{\gamma }_2}^{\quad \underline{\delta }%
_1}+T_{\qquad \underline{\alpha }_1\underline{\alpha }_2\underline{\delta }_1%
\underline{\gamma }_2}^{\underline{\tau }_1\underline{\tau
}_2}~\gamma
_{\quad \underline{\tau }_1\underline{\tau }_2\underline{\gamma }_1}^{%
\underline{\delta }_1}  \nonumber \\
&&+X_{\underline{\gamma }_2~\underline{\alpha }_1\underline{\alpha }_2%
\underline{\delta }_1\underline{\gamma }_2}^{\quad \underline{\delta }%
_2}+T_{\qquad \underline{\alpha }_1\underline{\alpha }_2\underline{\gamma }_1%
\underline{\delta }_2~}^{\underline{\tau }_1\underline{\tau
}_2}\gamma
_{\quad \underline{\tau }_1\underline{\tau }_2\underline{\gamma }_2}^{%
\underline{\delta }_2}-  \nonumber \\
&&\frac 12\varepsilon _{\underline{\gamma }_1\underline{\alpha }%
_1}\varepsilon _{\underline{\gamma }_2\underline{\alpha }_2}[X_{\quad ~%
\underline{~\beta }_1\quad \underline{\mu }_1\underline{\beta }_2}^{%
\underline{\beta }_1\underline{\mu }_1~~\underline{\beta }_2}+T_{\qquad ~~%
\underline{\beta }_2\underline{\mu }_1}^{\underline{\tau }_1\underline{\tau }%
_2\underline{\beta }_1\quad ~\underline{\beta }_2}~\gamma _{\quad \underline{%
\tau }_1\underline{\tau }_2\underline{\beta }_1}^{\underline{\mu
}_1}+
\nonumber \\
&&X_{\qquad \quad \underline{\beta }_2\underline{\mu }_2\underline{\nu }_1}^{%
\underline{\beta }_2\underline{\mu }_2\underline{\nu
}_1}+T_{\qquad
\underline{\beta }_1\quad ~\underline{\delta }_2~}^{\underline{\tau }_1%
\underline{\tau }_2~~\underline{\beta }_2\underline{\beta
}_1}\gamma _{\quad
\underline{\tau }_1\underline{\tau }_2\underline{\beta }_2}^{\underline{%
\delta }_2}]  \label{2.86a}
\end{eqnarray}
and
\begin{eqnarray}
& &\Phi _{<\gamma ><\alpha >} =\Phi _{\underline{\gamma }_1\underline{\gamma }%
_2\underline{\alpha }_1\underline{\alpha }_2}=\frac
1{2(n+m_1+...+m_z)}\varepsilon _{\underline{\gamma }_1\underline{\alpha }%
_1}\varepsilon _{\underline{\gamma }_2\underline{\alpha }_2}[X_{\quad ~%
\underline{~\beta }_1\quad \underline{\mu }_1\underline{\beta }_2}^{%
\underline{\beta }_1\underline{\mu }_1~~\underline{\beta }_2}+  \nonumber \\
&&T_{\qquad ~~\underline{\beta }_2\underline{\mu }_1}^{\underline{\tau }_1%
\underline{\tau }_2\underline{\beta }_1\quad ~\underline{\beta
}_2}~\gamma
_{\quad \underline{\tau }_1\underline{\tau }_2\underline{\beta }_1}^{%
\underline{\mu }_1}+X_{\qquad \quad \underline{\beta }_2\underline{\mu }_2%
\underline{\nu }_1}^{\underline{\beta }_2\underline{\mu }_2\underline{\nu }%
_1}+T_{\qquad \underline{\beta }_1\quad ~\underline{\delta }_2~}^{\underline{%
\tau }_1\underline{\tau }_2~~\underline{\beta }_2\underline{\beta
}_1}\gamma
_{\quad \underline{\tau }_1\underline{\tau }_2\underline{\beta }_2}^{%
\underline{\delta }_2}]-  \nonumber \\
&&\frac 12[X_{\underline{\gamma }_1~\underline{\alpha }_1\underline{\alpha }%
_2\underline{\delta }_1\underline{\gamma }_2}^{\quad \underline{\delta }%
_1}+T_{\qquad \underline{\alpha }_1\underline{\alpha }_2\underline{\delta }_1%
\underline{\gamma }_2}^{\underline{\tau }_1\underline{\tau
}_2}~\gamma
_{\quad \underline{\tau }_1\underline{\tau }_2\underline{\gamma }_1}^{%
\underline{\delta }_1}+  \nonumber \\
&&X_{\underline{\gamma }_2~\underline{\alpha }_1\underline{\alpha }_2%
\underline{\delta }_1\underline{\gamma }_2}^{\quad \underline{\delta }%
_2}+T_{\qquad \underline{\alpha }_1\underline{\alpha }_2\underline{\gamma }_1%
\underline{\delta }_2~}^{\underline{\tau }_1\underline{\tau
}_2}\gamma
_{\quad \underline{\tau }_1\underline{\tau }_2\underline{\gamma }_2}^{%
\underline{\delta }_2}].  \label{2.87a}
\end{eqnarray}

The components of the conformal Weyl d-spinor can be computed by
putting d-spinor values of the curvature (\ref{2.83a}) and Ricci
(\ref{2.84a}) d-tensors into corresponding expression for the
d-tensor (\ref{2.33}). We omit this calculus in this work.

\chapter{ Ha-Spinors and Field Interactions}

The problem of formulation gravitational and gauge field
equations on different types of locally anisotropic spaces is
considered, for instance, in \cite{ma94,bej,asa88} and \cite{vg}.
In this Chapter we shall introduce the basic field equations for
gravitational and matter field la-interactions in a generalized
form for generic higher order anisotropic spaces.

\section{Scalar field ha--interactions}

Let $\varphi \left( u\right) =(\varphi _1\left( u\right) ,\varphi
_2\left( u\right) \dot{,}...,\varphi _k\left( u\right) )$ be a
complex k-component scalar field of mass $\mu $ on ha-space
$\mathcal{E}^{<z>}.$ The d-covariant generalization of the
conformally invariant (in the massless case) scalar field
equation \cite{penr1,penr2} can be defined by using the d'Alambert
locally anisotropic operator \cite{ana94,vst96} $\Box =D^{<\alpha
>}D_{<\alpha >}$, where $D_{<\alpha >}$ is a d-covariant derivation on $%
\mathcal{E}^{<z>}$  and constructed, for simplicity, by using
Christoffel d--symbols  (all formulas for field equations and
conservation values can be deformed by using corresponding
deformations d--tensors $P_{<\beta ><\gamma
>}^{<\alpha >}$ from the Cristoffel d--symbols, or the canonical
d--connection to a general d-connection into consideration):

\begin{equation}
(\Box +\frac{n_E-2}{4(n_E-1)}\overleftarrow{R}+\mu ^2)\varphi
\left( u\right) =0,  \label{2.88a}
\end{equation}
where $n_E=n+m_1+...+m_z.$We must change d-covariant derivation
$D_{<\alpha
>}$ into $^{\diamond }D_{<\alpha >}=D_{<\alpha >}+ieA_{<\alpha >}$ and take
into account the d-vector current
\[
J_{<\alpha >}^{(0)}\left( u\right) =i(\left( \overline{\varphi
}\left(
u\right) D_{<\alpha >}\varphi \left( u\right) -D_{<\alpha >}\overline{%
\varphi }\left( u\right) )\varphi \left( u\right) \right)
\]
if interactions between locally anisotropic electromagnetic field
( d-vector potential $A_{<\alpha >}$ ), where $e$ is the
electromagnetic constant, and charged scalar field $\varphi $ are
considered. The equations (\ref{2.88a}) are (locally adapted to
the N-connection structure) Euler equations for the Lagrangian
\begin{eqnarray}
& & \mathcal{L}^{(0)}\left( u\right) = \label{2.89a}\\ & &
 \sqrt{|g|}\left[
g^{<\alpha> <\beta
>}\delta _{<\alpha >}\overline{\varphi }\left( u\right) \delta _{<\beta
>}\varphi \left( u\right) -\left( \mu ^2+\frac{n_E-2}{4(n_E-1)}\right)
\overline{\varphi }\left( u\right) \varphi \left( u\right)
\right] ,
 \nonumber
\end{eqnarray}
where $|g|=detg_{<\alpha ><\beta >}.$

The locally adapted variations of the action with Lagrangian
(\ref{2.89a}) on variables $\varphi \left( u\right) $ and
$\overline{\varphi }\left( u\right) $ leads to the locally
anisotropic generalization of the energy-momentum tensor,
\begin{eqnarray}
E_{<\alpha ><\beta >}^{(0,can)}\left( u\right)  &=&\delta _{<\alpha >}%
\overline{\varphi }\left( u\right) \delta _{<\beta >}\varphi
\left( u\right)
+  \label{2.90a} \\
&&\delta _{<\beta >}\overline{\varphi }\left( u\right) \delta
_{<\alpha
>}\varphi \left( u\right) -\frac 1{\sqrt{|g|}}g_{<\alpha ><\beta >}\mathcal{L%
}^{(0)}\left( u\right) ,  \nonumber
\end{eqnarray}
and a similar variation on the components of a d-metric
(\ref{dmetrichcv}) leads to a symmetric metric energy-momentum
d-tensor,
\begin{eqnarray}
 & & E_{<\alpha ><\beta >}^{(0)}\left( u\right)  =E_{(<\alpha ><\beta
>)}^{(0,can)}\left( u\right) -  \label{2.91a} \\
&&\frac{n_E-2}{2(n_E-1)}\left[ R_{(<\alpha ><\beta >)}+D_{(<\alpha
>}D_{<\beta >)}-g_{<\alpha ><\beta >}\Box \right] \overline{\varphi }\left(
u\right) \varphi \left( u\right) .  \nonumber
\end{eqnarray}
Here we note that we can obtain a nonsymmetric energy-momentum
d-tensor if we firstly vary on $G_{<\alpha ><\beta >}$ and than
impose the constraint of compatibility with the N-connection
structure. We also conclude that the existence of a N-connection
in dv-bundle $\mathcal{E}^{<z>}$ results in a nonequivalence of
energy-momentum d-tensors (\ref{2.90a}) and (\ref{2.91a}),
nonsymmetry of the Ricci tensor, nonvanishing of the d-covariant
derivation of the Einstein d-tensor, $D_{<\alpha
>}\overleftarrow{G}^{<\alpha ><\beta
>}\neq 0$ and, in consequence, a corresponding breaking of conservation laws
on higher order anisotropic spaces when $D_{<\alpha >}E^{<\alpha
><\beta
>}\neq 0\,$ . The problem of formulation of conservation laws on locally
anisotropic spaces is discussed in details and two variants of
its solution (by using nearly autoparallel maps and tensor
integral formalism on locally anisotropic and higher order
multispaces) are proposed in \cite{vst96}.

In this Chapter we  present only straightforward generalizations
of field equations and necessary formulas for energy-momentum
d-tensors of matter fields on $\mathcal{E}^{<z>}$ considering
that it is naturally that the conservation laws (usually being
consequences of global, local and/or intrinsic symmetries of the
fundamental space-time and of the type of field interactions)
have to be broken on locally anisotropic spaces.

\section{ Proca equations on ha--spaces} \index{Proca}

Let consider a d-vector field $\varphi _{<\alpha >}\left(
u\right) $ with mass $\mu ^2$ (locally anisotropic Proca field )
interacting with exterior la-gravitational field. From the
Lagrangian
\begin{equation}
\mathcal{L}^{(1)}\left( u\right) =\sqrt{\left| g\right| }\left[ -\frac 12{%
\overline{f}}_{<\alpha ><\beta >}\left( u\right) f^{<\alpha
><\beta >}\left( u\right) +\mu ^2{\overline{\varphi }}_{<\alpha
>}\left( u\right) \varphi ^{<\alpha >}\left( u\right) \right] ,
\label{2.92a}
\end{equation}
where $f_{<\alpha ><\beta >}=D_{<\alpha >}\varphi _{<\beta
>}-D_{<\beta
>}\varphi _{<\alpha >},$ one follows the Proca equations on higher order
anisotropic spaces
\begin{equation}
D_{<\alpha >}f^{<\alpha ><\beta >}\left( u\right) +\mu ^2\varphi
^{<\beta
>}\left( u\right) =0.  \label{2.93a}
\end{equation}
Equations (\ref{2.93a}) are a first type constraints for $\beta
=0.$ Acting with $D_{<\alpha >}$ on (\ref{2.93a}), for $\mu \neq
0$ we obtain second type constraints
\begin{equation}
D_{<\alpha >}\varphi ^{<\alpha >}\left( u\right) =0. \label{2.94a}
\end{equation}

Putting (\ref{2.94a}) into (\ref{2.93a}) we obtain second order
field equations with respect to $\varphi _{<\alpha >}$ :
\begin{equation}
\Box \varphi _{<\alpha >}\left( u\right) +R_{<\alpha ><\beta
>}\varphi ^{<\beta >}\left( u\right) +\mu ^2\varphi _{<\alpha
>}\left( u\right) =0. \label{2.95a}
\end{equation}
The energy-momentum d-tensor and d-vector current following from the (%
\ref{2.95a}) can be written as
\begin{eqnarray*}
E_{<\alpha ><\beta >}^{(1)}\left( u\right)  &=&-g^{<\varepsilon
><\tau
>}\left( {\overline{f}}_{<\beta ><\tau >}f_{<\alpha ><\varepsilon >}+{%
\overline{f}}_{<\alpha ><\varepsilon >}f_{<\beta ><\tau >}\right)  \\
&&+\mu ^2\left( {\overline{\varphi }}_{<\alpha >}\varphi _{<\beta >}+{%
\overline{\varphi }}_{<\beta >}\varphi _{<\alpha >}\right)
-\frac{g_{<\alpha
><\beta >}}{\sqrt{\left| g\right| }}\mathcal{L}^{(1)}\left( u\right)
\end{eqnarray*}
and
\[
J_{<\alpha >}^{\left( 1\right) }\left( u\right) =i\left( {\overline{f}}%
_{<\alpha ><\beta >}\left( u\right) \varphi ^{<\beta >}\left( u\right) -{%
\overline{\varphi }}^{<\beta >}\left( u\right) f_{<\alpha ><\beta
>}\left( u\right) \right) .
\]

For $\mu =0$ the d-tensor $f_{<\alpha ><\beta >}$ and the Lagrangian (%
\ref{2.92a}) are invariant with respect to locally anisotropic
gauge transforms of type
\[
\varphi _{<\alpha >}\left( u\right) \rightarrow \varphi _{<\alpha
>}\left( u\right) +\delta _{<\alpha >}\Lambda \left( u\right) ,
\]
where $\Lambda \left( u\right) $ is a d-differentiable scalar
function, and we obtain a locally anisot\-rop\-ic variant of
Maxwell theory.

\section{  Higher order anisotropic Dirac equations}

Let denote the Dirac d--spinor field on $\mathcal{E}^{<z>}$ as
$\psi \left( u\right) =\left( \psi ^{\underline{\alpha }}\left(
u\right) \right) $ and consider as the generalized Lorentz
transforms the group of automorphysm of
the metric $G_{<\widehat{\alpha }><\widehat{\beta }>}$ (see (\ref{dmetrichcv}%
)).The d--covariant derivation of field $\psi \left( u\right) $
is written as
\begin{equation}
\overrightarrow{\bigtriangledown _{<\alpha >}}\psi =\left[ \delta
_{<\alpha
>}+\frac 14C_{\widehat{\alpha }\widehat{\beta }\widehat{\gamma }}\left(
u\right) ~l_{<\alpha >}^{\widehat{\alpha }}\left( u\right) \sigma ^{\widehat{%
\beta }}\sigma ^{\widehat{\gamma }}\right] \psi ,  \label{2.96a}
\end{equation}
where coefficients $C_{\widehat{\alpha }\widehat{\beta }\widehat{\gamma }%
}=\left( D_{<\gamma >}l_{\widehat{\alpha }}^{<\alpha >}\right) l_{\widehat{%
\beta }<\alpha >}l_{\widehat{\gamma }}^{<\gamma >}$ generalize
for ha-spaces
the corresponding Ricci coefficients on Riemannian spaces \cite{foc}. Using $%
\sigma $-objects $\sigma ^{<\alpha >}\left( u\right) $ (see
(\ref{2.44a}) and (\ref{2.60a})--(\ref{2.62a})) we define the
Dirac equations on ha--spaces:
\[
(i\sigma ^{<\alpha >}\left( u\right)
\overrightarrow{\bigtriangledown _{<\alpha >}}-\mu )\psi =0,
\]
which are the Euler equations for the Lagrangian
\begin{eqnarray}
\mathcal{L}^{(1/2)}\left( u\right)  &=&\sqrt{\left| g\right|
}\{[\psi
^{+}\left( u\right) \sigma ^{<\alpha >}\left( u\right) \overrightarrow{%
\bigtriangledown _{<\alpha >}}\psi \left( u\right) -  \label{2.97a} \\
&&(\overrightarrow{\bigtriangledown _{<\alpha >}}\psi ^{+}\left(
u\right) )\sigma ^{<\alpha >}\left( u\right) \psi \left( u\right)
]-\mu \psi ^{+}\left( u\right) \psi \left( u\right) \},  \nonumber
\end{eqnarray}
where $\psi ^{+}\left( u\right) $ is the complex conjugation and
transposition of the column $\psi \left( u\right) .$

From (\ref{2.97a}) we obtain the d--metric energy-momentum
d-tensor
\begin{eqnarray*}
& & E_{<\alpha ><\beta >}^{(1/2)}= \frac i4[\psi ^{+}\left(
u\right) \sigma _{<\alpha >}\left( u\right)
\overrightarrow{\bigtriangledown _{<\beta >}}\psi \left( u\right)
+\psi ^{+}\left( u\right) \sigma _{<\beta
>}\left( u\right) \overrightarrow{\bigtriangledown _{<\alpha >}}\psi \left(
u\right)  \\ & & -(\overrightarrow{\bigtriangledown _{<\alpha
>}}\psi ^{+}\left( u\right)
)\sigma _{<\beta >}\left( u\right) \psi \left( u\right) -(\overrightarrow{%
\bigtriangledown _{<\beta >}}\psi ^{+}\left( u\right) )\sigma
_{<\alpha
>}\left( u\right) \psi \left( u\right) ]
\end{eqnarray*}
and the d-vector source
\[
J_{<\alpha >}^{(1/2)}\left( u\right) =\psi ^{+}\left( u\right)
\sigma _{<\alpha >}\left( u\right) \psi \left( u\right) .
\]
We emphasize that locally anisotropic interactions with exterior
gauge fields can be introduced by changing the higher order
anisotropic partial derivation from (\ref{2.96a}) in this manner:
\[
\delta _\alpha \rightarrow \delta _\alpha +ie^{\star }B_\alpha ,
\]
where $e^{\star }$ and $B_\alpha $ are respectively the constant
d-vector potential of locally anisotropic gauge interactions on
higher order anisotropic spaces (see \cite{vg} and the next
section).

\section{ D--spinor Yang--Mills fields} \index{Yang--Mills}

We consider a dv--bundle $\mathcal{B}_E,~\pi _B:\mathcal{B\rightarrow E}%
^{<z>}$ on ha--space $\mathcal{E}^{<z>}.\mathcal{\,}$Additionally
to d-tensor and d-spinor indices we shall use capital Greek
letters, $\Phi ,\Upsilon ,$ $\Xi ,\Psi ,...$ for fibre (of this
bundle) indices (see details in \cite{penr1,penr2} for the case
when the base space of the v-bundle $\pi
_B$ is a locally isotropic space-time). Let $\underline{\bigtriangledown }%
_{<\alpha >}$ be, for simplicity, a torsionless, linear connection in $%
\mathcal{B}_E$ satisfying conditions:
\begin{eqnarray*}
& &\underline{\bigtriangledown }_{<\alpha >} :{\ \Upsilon }^\Theta
\rightarrow {\ \Upsilon }_{<\alpha >}^\Theta \quad
 \left[ \mbox{or }{\ \Xi }%
^\Theta \rightarrow {\ \Xi }_{<\alpha >}^\Theta \right] , \\
& & \underline{\bigtriangledown }_{<\alpha >}\left( \lambda
^\Theta +\nu ^\Theta
\right) =\underline{\bigtriangledown }_{<\alpha >}\lambda ^\Theta +%
\underline{\bigtriangledown }_{<\alpha >}\nu ^\Theta ,\\
& & \underline{\bigtriangledown }_{<\alpha >}~(f\lambda ^\Theta
)=\lambda
^\Theta \underline{\bigtriangledown }_{<\alpha >}f+f\underline{%
\bigtriangledown }_{<\alpha >}\lambda ^\Theta ,\quad f\in {\ \Upsilon }%
^\Theta ~[\mbox{or }{\ \Xi }^\Theta ],
\end{eqnarray*}
where by ${\ \Upsilon }^\Theta ~\left( {\ \Xi }^\Theta \right) $
we denote the module of sections of the real (complex) v--bundle
$\mathcal{B}_E$
provided with the abstract index $\Theta .$ The curvature of connection $%
\underline{\bigtriangledown }_{<\alpha >}$ is defined as
\[
K_{<\alpha ><\beta >\Omega }^{\qquad \Theta }\lambda ^\Omega
=\left(
\underline{\bigtriangledown }_{<\alpha >}\underline{\bigtriangledown }%
_{<\beta >}-\underline{\bigtriangledown }_{<\beta >}\underline{%
\bigtriangledown }_{<\alpha >}\right) \lambda ^\Theta .
\]

For Yang-Mills fields as a rule one considers that
$\mathcal{B}_E$ is enabled with a unitary (complex) structure
(complex conjugation changes mutually the upper and lower Greek
indices). It is useful to introduce instead of $K_{<\alpha
><\beta >\Omega }^{\qquad \Theta }$ a Hermitian matrix
$F_{<\alpha ><\beta >\Omega }^{\qquad \Theta }=i$ $K_{<\alpha
><\beta
>\Omega }^{\qquad \Theta }$ connected with components of the Yang-Mills
d-vector potential $B_{<\alpha >\Xi }^{\quad \Phi }$ according
the formula:

\begin{equation}
\frac 12F_{<\alpha ><\beta >\Xi }^{\qquad \Phi
}=\underline{\bigtriangledown }_{[<\alpha >}B_{<\beta >]\Xi
}^{\quad \Phi }-iB_{[<\alpha >|\Lambda |}^{\quad \Phi }B_{<\beta
>]\Xi }^{\quad \Lambda },  \label{2.98a}
\end{equation}
where the locally anisotropic space indices commute with capital
Greek indices. The gauge transforms are written in the form:

\begin{eqnarray*}
B_{<\alpha >\Theta }^{\quad \Phi } &\mapsto &B_{<\alpha >\widehat{\Theta }%
}^{\quad \widehat{\Phi }}=B_{<\alpha >\Theta }^{\quad \Phi
}~s_\Phi ^{\quad \widehat{\Phi }}~q_{\widehat{\Theta }}^{\quad
\Theta }+is_\Theta ^{\quad \widehat{\Phi
}}\underline{\bigtriangledown }_{<\alpha >}~q_{\widehat{\Theta
}}^{\quad \Theta }, \\
F_{<\alpha ><\beta >\Xi }^{\qquad \Phi } &\mapsto &F_{<\alpha ><\beta >%
\widehat{\Xi }}^{\qquad \widehat{\Phi }}=F_{<\alpha ><\beta >\Xi
}^{\qquad \Phi }s_\Phi ^{\quad \widehat{\Phi }}q_{\widehat{\Xi
}}^{\quad \Xi },
\end{eqnarray*}
where matrices $s_\Phi ^{\quad \widehat{\Phi }}$ and $q_{\widehat{\Xi }%
}^{\quad \Xi }$ are mutually inverse (Hermitian conjugated in the
unitary case). The Yang-Mills equations on torsionless locally
anisotropic spaces \cite{vg} (see details in the next Section)
are written in this form:
\begin{eqnarray}
\underline{\bigtriangledown }^{<\alpha >}F_{<\alpha ><\beta
>\Theta
}^{\qquad \Psi } &=&J_{<\beta >\ \Theta }^{\qquad \Psi },  \label{2.99a} \\
\underline{\bigtriangledown }_{[<\alpha >}F_{<\beta ><\gamma
>]\Theta }^{\qquad \Xi } &=&0.  \nonumber
\end{eqnarray}
We must introduce deformations of connection of type $\underline{%
\bigtriangledown }_\alpha ^{\star }~\longrightarrow \underline{%
\bigtriangledown }_\alpha +P_\alpha ,$ (the deformation d-tensor
$P_\alpha $ is induced by the torsion in dv-bundle
$\mathcal{B}_E)$ into the definition
of the curvature of gauge ha--fields (\ref{2.98a}) and motion equations (%
\ref{2.99a})  if interactions are modeled on a generic higher
order anisotropic space.

\section{D--spinor Einstein--Cartan Theory} \index{Einstein--Cartan}

The Einstein equations in some models of higher order anisotropic
supergravity have been considered in \cite{vlasg,vbook}. Here we
note that the Einstein equations and conservation laws on
v--bundles provided with N-connection structures were studied in
detail in \cite {ma87,ma94,ana86,ana87,vodg,voa,vcl96}. In Ref.
\cite{vg} we proved that the locally anisotropic gravity can be
formulated in a gauge like manner and analyzed the conditions
when the Einstein gravitational locally anisotropic field
equations are equivalent to a corresponding form of Yang-Mills
equations. Our aim here is to write the higher order anisotropic
gravitational field equations in a form more convenient for
theirs equivalent reformulation in higher order anisotropic
spinor variables.

\subsection{Einstein ha--equations}

We define d-tensor $\Phi _{<\alpha ><\beta >}$ as to satisfy
conditions
\[
-2\Phi _{<\alpha ><\beta >}\doteq R_{<\alpha ><\beta >}-\frac
1{n+m_1+...+m_z}\overleftarrow{R}g_{<\alpha ><\beta >}
\]
which is the torsionless part of the Ricci tensor for locally
isotropic spaces \cite{penr1,penr2}, i.e. $\Phi _{<\alpha
>}^{~~<\alpha >}\doteq 0$.\ The Einstein equations on higher
order anisotrop\-ic spaces
\begin{equation}
\overleftarrow{G}_{<\alpha ><\beta >}+\lambda g_{<\alpha ><\beta
>}=\kappa E_{<\alpha ><\beta >},  \label{2.34a}
\end{equation}
where
\[
\overleftarrow{G}_{<\alpha ><\beta >}=R_{<\alpha ><\beta >}-\frac 12%
\overleftarrow{R}g_{<\alpha ><\beta >}
\]
is the Einstein d--tensor, $\lambda $ and $\kappa $ are
correspondingly the cosmological and gravitational constants and
by $E_{<\alpha ><\beta >}$ is denoted the locally anisotropic
energy--momentum d--tensor, can be rewritten in equivalent form:
\begin{equation}
\Phi _{<\alpha ><\beta >}=-\frac \kappa 2(E_{<\alpha ><\beta
>}-\frac 1{n+m_1+...+m_z}E_{<\tau >}^{~<\tau >}~g_{<\alpha
><\beta >}).  \label{2.35a}
\end{equation}

Because ha--spaces generally have nonzero torsions we shall add to (%
\ref{2.35a}) (equivalently to (\ref{2.34a})) a system of
algebraic d--field equations with the source $S_{~<\beta ><\gamma
>}^{<\alpha >}$ being the locally anisotropic spin density of
matter (if we consider a variant of higher order anisotropic
Einstein--Cartan theory ):
\begin{equation}
T_{~<\alpha ><\beta >}^{<\gamma >}+2\delta _{~[<\alpha >}^{<\gamma
>}T_{~<\beta >]<\delta >}^{<\delta >}=\kappa S_{~<\alpha ><\beta
>.}^{<\gamma >}  \label{2.36a}
\end{equation}
From (\ref{2.36a}) one follows the conservation law of higher
order anisotropic spin matter:
\[
\bigtriangledown _{<\gamma >}S_{~<\alpha ><\beta >}^{<\gamma
>}-T_{~<\delta
><\gamma >}^{<\delta >}S_{~<\alpha ><\beta >}^{<\gamma >}=E_{<\beta ><\alpha
>}-E_{<\alpha ><\beta >}.
\]

\subsection{Einstein--Cartan d--equations}

Now we can write out the field equations of the Einstein--Cartan
theory in the d-spinor form. So, for the Einstein equations
(\ref{2.34}) we have

\[
\overleftarrow{G}_{\underline{\gamma }_1\underline{\gamma }_2\underline{%
\alpha }_1\underline{\alpha }_2}+\lambda \varepsilon _{\underline{\gamma }_1%
\underline{\alpha }_1}\varepsilon _{\underline{\gamma }_2\underline{\alpha }%
_2}=\kappa E_{\underline{\gamma }_1\underline{\gamma }_2\underline{\alpha }_1%
\underline{\alpha }_2},
\]
with $\overleftarrow{G}_{\underline{\gamma }_1\underline{\gamma }_2%
\underline{\alpha }_1\underline{\alpha }_2}$ from (\ref{2.86a}),
or, by using the d-tensor (\ref{2.87a}),

\[
\Phi _{\underline{\gamma }_1\underline{\gamma }_2\underline{\alpha }_1%
\underline{\alpha }_2}+(\frac{\overleftarrow{R}}4-\frac \lambda
2)\varepsilon _{\underline{\gamma }_1\underline{\alpha }_1}\varepsilon _{%
\underline{\gamma }_2\underline{\alpha }_2}=-\frac \kappa 2E_{\underline{%
\gamma }_1\underline{\gamma }_2\underline{\alpha
}_1\underline{\alpha }_2},
\]
which are the d-spinor equivalent of the equations (\ref{2.35a}).
These equations must be completed by the algebraic equations
(\ref{2.36a}) for the d-torsion and d-spin density with d-tensor
indices changed into corresponding d--spinor ones.

\subsection{Higher order an\-i\-sot\-rop\-ic gravitons}

Let a massless d-tensor field $h_{<\alpha ><\beta >}\left(
u\right) $ is interpreted as a small perturbation of the locally
anisotropic background metric d-field $g_{<\alpha ><\beta
>}\left( u\right) .$ Considering, for simplicity, a torsionless
background we have locally anisotropic Fierz--Pauli equations
\[
\Box h_{<\alpha ><\beta >}\left( u\right) +2R_{<\tau ><\alpha
><\beta ><\nu
>}\left( u\right) ~h^{<\tau ><\nu >}\left( u\right) =0
\]
and d--gauge conditions
\[
D_{<\alpha >}h_{<\beta >}^{<\alpha >}\left( u\right) =0,\quad
h\left( u\right) \equiv h_{<\beta >}^{<\alpha >}(u)=0,
\]
where $R_{<\tau ><\alpha ><\beta ><\nu >}\left( u\right) $ is
curvature d-tensor of the locally anisotropic background space
(these formulae can be obtained by using a perturbation formalism
with respect to $h_{<\alpha
><\beta >}\left( u\right) $ developed in \cite{gri}; in our case we must
take into account the distinguishing of geometrical objects and
operators on higher order anisotropic spaces).

Finally, we remark that all presented geometric constructions
contain those elaborated for generalized Lagrange spaces
\cite{ma87,ma94} (for which a tangent bundle $TM$ is considered
instead of a v-bundle $\mathcal{E}^{<z>}$ ) and for constructions
on the so called osculator bundles with different prolongations
and extensions of Finsler and Lagrange metrics \cite{mirata}. We
also note that the higher order Lagrange (Finsler) geometry is
characterized by a metric of type (dmetrichcv) with components
parametized
as $g_{ij}=\frac 12\frac{\partial ^2\mathcal{L}}{\partial y^i\partial y^j}$ $%
\left( g_{ij}=\frac 12\frac{\partial ^2\Lambda ^2}{\partial y^i\partial y^j}%
\right) $ and $h_{a_pb_p}=g_{ij},$ where $\mathcal{L=L}$ $%
(x,y_{(1)},y_{(2)},....,y_{(z)})$ $\left( \Lambda =\Lambda \left(
x,y_{(1)},y_{(2)},....,y_{(z)}\right) \right) $ is a Lagrangian $\left( %
\mbox{Finsler metric}\right) $ on $TM^{(z)}$ (see details in \cite
{ma87,ma94,mat,bej}).






\part{Finsler Geometry and Spinor Variables}

\chapter{Metrics Depending on Spinor Variables}   \index{spinor variables}

\section{Lorentz Transformation}

We present the transformation character of the connection, the
nonlinear connection and the spin connection coefficients with
respect to local Lorentz transformations which depend on spinor
variables, vector variables as well as coordinates.

For any quantities which transform as
\begin{equation}
f\left( x,y,\xi ,\overline{\xi }\right) \rightarrow f^{\prime
}\left(
x,y,\xi ^{\prime },\overline{\xi }^{\prime }\right) =U\left( x,y,\xi ,%
\overline{\xi }\right)  \label{9.2.1}
\end{equation}
their derivatives with respect to $x^i,y^i,\xi _\alpha $ and $\overline{\xi }%
^\alpha $ under Lorentz transformations
\begin{equation}
x^{i^{\prime }}=x^i,y^{i^{\prime }}=y^i,\xi _\alpha ^{\prime
}=\Lambda _{\ \alpha }^\beta \xi _\beta ,\overline{\xi }^{\prime
\alpha }=\Lambda _{~\beta }^{-1\alpha }\overline{\xi }^\beta
\label{9.2.2}
\end{equation}
will be given as follows
\begin{eqnarray}
a)\quad \frac{\partial U}{\partial x^\lambda } &=&\frac{\partial f^{\prime }%
}{\partial x^\lambda }+\frac{\partial f^{\prime }}{\partial \xi
_\alpha
^{\prime }}\frac{\partial \Lambda _{\ \alpha }^\beta }{\partial x^\lambda }%
\xi _\beta +\frac{\partial f^{\prime }}{\partial \overline{\xi
}^{\prime
\alpha }}\frac{\partial \Lambda _{~\beta }^{-1\alpha }}{\partial x^\lambda }%
\xi _\beta ,  \notag \\
b)\quad \frac{\partial U}{\partial \xi _\alpha }
&=&\frac{\partial f^{\prime }}{\partial \xi _\beta ^{\prime
}}\Lambda _{\ \beta }^\alpha +\frac{\partial f^{\prime
}}{\partial \xi _\beta ^{\prime }}\frac{\partial \Lambda _{\ \beta
}^\gamma }{\partial \xi _\alpha }\xi _\gamma +\frac{\partial f^{\prime }}{%
\partial \overline{\xi }^{\prime \beta }}\frac{\partial \Lambda _{~\gamma
}^{-1\beta }}{\partial \xi _\alpha }\overline{\xi }^\gamma ,
\label{9.2.3}
\\
c)\quad \frac{\partial U}{\partial \overline{\xi }^\alpha } &=&\frac{%
\partial f^{\prime }}{\partial \overline{\xi }^{\prime \alpha }}\Lambda
_{~\alpha }^{-1\beta }+\frac{\partial f^{\prime }}{\partial \xi
_\beta
^{\prime }}\frac{\partial \Lambda _{\ \beta }^\gamma }{\partial \overline{%
\xi }^\alpha }\xi _\gamma +\frac{\partial f^{\prime }}{\partial \overline{%
\xi }^{\prime \beta }}\frac{\partial \Lambda _{~\gamma }^{-1\beta }}{%
\partial \overline{\xi }^\alpha }\overline{\xi }^\gamma ,  \notag \\
c)\quad \frac{\partial U}{\partial y^\lambda } &=&\frac{\partial f^{\prime }%
}{\partial y^\lambda }+\frac{\partial f^{\prime }}{\partial \xi
_\beta
^{\prime }}\frac{\partial \Lambda _{\ \beta }^\gamma }{\partial y^\lambda }%
\xi _\gamma +\frac{\partial f^{\prime }}{\partial \overline{\xi
}^{\prime
\beta }}\frac{\partial \Lambda _{~\gamma }^{-1\beta }}{\partial y^\lambda }%
\overline{\xi }^\gamma .  \notag
\end{eqnarray}

Taking into account that (2.23) of \cite{ot3}, namely:
\begin{eqnarray}
\frac{\partial ^{\lbrack \ast ]}}{\partial x^{\lambda }} &=&\left( \frac{%
\partial }{\partial x^{\lambda }}+N_{\alpha \lambda }\frac{\partial }{%
\partial \xi _{\alpha }}+\overline{N}_{\lambda }^{\alpha }\frac{\partial }{%
\partial \overline{\xi }^{\alpha }}\right)  \notag \\
&&-\left( \Gamma _{\tau \lambda }^{k}+\overline{C}_{\tau
}^{\kappa \alpha }N_{\alpha \lambda }+\overline{N}_{\lambda
}^{\alpha }C_{\tau \alpha
}^{\kappa }\right) y^{\tau }\frac{\partial }{\partial y^{\kappa }}  \notag \\
&=&\frac{\partial ^{\lbrack \ast ]}}{\partial x^{\lambda
}}-\left( \Gamma _{\tau \lambda }^{(\ast )k}y^{\tau }\right)
\frac{\partial }{\partial y^{\kappa }},  \label{9.2.4}
\end{eqnarray}
(where the nonlinear connection coefficients $N_{\alpha \lambda }$ and $%
\overline{N}_{\lambda }^{\alpha }$ are given in \cite{ot1}), we substitute (%
\ref{9.2.3}) in (\ref{9.2.4}), then the nonlinear connection
coefficients have to be transformed for Lorentz scalar quantities
as
\begin{eqnarray}
a)\quad N_{\alpha \lambda }^{\prime } &=&N_{\beta \lambda
}\Lambda _{\ \alpha }^{\beta }+\frac{\partial ^{\lbrack \ast
]}\Lambda _{\ \alpha
}^{\beta }}{\partial x^{\lambda }}\xi _{\beta },  \label{9.2.5} \\
a^{\prime })\quad \overline{N}_{\lambda }^{\prime \alpha } &=&\overline{N}%
_{\lambda }^{\beta }\Lambda _{~\beta }^{-1\alpha }+\frac{\partial
^{\lbrack
\ast ]}\Lambda _{~\beta }^{-1\alpha }}{\partial x^{\lambda }}\overline{\xi }%
^{\beta }.  \notag
\end{eqnarray}
In the above mentioned (\ref{9.2.5}) a), a') the relation
$\partial ^{\lbrack \ast ]}/\partial x^{\lambda }=\partial
^{\lbrack \ast ]\prime }/\partial x^{\lambda }$ was used for
$[\ast ]$--differential operators. For the calculation of the
transformation character of nonlinear connection
coefficients $n_{\alpha \lambda },\widetilde{n}_{\lambda }^{\alpha },%
\widetilde{n}_{\lambda }^{0\alpha },\widetilde{n}_{0}^{\beta
\alpha },n_{\beta \alpha }^{0},n_{0\alpha }^{\beta }$ are used
the relations
\begin{equation*}
\frac{\partial ^{\lbrack \ast ]}}{\partial \xi _{\alpha
}}=\Lambda _{\ \beta
}^{\alpha }\frac{\partial ^{\lbrack \ast ]\prime }}{\partial \xi _{\beta }},%
\frac{\partial ^{\lbrack \ast ]}}{\partial \xi _{\alpha
}}=\Lambda _{~\beta
}^{-1\alpha }\frac{\partial ^{\lbrack \ast ]\prime }}{\partial \xi ^{\alpha }%
},\frac{\partial ^{\lbrack \ast ]}}{\partial y^{\lambda
}}=\frac{\partial ^{\lbrack \ast ]\prime }}{\partial y^{\prime
\lambda }}.
\end{equation*}
Also by means of (2.23) b), c), d) of \cite{ot3} and
(\ref{9.2.3}) we obtain
\begin{eqnarray*}
b)\quad n_{\beta \lambda }^{\prime } &=&\Lambda _{\beta }^{\alpha
}n_{\alpha
\lambda }+\frac{\partial ^{\lbrack \ast ]}\Lambda _{\ \beta }^{\gamma }}{%
\partial y^{\lambda }}\xi _{\gamma }, \\
b^{\prime })\quad \widetilde{n}_{\lambda }^{\prime \beta }
&=&\Lambda _{~\alpha }^{-1\beta }\widetilde{n}_{\lambda }^{\alpha
}+\frac{\partial
^{\lbrack \ast ]}\Lambda _{~\gamma }^{-1\beta }}{\partial y^{\lambda }}%
\overline{\xi }^{\gamma }, \\
c)\quad \widetilde{n}_{\delta }^{\prime 0\beta } &=&\Lambda
_{~\alpha
}^{-1\beta }\left( \Lambda _{\ \delta }^{\varepsilon }\widetilde{n}%
_{\varepsilon }^{0\beta }+\frac{\partial ^{\lbrack \ast ]}\Lambda
_{\ \delta
}^{\gamma }}{\partial \xi _{\alpha }}\xi _{\gamma }\right) , \\
c^{\prime })\quad \widetilde{n}_{0}^{\prime \delta \beta }
&=&\Lambda
_{~\alpha }^{-1\beta }\left( \Lambda _{~\varepsilon }^{-1\delta }\widetilde{n%
}_{0}^{\varepsilon \alpha }+\frac{\partial ^{\lbrack \ast
]}\Lambda _{~\gamma }^{-1\delta }}{\partial \xi _{\alpha
}}\overline{\xi }^{\gamma
}\right) , \\
d)\quad n_{\beta \alpha }^{\prime 0} &=&\Lambda _{\ \alpha
}^{\delta }\left( \Lambda _{\ \beta }^{\gamma }n_{\gamma \alpha
}^{0}+\frac{\partial ^{\lbrack \ast ]}\Lambda _{\ \beta }^{\gamma
}}{\partial \overline{\xi }^{\delta }}\xi
_{\gamma }\right) , \\
d^{\prime })\quad n_{0\alpha }^{\prime \beta } &=&\Lambda _{\
\alpha
}^{\delta }\left( n_{0\delta }^{\gamma }\Lambda _{~\gamma }^{-1\beta }+%
\overline{\xi }^{\gamma }\frac{\partial ^{\lbrack \ast ]}\Lambda
_{~\gamma }^{-1\beta }}{\partial \overline{\xi }^{\delta
}}_{\gamma }\right) .
\end{eqnarray*}
Consequently, $[\ast ]$--derivatives of the quantities
{\ref{9.2.1}) will satisfy the following relations:
\begin{eqnarray}
a)\quad \frac{\partial ^{\lbrack \ast ]}U}{\partial x^{\lambda }} &=&\frac{%
\partial f^{\prime }}{\partial x^{\lambda }}+N_{\alpha \lambda }^{\prime }%
\frac{\partial f^{\prime }}{\partial \xi _{\alpha }^{\prime }}+\overline{N}%
_{\lambda }^{\prime \alpha }\frac{\partial f^{\prime }}{\partial \overline{%
\xi }^{\prime \alpha }}-\Gamma _{\tau \lambda }^{(\ast )\kappa
}y^{\prime \tau }\frac{\partial f^{\prime }}{\partial y^{\prime
\kappa }},
\label{9.2.6} \\
b)\quad \frac{\partial ^{\lbrack \ast ]}U}{\partial y^{\lambda }} &=&\frac{%
\partial f^{\prime }}{\partial y^{\prime \lambda }}+n_{\alpha \lambda
}^{\prime }\frac{\partial f^{\prime }}{\partial \xi _{\alpha }^{\prime }}+%
\overline{n}_{\lambda }^{\prime \alpha }\frac{\partial f^{\prime
}}{\partial \overline{\xi }^{\prime \alpha }}-C_{\tau \lambda
}^{(\ast )\kappa }y^{\prime \tau }\frac{\partial f^{\prime
}}{\partial y^{\prime \kappa }},
\notag \\
c)\quad \Lambda _{~\alpha }^{-1\beta }\frac{\partial ^{\lbrack \ast ]}U}{%
\partial \xi _{\beta }} &=&\frac{\partial f^{\prime }}{\partial \xi _{\alpha
}^{\prime }}+\widetilde{n}_{\beta }^{\prime 0\alpha
}\frac{\partial f^{\prime }}{\partial \xi _{\beta }^{\prime
}}+\widetilde{n}_{0}^{\prime \beta \alpha }\frac{\partial
f^{\prime }}{\partial \overline{\xi }^{\prime
\beta }}-\overline{C}_{\tau }^{\prime (\ast )\kappa \alpha }y^{\prime \tau }%
\frac{\partial f^{\prime }}{\partial y^{\prime \kappa }},  \notag \\
d)\quad \Lambda _{~\alpha }^{\beta }\frac{\partial ^{\lbrack \ast ]}U}{%
\partial \overline{\xi }^{\beta }} &=&\frac{\partial f^{\prime }}{\partial
\overline{\xi }^{\prime \alpha }}+\widetilde{n}_{\beta \alpha }^{\prime 0}%
\frac{\partial f^{\prime }}{\partial \xi _{\beta }^{\prime }}+\widetilde{n}%
_{0\alpha }^{\prime \beta }\frac{\partial f^{\prime }}{\partial \overline{%
\xi }^{\beta }}-C_{\tau }^{\prime (\ast )\kappa \alpha }y^{\prime \tau }%
\frac{\partial f^{\prime }}{\partial y^{\prime \kappa }}.  \notag
\end{eqnarray}
We have Lorentz--scalar quantities
\begin{equation}
f^{\prime }\left( x,y,\xi ^{\prime },\overline{\xi }^{\prime
}\right) =f\left( x,y,\xi ,\overline{\xi }\right) ,  \label{9.2.7}
\end{equation}
then, the
\begin{equation*}
\frac{\partial ^{\lbrack \ast ]}f}{\partial x^{\lambda
}},\frac{\partial
^{\lbrack \ast ]}f}{\partial y^{\lambda }},\frac{\partial ^{\lbrack \ast ]}f%
}{\partial \xi _{\alpha }},\frac{\partial ^{\lbrack \ast
]}f}{\partial \overline{\xi }_{\alpha }}
\end{equation*}
are transformed as Lorentz--scalar and spinors adjoint to each
other, respectively. Consequently $[\ast ]$--differentiation are
covariant differential operators for Lorentz--scalar quantities.
The spin connection coefficients $\omega _{ab\lambda }^{[\ast
]},\theta _{ab\lambda }^{[\ast ]},\theta _{ab}^{[\ast ]\beta },$
$\theta _{ab\beta }^{[\ast ]}$ will be transformed by Lorentz
transformations as follows: }

We consider the relation (3.23) a) of \cite{ot3}, namely:
\begin{eqnarray}
\omega _{ab\lambda }^{[*]} &=&\left( \frac{\partial ^{[*]}h_a^\mu
}{\partial x^\lambda }+\Gamma _{\nu \lambda }^{(*)\mu }h_a^\nu
\right) h_{\mu b},
\label{9.2.8} \\
\omega _{ab\lambda }^{[*]\prime } &=&\left( \frac{\partial
^{[*]}h_a^{\prime \mu }}{\partial x^\lambda }+\Gamma _{\nu
\lambda }^{(*)^{\prime }\mu
}h_a^{\prime \nu }\right) h_{\mu b}^{\prime },  \notag \\
\Gamma _{\nu \lambda }^{(*)\mu } &=&\Gamma _{\nu \kappa \lambda
}^{(*)}g^{\kappa \mu }  \notag
\end{eqnarray}
also for the tetrads $h_a^{\prime \mu }$ and $h_b^{\prime \mu }$
valid the relation $h_a^{\prime \mu }=L_a^bh_b^{\prime \mu }$
(4.1) of \cite {ot1}), then taking into account (\ref{9.2.8}) we
take the transformation
formula of spin connection coefficients $\omega _{ab\lambda }^{[*]},$%
\begin{eqnarray}
a)\quad \omega _{ab\lambda }^{[*]\prime } &=&L_a^cL_b^d\omega
_{cd\lambda }^{[*]}+\frac{\partial ^{[*]}L_a^c}{\partial
x^\lambda }h_{cd}L_b^d,
\label{9.2.11} \\
b)\quad \theta _{ab\lambda }^{[*]\prime } &=&L_a^cL_b^d\theta
_{cd\lambda }^{[*]}+\frac{\partial ^{[*]}L_a^c}{\partial
y^\lambda }n_{cd}L_b^d,  \notag
\\
c)\quad \theta _{ab}^{[*]\prime \beta } &=&\Lambda _{~\gamma
}^{-1\beta }
\left[ \theta _{cd}^{[*]\gamma }L_a^cL_b^d+\frac{\partial ^{[*]}L_a^c}{%
\partial \xi _\gamma }L_b^dn_{dc}\right] ,  \notag \\
d)\quad \theta _{ab\beta }^{[*]\prime } &=&\Lambda _{~\beta
}^\gamma \left[ \theta _{cd\gamma
}^{[*]}L_a^cL_b^d+\frac{\partial ^{[*]}L_a^c}{\partial
\overline{\xi }^\gamma }L_b^dn_{dc}\right] ,  \notag
\end{eqnarray}
where the connection coefficients $\Gamma _{\nu \lambda }^{(*)\mu
},\Gamma _{\nu \lambda }^{(*)^{\prime }\mu }$ are Lorentz--scalar
and similar
procedures are considered for the transformed connection coefficients of $%
\theta _{ab\lambda }^{[*]},\theta _{ab}^{[*]\beta },\theta
_{ab\beta }^{[*]}, $ using the relations (3.23) b),c), of
\cite{ot3}.

Next, we shall derive the transformation character of the spin
connection coefficients $\left( \Gamma _{\tau \lambda
}^{(*)\kappa },C_{\tau \lambda }^{(*)\kappa },\overline{C}_\tau
^{\prime (*)\kappa \alpha },C_\tau ^{\prime (*)\kappa \alpha
}\right) $ under Lorentz transformations. If we take the relation
(3.6) a) of \cite{ot3},
\begin{equation}
N_{\tau \lambda }=\Gamma _{\tau \lambda }^{(*)\kappa }\xi _\kappa
\label{9.2.12}
\end{equation}
and
\begin{equation}
N_{\tau \lambda }^{\prime }=\Gamma _{\tau \lambda }^{(*)^{\prime
}\kappa }\xi _\kappa ^{\prime },  \label{9.2.13}
\end{equation}
and we substitute (\ref{9.2.5}) a) in (\ref{9.2.13}), then we get
the required transformation formula,
\begin{eqnarray}
a)\quad \Gamma _{\alpha \lambda }^{(*)^{\prime }\delta }
&=&\Lambda _\varepsilon ^{-1\delta }\Lambda _\alpha ^\beta \Gamma
_{\beta \lambda
}^{(*)\varepsilon }+\frac{\partial ^{[*]}\Lambda _\alpha ^\varepsilon }{%
\partial x^\lambda }\Lambda _\varepsilon ^{-1\delta },  \label{9.2.14} \\
b)\quad C_{\alpha \lambda }^{(*)^{\prime }\delta } &=&\Lambda
_\varepsilon
^{-1\delta }\Lambda _\alpha ^\beta C_{\beta \lambda }^{(*)\varepsilon }+%
\frac{\partial ^{[*]}\Lambda _\alpha ^\varepsilon }{\partial y^\lambda }%
\Lambda _\varepsilon ^{-1\delta },  \notag \\
c)\quad \widetilde{C}_\varepsilon ^{(*)\delta \rho } &=&\left[
\Lambda _\varepsilon ^\gamma \widetilde{C}_\gamma ^{(*)\beta
\alpha }\Lambda _\beta ^{-1\delta }+\frac{\partial ^{[*]}\Lambda
_\alpha ^\varepsilon }{\partial \xi _\alpha }\Lambda _\gamma
^{-1\delta }\right] \Lambda _\alpha ^{-1\rho },
\notag \\
d)\quad C_{\varepsilon \rho }^{(*)^{\prime }\delta } &=&\Lambda
_\rho ^\alpha \left[ \Lambda _\varepsilon ^\gamma C_{\gamma
\alpha }^{(*)\beta }\Lambda _\beta ^{-1\delta }+\frac{\partial
^{[*]}\Lambda _\varepsilon
^\gamma }{\partial \overline{\xi }^\alpha }\Lambda _\gamma ^{-1\delta }%
\right] .  \notag
\end{eqnarray}

Finally, from (3.20) of \cite{ot3} and (\ref{9.2.5}),
(\ref{9.2.14}), arbitrary terms $a_\lambda ,$ $b_\lambda,$
$\overline{\beta }^\alpha ,\beta _\alpha $ are transformed as
follows
\begin{eqnarray*}
a_\lambda &=&a_\lambda ^{^{\prime }}+\overline{\beta }^{\prime
\alpha
}\left( \frac{\partial \Lambda _{~\beta }^\alpha }{\partial x^\lambda }%
\right) \xi _\beta +\overline{\xi }^\beta \left( \frac{\partial
\Lambda _{~\beta }^{-1\alpha }}{\partial x^\lambda }\right) \beta
_\alpha ^{\prime },
\\
b_\lambda &=&b_\lambda ^{^{\prime }}+\overline{\beta }^{\prime
\beta }\left( \frac{\partial \Lambda _{~\beta }^\gamma }{\partial
y^\lambda }\right) \xi _\gamma +\overline{\xi }^\gamma \left(
\frac{\partial ^{[*]}\Lambda _{~\gamma }^{-1\beta }}{\partial
y^\lambda }\right) \beta _\beta ^{\prime },
\\
\overline{\beta }^\alpha &=&\overline{\beta }^{\prime \gamma
}\Lambda _{~\gamma }^\alpha +\overline{\beta }^{\prime \gamma
}\left( \frac{\partial
\Lambda _{~\beta }^\gamma }{\partial \xi _\alpha }\right) \xi _\varepsilon +%
\overline{\xi }^\varepsilon \left( \frac{\partial ^{[*]}\Lambda
_{~\varepsilon }^{-1\gamma }}{\partial y^\lambda }\right) \beta
_\gamma
^{\prime }, \\
\beta _\alpha &=&\Lambda _{~\alpha }^{-1\gamma }\beta _\gamma ^{\prime }+%
\overline{\beta }^{\prime \gamma }\left( \frac{\partial \Lambda
_{~\gamma
}^\gamma }{\partial \overline{\xi }^\alpha }\right) \xi _\varepsilon +%
\overline{\xi }^\varepsilon \left( \frac{\partial ^{[*]}\Lambda
_{~\varepsilon }^{-1\gamma }}{\partial \overline{\xi }^\alpha
}\right) \beta _\gamma ^{\prime }.
\end{eqnarray*}

\section{Curvature}             \index{curvature}

In this section we shall present the form of the curvature of the
above--mentioned spaces. There must exist ten kinds of curvature
tensors
corresponding to four kind of covariant derivatives with respect to $%
x^i,y^\lambda ,\xi _\alpha ,\overline{\xi }^\alpha ,$
(coordinates, vector variables, spinor variables).

If we denote with $M,n$ the number of curvatures and the kind of
covariant derivatives, then we have generally, $N=n(n+1)/2.$ In
our case $N=10,n=4.$ Like in \cite{ot1} (paragraph 5), here, they
appear three different expressions of the above--mentioned ten
curvature tensors which are closely related to each other. The
relation between ten curvature tensors $T_{\nu XY}^\mu $ and ten
spin--curvature tensors $T_{abXY}$ will be the following:
\begin{equation}
T_{abXY}=T_{\nu XY}^\mu h_a^\nu h_{\mu b}  \label{9.3.1}
\end{equation}
which arises from integrability conditions of the partial
differential equations (cf. (3.22) of \cite{ot1}).

The curvature tensors which are calculated below come from the
Ricci
identities \cite{run,mat}, as well as the commutation formula of the $[*]$%
--differential operators $\partial ^{[*]}/\partial x^\lambda $
and $\partial ^{[*]\prime }/\partial y^\lambda .$

The curvature tensors $T_{\nu XY}^{\mu }$ are defined as follows
\begin{eqnarray*}
R_{\nu \lambda \kappa }^{\mu } &=&\frac{\partial ^{\lbrack \ast
]}\Gamma _{\nu \lambda }^{(\ast )\mu }}{\partial x^{\kappa
}}-\frac{\partial
^{\lbrack \ast ]}\Gamma _{\nu \kappa }^{(\ast )\mu }}{\partial x^{\lambda }}%
+\Gamma _{\nu \lambda }^{(\ast )\tau }\Gamma _{\tau \kappa
}^{(\ast )\mu }-\Gamma _{\nu \kappa }^{(\ast )\tau }\Gamma _{\tau
\lambda }^{(\ast )\mu }
\\
&&-\left( A_{\gamma \lambda \kappa }^{[\ast ]}\overline{C}_{\nu
}^{[\ast
]\mu \gamma }+\widehat{A}_{\lambda \kappa }^{[\ast ]\gamma }\overline{C}%
_{\nu \gamma }^{[\ast ]\mu }+\breve{A}_{\lambda \kappa }^{[\ast
]\tau }C_{\nu \tau }^{[\ast ]\mu }\right) ,
\end{eqnarray*}
where $A_{\gamma \lambda \kappa }^{[\ast ]},\widehat{A}_{\lambda
\kappa }^{[\ast ]\gamma },\breve{A}_{\lambda \kappa }^{[\ast
]\tau }$ are given by
\begin{eqnarray*}
A_{\gamma \lambda \kappa }^{[\ast ]} &=&A_{\gamma \lambda \kappa
}-C_{\gamma }^{0\xi }A_{\xi \lambda \kappa }-\hat{A}_{\lambda
\kappa }^{\xi }C_{\gamma
\xi }^{0}-\breve{A}_{\lambda \kappa }^{\xi }C_{\gamma \xi }^{0}, \\
\widehat{A}_{\lambda \kappa }^{[\ast ]\gamma } &=&A_{\lambda
\kappa }^{[\ast
]\gamma }+\tilde{C}_{0}^{\gamma \xi }A_{\xi \lambda \kappa }-\hat{A}%
_{\lambda \kappa }^{\xi }C_{0\xi }^{\gamma }+\breve{A}_{\lambda
\kappa
}^{\xi }C_{\gamma \xi }^{0}, \\
\breve{A}_{\lambda \kappa }^{[\ast ]\rho } &=&\breve{A}_{\lambda
\kappa }^{\rho }+\left( \bar{C}_{\tau }^{\rho \xi }y^{\tau
}\right) A_{\xi \lambda \kappa }+\hat{A}_{\lambda \kappa }^{\xi
}\left( C_{\tau \xi }^{\rho }y^{\tau
}\right) , \\
A_{\gamma \lambda \kappa } &=&\frac{\partial ^{\lbrack \ast
]}N_{\gamma \lambda }}{\partial x^{\kappa }}-\frac{\partial
^{\lbrack \ast ]}N_{\gamma
\kappa }}{\partial x^{\lambda }}, \\
\quad \hat{A}_{\lambda \kappa }^{\gamma } &=&\frac{\partial ^{\lbrack \ast ]}%
\bar{N}_{\lambda }^{\gamma }}{\partial x^{\kappa
}}-\frac{\partial ^{\lbrack
\ast ]}\bar{N}_{\kappa }^{\gamma }}{\partial x^{\lambda }}, \\
\breve{A}_{\lambda \kappa }^{[\ast ]\rho } &=&\frac{\partial
^{\lbrack \ast
]}}{\partial x^{\kappa }}\left[ -\left( \Gamma _{\tau \lambda }^{\rho }+\bar{%
C}_{\tau }^{\rho \alpha }N_{\alpha \lambda }+\bar{N}_{\lambda
}^{\alpha
}C_{\tau \alpha }^{\rho }\right) y^{\tau }\right] \\
&&-\frac{\partial ^{\lbrack \ast ]}}{\partial x^{\lambda }}\left[
-\left( \Gamma _{\tau \kappa }^{\rho }+\bar{C}_{\tau }^{\rho
\alpha }N_{\alpha
\kappa }+\bar{N}_{\kappa }^{\alpha }C_{\tau \alpha }^{\rho }\right) y^{\tau }%
\right] .
\end{eqnarray*}

Similarly, the curvatures $P_{\nu \lambda \alpha }^{\mu }$ and
$W_{\nu \lambda \alpha }^{\mu }$ can be defined as follows
\begin{eqnarray*}
P_{\nu \lambda \alpha }^{\mu } &=&\frac{\partial ^{\lbrack \ast
]}\Gamma _{\nu \lambda }^{(\ast )\mu }}{\partial \xi ^{\alpha
}}-\frac{\partial
^{\lbrack \ast ]}C_{\nu \alpha }^{(\ast )\mu }}{\partial x^{\lambda }}%
+\Gamma _{\nu \lambda }^{(\ast )\tau }C_{\tau \alpha }^{(\ast
)\mu }-\Gamma
_{\tau \lambda }^{(\ast )\mu }C_{\nu \alpha }^{(\ast )\tau } \\
&&-\left( E_{\gamma \lambda \alpha }^{[\ast ]}\overline{C}_{\nu
}^{[\ast
]\mu \gamma }+\widehat{E}_{\lambda \alpha }^{[\ast ]\gamma }\overline{C}%
_{\nu \gamma }^{[\ast ]\mu }+\check{E}_{\lambda \kappa }^{[\ast
]\tau
}C_{\nu \tau }^{[\ast ]\mu }\right) , \\
R_{\nu \lambda \kappa }^{\mu } &=&\frac{\partial ^{\lbrack \ast
]}C_{\nu \lambda }^{(\ast )\mu }}{\partial x^{\kappa
}}-\frac{\partial ^{\lbrack \ast ]}\Gamma _{\nu \kappa }^{(\ast
)\mu }}{\partial y^{\lambda }}+\Gamma _{\nu \lambda }^{(\ast
)i}C_{i\kappa }^{(\ast )\mu }-\Gamma _{i\kappa }^{(\ast
)\mu }C_{\tau \lambda }^{(\ast )i} \\
&&-\left( D_{\gamma \lambda \kappa }^{[\ast ]}C_{\nu }^{[\ast ]\mu \gamma }+%
\widehat{D}_{\lambda \kappa }^{[\ast ]\gamma }C_{\nu \gamma }^{[\ast ]\mu }+%
\check{D}_{\lambda \kappa }^{[\ast ]\tau }C_{\nu \tau }^{[\ast
]\mu }\right) .
\end{eqnarray*}
The quantities $E_{\gamma \lambda \alpha }^{[\ast
]},\widehat{E}_{\lambda
\alpha }^{[\ast ]\gamma },\check{E}_{\lambda \kappa }^{[\ast ]\tau }$ and $%
D_{\gamma \lambda \kappa }^{[\ast ]},\widehat{D}_{\lambda \kappa
}^{[\ast ]\gamma },\check{D}_{\lambda \kappa }^{[\ast ]\tau }$
are defined respectively to $A_{\gamma \lambda \kappa }^{[\ast
]},\widehat{A}_{\lambda \kappa }^{[\ast ]\gamma
},\breve{A}_{\lambda \kappa }^{[\ast ]\tau }.$ A a matter of fact
the expressions are too big to be presented for all ten curvature
tensors, we prefer to give an algorithm for the general case,
presenting the following the Table \ref{Table 1} of symbols for
nonlinear connection.

{\small
\begin{table*}[h]
\begin{center}
{\small
\begin{tabular}{|c|c|c|c|c|}
\hline {\
\begin{tabular}{l}
{coordinate} \\
vector \\
spinors
\end{tabular}
} & {\
\begin{tabular}{l}
{connection} \\
coefficients
\end{tabular}
} & $\mathbf{N}_{\gamma Y}$ & $\mathbf{\hat{N}}_X^Y$ & $- \mathbf{\bar{N}}%
_X^k $ \\ \hline
$x^i$ & $\Gamma _{\nu \lambda }^{(*)\mu }$ & $N_{\alpha \lambda }$ & $\bar{N}%
_\lambda ^\alpha $ & $\left( \Gamma _{\tau \lambda }^\rho
+\bar{C}_\tau ^{\rho \alpha }N_{\alpha \lambda }+\bar{N}_\lambda
^\alpha C_{\tau \alpha
}^\rho \right) y^\tau $ \\
$y^\lambda $ & $C_{\nu \alpha }^{(*)\mu }$ & $n_{\alpha \lambda }$ & $\bar{n}%
_\lambda ^\alpha $ & $\left( C_{\tau \lambda }^\rho +\bar{C}_\tau
^{\rho \alpha }n_{\alpha \lambda }+\bar{n}_\lambda ^\alpha
C_{\tau \alpha }^\rho
\right) y^\tau $ \\
$\xi _\alpha $ & $\overline{C}_\nu ^{[*]\mu \gamma }$ & $n_\beta
^{0\alpha }$
& $\bar{n}_0^{\beta \alpha }$ & $\left( \bar{C}_\tau ^{\rho \alpha }+\bar{C}%
_\tau ^{\rho \alpha }n_\beta ^{0\alpha }+\bar{n}_0^{\beta \alpha
}C_{\tau
\beta }^\rho \right) y^\tau $ \\
$\overline{\xi }^\alpha $ & $C_{\nu \lambda }^{(*)\mu }$ &
$n_{\beta \alpha }^0$ & $n_{0\alpha }^\beta $ & $\left( C_{\tau
\alpha }^\rho +\bar{C}_\tau ^{\rho \beta }n_{\beta \alpha
}^0+n_{0\alpha }^\beta C_{\tau \beta }^\rho \right) y^\tau $ \\
\hline
\end{tabular}
}
\end{center}
\caption{Nonlinear connections} \label{Table 1}
\end{table*}
} \index{nonlinear connection}

In general for each of the ten curvature tensors, we have
\begin{eqnarray}
\mathbf{T}_{\nu XY}^{\mu } &=&\frac{\partial ^{\lbrack \ast ]}Con\mathbf{X}%
_{\nu X}^{\mu }}{\partial Y}-\frac{\partial ^{\lbrack \ast ]}Con\mathbf{Y}%
_{\nu X}^{\mu }}{\partial X}  \label{9.3.2} \\
&&+Con\mathbf{X}_{\nu X}^{\mu }Con\mathbf{Y}_{\tau Y}^{\mu }-Con\mathbf{Y}%
_{\nu Y}^{\mu }Con\mathbf{X}_{\tau X}^{\mu }  \notag \\
&&-\left( \mathbf{A}_{\gamma XY}^{[\ast ]}\overline{C}_{\nu
}^{[\ast ]\mu \gamma }+\widehat{\mathbf{A}}_{XY}^{[\ast ]\gamma
}C_{\nu \gamma }^{[\ast ]\mu }+\mathbf{\breve{A}}_{XY}^{[\ast
]\tau }C_{\nu \tau }^{[\ast ]\mu }\right) ,  \notag
\end{eqnarray}
where the coefficients are given by
\begin{eqnarray*}
\mathbf{A}_{\gamma XY}^{[\ast ]} &=&\mathbf{A}_{\gamma \lambda
\kappa }-C_{\gamma }^{0\xi }\mathbf{A}_{\xi
XY}-\mathbf{\hat{A}}_{XY}^{\xi
}C_{\gamma \xi }^{0}-\mathbf{\breve{A}}_{XY}^{\xi }C_{\gamma \xi }^{0}, \\
\widehat{\mathbf{A}}_{XY}^{[\ast ]\gamma }
&=&\mathbf{A}_{XY}^{[\ast ]\gamma }+\tilde{C}_{0}^{\gamma \xi
}\mathbf{A}_{\xi XY}-\mathbf{\hat{A}}_{XY}^{\xi
}C_{0\xi }^{\gamma }+\mathbf{\breve{A}}_{XY}^{\xi }C_{\gamma \xi }^{0}, \\
\mathbf{\breve{A}}_{XY}^{[\ast ]\rho }
&=&\mathbf{\breve{A}}_{XY}^{\rho
}+\left( \bar{C}_{\tau }^{\rho \xi }y^{\tau }\right) \mathbf{A}_{\xi XY}+%
\mathbf{\hat{A}}_{XY}^{\xi }\left( C_{\tau \xi }^{\rho }y^{\tau }\right) , \\
\mathbf{A}_{\gamma XY} &=&\frac{\partial ^{\lbrack \ast
]}\mathbf{N}_{\gamma
X}}{\partial Y}-\frac{\partial ^{\lbrack \ast ]}\mathbf{N}_{\gamma Y}}{%
\partial X}, \\
\mathbf{\hat{A}}_{XY}^{\gamma } &=&\frac{\partial ^{\lbrack \ast ]}\mathbf{%
\bar{N}}_{X}^{\gamma }}{\partial Y}-\frac{\partial ^{\lbrack \ast ]}\mathbf{%
\bar{N}}_{Y}^{\gamma }}{\partial X}, \\
\mathbf{\breve{A}}_{\lambda \kappa }^{[\ast ]\rho }
&=&\frac{\partial ^{\lbrack \ast ]}\mathbf{\check{N}}_{X}^{\rho
}}{\partial Y}-\frac{\partial ^{\lbrack \ast
]}\mathbf{\check{N}}_{Y}^{\rho }}{\partial X},
\end{eqnarray*}
$Con\mathbf{X}_{\nu X}^{\mu }$ represent the connection
coefficients $\left(
\Gamma _{\nu \kappa }^{(\ast )\mu },C_{\nu \gamma }^{[\ast ]\mu },\overline{C%
}_{\nu }^{[\ast ]\mu \gamma },C_{\nu \alpha }^{[\ast ]\mu
}\right) .$ We can write down all ten curvatures using the
algorithm presented the above and adopt the following symbolism:

We can write down all the spin--curvature tensors using the
symbolism of
Table \ref{Table 2} with appropriate indices. The spin curvature tensros $%
T_{abXY}$ are defined in (\ref{9.3.3}). According the Tables
\ref{Table 2} and \ref{Table 3} our general formula becomes
\begin{eqnarray}
T_{abXY} &=&\frac{\partial ^{[*]}sp.ConX_{abX}}{\partial
Y}-\frac{\partial
^{[*]}sp.ConX_{abY}}{\partial X}  \label{9.3.3} \\
&&+sp.ConX_{acX}\ sp.ConY_{bY}^c-sp.ConY_{acY}\ sp.ConX_{bX}^c  \notag \\
&&-\left( \mathbf{A}_{\gamma XY}^{[*]}\ \theta _{ab}^{[*]\gamma }+\widehat{%
\mathbf{A}}_{XY}^{[*]\gamma }\theta _{ab\gamma }^{[*]}+\mathbf{\breve{A}}%
_{XY}^{[*]\tau }\theta _{ab\tau }^{[*]}\right) ,  \notag
\end{eqnarray}
where $sp.ConX_{abX}$ represent the spin connection coefficients
$\omega
_{ab\lambda }^{[*]},$ $\theta _{ab\lambda }^{[*]},$ $\overline{\theta }%
_{ab}^{[*]\alpha },$ $\theta _{ab\lambda }^{[*]}$ with before defined $%
\mathbf{A}_{\gamma XY}^{[*]},\widehat{\mathbf{A}}_{XY}^{[*]\gamma }, \mathbf{%
\breve{A}}_{XY}^{[*]\tau }.$

These spinor--curvature tensors will also appear in Ricci'
formulae for a Lorentz vector field. To examine the
transformation character of the curvature tensors it is
convenient to divide them into the parts $T_{\nu
XY}^{(0)\mu }$ and $T_{\nu XY}^{(1)\mu },$%
\begin{equation*}
T_{\nu XY}^\mu =T_{\nu XY}^{(0)\mu }-T_{\nu XY}^{(1)\mu },
\end{equation*}
where
\begin{eqnarray*}
T_{\nu XY}^{(0)\mu } &=&\frac{\partial ^{[*]}ConX_{\nu X}^\mu }{\partial Y}-%
\frac{\partial ^{[*]}ConY_{\nu Y}^\mu }{\partial X} \\
&&+ConX_{\nu X}^\tau ConY_{\tau Y}^\mu -ConY_{\nu Y}^\mu
ConX_{\tau X}^\mu ,
\\
T_{\nu XY}^{(1)\mu } &=&\mathbf{A}_{\gamma
XY}^{[*]}\overline{C}_\nu ^{[*]\mu \gamma
}+\widehat{\mathbf{A}}_{XY}^{[*]\gamma }C_{\nu \gamma }^{[*]\mu
}+\mathbf{\breve{A}}_{XY}^{[*]\tau }C_{\nu \tau }^{[*]\mu }.
\end{eqnarray*}
The curvature tensors $T_{\nu XY}^{(1)\mu }$ are expected to have
the same transformation character as $T_{\nu XY}^{(0)\mu }$ and
$T_{\nu XY}^\mu $ and are confirmed to transform as tensors or
spinors under general coordinate transformations and local
Lorentz transformations by formulae (\ref{9.2.3}), (\ref{9.2.5})
and (\ref{9.2.14}). The arbitrary terms of spin connection
coefficients are contained only in the parts $T_{\nu XY}^{(1)\mu
}$, the arbitrariness disappear completely by virtue of the
homogeneity of $\Gamma _{\nu \kappa }^{(*)\mu },C_{\nu \gamma
}^{[*]\mu },\overline{C}_\nu ^{[*]\mu \gamma },C_{\nu \alpha
}^{[*]\mu }.$ Therefore, $T_{\nu XY}^\mu $ as well as $T_{\nu
XY}^{(1)\mu }$ are defined unambiguously. The following conditions
are imposed on $T_{\nu XY}^{(0)\mu }$ and $T_{\nu XY}^{(1)\mu }$
and, therefore, on $T_{\nu XY}^\mu .$

\begin{table*}[h]
\begin{center}
\begin{tabular}{|c|c|c|c|c|c|c|}
\hline
$X-Y$ & $T_{\nu XY}^\mu $ & $T_{\varepsilon XY}^\delta $ & $T_{XY}$ & $%
\mathbf{A}^{[*]}$ & $\widehat{\mathbf{A}}^{[*]}$ &
$\mathbf{\breve{A}}^{[*]}$
\\ \hline
$x-x$ & $R$ & $X$ & $\varphi $ & $A$ & $\hat{A}$ & $\check{A}$ \\
$x-\xi $ & $P$ & $\overline{\Xi }$ & $\Psi $ & $E$ & $\hat{E}$ &
$\check{E}$
\\
$x-\bar{\xi}$ & $\overline{P}$ & $\Xi $ & $\overline{\Psi }$ &
$F$ & $\hat{F}
$ & $\check{F}$ \\
$x-y$ & $W$ & $\Psi $ & $x$ & $D$ & $\hat{D}$ & $\check{D}$ \\
$\xi -\xi $ & $\overline{Q}$ & $O$ & $\rho $ & $B$ & $\hat{B}$ &
$\check{B}$
\\
$\xi -\bar{\xi}$ & $S$ & $K$ & $\mu $ & $V$ & $\hat{V}$ & $\check{V}$ \\
$\xi -y$ & $\Omega $ & $\widetilde{U}$ & $\nu $ & $G$ & $\hat{G}$ & $\check{G%
}$ \\
$\bar{\xi}-\bar{\xi}$ & $Q$ & $O$ & $\overline{\rho }$ & $J$ & $\hat{J}$ & $%
\check{J}$ \\
$\bar{\xi}-y$ & $\overline{\Omega }$ & $U$ & $\overline{\nu }$ & $\Phi $ & $%
\hat{\Phi}$ & $\check{\Phi}$ \\
$y-y$ & $Z$ & $Y$ & $\upsilon $ & $H$ & $\hat{H}$ & $\check{H}$
\\ \hline
\end{tabular}
\end{center}
\caption{Curvatures} \label{Table 2}
\end{table*}

Contractions of $\overline{\xi },\xi ,y^\lambda $ with the
curvature tensors give the following:
\begin{eqnarray}
\overline{\xi }^\alpha T_{(\overline{\xi }^\alpha ,x^\lambda )}^{(\ )} &=&0,~%
\overline{\xi }^\alpha T_{(\overline{\xi }^\alpha ,\xi _\alpha
)}^{(\ )}=0,
\label{9.3.4} \\
\overline{\xi }^\alpha T_{(\overline{\xi }^\alpha ,y^\lambda )}^{(\ )} &=&0,~%
\overline{\xi }^\alpha T_{(\overline{\xi }^\alpha ,\overline{\xi
}^\alpha )}^{(\ )}=0,  \notag
\end{eqnarray}

The above mentioned structures and properties of curvature
tensors $T_{\nu XY}^\mu $ are transformed to those of
spin--curvature tensors $T_{abXY}$ through the relations
(\ref{9.3.1}). Also, the integrability conditions of the partial
differential equations of Ricci formulae for a spinor field, led
to another spin--curvature tensors $T_{\varepsilon XY}^\delta $
which are related to $T_{abXY}$ by the relation of
\begin{equation*}
T_{\varepsilon XY}^\delta =\frac 12T_{abXY}~\left( S^{ab}\right)
_\varepsilon ^\delta +iT_{XY}~I_\varepsilon ^\delta ,
\end{equation*}
where $I_\varepsilon ^\delta $ is the unit matrix, $T_{\varepsilon
XY}^\delta ,T_{XY}$ are defined by (\ref{9.3.5}) and (\ref{9.3.6})
respectively, $T_{abXY}$ are given by (\ref{9.3.3}) and $S^{ab}$
and (3.18) of \cite{ot3}.

\begin{table}[h]
\begin{tabular}{|c|c|c|c|}
\hline
\begin{tabular}{l}
coordinate \\
vector \\
spinors
\end{tabular}
&
\begin{tabular}{l}
Spin connection \\
coefficients 1
\end{tabular}
&
\begin{tabular}{l}
Spin connection \\
coefficients 2
\end{tabular}
& coef. Xx \\ \hline $x^{\lambda }$ & $\omega _{ab\lambda
}^{[\ast ]}$ & $\Gamma _{\nu k}^{(\ast
)\mu }$ & $a_{\lambda }^{[\ast ]}$ \\
$y^{\lambda }$ & $\theta _{ab\lambda }^{[\ast ]}$ & $C_{\nu
\gamma }^{(\ast
)\mu }$ & $b_{\lambda }^{[\ast ]}$ \\
$\xi _{\alpha }$ & $\overline{\theta }_{ab}^{[\ast ]\alpha }$ & $\overline{C}%
_{\nu }^{(\ast )\mu \gamma }$ & $\overline{\beta }^{[\ast ]\alpha }$ \\
$\overline{\xi }^{\alpha }$ & $\theta _{ab\lambda }^{[\ast ]}$ &
$C_{\nu \alpha }^{(\ast )\mu }$ & $\beta _{\alpha }^{[\ast ]}$ \\
\hline
\end{tabular}
\caption{Spin Connections} \label{Table 3}
\end{table}

Again, in order to present the spin--curvature tensors
$T_{\varepsilon XY}^\delta $ we are going to use an algorithm
along with appropriate columns in the Tables \ref{Table 2} and
\ref{Table 3}. The general formula is
\begin{eqnarray}
T_{\varepsilon XY}^\delta &=&\frac{\partial
^{[*]}sp.ConX_{\varepsilon X}^\delta }{\partial Y}-\frac{\partial
^{[*]}sp.ConY_{\varepsilon Y}^\delta
}{\partial X}  \label{9.3.5} \\
&&+sp.ConX_{\varepsilon X}^j\ sp.ConY_{jY}^\delta
-sp.ConY_{\varepsilon
Y}^j\ sp.ConX_{jX}^\delta  \notag \\
&&-\left( \mathbf{A}_{jXY}^{[*]}\ \widetilde{C}_\varepsilon ^{[*]\delta j}+%
\widehat{\mathbf{A}}_{XY}^{[*]j}C_\varepsilon ^{[*]\delta j}+\mathbf{\breve{A%
}}_{XY}^{[*]\tau }C_{\varepsilon \tau }^{[*]\delta }\right) ,
\notag
\end{eqnarray}
where $sp.ConX_{\varepsilon X}^j$ represent the spin connection
coefficients and $\mathbf{A}^{[*]}$ are defined as before.

The spin--curvature tensors $T_{XY}$ consisting of the arbitrary terms of $%
\Gamma _{\nu \kappa }^{(*)\mu },$ $C_{\nu \gamma }^{[*]\mu },$ $\overline{C}%
_\nu ^{[*]\mu \gamma },$ $C_{\nu \alpha}^{[*]\mu} $ are defined
as follows
\begin{eqnarray}
T_{XY} &=&\frac{\partial ^{[*]}coefX_X}{\partial Y}-\frac{\partial
^{[*]}coefY_Y}{\partial X}  \label{9.3.6} \\
&&+i\left( coefX_X~coefY_Y-coefY_Y\ coefX_X\right)  \notag \\
&&-\left( \mathbf{A}_{\gamma XY}^{[*]}\ \overline{\beta }_{XY}^{[*]\gamma }+%
\widehat{\mathbf{A}}_{XY}^{[*]\gamma }\beta _{XY}^{[*]\gamma }+\mathbf{%
\breve{A}}_{XY}^{[*]\tau }b_\tau ^{[*]}\right) ,  \notag
\end{eqnarray}
where the $coefX_X$ are defined in Table \ref{Table 3}. If we
want to write down all ten spin--curvature tensors $T_{XY}$ we
must use the corresponding column in Table \ref{Table 2}. These
objects are defined uniquely on account of the conditions (3.27)
or (3.28) of \cite{ot3} and the homogeneity
properties of $\Gamma _{\nu \kappa }^{(*)\mu },C_{\nu \gamma }^{[*]\mu },%
\overline{C}_\nu ^{[*]\mu \gamma },C_{\nu \alpha }^{[*]\mu }. $
There are imposed on $T_{XY}$ conditions similar to
(\ref{9.3.4}): that is
contractions of $\overline{\xi },\xi ,y^\lambda $ with the spin--curvature $%
T_{XY}$ results
\begin{equation}
\overline{\xi }^\alpha \Psi _{\lambda \alpha }=0.  \label{9.3.7}
\end{equation}
Now, from (\ref{9.3.1}), (\ref{9.3.5}) together with
(\ref{9.3.4}), (\ref {9.3.7}), it is easily shown that the
similar conditions to (\ref{9.3.2}) on $T_{\nu XY}^\mu $ must be
imposed on $T_{\varepsilon XY}^\delta .$


\chapter{Field Equations in Spinor Variables} \index{spinor variables}

\section{Introduction}

The introduction of a metric $g_{\mu \nu }(x,\omega )$ that
depends on the position variables $x$ as well as on the spinor
variables $\omega $ assigns a non-Riemannian structure to the
space and provides it with torsion. This procedure enables the
construction of a non-localized (bi--local) gravitational field,
identical to the one proposed by Yukawa \cite{yuk} that allows a
more general description of elementary particles. Further
arguments have been developed by some other authors
\cite{ik1,ot1,t1}. In our context each point of the space-time is
characterized by the influence of two fields: an external one
which is the conventional field in Einstein`s sense, and an
internal one due to the introduction of the spinor variables.
These fields are expected to play the role of a geometrical
unification of the fields. If $\omega $ is represented by a
vector $y$, then we work in the Finslerian framework
\cite{asa,ik2,ma94}. A more general case of the gauge approach in
the framework of Finsler and Lagrange geometry has been studied
e.g. in \cite{asa88a,asa89a,beja,ma94,mb,mtbr}.

In the following, we consider a space-time and we denote its
metric tensor by
\begin{equation*}
g_{\mu \nu }(Z^{M}),
\end{equation*}
(here $Z^{M}=(x^{\mu },\xi _{\alpha },\bar{\xi}^{\alpha }),x^{\mu
},\xi _{\alpha },\bar{\xi}^{\alpha }$ represent the position and
the 4-spinor variables $\bar{\xi}$ denotes the Dirac conjugate of
$\xi $) \cite{t1}. With the Greek letters $\lambda ,\mu ,\nu $
and $\alpha ,\beta ,\gamma $ we
denote the space-time indices and the spinor indices, also Latin letters $%
\alpha ,b,c$ are used for the Lorenz (flat) indices. The
(*)-differential operators $\partial _{M}^{(\ast )}$ are defined
as
\begin{equation}
\partial _{M}^{(\ast )}=\frac{\partial ^{(\ast )}}{\partial Z^{M}}=\biggl(%
\frac{\partial ^{(\ast )}}{\partial x^{\mu }},\frac{\partial ^{(\ast )}}{%
\partial \xi _{\alpha }},\frac{\partial ^{(\ast )}}{\partial \bar{\xi}%
^{\alpha }}\biggr),  \label{10.1.1}
\end{equation}
with
\begin{eqnarray*}
\frac{\partial ^{(\ast )}}{\partial x^{\lambda }}&=&\frac{\partial }{%
\partial x^{\lambda }}+N_{\alpha \lambda }\frac{\partial }{\partial \xi
^{\alpha }}+\overline{N}_{\lambda }^{\alpha }\frac{\partial }{\partial \bar{%
\xi}^{\alpha }}, \\
\frac{\partial ^{(\ast )}}{\partial \xi _{\alpha }}& =& \frac{\partial }{%
\partial \xi _{\alpha }}+\tilde{\eta}_{\beta }^{0\alpha }\frac{\partial }{%
\partial \xi _{\alpha }}+\tilde{\eta}_{0}^{\beta \alpha }\frac{\partial }{%
\partial \bar{\xi}^{\beta }}, \\
\frac{\partial ^{(\ast )}}{\partial \bar{\xi}^{\alpha }}&=&\frac{\partial }{%
\partial \bar{\xi}^{\alpha }}+\eta _{\beta \alpha }^{0}\frac{\partial }{%
\partial \xi _{\beta }}+\eta _{0\alpha }^{\beta }\frac{\partial }{\partial
\bar{\xi}^{\beta }},
\end{eqnarray*}
here $N_{\alpha \lambda },\overline{N}_{\lambda }^{\alpha },\tilde{\eta}%
_{\beta }^{0\alpha },\tilde{\beta \alpha }_{0},\eta _{\beta \alpha
}^{0},\eta _{0\alpha }^{\beta }$ are the nonlinear connections
\cite{ot1}.

In our study, field equations are obtained from a Lagrangian
density of the form
\begin{equation}
L(\Psi ^{(A)},\partial _M^{(*)}\Psi ^{(A)}),  \label{10.1.2}
\end{equation}
here $\Psi ^{(A)}$ is the set
\begin{equation*}
\Psi ^{(A)}=\{h_\mu ^a(x,\xi ,\bar{\xi}),\omega _\mu ^{(*)ab}(x,\xi ,\bar{\xi%
}),\theta _\alpha ^{(*)ab}(x,\xi
,\bar{\xi}),\bar{\theta}^{(*)ab\alpha }(x,\xi ,\bar{\xi})\}.
\end{equation*}
Thus $L$ is a function of the tetrad field, of the spin connection
coefficients and of their (*)-derivatives. the variables $h,\omega
^{(*)},\theta ^{(*)},\bar{\theta}^{(*)}$ are considered as
independent.

It is known that gravity can be described by the tetrad field and
the Lorenz connection coefficients \cite{ramon}. The variation of
the Palatini action
 \index{Palatini}
with respect to $h$ and $\omega $ yields a set of two equations:
\begin{eqnarray}
R_\mu ^a-\frac 12Rh_\mu ^a & =& 0\qquad (a)  \notag \\
D_\mu [h(h_a^\nu h_b^\mu -h_b^\nu h_a^\mu )] & =0 & \qquad (b)
\label{10.1.3}
\end{eqnarray}
$R_\mu ^a$ is the determinant of the tetrad $h_\mu ^a$ and $D_\mu
$ is the gauge covariant derivative
\begin{equation*}
D_\mu =\partial _\mu +\sum \omega _\mu ,
\end{equation*}
where the sum is taken over all Lorentz indices.

In spaces whose metric tensor depends on spinor variables, an
analogous method can be applied, but instead of one connection we
have three connections:
\begin{equation*}
\omega _{\mu }^{(\ast )}(x,\xi ,\bar{\xi}),\qquad \theta
_{a}^{(\ast )}(x,\xi ,\bar{\xi}),\qquad \bar{\theta}^{(\ast
)\alpha }(x,\xi ,\bar{\xi}).
\end{equation*}
So we choose a Lagrangian density of the form (\ref{10.1.2}) from
which four equations are obtained. The analogous gauge covariant
derivatives of $D_{\mu }$ appear naturally as
\begin{eqnarray}
a)~~D_{\mu }^{(\ast )}& = & \partial _{\mu }^{(\ast )}+\sum \omega
_{\mu}^{(\ast )},\qquad  \notag \\
b)~D_{\alpha }^{(\ast )}& = &\partial _{\alpha }^{(\ast )}+\sum
\theta
_{\alpha }^{(\ast )},\qquad  \notag \\
c)~~D^{(\ast )\alpha }& =&\partial ^{(\ast )\alpha }+\sum \bar{\theta}%
^{(\ast )\alpha }.\qquad  \label{10.1.4}
\end{eqnarray}
Transformation laws of the connection coefficients $\omega
_{ab\lambda
}^{(\ast )}(x,\xi ,\bar{\xi}),\theta _{ab\alpha }^{(\ast )}(x,\xi ,\bar{\xi}%
) $ and $\bar{\theta}_{ab}^{(\ast )\alpha }(x,\xi ,\bar{\xi})$
under local Lorenz transformations are the expected
transformation laws for the gauge potentials \cite{ramon}
\begin{eqnarray}
a)~\omega _{ab\lambda }^{(\ast )}& = & L_{a}^{c}L_{b}^{d}\omega
_{cd\lambda
}^{(\ast )}+\frac{\partial ^{(\ast )}L_{\alpha }^{c}}{\partial x^{\lambda }}%
L_{bc},\qquad  \notag \\
b)~\bar{\theta}_{ab}^{(\ast )^{\prime }a}& = & \biggl[L_{a}^{c}L_{b}^{d}\bar{%
\theta}_{cd}^{(\ast )\beta }+\frac{\partial ^{(\ast )}L_{\alpha }^{c}}{%
\partial \xi _{\beta }}L_{bc}\biggr](\Lambda ^{-1})_{\beta }^{\alpha },\qquad
\notag \\
c)~\theta _{ab\alpha }^{(\ast )^{\prime }}& = & \Lambda _{\alpha }^{\beta }%
\biggl[L_{a}^{c}L_{b}^{d}\theta _{cd\beta }^{(\ast
)}+\frac{\partial ^{(\ast )}L_{\alpha }^{c}}{\partial
\bar{\xi}_{\beta }}L_{bc}\biggr].\qquad \label{10.1.5}
\end{eqnarray}
The matrices $L$ and $\Lambda $ belong to the vector and spinor
representations of the Lorentz group, respectively.

\section{Derivation of the field equations}

The field equations will be the Euler-Lagrange equations for a
given Lagrangian. We postulate the explicit form of the
Lagrangian density
\begin{equation}
L(\Psi ^{(A)},\partial _{M}^{(\ast )}\psi ^{(A)}).  \label{10.2.1}
\end{equation}
But first we observe that the metric tensor $g_{\mu \nu }$ and the tetrad $%
h_{\mu }^{a}$ are related by (cf. \cite{ramon})
\begin{equation*}
a)~g_{\mu \nu }(x,\xi ,\bar{\xi})=h_{\mu }^{a}h_{\nu }^{b}\eta
_{ab},\qquad
\end{equation*}
\begin{equation}
b)~g^{\mu \nu }(x,\xi ,\bar{\xi})=h_{a}^{\mu }h_{b}^{\nu }\eta
^{ab},\qquad \label{10.2.2}
\end{equation}
where $\eta _{ab}$ is the Minkowski metric tensor and it is of the form%
\newline
$diag(+1,-1,-1,-1)$. From the relations \cite{W}:
\begin{equation}
a)\quad g=-h^{2},\qquad b)\quad dg=gg^{\mu \nu }dg_{\mu \nu },
\label{10.2.3}
\end{equation}
and using (\ref{10.2.2}), we get
\begin{equation}
\frac{\partial h}{\partial h_{a}^{\mu }}=-\frac{1}{2}hh_{\mu
}^{a}, \label{10.2.4}
\end{equation}
where $g=$det$(g_{\mu \nu })$.

Now we postulate the Lagrangian density in the form
\begin{equation}
L=h(R+P+Q+S),  \label{10.2.5}
\end{equation}
where $R,P,Q,S$ are the scalar curvatures obtained by contraction
of the spin curvature tensors:
\begin{eqnarray}
R &=&h_{a}^{\mu }h_{b}^{\nu }R_{\mu \nu }^{ab},\ P=h_{a}^{\mu
}h_{b}^{\nu
}P_{c\mu \alpha }^{a}\overline{P}_{\nu }^{bc\alpha },  \label{10.2.6} \\
Q &=&Q_{ab\beta \alpha }\widetilde{Q}^{ab\beta \alpha },\
S=S_{ab\beta }^{\alpha }S_{\alpha }^{ab\beta }.  \notag
\end{eqnarray}

The spin curvature tensors are given by the components
\begin{eqnarray}
P_{\lambda \alpha }^{ab} &=&\frac{\partial ^{(\ast )}\omega
_{\lambda }^{(\ast )ab}}{\partial \bar{\xi}^{\alpha
}}-\frac{\partial ^{(\ast )}\theta
_{\alpha }^{(\ast )ab}}{\partial x^{\lambda }}  \label{10.2.7} \\
&&+\omega _{c\lambda }^{(\ast )a}\theta _{\alpha }^{(\ast
)cb}-\theta _{c\alpha }^{(\ast )a}\omega _{\lambda }^{(\ast
)cb}-(\bar{\theta}^{ab\beta }E_{\beta \lambda \alpha }+F_{\lambda
\alpha }^{\beta }\theta _{\beta }^{ab}),  \notag
\end{eqnarray}
\begin{eqnarray*}
\overline{P}_{\lambda }^{ab\alpha } &=&\frac{\partial ^{(\ast
)}\omega
_{\lambda }^{(\ast )ab}}{\partial \xi _{\alpha }}-\frac{\partial ^{(\ast )}%
\bar{\theta}^{(\ast )ab\alpha }}{\partial x^{\lambda }}  \notag \\
&&+\omega _{c\lambda }^{(\ast )a}\bar{\theta}^{(\ast )a\alpha b}-\bar{\theta}%
_{c}^{(\ast )a\alpha }\omega _{\lambda }^{(\ast )cb}-(\bar{\theta}^{ab\beta }%
\widetilde{F}_{\beta \lambda }^{\alpha }+\widetilde{E}_{\lambda
}^{\beta
\alpha }\theta _{\beta }^{ab}),  \notag \\
S_{\beta }^{ab\alpha } &=&\frac{\partial ^{(\ast )}\omega _{\beta
}^{(\ast )ab}}{\partial \xi _{\alpha }}-\frac{\partial ^{(\ast
)}\bar{\theta}^{(\ast
)ab\alpha }}{\partial \bar{\xi}^{\beta }}  \notag \\
&&+\theta _{c\beta }^{(\ast )a}\bar{\theta}^{(\ast )c\alpha b}-\bar{\theta}%
_{c}^{(\ast )a\alpha }\theta _{\beta }^{(\ast )cb}-(\theta
_{\gamma }^{ab}G_{\gamma \beta }^{\alpha }+H_{\beta }^{\gamma
\alpha }\theta _{\gamma
}^{ab}),  \notag \\
R_{\mu \nu }^{ab} &=&\frac{\partial ^{(\ast )}\omega _{\mu }^{(\ast )ab}}{%
\partial x^{\nu }}-\frac{\partial ^{(\ast )}\theta _{\nu }^{(\ast )ab}}{%
\partial x^{\mu }}  \notag \\
&&+\omega _{\mu }^{(\ast )ac}\omega _{c\nu }^{(\ast )b}-\omega
_{\nu }^{(\ast )ac}\omega _{c\mu }^{(\ast
)b}-(\bar{\theta}^{ab\beta }A_{\beta \mu
\nu }+\overline{A}_{\mu \nu }^{\beta }\theta _{\beta }^{ab}),  \notag \\
\widetilde{Q}_{ab}^{\beta \alpha } &=&\frac{\partial ^{(\ast )}\bar{\theta}%
_{ab}^{(\ast )\beta }}{\partial \xi _{\alpha }}-\frac{\partial ^{(\ast )}%
\bar{\theta}_{ab}^{(\ast )\alpha }}{\partial \xi _{\beta }}  \notag \\
&&+\bar{\theta}_{ac}^{(\ast )\beta }\theta _{b}^{(\ast )ca}-\bar{\theta}%
_{ac}^{(\ast )\alpha }\bar{\theta}_{b}^{(\ast )c\beta }-(\bar{\theta}%
_{ab}^{\gamma }\widetilde{K}_{\gamma }^{\beta \alpha
}+\widetilde{J}^{\gamma
\beta \alpha }\theta _{ab\gamma }),  \notag \\
Q_{\beta \alpha }^{ab} &=&\frac{\partial ^{(\ast )}\theta _{\beta
}^{(\ast )ab}}{\partial \bar{\xi}^{\alpha }}-\frac{\partial
^{(\ast )}\theta _{\alpha
}^{(\ast )ab}}{\partial x^{\lambda }}  \notag \\
&&+\theta _{c\beta }^{(\ast )a}\theta _{b\alpha }^{(\ast
)c}-\theta _{c\alpha }^{(\ast )a}\theta _{\beta }^{(\ast
)cb}-(\theta ^{ab\gamma }J_{\gamma \beta \alpha }+K_{\beta \alpha
}^{\gamma }\theta _{\gamma }^{ab}), \notag
\end{eqnarray*}
where the coefficients are defined
\begin{eqnarray*}
A_{\beta \mu \nu } &=&\frac{\partial ^{(\ast )}N_{\beta \mu
}}{\partial
x^{\nu }}-\frac{\partial ^{(\ast )}N_{\beta \nu }}{\partial x^{\mu }},%
\overline{A}_{\mu \nu }^{\beta }=\frac{\partial ^{(\ast
)}\overline{N}_{\mu }^{\beta }}{\partial x^{\nu }}-\frac{\partial
^{(\ast )}\overline{N}_{\nu
}^{\beta }}{\partial x^{\mu }}, \\
E_{\beta \lambda \alpha } &=&\frac{\partial ^{(\ast )}N_{\beta \lambda }}{%
\partial \bar{\xi}^{\alpha }}-\frac{\partial ^{(\ast )}\eta _{\beta \alpha
}^{0}}{\partial x^{\lambda }},F_{\lambda \alpha }^{\beta
}=\frac{\partial
^{(\ast )}\overline{N}_{\lambda }^{\beta }}{\partial \bar{\xi}^{\alpha }}-%
\frac{\partial ^{(\ast )}\eta _{0\alpha }^{\beta }}{\partial
x^{\lambda }},
\\
\widetilde{F}_{\beta \lambda }^{\alpha } &=&\frac{\partial ^{(\ast
)}N_{\beta \lambda }}{\partial \xi _{\alpha }}-\frac{\partial ^{(\ast )}%
\tilde{\eta}_{\beta }^{0\alpha }}{\partial x^{\lambda }},\widetilde{E}%
_{\lambda }^{\beta \alpha }=\frac{\partial ^{(\ast
)}\overline{N}_{\lambda
}^{\beta }}{\partial \xi _{\alpha }}-\frac{\partial ^{(\ast )}\tilde{\eta}%
_{0}^{\beta \alpha }}{\partial x^{\lambda }}, \\
G_{\gamma \beta }^{\alpha } &=&\frac{\partial ^{(\ast )}\eta
_{\gamma \beta
}^{0}}{\partial \xi _{\alpha }}-\frac{\partial ^{(\ast )}\tilde{\eta}%
_{\gamma }^{0\alpha }}{\partial \bar{\xi}^{\beta }},H_{\beta
}^{\gamma \alpha }=\frac{\partial ^{(\ast )}\eta _{0\beta
}^{\gamma }}{\partial \xi
_{\alpha }}-\frac{\partial ^{(\ast )}\tilde{\eta}_{0}^{\gamma \alpha }}{%
\partial \bar{\xi}^{\beta }}, \\
J_{\gamma \beta \alpha } &=&\frac{\partial ^{(\ast )}\eta
_{\gamma \beta }^{0}}{\partial \bar{\xi}^{\alpha
}}-\frac{\partial ^{(\ast )}\eta _{\gamma
\alpha }^{0}}{\partial \bar{\xi}^{\beta }},K_{\beta \alpha }^{\gamma }=\frac{%
\partial ^{(\ast )}\eta _{0\beta }^{\gamma }}{\partial \bar{\xi}^{\alpha }}-%
\frac{\partial ^{(\ast )}\eta _{0\alpha }^{\gamma }}{\partial \bar{\xi}%
^{\beta }}, \\
\widetilde{K}_{\gamma }^{\beta \alpha } &=&\frac{\partial ^{(\ast
)}\eta
_{\gamma }^{0\beta }}{\partial \xi _{\alpha }}-\frac{\partial ^{(\ast )}%
\tilde{\eta}_{\gamma }^{0\alpha }}{\partial \xi _{\beta }},\widetilde{J}%
^{\gamma \beta \alpha }=\frac{\partial ^{(\ast
)}\tilde{\eta}_{0}^{\gamma
\beta }}{\partial \xi _{\alpha }}-\frac{\partial ^{(\ast )}\tilde{\eta}%
_{0}^{\gamma \alpha }}{\partial \xi _{\beta }}.
\end{eqnarray*}

The Lagrangian (\ref{10.2.5}) is the only possible scalar that
can be made from the curvature tensors (\ref{10.2.7}) and it must
be the sum of the first-order quantity $R$ and the second--order
quantities $P,Q$ and $S$. The mixing of the quantities of
different order is not impossible. It is known
that the Einstein-Maxwell Lagrangian is the sum of the first-order quantity $%
R$ and the second-order quantity $F_{\mu \nu }F^{\mu \nu }$. So,
our Lagrangian (\ref{10.2.5}) is physically acceptable.

The Euler--Lagrange equations for the objects
\begin{equation*}
\Psi ^{(A)}=\{h^{\mu },\omega _{\mu }^{(\ast )},\theta _{\alpha }^{(\ast )},%
\bar{\theta}^{(\ast )\alpha }\}
\end{equation*}
are of the form
\begin{equation}
\partial _{M}^{(\ast )}\biggl(\frac{\partial L}{\partial (\partial
_{M}^{(\ast )}\Psi ^{(A)})}\biggr)-\frac{\partial L}{\partial
\Psi ^{(A)}}=0, \label{10.2.8}
\end{equation}
where $\partial _{M}^{(\ast )}$ was defined in (\ref{10.1.1}).
From the variation of $L$ with respect to the tetrad we have
\begin{equation}
\frac{\partial L}{\partial h_{b}^{\nu }}=0.  \label{10.2.9}
\end{equation}
Taking into account (\ref{10.2.3}), and (\ref{10.2.4}) we get the
equation
\begin{equation}
H_{\nu }^{b}-\frac{1}{2}h_{\nu }^{b}=0,  \label{10.2.10}
\end{equation}
where
\begin{equation}
H_{\nu }^{b}=R_{\nu }^{b}+P_{\nu }^{b}=h_{a}^{\mu }R_{\mu \nu
}^{ab}+h_{a}^{\mu }P_{c\mu \alpha }^{a}\overline{P}_{\nu
}^{bc\alpha }, \label{10.2.11}
\end{equation}
and
\begin{equation}
H=R+P.  \label{10.2.12}
\end{equation}
From the variation of $L$ with respect to $\omega _{\mu }^{(\ast
)}$ we get
\begin{eqnarray}
\partial _{\mu }^{(\ast )}\frac{\partial L}{\partial (\partial _{\mu
}^{(\ast )}\omega _{\nu }^{(\ast )ab})}+\partial ^{(\ast )\alpha }\frac{%
\partial L}{\partial (\partial ^{(\ast )\alpha }\omega _{\nu }^{(\ast )ab})}%
+ &&  \label{10.2.13} \\
\partial _{\alpha }^{(\ast )}\frac{\partial L}{\partial (\partial _{\alpha
}^{(\ast )}\omega _{\nu }^{(\ast )ab})}-\frac{\partial
L}{\partial \omega _{\nu }^{(\ast )ab}} &=&0.  \notag
\end{eqnarray}
The spin-connection coefficients $\omega _{\nu }^{(\ast )ab}$ are
contained in $R$ and $P$:
\begin{equation*}
h(R+P)=hh_{a}^{\mu }h_{b}^{\nu }(R_{\mu \nu }^{ab}+P_{c\mu \alpha }^{a}%
\overline{P}_{\nu }^{bc\alpha }).
\end{equation*}
From relation (\ref{10.2.13}) we get the following variation of
the term $hR$ with respect to $\omega _{\mu }^{(\ast )}$:
\begin{eqnarray}
&&\partial _{\mu }^{(\ast )}\frac{\partial (hR)}{\partial
(\partial _{\mu
}^{(\ast )}\omega _{\nu }^{(\ast )ab})}+\partial ^{(\ast )\alpha }\frac{%
\partial (hR)}{\partial (\partial ^{(\ast )\alpha }\omega _{\nu }^{(\ast
)ab})}+  \notag \\
&&\partial _{\alpha }^{(\ast )}\frac{\partial (hR)}{\partial
(\partial
_{\alpha }^{(\ast )}\omega _{\nu }^{(\ast )ab})}-\frac{\partial (hR)}{%
\partial \omega _{\nu }^{(\ast )ab}}.  \label{10.2.14}
\end{eqnarray}
By a direct calculation, the first term of (\ref{10.2.14}) can be
written as
\begin{equation*}
\partial _{\mu }^{(\ast )}[h(h_{a}^{\nu }h_{b}^{\mu }-h_{a}^{\mu }h_{b}^{\nu
})].
\end{equation*}
The second and the third terms of (\ref{10.2.14}) are equal to
zero. The fourth term equals
\begin{equation}
h(h_{c}^{\nu }h_{b}^{\mu }-h_{b}^{\nu }h_{c}^{\mu })\omega _{a\mu
}^{(\ast )c}+h(h_{c}^{\nu }h_{a}^{\mu }-h_{a}^{\nu }h_{c}^{\mu
})\omega _{b\mu }^{(\ast )c}.  \label{(2.15)}
\end{equation}
Consequently, the first and the fourth terms can be rewritten as
\begin{equation}
D_{\mu }^{(\ast )}[h(h_{a}^{\nu }h_{b}^{\mu }-h_{b}^{\nu
}h_{a}^{\mu })], \label{10.2.16}
\end{equation}
where we have used the gauge covariant derivative $D_{\mu }^{(\ast )}$ from (%
\ref{10.1.4}). Contribution from the P-part is equal to
\begin{eqnarray}
\partial _{\mu }^{(\ast )}\frac{\partial (hP)}{\partial (\partial _{\mu
}^{(\ast )}\omega _{\nu }^{(\ast )ab})}+\partial ^{(\ast )\alpha }\frac{%
\partial (hP)}{\partial (\partial ^{(\ast )\alpha }\omega _{\nu }^{(\ast
)ab})}+ &&  \notag \\
\partial _{\alpha }^{(\ast )}\frac{\partial (hP)}{\partial (\partial
_{\alpha }^{(\ast )}\omega _{\nu }^{(\ast )ab})}-\frac{\partial (hP)}{%
\partial \omega _{\nu }^{(\ast )ab}}. &&  \label{10.2.17}
\end{eqnarray}
The first term of (\ref{10.2.17}) is equal to zero.The second and
the third terms can be written as
\begin{gather}
\partial ^{(\ast )\alpha }(hh_{c}^{\mu }h_{a}^{\nu }P_{b\mu \alpha }^{c}),
\label{10.2.18} \\
\partial _{\alpha }^{(\ast )}(hh_{c}^{\mu }h_{a}^{\nu }\overline{P}_{b\mu
}^{c\alpha }),  \label{10.2.19}
\end{gather}
respectively. The fourth term may be written as
\begin{equation}
hh_{a}^{\nu }h_{l}^{\mu }\theta _{bc\alpha }^{(\ast
)}\overline{P}_{\mu
}^{cl\alpha }-hh_{l}^{\nu }h_{k}^{\mu }\theta _{a\alpha }^{(\ast )l}%
\overline{P}_{b\mu }^{k\alpha }-hh_{l}^{\mu }h_{a}^{\nu }P_{c\mu \alpha }^{l}%
\bar{\theta}_{b}^{(\ast )c\alpha }-hh_{k}^{\nu }h_{l}^{\mu
}P_{b\mu \alpha }^{l}\bar{\theta}_{a}^{(\ast )k\alpha }.
\label{10.2.20}
\end{equation}
The sum of (\ref{10.2.18}), (\ref{10.2.19}) and (\ref{10.2.20})
is equal to
\begin{equation}
D_{\alpha }^{(\ast )}(hh_{a}^{\nu }h_{l}^{\mu }\overline{P}_{b\mu
}^{l\alpha })+D^{(\ast )\alpha }(hh_{a}^{\nu }h_{l}^{\mu }P_{b\mu
}^{l\alpha }). \label{10.2.21}
\end{equation}
So, (\ref{10.2.17}) is written in the form
\begin{equation}
D_{\mu }^{(\ast )}[h(h_{a}^{\nu }h_{b}^{\mu }-h_{b}^{\nu
}h_{a}^{\mu })]+D_{\alpha }^{(\ast )}(hh_{a}^{\nu }h_{l}^{\mu
}\overline{P}_{b\mu }^{l\alpha })+D^{(\ast )\alpha }(hh_{a}^{\nu
}h_{l}^{\mu }\overline{P}_{b\mu }^{l\alpha })=0.  \label{10.2.22}
\end{equation}
Taking the variation of $L$ with respect to $\theta _{\alpha
}^{(\ast )}$ we have contributions from $(P+Q+S)$. The field
equation is
\begin{eqnarray}
\partial _{\mu }^{(\ast )}\frac{\partial (hL)}{\partial (\partial _{\mu
}^{(\ast )}\theta _{\nu }^{(\ast )ab})}+\partial ^{(\ast )\alpha }\frac{%
\partial (hL)}{\partial (\partial ^{(\ast )\alpha }\theta _{\nu }^{(\ast
)ab})} &&  \label{10.2.23} \\
+\partial _{\alpha }^{(\ast )}\frac{\partial (hL)}{\partial
(\partial
_{\alpha }^{(\ast )}\theta _{\nu }^{(\ast )ab})}-\frac{\partial (hL)}{%
\partial \theta _{\nu }^{(\ast )ab}} &=&0.  \notag
\end{eqnarray}
We proceed in the same way as before. The contribution from the
$hP$ term is
\begin{equation}
-D_{\mu }^{(\ast )}(hh_{a}^{\mu }h_{c}^{\nu }\overline{P}_{b\nu
}^{ca}). \label{10.2.24}
\end{equation}
The contribution from the $hQ$ term gives
\begin{equation}
D_{\beta }^{(\ast )}(2h\widetilde{Q}_{ab}^{[\alpha \beta ]}).
\label{10.2.25}
\end{equation}
Similarly, the $hS$ term yields
\begin{equation}
2D^{(\ast )\beta }(hS_{ab\beta }^{\alpha }).  \label{10.2.26}
\end{equation}
So, the third equation is written in the form
\begin{equation}
D_{\mu }^{(\ast )}(hh_{a}^{\mu }h_{c}^{\nu }\overline{P}_{b\nu
}^{ca})-D_{\beta }^{(\ast )}(2h\widetilde{Q}_{ab}^{[\alpha \beta
]})-2D^{(\ast )\beta }(hS_{ab\beta }^{\alpha })=0. \label{10.2.27}
\end{equation}
Finally, the variation with respect to $\theta ^{(\ast )\alpha }$
yields the equation ''conjugate'' to (\ref{10.2.27})
\begin{equation}
D_{\mu }^{(\ast )}(hh_{c}^{\mu }h_{a}^{\nu }P_{b\nu \alpha
}^{c})-D^{(\ast )\beta }(2hQ_{ab[\alpha \beta ]})-2D_{\beta
}^{(\ast )}(hS_{ab\alpha }^{\beta })=0.  \label{10.2.28}
\end{equation}

\section{Generalized Conformally Flat Spaces}

In this Section we study the form of the spin-connection
coefficients, spin-curvature tensors, and the field equations for
generalized conformally
flat spaces (GCFS) $(M,g_{\mu \nu }(x,\xi ,\bar{\xi})=e^{2\sigma (x,\xi ,%
\bar{\xi})}\eta _{\mu \nu })$, where $\eta _{\mu \nu }$
represents the
Lo\-renz metric tensor $\eta _{\mu \nu }=$diag$(+,-,-,-)$, and $\xi ,\bar{\xi%
}$ are internal variables. The case of conformally related
metrics of the Riemannian and the generalized Lagrange spaces has
been extensively studied in \cite{mb,mtbr}. It is remarkable that
in the above mentioned GCFS spaces, some spin--connection and
spin-curvature tensors vanish.

As pointed out in \cite{ot1}, the absolute differential $DV^\mu $
of a vector field $V^\mu (x,\xi .\bar{\xi})$ is expressed in
terms of the coefficients
\begin{equation}
\{\Gamma _{\nu \lambda }^\mu ,\overline{C}_\nu ^{\mu \alpha
},C_{\nu \alpha }^\mu \}.  \label{10.3.1}
\end{equation}
Considering the absolute differentials of the spinor variables $\xi _\alpha ,%
\bar{\xi}^\alpha $:
\begin{gather*}
D\xi _\alpha =d\xi _\alpha -N_{\alpha \lambda }dx^\lambda -\tilde{\eta}%
_\alpha ^{0\beta }D\xi _\beta -D\bar{\xi}^\beta \eta _{\alpha \beta }^0, \\
D\bar{\xi}^\alpha =d\bar{\xi}^\alpha -\overline{N}_\lambda ^\alpha
dx^\lambda -\tilde{\eta}_0^{\alpha \beta }D\xi _\beta
-D\bar{\xi}^\beta \eta _{0\beta }^\alpha ,
\end{gather*}
which depend on the nonlinear connections:
\begin{equation}
\{N_{\alpha \lambda },\overline{N}_\lambda ^\alpha
,\tilde{\eta}_\alpha ^{0\beta },\tilde{\eta}_0^{\alpha \beta
}\eta _{\alpha \beta }^0,\eta _{0\beta }^\alpha \}, \label{10.3.2}
\end{equation}
and expressing $DV^\mu $ in terms of $dx^\lambda $,$D\xi _\alpha $,$D\bar{\xi%
}^\alpha $, we obtain the connection coefficients
\begin{equation}
\{\Gamma _{\nu \alpha }^{(*)\mu },\overline{C}_\nu ^{(*)\mu
\alpha },C_{\nu \alpha }^{(*)\mu }\}  \label{10.3.3}
\end{equation}
related to the coefficients (\ref{10.3.1}) via the non-linear connections (%
\ref{10.3.2}) \cite{ot1}.

By imposing the postulates of the length preservation for the
parallel vector fields and symmetry of the derived coefficients
\begin{equation}
\{\Gamma _{\nu \mu \lambda }^{(*)},\overline{C}_{\nu \mu }^\alpha
,C_{\nu \mu \alpha }\}  \label{10.3.4}
\end{equation}
in the first two tensor indices, we have the relations:
\begin{gather}
\Gamma _{\nu \mu \lambda }^{(*)}=\frac 12\biggl(\frac{\partial
^{(*)}g_{\mu \nu }}{\partial x^\lambda }-\frac{\partial
^{(*)}g_{\nu \lambda }}{\partial
x^\mu }\biggr),  \notag \\
\overline{C}_{\nu \mu }^\alpha =\frac 12\frac{\partial g_{\nu \mu }}{%
\partial \xi _\alpha },\qquad C_{\nu \mu \alpha }=\frac 12\frac{\partial
g_{\nu \mu }}{\partial \bar{\xi}^\alpha },  \label{10.3.5}
\end{gather}
where $\tau _{\{\mu \nu \}}=\tau _{\mu \nu }+\tau _{\nu \mu }$.

\begin{theorem}
\label{th10.1} For the GCFS spaces we infer the following:\newline
(a) The coefficients (\ref{10.3.4}) have the explicit form
\begin{equation}
\Gamma _{\nu \mu \lambda }^{*}=e^{2\sigma }(\eta _\mu \{_\nu
\sigma _\lambda ^{*}\}-\eta _{\nu \lambda }\sigma _\mu
^{*}),\quad \overline{C}_\nu ^{\mu
\alpha }=\delta _\nu ^\mu \sigma ^\alpha ,\quad C_{\nu \alpha }^\mu =\bar{%
\sigma}_\alpha \delta _\nu ^\mu ,  \label{10.3.6}
\end{equation}
where $\sigma ^\alpha =\partial \sigma /\partial \xi _\alpha ,\bar{\sigma}%
_\alpha =\partial \sigma /\partial \bar{\xi}^\alpha ,\sigma
_\lambda ^{*}=\partial ^{*}\sigma /\partial x^\lambda $ are the
derivation operators of scalar fields involving the coefficients
(\ref{10.3.2}).\newline (b) The following relations hold:
\begin{gather}
\Gamma _{\nu \lambda }^\mu =\Gamma _{\nu \lambda }^{*\mu }-\delta
_\nu ^\mu \sigma ^\alpha N_{\alpha \lambda }-\delta _\nu ^\mu
\bar{\sigma}_\alpha
\overline{N}_\lambda ^\alpha ,  \notag \\
\overline{C}_\nu ^{*\mu \alpha }=\overline{C}_\nu ^{\mu \alpha
}+\delta _\nu
^\mu \sigma ^\beta \tilde{\eta}_\beta ^{0\alpha }+\delta _\nu ^\mu \bar{%
\sigma}_\beta \tilde{\eta}_0^{\beta \alpha },  \label{10.3.7} \\
C_{\nu \alpha }^{*\mu }=C_{\nu \alpha }^\mu +\delta _\nu ^\mu
\sigma ^\beta \eta _{\beta \alpha }^0+\delta _\nu ^\mu \eta
_{0\alpha }^\beta .  \notag
\end{gather}
\end{theorem}

\begin{proof}  Computational, using the consequences  (\ref{10.3.5})
 of the above postulates and identifying the absolute differentials
 expressed in terms of (\ref{10.3.1})  and (\ref{10.3.3}) .
\end{proof}

Considering the absolute differentials of a Dirac spinor field $\psi (x,\xi ,%
\bar{\xi})$ and of its adjoint $\bar{\psi}(x,\xi ,\bar{\xi})$ we
have the coefficients    \index{spinor}
\begin{equation}
\{\Gamma _{\gamma \lambda }^\beta ,\widetilde{C}_\gamma ^{\beta
\alpha },C_{\gamma \alpha }^\beta \}.  \label{10.3.8}
\end{equation}
Expressing $D\psi $ and $D\bar{\psi}$ in terms of $dx^\lambda
,D\xi _\alpha ,D\bar{\xi}^\alpha $, we are led to the
spin-connection coefficients I:
\begin{equation}
\{\Gamma _{\gamma \lambda }^{*\beta },\widetilde{C}_\gamma
^{*\beta \alpha },C_{\gamma \alpha }^{*\beta }\}  \label{10.3.9}
\end{equation}
connected to (\ref{10.3.8}) \cite{ot1}. In a similar manner, the
absolute differential of a Lorenz vector $V^a(x,\xi ,\bar{\xi})$
produces the coefficients
\begin{equation}
\{\omega _{ba\lambda },\bar{\theta}_{ba}^\alpha ,\theta
_{ba\alpha }\}, \label{10.3.10}
\end{equation}
where the raising and lowering of the indices $a,b,\ldots
=1,\ldots ,4$ are performed by means of $\eta _{ab}$, and also
the spin-connection coefficients II:
\begin{equation}
\{\omega _{ba\lambda }^{*},\bar{\theta}_{ba}^{*\alpha },\theta
_{ba\alpha }^{*}\}  \label{10.3.11}
\end{equation}
related to the coefficients (\ref{10.3.10}) and (\ref{10.3.3})
\cite{ot1}. Similarly to the previous work of Takano and Ono
\cite{ot1}, we shall postulate the invariance of length of the
parallel Lorentz vector fields, and the vanishing of the absolute
differentials and covariant derivatives of the tetrads $h_\alpha
^\mu $,which involve the connection coefficients (\ref {10.3.3})
and (\ref{10.3.11}).

In the GCFS, the tetrads are given by $h_{\mu }^{a}(x,\xi ,\bar{\xi}%
)=e^{\sigma (x,\xi ,\bar{\xi})}\delta _{\mu }^{a}$ and lead to
the dual
entities $h_{a}^{\mu }(x,\xi ,\bar{\xi})=e^{-\sigma (x,\xi ,\bar{\xi}%
)}\delta _{a}^{\mu }$. In general, the above postulates produce
the relations:
\begin{gather}
\omega _{ab\lambda }=\left( \frac{\partial h_{a}^{\mu }}{\partial
x^{\lambda
}}+\Gamma _{\nu \lambda }^{\mu }h_{a}^{\nu }\right) h_{\mu b},  \notag \\
\bar{\theta}_{ab}^{\alpha }=\left( \frac{\partial h_{a}^{\mu
}}{\partial \xi _{\alpha }}+\overline{C}_{\nu }^{\mu \alpha
}h_{a}^{\nu }\right) h_{\mu b},
\label{10.3.12} \\
\theta _{ab\alpha }=\left( \frac{\partial h_{a}^{\mu }}{\partial \bar{\xi}%
^{\alpha }}+C_{\nu \alpha }^{\mu }h_{a}^{\nu }\right) h_{\mu b},  \notag \\
\omega _{ab\lambda }^{\ast }=\left( \frac{\partial ^{\ast }h_{a}^{\mu }}{%
\partial x^{\lambda }}+\Gamma _{\nu \lambda }^{\ast \mu }h_{a}^{\nu }\right)
h_{\mu b}.  \label{10.3.13}
\end{gather}
For the GCFS case we are led to

\begin{theorem}
\label{th10.2} The spin--connection coefficients (II) and the
coefficients\\
 (\ref{10.3.10}) are subject to
\begin{gather}
\omega _{ba\lambda }=h_{\mu a}\Gamma _{b\lambda }^{\mu }-\sigma
_{\lambda }\eta _{ba},\quad \bar{\theta}_{ab}^{\alpha }=0,\quad
\theta _{ab\alpha }=0,
\label{10.3.14} \\
\omega _{ba\lambda }^{\ast }=\eta _{\lambda (a}\sigma _{b)}^{\ast
},\quad \bar{\theta}_{ab}^{\ast \alpha }=0,\quad \theta
_{ab\alpha }^{\ast }=0,
\label{10.3.15} \\
\omega _{ba\lambda }^{\ast }=\omega _{ba\lambda }, \label{10.3.16}
\end{gather}
where $h_{\mu a}=e^{\sigma }\eta _{\mu a}$ and
$T_{(ab)}=T_{ab}-T_{ba}$.
\end{theorem}

\begin{proof}  Relations (\ref{10.3.12})
 imply (\ref{10.3.14}) ; (\ref{10.3.13}) and
 $$\omega^{*}_{b a \lambda}=\omega_{b a \lambda}+
\theta^{\beta}_{b a}N_{\beta \lambda}+
\overline{N}^{\beta}_{\lambda}\theta_{b a \beta}$$
 produce (\ref{10.3.16})  and
\begin{equation*} \bar{\theta}^{* \alpha}_{b a}=
\bar{\theta}^{\alpha}_{b a}+ \bar{\theta}^{\beta}_{b
a}\tilde{\eta}^{0 \alpha}_{\beta}+ \tilde{\eta}^{\beta
\alpha}_{0}\bar{\theta}_{b a \beta}, \quad \theta^{*}_{b a
\alpha}=\theta_{b a \alpha}+ \bar{\theta}^{\beta}_{b
a}\eta^{0}_{\beta \alpha}+ \eta^{\beta}_{0 \alpha}\theta_{b a
\beta}.
\end{equation*}
So,  we infer (\ref{10.3.14})  and (\ref{10.3.15}) .
\end{proof}

The connections (\ref{10.3.3}) and (\ref{10.3.9}) give rise to 8
curvature tensors as described in (5.2) of \cite{ot1}. But also
the spin-connections (II) connected to (\ref{10.3.3}) lead to six
spin-curvature tensors (\ref {10.2.7})
\begin{equation}
\{R_{ab\lambda \mu },P_{ab\lambda \alpha
},\overline{P}_{ab\lambda }^\alpha ,S_{ab\beta }^\alpha
,Q_{ab\beta \alpha },\widetilde{Q}_{ab}^{\beta \alpha }\}.
\label{10.3.17}
\end{equation}
Taking into account Theorems \ref{th10.1} and \ref{th10.2} we can
express these tensors as follows.

\begin{theorem}
\label{th10.3} In the GCFS spaces the spin-curvature tensors are
given by
\begin{gather}
R_{ab\lambda \mu }=\eta _{\lambda (b}\sigma _{\mu a)}^{*}+\eta
_{\mu (a}\sigma _{\lambda b)}^{*}+\eta _{\mu (b}\sigma _\lambda
^{*}\sigma
_{a)}^{*}  \label{10.3.18} \\
+\eta _{\lambda (a}\sigma _\mu ^{*}\sigma _{b)}^{*}+\eta _{(\mu
a}\eta
_{\lambda )b}\eta ^{cd}\sigma _c^{*}\sigma _d^{*},  \notag \\
P_{ab\lambda \alpha }=\eta _{\lambda (b}\sigma _{\alpha
a)}^{*},\qquad \overline{P}_{ab\lambda }^\alpha =\eta _{\lambda
(b}\sigma _{a)}^{*\alpha },
\notag \\
S_{ab\beta }^\alpha =0,\quad Q_{ab\beta \alpha }=0,\quad \widetilde{Q}%
_{ab}^{\beta \alpha }=0,  \label{10.3.19}
\end{gather}
where $\sigma _{xy}^{*}=\partial ^{*2}\sigma /\partial
x^x\partial x^y;\quad x,y=\{\lambda ,\alpha ,a\}$ and $\eta
_{\lambda (b}\sigma _{\mu a)}^{*}=\eta _{\lambda b}\sigma _{\mu
a}^{*\alpha }-\eta _{\lambda a}\sigma _{\mu b}^{*}$.
\end{theorem}

\begin{proof} Relations (\ref{10.3.19})  are
directly implied by (\ref{10.3.15})  and
 (\ref{10.3.16}). (\ref{10.3.6})  leads to (\ref{10.3.18})  after a
straightforward calculation. Also, using Theorem \ref{th10.2}, we
infer that
\begin{equation}\label{(10.3.20)}  P_{a b \lambda \alpha}=\omega^{*}_{a b \lambda , 
\alpha}, \qquad  \overline{P}_{a b \lambda \alpha}=\omega^{*}_{a b
\lambda , \alpha},
\end{equation}
where
\begin{equation*} \omega^{*}_{a b \lambda , \alpha}=\frac{\partial^{*}\omega_{a b 
\lambda}}{\partial \bar{\xi}^{\alpha}}, \qquad  \omega^{*}_{a b
\lambda , \alpha}=\frac{\partial^{*}\omega_{a b \lambda}}{\partial
\xi_{\alpha}}.
\end{equation*}
Then (\ref{10.3.20})   leads to (\ref{10.3.18}) and
 (\ref{10.3.19}).
\end{proof}
Relations (\ref{10.3.19}) are directly implied by
(\ref{10.3.14})--(\ref {10.3.16}). The relations (\ref{10.3.6})
leads to (\ref{10.3.18}) after a straightforward calculations.
Also, using Theorem \ref{th10.3}, we infer that
\begin{equation}
P_{ab\lambda \alpha }=\frac{\partial ^{*}\omega _{ab\lambda
}}{\partial
\overline{\xi }^\alpha },\quad \overline{P}_{ab\lambda \alpha }=\frac{%
\partial ^{*}\omega _{ab\lambda }}{\partial \xi _\alpha }.  \label{10.3.20}
\end{equation}
Then (\ref{10.3.15}) leads to (\ref{10.3.18}) and (\ref{10.3.19}).

As a consequence of this theorem we state the following

\begin{corollary}
\label{cor10.1} In the GCFS space $(M,g_{\mu \nu })$ the Ricci
tensor
 \index{Ricci tensor}  fields have the form
\begin{gather}
R_\mu ^d=e^{-\sigma }(2\eta ^{bd}\sigma _\mu ^{*}\sigma
_b^{*}-2\eta ^{bd}\sigma _{\mu b}^{*}-\delta _\mu ^d\eta
^{a\lambda }\sigma _{\lambda
a}^{*}-2\delta _\mu ^d\eta ^{ef}\sigma _c^{*}\sigma _f^{*}),  \notag \\
P_\nu ^b=-3e^{-\sigma }(\eta ^{bc}\sigma _{\alpha c}^{*}\sigma
_\nu ^{*\alpha }-\sigma _{\alpha \nu }^{*}\sigma _e^{*\alpha
}\eta ^{eb}). \label{10.3.21}
\end{gather}
\end{corollary}

\begin{proof} Using Theorem \ref{th10.3}
 $R^{d}_{\mu}=h^{\lambda}_{c}R^{c d}_{\lambda \mu},P^{b}_{\nu} =
h^{\mu}_{a}P^{a}_{c \mu \alpha}\overline{P}^{b c \alpha}_{\nu}$ we
obtain relations (\ref{10.3.21}) .
\end{proof}
\emph{Remark} (1) It follows that the scalar curvature takes the
form
\begin{equation}
R=R_\mu ^dh_d^\mu =-6e^{-2\sigma }(\eta ^{bd}\sigma _{db}^{*}+\eta
^{ef}\sigma _e^{*}\sigma _f^{*}).  \label{10.3.22}
\end{equation}
Furthermore, it can be easily seen that
\begin{equation}
P=P_\nu ^bh_b^\nu \equiv 0.  \label{10.3.23}
\end{equation}

As we have previously remarked, the scalar curvature fields
\begin{equation*}
Q=Q_{ab\beta \alpha }\widetilde{Q}^{ab\beta \alpha } \mbox{ and }
S=S_{ab\alpha \beta }S^{ab\alpha \beta }
\end{equation*}
vanish identically. Then the employed Lagrangian density
(\ref{10.2.5})
\begin{equation*}
L=h(R+P+Q+S),\qquad det(g_{\mu \nu })=-h^2,
\end{equation*}
reduces to $L=e^\sigma (R+P)$ and depends on the fields\newline
$\varphi \in \{h_\nu ^b,\omega _{ab\lambda }^{*},\theta _{ab\alpha }^{*},%
\bar{\theta}_{ab}^{*\alpha }\}$. The Euler-Lagrange equations
\begin{equation}
\partial _M^{*}\biggl(\frac{\partial L}{\partial (\partial _M^{*}\varphi )}%
\biggr)-\frac{\partial L}{\partial \varphi }=0  \label{10.3.24}
\end{equation}
for these fields produce the field equations (\ref{10.2.10}), (\ref{10.2.22}%
), (\ref{10.2.27}) and (\ref{10.2.28}).

We shall obtain their form for the GCFS as follows.

\begin{theorem}
\label{th10.4}The field equations for the GCFS are
\begin{eqnarray*}
\delta _\mu ^d\eta ^{ef}(2\sigma _{ef}^{*}-\sigma _e^{*}\sigma
_f^{*})+2\eta
^{bd}(\sigma _{\mu b}^{*}-\sigma _\mu ^{*}\sigma _b^{*}) && \\
+3\eta ^{ed}\sigma _{\alpha \mu }^{*}\sigma _e^{*\alpha }-3\eta
^{dc}\sigma
_{\alpha c}^{*}\sigma _\mu ^{*\alpha } &=&0,\qquad (F1) \\
\sigma _{(b}^{*}\delta _{a)}^\nu -3\sigma _\alpha ^{*}\sigma
_{(b}^{*\alpha }\delta _{a)}^\nu -3\delta _a^\nu \sigma _{\alpha
b}^{*\alpha } &=&0,\qquad
(F2) \\
2\sigma _a^{*}\sigma _{a\alpha \beta }^{*} &=&0,\qquad (F3) \\
2\sigma _\mu ^{*}\eta ^{\mu d}\eta _{\alpha b}\sigma _{\alpha
d}^{*}-2\sigma _a^{*}\sigma _{\alpha b}^{*}+\eta ^{\mu d}\eta
_{ab}\sigma _{\mu \alpha d}^{*}-\sigma _{a\alpha b}^{*}
&=&0,\qquad (F4)
\end{eqnarray*}
\end{theorem}

where we have put $\sigma^{* \beta}_{\alpha
b}=\partial^{*3}\sigma /
\partial \xi_{\beta}\partial \bar{\xi}^{\alpha}\partial x^{b}$.

\begin{proof} By virtue of relations (\ref{10.2.10})
 and (\ref{10.2.11}), and using Corollary \ref{cor10.1}
 and Remark (1), we get (F1).

Considering Theorem \ref{th10.2} we infer that
$D^{*}_{\alpha}=\partial^{*}_{\alpha}$ and
$D^{*\alpha}=\partial^{*\alpha}$. Also from $\omega^{(*)}_{\alpha
b \lambda}=\omega_{\alpha b \lambda}=-\omega_{b a \lambda}$, we
derive $D^{*}_{\mu}=\partial^{*}_{\mu}$. Taking into account
 (\ref{10.2.22}) , we obtain relation 
(F2) by a straightforward computation. Also, by means of Theorem
\ref{th10.3}  and noticing that $\overline{P}^{\mu \alpha}_{b
\mu}=-3 \sigma^{* \alpha}_{b}$, after substituting to
(\ref{10.2.27}), we infer (F3). Finally, from (\ref{10.2.28})  we
derive (F4).
\end{proof}

\section{Geodesics and geodesic deviation} \index{geodesic deviation}

We shall now give the form of geodesics in spaces with the $g_{\mu
\nu}(x,\xi,\bar{\xi})$ metric.

A curve $c$ in a space $(M,g_{\mu \nu}(x,\xi,\bar{\xi}))$ is
defined as a
smooth mapping $c: I \rightarrow U \subset M: t \rightarrow (x(t),\xi(t),%
\bar{\xi}(t))$, where $U$ is an open set of $M$ and $t$ is an
arbitrary parameter.

\begin{definition}
A curve $c$ is a geodesic if it satisfies the set of equations:
\begin{align}
\frac{D\dot{x}^\mu }{ds}& \equiv \frac{d^2x^\mu
}{ds^2}+\dot{x}^\nu (\Gamma _{\nu \lambda x}^\mu \dot{x}^\lambda
+\overline{C}_\nu ^{\mu \alpha }\xi
_\alpha +C_{\nu \alpha }^\mu \bar{\xi}^\alpha )=0,\qquad (a)  \nonumber \\
\frac{D^2\xi _\alpha }{ds^2}& \equiv \frac D{ds}[\dot{\xi}_\alpha
-\xi
_\gamma (\Gamma _{\alpha \lambda }^\gamma \dot{x}^\lambda +\widetilde{C}%
_\alpha ^{\gamma \beta }\dot{\xi}_\beta +C_{\alpha \beta }^\gamma \bar{\xi}%
^\beta )]=0,\qquad (b)  \nonumber \\
\frac{D^2\bar{\xi}^\alpha }{ds^2}& \equiv \frac D{ds}[\bar{\xi}^\alpha +\bar{%
\xi}^\gamma (\Gamma _{\gamma \lambda }^\alpha \dot{x}^\lambda +\widetilde{C}%
_\gamma ^{\alpha \beta }\dot{\xi}_\beta +C_{\gamma \beta }^\alpha \bar{\xi}%
^\beta )]=0,\qquad (c)  \label{10.4.1}
\end{align}
where $\dot{x}^\mu =dx^\mu /ds,\dot{\xi}_\alpha =d\xi _\alpha /ds,\bar{\xi}%
^\alpha =d\bar{\xi}^\alpha /ds$, and the coefficients $\Gamma
_{\nu \lambda }^\mu ,$ $\Gamma _{\alpha \lambda }^\gamma ,$
$\overline{C}_\nu ^{\mu \alpha }$, $\widetilde{C}_\gamma ^{\alpha
\beta },C_{\nu \alpha }^\mu ,C_{\alpha \beta }^\gamma $ satisfy
the postulates imposed by Y. Takano and T. Ono \cite {ot1}.
\end{definition}

\begin{proposition}
(a) If $\overline{C}_\nu ^{\mu \alpha }=0$ and $C_{\nu \alpha
}^\mu =0$,
then $\Gamma _{\nu \lambda }^\mu =\Gamma _{\lambda \nu }^\mu $ and relation (%
\ref{10.4.1}) becomes
\begin{equation}
\frac{d^2x^\mu }{ds^2}+\Gamma _{\nu \lambda }^\mu (x,\xi (x),\bar{\xi}(x))%
\frac{dx^\nu }{ds}\frac{dx^\lambda }{ds}=0.  \label{10.4.2}
\end{equation}
(b) For the GCFS, equation (\ref{10.4.1}) has the form
\begin{equation}
\frac{d^2x^\mu }{ds^2}+\Gamma _{\nu \lambda }^\mu \dot{x}^\nu \dot{x}%
^\lambda +\dot{x}^\mu (\sigma ^\alpha \dot{xi}_\alpha
+\bar{\sigma}_\alpha \bar{\xi}^\alpha )=0.  \label{10.4.3}
\end{equation}
In this case $\overline{C}_\nu ^{\mu \alpha }=0,C_{\nu \alpha
}^\mu =0$ hold true iff $\sigma ^\alpha =\bar{\sigma}_\alpha =0$,
i.e., for $\sigma $ depending only on $x$.
\end{proposition}

\begin{proof}Equations (\ref{10.4.2})  and (\ref{10.4.3})
 are consequences of Definition (\ref{10.4.1}) (a)
and relations (\ref{10.3.6}) .
\end{proof}

\emph{Remark} (2): The spinor parts of equations (\ref{10.4.2})
and (\ref {10.4.3}) also write as
\begin{gather}
\ddot{\xi}_\alpha -\dot{\xi}_\gamma T_\alpha ^\gamma -\xi _\gamma \dot{T}%
_\alpha ^\gamma -(\dot{\xi}_\gamma -\xi _\delta T_\gamma ^\delta
)T_\alpha
^\gamma =0,  \label{10.4.4} \\
\bar{\xi}^\alpha +\bar{\xi}^\alpha \overline{T}_\alpha ^\gamma +\bar{\xi}%
^\gamma \overline{T}_\gamma ^\alpha +(\bar{\xi}^\gamma
+\bar{\xi}^\delta \overline{T}_\delta ^\gamma )T_\gamma ^\alpha
=0,  \nonumber
\end{gather}
where
\[
T_\alpha ^\gamma \equiv \Gamma _{\alpha \lambda }^\gamma \dot{x}^\lambda +%
\widetilde{C}_\alpha ^{\gamma \beta }\dot{\xi}_\beta +C_{\alpha
\beta }^\gamma \bar{\xi}^\beta =\overline{T}_\alpha ^\gamma .
\]

Having the equations of geodesics, it remains to derive the
equations of geodesic deviation of our spaces. This geodesic
deviation can be given a physical meaning if we consider two very
close geodesic curves and the curvature tensor is Riemannian.

In the general case of GCFS, the spinor variables are independent
of the position, so it is difficult to convey a physical meaning
to the equations of geodesic deviation. For this reason it is
convenient to study the deviation of the geodesics in the case
where the spinor field $\xi _\alpha =\xi _\alpha (x^\mu )$(and
$\bar{\xi}^\alpha =\bar{\xi}^\alpha (x^\mu )$) is defined on the
manifold. This spinor field associates a spinor -and its
conjugate-to every point of the space-time.

In this case, from Proposition (1) and relation (\ref{10.4.2}) the
Christoffel symbols $\Gamma _{\nu \lambda }^\mu $ are symmetric
in the lower indices and the equation of geodesics is similar to
the Riemannian one, except that the connection coefficients have
the additional dependence on the spinors $\xi _\alpha (x^\mu
),\bar{\xi}^\alpha (x^\mu )$. Thus our approach is more general.
The equation of geodesic deviation in our case is given by
\begin{equation}
\frac{D^2\zeta ^\lambda }{ds^2}+R_{\mu \nu \varrho }^\lambda \frac{dx^\mu }{%
ds}\frac{d\zeta ^\nu }{ds}\frac{dx^\varrho }{ds}=0. \label{10.4.5}
\end{equation}
The above curvature tensor $R_{\mu \nu \varrho }^\lambda (x,\xi (x),\bar{\xi}%
(x))$ has a modified Riemannian form. This equation has additional
contributions from the spinor parts which enter the curvature
tensor $R$ and the covariant derivative. In (\ref{10.4.5}) $\zeta
^\mu $ denotes the deviation vector, and $s$ the arc length.

For the GCFS, the deviation equation has the above form, where
the curvature tensor depends on the function $\sigma (x,\xi
(x),\bar{\xi}(x))$ and its derivatives, as we have proved in
Theorem \ref{th10.3}, relation (\ref {10.3.18}). After a direct
calculation from (\ref{10.4.5}) and
\begin{equation}
R_{\mu \nu \varrho }^\lambda =R_{ab\mu \varrho }h^{b\lambda
}h_\nu ^a, \label{10.4.6}
\end{equation}
where $h^{b\lambda }=e^{-\sigma }\eta ^{b\lambda },h_\nu
^a=e^\sigma \delta
_\nu ^a$, we get the equation of geodesic deviation for the GCFS, with $%
\overline{C}_\nu ^{\mu \alpha }=0,C_{\nu \alpha }^\mu =0$, in the
form
\begin{multline}
\frac{D^2\zeta ^2}{ds^2}+(\delta _{(\mu }^\lambda \sigma
_{\varrho )\nu }^{*}+\eta _{\nu (\varrho }\sigma _{\mu
)b}^{*}\eta ^{b\lambda }+\delta
_{(\varrho }^\lambda \sigma _{\mu )}^{*}\sigma _\nu ^{*}+  \label{10.4.7} \\
+\eta _{\nu (\mu }\sigma _{\varrho )}^{*}\sigma _b^{*}h^{b\lambda
}+\eta
_{\nu (\varrho }\delta _{\mu )}^\lambda \eta ^{cd}\sigma _c^{*}\sigma _d^{*})%
\frac{dx^\mu }{ds}\frac{d\zeta ^\nu }{ds}\frac{dx^\varrho
}{ds}=0.  \nonumber
\end{multline}

\section{Conclusions}

(a) We derived the gravitational field equations in spaces whose
metric tensor depends on spinor variables. Equations
(\ref{10.2.10}) and (\ref {10.2.22}) are generalizations of the
conventional equations (\ref{10.1.3}) a) and (\ref{10.1.3}) b).
They are reduced to equations (\ref{10.1.3}) a) and
(\ref{10.1.3}) b) when the coefficients
\[
(\omega _\mu ^{(*)},\theta _\alpha ^{(*)},\bar{\theta}^{(*)\alpha
})\rightarrow (\omega _\mu ).
\]
Relations (\ref{10.2.27}) and (\ref{10.2.28}) give rise to new
results.

(b) Equations (F1)-(F4) represent the field equations on the
GCFS\newline $(M,g_{\mu \nu }(x,\xi ,\bar{\xi}))$. The solutions
of these equations are
the subject of further 
concern. They represent an application of the gauge approach, for
spaces
with the metric $g(x,\xi ,\bar{\xi})$, studied by two of the authors in %
\cite{sm1,sm2}.

(c) The vanishing of the curvatures $S_{ab\beta }^\alpha
,Q_{ab\beta \alpha },\widetilde{Q}_{ab}^{\beta \alpha }$ (Theorem
\ref{th10.3}), reduces the 6
spin curvatures of the theory of Y. Takano and T. Ono to the three ones $%
R_{ab\lambda \mu },P_{ab\lambda \alpha },\overline{P}_{ab\lambda
}^\alpha $. This simplifies considerably the study of the
generalized conformally flat spaces.

\chapter{Gauge Gravity Over Spinor Bundles}       \index{gauge gravity}
\label{ChapterGauge}

\section{Introduction}

The concept of the nonlocalized field theory has already been
developed in recent years by Japanese authors (see, for instance,
\cite{ik3}) in order to provide a unified description of
elementary particles. In this approach, the internal variable is
replaced by a spinor $\omega =(\xi ,{\bar{\xi}})$ ($\xi $ and its
conjugate $\bar{\xi}$ are considered as independent variables).

The description of gravity through the introduction of variables
$\omega _\mu ^{ab}(x)$ as a gravitational potential (Lorentz
connection coefficients) was proposed originally by Utiyama
\cite{utiyama,carmeli}. He considered the Lorentz group as a
local transformation group. The gravitational field is described
by the tetrad $h_\mu ^a(x)$ viewed as independent variables. With
the help of these variables we may pass from a general system of
coordinates to a local Lorentz ones.

The Einstein equation were derived in the context of Utiyama's
approach, but this was not satisfactory because of the
arbitrariness of the elements introduced. Later T. Kibble
\cite{hermann,ik3,kibble} introduced a gauge approach which
enables the introduction of all gravitational variables. To
achieve this goal it is important to use the Poincar\'{e} group
(i.e. a group consisting of rotations, boosts and translations).

This group first assigns an exact meaning to the terms:
``momentum'', ``energy'', ``mass'' and ``spin'' used to determine
characteristics of elementary particles. On the other hand, it is
a gauge acting locally in the space-time. Thus, we may perform
Poincar\'{e} transformations for a physical approach. Hence by
treating the Poincar\'{e} group as a local group, we introduce
the fundamental 1-form field $\rho _\mu $ taking in the Lie
algebra of the Poincar\'{e} group.

In our present study the basic idea is to consider a spinor
bundle with a base manifold $M$ of a metric tensor $g_{\mu \nu
}(x,\xi ,{\bar{\xi}})$ that
depends on the position coordinates $x^k$ and the spinor (Dirac) variables $%
(\xi _{{\alpha }},{\bar{\xi}}^{{\alpha }})\in {\mathbb C}^4\times {\mathbb C}%
^4$, where ${\bar{\xi}}^\alpha $ is the adjoint of $\xi _\alpha
$, an independent variable, similar to the one proposed by
Y.~Takano \cite{t1},
and Y.~Takano and T.~Ono \cite{ot1,ot2,ot3,ono}. The spinor bundle $%
S^{(1)}(M)$ is constructed from one of the principal fiber
bundles with a fiber: $F={\mathbb C}^4$.

Each fiber is diffeomorphic with one proper Lorentz group (which
is produced
by Lorentz transformations) and it entail a principal bundle $SL(4,{\mathbb C%
} )$ over $M$, ($SL(4,{\mathbb C} )$ consists of the group of
ratations and boosts of unit determinant acting on a
four-dimensional complex space, which is reducible to $(SL (2,
{\mathbb C} )$).

The consideration of the $d$-connections that preserve the $(hv)$%
-distribution by the parallel translation (cf.\cite{ma94,mwi}, in
relation to the second order bundles $S^2(M)=M\times {\mathbb
C}^{2\cdot 4}$ enables us to use a more general group $G^{(2)}$
called a structured group of all rotations and translations that
is isomorphic to the Poincar\'{e} Lie algebra. Therefore, a
\emph{spinor} in $x\in M$ is an element of the spinor bundle
$S^{(2)}(M)$.
\[
(x^\mu ,\xi _\alpha ,{\bar{\xi}}^\alpha )\in S^{(2)}(M).
\]
A \emph{spinor field} is a section of $S^{(2)}(M)$.

Moreover, the fundamental gauge 1-form field mentioned above in
connection with the spaces that possess metric tensor $g_{\mu \nu
}(x,\xi ,{\bar{\xi}})$ will take a similar but more general form
than that proposed by other
authors \cite{menotti}. We shall define a nonlinear connection on $%
S^{(2)}(M) $ such as,
\[
T(S^{(2)}M)=H(S^{(2)}M)\oplus \mathcal{F}^{(1)}(S^{(2)}M)\oplus \mathcal{F}%
^{(2)}(S^{(2)}M),
\]
where $H$, $\mathcal{F}^{(1)}$, $\mathcal{F}^{(2)}$ represent the
horizontal, vertical, and normal distribution. In a local base,
for the horizontal distribution $H(S^{(2)}M)$ we have:
\[
\rho _\mu (x,\xi ,{\bar{\xi}})=\frac 12\omega _\mu
^{*ab}J_{ab}+h_\mu ^a(x,\xi ,{\bar{\xi}})P_a,
\]
where $J_{ab}$, $P_a$ are the generators of the four-dimensional
Poincar\'{e} group satisfying relations of the form:
\begin{eqnarray*}
&[J_{ab},J_{cd}] = n_{bc}J_{ad}-n_{bd}J_{ac}+n_{ad}J_{bc}-n_{ac}J_{bd}, \\
&[J_{ab},P_c] = n_{bc}P_a-n_{ac}P_b,\ \ \ [P_a,P_b]=0,\ \ \
J_{ab}+J_{ba}=0. &
\end{eqnarray*}

The quantities $\omega _\mu ^{(*)ab}$ represent the (Lorentz) spin
connection coefficients and are considered as given, $n_{ab}$ is
the metric for the local Lorentz spaces with signature
$(+---)$.\newline These are connected with $g_{\mu \nu }$ by
\[
g_{\mu \nu }h_a^\mu h_b^\nu =n_{ab},\ \ g^{\mu \nu
}=n^{ab}h_a^\mu h_b^\nu ,
\]
where $h_a^\nu $ represents the tetrads. Similarly, for the
vertical and
normal distributions $\mathcal{F}^{(1)}(S^{(2)}M)$, $\mathcal{F}%
^{(2)}(S^{(2)}M)$ the fundamental 1-forms $\zeta _\alpha $, ${\bar{\zeta}}%
^\alpha $ are given by
\begin{eqnarray*}
\zeta _\alpha &=&\frac 12\Theta _{{\alpha }}^{(*)ab}J_{ab}+\Psi _{{\alpha }%
}^aP_a, \\
{\bar{\zeta}}^{{\alpha }} &=&\frac 12{\bar{\Theta}}^{(*){\alpha }ab}J_{ab}+{%
\bar{\Psi}}^{{\alpha }a}P_a,
\end{eqnarray*}
where ${\bar{\psi}}^{{\alpha }a}$, $\psi _{{\alpha }}^a$ are the
spin tetrad coefficients, and $\Theta _{{\alpha }}^{(*)ab}$,
${\bar{\Theta}}^{(*){\alpha }ab}$ are the given spin connection
coefficients which are determined in such a way that the absolute
differential and the covariant derivatives of the metric tensor
$g_{\mu \nu }(x,\xi ,{\bar{\xi}})$ vanish identically.

We use the Greek letters ${\lambda} ,\mu ,\nu \ldots$ for
space-time
indices, ${\lambda}, {\beta}, {\gamma}$ for spinors, and the Latin letters $%
a,b,c,\ldots$ for the Lorentz indices.

The general transformations of coordinates on $S^{(2)}(M)$ are:
\begin{eqnarray}
x^{\prime \mu} =x^{\prime \mu}(x^{\nu}), \ \ \ \xi^{\prime}_{{\alpha}%
}=\xi^{\prime}_{{\alpha}}(\xi_{{\beta}}, {\bar \xi}^{{\gamma}}),
\ \ \ {\bar \xi}^{\prime {\alpha}}={\bar \xi}^{\prime
{\alpha}}({\bar \xi}^{{\beta}} ,\xi_{{\gamma}}).
\end{eqnarray}

\section{Connections}

We define the following gauge covariant derivatives
\begin{eqnarray*}
D_\mu ^{(*)} &=&\frac \delta {\delta x^\mu }+\frac 12{\omega }_\mu
^{(*)ab}J_{ab}, \\
D^{(*){\alpha }} &=&\frac \delta {\delta \xi _\alpha }+\frac 12{\bar{\Theta}}%
^{(*){\alpha }ab}J_{ab}, \\
D^{(*){\alpha }} &=&\frac \delta {\delta \xi ^{{\alpha }}}+\frac 12\Theta _{{%
\alpha }}^{(*)ab}J_{ab},
\end{eqnarray*}
where
\begin{eqnarray*}
\frac \delta {\delta x^\mu } &=& \frac \partial {\partial x^\mu }+N_{{\alpha }%
\mu }\frac \partial {\partial \xi _\alpha }-{\bar{N}}_\mu
^{{\alpha }}\frac\partial {\partial {\bar{\xi}}_\alpha },\\
 \frac \delta {\delta \xi _{{\alpha }}} &=& \frac \partial {\partial \xi _{{\alpha
}}}-{\tilde{N}}_0^{{\alpha }{\beta }}\frac \partial {\partial
{\bar{\xi}}^\beta }.
\end{eqnarray*}
$N_{{\alpha }{\lambda }}$, ${\bar{N}}_{{\lambda }}^{{\alpha }}$, ${\tilde{N}}%
_0^{{\alpha }{\beta }}$ are the nonlinear connections which we
shall define below.

The covariant derivatives of the metric tensor $g_{\mu \nu }$ are
all zero:
\[
D_\mu ^{(*)}g_{{\kappa }{\lambda }}=0,\ \ D^{(*){\alpha }}g_{{\kappa }{%
\lambda }}=0,\ \ D_{{\alpha }}^{(*)}g_{{\kappa }{\lambda }}=0.
\]
The space-time frame $\delta /\delta x^\mu $ and the local Lorentz frame $%
\delta /\delta x^a$ are connected with
\[
\frac \delta {\delta x^\mu }=h_\mu ^a\frac \delta {\delta x^a}.
\]
Similarly, the spin-tetrad coefficients $\psi _{{\alpha }}^a$ and adjoint ${%
\bar{\psi}}^{{\alpha }a}$ connect the spin frames, $\partial /\partial \xi _{%
{\alpha }}$, $\partial /\partial {\bar{\xi}}^\alpha $ with
$\partial /\partial x^a$:
\[
\frac \partial {\partial \xi _{{\alpha }}}={\bar{\psi}}^{{\alpha
}a}\frac
\partial {\partial x^a},
\]
\[
\frac \partial {\partial {\bar{\xi}}^{{\alpha }}}=\psi _{{\alpha
}}^a\frac
\partial {\partial x^a}.
\]
The absolute differential of an arbitrary contravariant vector
$X^\nu $ is given by
\[
DX^\nu =(D_\mu ^{(*)}dx^\mu +D^{(*){\alpha }}X^\nu )d\xi _{{\alpha }}+(D_{{%
\alpha }}^{(*)}X^\nu )d{\bar{\xi}}^{{\alpha }}.
\]

\subsection{Nonlinear connections} \index{nonlinear connection}

We give the nonlinear connections $N=\{N_{{\beta }\mu },{\tilde{N}}_{{\beta }%
}^{0{\alpha }},N_{{\alpha }{\beta }}^0,{\bar{N}}_\mu ^{{\beta }},{\tilde{N}}%
_0^{{\beta }{\alpha }},N_{0{\alpha }}^{{\beta }}\}$ in the
framework of our consideration in the following form:
\begin{eqnarray}
N_{{\beta }\mu } &=&\frac 12{\omega }_\mu ^{(*)ab}J_{ab}\xi _{{\beta }%
},\quad {\tilde{N}}_{{\beta }}^{0{\alpha }}=\frac 12{\bar{\Theta}}^{(*){%
\alpha }ab}J_{ab}\xi _{{\beta }},  \label{11.2.5} \\
N_{{\alpha }{\beta }}^0 &=&\frac 12\Theta _{{\alpha }}^{(*)ab}J_{ab}\xi _{{%
\beta }},\quad {\bar{N}}_\mu ^{{\beta }}=-\frac 12{\omega }_\mu
^{(*)ab}J_{ab}{\bar{\xi}}^{{\beta }},  \nonumber \\
{\tilde{N}}_0^{{\alpha }{\beta }} &=&-\frac 12{\bar{\Theta}}^{(*){\alpha }%
ab}J_{ab}{\bar{\xi}}^{{\beta }},\quad N_{0{\alpha }}^{{\beta
}}=-\frac 12\Theta _{{\alpha }}^{(*)ab}J_{ab}{\bar{\xi}}^{{\beta
}}.  \nonumber
\end{eqnarray}

The differentials of $D\xi _{{\alpha }}$, $D{\bar{\xi}}^{{\alpha
}}$ can be written, after the relations (\ref{11.2.5}), in the
form:
\begin{eqnarray}
D\xi _{{\beta }} &=&d\xi _{{\beta }}+N_{{\alpha }{\beta }}^0d{\bar{\xi}}_{{%
\alpha }}+{\tilde{N}}_{{\beta }}^{0{\alpha }}d\xi _{{\alpha }}+N_{{\beta }%
\mu }dx^\mu ,  \label{11.2.6} \\
D{\bar{\xi}}^{{\beta }} &=&d{\bar{\xi}}^{{\beta }}+N_{0{\alpha }}^{{\beta }}d%
{\bar{\xi}}^{{\alpha }}-{\tilde{N}}_0^{{\beta }{\alpha }}d\xi _{{\alpha }}-{%
\tilde{N}}_\mu ^{{\beta }}dx^\mu ,  \nonumber
\end{eqnarray}
The metric in the second order tangent bundle is given by the
relation
\[
G=g_{{\kappa }{\lambda }}dx^{{\kappa }}dx^{{\lambda
}}+g_{ij}\delta y^i\delta y^j+g_{{\alpha }{\beta }}\delta
u^{{\alpha }}\delta u^{{\beta }},
\]
and the adapted frame
\[
\frac \partial {\partial Z^A}=\left( \frac \delta {\delta
x^\lambda }=\frac
\partial {\partial x^\lambda }-N_{{\lambda }}^i\frac \partial {\partial
y^i}-M_{{\lambda }}^{{\alpha }}\frac \partial {\partial
u^{{\alpha }}},\frac \delta {\delta y^i},\frac \partial {\partial
u^{{\alpha }}}\right)
\]
where $\delta /\delta y^i=\partial /\partial y^i-L_i^{{\alpha
}}\partial /\partial u^{{\alpha }}$.

Furthermore, the dual frame is
\[
\delta Z^A=(dx^{{\kappa }},\delta y^i+N_{{\lambda
}}^idx^{{\lambda }},\delta
u^{{\alpha }}=du^{{\alpha }}+L_i^{{\alpha }}dy^i+M_{{\lambda }}^{{\alpha }%
}dx^{{\lambda }}).
\]
The metrical structure in the bundle will be defined as follows:
\[
G=g_{\mu \nu }(x,\xi ,{\bar{\xi}})dx^\mu dx^\nu +g_{{\alpha
}{\beta }}(x,\xi
,{\bar{\xi}})D{\bar{\xi}}^{{\alpha }}D{\bar{\xi}}^{*{\beta }}+g^{{\alpha }{%
\beta }}D\xi _{{\alpha }}D\xi _{{\beta }}^{*}.
\]
an analogy with the previous adapted frame, a local adapted frame
on a spinor bundle $S^{(2)}(M)$ will be defined as
\begin{eqnarray*}
\left( \frac \partial {\partial \zeta ^A}\right) &=&\left\{ \frac
\delta {\delta x^{{\lambda }}},\frac \delta {\delta \xi _{{\alpha
}}},\frac \delta
{\delta {\bar{\xi}}^{{\alpha }}}\right\} , \\
\frac \delta {\delta x^{{\lambda }}} &=&\frac \partial {\partial
x^{{\lambda
}}}+N_{{\alpha }{\lambda }}\frac \partial {\partial \xi _{{\alpha }}}-{\bar{N%
}}_{{\lambda }}^{{\alpha }}\frac \partial {\partial
{\bar{\xi}}^{{\alpha }}},
\\
  \frac \delta {\delta \xi _{{\alpha }}} &=&\frac \partial {\partial \xi _{%
{\alpha }}}-{\tilde{N}}_0^{{\beta }{\alpha }}\frac \partial {\partial {\bar{%
\xi}}^{{\beta }}},
\end{eqnarray*}
and
\[
\delta \zeta ^A=\{dx^{{\kappa }},D\xi _{{\beta
}},D{\bar{\xi}}^\beta \},
\]
where the expressions $D\xi _{{\beta }},D{\bar{\xi}}^{{\beta }}$
are given
by (\ref{11.2.6}). If we consider the connection coefficients ${\Gamma }%
_{BC}^A$ given in the general case, then in the total space
$S^{(2)}(M)$ we have
\[
{\Gamma }_{BC}^A=\{{\Gamma }_{\nu \rho }^{(*)\mu },C_{\nu {\alpha }}^\mu ,{%
\bar{C}}_\nu ^{\mu {\alpha }},{\bar{\Gamma}}_{{\beta }{\lambda }}^{(*){%
\gamma }},C_{{\beta }}^{{\gamma }{\alpha }},{\tilde{C}}_{{\beta }}^{{\gamma }%
{\alpha }},{\tilde{C}}_{{\alpha }{\beta }}^{{\gamma }},{\Gamma }_{{\alpha }{%
\lambda }}^{(*){\beta }},C_{{\alpha }{\beta }}^{{\gamma }}\}.
\]

Considering that the connections are $d$--connections
\cite{ma94,mwi} in an adapt\-ed base, we get the following
relations
\[
D_{\partial /\partial x^C}\frac \partial {\partial x^B}={\Gamma }%
_{BC}^A\frac \partial {\partial x^A},
\]
or, in explicit form,
\newpage
\begin{eqnarray*}
D_{\delta /\delta x^\rho }\frac \delta {\delta x^\nu } &=&{\Gamma
}_{\nu
\rho }^{(*)\mu }\frac \delta {\delta x^\mu },\quad D_{\partial /\partial {%
\bar{\xi}}^{{\alpha }}}\frac \delta {\delta x^\nu }=C_{\nu
{\alpha }}^\mu
\frac \delta {\delta x^\mu }, \\
D_{\delta /\delta \xi _{{\alpha }}}\frac \delta {\delta x^\nu } &=&{\bar{C}}%
_\nu ^{\mu {\alpha }}\frac \delta {\delta x^\mu },\quad
D_{\partial
/\partial {\bar{\xi}}^{{\alpha }}}\frac \delta {\delta x^\nu }={\Gamma }%
_{\nu {\alpha }}^{(*){\gamma }}\frac \delta {\delta \xi ^{{\gamma }}}, \\
D_{\delta /\delta x^{{\alpha }}}\frac \delta {\delta x_{{\beta }}} &=&{\bar{{%
\Gamma }}}_{{\lambda }{\gamma }}^{(*){\beta }}\frac \delta {\delta \xi _{{%
\gamma }}},\quad D_{\delta /\delta \xi _{{\alpha }}}\frac
\partial {\partial \xi ^{{\beta }}}=C_{{\beta }}^{{\gamma
}{\alpha }}\frac \partial {\partial
\xi ^{{\gamma }}}, \\
D_{\delta /\delta \xi _{{\alpha }}}\frac \partial {\partial \xi
^{{\beta }}}
&=&C_{{\beta }{\alpha }}^{{\gamma }}\frac \partial {\partial \xi ^{{\gamma }%
}},\quad D_{\partial /\partial {\bar{\xi}}^{{\alpha }}}\frac
\delta {\delta \xi _{{\beta }}}={\tilde{C}}_{{\alpha }{\gamma
}}^{{\beta }}\frac \delta
{\delta \xi _{{\gamma }}}, \\
D_{\delta /\delta \xi _{{\alpha }}}\frac \delta {\delta \xi _{{\beta }}} &=&{%
\tilde{C}}_{{\gamma }}^{{\beta }{\alpha }}\frac \delta {\delta
\xi _{{\gamma }}}.
\end{eqnarray*}

The covariant differentiation of tensor and spin-tensors of
arbitrary rank may be classified into three types:
\begin{eqnarray*}
{\bigtriangledown }_{{\lambda }}T_{\nu \ldots }^{\mu \ldots }& =&\frac{%
\delta T_{\nu \ldots }^{\mu \ldots }}{\delta x^{{\lambda }}}+{\Gamma }_{{%
\kappa }{\lambda }}^{(*)\mu }T_{\nu \ldots }^{{\kappa }\ldots }+\cdots -{%
\Gamma }_{\nu {\lambda }}^{(*){\kappa }}T_{{\kappa }\ldots }^{\mu
\ldots },
\\
{\bigtriangledown }^{{\alpha }}T_{\nu \ldots }^{\mu \ldots }&
=&\frac{\delta
T_{\nu \ldots }^{\mu \ldots }}{\delta \xi _{{\alpha }}}+{\bar{C}}_{{\kappa }%
}^{(*)\mu {\alpha }}T_{\nu \ldots }^{{\kappa }\ldots }+\cdots
-{\bar{C}}_\nu
^{(*){\kappa }{\alpha }}T_{{\kappa }\ldots }^{\mu \ldots }, \\
{\bigtriangledown }_{{\alpha }}T_{\nu \ldots }^{\mu \ldots }& =&\frac{%
\partial T_{\nu \ldots }^{\mu \ldots }}{\partial \xi ^{{\alpha }}}+C_{{%
\kappa }{\alpha }}^{(*)\mu }T_{\nu \ldots }^{{\kappa }\ldots
}+\cdots
-C_{\nu {\alpha }}^{(*){\kappa }}T_{{\kappa }\ldots }^{\mu \ldots }, \\
{\bigtriangledown }_{{\lambda }}\Phi _{{\beta }\ldots }^{{\alpha
} \ldots }&
= & \frac{\delta \Phi _{{\beta }\ldots }^{{\alpha }\ldots }}{\delta x^{{%
\lambda }}}-{\Gamma }_{{\beta }{\lambda }}^{(*){\gamma }}\Phi _{{\gamma }%
\ldots }^{{\alpha }\ldots }-\cdots +\Phi _{{\beta }\ldots
}^{{\gamma }\ldots
}{\Gamma }_{{\gamma }{\lambda }}^{(*){\alpha }}+\ldots , \\
{\bigtriangledown }^\delta \Phi _{{\beta }\ldots }^{{\alpha
}\ldots }& = &
\frac{\delta \Phi _{{\beta }\ldots }^{{\alpha }\ldots }}{\delta \xi _\delta }%
-{\tilde{C}}_{{\beta }}^{(*){\gamma }\delta }\Phi _{{\gamma }\ldots }^{{%
\alpha }\ldots }-\cdots +\Phi _{{\beta }\ldots }^{{\gamma }\ldots }{\tilde{C}%
}_{{\gamma }}^{(*){\alpha }\delta }+\ldots , \\
{\bigtriangledown }_\delta \Phi _{{\beta }\ldots }^{{\alpha
}\ldots }& = & \frac{\partial \Phi _{{\beta }\ldots }^{{\alpha
}\ldots }}{\partial \xi
^\delta }-C_{{\beta }\delta }^{(*){\gamma }}\Phi _{{\gamma }\ldots }^{{%
\alpha }\ldots }-\cdots +\Phi _{{\beta }\ldots }^{{\gamma }\ldots }C_{{%
\gamma }\delta }^{(*){\alpha }}+\ldots , \\
{\bigtriangledown }_\mu ^{(*)}V_{c\ldots }^{{\alpha }\ldots }& = & \frac{%
\delta V_{c\ldots }^{{\alpha }\ldots }}{\delta x^\mu }+\omega _{\mu b}^{(*){%
\alpha }}V_{c\ldots }^{b\ldots }+\cdots -\omega _{\mu c}^{(*)b}V_{b\ldots }^{%
{\alpha }\ldots }, \\
{\bigtriangledown }^{(*){\alpha }}V_{c\ldots }^{{\alpha }\ldots }& = & \frac{%
\delta V_{c\ldots }^{{\alpha }\ldots }}{\delta \xi _{{\alpha }}}+{\bar{\Theta%
}}_b^{(*)a{\alpha }}V_{c\ldots }^{b\ldots }+\cdots -{\bar{\Theta}}_c^{(*){%
\alpha }b}V_{b\ldots }^{{\alpha }\ldots }, \\
{\bigtriangledown }_{{\alpha }}^{(*)}V_{c\ldots }^{{\alpha
}\ldots }& =&
\frac{\partial V_{c\ldots }^{{\alpha }\ldots }}{\delta \xi ^{{\alpha }}}%
+\Theta _{{\alpha }b}^{(*)a}V_{c\ldots }^{b\ldots }+\cdots
-\Theta _{{\alpha }c}^{(*)b}V_{b\ldots }^{{\alpha }\ldots }.
\end{eqnarray*}

\subsection{Lorentz transformation}        \index{Lorentz transform}

We can get the Lorentz transformations of linear connections
${\omega }_\nu ^{(*)ab}$, $\bar{\Theta}^{(*){\beta }ab}$, $\Theta
_{{\beta }}^{(*)ab}$ in the following form:
\begin{eqnarray*}
{\omega }^{\prime }{}_\mu ^{(*)ab} &=&L_c^aL_d^b{\omega }_\mu ^{(*)cd}+\frac{%
\delta L_c^a}{\delta x^\mu }L_d^bn^{cd}, \\
{\bar{\Theta}}^{(*)^{\prime }{\alpha }ab} &=&\left[ L_c^aL_d^b{\bar{\Theta}}%
^{(*){\beta }cd}+\frac{\delta L_c^a}{\delta \xi _{{\beta }}}%
L_d^bn^{cd}\right] {\Lambda }_{{\beta }}^{-1{\alpha }}, \\
\Theta _{{\alpha }}^{(*)^{\prime }ab} &=&{\Lambda }_{{\alpha }}^{{\beta }%
}\left[ L_c^aL_d^b{\bar{\Theta}}_{{\beta }}^{(*)cd}+\frac{\partial L_c^a}{%
\partial \xi ^{{\beta }}}L_d^bn^{cd}\right] ,
\end{eqnarray*}
Similarly, the Lorentz transformation law of nonlinear connection
is given by:
\begin{eqnarray*}
{\tilde N}_{{\beta} \mu}&=& \frac{1}{2}{\omega}^{(*)ab}_{\mu}J_{ab}\xi_{{%
\alpha}}L^{{\alpha}}_{{\beta}}+ \frac{1}{2}n^{cd}\frac{\delta L^{a}_{c}}{%
\delta x^{\mu}}L^{b}_{d}
J_{ab}{\Lambda}^{{\alpha}}_{{\beta}}\xi_{{\alpha}}
\\
&=& N_{{\alpha} \mu} {\Lambda}^{{\alpha}}_{{\beta}} + \frac{1}{2}n^{cd}\frac{%
\delta L^{a}_{c}}{\delta x^{\mu}}J^{\prime}_{ab} {\Lambda}^{{\alpha}}_{{\beta%
}}\xi_{{\alpha}},
\end{eqnarray*}
where
\begin{eqnarray*}
{\tilde N}^{0{\alpha}}_{{\beta}}&=& \left[{\tilde N}^{0\delta}_{{\gamma}} {%
\Lambda}^{{\gamma}}_{{\beta}} + \frac{1}{2}n^{cd}\frac{\delta L^{a}_{c}} {%
\delta \xi_{\delta}}L^{b}_{d}J^{\prime}_{ab}{\Lambda}^{{\gamma}}_{{\beta}%
}\xi_{{\gamma}} \right] {\Lambda}^{-1{\alpha}}_{\delta}, \\
{\tilde N}^{0}_{{\alpha} {\beta}}&=&
{\Lambda}^{\delta}_{{\alpha}}\left[
N^{0}_{{\gamma} \delta} {\Lambda}^{{\gamma}}_{{\beta}} + \frac{1}{2}n^{cd}%
\frac{\partial L^{a}_{c}} {\partial \xi^{\delta}}L^{b}_{d}J^{\prime}_{ab} {%
\Lambda}^{{\gamma}}_{{\beta}}\xi_{{\gamma}} \right], \\
{\bar N}^{{\beta}}_{\mu}&=& N^{{\alpha}}_{\mu} {\Lambda}^{1-{\beta}}_{{\alpha%
}} - \frac{1}{2}n^{cd}\frac{\delta L^{a}_{c}}{\delta x_{\mu}}
L^{b}_{d}J^{\prime}_{ab}{\bar \xi}^{{\gamma}}{\Lambda}^{-1{\beta}}_{{\gamma}%
}, \\
{\tilde N}^{{\alpha} {\beta}}_{0}&=& \left[{\tilde N}^{{\gamma} \delta}_{0} {%
\Lambda}^{-1 {\beta}}_{{\gamma}} -\frac{1}{2}n^{cd}\frac{\delta L^{a}_{c}} {%
\delta \xi_{\delta}}L^{b}_{d}J^{\prime}_{ab}{\Lambda}^{-1 {\beta}}_{{\gamma}%
}\right] {\Lambda}^{-1{\alpha}}_{\delta}, \\
{\tilde N}^{{\beta}}_{0{\alpha}}&=& {\Lambda}^{\delta}_{{\alpha}}\left[ N^{{%
\gamma}}_{0 \delta} {\Lambda}^{-{\beta}}_{{\gamma}} - \frac{1}{2}n^{cd}\frac{%
\partial L^{a}_{c}} {\partial\xi^{\delta}} L^{b}_{d}J^{\prime}_{ab}{\Lambda}%
^{{\gamma}}_{{\beta}}{\bar \xi}^{{\gamma}}
{\Lambda}^{-1{\beta}}_{{\gamma}} \right],
\end{eqnarray*}
where $J_{ab}^{\prime }=L_{\hat{a}}^cL_b^dJ_{cd}$.

\section{Curvatures and torsions}  \index{curvature} \index{torsion}

From the covariant derivatives $D_\mu ^{(*)}$, $D^{(*){\alpha }}$, $D_{{%
\alpha }}^{(*)}$ we get six curvatures and torsions:
\begin{eqnarray*}
a)\quad \left[ D_\mu ^{(*)},D_\nu ^{(*)}\right]  &=&D_\mu
^{(*)}D_\nu ^{(*)}-D_\nu ^{(*)}D_\mu ^{(*)}=R_{\mu \nu
}^aP_a+\frac 12R_{\mu \nu
}^{ab}J_{ab}, \\
R_{\mu \nu }^a &=&\frac{\delta h_\mu ^a}{\delta x^\nu }-\frac{\delta h_\nu ^a%
}{\delta x^\mu }+{\omega }_{\mu b}^{(*)a}h_\nu ^b-{\omega }_{\nu
b}^{(*)a}h_\mu ^b, \\
R_{\mu \nu }^{ab} &=&\frac{\delta {\omega }_\mu ^{(*)ab}}{\delta x^\nu }-%
\frac{\delta {\omega }_\nu ^{(*)ab}}{\delta x^\mu }+{\omega }_\mu
^{(*)a\rho }{\omega }_{\nu \rho }^{(*)b}-{\omega }{^{(}*)\rho
a}_\nu {\omega }_{\mu
\rho }^{(*)b}, \\
b)\quad \left[ D_\mu ^{(*)},D_{{\alpha }}^{(*)}\right]  &=&P_{\mu {\alpha }%
}^aP_a+\frac 12P_{\mu {\alpha }}^{ab}J_{ab}, \\
P_{\mu {\alpha }}^{ab} &=&\frac{\delta \theta _{{\alpha
}}^{(*)ab}}{\delta
x^\nu }-\frac{\partial {\omega }_\mu ^{(*)ab}}{\delta {\bar{\xi}}^{{\alpha }}%
}+\Theta _{{\alpha }c}^{(*)b}{\omega }_\mu ^{(*)a}-\Theta _{{\alpha }%
c}^{(*)a}{\omega }_\mu ^{(*)cb}, \\
P_{\mu {\alpha }}^a &=&\frac{\delta \psi _{{\alpha }}^a}{\delta x^\mu }-%
\frac{\partial h_\mu ^a}{\delta {\bar{\xi}}^{{\alpha }}}+{\omega
}_{\mu
c}^{(*)a}\psi _{{\alpha }}^c-\Theta {^{(}*)a}_{{\alpha }c}h_\mu ^c, \\
c)\quad \left[ D_\mu ^{(*)},D_{{\alpha }}^{(*)}\right]  &=&{\bar{P}}_\mu ^{a{%
\alpha }}P_a+\frac 12{\bar{P}}_\mu ^{ab{\alpha }}J_{ab}, \\
{\bar{P}}_\mu ^{ab{\alpha }} &=&\frac{\delta \Theta _{{\alpha
}}^{(*){\alpha
}ab}}{\delta x^\mu }-\frac{\delta {\omega }_\mu ^{(*)ab}}{\delta \xi _{{%
\alpha }}}+{\bar{\Theta}}_c^{(*)ab}{\omega }_\mu ^{(*)ac}-{\bar{\Theta}}%
_c^{(*){\alpha }a}{\omega }_\mu ^{(*)cb}, \\
{\bar{P}}_{\mu {\alpha }}^{a{\alpha }} &=&\frac{\delta
{\bar{\psi}}_{{\alpha
}}^{{\alpha }a}}{\delta x^\mu }-\frac{\delta h_\mu ^a}{\delta \xi ^{{\alpha }%
}}+{\omega }_{\mu c}^{(*)a}{\bar{\psi}}^{c{\alpha }}-{\bar{\Theta}}_c^{(*){%
\alpha }a}h_\mu ^c, \\
d)\quad \left[ D_{{\alpha }}^{(*)},D^{(*){\beta }}\right]  &=&S_{{\alpha }}^{%
{\beta }a}P_a+\frac 12S_{{\alpha }}^{ab{\beta }}J_{ab}, \\
S_{{\alpha }}^{{\beta }a} &=&\frac{\delta {\bar{\psi}}^{{\beta }a}}{\delta {%
\bar{\xi}}_{{\alpha }}}-\frac{\delta \psi _{{\alpha }}^a}{\delta \xi _{{%
\beta }}}+{\bar{\Theta}}^{(*){\beta }ba}\psi _{ab}-\Theta _{{\alpha }%
}^{(*)ab}{\bar{\psi}}_b^{{\beta }}, \\
S_{{\alpha }}^{ab{\beta }} &=&\frac{\partial {\bar{\Theta}}^{{\beta }ab}}{%
\partial \xi ^{{\alpha }}}-\frac{\partial \Theta _{{\alpha }}^{(*)ab}}{%
\delta \xi _{{\beta }}}+\Theta _{{\alpha
}c}^{(*)a}{\bar{\Theta}}^{(*){\beta
}cb}-\Theta {^{(}*)b}_{ac}{\bar{\Theta}}^{{\beta }ca}, \\
e)\quad \left[ D_{{\alpha }}^{(*)},D_{{\beta }}^{(*)}\right]  &=&Q_{{\alpha }%
}^aP_a+\frac 12Q_{{\alpha }{\beta }}^{ab}J_{ab}, \\
Q_{{\alpha }{\beta }}^a &=&\frac{\partial \psi _{{\beta }}^a}{\partial {\bar{%
\xi}}^{{\alpha }}}-\frac{\partial \psi _{{\alpha }}^a}{\partial {\bar{\xi}}^{%
{\beta }}}+\Theta _{{\beta }}^{(*)ba}\psi _{ab}-\Theta _{{\alpha }%
}^{(*)ab}\psi _{{\beta }b}, \\
Q_{{\alpha }{\beta }}^{ab} &=&\frac{\partial \theta _{{\alpha }}^{(*)ab}}{%
\partial {\bar{\xi}}^{{\alpha }}}-\frac{\partial \theta _{{\alpha }}^{(*)ab}%
}{\partial {\bar{\xi}}^{{\beta }}}+\Theta _{{\alpha }c}^{(*)a}\Theta _{{%
\beta }}^{(*)cb}-\Theta {^{(}*)b}_{{\alpha }c}\Theta _{{\beta
}}^{(*)ca},
\end{eqnarray*}

\begin{eqnarray*}
f)\quad \left[ D^{(*){\alpha }},D^{(*){\beta }}\right]  &=&{\tilde{Q}}^{{%
\alpha }{\beta }a}P_a+\frac 12{\tilde{Q}}^{ab{\alpha }{\beta }}J_{ab}, \\
{\tilde{Q}}^{{\alpha }{\beta }a} &=&\frac{\delta \psi _{{\beta
}}^a}{\delta
\xi _{{\alpha }}}-\frac{\delta \psi _{{\alpha }}^a}{\partial \xi _{{\beta }}}%
+{\bar{\Theta}}^{(*){\beta }ba}{\bar{\psi}}_b^{{\alpha }}-{\bar{\Theta}}^{(*)%
{\alpha }ba}{\bar{\psi}}_b^{{\beta }}, \\
{\tilde{Q}}^{ab{\alpha }{\beta }} &=&\frac{\delta {\bar{\theta}}^{{\beta }ab}%
}{\delta \xi _{{\alpha }}}-\frac{\delta {\bar{\theta}}^{{\alpha }ab}}{%
\partial \xi _{{\beta }}}+{\bar{\Theta}}^{(*){\beta }ab}{\bar{\Theta}}_c^{(*)%
{\alpha }a}-{\bar{\Theta}}_c^{(*){\alpha
}b}{\bar{\Theta}}^{{\beta }ca},
\end{eqnarray*}

\section{Field equations}          \index{field equations}

We derive the field equations using the spin--tetrad frames
\index{spin--tetrad} in the Lagrangian
form: $\mathcal{L}(h,{\omega }^{(*)},\psi ,\Theta ^{(*)},{\bar{\psi}},{\bar{%
\Theta}}^{(*)})$. The method of derivation of equations is
similar to Palatini's one.

We get the Lagrangian
\[
\mathcal{L}(h,{\omega }^{(*)},\psi ,\Theta ^{(*)},{\bar{\psi}},{\bar{\Theta}}%
^{(*)})
\]
or
\begin{equation}
\mathcal{L}\psi ^A,\delta _M\psi
^A=h(R+P+{\bar{P}}+S+Q+{\tilde{Q}}), \label{11.4.1}
\end{equation}
where
\[
\psi ^A=\left( h_\mu ^a(x,\xi ,{\bar{\xi}}),{\omega }_\mu ^{(*)ab}(x,\xi ,{%
\bar{\xi}}),\psi _{{\alpha }}^a(x,\xi
,{\bar{\xi}}),{\bar{\psi}}_{(\ldots )}^{{\alpha }a},\Theta
_{{\alpha }(\ldots )}^{(*)ab},{\bar{\Theta}}_{(\ldots
)}^{(*){\alpha }ab}\right) ,
\]
\[
\delta _M=\frac \delta {\delta z^M}=\left( \frac \delta {\delta
x^m},\frac
\delta {\delta \xi _{{\alpha }}},\frac \delta {\delta {\bar{\xi}}^{{\alpha }%
}}\right) ,\qquad z^M=(x^\mu ,\xi _{{\alpha
}},{\bar{\xi}}^{{\alpha }}),
\]
\begin{eqnarray*}
R & =& h_a^\mu h_b^\nu R_{\mu \nu }^{ab}, \\
P & =& h_a^\mu {\bar{\psi}}_b^{{\alpha }}P_{\mu {\alpha }}^{ab},\qquad {\bar{P}}%
=h_a^\mu {\bar{\psi}}_{{\alpha }b}{\bar{P}}_\mu ^{ab{\alpha }}, \\
Q & =& Q_{{\alpha }{\beta }}^{ab}{\bar{\psi}}_a^{{\alpha }}{\bar{\psi}}_b^{{%
\beta }},\qquad {\tilde{Q}}={\tilde{Q}}^{ab{\alpha }{\beta }}\psi _{{\alpha }%
a}\psi _{{\beta }b}, \\
S & =& {\bar{\psi}}_a^{{\alpha }}\psi _{{\beta }b}S_{{\alpha
}}^{ab{\beta }}.
\end{eqnarray*}
The Euler--Lagrange equations are written in the form:
\[
\frac{\delta \mathcal{L}}{\delta z^M}=\frac{\partial
\mathcal{L}}{\partial (\delta _M\psi ^{(A)})}-\frac{\partial
\mathcal{L}}{\partial \psi ^{(A)}}=0.
\]
From the relation (\ref{11.4.1}), the variation of $\mathcal{L}$
with respect to $h^\nu {b}$ yields the equations
\begin{eqnarray*}
(R_\mu ^a+P_\mu ^a+{\bar{P}}_\mu ^a)-\frac 12(R+P+{\bar{P}})h_\mu ^a &=&0, \\
H_\mu ^a-\frac 12Hh_\mu ^a &=&0,
\end{eqnarray*}
where
\[
P_\mu ^a={\bar{\psi}}_b^{{\alpha }}P_{\mu {\alpha }}^{ab},\quad {\bar{P}}%
_\mu ^a=\psi _{ab}{\bar{P}}_\mu ^{ab{\alpha }},\quad R_\mu
^a=h_b^\nu R_{\mu \nu }^{ab},
\]
and
\[
H_\mu ^a=R_\mu ^a+P_\mu ^a+{\bar{P}}_\mu ^a,\quad H=R+P+{\bar{P}}.
\]

From the variation of $\mathcal{L}$ with respect to ${\omega
}_\mu ^{(*)ab}$
\begin{eqnarray*}
\frac \delta {\delta x^\mu }\left( \frac{\partial
\mathcal{L}}{\partial \left( \frac \delta {\delta x^\mu }{\omega
}_\nu ^{(*)ab}\right) }\right) +\frac \delta {\delta \xi
_{{\alpha }}}\frac{\partial \mathcal{L}}{\partial
\left( \frac{\delta {\omega }_\nu ^{(*)ab}}{\delta \xi _{{\alpha }}}\right) }%
  & &\\
+\frac \delta {\delta {\bar{\xi}}^{{\alpha }}}\left( \frac{\partial \mathcal{%
L}}{\partial \left( \frac \delta {\delta {\bar{\xi}}^{{\alpha }}}{\omega }%
_\nu ^{(*)ab}\right) }\right) -\frac{\partial \mathcal{L}}{\partial {\omega }%
_\nu ^{(*)ab}}&=&0,\
\end{eqnarray*}
we get
\begin{eqnarray*}
D_\mu ^{(*)}[h(h_a^\nu h_b^\mu -h_b^\nu h_a^\mu )]+D_{{\alpha }%
}^{(*)}[h(h_a^\nu {\bar{\psi}}_b^{{\alpha }}-h_a^\nu
{\bar{\psi}}_a^{{\alpha
}})] \\
+D^{(*){\alpha }}[h(h_a^\nu \psi _{{\alpha }b}-h_b^\nu \psi _{{\alpha }%
}^a)]&=&0.
\end{eqnarray*}

The variations with respect to $\Theta _{{\alpha }}^{(*)ab}$, ${\bar{\Theta}}%
^{(*){\alpha }ab}$ yield the relation
\[
\frac \delta {\delta x^\mu }\left( \frac{\partial
\mathcal{L}}{\partial \left( \frac{\delta {\Omega }^{(*)}}{\delta
x^\mu }\right) }\right) +\frac \delta {\delta \xi _{{\alpha
}}}\frac{\partial \mathcal{L}}{\partial \left( \frac{\delta
{\Omega }^{(*)}}{\delta \xi _{{\alpha }}}\right) }+\frac \delta
{\delta {\bar{\xi}}^{{\alpha }}}\left( \frac{\partial
\mathcal{L}}{\partial
\left( \frac{\delta {\Omega }^{(*)}}{\delta {\bar{\xi}}^{{\alpha }}}\right) }%
\right) -\frac{\partial \mathcal{L}}{\partial {\Omega }^{(*)}}=0,\
\]
with
\[
{\Omega }^{(*)}=\left\{ \Theta _{{\alpha }}^{(*)ab},{\bar{\Theta}}^{(*){%
\alpha }ab}\right\}
\]
which gives us the equations:
\begin{eqnarray*}
D_\mu ^{(*)}(hh_a^\mu {\bar{\psi}}_a^{{\alpha }})-D_{{\beta }}^{(*)}(2h{\bar{%
\psi}}_a^{{\alpha }}{\bar{\psi}}_b^{{\beta }})-2D^{(*){\beta }}(h{\bar{\psi}}%
_a^{{\alpha }}\psi _{{\beta }b}) &=&0, \\
D_\mu ^{(*)}(hh_a^\mu \psi _{b{\alpha }})-2D_{{\beta }}^{(*)}(h\psi _{a{%
\alpha }}{\bar{\psi}}_b^{{\beta }})-D^{(*){\beta }}(2h\psi
_{a{\alpha }}\psi _{b{\beta }}) &=&0.
\end{eqnarray*}
Finally, the variation of $\mathcal{L}$ with respect to the
spin-tetrad coefficients ${\bar{\psi}}_a^{{\alpha }}$, $\psi
^{{\alpha }a}$ derives the equations:
\begin{eqnarray*}
Q_{{\alpha }{\beta }}^{ab}{\bar{\psi}}_b^{{\beta }}+\frac 12S_{{\alpha }}^{ab%
{\beta }}\psi _{{\beta }b}+\frac 12P_{\mu {\alpha }}^{ba}h_b^\mu  &=&0, \\
{\tilde{Q}}^{a{\alpha }}-\frac 12(S^{a{\alpha
}}+{\bar{P}}^{a{\alpha }}) &=&0.
\end{eqnarray*}

\section{Bianchi identities}      \index{Bianchi identities}

From Jacobi identities,
\[
\mathcal{Q}_{(XYZ)}\left[ D_X^{(*)},[D_Y^{(*)},D_Z^{(*)}]\right]
=0,
\]
we get $18(3\times 6)$ relations of different types. For each
relation we derive two identities, namely 36 ones in total.
Taking into account that
\[
D_\mu ^{(*)}=\frac \delta {\delta x^\mu }+\frac 12{\omega }_\mu
^{(*)ab}J_{ab},
\]
where
\begin{eqnarray*}
\frac \delta {\delta x^\mu } &=&\frac \partial {\partial x^\mu }-N_{\mu {%
\alpha }}\frac \partial {\partial \xi _{{\alpha }}}-{\bar{N}}_\mu ^{{\alpha }%
}\frac \partial {\partial {\bar{\xi}}^{{\alpha }}} \\
&=&h_\mu ^aP_a-N_{\mu {\alpha }}{\bar{\Psi}}^{{\alpha }a}P_a-{\bar{N}}_\mu ^{%
{\alpha }}\Psi _{{\alpha }}^aP_a=A_\mu ^aP_a, \\
A_\mu ^a &=&h_\mu ^a-N_{\mu {\alpha }}{\bar{\psi}}^{{\alpha }a}-{\bar{N}}%
_\mu ^{{\alpha }},\quad P_a=\frac \partial {\partial x^a},
\end{eqnarray*}
we can get
\begin{eqnarray}
\left[ D_\mu ^{(*)},[D_{{\kappa }}^{(*)},D_{{\lambda
}}^{(*)}]\right] &=&\left[ A_\mu ^cP_c,\frac 12R_{{\kappa
}{\lambda }}^{ab}J_{ab}\right]
+[A_\mu ^cP_c,R_{{\kappa }{\lambda }}^aP_a]  \label{11.5.1} \\
&&+\frac 12{\omega }_\mu ^{(*)ab}R_{{\kappa }{\lambda }%
}^{cd}[J_{ab},J_{cd}]+\frac 12{\omega }_\mu ^{(*)ab}R_{{\kappa }{\lambda }%
}^c[J_{ab},P_c].  \nonumber
\end{eqnarray}

The first term of the right hand side of (\ref{11.5.1}) by
straightforward calculations is written in the form
\[
\left[ A_\mu ^cP_c,\frac 12R_{{\kappa }{\lambda
}}^{ab}J_{ab}\right] =\frac
12\frac{\delta R_{{\kappa }{\lambda }}^{ab}}{\delta x^\mu }J_{ab}+R_{b{%
\kappa }{\lambda }}^aA_\mu ^bP_a.
\]

Similarly, the second, third, and fourth terms of (\ref{11.5.1})
yield the relations
\[
\lbrack A_\mu ^cP_c,R_{{\kappa }{\lambda }}^aP_a]=\frac{\delta R_{{\kappa }{%
\lambda }}^a}{\delta x^\mu }P_a+A_\mu ^cR_{{\kappa }{\lambda }}^a[P_c,P_a]=%
\frac{\delta R_{{\kappa }{\lambda }}^a}{\delta x^\mu }P_a,
\]
where we used the fact that $[P_c,P_a]=0$ Also
\begin{eqnarray*}
\frac 14{\omega }_\mu ^{(*)ab}R_{{\kappa }{\lambda }}^{cd}[J_{ab},J_{cd}] &=&%
{\omega }_\mu ^{(*)ac}R_{c{\kappa }{\lambda }}^bJ_{ab}, \\
\frac 12{\omega }_\mu ^{(*)ab}R_{{\kappa }{\lambda }}^c[J_{ab},P_c] &=&{%
\omega }_\mu ^{(*)ac}R_{{\kappa }{\lambda }}^bP_a,
\end{eqnarray*}
so the relation (\ref{11.5.1}) is written as
\begin{eqnarray*}
\left[ D_\mu ^{(*)},[D_{{\kappa }}^{(*)},D_{{\lambda
}}^{(*)}]\right]
&=&\left( \frac 12\frac{\delta R_{{\kappa }{\lambda }}^{ab}}{\delta x^\mu }+{%
\omega }_\mu ^{(*)ac}R_{c{\kappa }{\lambda }}^b\right)  \\
&&+J_{ab}+\left( \frac{\delta R_{{\kappa }{\lambda }}^a}{\delta x^\mu }+R_{b{%
\kappa }{\lambda }}^aA_\mu ^b+R_{{\kappa }{\lambda }}^c{\omega
}_{\mu b}^{(*)a}\right) P_a.
\end{eqnarray*}
Defining
\begin{eqnarray}
D_\mu R_{{\kappa }{\lambda }}^{ab} &=&\frac 12\frac{\delta R_{{\kappa }{%
\lambda }}^{ab}}{\delta x^\mu }+{\omega }_\mu ^{(*)ac}R_{c{\kappa }{\lambda }%
}^b, \\
D_\mu R_{{\kappa }{\lambda }}^a &=&\frac 12\frac{\delta
R_{{\kappa }{\lambda }}^a}{\delta x^\mu }+A_\mu ^b+R_{{\kappa
}{\lambda }}^c{\omega }_{\mu b}^{(*)a},
\end{eqnarray}
we have the relations:
\begin{eqnarray*}
&&D_\mu R_{{\kappa }{\lambda }}^{ab}+D_\kappa R_{{\lambda }\mu }^{ab}+D_{{%
\lambda }}R_{\mu {\kappa }}^{ab}=0, \\
&&D_\mu R_{{\kappa }{\lambda }}^a+D_\kappa R_{{\lambda }\mu }^a+D_{{\lambda }%
}R_{\mu {\kappa }}^a=0.
\end{eqnarray*}
In the similar way, from
\[
\mathcal{Q}_{({\alpha }{\beta }{\gamma })}\left[ D_\alpha ^{(*)},[D_{{\beta }%
}^{(*)},D_{{\gamma }}^{(*)}]\right] =0
\]
we get for the $Q$-curvature and torsion the identities below:
\[
D_\alpha Q_{{\beta }{\gamma }}^{ab}+D_{{\beta }}Q_{{\gamma }{\alpha }%
}^{ab}+D_\gamma Q_{{\alpha }{\beta }}^{ab}=0
\]
and
\[
D_\alpha Q_{{\beta }{\gamma }}^a+D_{{\beta }}Q_{{\gamma }{\alpha }}^a+D_{{%
\gamma }}Q_{{\alpha }{\beta }}^a=0,
\]
where we put
\begin{eqnarray*}
D_{{\alpha }}Q_{{\beta }{\gamma }}^{ab} &=&\frac 12\frac{\partial Q_{{\beta }%
{\gamma }}^{ab}}{\partial {\bar{\xi}}^{{\alpha }}}+\Theta _{{\alpha }%
}^{(*)ac}Q_{c{\beta }{\gamma }}^b, \\
D_{{\alpha }}Q_{{\beta }{\gamma }}^a &=&\frac{\partial Q_{{\beta }{\gamma }%
}^a}{\partial {\bar{\xi}}^{{\alpha }}}+Q_{b{\beta }{\gamma }}^a\Psi _{{%
\alpha }}^b+Q_{{\beta }{\gamma }}^b\Theta _{{\alpha }b}^{(*)a}.
\end{eqnarray*}

\section{Yang-Mills fields}   \index{Yang--Mills fields}

In this section, we study Yang-Mills fields and we derive the
generalized Yang-Mills equations in the framework of our
approach. In such a case we consider a vector field $A$
\begin{equation}
F_{\mu \nu }=D_\mu A_\nu -D_\nu A_\mu +i[A_\mu ,A_\nu ]
\label{11.6.2}
\end{equation}
represents the Yang-Mills field, $A_\mu $ is given by
\begin{equation}
A_\mu =A_\mu ^i\tau _i,\qquad [\tau _i,\tau _j]=C_{ij}^k\tau _k,
\label{11.6.3}
\end{equation}
the elements $\tau _i$ are the generators which satisfy the
commutation relations of the Lie algebra, and $D_\mu $ represent
the gauge covariant derivatives.

Using (\ref{11.6.2}), (\ref{11.6.3}) of the matrices $A_\mu $ we
find that
\[
F_{\mu \nu }=F_{\mu \nu }^i\tau _i,
\]
where the field strengths are given by
\[
F_{\mu \nu }^k=D_\mu A_\nu ^k-D_\nu A_\mu ^k+iA_\mu ^iA_\nu
^jC_{ij}^k.
\]
Moreover, the generalized gauge field is defined by the quantities $%
F_{XY},X,Y=\{\mu ,\nu ,{\alpha },{\beta }\}$, that is
\begin{eqnarray*}
\lbrack {\tilde{D}}_\mu ,{\tilde{D}}_{{\alpha }}] &=&[D_\mu ,D_{{\alpha }%
}]+iF_{\mu {\alpha }}, \\
\lbrack {\tilde{D}}_\mu ,{\tilde{D}}^{{\alpha }}] &=&[D_\mu ,D^{{\alpha }}]+i%
{\bar{F}}_\mu ^{{\alpha }}, \\
\lbrack {\tilde{D}}_{{\alpha }},{\tilde{D}}_{{\beta }}] &=&[D_{{\alpha }},D_{%
{\beta }}]+iF_{{\alpha }}^{{\beta }}, \\
\lbrack {\tilde{D}}_{{\alpha }},{\tilde{D}}^{{\beta }}] &=&[D_{{\alpha }},D^{%
{\beta }}]+iF_{{\alpha }{\beta }}, \\
\lbrack {\tilde{D}}^{{\alpha }},{\tilde{D}}^{{\beta }}] &=&[D^{{\alpha }},D^{%
{\beta }}]+iF^{{\alpha }{\beta }},
\end{eqnarray*}
with
\begin{eqnarray*}
F_{\mu {\alpha }} &=&D_\mu A_{{\alpha }}-D_{{\alpha }}A_\mu +i[A_\mu ,A_{{%
\alpha }}], \\
{\bar{F}}_\mu ^{{\alpha }} &=&D_\mu {\bar{A}}^{{\alpha
}}-{\bar{D}}^{{\alpha
}}A_\mu +i[A_\mu ,{\bar{A}}^{{\alpha }}], \\
F_{{\alpha }}^{{\beta }} &=&D_{{\alpha }}{\bar{A}}^{{\beta }}-{\bar{D}}^{{%
\beta }}A_{{\alpha }}+i[A_{{\alpha }},{\bar{A}}^{{\beta }}], \\
F_{{\alpha }{\beta }} &=&D_{{\alpha }}A_{{\beta }}-D_{{\beta }}A_{{\alpha }%
}+i[A_{{\alpha }},A_{{\beta }}], \\
{\bar{F}}^{{\alpha }{\beta }} &=&{\bar{D}}^{{\alpha }}{\bar{A}}^{{\beta }}-{%
\bar{D}}^{{\beta }}{\bar{A}}^{{\alpha }}+i[{\bar{A}}^{{\alpha }},{\bar{A}}^{{%
\beta }}].
\end{eqnarray*}
In our space $S^{(*)}(M)$ the Yang-Mills generalized action can
be written in the form
\begin{equation}
S_{GF}=\int d^4xd^4\xi d^4{\bar{\xi}}h(trF_{\mu \nu }F^{\mu \nu }+trF_{\mu {%
\alpha }}{\bar{F}}^{\mu {\alpha }}+trF_{{\alpha }{\beta
}}{\bar{F}}^{{\alpha }{\beta }}+trF_{{\alpha }}^{{\beta
}}F_{{\beta }}^{{\alpha }}), \label{11.6.8}
\end{equation}
where $F_{\mu \nu }$ represent the internal quantities in the
base manifold,
$F_{{\alpha }}^\mu $ the field in the tensor bundle and $F_{{\alpha }{\beta }%
}$ the internal quantities in the internal space.

In order to derive the generalized Yang-Mills equations we get the
Lagrangian
\[
\mathcal{L}_{YM}(A_X,D_XA_Y),
\]
where $A_X=\{A_\mu ,A_\alpha ,A^\beta \}$ and $D_XA_Y$ represent
\[
D_XA_Y=\{D_\mu A_\nu ,D_\alpha A_\nu ,{\bar{D}}^\alpha ,D_\alpha A_\beta ,{%
\bar{D}}^\alpha A_\beta ,D_\mu A_\alpha ,D_\mu A^\alpha \}.
\]

Varying the action (\ref{11.6.8}) and taking into account the
Euler-Lagrange equations
\begin{equation}
D_X\left( \frac{\partial \mathcal{L}_{YM}}{\partial (D_XA_Y)}\right) -\frac{%
\partial \mathcal{L}_{YM}}{\partial A_Y}=0,  \label{11.6.10}
\end{equation}
obtain the generalized Yang-Mills equations in the following form:
\begin{eqnarray*}
{\tilde{D}}^\mu F_{\mu \nu }+{\tilde{D}}^{{\alpha }}F_{{\alpha }\nu }+{%
\tilde{D}}_{{\alpha }}{\bar{F}}_\nu ^{{\alpha }} &=&0, \\
{\tilde{D}}_\mu F^{\mu {\beta }}+{\tilde{D}}_{{\alpha }}F^{{\alpha }{\beta }%
}+{\tilde{D}}^{{\alpha }}F_{{\alpha }}^{{\beta }} &=&0, \\
{\tilde{D}}_\mu F_{{\beta }}^\mu +{\tilde{D}}_{{\alpha }}F_{{\beta }}^{{%
\alpha }}+{\tilde{D}}^{{\alpha }}F_{{\alpha }{\beta }} &=&0,
\end{eqnarray*}
we used the trace properties of the operators $\tau _{{\alpha }}$
with the normalization condition
\[
tr(\tau ^{{\alpha }}\tau ^{{\beta }})=\frac 12\delta ^{{\alpha
}{\beta }}.
\]

\section{Yang-Mills-Higgs field}                       \index{Higgs}

In this last Section we shall give the form of Yang-Mills-Higgs
field in a sufficiently generalized form. The usual case has been
studied with the appropriate Lagrangian $\mathcal{L}$ ... the
corresponding Euler--Lagrange equations.

Here, we get a scalar field $\phi $ of mass $m$ which is valuated
in the Lie algebra $\mathcal{G}$ of consideration and is defined
by
\[
\phi :M^{(4)}\times {\mathbb C}^4\times {\mathbb C}^4\to
\mathcal{G}
\]
\[
\phi (x^\mu ,\xi _{{\alpha }},{\bar{\xi}}^{{\alpha }}\in
\mathcal{G}.
\]
... is in adjoint representations, its covariant derivatives are
given by
\[
{\tilde{D}}_\mu \phi =D_\mu \phi +[A_\mu ,\phi
],{\tilde{D}}_{{\alpha }}\phi
=D_{{\alpha }}\phi +[A_{{\alpha }},\phi ],{\tilde{D}}^{{\alpha }}\phi =D^{{%
\alpha }}\phi +[A^{{\alpha }},\phi ].
\]
The first of these relations, after taking into account
(\ref{11.6.3}), becomes
\begin{equation}
{\tilde{D}}_\mu \phi =D_\mu \phi +A_\mu ^{{\alpha }}\phi ^bC_{{\alpha }%
c}^c\tau _b;  \label{11.7.2}
\end{equation}
for ${\tilde{D}}_{{\alpha }}\phi $, ${\tilde{D}}^{{\alpha }}\phi
$ similar relations are produced.

The generalized Lagrangian is given by the following form:
\[
\mathcal{L}=\mathcal{L}_{YM}-\frac 12tr({\tilde{D}}_\mu \phi )-\frac 12tr({%
\tilde{D}}_{{\alpha }}\phi )({\tilde{D}}^{{\alpha }}\phi )+\frac
12m^2tr\phi ^2.
\]
Using (\ref{11.7.2}) and getting (\ref{11.6.10}) for this Lagrangian $%
\mathcal{L}$, the generalized Yang-Mills-Higgs equations are as
follows:
\begin{eqnarray*}
{\tilde{D}}^\mu F_{\mu \nu }+{\tilde{D}}^{{\alpha }}F_{{\alpha }\nu }+{%
\tilde{D}}_{{\alpha }}F_\nu ^{{\alpha }}+[\phi ,{\tilde{D}}_\nu
\phi ] &=&0,
\\
{\tilde{D}}_\mu F^{\mu {\beta }}+{\tilde{D}}_{{\alpha }}F^{{\alpha }{\beta }%
}+{\tilde{D}}^{{\alpha }}F_{{\alpha }}^{{\beta }}+[\phi
,{\tilde{D}}^{{\beta
}}\phi ] &=&0, \\
{\tilde{D}}_\mu F_{{\beta }}^\mu +{\tilde{D}}_{{\alpha }}F_{{\beta }}^{{%
\alpha }}+{\tilde{D}}^{{\alpha }}F_{{\alpha }{\beta }}+[\phi ,{\tilde{D}}_{{%
\beta }}\phi ] &=&0.
\end{eqnarray*}

These equations defines a Poincare like gravity theory on spaces
where the metric tensor $g_{\mu \nu }\left( x,\omega \right) $
depends on internal independent variables
 $\omega =\left(\xi,\bar{\xi}\right) .$

\chapter[Spinors on Internal Deformed Systems]{Spinor
 Bundle on Internal Deformed Systems}

\section{Introduction}

It was formulated \cite{st1,sm3} the concept of a spinor bundle
$S^{(2)}M$ and its relation to the Poincar$\grave{e}$ group. This
group, consisting of the set of rotations, boosts and
translations, gives an exact meaning to the terms: ``momentum'',
``energy'', ``mass'', and ``spin'' and is used to determine
characteristics of the elementary particles. Also, it is a gauge,
acting locally in the space-time. Hence we may perform
Poincar$\grave{e}$ transformations for a physical approach. In
that study we considered a base manifold $(M,g_{\mu \nu }(x,\xi
,\bar{\xi}))$ where the metric tensor depends on the position
coordinates and the spinor (Dirac) variables $(\xi _\alpha
,\bar{\xi ^\alpha })\in C^4\times C^4$. A spinor bundle
$S^{(1)}(M)$
can be constructed from one of the principal fiber bundles with fiber $F=C^4$%
. Each fiber is diffeomorphic with one proper Lorentz group.

In our study we apply an analogous method as in the theory of
deformed
bundles developed in \cite{si1}, for the case of a \textbf{spinor bundle} $%
S^{(2)}M=M\times C^{4\cdot 2}$ \index{spinor bundle}  in connection with a deformed \textbf{%
internal fiber} $R.$ Namely our space has the form $S^{(2)}M\times
R$. The consideration of Miron $d$ - connections \cite{ma94},
which preserve the $h-$ and $v-$distributions is of vital
importance in our approach, as in the
previous work. This standpoint enables us to use a more general group $%
G^{(2)}$, called the structural group of all rotations and
translations, that is isomorphic to the Poincar$\grave{e}$ Lie
algebra. A \textbf{spinor} is an element of the spinor bundle
$S^{(2)}(M)\times R$ where $R$ represents the \textbf{internal
fiber of deformation.}  \index{deformation} \index{internal fiber}
 The local variables are in this case
\[
(x^\mu ,\xi _\alpha ,\bar{\xi}^\alpha ,\lambda )\in S^{(2)}(M)\times R=%
\tilde{S}^{(2)}(M),\lambda \in R.
\]

The non-linear connection on $\tilde{S}^{(2)}(M)$ is defined
analogously, as for the vector bundles of order two
\cite{mat1,muat}
\[
T(\tilde{S}^{(2)}{M})=H(\tilde{S}^2M)\oplus \mathcal{F}^{(1)}(\tilde{S}^{(2)}%
{M})\oplus \mathcal{F}^{(2)}(\tilde{S}^{(2)}{M})\oplus R,
\]
where $\mathcal{H}$,$\mathcal{F}^{(1)}$,$\mathcal{F}^{(2)}$,$R$
represent the horizontal, vertical, normal and deformation
distributions respectively.

The fundamental gauge $1$-form fields which take values from the
Lie algebra of the Poincar$\grave{e}$ group will have the
following form in the local bases of their corresponding
distributions
\begin{eqnarray}
\rho_{\mu}(x,\xi,\bar{\xi},\lambda)& =&
\frac{1}{2}\omega_{\mu}^{*ab} \mathcal{%
J}_{ab} + h_{\mu}^{a}(x,\xi,\bar{\xi},\lambda) P_{a}
\label{sx1}\\
\zeta_{\alpha} & = & \frac{1}{2} \theta_{\alpha}^{(*) a b} J_{ab}
+ \psi_{\alpha}^{a} P_{a}  \label{sx2} \\
 \bar{\zeta}^{\alpha}& = & \frac{1}{2} \bar{\theta}^{(*) \alpha a
b} J_{ab} + \bar{\psi}^{\alpha a} P_{a}  \label{sx3} \\
 \rho _o & = & \frac 12\omega _o^{ab}J_{ab}+L_o^aP_a \label{sx4}
\end{eqnarray}
where, $J_{ab},P_a$ are the generators of the four--dimensional
 Poincar$%
\grave{e}$ group, namely the angular momentum and linear
momentum, $\omega
_\mu ^{(*)ab}$ represent the Lorentz - spin connection coefficients, $\bar{%
\Psi}^{\alpha a},$ $\Psi _\alpha ^a$, $\theta _\alpha ^{(*)ab}$, $\bar{\theta%
}^{(*)\alpha ab}$ are the given spin-tetrad and spin - connection
coefficients, and $L_o^a$ deformed tetrad coefficients. We use
Greek letters $\lambda ,\mu ,$ $\nu ,\ldots $ for space-time
indices, $\alpha ,\beta ,\gamma ,\ldots $ for the spinor,
$a,b,c,\ldots $ for Lorentz ones, and the
index $(o)$ represents the deformed variable; $\lambda ,\alpha ,$ $%
a=1,\ldots ,4.$ The general transformations of coordinates on $\tilde{S}%
^{(2)}{M}$ are
\begin{equation}
{x^{\prime }}^\mu ={x^{\prime }}^\mu (x^\nu ),{\xi }_\alpha ^{\prime }={\xi }%
_\alpha ^{\prime }(\xi _\beta ,\bar{\xi}^\beta ),\bar{{\xi }^{\prime }}%
^\alpha =\bar{{\xi }^{\prime }}^\alpha (\bar{\xi}^\beta ,\xi _\beta ),{%
\lambda }^{\prime }=\lambda   \label{sx5}
\end{equation}


\section{Connections}

We define the following gauge covariant derivatives
\begin{eqnarray}
(a) \; \mathcal{D}_{\mu}^{(*)} & = & \frac{\delta}{\delta x^{\mu}}
+ \frac{1}{2} \omega_{\mu}^{(*)ab} \mathcal{J}_{ab} \qquad (b) \;
\mathcal{D}^{(*) \alpha} = \frac{\delta}{\delta \xi_{\alpha}} +
\frac{1}{2} \bar{\Theta}^{(*) \alpha ab} \mathcal{J}_{ab}
 \nonumber \\
(c) \; \mathcal{D}^{(*)}_{ \alpha} & = &
\frac{\partial}{\partial \bar{\xi}%
^{\alpha}} + \frac{1}{2} \Theta_{\alpha}^{(*)ab} \mathcal{J}_{ab}
 \nonumber \\
 (d) \; \mathcal{D}_{o}^{(*)}& =&  \frac{\partial}{\partial \lambda}
+ \omega_{o}^{ab} \mathcal{J}_{ab}  \label{sx6}
\end{eqnarray}
where,
\begin{eqnarray}
\frac{\delta}{\delta x^{\mu}} &=&
 \frac{\partial}{\partial x^{\mu}} + \mathcal{%
N}_{\alpha \mu}\frac{\partial}{\partial \xi_{\alpha}} - \bar{\mathcal{N}}%
_{\mu}^{\alpha}\frac{\partial}{\partial \bar{\xi}^{\alpha}} - \mathcal{N}%
_{\mu}^{o}\frac{\partial}{\partial \lambda} \nonumber \\
\frac{\delta}{\delta \xi_{\alpha}}& =&  \frac{\partial}{\partial
\xi_{\alpha}}
- \bar{\mathcal{N}}_{o}^{\alpha \beta}\frac{\partial}{\partial \bar{\xi}%
^{\beta}}, \frac{\partial}{\partial \lambda} = L_{o}^{\mu}\frac{\partial}{%
\partial x^{\mu}}.  \label{sx7}
\end{eqnarray}

The nonlinear connection coefficients are defined further. The
space-time, Lorentz, spin frames and the deformed frame are
connected by the relations
\begin{eqnarray}
(a) \; \frac{\delta}{\delta x^{\mu}} & = &
h_{\mu}^{a}\frac{\delta}{\delta x^{a}} \nonumber \\
 (b) \; \frac{\partial}{\partial \xi}_{\alpha} & = &
\bar{\Psi}^{\alpha a}%
\frac{\partial}{\partial x^{a}}
 (b^{\prime}) \; \frac{\partial}{\partial \bar{\xi}^{\alpha}}  =
\Psi_{\alpha}^{a}\frac{\partial}{\partial x^{a}} \nonumber \\
 (c) \; \frac{\partial}{\partial \lambda}
 & = & L_{o}^{\mu}\frac{\partial}{\partial
x^{\mu}}. \label{sx9}
\end{eqnarray}

The relation $(\ref{sx9}a)$ is a generalization of the well -
known principle of equivalence in the total space of the spinor
bundle $S^{(2)}M$. In addition, the relations
$(\ref{sx9}a,b,b^{\prime },c)$ represent a generalized form of
the equivalence principle, since the considered deformed spinor
bundle contains spinors as internal variables.

The absolute differential of an arbitrary contravariant vector $X^{\nu}$ in $%
\tilde S^{(2)}M$, has the form
\begin{equation}
\mathcal{D}X^{\nu} = (\mathcal{D}_{\mu}^{(*)}X^{\nu})dx^{\mu} +(\mathcal{D}%
^{(*) \alpha}X^{\nu})d\xi_{\alpha} + \newline
(\mathcal{D}_{\alpha}^{(*)}X^{\nu})d\bar{\xi}^{\alpha} + (\mathcal{D}%
_{o}^{(*)}X^{\nu})d\lambda  \label{sx10}
\end{equation}

The differentials $\mathcal{D}\xi_{\alpha}$, $\mathcal{D}\bar{\xi}^{\alpha}$%
, $\mathcal{D}\lambda$ can be written
\begin{eqnarray}
\mathcal{D}\xi _\beta &= & d\xi _\beta + \mathcal{N}_{\alpha
\beta}^od\bar{\xi}^\alpha + \tilde{\mathcal{N}}_\beta ^{o\alpha
}d\xi _\alpha + \mathcal{N}_{\beta \mu }dx^\mu \nonumber \\
\mathcal{D}\bar{\xi}^\beta &= & d\bar{\xi}^\beta
+\mathcal{N}_{o\alpha }^\beta d \bar{\xi}^\alpha -
\tilde{\mathcal{N}}_o^{\beta \alpha }d\xi _\alpha -
\bar{\mathcal{N}}_\mu ^\beta dx^\mu  \label{sx11}\\
\mathcal{D}_o\lambda & = &
d\lambda +\mathcal{N}_\kappa ^odx^\kappa -\tilde{%
\mathcal{N}}_o^\alpha d\xi _\alpha -\mathcal{N}_\alpha
^od\bar{\xi}^\alpha, \nonumber
\end{eqnarray}
 where
\[
\mathcal{N}=\{\mathcal{N}_{\alpha \beta }^o,\mathcal{N}_{\beta \mu },\tilde{%
\mathcal{N}}_\beta ^{o\alpha },\bar{\mathcal{N}}_\mu ^\beta ,\tilde{\mathcal{%
N}}_o^{\beta \alpha },\mathcal{N}_{o\alpha }^\beta ,\mathcal{N}_\kappa ^o,%
\tilde{\mathcal{N}}_o^\beta ,\mathcal{N}_\alpha ^o\}
\]
represent the coefficients of the nonlinear connection which are
given by
\begin{eqnarray}
\mathcal{N}_{\beta \mu} & = & \frac{1}{2} \omega^{(*)ab}_{\mu}
\mathcal{J}_{ab} \xi_{\beta}, \qquad
\bar{\mathcal{N}}_{\mu}^{\beta} = - \frac{1}{2}
\omega_{\mu}^{(*) ab} \mathcal{J}_{ab} \bar{\xi}^{\beta}, \mathcal{N}%
_{\mu}^{o} = \frac{1}{2} \omega_{o \mu}^{\alpha b}
\mathcal{J}_{ab}
 \nonumber \\
 \tilde{\mathcal{N}}_{\beta}^{o \alpha} & = & \frac{1}{2}
\bar{\theta}^{(*) \alpha ab} \mathcal{J}_{ab} \xi_{\beta}, \,
\tilde{\mathcal{N}}_{o}^{\alpha
\beta} = - \frac{1}{2} \bar{\theta}^{(*) \alpha ab} \mathcal{J}_{ab} \bar{\xi%
}^{\beta},\, \tilde{\mathcal{N}}_{o}^{\alpha} = \frac{1}{2}
\omega_{o}^{ab} \mathcal{J}_{ab} \bar{\xi}^{\alpha},
\label{sx12} \\
 \mathcal{N}_{\alpha \beta}^{o} & = & \frac{1}{2}
\theta_{\alpha}^{(*) ab}
\mathcal{J}_{ab} \xi_{\beta}, \, \mathcal{N}_{o \alpha}^{\beta} = - \frac{1}{%
2} {\theta}_{\alpha}^{(*)ab} \mathcal{J}_{ab} \bar{\xi}^{\beta}, \, \mathcal{%
N}_{\alpha}^{o} = \frac{1}{2} \omega_{o}^{ab} \mathcal{J}_{ab}
\xi_{\alpha}. \nonumber
\end{eqnarray}

The metrical structure in the deformed spinor bundle $\tilde
S^{(2)}M$ has the form
\begin{eqnarray}
G &= & g_{\mu \nu }(x,\xi ,\bar{\xi},\lambda )dx^\mu \otimes
dx^\nu +g_{\alpha \beta }(x,\xi ,\bar{\xi}, \lambda
)\mathcal{D}\bar{\xi}^\alpha \otimes \mathcal{D}\bar{\xi}^{*\beta
}+ \nonumber  \\
&&  +g^{\alpha \beta }(x,\xi ,\bar{\xi},\lambda )\mathcal{D}\xi
_\alpha \otimes
\mathcal{D}\xi _\beta ^{*}+g_{oo}(x,\xi ,\bar{\xi},\lambda )\mathcal{D}%
\lambda \otimes \mathcal{D}\lambda \label{sx13}
\end{eqnarray}
where '*' denotes Hermitean conjugation. The local adapted frame
is given by
\[
\frac \delta {\delta \zeta ^A}=\{\frac \delta {\delta x^\lambda
},\frac \delta {\delta \xi _\alpha },\frac \partial {\partial
\bar{\xi}^\alpha },\frac \partial {\partial \lambda }\}
\]
and the associated dual frame
\begin{equation}
\delta \zeta ^A=\{\mathcal{D}x^\kappa \equiv dx^\kappa
,\mathcal{D}\xi _\beta ,\mathcal{D}\bar{\xi}^\beta
,\mathcal{D}_o\lambda \},  \label{sx14}
\end{equation}
where the terms $\frac \delta {\delta x^\lambda },\frac \delta
{\delta \xi _\alpha },\mathcal{D}_o\lambda ,$
$\mathcal{D}x^\kappa ,\mathcal{D}\xi _\beta
,\mathcal{D}\bar{\xi}^\beta ,$ are provided by the relations (\ref
{sx7}), (\ref{sx10}), (\ref{sx11}).

The considered connection in $\tilde{S}^{(2)}(M)$ is a
d-connection, having with respect to the adapted basis the
coefficients
\begin{eqnarray}
\Gamma _{BC}^A &=&
\{\Gamma _{\nu \rho }^{(*)\mu },C_{\nu \alpha }^\mu ,\bar{C}%
_\nu ^{\mu \alpha },\Gamma _{\nu o}^{(*)\mu },\bar{\Gamma}_{\beta
\lambda }^{(*)\alpha },\tilde{C}_{\alpha \gamma }^\beta
,\tilde{C}_\gamma ^{\beta \alpha },\bar{\Gamma}_{o\gamma
}^{(*)\beta }, \nonumber \\
&&\Gamma _{\alpha \nu }^{(*)\beta },C_{\beta \alpha }^\gamma
,C_\beta ^{\gamma \alpha },C_{\beta o}^\alpha ,\Gamma _{o\mu
}^{(*)o},\bar{C}_o^{o\alpha },C_{o\alpha }^o,L_{oo}^o\}
\label{sx15}
\end{eqnarray}
defined by the generic relations
\begin{equation}
\mathcal{D}_{\frac{\delta}{\delta z^{C}}} \frac{\delta}{\delta
z^{B}} =
\Gamma_{BC}^{A} \frac{\delta}{\delta z^{A}},\qquad \frac{\delta}{\delta z^{A}%
} \in \left \{ \frac{\delta}{\delta x^{\mu}}, \frac{\delta}{\delta
\xi_{\alpha}}, \frac{\partial}{\partial \bar{\xi}^{\alpha}}, \frac{\partial}{%
\partial \lambda} \right\}.  \label{sx16}
\end{equation}


It preserves the distributions $\mathcal{H}$,$\mathcal{F}^{(1)}$,
$\mathcal{F}%
^{(2)}$,$R$, and its coefficients are defined by
\begin{eqnarray*}
\mathcal{D}_{\frac \delta {\delta x^\rho }}\frac \delta {\delta
x^\nu } &=& \Gamma _{\nu \rho }^{(*)\mu }
\frac \delta {\delta x^\mu },\qquad \mathcal{D%
}_{\frac \partial {\partial \bar{\xi}^\alpha }}\frac \delta
{\delta x^\nu }=C_{\nu \alpha }^\mu
\frac \delta {\delta x^\mu},\\
\mathcal{D}_{\frac \delta {\delta \xi _\alpha }} \frac \delta
{\delta x^\nu } &=& \bar{C}_\nu ^{\mu \alpha }\frac \delta
{\delta x^\mu },\qquad
 \mathcal{D}_{\frac \partial {\partial \lambda }}
 \frac \delta {\delta x^\nu}=\Gamma _{\nu o}^{(*)\mu }
 \frac \delta {\delta x^\mu }, \\
 \mathcal{D}_{\frac \delta {\delta x^\lambda }}\frac \delta
{\delta \xi _\alpha } &= &\bar{\Gamma}_{\beta \lambda }^{(*)\alpha
}\frac \delta {\delta \xi _\beta },\qquad \mathcal{D}_{\frac
\partial {\partial \bar{\xi}_\alpha }}\frac \delta {\delta \xi
_\beta }=\tilde{C}_{\alpha \gamma }^\beta \frac \delta {\delta
\xi _\gamma }, \\
\mathcal{D}_{\frac \delta {\delta \xi _\alpha }}\frac \delta
{\delta \xi _\beta } &=& \tilde{C}_\gamma ^{\beta \alpha }\frac
\delta {\delta \xi _\gamma },\qquad \qquad \mathcal{D}_{\frac
\partial {\partial \lambda }}\frac \delta {\delta \xi _\beta
}=\bar{\Gamma}_{o\gamma }^{(*)\beta }\frac \delta {\delta \xi
_\gamma }, \\
\mathcal{D}_{\frac \delta {\delta x^\nu }} {\frac \partial
{\partial \bar{\xi}^\alpha }}&=&
\Gamma _{\alpha \nu }^{(*)\beta }\frac \partial {\partial \bar{\xi%
}^\beta },\qquad \mathcal{D}_{\frac \partial {\partial
{\bar{\xi}}^\alpha }}\frac \partial {\partial \bar{\xi}^\beta
}=C_{\beta \alpha }^\gamma \frac
\partial {\partial \bar{\xi}^\gamma }, \\
 \mathcal{D}_{\frac \delta {\delta \xi _\alpha }}\frac \partial
{\partial \bar{\xi}^\beta } &=&
C_\beta ^{\gamma \alpha }\frac \partial {\partial \bar{\xi}%
^\gamma },\qquad \mathcal{D}_{\frac \partial {\partial \lambda
}}\frac
\partial {\partial \bar{\xi}^\beta }=C_{\beta o}^\alpha \frac \partial
{\partial \bar{\xi}^\alpha }, \\
\mathcal{D}_{\frac \delta {\delta x^\mu }} {\frac \partial
{\partial \lambda }}&=& \Gamma _{o\mu }^{(*)o}\frac \partial
{\partial \lambda },
\qquad \mathcal{D}%
_{\frac \delta {\delta \xi _\alpha }}\frac \partial {\partial
\lambda } = \bar{C}_o^{o\alpha }\frac \partial {\partial \lambda
} \mathcal{D}_{\frac \partial {\partial \bar{\xi}^\alpha }}\frac
\partial {\partial \lambda } = C_{o\alpha }^o\frac \partial
{\partial \lambda }, \\
\mathcal{D}_{\frac \partial {\partial \lambda }}\frac \partial
{\partial \lambda } &= & L_{oo}^o\frac
\partial {\partial \lambda }.
\end{eqnarray*}

The covariant differentiation of tensors, spin-tensors and
Lorentz - type tensors of arbitrary rank is defined as follows:
\begin{eqnarray}
 \bigtriangledown_{\kappa} T_{\nu \ldots}^{\mu
\ldots} &=& \frac{\delta T_{\nu \ldots}^{\mu \ldots}}{\delta
x^{\kappa}} + \Gamma_{\rho \kappa}^{(*) \mu} T_{\nu \ldots}^{\rho
\ldots} + \ldots - \Gamma_{\nu \kappa}^{(*) \rho} T_{\rho
\ldots}^{\mu
\ldots} \nonumber \\
 \bigtriangledown^{\alpha} T_{\nu \ldots}^{\mu \ldots} &=&
\frac{\delta T_{\nu \ldots}^{\mu \ldots}}{\delta \xi_{\alpha}} +
\bar{C}_{\kappa}^{\mu \alpha} T_{\nu \ldots}^{\kappa \ldots} +
\ldots - \bar{C}_{\nu}^{\kappa \alpha} T_{\kappa \ldots}^{\mu
\ldots} \nonumber \\
 \bigtriangledown_{\alpha} T_{\nu \ldots}^{\mu \ldots} &=&
\frac{\partial T_{\nu \ldots}^{\mu \ldots}}{\partial
\bar{\xi}^{\alpha}} + C_{\kappa \alpha}^{\mu} T_{\nu
\ldots}^{\kappa \ldots} + \ldots - \bar{C}_{\nu \alpha}^{\kappa}
T_{\kappa \ldots}^{\mu \ldots}\nonumber \\
 \bigtriangledown_{o} T_{\nu \ldots}^{\mu \ldots} & = &
\frac{\partial T_{\nu \ldots}^{\mu \ldots}}{\partial \lambda} +
\Gamma_{\kappa o}^{(*) \mu} T_{\nu \ldots}^{\kappa \ldots} -
\ldots + \Gamma_{\nu o}^{(*) \kappa} T_{\kappa \ldots}^{\mu
\ldots} \nonumber \\
\bigtriangledown_{\kappa} \Phi_{\beta \ldots}^{\alpha \ldots} &=&
\frac{\delta
\Phi_{\beta \ldots}^{\alpha \ldots}}{\delta x^{\kappa}} - \bar{\Gamma}%
_{\beta \kappa}^{(*) \gamma} \Phi_{\gamma \ldots}^{\alpha \ldots}
- \ldots + \Phi_{\beta\ldots}^{\gamma\ldots} \bar{\Gamma}_{\gamma
\kappa}^{(*) \alpha \ldots}  \label{sx17} \\
 \bigtriangledown^{\delta} \Phi_{\beta\ldots}^{\alpha\ldots} &=&
\frac{\delta
\Phi_{\beta \ldots}^{\alpha \ldots}}{\delta \xi_{\delta}} - \tilde{C}%
_{\beta}^{\gamma \delta} \Phi_{\gamma \ldots}^{\alpha \ldots} -
\ldots + \Phi_{\beta\ldots}^{\gamma\ldots}
\tilde{C}_{\gamma}^{\alpha \delta}
 \nonumber
 \end{eqnarray}
 \begin{eqnarray}
 \bigtriangledown_{\delta} \Phi_{\beta\ldots}^{\alpha\ldots} &=&
\frac{\partial \Phi_{\beta \ldots}^{\alpha \ldots}}{\partial
\bar{\xi}^{\delta}} - C_{\beta \delta}^{\gamma} \Phi_{\gamma
\ldots}^{\alpha \ldots} - \ldots +
\Phi_{\beta\ldots}^{\gamma\ldots} C_{\gamma \delta}^{\alpha}
\nonumber \\
 \bigtriangledown_{o} \Phi_{\beta\ldots}^{\alpha\ldots} &=&
\frac{\partial \Phi_{\beta \ldots}^{\alpha \ldots}}{\partial
\lambda} - \bar{\Gamma}_{o \beta}^{(*) \gamma} \Phi_{\gamma
\ldots}^{\alpha \ldots} - \ldots +
\Phi_{\beta\ldots}^{\gamma\ldots} \Gamma_{o \gamma}^{(*) \alpha}
 \nonumber \\
 \bigtriangledown_{\mu}^{(*)} V_{c \ldots}^{a \ldots} & = &
\frac{\delta V_{c \ldots}^{a \ldots}}{\delta x^{\mu}} +
\omega_{\mu b}^{(*) a} V_{c \ldots}^{b \ldots} + \ldots -
\omega_{\mu c}^{(*) b} V_{b \ldots}^{a \ldots} \nonumber \\
 \bigtriangledown^{(*) \alpha} V_{c \ldots}^{a \ldots} &=&
\frac{\delta V_{c \ldots}^{a \ldots}}{\delta \xi_{\alpha}} +
\ldots + \bar{\theta}_{\alpha b}^{(*) a} V_{c \ldots}^{b \ldots}
+ \ldots - \theta_{\alpha c}^{(*) b} V_{b \ldots}^{a \ldots}
\nonumber \\
 \bigtriangledown_{\alpha}^{(*)} V_{c \ldots}^{a \ldots} &=&
\frac{\partial V_{c \ldots}^{a \ldots}}{\partial
\bar{\xi}_{\alpha}} + \theta_{\alpha b}^{(*) a} V_{c \ldots}^{b
\ldots} + \ldots - \theta_{\alpha c}^{(*) b} V_{b \ldots}^{a
\ldots} \nonumber \\
 \bigtriangledown_{o}^{(*)} V_{c \ldots}^{a \ldots} & =&
\frac{\partial V_{c \ldots}^{a \ldots}}{\partial \lambda} +
\omega_{o b}^{(*) a} V_{c \ldots}^{b \ldots} + \ldots -
\omega_{oc}^{(*) b} V_{b \ldots}^{a \ldots}. \nonumber
\end{eqnarray}
The covariant derivatives of the metric tensor $g_{\mu \nu}$ are
postulated to be zero:
\begin{equation}
\mathcal{D}_{\mu}^{(*)}g_{\kappa \lambda} = 0, \mathcal{D}^{(*)
\alpha}g_{\kappa \lambda} = 0,
\mathcal{D}_{\alpha}^{(*)}g_{\kappa \lambda} = 0,
\mathcal{D}_{(o)}^{(*)}g_{\kappa \lambda} = 0.  \label{sx8}
\end{equation}

\section{Curvatures and Torsions} \index{curvature} \index{torsion}

From the relations (\ref{sx6}) we obtain the curvatures and
torsions of the space $\tilde S^{(2)}M$
\begin{equation}
\lbrack \mathcal{D}_\mu ^{(*)},\mathcal{D}_\nu ^{(*)}]=\mathcal{D}_\mu ^{(*)}%
\mathcal{D}_\nu ^{(*)}-\mathcal{D}_\nu ^{(*)}\mathcal{D}_\mu
^{(*)}=R_{\mu \nu }^aP_a+\frac 12R_{\mu \nu }^aJ_{ab} \label{sx18}
\end{equation}
with their explicit expressions given by
\begin{eqnarray*}
R_{\mu \nu }^a &=&\frac{\delta h_\mu ^a}{\delta x^\nu }-\frac{\delta h_\nu ^a%
}{\delta x^\mu }+\omega _{\mu b}^{(*)a}h_\nu ^b-\omega _{\nu
b}^{(*)a}h_\mu
^b, \nonumber \\
R_{\mu \nu }^{ab} &=&\frac{\delta \omega _\mu ^{(*)ab}}{\delta x^\nu }-\frac{%
\delta \omega _\nu ^{(*)ab}}{\delta x^\mu }+\omega _\mu
^{(*)a\rho }\omega _{\nu \rho }^{(*)b}-\omega _\nu ^{(*)\rho
a}\omega _{\mu \rho }^{(*)b}, \nonumber \\
R_{\mu \nu }^{ab} &=&\frac{\delta \omega _\mu ^{(*)ab}}{\delta x^\nu }-\frac{%
\delta \omega _\nu ^{(*)ab}}{\delta x^\mu }+\omega _\mu
^{(*)a\rho }\omega _{\nu \rho }^{(*)b}-\omega _\nu ^{(*)\rho
a}\omega _{\mu \rho }^{(*)b},
\end{eqnarray*}

\begin{equation}
\lbrack {D}_\mu ^{(*)}, {D}_\alpha ^{(*)} \rbrack =  P_{\mu
\alpha }^aP_a+\frac 12P_{\mu \alpha }^{ab}J_{ab}
  \label{sx19}
\end{equation}

\begin{eqnarray*}
P_{\mu \alpha }^{ab} &=&\frac{\delta \theta _\alpha ^{(*)ab}}{\delta x^\mu }-%
\frac{\partial \omega _\mu ^{(*)ab}}{\partial \bar{\xi}^\alpha
}+\theta _{\alpha c}^{(*)b}\omega _\mu ^{(*)ac}-\theta _{\alpha
c}^{(*)a}\omega _\mu ^{(*)cb}, \nonumber \\
P_{\mu \alpha }^a &=&\frac{\delta \psi _\alpha ^a}{\delta x^\mu }-\frac{%
\partial h_\mu ^a}{\partial \bar{\xi}^\alpha }+\omega _{\mu c}^{(*)a}\psi
_\alpha ^c-\theta _{\alpha c}^{(*)a}h_\mu ^c, \nonumber \\
P_{\mu \alpha }^a &=&\frac{\delta \psi _\alpha ^a}{\delta x^\mu }-\frac{%
\partial h_\mu ^a}{\partial \bar{\xi}^\alpha }+\omega _{\mu c}^{(*)a}\psi
_\alpha ^c-\theta _{\alpha c}^{(*)a}h_\mu ^c. \nonumber
\end{eqnarray*}

Similarly to \cite{sm3}, the other four curvatures and torsions
result from the commutation relations
\begin{eqnarray}
\lbrack {D}_{\mu}^{(*)} , {D}^{\alpha} \rbrack & =&
\bar{P}_{\mu}^{a \alpha} P_{a} + \frac{1}{2}
\bar{P}_{\mu}^{ab\alpha} J_{ab}  \label{sx20}\\
 \lbrack {D}_{\alpha}^{(*)} , {D}^{(*) \beta}
 \rbrack &=&
S_{\alpha}^{\beta a} P_{a} + \frac{1}{2} S_{\alpha}^{ab \beta}
J_{ab}  \label{sx21}\\
\lbrack {D}_{\alpha}^{(*)}, {D}_{\beta}^{(*)} \rbrack &=&
Q_{\alpha \beta}^{a} P_{a} + \frac{1}{2} Q_{\alpha \beta}^{ab}
J_{ab}  \label{sx22} \\
\lbrack {D}^{(*) \alpha} , {D}^{(*) \beta} \rbrack &=&
\tilde{Q}^{\alpha \beta a} P_{a} + \frac{1}{2} \tilde{Q}^{ab
\alpha \beta} J_{ab}.  \label{sx23}
\end{eqnarray}

The contribution of the $\lambda$ - covariant derivative $\mathcal{D}%
_{o}^{(*)}$ provides us the following curvatures and torsions

\begin{equation}
\lbrack \mathcal{D}_o^{(*)},\mathcal{D}_\mu ^{(*)}]=R_{o\mu
}^aP_a+\frac 12R_{o\mu }^{ab}J_{ab}  \label{sx24}
\end{equation}
\begin{eqnarray*}
R_{o\mu }^a &=&\frac{\delta L_\mu ^a}{\delta \lambda }-\frac{\delta h_o^a}{%
\delta x^\mu }+\omega _{\mu b}^{(*)a}L_o^b-\omega _{bo}^{(*)a}h_\mu ^b, \\
R_{o\mu }^{ab} &=&\frac{\delta \omega _\mu ^{(*)ab}}{\delta \lambda }-\frac{%
\delta \omega _o^{(*)ab}}{\delta x^\mu }+\omega _\mu ^{(*)a\rho
}\omega
_{o\rho }^{(*)b}-\omega _o^{(*)\rho a}\omega _{\mu \rho }^{(*)b}, \\
R_{o\mu }^{ab} &=&\frac{\delta \omega _\mu ^{(*)ab}}{\delta \lambda }-\frac{%
\delta \omega _o^{(*)ab}}{\delta x^\mu }+\omega _\mu ^{(*)a\rho
}\omega _{o\rho }^{(*)b}-\omega _o^{(*)\rho a}\omega _{\mu \rho
}^{(*)b},
\end{eqnarray*}

\begin{equation}
\lbrack \mathcal{D}_o^{(*)},\mathcal{D}_o^{(*)}]=0,\qquad
R_{oo}^{ab}=0,\qquad R_{oo}^a=0.  \label{sx25}
\end{equation}

\begin{equation}
\lbrack \mathcal{D}_o^{(*)},\mathcal{D}^{(*)\alpha
}]=\bar{P}_o^{a\alpha }P_a+\frac 12\bar{P}_o^{ab\alpha }J_{ab}
\label{sx26}
\end{equation}
\begin{eqnarray*}
\bar{P}_o^{a\alpha } &=&\frac{\partial \bar{\psi}^{a\alpha
}}{\partial
\lambda }-\frac{\delta L_o^a}{\delta \xi _\alpha }+\omega _{oc}^{(*)a}\bar{%
\psi}^{\alpha a}-\bar{\theta}_c^{(*)a\alpha }L_o^c, \\
\bar{P}_o^{ab\alpha } &=&\frac{\partial \bar{\theta}^{(*)\alpha ab}}{%
\partial \lambda }-\frac{\delta \omega _o^{(*)ab}}{\delta \xi _\alpha }+\bar{%
\theta}_c^{(*)ab}\omega _\mu ^{(*)ac}-\bar{\theta}_c^{(*)a\alpha
}\omega
_o^{(*)ab}, \\
\bar{P}_o^{ab\alpha } &=&\frac{\partial \bar{\theta}^{(*)\alpha ab}}{%
\partial \lambda }-\frac{\delta \omega _o^{(*)ab}}{\delta \xi _\alpha }+\bar{%
\theta}_c^{(*)ab}\omega _\mu ^{(*)ac}-\bar{\theta}_c^{(*)a\alpha
}\omega _o^{(*)ab},
\end{eqnarray*}

\begin{equation}
\lbrack \mathcal{D}_o^{(*)},\mathcal{D}_\alpha ^{(*)}]=P_{o\alpha
}^aP_a+\frac 12P_{o\alpha }^{ab}J_{ab},  \label{sx27}
\end{equation}
\begin{eqnarray*}
P_{o\alpha }^a &=&\frac{\partial \psi _\alpha ^a}{\partial \lambda }-\frac{%
\partial L_o^a}{\partial \bar{\xi}^\alpha }+\omega _{oc}^{(*)a}\psi _\alpha
^c-\theta _{\alpha c}^{(*)a}L_o^c, \\
P_{o\alpha }^{ab} &=&\frac{\partial \theta _\alpha
^{(*)ab}}{\partial \lambda }-\frac{\delta \omega
_o^{(*)ab}}{\delta \bar{\xi}^\alpha }+\theta
_{ac}^{(*)b}\omega _o^{(*)ac}-\theta _{\alpha c}^{(*)a}\omega _o^{(*)cb}, \\
P_{o\alpha }^{ab} &=&\frac{\partial \theta _\alpha
^{(*)ab}}{\partial \lambda }-\frac{\delta \omega
_o^{(*)ab}}{\delta \bar{\xi}^\alpha }+\theta _{ac}^{(*)b}\omega
_o^{(*)ac}-\theta _{\alpha c}^{(*)a}\omega _o^{(*)cb}.
\end{eqnarray*}

\section{Field Equations}             \index{Field equations}

In the following, we derive by means of the Palatini method the
field equations, using a Lagrangian of the form
\begin{equation}
\mathcal{L}=h(R+P+\bar{P}+Q+\tilde{Q}+R_o+\bar{P}_o+P{o})
\label{sx28}
\end{equation}
which depends on the tetrads and on the connection coefficients,
\[
\mathcal{L}(\kappa ^A,\delta _M\kappa ^A)=\mathcal{L}(h,\omega ^{(*)},\psi ,%
\bar{\psi},\theta ^{(*)},\bar{\theta}^{(*)},\omega _o^{(*)})
\]
where,
 \begin{eqnarray}
&& \kappa^{A} \in \{ (h_{\mu}^{a} (z),\omega_{\mu}^{(*)ab}(z),
\psi_{\alpha}^{a} (z),\bar{\psi}^{\alpha a} (z),
\theta_{\alpha}^{(*) ab} (z),\bar{\theta}^{(*) \alpha ab} (z),
\omega_{o}^{(*) ab} (z))\}, \nonumber \\
&&\delta_{M} =  \frac{\delta}{\delta z^{M}} \in \left \{
\frac{\delta}{\delta
x^{\mu}}, \frac{\delta}{\delta \xi_{\alpha}}, \frac{\partial}{\partial \bar{%
\xi}^{\alpha}}, \frac{\partial}{\partial \lambda} \right\}, \
z=(z^{M}) = (x^{\mu}, \xi_{\alpha}, \bar{\xi}^{\alpha},\lambda)
\label{sx29}
\end{eqnarray}
and
\[
\left\{
\begin{array}{lll}
R = h_{a}^{\mu} h_{c}^{\kappa} R_{\mu \kappa}^{ c}, & P = h_{
}^{\mu}
\psi_{c}^{\alpha} P_{\mu \alpha}^{ c}, & \bar{P} = h_{ }^{\mu} \bar{\psi}%
_{\alpha c} \bar{P}_{\mu}^{ c \alpha}, \\
&  &  \\
Q = Q_{\alpha \beta}^{ab} \bar{\psi}_{a}^{\alpha}
\bar{\psi}_{b}^{\beta}, & \tilde{Q}=\tilde{Q}^{ab \alpha \beta} =
\psi_{\alpha a} \psi_{\beta b}, & S
= \bar{\psi}_{a}^{\alpha} \psi_{\beta b} S_{\alpha}^{ab \beta}, \\
&  &  \\
R_{o} = L_{\kappa}^{o} h_{c}^{\mu} h_{a}^{\kappa} R_{o \mu}^{ac}, & \bar{P}%
_{o} = L_{\kappa}^{o} h_{a}^{\kappa} \bar{\psi}_{\alpha c}
\bar{P}_{o}^{ac \alpha}, & P_{o} = L_{\kappa}^{o} h_{a}^{\kappa}
\bar{\psi}_{c}^{\alpha}P_{o
\alpha}^{ac}. \\
&  &
\end{array}
\right.
\]

The Euler-Lagrange equations are generally given by
\begin{equation}
\frac{\delta\mathcal{L}}{\delta {\kappa}^{(A)}} = \partial_M \left ( \frac{%
\partial\mathcal{L}}{\partial(\partial_{M} \kappa^{(A)})} \right ) - \frac{%
\partial \mathcal{L}} {\partial \kappa^{(A)}}= 0,  \label{sx31}
\end{equation}
with $\partial_{M} = \frac{\partial}{\partial z^{M}} \in \left \{ \frac{%
\partial}{\partial x^{\mu}}, \frac{\partial}{\delta \xi_{\alpha}},\frac{%
\partial}{\partial \bar{\xi}^{\alpha}}, \frac{\partial}{\partial\lambda}
\right\}.$

From the relation (\ref{sx31}), the variation of $\mathcal{L}$
with respect to the tetrads $h_b^\nu $ gives us the first field
equation
\begin{equation}
\partial _\mu \frac{\partial \mathcal{L}}{\partial (\partial _\mu h_b^\nu )}%
+\partial ^\alpha \frac{\partial \mathcal{L}}{\partial (\partial
^\alpha h_b^\nu )}+\partial _\alpha \frac{\partial
\mathcal{L}}{\partial (\partial _\alpha h_b^\nu )}+\partial
_o\frac{\partial \mathcal{L}}{\partial (\partial _oh_b^\nu
)}-\frac{\partial \mathcal{L}}{\partial h_b^\nu }=0,  \label{sx32}
\end{equation}
where we denoted $\partial _o=\frac \partial {\partial \lambda
}$. Finally, after some calculations we get,
\begin{equation}
\tilde{H}_\nu ^b-\frac 12h_\nu ^b\tilde{H}=0,  \label{sx33}
\end{equation}
where we put,
\begin{eqnarray}
\tilde{H}&=&R+P+\bar{P}+Q+\tilde{Q}+S+R_o+\bar{P}_o+P_o,
\nonumber \\
\tilde{H}_\nu ^b &=& 2R_\nu ^b+P_\nu ^b+\bar{P}_\nu ^b+R_{o\nu
}^b+\bar{P}_{o\nu }^b+P_{(o)\nu }^b, \nonumber \\
&& \left\{
\begin{array}{lll}
R_\nu ^b=h_c^\kappa R_{\nu \kappa }^{bc}, & P_\nu
^b=\bar{\psi}_c^\alpha P_{\nu \alpha }^{bc}, & R_{o\nu
}^b=L_\kappa ^oh_{}^\kappa R_{o\nu }^b+L_\nu
^oh_c^\mu R_{o\mu }^{bc},  \\
&  &  \\
\bar{P}_\nu ^b=\psi _{\alpha c}\bar{P}_\nu ^{bc\alpha }, &
\bar{P}_{o\nu }^b=L_\nu ^o\bar{\psi}_{\alpha
c}\bar{P}_o^{bc\alpha }, & P_{o\nu }^b=L_\nu
^o\bar{\psi}_c^\alpha P_{o\alpha }^{bc} \\
&  &
\end{array}
\right.  \nonumber
\end{eqnarray}
The equation (\ref{sx33}) is the \textbf{Einstein equation for
empty space}, in the framework of our consideration. Also, the
variation of $\mathcal{L}$ with respect to $\omega _\mu ^{(*)ab}$
gives,
\begin{equation}
\partial _\nu \frac{\partial \mathcal{L}}{\partial (\partial _\nu \omega
_\mu ^{(*)ab})}+\partial ^\alpha \frac{\partial
\mathcal{L}}{\partial (\partial ^\alpha \omega _\mu
^{(*)ab})}+\partial _\alpha \frac{\partial
\mathcal{L}}{\partial (\partial _\alpha \omega _\mu ^{(*)ab})}-\frac{%
\partial \mathcal{L}}{\partial \omega _\mu ^{(*)ab}}=0.  \label{sx34}
\end{equation}
From this relation we get the second field equation in the
following form,
\[
\partial _\nu [h(h_a^\mu h_b^\nu -h_b^\mu h_a^\nu )]-\partial ^\alpha
[hh_a^\mu \psi _{\alpha b}]-\partial _\alpha [hh_a^\mu
\bar{\psi}_b^\alpha ]-\partial _o[hL_a^oh_b^\mu ]-
\]
\[
-h[\omega _{\kappa (a}^{(*)d}h_{b)}^{(\kappa }h_d^{\mu )}+h_a^\mu (\bar{\psi}%
_c^\alpha \theta _{\alpha b}^{(*)c}+\psi _{\alpha c}\bar{\theta}%
_b^{(*)\alpha c})+
\]
\begin{equation}
+h_d^\mu (\bar{\psi}_b^\alpha \theta _{\alpha a}^{(*)d}+\psi
_{\alpha b}\theta _a^{(*)\alpha d})-L_\kappa ^o\omega
_{ob}^{(*)c}h_{[c}^\mu h_{a]}^\kappa =0,  \label{sx35}
\end{equation}
where the parantheses () and [] are used to denote symmetrization
and antisymmetrization respectively.

The variation of $\mathcal{L}$ with respect to
$\psi_{\alpha}^{a}$ provides the field equation

\begin{equation}
\partial _\nu \frac{\partial \mathcal{L}}{\partial (\partial _\nu \psi
_\alpha ^a)}+\partial ^\beta \frac{\partial \mathcal{L}}{\partial
(\partial
^\beta \psi _\alpha ^a)}+\partial _\beta \frac{\partial \mathcal{L}}{%
\partial (\partial _\beta \psi _\alpha ^a)}+\partial _o\frac{\partial
\mathcal{L}}{\partial (\partial _o\psi _\alpha ^a)}-\frac{\partial \mathcal{L%
}}{\partial \psi _\alpha ^a}=0,  \label{sx36}
\end{equation}
having the explicit form
\begin{equation}
\frac 12\bar{\psi}_c^\beta S_{\beta a}^{c\alpha }+\frac 12P_{\mu
\alpha }^{ba}h_b^\mu +\tilde{Q}_a^{d\gamma \alpha }\psi _{\gamma
d}=0.  \label{sx37}
\end{equation}

From the variation with respect to $\bar{\psi}^{\alpha a}$
\begin{equation}
\partial _\nu \frac{\partial \mathcal{L}}{\partial (\partial _\nu \bar{\psi}%
^{\alpha a})}+\partial ^\beta \frac{\partial
\mathcal{L}}{\partial (\partial
^\beta \bar{\psi}^{\alpha a})}+\partial _\beta \frac{\partial \mathcal{L}}{%
\partial (\partial _\beta \bar{\psi}^{\alpha a})}+\partial _o\frac{\partial
\mathcal{L}}{\partial (\partial _o\bar{\psi}^{\alpha
a})}-\frac{\partial \mathcal{L}}{\partial \bar{\psi}^{\alpha
a}}=0,  \label{sx38}
\end{equation}
we get the fourth field equation
\begin{equation}
\frac 12\bar{P}_\mu ^{ba\alpha }h_b^\mu +\frac 12\psi _{\beta
b}S_{a\alpha }^{b\beta }+\bar{\psi}_d^\gamma Q_{a\gamma \alpha
}^d=0.  \label{sx39}
\end{equation}

Finally, we write down the other three field equations which are
derived from the variation of $\mathcal{L}$ with respect to the
connection
coefficients $\theta _\alpha ^{(*)ab}$, $\bar{\theta}^{(*)\alpha ab}$ and $%
\omega _{(o)}^{(*)ab}$
\begin{equation}
\partial _\mu (\frac{\partial \mathcal{L}}{\partial (\partial _\mu \Omega
^{(\star )})})+\partial ^\beta (\frac{\partial
\mathcal{L}}{\partial (\partial ^\beta \Omega ^{(\star
)})})+\partial _\beta (\frac{\partial
\mathcal{L}}{\partial (\partial _\beta \Omega ^{(\star )})})+\partial _o(%
\frac{\partial \mathcal{L}}{\partial (\partial _o\Omega ^{(\star
)})})=0 \label{sx40}
\end{equation}
with $\Omega ^{(*)}\in \{\theta _\alpha
^{(*)ab},\bar{\theta}^{(*)\alpha ab},\omega _{(o)}^{(*)ab}\}$,
which have the corresponding detailed forms
\begin{eqnarray}
\partial _\nu (hh_a^\nu \psi _b^\alpha )+\partial _\beta (h\psi _a^{(\beta }%
\bar{\psi}_b^{\alpha )})-\partial ^\beta (h\bar{\psi}_a^\alpha
\psi _{\beta b})+\partial _o(hL_\kappa ^oh_a^\kappa
\bar{\psi}_b^\alpha )- &&
\label{sx41} \\
\,\,-h[\omega _{\mu b}^{(*)d}h_{(d}^\mu \psi _{a)}^\alpha +\bar{\psi}%
_d^{[\gamma }\bar{\psi}_{(a}^{\alpha ]}\theta _{b)\gamma
}^{(*)d}+\theta _b^{(*)\beta d}\bar{\psi}_{(a}^\alpha \psi
_{d)\beta }+L_\kappa ^o\omega
_{ob}^{(*)}h_{(d}^\kappa \bar{\psi}_{a)}^\alpha ] &=&0,  \nonumber \\
\,-h[\omega _{\mu b}^{(*)d}h_{(d}^\mu \psi _{a)}^\alpha +\bar{\psi}%
_d^{[\gamma }\bar{\psi}_{(a}^{\alpha ]}\theta _{b)\gamma
}^{(*)d}+\theta _b^{(*)\beta d}\bar{\psi}_{(a}^\alpha \psi
_{d)\beta }+L_\kappa ^o\omega _{ob}^{(*)}h_{(d}^\kappa
\bar{\psi}_{a)}^\alpha ] &=&0  \nonumber
\end{eqnarray}
\begin{equation}
\left\{
\begin{array}{lll}
\partial _\nu (hh_a^\nu \bar{\psi}_{\alpha b})+ & \partial ^\beta (h\psi
_{a[\beta }\psi _{\alpha ]b})+ & \partial _\beta
(h\bar{\psi}_a^\beta \psi
_{\alpha b})+ \\
&  &  \\
\partial _o(hL_\kappa ^oh_a^\kappa \psi _{\alpha b})- & h[\omega _{b\mu
}^{(*)d}h_{[d}^\mu \bar{\psi}_{a]\alpha }+ & \theta _{\beta a}^{(*)d}\bar{%
\psi}_{[d}^\beta \psi _{b]\alpha }+ \\
&  &  \\
2\theta _a^{(*)\gamma d}\psi _{\gamma [d}\psi _{b]\alpha }+ &
L_\kappa ^o(h_d^\kappa \bar{\psi}_{\alpha a}\omega _{ob}^{(*)d}-
& \bar{\psi}_{\alpha d}h_b^\kappa \omega _{oa}^{(*)d})]=0
\end{array}
\right.   \label{sx42}
\end{equation}
\begin{equation}
\left\{
\begin{array}{lll}
hL_\kappa ^o[h_a^\mu h_d^\kappa \omega _{\mu b}^{(*)d}- & h_d^\mu
h_b^\kappa \omega _{\mu a}^{(*)d}+ & h_a^\kappa
\bar{\psi}_c^\alpha \theta _{\alpha
b}^{(*)c}- \\
&  &  \\
\theta _{\alpha a}^{(*)d}\bar{\psi}_b^\alpha h_d^\kappa + & h_a^\kappa \bar{%
\psi}_{\alpha c}\bar{\theta}_b^{(*)\alpha c}- & \bar{\theta}_a^{(*)\alpha d}%
\bar{\psi}_{\alpha b}h_d^\kappa ]+ \\
&  &  \\
\partial _\nu [hL_\kappa ^oh_b^\nu h_\alpha ^\kappa ]+ & \partial _\beta
[hL_\kappa ^oh_a^\kappa \bar{\psi}_b^\beta ]+ & \partial ^\beta
[hL_\kappa ^oh_a^\kappa \bar{\psi}_{\beta b}]=0
\end{array}
\right.   \label{sx43}
\end{equation}

\chapter[Bianchi Identities and Deformed Bundles]{Bianchi Identities,
Gauge--Higgs Fields and Deformed Bundles} \label{ChapterBianchi}

\section{Introduction}

In the works \cite{sm3},\cite{sbmp1} the concepts of spinor bundle ${S}%
^{(2)}M$ as well as of deformed spinor bundle $\tilde{S}^{(2)}M$
of order two, were intorduced in the framework of a geometrical
generalization of the proper spinor bundles as they have been
studied from different authors e.g.
\cite{carmelli},\cite{wald},\cite{ward}.

The study of fundamental geometrical subjects as well as the
gauge covariant
derivatives, connections field equations e.t.c. in a deformed spinor bundle $%
\tilde{S}^{(2)}M$, has been developed in a sufficiently
generalized approach. \cite{sbmp1} In these spaces the internal
variables or the gauge variables of fibration have been
substituted by the internal (Dirac) variables $\omega =(\xi
,\bar{\xi})$. In addition, another central point of our
consideration is that of the internal fibres $C^4$.

The initial spinor bundle $(S^{(2)}M,\pi ,F),\,\pi
:S^{(2)}M\rightarrow M$
was constructed from the one of the principal fibre bundles with fibre $%
F=C^4 $ ($C^4$ denotes the complex space) and $M$ the base
manifold of space-time events of signature $(+,-,-,-)$. A spinor
in $x\in M$ is an element of the spinor bundle $S^{(2)}M$
\cite{sm3},
\[
(x^{\mu},{\xi}_{\alpha},\bar{\xi}^\alpha) \in S^{(2)}M.
\]

A spinor field is section of $S^{(2)}M$. A generalization of the
spinor bundle $S^{(2)}M$ in an internal deformed system, has been
given in the work \cite{sbmp1} .The form of this bundle
determined as
\[
\tilde{S}^{(2)}M=\tilde{S}^{(2)}M\times R
\]
where $R$ represents the internal on dimension fibre of
deformation. The metrical structure in the deformed spinor bundle
$\tilde{S}^{(2)}M$ has the form:
\begin{eqnarray}
 G & = & g_{\mu\nu}(x,\xi,\bar{\xi})dx^{\mu}\otimes dx^{\nu} +
g_{\alpha\beta}(x,\xi,\bar{\xi},\lambda)
D\bar{\xi}^{\alpha}\otimes D\bar{\xi}^{\ast \beta}+ \nonumber \\
& & +g^{\alpha\beta}(x,\xi,\bar{\xi},\lambda)
D{\xi}_{\alpha}\otimes D{\xi}^{\ast}_{\beta}+
g_{0,0}(x,\xi,\bar{\xi},\lambda) D\lambda \otimes D\lambda
\label{eq1.1}
\end{eqnarray}
where $`` \ast \textquotedblright$ denotes Hermitean conjugation.
The local adapted frame is given by:
\begin{equation}
\frac \delta {\delta {\zeta }^A}=\{\frac \delta {\delta
{x}^\lambda },\frac \delta {\delta {\xi }_\alpha },\frac \delta
{\delta \bar{\xi}^\alpha },\frac
\partial {\partial \lambda }\}  \label{eq1.2}
\end{equation}
and the associated dual frame:
\begin{equation}
\delta {\zeta }^A=\{D{x}^K\equiv dx^K,D{\xi }_\beta
D\bar{\xi}_\beta ,D_{0\lambda }\}  \label{eq1.3}
\end{equation}
where the terms $\frac \delta {\delta {\chi }^\lambda },\frac
\delta {\delta {\xi }_\alpha }D_{0\lambda },D{x}^K,D{\xi }_\beta
,D\bar{\xi}^\beta $ are provided by the relations (6)-(7) of
\cite{sbmp1}.

The considered connection in $\tilde{S}^{(2)}M$ is a
$d$-connection \cite {ma94} having with respect to the adapted
basis the coefficients(cf. \cite {sbmp1} ).
\begin{eqnarray}
{\Gamma}^{A}_{BC} &=& \{ {\Gamma}^{(\ast)\mu}_{\nu\rho},
C^{\mu}_{\nu\alpha}, \bar{C}^{\mu\alpha}_{\nu},
{\Gamma}^{(\ast)\mu}_{\nu 0}, \bar{\Gamma}^{(\ast)\alpha}_{\beta
\lambda},
\tilde{C}^{\beta}_{\alpha\gamma},\tilde{C}^{\beta\alpha}_{\gamma},
\nonumber \\
{} & & \tilde{\Gamma}^{(\ast)\beta}_{\alpha \gamma},
{\Gamma}^{(\ast)\beta}_{\alpha\nu}, C^{\gamma}_{\beta\alpha},
C^{\gamma\alpha}_{\beta}, C^{\alpha}_{\beta 0},
{\Gamma}^{(\ast)0}_{0 \mu}, \bar{C}^{0\alpha}_{)},
C^{0}_{0\alpha}, L^{0}_{00} \}. \label{eq1.4}
\end{eqnarray}

The metric $G$ of relation (\ref{eq1.1}) could be considered as a
definite physical application like the one given by R. Miron and
G. Atanasiu for Lagrange spaces \cite{mat1} for the case of
spinor bundles of order two.
According to our approach on $\tilde{S}^{(2)}M$ the internal variables $\xi $%
, $\bar{\xi}$ play a crucial role similar to the variables
$y^{(1)},y^{(2)}$ of the vector bundles of order two.

The non-linear connection on $\tilde{S}^{(2)}M$ is defined
analogously to the vector bundles at order two (cf. \cite{mat1} )
but in a gauge covariant form:
\begin{equation}
T(\tilde{S}^{(2)}M)=H(\tilde{S}^{(2)}M)\oplus F^{(1)}(\tilde{S}%
^{(2)}M)\oplus F^{(2)}(\tilde{S}^{(2)}M)\oplus R  \label{eq1.5 }
\end{equation}
where $H,F^{(1)},F^{(2)},R$ represent the horizontal vertical
normal and deformation distributions respectively.

In the following we study the Bianchi identities and
Yang-Mills-Higgs fields on $\tilde{S}^{(2)}M$ bundle..

\section{Bianchi Identities}             \index{Bianchi identities}

In order to study the Bianchi Identities (kinematic constraints)
it is necessary to use the Jacobi identities:
\begin{equation}
S_{(XYZ)}[\tilde{D}_X^{(*)},[\tilde{D}_Y^{(*)},\tilde{D}_Z^{(*)}]]=0
\label{eq1.6}
\end{equation}

There are forty-eight Bianchi relations derived from twenty-four
different types of Jacobi identities. Two of these relations are
identically zero. Therefore remain forty-six Bianchi relations.
We will give now some characteristic cases of the Bianchi
identities.

Similarly to our previous work \cite{sbmp1}, the gauge covariant
derivative will take the form
\begin{equation}
\tilde{D}_\mu ^{(*)}=\frac{\tilde{\delta}}{\delta x^\mu }+\frac 12{\omega }%
_\mu ^{(*)ab}J_{ab}  \label{eq1.7}
\end{equation}
here
\[
\frac{\tilde{\delta}}{\delta x ^{\mu}} = \frac{\partial}{\partial
{x}^{\mu}}
- N_{\alpha\mu}\frac{\partial}{\partial {\xi}_{\alpha}} - \bar{N}%
^{\alpha}_{\mu} \frac{\partial}{\partial \bar{\xi}^{\alpha}}- \bar{N}%
^{0}_{\mu} \frac{\partial}{\partial \lambda}
\]
or
\[
\frac{\tilde{\delta}}{\delta x ^{\mu}} = \tilde{A}^{a}_{\mu} P_{a}
\]
with
\begin{eqnarray*}
\tilde{A}_\mu ^a &=&A_\mu ^a-N_\mu ^0L_0^kL_k^a,~P_a=\frac
\partial
{\partial x^a}, \\
A_\mu ^a &=&h_\mu ^a-N_{\alpha \mu }\bar{\psi}^{\alpha
a}-\bar{N}_\mu ^\alpha {\psi }_\alpha ^a.
\end{eqnarray*}

After some calculations we get:
\begin{eqnarray}
 [ \tilde{D}^{(\ast)}_{\mu},
 [\tilde{D}^{(\ast)}_{\kappa}, \tilde{D}^{(\ast)}_{\lambda} ] ] &=&
  (\frac{\delta \tilde{R}^{a}_{\kappa\lambda}}{\partial x^{\mu}} +
\tilde{R}^{a}_{b\kappa\lambda} A^{b}_{\mu} + {\omega}^{(\ast)a
b}_{\mu c} \tilde{R}^{c}_{\kappa\lambda} ) P_{a} \nonumber \\
{ }& &  +(\frac{1}{2}\frac{\delta \tilde{R}^{c
e}_{\kappa\lambda}}{\partial x^{\mu}} + {\omega}^{(\ast)c
d}_{\mu} \tilde{R}^{e}_{d\kappa\lambda}) J_{c e} \label{eq1.8}
\end{eqnarray}
and ${\omega }_\mu ^{(*)ab}$ represent the Lorentz-spin connection
coefficients. We define also:
\begin{eqnarray}
\tilde{D}_\mu \tilde{R}_{\kappa \lambda }^{ce} &=&\frac 12\frac{\tilde{\delta%
}\tilde{R}_{\kappa \lambda }^{ce}}{\partial x^\mu }+{\omega }_\mu ^{(*)cd}%
\tilde{R}_{d\kappa \lambda }^e  \label{eq1.9} \\
\tilde{D}_\mu ^e\tilde{R}_{\kappa \lambda }^a &=&\frac{\tilde{\delta}\tilde{R%
}_{\kappa \lambda }^\alpha }{\delta x^\mu }+\tilde{R}_{b\kappa
\lambda }^\alpha \tilde{A}_\mu ^b+{\omega }_{\mu
c}^{(*)ab}\tilde{R}_{\kappa \lambda }^c  \label{eq1.10}
\end{eqnarray}

By cyclic permutation of the independent generators $J_{ce},P_a$
we get the following Bianchi identities,

\begin{equation}
\tilde{D}_\mu \tilde{R}_{\kappa \lambda }^a+\tilde{D}_\kappa \tilde{R}%
_{\lambda \mu }^a+\tilde{D}_\lambda \tilde{R}_{\mu \kappa }^a=0
\label{eq1.11}
\end{equation}

\begin{equation}
\tilde{D}_\mu \tilde{R}_{\kappa \lambda }^{ce}+\tilde{D}_\kappa \tilde{R}%
_{\lambda \mu }^{ce}+\tilde{D}_\lambda \tilde{R}_{\mu \kappa
}^{ce}=0 \label{eq1.12}
\end{equation}

Using the Jacobi identities $Q_(\alpha,\beta,\gamma) [ \tilde{D}%
^{(\ast)}_{\alpha}, [ \tilde{D}^{(\ast)}_{\beta}, \tilde{D}%
^{(\ast)}_{\gamma} ] ] = 0 $ the Bianchi identities with respect
to spinor quantities produce the relations,

\begin{eqnarray}
\tilde{D}_\alpha {Q}_{\beta \gamma }^{ab}+\tilde{D}_\beta
{Q}_{\gamma \alpha
}^{ab}+\tilde{D}_\gamma {Q}_{\alpha \beta }^{ab} &=&0,  \label{eq1.13} \\
\tilde{D}_\alpha {Q}_{\beta \gamma }^a+\tilde{D}_\beta
{Q}_{\gamma \alpha }^a+\tilde{D}_\gamma {Q}_{\alpha \beta }^a
&=&0.  \label{eq1.14}
\end{eqnarray}

The new Jacobi identity, due to $\lambda$, has the form

\begin{equation}
\lbrack {D}_0^{(*)},[{D}_0^{(*)},{D}_0^{(*)}]]=0  \label{eq1.15}
\end{equation}
which yields us no Bianchi identity.

Bianchi identities of mixed type give us the kinematic constraint
which encompass space-time, spinors and deformed gauge covariant
derivatives. In that case from the Jacobi identities
\[
Q_{\mu \alpha 0} [ \tilde{D}^{(\ast)}_{\mu}, [
\tilde{D}^{(\ast)}_{\alpha}, \tilde{D}^{(\ast)}_{0} ] ] = 0
\]
we get the relations
\begin{eqnarray}
 & & [ \tilde{D}^{(\ast)}_{\mu}, [
\tilde{D}^{(\ast)}_{\alpha}, \tilde{D}^{(\ast)}_{0} ] ] = (
\frac{\tilde{\delta} \tilde{P}^{d}_{0\alpha}}{\delta x^{\mu}} +
\tilde{P}^{d}_{c 0\alpha} A^{e}_{\mu} + {\omega}^{(\ast)d}_{\mu a}
\tilde{P}^{a}_{0 \alpha} ) P_{d} \nonumber \\
& & +(\frac{1}{2}\frac{\tilde{\delta} \tilde{P}^{c d}_{0
\alpha}}{\delta x^{\mu}} + {\omega}^{(\ast)c}_{\mu\alpha}
\tilde{P}^{a d}_{0\alpha}) J_{c d}   \label{eq1.16}
\\
& & [ \tilde{D}^{(\ast)}_{\alpha}, [ \tilde{D}^{(\ast)}_{0},
\tilde{D}^{(\ast)}_{\mu} ] ]  =  ( \frac{\partial
\tilde{P}^{d}_{\mu 0}}{\partial \bar{\xi}^{\alpha}} +
\tilde{P}^{d}_{c \mu 0} A^{c}_{\alpha} + \ast
{\omega}^{(\ast)d}_{\alpha a} \tilde{P}^{a}_{\mu 0} ) P_{d}
 \nonumber \\
 & & +(\frac{1}{2}\frac{\partial \tilde{P}^{c d}_{\mu 0}}{\partial
\bar{\xi}^{\alpha}} + {\omega}^{(\ast)c}_{\alpha a} \tilde{P}^{a
d}_{\mu 0}) J_{c d}   \label{eq1.17}
\\
 & & [ \tilde{D}^{(\ast)}_{0}, [ \tilde{D}^{(\ast)}_{\mu},
\tilde{D}^{(\ast)}_{\alpha} ] ]  =  ( \frac{\partial
\tilde{P}^{d}_{\alpha \mu}}{\partial \lambda} + \tilde{P}^{d}_{c
\alpha \mu} A^{c}_{0} + {\omega}^{(\ast)d}_{0 a}
\tilde{P}^{a}_{\alpha \mu} ) P_{d}
 \nonumber \\
 & & +(\frac{1}{2}\frac{\partial \tilde{P}^{c d}_{\alpha
\mu}}{\partial \lambda} + {\omega}^{(\ast)c}_{0 a} \tilde{P}^{a
d}_{\alpha \mu}) J_{c d}
  \label{eq1.18}
\end{eqnarray}
where,
\begin{eqnarray*}
\tilde{D}_\mu ^{(*)} &=&\frac{\tilde{\delta}}{\delta {x}^\mu }+\frac 12{%
\omega }_\mu ^{(*)ab}J_{ab},~\tilde{D}_\alpha ^{(*)}=\frac
\partial
{\partial \bar{\xi}^\alpha }+\frac 12{\Theta }_\alpha ^{(*)ab}J_{ab}, \\
\tilde{D}_0^{(*)} &=&\frac \partial {\partial \lambda }+{\omega }%
_0^{ab}J_{ab},\frac \partial {\partial \lambda }=L_0^\mu h_\mu
^aP_a,\frac
\partial {\partial \bar{\xi}^\alpha }={\psi }_\alpha ^aP_a
\end{eqnarray*}

Now we put,
\begin{eqnarray}
\tilde{D}_\mu \tilde{P}_{0\alpha }^d &=&\frac{\tilde{\delta}\tilde{P}%
_{0\alpha }^d}{\delta x^\mu }+\tilde{P}_{c\mu 0}^dA_\mu
^c+{\omega }_{\alpha
a}^{(*)d}\tilde{P}_{\mu 0}^a,  \label{eq1.19} \\
\tilde{D}_\alpha \tilde{P}_{\mu 0}^d &=&\frac{\partial \tilde{P}_{\mu 0}^d}{%
\partial \bar{\xi}^\alpha }+\tilde{P}_{c\mu 0}^dA_\mu ^c+{\omega }_{\alpha
a}^{(*)d}\tilde{P}_{\mu 0}^a,  \label{eq1.20} \\
\tilde{D}_0\tilde{P}_{\alpha \mu }^d &=&\frac{\partial
\tilde{P}_{\alpha \mu
}^d}{\partial \lambda }+\tilde{P}_{c\alpha \mu }^dA_0^c+{\omega }_{0a}^{(*)d}%
\tilde{P}_{\alpha \mu }^a.  \label{eq1.21}
\end{eqnarray}
in virtue of (\ref{eq1.14}), (\ref{eq1.15}) and (\ref{eq1.16}) we
get the identity
\begin{equation}
\tilde{D}_\mu \tilde{P}_{0\alpha }^d+\tilde{D}_\alpha \tilde{P}_{\mu 0}^d+%
\tilde{D}_0\tilde{P}_{\alpha \mu }^d=0  \label{eq1.22}
\end{equation}

Similarly we define
\begin{eqnarray}
\tilde{D}_\mu \tilde{P}_{0\alpha }^{cd} &=&\frac 12\frac{\partial \tilde{P}%
_{0\alpha }^{cd}}{\partial x^\mu }+{\omega }_{\mu a}^{(*)c}\tilde{P}%
_{0\alpha }^{ad},  \label{eq1.23} \\
\tilde{D}_\alpha \tilde{P}_{\mu 0}^{cd} &=&\frac 12\frac{\partial \tilde{P}%
_{\mu 0}^{cd}}{\partial \bar{\xi}^\alpha }+{\omega }_{\alpha a}^{(*)c}\tilde{%
P}_{\mu 0}^{ad},  \label{eq1.24} \\
\tilde{D}_0\tilde{P}_{\alpha \mu }^{cd} &=&\frac 12\frac{\partial \tilde{P}%
_{\alpha \mu }^{cd}}{\partial \lambda }+{\omega }_{0\alpha }^{(*)c}\tilde{P}%
_{\alpha \mu }^{ad}.  \label{eq1.25}
\end{eqnarray}

From (\ref{eq1.18})--(\ref{eq1.20}) we get
\begin{equation}
\tilde{D}_\mu \tilde{P}_{0\alpha }^{cd}+\tilde{D}_\alpha
\tilde{P}_{\mu 0}^{cd}+\tilde{D}_0\tilde{P}_{\alpha \mu }^{cd}=0
\label{eq1.26}
\end{equation}


\section{Yang-Mills-Higgs equations.}   \index{Higgs}

The study of Yang-Mills-Higgs equations within the framework of
the geometrical structure of $\tilde{S}^{(2)}(M)$-bundle that
contains the one-dimensional fibre as an internal deformed system
can characterize the Higgs field which is studied in the
elementary particle physics. In our description we are allowed to
choose a scalar from the internal deformed fibre of the spinor
bundle $\tilde{S}^{(2)}(M)$. Its contribution to the Lagrangian
density provides us with the generated Yang-Mills-Higgs equations.

In the following we define a gauge potential $\tilde{A}=(A_{\mu}, A_{\alpha},%
\bar{A}^{\alpha},\varphi)$ with space-time and spinor components, $%
\varphi:R\longrightarrow g$ which takes its values in a Lie
Algebra $g$.
\begin{eqnarray*}
&& \tilde{A}:\tilde{S}(M)\longrightarrow g \\
&&\tilde{A}_{X} = A^{i}_{X}{\tau}_{i}, [{\tau}_{i},{\tau}_{j}] = C^{k}_{ij}{%
\tau}_{k} \\
&& \tilde{A}_{X} = \{ A_{\mu}, A_{\alpha},\bar{A}^{\alpha},\varphi
\} \end{eqnarray*} where the elements ${\tau}_{i}$ are the
components which satisfy the commutation relations of the Lie
algebra. Then $\tilde{A}$ is called a {\ g-valued spinor gauge
potential}.

We can define the gauge covariant derivatives:
\begin{eqnarray}
 \hat{D}_{\mu} & = & \tilde{D}_{\mu}+ i A_{\mu} \nonumber \\
\hat{D}_{\alpha} &=& \tilde{D}_{\alpha}+ i A_{\alpha} \nonumber \\
\hat{D}^{\alpha} = \tilde{\bar{D}}^{\alpha}+ i A_{\mu}
 \label{eq2.1 }
\end{eqnarray}

In virtue of the preceding relations we get the following theorem:
\begin{theorem}
The commutators of gauge covariant derivatives on a
\\ $\tilde{S}^{(2)}M$ deformed bundle are given by the relations:
\begin{eqnarray}
 &a)\quad & [\hat{D}_{\mu},\hat{D}_{\nu}] = [
\tilde{D}_{\mu},\tilde{D}_{\nu}] + i \tilde{F}_{\mu \nu} \nonumber \\
& b)\quad & [\hat{D}_{\mu},\hat{\bar{D}}^{\alpha}] = [
\tilde{D}_{\mu},\tilde{\bar{D}}^{\alpha}] + i
\tilde{F}^{\alpha}_{\mu}
 \nonumber \\
& c)\quad & [\hat{D}_{\alpha},\hat{\bar{D}}^{\beta}] = [
\tilde{D}_{\alpha},\tilde{\bar{D}}^{\beta}] + i
\tilde{F}^{\beta}_{\alpha} \nonumber \\
& d)\quad & [\hat{D}_{\alpha},\hat{\bar{D}}_{\beta}] = [
\tilde{D}_{\alpha},\tilde{\bar{D}}_{\beta}] + i \tilde{F}_{\alpha
\beta}
 \nonumber \\
&e)\quad & [\hat{D}_{\mu},\hat{\bar{D}}_{\alpha}] = [
\tilde{D}_{\mu},\tilde{\bar{D}}_{\alpha}] + i \tilde{F}_{\mu
\alpha}
\nonumber \\
& f)\quad & [\hat{\bar{D}}^{\alpha},\hat{\bar{D}}^{\beta}] = [
\tilde{\bar{D}}^{\alpha},\tilde{\bar{D}}^{\beta}] + i
\tilde{\bar{F}}^{\alpha \beta}  \label{eq2.2}
\end{eqnarray}
The curvature two-forms
$\tilde{F}_{XY},\tilde{F}^{XY},F_Y^X,X,Y=\{\alpha ,\beta ,\mu
,\nu \}$ are the g-valued field strengths on $\tilde{S}^{(2)}M$
and they have the following form:
\begin{eqnarray}
 \tilde{F}_{\mu\nu} &=& \tilde{D}_{\mu}A_{\nu}-
\tilde{D}_{\nu}A_{\mu} + i [ A_{\mu},A_{\nu} ] \nonumber \\
\tilde{F}_{\mu\alpha} &=& \tilde{D}_{\mu}A_{\alpha}-
\tilde{D}_{\alpha}A_{\mu} + i [ A_{\alpha},A_{\mu} ]\nonumber  \\
\tilde{\bar{F}}^{\beta}_{\alpha} &=&
\tilde{D}_{\alpha}\bar{A}^{\beta}-
\tilde{\bar{D}}^{\beta}\bar{A}_{\alpha} + i [
A_{\alpha},\bar{A}^{\beta} ] \nonumber  \\
\tilde{\bar{F}}^{\alpha}_{\mu} &= &
\tilde{D}_{\mu}\bar{A}^{\alpha}-
\tilde{\bar{D}}_{\alpha}{A}_{\mu} + i [ A_{\mu},\bar{A}^{\alpha}
] \nonumber  \\
\tilde{F}_{\alpha\beta} &= & \tilde{D}_{\alpha}A_{\beta}-
\tilde{D}_{\beta}A_{\alpha} + i [ A_{\alpha},A_{\beta} ]
 \nonumber \\
\tilde{\bar{F}}^{\alpha\beta} &= &
\tilde{\bar{D}}^{\alpha}\bar{A}^{\beta}-
\tilde{\bar{D}}^{\beta}\bar{A}_{\alpha} + i [
\bar{A}^{\alpha},\bar{A}^{\beta} ]   \label{eq2.3}
\end{eqnarray}
\end{theorem}

The appropriate Lagrangian density of Yang-Mills(Higgs) can be
written in the form
\begin{eqnarray}
 \tilde{L} &=& tr(\tilde{F}_{\mu \nu} \tilde{F}^{\mu \nu} +
tr(\tilde{F}_{\mu \alpha} \tilde{\bar{F}}^{\mu \alpha} +
tr(\tilde{F}_{\alpha \beta} \tilde{\bar{F}}^{\alpha \beta} +
tr(\tilde{\bar{F}}^{\beta}_{\alpha}
\tilde{\bar{F}}^{\alpha}_{\beta} \nonumber \\
& & + \frac{1}{2}{m}^2{\varphi}^2 - \frac{1}{2}
tr[(\hat{D}_{\mu}\varphi)(\hat{D}^{\mu}\varphi)] - \frac{1}{2}
tr[(\hat{D}_{\alpha}\varphi)(\hat{D}^{\mu}\varphi)]
 \label{eq2.4}
\end{eqnarray}

In our case the Yang-Mills(Higgs) generalized action can be
written in the form
\begin{equation}
\tilde{\mathcal{I}}_{YM(H)}=\int \tilde{\mathcal{L}}d^4xd^4\xi d^4\bar{\xi}d{%
\lambda }  \label{eq2.5}
\end{equation}

From the Euler-Lagrange equations
\begin{equation}
\frac{\delta \tilde{L}}{\delta A_Y}=\tilde{D}_X(\frac{\partial \tilde{L}}{%
\partial (\tilde{D}_XA_Y)})-\frac{\partial \tilde{L}}{\partial A_Y}=0
\label{eq2.6}
\end{equation}
the variation of $\tilde{L}$ with respect to $A_{\lambda}$ is
\begin{equation}
\tilde{D}_k(\frac{\partial \tilde{L}}{\partial (\tilde{D}_kA_\lambda )})+%
\tilde{D}_\beta (\frac{\partial \tilde{L}}{\partial
(\tilde{D}_\beta
A_\lambda )})+\bar{D}^\beta (\frac{\partial \tilde{L}}{\partial (\bar{D}%
_\beta A_\lambda )})-\frac{\partial \tilde{L}}{\partial A_\lambda
}=0 \label{eq2.7}
\end{equation}
and it will give us after some straightforward calculations the
equation:
\begin{equation}
\hat{D}_k\tilde{F}^{k\lambda }+\hat{D}_\beta \bar{F}^{\lambda \beta }+\bar{D}%
^\beta \tilde{F}_\beta ^\lambda +[\varphi ,\hat{D}^\lambda
\varphi ]=0 \label{eq2.8}
\end{equation}

Similarly from the variation of $\tilde{L}$ with respect to $A_\alpha $ and $%
\tilde{A}^\beta $ we associate the equations:
\begin{eqnarray}
\hat{D}_k\bar{F}^{k\gamma }+\hat{D}_\delta \bar{F}_\delta ^\gamma +\bar{D}%
^\delta \bar{F}_\delta ^\gamma +[\varphi ,\hat{D}^\gamma \varphi
] &=&0,
\label{eq2.9} \\
\hat{D}_k\tilde{F}_\gamma ^k+\hat{D}_\delta \tilde{F}_\gamma ^\delta +\bar{D}%
^\delta \tilde{F}_{\delta \gamma }+[\varphi ,\bar{D}_\gamma
\varphi ] &=&0. \label{eq2.10}
\end{eqnarray}

So we can state the following theorem:
\begin{theorem}
The Yang-Mills-Higgs equations of $\tilde{S}^{(2)}M$-bundle are
given by the relations (\ref{eq2.4})--(\ref{eq2.5}).
\end{theorem}

\section{Field Equations of an Internal Deformed System}
\index{deformed system}

 Considering on the deformed spinor bundle $S^{(2)}M\times R$ a
nonlinear connection and a gauge d-connection, the authors obtain
the equivalence principle and the explicit expressions of the
field equations corresponding to a Utiyama gauge invariant
Lagrangian density produced by the corresponding scalars of
curvature; these results extend the corresponding ones for
$S^{(2)}M$. \vskip 0.2cm

The concept of a spinor bundle $S^{(2)}M$ and its relation to the Poincar$%
\grave{e}$ group were introduced in \cite{st1,sm3}. This group,
consisting of the set of rotations, boosts and translations,
gives an exact meaning to the terms: ``momentum'', ``energy'',
``mass'', and ``spin'' and is used to determine characteristics
of the elementary particles; also, it is a gauge, acting locally
in the space-time. Hence we may perform Poincar$\grave{e}$
transformations for a physical approach. In \cite{st1}, the
metric tensor of the base manifold $(M,g_{\mu \nu }(x,\xi
,\bar{\xi}))$, depends on the
position coordinates and on the spinor (Dirac) variables $(\xi _\alpha ,\bar{%
\xi ^\alpha })\in C^4\times C^4$. A spinor bundle $S^{(1)}(M)$
can be constructed from one of the principal fiber bundles with
fiber $F=C^4$. Each fiber is diffeomorphic with one proper
Lorentz group. In this study we apply for the space
$S^{(2)}M\times R$ an analogous method as in the theory of
deformed bundles developed in [6], for the case of a \textbf{spinor bundle} $%
S^{(2)}M=M\times C^{4\cdot 2}$ in connection with a deformed \textbf{%
internal fibre} $R.$ The consideration of Miron $d$ - connections \cite{ma94}%
, which preserve the $h-$ and $v-$distributions is essential in
our approach, as in the previous work: this standpoint enables
using a more general group $G^{(2)}$, called the structural group
of all rotations and translations, that is isomorphic to the
Poincar$\grave{e}$ Lie algebra. A \textbf{spinor} is an element
of the spinor bundle $S^{(2)}(M)\times R$ where $R$ represents
the \textbf{internal fibre of deformation.} The local variables
are in this case
\[
(x^\mu ,\xi _\alpha ,\bar{\xi}^\alpha ,\lambda )\in S^{(2)}(M)\times R=%
\tilde{S}^{(2)}(M),\lambda \in R.
\]
The non-linear connection on $\tilde{S}^{(2)}(M)$ is defined
analogously, as for the vector bundles of order two
\cite{mat1,muat}
\[
T(\tilde{S}^{(2)}{M})=H(\tilde{S}^2M)\oplus \mathcal{F}^{(1)}(\tilde{S}^{(2)}%
{M})\oplus \mathcal{F}^{(2)}(\tilde{S}^{(2)}{M})\oplus R,
\]
where $\mathcal{H}$,$\mathcal{F}^{(1)}$,$\mathcal{F}^{(2)}$,$R$
represent the horizontal, vertical, normal and deformation
distributions respectively.

We introduce the fundamental gauge $1$-form fields which take
values from the Lie algebra of the Poincar$\grave{e}$ group and
denote by $J_{ab},P_a$ the generators of the four-dimensional
Poincar$\grave{e}$ group (namely the angular momentum and linear
momentum), by $\omega _\mu ^{(*)ab}$ - the Lorentz - spin
connection coefficients, $\bar{\Psi}^{\alpha a},$ $\Psi _\alpha
^a$, $\theta _\alpha ^{(*)ab}$, $\bar{\theta}^{(*)\alpha ab}$ -
the spin-tetrad and spin - connection coefficients, and $L_o^a$ -
the deformed tetrad coefficients. We use Greek letters $\lambda
,\mu ,$ $\nu ,\ldots $
for space-time indices, $\alpha ,\beta ,\gamma ,\ldots $ for the spinor, $%
a,b,c,\ldots $ for Lorentz ones, and the index $(o)$ represents
the deformed variable; $\lambda ,\alpha ,$ $a=1,\ldots ,4.$ The
general transformations of coordinates on $\tilde{S}^{(2)}{M}$ are
\begin{equation}
{x^{\prime }}^\mu ={x^{\prime }}^\mu (x^\nu ),{\xi }_\alpha ^{\prime }={\xi }%
_\alpha ^{\prime }(\xi _\beta ,\bar{\xi}^\beta ),\bar{{\xi }^{\prime }}%
^\alpha =\bar{{\xi }^{\prime }}^\alpha (\bar{\xi}^\beta ,\xi _\beta ),{%
\lambda }^{\prime }=\lambda  \nonumber 
\end{equation}

Like in \cite{st1,sm3} we define the following gauge covariant
derivatives, including the new derivative corresponding to the
deformation-parameter.

The space-time, Lorentz, spin frames and the deformed frame are
shown to be connected by a set of the relations which generalize
the well - known principle of equivalence in the total space of
the spinor bundle $S^{(2)}M$.

The deformed spinor bundle $\tilde S^{(2)}M$ is endowed with a
metrical structure.The considered connection in
$\tilde{S}^{(2)}(M)$ is a
d-connection; it preserves the distributions $\mathcal{H}$,$\mathcal{F}%
^{(1)} $,$\mathcal{F}^{(2)}$,$R$, and is assumed to be metrical.

The covariant differentiation of tensors, spin-tensors and
Lorentz - type tensors of arbitary rank is defined as in
\cite{st1,sm3}; also, are present the supplementary derivation
laws relative to the deformation component.

From the anticommutation relations of the adapted basis, we
obtain the curvatures and torsions of the space $\tilde S^{(2)}M$
\[
R_{\mu \nu }^a,R_{\mu \nu }^{ab},P_{\mu \alpha }^{ab},P_{\mu
\alpha }^a
\]
and, similarly to \cite{sm3}, other four curvatures and torsions.
The contribution of the $\lambda $ - covariant derivative
$\mathcal{D}_o^{(*)}$ provides us the following curvatures and
torsions
\[
R_{o \mu}^{a},R_{o \mu}^{ab},R_{oo}^{ab} = 0,R_{oo}^{a} = 0,
\bar{P}_{o}^{a \alpha},\bar{P}_{o}^{ab \alpha},P_{o \alpha}^{a},
P_{o \alpha}^{ab}.
\]

In the following, are derived the field equations, by means of
the Palatini method, using a Lagrangian of the form
\begin{equation}
\mathcal{L}=h(R+P+\bar{P}+Q+\tilde{Q}+R_o+\bar{P}_o+P{o})
 \nonumber 
\end{equation}
which depends on the tetrads and on the connection coefficients,
\[
\mathcal{L}(\kappa ^A,\delta _M\kappa ^A)=\mathcal{L}(h,\omega ^{(*)},\psi ,%
\bar{\psi},\theta ^{(*)},\bar{\theta}^{(*)},\omega _o^{(*)})
\]
where,
\[
\kappa ^A\in \{(h_\mu ^a(z),\omega _\mu ^{(*)ab}(z),\psi _\alpha ^a(z),\bar{%
\psi}^{\alpha a}(z),\theta _\alpha
^{(*)ab}(z),\bar{\theta}^{(*)\alpha ab}(z),\omega
_o^{(*)ab}(z))\},
\]
\begin{equation}
\delta _M=\frac \delta {\delta z^M}\in \left\{ \frac \delta
{\delta x^\mu
},\frac \delta {\delta \xi _\alpha },\frac \partial {\partial \bar{\xi}%
^\alpha },\frac \partial {\partial \lambda }\right\} ,\qquad
z=(z^M)=(x^\mu ,\xi _\alpha ,\bar{\xi}^\alpha ,\lambda ) \nonumber
\end{equation}
and
\[
\left\{
\begin{array}{lll}
R=h_a^\mu h_c^\kappa R_{\mu \kappa }^c, & P=h_{}^\mu \psi
_c^\alpha P_{\mu \alpha }^c, & \bar{P}=h_{}^\mu
\bar{\psi}_{\alpha c}\bar{P}_\mu ^{c\alpha },
\\
&  &  \\
Q=Q_{\alpha \beta }^{ab}\bar{\psi}_a^\alpha \bar{\psi}_b^\beta , & \tilde{Q}=%
\tilde{Q}^{ab\alpha \beta }=\psi _{\alpha a}\psi _{\beta b}, & S=\bar{\psi}%
_a^\alpha \psi _{\beta b}S_\alpha ^{ab\beta }, \\
&  &  \\
R_o=L_\kappa ^oh_c^\mu h_a^\kappa R_{o\mu }^{ac}, &
\bar{P}_o=L_\kappa ^oh_a^\kappa \bar{\psi}_{\alpha
c}\bar{P}_o^{ac\alpha }, & P_o=L_\kappa
^oh_a^\kappa \bar{\psi}_c^\alpha P_{o\alpha }^{ac}. \\
&  &
\end{array}
\right.
\]

The Euler-Lagrange equations are generally given by
\begin{equation}
\frac{\delta\mathcal{L}}{\delta {\kappa}^{(A)}} = \partial_M \left ( \frac{%
\partial\mathcal{L}}{\partial(\partial_{M} \kappa^{(A)})} \right ) - \frac{%
\partial \mathcal{L}} {\partial \kappa^{(A)}}= 0,
 \nonumber 
\end{equation}
with $\partial_{M} = \frac{\partial}{\partial z^{M}} \in \left \{ \frac{%
\partial}{\partial x^{\mu}}, \frac{\partial}{\delta \xi_{\alpha}},\frac{%
\partial}{\partial \bar{\xi}^{\alpha}}, \frac{\partial}{\partial\lambda}
\right\}.$

The variation of $\mathcal{L}$ with respect to the tetrads
$h_b^\nu $ gives us the first field equation
\begin{equation}
\tilde{H}_\nu ^b-\frac 12h_\nu ^b\tilde{H}=0,
 \nonumber  \label{asx33}
\end{equation}
where we put,
\begin{eqnarray*}
\tilde{H} &=&R+P+\bar{P}+Q+\tilde{Q}+S+R_o+\bar{P}_o+P_o, \\
\tilde{H}_\nu ^b &=&2R_\nu ^b+P_\nu ^b+\bar{P}_\nu ^b+R_{o\nu }^b+\bar{P}%
_{o\nu }^b+P_{(o)\nu }^b, \\
R_\nu ^b &=&h_c^\kappa R_{\nu \kappa }^{bc},\qquad P_\nu ^b=\bar{\psi}%
_c^\alpha P_{\nu \alpha }^{bc},\qquad R_{o\nu }^b=L_\kappa
^oh_{}^\kappa
R_{o\nu }^b+L_\nu ^oh_c^\mu R_{o\mu }^{bc}, \\
\bar{P}_\nu ^b &=&\psi _{\alpha c}\bar{P}_\nu ^{bc\alpha },\qquad \bar{P}%
_{o\nu }^b=L_\nu ^o\bar{\psi}_{\alpha c}\bar{P}_o^{bc\alpha
},\qquad P_{o\nu }^b=L_\nu ^o\bar{\psi}_c^\alpha P_{o\alpha
}^{bc}.
\end{eqnarray*}
The equation (\ref{asx33}) is the \textbf{Einstein equation for
empty space}, in the framework of our consideration. Also, the
variation of $\mathcal{L}$ with respect to $\omega _\mu ^{(*)ab}$
gives the second field equation
\begin{eqnarray}
\partial _\nu [h(h_a^\mu h_b^\nu -h_b^\mu h_a^\nu )]-\partial ^\alpha
[hh_a^\mu \psi _{\alpha b}]-\partial _\alpha [hh_a^\mu
\bar{\psi}_b^\alpha
]-\partial _o[hL_a^oh_b^\mu ]- &&  \nonumber \\
-h[\omega _{\kappa (a}^{(*)d}h_{b)}^{(\kappa }h_d^{\mu )}+h_a^\mu (\bar{\psi}%
_c^\alpha \theta _{\alpha b}^{(*)c}+\psi _{\alpha c}\bar{\theta}%
_b^{(*)\alpha c})+ &&  \nonumber \\
+h_d^\mu (\bar{\psi}_b^\alpha \theta _{\alpha a}^{(*)d}+\psi
_{\alpha b}\theta _a^{(*)\alpha d})-L_\kappa ^o\omega
_{ob}^{(*)c}h_{[c}^\mu h_{a]}^\kappa &=&0,
 \nonumber 
\end{eqnarray}
where the parantheses (... ) and [... ] are used to denote
symmetrization and antisymmetrization respectively.

The variation of $\mathcal{L}$ with respect to
$\psi_{\alpha}^{a}$ provides the field equation
\begin{equation}
\frac{1}{2} \bar{\psi}_{c}^{\beta} S_{\beta a}^{c \alpha} +
\frac{1}{2} P_{\mu \alpha}^{b a} h_{b}^{\mu} + \tilde{Q}_{a}^{d
\gamma \alpha}
\psi_{\gamma d} = 0.  \nonumber 
\end{equation}

From the variation with respect to $\bar{\psi}^{\alpha a}$ we get
the fourth field equation
\begin{equation}
\frac{1}{2} \bar{P}_{\mu}^{ba \alpha} h_{b}^{\mu} + \frac{1}{2}
\psi_{\beta b} S_{a \alpha}^{b \beta} + \bar{\psi}_{d}^{\gamma}
Q_{a \gamma \alpha}^{d}
= 0.  \nonumber 
\end{equation}

Finally, are obtained the explicit expressions of the other three
field equations, by means of the variation of $\mathcal{L}$ with
respect to the connection coefficients $\theta _\alpha ^{(*)ab}$,
$\bar{\theta}^{(*)\alpha ab}$ and $\omega _{(o)}^{(*)ab}$.

\chapter[Tensor and Spinor Equivalence]{Tensor and Spinor Equivalence on
Generalized Metric Tangent Bundles}

\section{Introduction}

The theory of spinors on pseudo-Riemannian spaces has been
recognized by many authors, e.g. \cite{penr1,carmelli,ward} for
the important role it has played from the mathematical and
physical point of view.

The spinors that we are dealing with here, are associated with the group $%
SL(2,C)$. In particular $SL(2,C)$ acts on $C^2$. Each elements of
$C^2$ represents a two-component spinor. This group is the
covering group of the Lorentz group in which the tensors are
described \cite{carmelli}. The correspondence between spinors and
tensors is achieved by means of mixed quantities initially
introduced by Infeld and Van der Waerden.

The correspondence of tensors and spinors establishes a
homomoerhism between the Lorentz group and the covering group
$SL(2,C)$.

In the following, we give some important relations between
spinors and tensors on a general manifold of metric $g_{\mu \nu}$.

Let ${\sigma }:S\otimes {\bar{S}}\to V^4$ be a homomorphism
between spinor spaces $S,{\bar{S}}$ and four-vectors belonging to
the $V^4$ space, then the
components of ${\sigma }$, which are called the \emph{Pauli-spinor matrices}%
, are given by
\begin{eqnarray}
{\sigma }_{AB^{\prime }}^0 &=&\frac 1{\sqrt{2}}\left(
\begin{array}{cc}
1 & 0 \\
0 & 1
\end{array}
\right) ,\qquad {\sigma }_{AB^{\prime }}^1=\frac 1{\sqrt{2}}\left(
\begin{array}{cc}
0 & 1 \\
1 & 0
\end{array}
\right) ,  \label{15.1} \\
{\sigma }_{AB^{\prime }}^2 &=&\frac 1{\sqrt{2}}\left(
\begin{array}{cc}
0 & i \\
-i & 0
\end{array}
\right) ,\qquad {\sigma }_{AB^{\prime }}^3=\frac 1{\sqrt{2}}\left(
\begin{array}{cc}
& 0 \\
0 & -1
\end{array}
\right)  \nonumber
\end{eqnarray}
The hermitian spinorial equivalent notation of ${\sigma
}_{AB^{\prime }}^\mu $ is given by ${\sigma }_{AB^{\prime }}^\mu
={\bar{\sigma}}_{BA^{\prime }}^\mu ={\bar{\sigma}}_{B^{\prime
}A}^\mu $. Greek letters $\mu ,\nu ,\cdots $ represent the usual
space-time indices taking the values $0,1,2,3$ and the Roman
capital indices $A,B,A^{\prime },B^{\prime }$ are the spinor
indices taking the values $0,1$. The tensor indices are raised
and lowered by means of the metric tensor, whereas the raising
and lowering of spinor indices is
given by the \emph{spinor metric tensors} ${\varepsilon }_{AC}$, ${%
\varepsilon }_{B^{\prime }C^{\prime }}$ which are of
skew-symmetric form. Thus, for two spinors $\xi ^A$,
$n^{A^{\prime }}$ we have the relations, moreover we have,
\[
\xi ^An_A=\xi ^An^B{\varepsilon }_{BA}=-\xi ^A{\varepsilon
}_{AB}n^B=-\xi _Bn^B.
\]
For a real vector $V_\mu $ its spinor equivalent it
\begin{equation}
V_{AB^{\prime }}=V_\mu {\sigma }_{AB^{\prime }}^\mu , \label{15.3}
\end{equation}
where ${\sigma }_{AB^{\prime }}^\mu $ are given by the relation (\ref{15.1}%
). Also, the following formulas are satisfied,
\[
{\sigma }_{AB^{\prime }}^\mu {\sigma }^{\nu AB^{\prime }}=g^{\mu
\nu },\quad {\sigma }_{AB^{\prime }}^\mu {\sigma }_\nu
^{AB^{\prime }}=\delta _\nu ^\mu .
\]
The spinor equivalent of a tensor $T_{\mu \nu }$ is given by
\[
T_{\mu \nu }={\sigma }_\mu ^{AB^{\prime }}{\sigma }_\nu
^{CD^{\prime }}T_{AB^{\prime }CD^{\prime }}
\]
and the tensor corresponding to the spinor $T_{AB^{\prime
}CD^{\prime }}$ is,
\[
T_{AB^{\prime }CD^{\prime }}={\sigma }_{AB^{\prime }}^\mu {\sigma }%
_{CD^{\prime }}^\nu T_{\mu \nu }.
\]

The relationship between the matrices ${\sigma }^\nu $ and the
geometric tensor $g_{\mu \nu }$, as well as its spinor equivalent
are
\begin{eqnarray}
g_{\mu \nu }{\sigma }_{AB^{\prime }}^\mu {\sigma }_{CD^{\prime }}^\nu &=&{%
\varepsilon }_{AC}{\varepsilon }_{B^{\prime }D^{\prime }},~  \label{15.6} \\
g_{AB^{\prime }CD^{\prime }} &=&{\sigma }_{AB^{\prime }}^\mu {\sigma }%
_{CD^{\prime }}^\nu g_{\mu \nu }={\varepsilon }_{AC}{\varepsilon }_{{%
B^{\prime }D^{\prime }}},  \nonumber \\
g^{AB^{\prime }CD^{\prime }} &=&{\sigma }_\mu ^{AB^{\prime
}}{\sigma }_\nu ^{CD^{\prime }}g^{\mu \nu }={\varepsilon
}^{AC}{\varepsilon }^{B^{\prime }D^{\prime }}.  \nonumber
\end{eqnarray}

The complex conjugation of the spinor $S_{AB^{\prime }}$ is
\[
\overline{S_{AB^{\prime }}}={\bar{S}}_{A^{\prime }B}.
\]
Furthermore, for a real vector $V_\mu $ the spinor hermitian
equivalence yields ${\bar{V}}_{B^{\prime }A}=V_{AB^{\prime }}$.
If a vector $y^k$ is a null-vector,
\begin{equation}
y^ky_k=g_{k{\lambda }}y^ky^{{\lambda }}=0,  \label{15.8}
\end{equation}
then its spinor equivalent will take the form
\begin{equation}
y^k={\sigma }_{AB^{\prime }}^k{\theta
}^A{\bar{\theta}}^{B^{\prime }}, \label{15.9}
\end{equation}
where,${\theta }^A,{\bar{\theta}}^{B^{\prime }}$ represents the
two-component spinors of $SL(2,C)$ group.

In the Riemannian space, the covariant derivative of
$x$-dependent spinors will take the form
\begin{eqnarray*}
D_\mu \xi ^A &=&\frac{\partial \xi ^A}{\partial x^\mu }+L_{B\mu
}^A\xi
^B,~D_\mu {\bar{\xi}}^{A^{\prime }}=\frac{\partial {\bar{\xi}}^{A^{\prime }}%
}{\partial x^\mu }+{\bar{L}}_{B^{\prime }\mu }^{A^{\prime }}{\bar{\xi}}%
^{B^{\prime }}, \\
D_\mu \xi _A &=&\frac{\partial \xi _A}{\partial x^\mu }-L_{A\mu
}^B\xi
_B,~D_\mu {\bar{\xi}}_{A^{\prime }}=\frac{\partial {\bar{\xi}}_{A^{\prime }}%
}{\partial x^\mu }+{\bar{L}}_{A^{\prime }\mu }^{B^{\prime }}{\bar{\xi}}%
_{B^{\prime }},
\end{eqnarray*}
where $\xi ^A,{\bar{\xi}}^{A^{\prime }},{\bar{\xi}}_{A^{\prime }}$
represents two--components spinors and $L_{B\mu
}^A,{\bar{L}}_{B^{\prime }\mu }^{A^{\prime }}$ are the spinor
affine connections. In the case that we have spinors with two
indices, the covariant derivative will be in the form
\[
D_\mu \xi ^{AB^{\prime }}=\frac{\partial \xi ^{AB^{\prime
}}}{\partial \xi ^\mu }+L_{C_\mu }^A\xi ^{CB^{\prime
}}+{\bar{L}}_{C_\mu ^{\prime }}^{B^{\prime }}\xi ^{AC^{\prime }}.
\]
Applying this formula to the spinor metric tensors ${\varepsilon }_{AC},{%
\varepsilon }_{B^{\prime }C^{\prime }}$ we get
\begin{equation}
D_\mu {\varepsilon }_{AB}=\frac{\partial {\varepsilon
}_{AB}}{\partial x^\mu }-L_{A_\mu }^C{\varepsilon }_{CB}-L_{B_\mu
}^C{\varepsilon }_{AC}. \label{15.12}
\end{equation}
If
\[
D_\mu {\varepsilon }_{AB}=0,
\]
we shall say that the spinor connection coefficients $L_{B_\mu
}^A$ are \emph{metrical} together with the relations
\begin{equation}
D_\mu {\sigma }_{AB^{\prime }}^\nu =0,\quad D_\mu {\varepsilon }%
^{AB}=0,\quad D_\mu {\varepsilon }_{A^{\prime }B^{\prime }}=0,\quad D_\mu {%
\varepsilon }^{A^{\prime }B^{\prime }}=0.  \label{15.13}
\end{equation}
From the relation (\ref{15.12}) we immediately obtain
\[
L_{BA_\mu }=L_{AB_\mu }
\]
where we used the relation
\[
L_{AB_\mu }=L_{B_\mu }^C{\varepsilon }_{CA}.
\]
Also from the relation (\ref{15.13}) we have
\[
D_\mu {\sigma }_{AB^{\prime }}^\nu =\partial _\mu {\sigma
}_{AB^{\prime }}^\nu +L_{\mu \rho }^\nu {\sigma }_{AB^{\prime
}}^\rho -L_{A_\mu }^C{\sigma
}_{CB^{\prime }}^\nu -{\bar{L}}_{B_\mu ^{\prime }}^{D^{\prime }}{\sigma }%
_{AD^{\prime }}^\nu =0.
\]

\section[Generalization Spinor--Tensor Equivalents]{Generalization of
the Equivalent of Two Component--Spinors with Tensors}

The above mentioned well-known procedure for $SL(2,C)$ group
between spinors and tensors in a pseudo-Riemannian space-time can
be applied to more generalized metric spaces or bundles. For
example G. Asanov \cite{asa} applied this method for Finsler
spaces (FS), where the two-component spinors $n(x,y)$ depend on
the position and direction variables or $n(x^i,z^{{\alpha }}$,
with $z{\alpha }$ a scalar for a gauge approach. Concerning this
approach some results were given relatively to the gauge covariant
derivative of spinors and the Finslerian tetrad. In our present
study we give the relation between spinors of $SL(2,C)$ group and
tensors in the framework of Lagrabge spaces $(LS)$.

The expansion for the covariant derivatives, connections
non-linear connections, torsions and curvatures are the main
purpose of our approach.

In the following, we shall study the case that the vectors of
$LS$ are null-vectors and consequently fulfill the relation
(\ref{15.9}). In Finsler type space-time the metric tensor
$g_{ij}(x,y)$ depends on the position and directional variables,
where the vector $y$ may be identified with the frame velocity
(\cite{asa} ch. t). So, a vector $v^i$ will be called \emph{null}
if
\begin{equation}
g_{ij}(x,v)v^iv^j=0.  \label{15.15}
\end{equation}
In this case there is no unique solution for the light-cone \cite
{ishikawa,beil}. The problem of causality is solved considering
the velocity as a parameter and the motion of a particle in
Finsler space is described by a pair $(x,y)$. The metric form in
such a case will be given by
\[
ds^2=g_{ij}(x,v)dx^idx^j.
\]
When a particle is moving in the tangent bundle of a Finsler
(Lagrange) space-time its line-element will be given by
\begin{eqnarray}
\qquad d{\sigma }^2 &=&G_{ab}dx^adx^b  \label{15.16} \\
&=&g_{ij}^{(0)}(x,y)dx^idx^j+g_{{\alpha }{\beta }}^{(1)}(x,y)\delta y^{{%
\alpha }}\delta y^{{\beta }},\ \ \left( y^{{\alpha }}=\frac{dx^{{\alpha }}}{%
dt}\right) ,  \nonumber
\end{eqnarray}
where the indices $i,j$ and ${\alpha },{\beta }$ taking the
values $1,2,3,4$ and
\[
\delta y^{{\alpha }}=dy^{{\alpha }}+\mathcal{N}_j^{{\alpha }}dx^j.
\]
Thus we have
\begin{theorem}
The null-geodesic condition (\ref{15.15}) is satisfied for a
particle moving in the tangent bundle of Finsler space-time of
metric $d{\sigma }^2$ (\ref {15.16}) with the assumption, the
velocity $v$ is taken as a parameter of the absolute parallelism
\[
\delta y^{{\alpha }}=0.
\]
{\ The previous treatment of null-vectors in Finsler spaces can
also be considered for Lagrange spaces involving Lagrangians
which are not
homogeneous \cite{ma94,beil}. The introduction of spinors ${\theta },{\bar{%
\theta}}$ of the covering group $SL(2,C)$ in the metric tensor $g(x,{\theta }%
,{\bar{\theta}})$ under the correspondence between spinors and
tensors in $LS $,
\[
(x,y)\to (x,V_{AB^{\prime }}\to (x,{\theta
}^A,{\bar{\theta}}^{A^{\prime }})
\]
preserves the anisotropy of space with torsions. in this case all
objects
depend on the position and spinors, e.g. the Pauli matrices ${\tilde{\sigma}}%
_{AA^{\prime }}^i(x,{\theta },{\bar{\theta}})$. Such an approach
can be developed for a second-order spinor bundle applying the
method analogous to \cite{si1}. In virtue of relation
(\ref{15.8}), a null vector in spinor form can be characterized by
\begin{equation}
g_{AA^{\prime }BB^{\prime }}{\theta }^A{\bar{\theta}}^{A^{\prime }}{\theta }%
^B{\bar{\theta}}^{B^{\prime }}={\tilde{\sigma}}_{AA^{\prime }}^i{\tilde{%
\sigma}}_{BB^{\prime }}^jg_{ij}{\theta }^A{\bar{\theta}}^{A^{\prime }}{%
\theta }^B{\bar{\theta}}^{B^{\prime }}=0.  \label{15.18}
\end{equation}
}
\end{theorem}

\begin{proposition}
In a tangent bundle of metric (Finsler, Lagrange)
\[
G=g_{ij}(x,y)dx^idx^j+h_{ab}(x,y)\delta y^a\delta y^b,
\]
if the vector $y$ is a null, then the corresponding spinor metric
of the bundle will be given in the form
\begin{equation}
G=g_{AA^{\prime }BB^{\prime }}d{\theta }^Ad{\bar{\theta}}^{A^{\prime }}d{%
\theta }^Bd{\bar{\theta}}^{B^{\prime }}+h_{AA^{\prime }BB^{\prime }}\delta ({%
\theta }^B{\bar{\theta}}^{B^{\prime }})\delta ({\theta }^A{\bar{\theta}}%
^{A^{\prime }})  \label{15.19}
\end{equation}
or equivalently
\[
G=g_{AA^{\prime }BB^{\prime }}d{\theta }^Ad{\bar{\theta}}^{A^{\prime }}d{%
\theta }^Bd{\bar{\theta}}^{B^{\prime }}+h_{AA^{\prime }BB^{\prime
}}\delta y^{AA^{\prime }}\delta y^{BB^{\prime }},
\]
where $y^{AA^{\prime }}={\theta }^A{\bar{\theta}}^{A^{\prime }}$,
when $y$ is null vector (cf. \cite{carmelli}).
\end{proposition}

\textbf{Proof.} The relation (\ref{15.19}) is obvious by virtue
of (\ref {15.6}) and (\ref{15.9}).

\medskip \textbf{Remark.} A generalized spinor can be considered as the
equare root of a Finsler (Lagrange) null vector.

\section{Adapted Frames and Linear Connec\-ti\-ons}

In the general case of a $LS$, the spinor equivalent to the
metric tensor
\[
g_{ij}=\frac{\partial ^2L}{\partial y^i\partial y^j},\ \ L=\frac
12F^2
\]
is given by
\[
g_{ij}={\tilde{\sigma}}_i^{AA^{\prime
}}{\tilde{\sigma}}_j^{BB^{\prime }}g_{AA^{\prime }BB^{\prime }}.
\]
The corresponding Lagrangian will be ${\bar{L}}:M\times C^2\times
C^2\to R$, with the property ${\bar{L}}(x,{\theta
},{\bar{\theta}})=L(x,y)$, where $L$ represents the Lagrangian in
a Lagrange space. We can adopt the spinor equivalent form of the
adapted frames and their duals in a $LS$,
\[
\left( \frac \delta {\delta x^\mu },\frac \partial {\partial
y^i}\right) \to
\left( \frac \delta {\delta x^\mu },\frac \partial {\partial {\theta }%
^A},\frac \partial {\partial {\bar{\theta}}^{A^{\prime }}}\right)
,\qquad
(dx^\mu ,\delta y^i)\to (dx^\mu ,\delta {\theta }^A,\delta {\bar{\theta}}%
^{A^{\prime }})
\]
as well as the spinor counterpart of the non-linear connection $\mathcal{N}%
_\mu ^i$ of a $LS$,
\[
\mathcal{N}_\mu ^i\to (N_\mu ^A,{\bar{N}}_\mu ^{A^{\prime }}).
\]
The geometrical objects $\delta {\theta }^A,\delta
{\bar{\theta}}^{A^{\prime }}$ are given by
\begin{equation}
\delta {\theta }^A=d{\theta }^A+N_\mu ^Adx^\mu ,\qquad \delta {\bar{\theta}}%
^{A^{\prime }}=d{\bar{\theta}}^{A^{\prime }}+{\bar{N}}_\mu
^{A^{\prime }}dx^\mu .  \label{15.21}
\end{equation}
In virtue of (\ref{15.3}), the bases $\partial _\mu ,\partial
_{AA^{\prime }} $ are related as follows
\begin{equation}
\partial _\mu ={\tilde{\sigma}}_\mu ^{AA^{\prime }}\partial _{AA^{\prime }},
\label{15.22}
\end{equation}
where $\partial _\mu =\frac \partial {\partial x^\mu }$ and
$\partial _{AA^{\prime }}=\frac \partial {\partial {\theta
}^A}\frac \partial {\partial {\bar{\theta}}^{A^{\prime }}}$.

\begin{theorem}
In a Lagrange space the spinor equivalent of the adapted basis
$(\delta /\delta x^\mu ,\partial /\partial y^{{\alpha }})$ and
its dual $(dx^\mu ,\delta y^{{\alpha }})$ are given by
\begin{align}
& \mbox{a)}\quad \frac \delta {\delta x^\mu }={\tilde{\sigma}}_\mu
^{AA^{\prime }}\partial _A\partial _{A^{\prime }}-N_\mu ^A\partial _A-{\bar{N%
}}_\mu ^{A^{\prime }}\partial _{A^{\prime }}  \label{15.23} \\
& \mbox{b)}\quad \partial _P{\tilde{\sigma}}_P^{AA^{\prime
}}=\partial
_{AA^{\prime }},\qquad P=\{i,{\alpha }\}  \nonumber \\
& \mbox{c)}\quad dx^\mu ={\tilde{\sigma}}_{AA^{\prime }}^\mu d{\theta }^Ad{%
\bar{\theta}}^{A^{\prime }}  \nonumber \\
& \mbox{d)}\quad \delta y^{{\alpha }}=({\bar{\theta}}^{A^{\prime }}d{\theta }%
^A+{\theta }^Ad{\bar{\theta}}^{A^{\prime }}){\tilde{\sigma}}_{AA^{\prime }}^{%
{\alpha }}+({\bar{\theta}}^{A^{\prime }}N_{{\gamma }}^A+{\theta }^A{\bar{N}}%
_{{\gamma }}^{A^{\prime }}){\tilde{\sigma}}_{AA^{\prime }}^{{\gamma }}d{%
\theta }^Bd{\theta }^{\bar{B}^{\prime }}  \nonumber
\end{align}
\end{theorem}

\textbf{Proof.}The relations (\ref{15.23}) are derived from
(\ref{15.21}) and (\ref{15.22}).

\begin{proposition}
The null-geodesic equation of spinor equivalence in a $LS$ or
$FS$ is given by
\begin{equation}
dy^{{\alpha }}={\tilde{\sigma}}_{AA^{\prime }}^{{\alpha }}({\bar{\theta}}%
^{A^{\prime }}d{\theta }^A+{\theta }^Ad{\bar{\theta}}^{A^{\prime
}}),\ \
N_j^{{\alpha }}={\tilde{\sigma}}_{AA^{\prime }}^{{\alpha }}({\bar{\theta}}%
^{A^{\prime }}N_j^A+{\theta }^A{\bar{N}}_j^{A^{\prime }}).
\label{15.24}
\end{equation}
\end{proposition}

\textbf{Proof.} The relation (\ref{15.24}) is obvious because of (\ref{15.23}%
) d).

\begin{proposition}
The null-geodesic equation of spinor equivalence in a $LS$ or
$FS$ is given by
\begin{equation}
{\bar{\theta}}^{A^{\prime }}d{\theta
}^A({\tilde{\sigma}}_{AA^{\prime }}^\mu
N_\mu ^Ad{\bar{\theta}}^{A^{\prime }}+1)+{\theta }^Ad{\bar{\theta}}%
^{A^{\prime }}({\tilde{\sigma}}_{AA^{\prime }}^{A^{\prime
}}{\bar{N}}_\mu ^{A^{\prime }}d{\theta }^A+1)=0.  \label{15.25}
\end{equation}
\end{proposition}

\textbf{Proof.} In virtue of relations (\ref{15.18}) and
(\ref{15.23}) c,d) we obtain the relation (\ref{15.25}).

Affine connections and affine spinor connections are defined in
the frames
of $LS$ by the following formulas 
\begin{align}
& D_{\delta /\delta x^\mu }\left( \frac \delta {\delta x^\nu
}\right) =L_{\nu \mu }^k\frac \delta {\delta x^k},\ \ D_{\delta
/\delta x^\mu }\left( \frac \partial {\partial {\theta
}^A}\right) =L_{A\mu }^B\frac \partial
{\partial {\theta }^B},  \nonumber \\
& D_{\delta /\delta x^\mu }\left( \frac \partial {\partial {\bar{\theta}}%
^{A^{\prime }}}\right) ={\bar{L}}_{A^{\prime }\mu }^{B^{\prime
}}\frac
\partial {\partial {\bar{\theta}}^{B^{\prime }}},\ \ D_{\partial /\partial {%
\theta }^A}\left( \frac \delta {\delta x^\mu }\right) =C_{\mu
A}^\nu \frac
\delta {\delta x^\nu },  \nonumber \\
& D_{\partial /\partial {\theta }^A}\left( \frac \partial {\partial {\bar{%
\theta}}^{B^{\prime }}}\right) =C_{B^{\prime }A}^{C^{\prime
}}\frac \partial
{\partial {\theta }^{C^{\prime }}},\ \ D_{\partial /\partial {\bar{\theta}}%
^{A^{\prime }}}\left( \frac \partial {\partial {\theta }^B}\right) ={\bar{C}}%
_{BA^{\prime }}^C\frac \partial {\partial {\theta }^C},  \nonumber \\
& D_{\partial /\partial {\bar{\theta}}^{A^{\prime }}}\left( \frac
\partial {\partial {\bar{\theta}}^{B^{\prime }}}\right)
={\bar{C}}_{B^{\prime }A^{\prime }}^{C^{\prime }}\frac \partial
{\partial {\bar{\theta}}^C},\ \
D_{\partial /\partial {\theta }^A}\left( \frac \partial {\partial {\theta }%
^B}\right) =C_{BA}^C\frac \partial {\partial {\theta }^C},  \nonumber \\
& D_{\partial /\partial {\bar{\theta}}^{A^{\prime }}}\left( \frac
\delta {\delta x^\mu }\right) ={\bar{C}}_{\mu A^{\prime }}^\nu
\frac \delta {\delta x^\nu }.  \label{15.26}
\end{align}

We can give the covariant derivatives of the higher order
generalized spinors
$\zeta^{AB^{\prime}}_{BA^{\prime}}(x,{\theta},{\bar {\theta}})$,
\begin{eqnarray}
&\bigtriangleup _\mu \zeta _{BA^{\prime }}^{AB^{\prime
}}=&\frac{\delta \zeta _{BA^{\prime }\ldots }^{AB^{\prime }\ldots
}}{\delta x^\mu }+L_{C\mu }^A\zeta _{BA^{\prime }\ldots
}^{CB^{\prime }\ldots }+{\bar{L}}_{C^{\prime }\mu }^{B^{\prime
}}\zeta _{BA^{\prime }\ldots }^{AC^{\prime }\ldots
}-L_{B\mu }^C\zeta _{CA^{\prime }\ldots }^{AB^{\prime }\ldots }-{\bar{L}}%
_{\mu A^{\prime }}^{C^{\prime }}\zeta _{BC^{\prime }\ldots
}^{AB^{\prime
}\ldots }  \nonumber \\
&\bigtriangleup _E\zeta _{BA^{\prime }}^{AB^{\prime
}}=&\frac{\partial \zeta
_{BA^{\prime }\ldots }^{AB^{\prime }\ldots }}{\partial {\theta }^E}%
+C_{CE}^A\zeta _{BA^{\prime }\ldots }^{CB^{\prime }\ldots }+{\bar{C}}%
_{C^{\prime }E}^{B^{\prime }}\zeta _{BA^{\prime }\ldots
}^{AC^{\prime
}\ldots }-C_{BE}^C\zeta _{CA^{\prime }\ldots }^{AB^{\prime }\ldots }-{\bar{C}%
}_{EA^{\prime }}^{C^{\prime }}\zeta _{BC^{\prime }\ldots
}^{AB^{\prime
}\ldots }  \nonumber \\
&\bigtriangleup _{Z^{\prime }}\zeta _{BA^{\prime }}^{AB^{\prime }}=&\frac{%
\partial \zeta _{BA^{\prime }\ldots }^{AB^{\prime }\ldots }}{\partial {\bar{%
\theta}}^{Z^{\prime }}}+{\bar{C}}_{CZ^{\prime }}^A\zeta
_{BA^{\prime }\ldots }^{CB^{\prime }\ldots }+{\bar{C}}_{C^{\prime
}Z^{\prime }}^{B^{\prime }}\zeta _{BA^{\prime }\ldots
}^{AC^{\prime }\ldots }-{\bar{C}}_{Z^{\prime }A^{\prime
}}^{C^{\prime }}\zeta _{BC^{\prime }\ldots }^{AB^{\prime }\ldots
}.  \label{15.27}
\end{eqnarray}

\begin{proposition}
If the connections defined by the relations (\ref{15.26}) are of
the Cartan-type, then the spinor equivalent relations are given by
\begin{eqnarray}
{\bar{\theta}}^{A^{\prime }}\frac{\delta {\theta }^A}{\delta x^k}+L_{Ck}^A{%
\theta }^C{\bar{\theta}}^{A^{\prime }}+{\theta }^A\frac{\delta {\bar{\theta}}%
^{A^{\prime }}}{\delta x^k}+{\bar{L}}_{C^{\prime }k}^{A^{\prime }}{\bar{%
\theta}}^{C^{\prime }}{\theta }^A &=&0,  \nonumber \\
({\tilde{\sigma}}_{{\beta }}^{AA^{\prime
}})^{-1}({\bar{\theta}}^{A^{\prime
}}\bigtriangleup _E{\theta }^A+{\theta }^A\bigtriangleup _E{\bar{\theta}}%
^{A^{\prime }}) &=&1,  \label{15.28} \\
({\tilde{\sigma}}_{{\gamma }}^{AA^{\prime
}})^{-1}({\bar{\theta}}^{A^{\prime
}}\bigtriangleup _Z{\theta }^A+{\theta }^A\bigtriangleup _{Z^{\prime }}{\bar{%
\theta}}^{A^{\prime }}) &=&1.  \nonumber
\end{eqnarray}
\end{proposition}

\textbf{Proof.} Applying the relations (\ref{15.27}) to a null
vector $y$
with the Cartan-type properties $y_{|k}^{{\alpha }}=0$ and $y^{{\alpha }%
}\mid _{{\beta }}=\delta _{{\beta }}^{{\alpha }}$ \cite{mwi,asa},
and taking into account the (\ref{15.3}) a), (\ref{15.9}) we
obtain the relations (\ref {15.28}). (As we have mentioned
previously the $y$-covariant derivative has corresponded to the
spinor covariant derivatives).

\section{Torsions and Curvatures}         \index{torsion} \index{curvature}

The spinor torsions \index{spinor torsions}  corresponding to the
torsions of $LS$ are given by an analogous method to that one we
derived in \cite{si1} for a deformed bundle. The torsion tensor
field $T$ of a $D$-connection is given by
\[
T(X,Y)=D_XY-D_YX-[X,Y]
\]
.

Relatively to an adapted frame we have the relations
\begin{align}
& \mbox{a)}\quad T\left( \frac \delta {\delta x^k},\frac \delta {\delta x^{{%
\lambda }}}\right) =T_{{\lambda }k}^\mu \frac \delta {\delta x^\mu }+T_{{%
\lambda }k}^A\frac \partial {\partial {\theta }^A}+{\bar{T}}_{{\lambda }%
k}^{A^{\prime }}\frac \partial {\partial
{\bar{\theta}}^{A^{\prime }}}
\nonumber \\
& \mbox{b)}\quad T\left( \frac \partial {\partial {\theta
}^A},\frac \delta {\delta x^\mu }\right) =T_{\mu A}^\nu \frac
\delta {\delta x^\nu }+T_{\mu A}^B\frac \delta {\delta {\theta
}^B}+{\bar{T}}_{\mu A}^{B^{\prime }}\frac
\partial {\partial {\bar{\theta}}^{B^{\prime }}}  \nonumber \\
& \mbox{c)}\quad T\left( \frac \partial {\partial
{\bar{\theta}}^{A^{\prime }}},\frac \delta {\delta x^\mu }\right)
=T_{\mu A^{\prime }}^\nu \frac \delta {\delta x^\nu }+T_{\mu
A^{\prime }}^B\frac \partial {\partial {\theta
}^B}+{\bar{T}}_{\mu A^{\prime }}^{B^{\prime }}\frac \partial {\partial {\bar{%
\theta}}^{B^{\prime }}}  \nonumber \\
& \mbox{d)}\quad T\left( \frac \partial {\partial {\theta
}^A},\frac
\partial {\partial {\bar{\theta}}^B}\right) =T_{BA}^\mu \frac \delta {\delta
x^\mu }+T_{BA}^C\frac \partial {\partial {\theta }^C}+{\bar{T}}%
_{BA}^{C^{\prime }}\frac \partial {\partial {\theta }^{C^{\prime
}}}
\label{15.29} \\
& \mbox{e)}\quad T\left( \frac \partial {\partial {\theta
}^A},\frac
\partial {\partial {\bar{\theta}}^{B^{\prime }}}\right) =T_{B^{\prime
}A}^\mu \frac \delta {\delta x^\mu }+T_{B^{\prime }A}^C\frac
\partial {\partial {\theta }^C}+{\bar{T}}_{B^{\prime
}A}^{C^{\prime }}\frac \partial
{\partial {\bar{\theta}}^{C^{\prime }}}  \nonumber \\
& \mbox{f)}\quad T\left( \frac \partial {\partial
{\bar{\theta}}^{A^{\prime }}},\frac \partial {\partial
{\bar{\theta}}^B}\right) =T_{BA^{\prime }}^\mu
\frac \delta {\delta x^\mu }+T_{BA^{\prime }}^C\frac \partial {\partial {%
\theta }^C}+{\bar{T}}_{B^{\prime }A}^{C^{\prime }}\frac \partial {\partial {%
\bar{\theta}}^{C^{\prime }}}  \nonumber \\
& \mbox{g)}\quad T\left( \frac \partial {\partial
{\bar{\theta}}^{A^{\prime }}},\frac \partial {\partial
{\bar{\theta}}^{B^{\prime }}}\right) =T_{B^{\prime }A^{\prime
}}^\mu \frac \delta {\delta x^\mu }+T_{B^{\prime }A^{\prime
}}^C\frac \partial {\partial {\theta }^C}+{\bar{T}}_{B^{\prime
}A^{\prime }}^{C^{\prime }}\frac \partial {\partial {\bar{\theta}}%
^{C^{\prime }}}.  \nonumber
\end{align}

The torsion (\ref{15.29}) a) can be written in the form %
\setcounter{equation}{29}
\begin{eqnarray}
&&D_{\delta /\delta x^\mu }\frac \delta {\delta x^\nu }-D_{\delta
/\delta x^\nu }\frac \delta {\delta x^\mu }-\left[ \frac \delta
{\delta x^\mu
},\frac \delta {\delta x^\nu }\right] =  \label{15.30} \\
&&L_{\nu \mu }^{{\lambda }}\frac \delta {\delta x^{{\lambda
}}}-L_{\mu \nu }^{{\lambda }}\frac \delta {\delta x^{{\lambda
}}}-R_{\mu \nu }^A\frac
\partial {\partial {\theta }^A}-V_{\mu \nu }^{A^{\prime }}\frac \partial
{\partial {\bar{\theta}}^{A^{\prime }}},  \nonumber
\end{eqnarray}
where the brackets have the form
\begin{equation}
\lbrack \delta /\delta x^\mu ,\delta /\delta x^\nu ]=R_{\mu \nu
}^A\frac
\partial {\partial {\theta }^A}+V_{\mu \nu }^{A^{\prime }}\frac \partial
{\partial {\bar{\theta}}^{A^{\prime }}},  \label{15.31}
\end{equation}
and $\delta /\delta x^\mu $, $R_{\mu \nu }^A$, $V_{\mu \nu
}^{A^{\prime }}$ are given by
\begin{eqnarray}
\frac \delta {\delta x^k} &=&\frac \partial {\partial x^k}-\mathcal{N}%
_k^A\frac \partial {\partial {\theta }^A}-{\bar{N}}_k^{A^{\prime
}}\frac
\partial {\partial {\bar{\theta}}^{A^{\prime }}},\   \label{15.32} \\
R_{\mu \nu }^A &=&\frac{\delta N_\mu ^A}{\delta x^\nu
}-\frac{\delta \mathcal{N}_\nu ^A}{\delta x^\mu },\ V_{\mu \nu
 }^{A^{\prime }}=\frac{\delta N^{A^{\prime }}}{\delta x^\nu
}-\frac{\delta N^{A^{\prime }}}{\delta x^\mu }. \nonumber
\end{eqnarray}
The terms $R_{\mu \nu }^A$, $V_{\mu \nu }^{A^{\prime }}$ represents \emph{%
the spinor-curvatures of non-linear connections} \index{spinor
curvature} $N_\nu ^A$, $N_\mu
^{A^{\prime }}$. In virtue of the relations (\ref{15.29}), (\ref{15.30}), (%
\ref{15.31}) we obtain
\[
T_{\mu \nu }^{{\lambda }}=L_{\mu \nu }^{{\lambda }}-L_{\nu \mu }^{{\lambda }%
},\ \ T_{\nu \mu }^A=-R_{\mu \nu }^A,\ \ {\bar{T}}_{\nu \mu
}^{A^{\prime }}=-V_{\mu \nu }^{A^{\prime }}.
\]
Similarly from the relations (\ref{15.29}) b)- g), comparing with
the torsion in the following form,
\[
T\left( \frac \delta {\delta Y^P},\frac \delta {\delta Y^Q}\right)
=D_{\delta /\delta Y^P}\frac \delta {\delta Y^Q}-D_{\delta
/\delta Y^Q}\frac \delta {\delta Y^P}-\left[ \frac \delta {\delta
Y^P},\frac \delta {\delta Y^Q}\right]
\]
we can obtain the relations 
\begin{eqnarray}
 T_{A\mu }^{{\lambda }}&=& C_{A\mu }^{{\lambda }},\ \ T_{B\mu }^A=\frac{%
\partial N_\mu ^A}{\partial {\theta }^B}-L_{B\mu }^A  \label{15.34} \\
 T_{A\mu }^{A^{\prime }}&=&-{\tilde{Y}}_{A\mu }^{A^{\prime }},\ \ T_{AB}^{{%
\lambda }}=T_{AA^{\prime }}^{{\lambda }}  \nonumber \\
 T_{AB}^l &= & C_{AB}^l-C_{BA}^l,\ \ {\bar{T}}_{AB}^{A^{\prime
}}=-R_{AB}^{A^{\prime }},\ \ T_{\mu A^{\prime }}^{{\lambda }}=-{\bar{C}}%
_{A^{\prime }\mu }^{{\lambda }}  \nonumber \\
 T_{\mu A^{\prime }}^A &=& -\frac{\partial N_\mu ^A}{\partial {\bar{\theta}}%
^{A^{\prime }}},\ \ T_{\mu B^{\prime }}^{A^{\prime }}=C_{\mu
B^{\prime }}^{A^{\prime }}=C_{B^{\prime }\mu }^{A^{\prime
}}-P_{\mu B}^{A^{\prime }}
\nonumber \\
T_{AA^{\prime }}^B & =&  -C_{AA^{\prime }}^B,\ \ T_{AB^{\prime
}}^{A^{\prime }}=C_{AB^{\prime }}^{A^{\prime }}-\frac{\partial
C_A^{A^{\prime }}}{\partial {\bar{\theta}}^{B^{\prime }}},
\nonumber
\end{eqnarray}
where we have put
\begin{eqnarray*}
\frac \delta {\delta Y^P} &=&\left\{ \frac \partial {\partial {\theta }%
^A},\frac \partial {\partial {\bar{\theta}}^{A^{\prime
}}}\right\} ,\frac \delta {\delta Y^Q}=\left\{ \frac \delta
{\delta x^\mu },\frac \partial
{\partial {\bar{\theta}}^{\mit\Delta }}\right\} , \\
\Delta &=&B,B^{\prime }\ \mbox{and}\ C_A^{A^{\prime }}=C_A^{A^{\prime }}{%
\theta }^B.
\end{eqnarray*}
So, we obtain the following:

\begin{proposition}
In the adapted basis of a generalized metric tangent bundle the
spinor equivalent of coefficients of the torsion $T$ of a
$D$-connection, are given by the relations
(\ref{15.32})-(\ref{15.34}).
\end{proposition}

\begin{proposition}
$D$-connection has no torsion if and only if all terms of the
relation (\ref {15.34}) are equal to zero.
\end{proposition}

The curvature tensor field $R$ of a $D$-connection has the form
\[
R(X,Y)Z=[D_X,D_Y]Z-D_{[X,Y]}Z\ \forall X,Y,Z\in \mathcal{X}(TM).
\]
The coefficients of the curvature tensor and the corresponding
spinor cur\-va\-tu\-re tensors in spinor bundle are given by
\begin{eqnarray}
R_{{\lambda }\nu \mu }^k &=&\frac{\delta L_{{\lambda }\mu }^k}{\delta x^\mu }%
-\frac{\delta L_{{\lambda }\mu }^k}{\delta x^\nu }+L_{{\lambda
}\nu }^\rho
L_{\rho \mu }^k-L_{{\lambda }\mu }^\rho L_{\rho \nu }^k-R_{\mu \nu }^AC_{A{%
\lambda }}^k-V_{\mu \nu }^{A^{\prime }}{\bar{C}}_{A^{\prime
}{\lambda }}^k
\label{15.35} \\
R_{A\nu \mu }^B &=&\frac{\delta L_{\nu A}^B}{\delta x^\mu
}-\frac{\delta L_{\mu A}^B}{\delta x^\nu }+L_{A\nu }^\rho L_{\rho
\mu }^B-L_{A\mu }^\rho
L_{\rho \nu }^B-R_{\mu \nu }^\rho L_{A\rho }^B-V_{\mu \nu }^{A^{\prime }}{%
\bar{C}}_{A^{\prime }A}^B  \nonumber \\
R_{A^{\prime }\nu \mu }^{B^{\prime }} &=&\frac{\delta
{\bar{L}}_{\nu A^{\prime }}^{B^{\prime }}}{\delta x^\mu
}-\frac{\delta {\bar{L}}_{\mu A^{\prime }}^{B^{\prime }}}{\delta
x^\nu }+{\bar{L}}_{\nu A^{\prime }}^{B^{\prime }}L_\mu
-L_{A^{\prime }\nu }^{B^{\prime }}L_\mu -R_{\mu \nu
}^AC_{A^{\prime }A}^{B^{\prime }}-V_{\mu \nu }^{D^{\prime }}{\bar{C}}%
_{D^{\prime }A^{\prime }}^{B^{\prime }}  \nonumber \\
P_{\nu \mu A}^k &=&\frac{\delta L_{\nu \mu }^k}{\delta {\theta }^A}-\frac{%
\delta C_{A\nu }^k}{\partial x^\mu }+L_{\nu \mu }^{{\lambda }}C_{A{\lambda }%
}^k-C_{A\nu }^{{\lambda }}L_{{\lambda }\mu }^k+\frac{\partial N_\mu ^E}{%
\partial {\theta }^A}C_{E\nu }^k+{\tilde{Y}}_{\mu AA^{\prime }\nu
}^{^{\prime }k}  \nonumber \\
P_{AB\mu }^l &=&\frac{\partial L_{A\mu }^l}{\partial {\theta }^B}-\frac{%
\delta C_{AB}^l}{\delta x^\mu }+L_{A\mu }^kC_{kB}^l-C_{AB}^kL_{k\mu }^l+%
\frac{\partial N_\mu ^n}{\partial {\theta
}^A}C_{Bn}^l+{\tilde{Y}}_{\mu
A}^{A^{\prime }}C_{A^{\prime }B}^l  \nonumber \\
P_{A^{\prime }A\mu }^{B^{\prime }} &=&\frac{\partial L_{A^{\prime
}\mu }^{B^{\prime }}}{\partial {\theta }^A}-\frac{\delta
C_{A^{\prime }A}^{B^{\prime }}}{\delta x^\mu }+L_{A^{\prime }\mu
}^BC_{BA}^{B^{\prime
}}-C_{AA^{\prime }}^BL_{B\mu }^{B^{\prime }}  \nonumber \\
&&+\frac{\delta N_\mu ^E}{\partial {\theta }^A}C_{A^{\prime
}E}^{B^{\prime }}+{\tilde{Y}}_{\mu A}^{E^{\prime
}}{\bar{C}}_{E^{\prime }A^{\prime }}^{B^{\prime }}\nonumber\\
S_{\mu AB}^k &=&\frac{\partial C_{\mu A}^k}{\partial {\theta }^B}-\frac{%
\partial C_{\mu B}^k}{\partial {\theta }^A}+C_{\mu A}^{{\lambda }}C_{{%
\lambda }B}^k-C_{\mu B}^{{\lambda }}C_{{\lambda
}A}^k-R_{AB}^{A^{\prime
}}C_{A^{\prime }{\lambda }}^k  \nonumber \\
S_{lAB}^m &=&\frac{\partial C_{lA}^m}{\partial {\theta
}^B}-\frac{\partial
C_{lB}^m}{\partial {\theta }^A}%
+C_{lA}^nC_{nB}^m-C_{lB}^nC_{nA}^m-R_{AB}^{A^{\prime
}}{\bar{C}}_{A^{\prime
}l}^m  \nonumber \\
S_{A^{\prime }AB}^{B^{\prime }} &=&\frac{\partial C_{A^{\prime
}A}^{B^{\prime }}}{\partial {\theta }^B}-\frac{\partial
C_{A^{\prime }B}^{B^{\prime }}}{\partial {\theta
}^A}+C_{A^{\prime }A}^{D^{\prime }}C_{D^{\prime }B}^{B^{\prime
}}-C_{A^{\prime }B}^{D^{\prime }}C_{D^{\prime }B}^{B^{\prime
}}-R_{AB}^{D^{\prime }}{\bar{C}}_{D^{\prime }A^{\prime
}}^{B^{\prime }}  \nonumber \\
I_{\nu A^{\prime }\mu }^k &=&\frac{\delta {\bar{C}}_{A^{\prime }\nu }^k}{%
\delta x^\mu }-\frac{\partial L_{\nu \mu }^k}{\partial {\bar{\theta}}%
^{A^{\prime }}}+C_{A^{\prime }\nu }^\rho L_{\rho \mu }^k-L_{\nu \mu }^\rho {%
\bar{C}}_{A^{\prime }\rho }^k-\frac{\partial N_\mu ^A}{\partial {\bar{\theta}%
}^{A^{\prime }}}C_{A\nu }^k-{\bar{L}}_{A^{\prime }\mu
}^{A^{\prime }}C_{A\nu
}^k  \nonumber \\
I_{AA^{\prime }\mu }^B &=&\frac{\delta C_{A^{\prime }A}^B}{\delta x^\mu }-%
\frac{\partial L_{A\mu }^B}{\partial {\bar{\theta}}^{A^{\prime }}}%
+C_{A^{\prime }A}^\rho L_{\rho \mu }^B-L_{A\mu }^\rho C_{A^{\prime }\rho }^B-%
\frac{\partial N_\mu ^\rho }{\partial {\bar{\theta}}^{A^{\prime
}}}C_{A\rho }^B-{\bar{L}}_{A^{\prime }\mu }^{D^{\prime
}}{\bar{C}}_{D^{\prime }A}^B
\nonumber \\
I_{A^{\prime }C^{\prime }\mu }^{B^{\prime }} &=&\frac{\delta {\bar{C}}%
_{A^{\prime }C^{\prime }}^{B^{\prime }}}{\delta x^\mu }-\frac{\partial {\bar{%
L}}_{A^{\prime }\mu }^{B^{\prime }}}{\partial {\bar{\theta}}^{C^{\prime }}}+{%
\bar{C}}_{A^{\prime }D^{\prime }}^{B^{\prime
}}{\bar{L}}_{C^{\prime }\mu
}^{D^{\prime }}-{\bar{L}}_{E^{\prime }\mu }^{B^{\prime }}{\bar{C}}%
_{A^{\prime }B^{\prime }}^{E^{\prime }} \nonumber  \\
&&-\frac{\partial N_\mu ^\rho }{\partial {\bar{\theta}}^{A^{\prime }}}{\bar{L%
}}_{C^{\prime }\rho }^{B^{\prime }}-{\bar{L}}_{A^{\prime }\mu }^{D^{\prime }}%
{\bar{C}}_{D^{\prime }C^{\prime }}^{B^{\prime }} \nonumber
\end{eqnarray}
\begin{eqnarray*}   J_{\nu A^{\prime }B}^k &=&\frac{\partial C_{A^{\prime }\nu }^k}{\partial {%
\theta }^B}-\frac{\partial C_{B\nu }^k}{\partial {\bar{\theta}}^{A^{\prime }}%
}+C_{A^{\prime }\nu }^\rho C_{B\rho }^k-C_{B\nu }^\rho
C_{A^{\prime }\rho
}^k-\frac{\partial L_B^{D^{\prime }}}{\partial {\bar{\theta}}^{A^{\prime }}}%
C_{D^{\prime }\nu }^k  \nonumber \\
J_{AA^{\prime }B}^{rho} &=&\frac{\partial C_{A^{\prime }B}^\rho }{\partial {%
\theta }^A}-\frac{\partial C_{AB}^\rho }{\partial
{\bar{\theta}}^{A^{\prime
}}}+C_{A^{\prime }A}^kC_{kB}^\rho -C_{AB}^kC_{A^{\prime }k}^\rho -\frac{%
\partial L_A^{D^{\prime }}}{\partial {\bar{\theta}}^{A^{\prime }}}%
C_{D^{\prime }B}^\rho   \nonumber \\
J_{A^{\prime }C^{\prime }A}^B &=&\frac{\partial
{\bar{C}}_{A^{\prime }C^{\prime }}^{B^{\prime }}}{\partial
{\theta }^A}-\frac{\partial
C_{A^{\prime }A}^{B^{\prime }}}{\partial {\bar{\theta}}^{A^{\prime }}}%
+C_{A^{\prime }D}^{B^{\prime }}C_{C^{\prime }A}^D-C_{A^{\prime
}E}^{B^{\prime }}C_{C^{\prime }A}^E-\frac{\partial L_A^{D^{\prime }}}{%
\partial {\bar{\theta}}^{A^{\prime }}}{\bar{C}}_{C^{\prime }D^{\prime
}}^{B^{\prime }}  \nonumber \\
K_{\mu A^{\prime }B^{\prime }}^\nu  &=&K_{AA^{\prime }B^{\prime
}}^B=K_{A^{\prime }C^{\prime }D^{\prime }}^{B^{\prime }}=0.
\nonumber
\end{eqnarray*}

So we have

\begin{theorem}
The coefficients of the curvatures of a $D$-connection are given
by the relation (\ref{15.35}).
\end{theorem}

\begin{theorem}
In a tangent bundle a $D$-connection has no curvature if and only
if all the coefficients (\ref{15.35}) of the cur\-va\-tures are
equal to zero.
\end{theorem}

Finally, we note that the gravitational field can be described by
virtue of the corresponding spinorial form of the metric tensor
equivalent to the spinor bundle. This will be the object of our
future study.
 \backmatter

\printindex

\end{document}